**Technical Design Report**
Final version
July 14th 2014

**Scientific Editor**
Luca Serafini

# Technical Design Report
## EuroGammaS proposal for the ELI-NP Gamma beam System
### With 73 tables and 230 figures

*O. Adriani, S. Albergo, D. Alesini, M. Anania, D. Angal-Kalinin, P. Antici, A. Bacci, R. Bedogni, M. Bellaveglia, C. Biscari, N. Bliss, R. Boni, M. Boscolo, F. Broggi, P. Cardarelli, K. Cassou, M. Castellano, L. Catani, I. Chaikovska, E. Chiadroni, R. Chiche, A. Cianchi, J. Clarke, A. Clozza, M. Coppola, A. Courjaud, C. Curatolo, O. Dadoun, N. Delerue, C. De Martinis, G. Di Domenico, E. Di Pasquale, G. Di Pirro, A. Drago, F. Druon, K. Dupraz, F. Egal, A. Esposito, F. Falcoz, B. Fell, M. Ferrario, L. Ficcadenti, P. Fichot, A. Gallo, M. Gambaccini, G. Gatti, P. Georges, A. Ghigo, A. Goulden, G. Graziani, D. Guibout, O. Guilbaud, M. Hanna, J. Herbert, T. Hovsepian, E. Iarocci, P. Iorio, S. Jamison, S. Kazamias, F. Labaye, L. Lancia, F. Marcellini, A. Martens, C. Maroli, B. Martlew, M. Marziani, G. Mazzitelli, P. McIntosh, M. Migliorati, A. Mostacci, A. Mueller, V. Nardone, E. Pace, D.T. Palmer, L. Palumbo, A. Pelorosso, F.X. Perin, G. Passaleva, L. Pellegrino, V. Petrillo, M. Pittman, G. Riboulet, R. Ricci, C. Ronsivalle, D. Ros, A. Rossi, L. Serafini, M. Serio, F. Sgamma, R. Smith, S. Smith, V. Soskov, B. Spataro, M. Statera, A. Stecchi, A. Stella, A. Stocchi, S. Tocci, P. Tomassini, S. Tomassini, A. Tricomi, C. Vaccarezza, A. Variola, M. Veltri, S. Vescovi, F. Villa, F. Wang, E. Yildiz, F. Zomer*

# Table of contents

# Tables

# Figures

## 0.1. Executive Summary

The machine described in this document is an advanced Source of up to 20 MeV Gamma Rays based on Compton back-scattering, i.e. collision of an intense high power laser beam and a high brightness electron beam with maximum kinetic energy of about 720 MeV. Fully equipped with collimation and characterization systems, in order to generate, form and fully measure the physical characteristics of the produced Gamma Ray beam. The quality, i.e. phase space density, of the two colliding beams will be such that the emitted Gamma ray beam is characterized by energy tunability, spectral density, bandwidth, polarization, divergence and brilliance compatible with the requested performances of the ELI-NP user facility, to be built in Romania as the Nuclear Physics oriented Pillar of the European Extreme Light Infrastructure. This document illustrates the Technical Design finally produced by the EuroGammaS Collaboration, after a thorough investigation of the machine expected performances within the constraints imposed by the ELI-NP tender for the Gamma Beam System (ELI-NP-GBS), in terms of available budget, deadlines for machine completion and performance achievement, compatibility with lay-out and characteristics of the planned civil engineering.

The machine hereby described may not be the ideal machine to maximize performances of the Gamma Beam System under general circumstances: different technical solutions could have been explored in a general, un-constrained technical and scientific investigation. As a matter of fact, the solution proposed in this document is the best compromise between the requested specifications for the Gamma Ray beam and the imposed constraints.

In the course of preparation of this document, several significant improvements have been carried out by the EuroGammaS technical team on many technical and scientific aspects of the machine, which brought to new, unprecedented achievements in fields like:

i. study and start-to-end simulation of multi-bunch high brightness electron beams colliding in low-recoil mode, quasi-elastic scattering regime, with high quality intense recirculated laser pulses, to generate very mono-chromatic low bandwidth bright Gamma Ray beams
ii. phase space density evolution during electron-photon beam collisions
iii. recirculation of J-class high quality lasers by newly conceived optical systems
iv. stabilization and control of highly collimated Gamma Ray beams
v. characterization of high Gamma photon density at high energy resolution
vi. advanced integration of dual interaction point Gamma Beam Systems

Many are the expertise from various fields that have been gathered within the EuroGammaS team in order to successfully conceive, analyze and design this machine as a whole integrated system driving such an advanced Compton Source of narrow bandwidth Gamma Rays.

a. quantum and classical electro-dynamics analysis of collective multi-particle electron photon collisions with phase space reconstruction
b. design of high brightness RF electron linacs as drivers for Free Electron Lasers
c. design of advanced high gradient RF Structures and systems

    d.     timing, synchronization, stability and control of multiple beam accelerators working with beams of electrons and photons at the 100's fs level of synchronization and time resolution
    e.     advanced electron and photon beam diagnostics for phase space reconstruction with single and multi-bunch/shot capability, as well as for beam orbit stabilization at collision point
    f.     ultra-stable optical cavities for high power laser beams interacting with electron accelerators
    g.     intense high quality and high power ps laser systems based on advanced new laser technologies
    h.     design, integration and mechanical analysis of complex systems involving ultra-high stabilization, vacuum, precise magnetic fields, requiring very high mechanical and thermal stabilization of long transport lines for RF power, laser beams, control and diagnostic signals, etc
    i.     ultra stable and effective collimating techniques for high energy Gamma Ray photon beams
    j.     challenging the characterization of Gamma Ray beams with unprecedented energy resolution to reconstruct the spectral distribution of very narrow bandwidths

As extensively described in the course of this document, the specifications set by the ELI-NP-GS tender aim at upgrading the present state of the art in Gamma Ray Systems for Nuclear Photonics by at least one order of magnitude: to address such a challenging task, all of the expertise and capabilities described above had to sinergically act together in pursuing, and successfully achieving, the design study reported in this document. Its content is organized as follows:

A general introduction takes the reader to consider the basics aspects of Thomson vs. Compton backscattering of an optical photon colliding with a highly relativistic electron with energy of several hundreds of MeV's. Scalings of the back-scattered photon energy in the Thomson regime (negligible electron recoil, elastic scattering) and in the small recoil Compton regime are presented, underlining various mechanisms of red shift in the emitted radiation, due to recoil, collision angle, scattering angle, etc. Then a simple description of beam collisions is presented exploiting the concept of Luminosity, typically used in high energy particle colliders: powerful scaling laws are derived to predict the performances of the Compton Source in terms of total number of scattered photons, bandwidth of the emitted radiation vs. the collimation angle, the beam quality (both electron and laser) and the incoming fluxes. These simple analytical laws allow quick but still accurate predictions of the characteristics of the Gamma Ray beam, quite useful either for a first order design of the machine and for a consistency check of the numerical simulations outputs.

In the first Chapter we illustrate a complete analysis of the theory of Thomson/Compton backscattering of electron and photon beams, both based on classical electro-dynamics and on QED. The unprecedented characteristics of the Gamma Ray beam to be delivered for ELI-NP-GBS, in particular concerning the small bandwidth, lower than 0.5%, and the high spectral density Sr, higher than $10^4$ photons/s.eV, set up a quite challenging and demanding task of evaluating the Gamma Ray beam properties in its 6D phase space distribution, in presence of small, but not negligible, effects due to the collective behavior of the laser beam of photons (bandwidth, phase front curvature, ponderomotive radiation pressure from intensity time-space modulations, $M^2$ quality degradation, average collision angle, polarization distribution, etc) as well as from non linearities arising in classical electron trajectory through the laser e.m. field, multi-photon scattering, etc. Analytical treatments and numerical modeling, either via classical e.m. treatment of the laser synchrotron radiation emission by the electrons crossing the laser field, or via Klein-Nishina formulation of the two-body electron-photon scattering, carried out via montecarlo calculations, are all sinergically applied in order to achieve a global description of the Compton Source performances. The three simulation tools, CAIN, TSST and our home-developed Quantum Model have been extensively used, as reported, to assess the predicted ELI-NP-GBS specifications minimizing the uncertainties inherent in each specific tool. This represents an unprecedented effort in modeling self-consistently and with start-to-end simulations a modern Compton Source based on electron linac.

In the second Chapter we describe the design of the electron linac capable to produce multi-bunch trains of electrons with maximum phase space density at the two Interaction Points foreseen for the machine: as a matter of fact, it is shown in Chapter 1 that the maximum spectral density for the Gamma Ray beam is achieved, for a given laser average power and intensity at the collision point, maximizing the phase space density of the electron bunch, which basically scales like the ratio between the bunch charge (number of electrons in the bunch) and the square of the rms normalized total projected transverse emittance of the electron beam. For a room temperature RF linac running at maximum rep rate of about 100 Hz, the specifications on the requested spectral density cannot be achieved with single bunch collisions (one electron bunch per RF pulse colliding with one laser pulse): this forces a design for the RF linac with capability to provide trains of bunches in each RF pulse, spaced by the same time interval needed to recirculate the laser pulse in a properly conceived and designed laser recirculator, in such a way that the same laser pulse will collide with all the electron bunches in the RF pulse, before being dumped. Our final optimization foresees trains of 32 electron bunches separated by 16 ns, distributed along a 0.5 µs RF pulse, repeated at a rep rate of 100 Hz. Through the Chapter we present our solutions on how to deal with electron beam quality (emittance, energy spread) preservation in single bunch mode (applying techniques mutuated by the design of FEL's based on RF linacs), as well as the proper control and minimization on how multi-bunch effects impact on phase space distribution and density of electron bunches following the leading one in the train (the so called Beam Break-Up instability), leading to the design of properly High-Order-Mode damped RF structures for the electron beam acceleration.

The third Chapter describes how the previously discussed design criteria have been concretely implemented in generating a real lay-out and component detailed design for the RF linear accelerator. From photo-injector to RF accelerator sections, from magnets to vacuum system and beam pipes, from RF power stations to wave-guide distribution system, all components of the machine are specifically described in all details. Special paragraphs are devoted to the beam diagnostics system, timing and synchronization and the control system: needless to say, without an advanced, highly performant and reliable diagnostics apparatus even the best designed machine couldn't be commissioned to achieve its full potentialities, and an efficiently interfaced and user friendly control system is the tool to accomplish this goal. A detailed description of the procedures to install an integrated machine, and pursuing a proper alignment strategy by evaluating ground vibrations and stability, as well as thermal stability requirements, is also given at the end of the Chapter.

In Chapter 4 the Photon Machine is described (as the Electron Machine is described in Chapter 3), basically constituted by a Yb:Yag J-class laser system delivering high quality ($M^2<1.2$) ps laser pulses that must be focused down to focal spot sizes of about 20-30 microns at the collision points. The high average power demand of such a 100 Hz laser system, joined to the high quality ps pulse requirements, makes it very challenging and advanced, needing to address the development of a new laser technology based on Yb:Yag, which, although already proven to be capable to reach these performances, must be implemented in a stable, reliable and user facility oriented laser system as the one needed for the ELI-NP-GBS. A detailed description of the laser amplification chain and laser transport lines and control to the IP's is given in this Chapter, together with a description also of the photocathode drive laser, based on Ti:Sa technology, responsible for the generation of the low emittance electron multi-bunch trains at the photocathode in the photo-injector.

One of the most advanced conceptual achievement and most challenging technological device to be implemented in the machine, i.e. the laser recirculator, is described in a paragraph of Chapter 4. The analysis of stability, synchronization, diagnostics and control of such an innovative and advanced optical device, shows the capability of achieving a Time Averaged Spectral Density (TASD) almost equal to the nominal product between the number of electron bunches in the train and the single bunch predicted performances on the Gamma Ray spectral density. The development of this newly conceived optical device deserves an attention and support as for a project in its own.

Chapter 5 focuses on the collimation systems and the measurement apparatus, expected to experimentally prove that the machine has reached its foreseen performances in terms of the characteristics of the Gamma Ray beam. Quite a lot of innovative techniques have been discussed and integrated in these two systems in order to achieve these two goals: forming the gamma Ray beam by accurate collimation, at the level of few tens of micro-radians, and measuring the photon flux and bandwidth at the level of $10^5$ photons per shot and 0.3% energy resolution in the full dynamic range of the Compton Source tunability energy domain (0.2-19.5 MeV).

The integration of the machine with the conventional facilities and the expected building lay-out and characteristics is discussed in Chapter 6.

Possible future upgrades, requiring an extension of the available budget, but potentially upgrading in a significant way the expected phase space density of the electron beam, are illustrated in Chapter 7.

Finally, in Chapter 8 we summarize the expected performances of the 3 beams to be delivered by the machine, the electron, the optical photon and the Gamma Ray beam. A complete Parameter Table, split in sub-tables for each sub-system, is reported, showing in details the evaluated performances. An assessment of the risk associated with installation, integration and commissioning of several key components of the machine is listed in a dedicated Risk Table. Several Appendixes are completing specific discussions with additional details on various components of the machine.

### 0.1.1. Acknowledgments

We would like to acknowledge the great help and assistance by Franck Brottier in writing and properly editing this document: his collaboration was really invaluable. As it's been the support and encouragement, throughout these last couple years of endeavoring the preparation of this document, from the President of INFN, Fernando Ferroni, to whom we address our gratitude.

## 0.2. Introduction

### 0.2.1. General Considerations

In this document we describe the result of the activity carried out by a recently formed European Collaboration, denominated EuroGammaS, for the conceivement and design of the Gamma beam System (ELI-NP-GS) to be built for ELI-NP, the Romanian Pillar of the European Extreme Light Infrastructure (ELI) dedicated to Nuclear Physics/Photonics (ELI-NP). This document illustrates the Technical Design finally produced by EuroGammaS for ELI-NP-GS: the machine proposed here will be based on a back-scattering Compton Source, in which a 700 MeV high brightness electron beam will scatter an intense high power optical laser beam, converting optical photons into energetic gamma ray photons up to 20 MeV. The beam of gamma ray photons will have, as described throughout this document, unprecedented quality, in particular in terms of small bandwidth and very high spectral density and peak brilliance, representing a substantial improvement of the present state of the art, with an expected step-up of the various beam performances by one or two orders of magnitude.

ELI-NP is being funded by the European Commission in order to build at the Măgurele site (outskirt of Bucharest, Romania) a Research Infrastructure aimed at exploring the scientific advanced potentials of a high intensity laser system (up to $10^{24}$ W/cm$^2$) joined with a high brilliance Gamma ray Beam System, in the field of Nuclear Physics and Nuclear Photonics. The aim is to make a jump in the available technologies of at least 2 order of magnitudes with respect to the present state of the art. This implies, as far as the Gamma Beam System is concerned, the capability to produce a beam with extremely advanced characteristics in energy, tunability, mono-chromaticity, collimation, brilliance, time rapidity, polarizability, etc. In such a way that many applications and nuclear physics studies not addressable nowadays might be pursued and succesfully achieved by exploitation of the gamma beam expected properties. The aim is to open the era of nuclear photonics and pursue advanced applications in the field of national security, nuclear waste treatment, nuclear medicine, as well as fundamental studies in nuclear physics dealing with the nucleus structure and the role of giant dipole resonances, of great relevance also for astrophysics studies.

As a consequence, some key specifications on the Gamma Beam System were set as quality markers and primary goals for the design of the machine.

The main specifications of such a Compton Gamma-ray Source are: photon energy tunable in the range 1-20 MeV, rms bandwidth smaller than 0.5% and spectral density lager than $10^4$ photons/sec·eV, with source spot sizes smaller than 100 microns and linear polarization of the gamma-ray beam larger than 95%. Not to forget the peak brilliance of the gamma ray beam, expected to be larger than $10^{20}$ photons/(sec·mm$^2$·mrad$^2$·0.1%) .

In order to meet these quite challenging specifications, an European collaboration eventually named EuroGammaS Association (hereinafter "EuroGammaS") has been set up in 2012 to prepare a Technical Design Report for a machine capable to deliver by September 2018 the gamma-ray beam to users in the ELI-NP research infrastructure. The collaboration is formed by the following Universities/Institutions: Istituto Nazionale di Fisica Nucleare and Universita' di Roma La Sapienza, in Italy, and IN2P3/CNRS, in France, together with European companies such as ACP, a company of the Amplitude Group, France; Alsyom, a company of the Alcen Group, France; Comeb, Italy; ScandiNova Systems, Sweden. The collaboration is supported by 8 other European companies and research institutions: ALBA, Spain; ASTeC, a center of STFC, UK; Cosylab, Slovenia; Danfysik, Denmark; Instrumentation technologies, Slovenia; M+W Group Italy, Italy; Menlo Systems, Germany; and Research Instruments, Germany. The rationale for such a collaboration, the legal, financial and other relevant aspects thereof have been detailed in separate documents, attached to the EuroGammaS bid documentation related to the public procurement contract (tender notice #148351) issued by the Horia Hulubei National Institute for Research and Development in Physics and Nuclear Engineering on December 5th 2013. The EuroGammaS Association was awarded the contract for the procurement of the ELI-NP High-Intensity Gamma-Beam System on March 6th 2014. The Technical structure of the machine envisioned for the Gamma Beam System will be described in this document.

In the following of this Introduction we briefly summarize the challenges, the Physics and Technological Issues set by the ELI-NP requirements on the Gamma beam, and the general choices taken about the solutions proposed to achieve the specifications within the requested milestones and the defined machine cost. The design criteria will be delineated and motivated on the basis of a simplified collider-like description of the Compton source, and the main Parameter Tables for the electron and laser beam will be illustrated at the end of this introductory section.

### 0.2.2. Thomson elastic vs. Compton inelastic back-scattering

The Thomson/Compton backscattering [1-[10] between a relativistic electron bunch and a high power laser pulse is the preferred technique to generate high brilliance gamma ray beams.

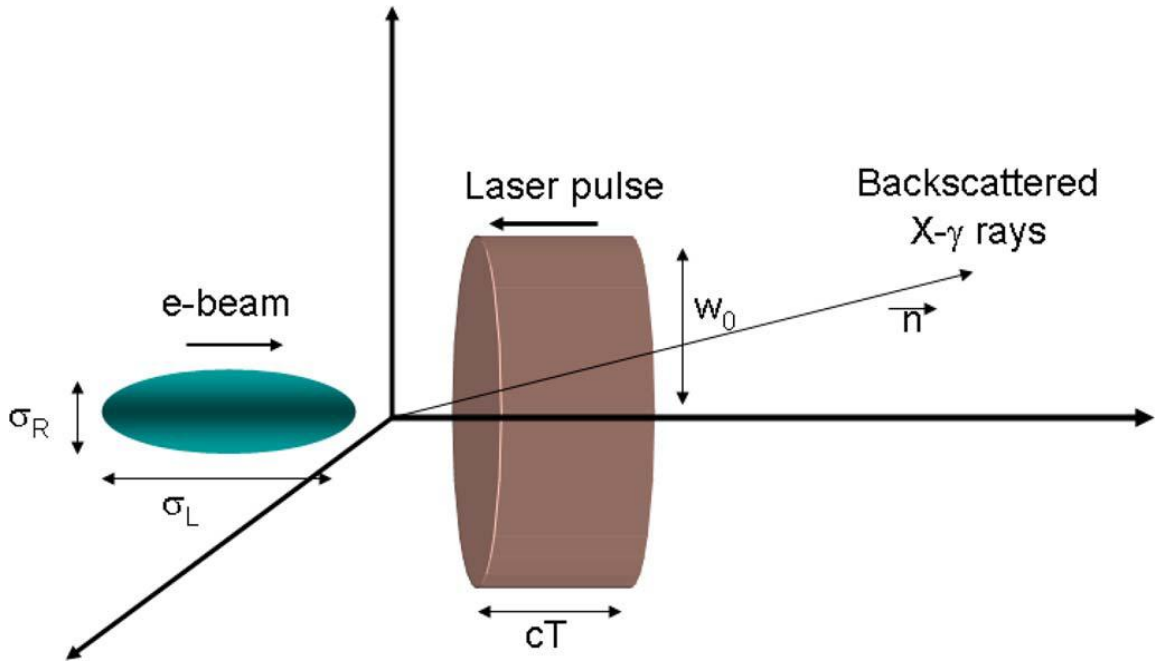

**Fig. 1.** Compton/Thomson backscattering geometry. The electron beam of longitudinal and transverse sizes σ_L and σ_R, respectively, is moving at a relativistic speed from left to right, colliding with a photon beam of waist size w_0 and duration T, thus emitting scattered radiation mainly in the direction of motion of the electron beam.

It has a number of advantages over the more conventional brehmstrahlung sources, which are based on impinging the electron beam onto a solid target of high Z material. It offers good and controllable monochromaticity, easy tunability, higher collimation and full control of the gamma photon polarization. It is of course a more challenging technique, since the requirements on the alignment, synchronization and phase space density of the two colliding beams are quite tight and represent a further generation of machines than the present one, based on brehmstrahlung sources.

In the collision between a laser beam and an electron beam there are two regimes: elastic Thomson scattering [1-8] and Compton inelastic scattering [9-11]. In order to maximize the energy of the scattered photon the back-scattering configuration is often used, as depicted in Fig. 1.. Here the two beams counter-propagate and collide almost head-on: in this configuration the energy of the back-scattered photons is maximum and their propagation is mainly in the direction of the electron beam. Indeed, for an electron of kinetic energy $T = (\gamma - 1)mc^2$, colliding with an optical photon of energy $h\nu_L$ which is transported in a laser beam of wavelength $\lambda_L$ ($\lambda_L \nu_L = c$), the energy $h\nu_\gamma^{TH}$ of the back-scatttered $X/\gamma$ photon is given by:

$$h\nu_\gamma^{TH} = h\nu_L \frac{4\gamma^2}{1 + \gamma^2 \vartheta^2 + \frac{\delta^2}{4} + \frac{a_{0p}^2}{2}} \qquad (1)$$

where $\vartheta$ is the angle of scattering (*i.e.* the angle between the direction of the incoming electron and the direction of the back-scattered photon), $\delta$ is the incidence angle of the incoming photon (again with respect to the direction of propagation of the incoming electron) and $a_{0p}$ is the dimensionless amplitude of the vector potential associated to the laser *e.m.* field, often called the laser parameter. In practical units $a_{0p}$ is given by: $a_{0p} = 4.3 \frac{\lambda_L}{w_0} \sqrt{\frac{U[J]}{\sigma_t[ps]}}$ . In all cases of interest to our EuroGammaS Project all angles $\vartheta, \delta$ and the laser parameter $a_{0p}$ are small quantities: typically $\vartheta$ is smaller than $2\cdot10^{-4}$ radians, $\delta$ smaller than $10^{-1}$ radians and $a_{0p}$ smaller than 0.05 (indeed $a_{0p} = 0.05$ when $\lambda_L = 0.5$ $\mu m$ and $w_0 = 25 \mu m$, with $U = 0.5 J$ and $\sigma_t = 1.5 ps$, a typical case).

Equation 1 has been derived by adopting a classical description of spontaneous synchrotron radiation emitted by the relativistic electron wiggling in the *e.m* field of the counter-propagating laser (assumed as a plane wave, no diffraction present). In this approximation weakly non-linear effects on the electron transverse motion are taken into account, and represented in the equation by the term scaling like $a_{0p}^2$, which accounts for radiation pressure related (or ponderomotive) effects. These non-linear effects are responsible for a small red-shift (decrease of the photon frequency) and also for emission of higher harmonics. Note that also the effects of having a non-zero collision angle $\delta$ (*i.e.* a non-exactly head-on collision) and a non-vanishing scattering angle $\vartheta$ result in small red-shifts of the back-scattered photon.

The Thomson approximation implies elastic scattering, *i.e.* zero recoil of the incoming electrons: as typical of synchotron radiation in undulators by relativistic electron beams, and even in Free Electron Laser scenarios, the loss of energy and momentum of the electrons emitting radiation is neglected and negligible, since the energy loss is in these scenarios much smaller than all the related bandwidths of interest (*e.g.* an Xray FEL where 12 keV photons are emitted by 12 GeV electrons).

The measure of the electron recoil, and of how much relevant this may be for the back-scattering photon energy determination, is given by a Compton recoil correction parameter $\Delta$, which represents the red-shift due to the inelastic electron recoil, and is given by (see [12] for an extensive discussion)

$$\Delta = \frac{4\gamma\, h\nu_L / mc^2}{1 + 2\gamma\, h\nu_L / mc^2} \qquad (2)$$

The quantity $\gamma h \nu_L$ represents the energy of the incoming optical photon as seen by the electron in its own rest frame (a factor $\gamma$ for the Lorentz transformation from laboratory frame to the rest frame of the electron):

clearly $\gamma h\nu_L/mc^2$ represents the electron recoil as seen in its rest frame when impinged by the up-converted $\gamma h\nu_L$ optical laser photon.

As a result, the true Compton back-scattered photon energy is given by:

$$h\nu_\gamma^{CO} = h\nu_\gamma^{TH}(1-\Delta) \qquad (3)$$

whenever the recoil parameter $\Delta$ is much smaller than 1.

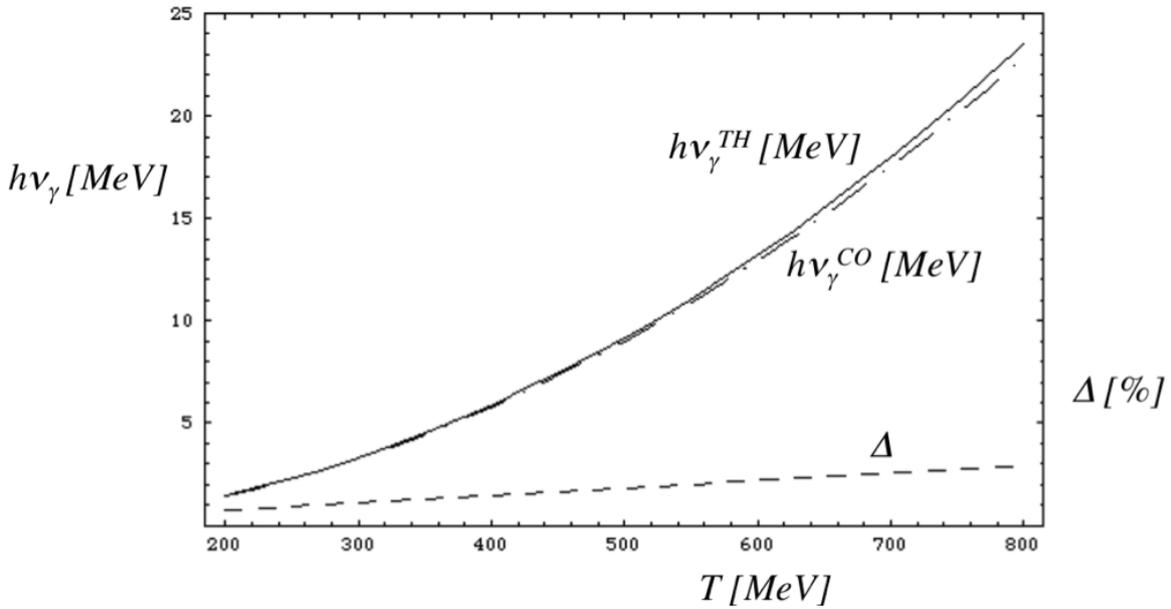

**Fig. 2.** Compton/Thomson backscattered photon energy plotted as function of the kinetic energy for the EuroGammaS scenario (including the weak non-linear effects of the laser carrying $a_{0p}=0.05$). The solid line gives the elastic Thomson approximation, while the dotted-dashed line represents the complete Comtpon calculation (left scale, in MeV). The deviation $\Delta$ due to inelastic Compton effects (i.e. the red-shift induced by the electron recoil) is given by dashed line (right scale, in %)

In practical cases of interest for our EuroGammaS Project, the electron kinetic energy are spanning the 200-720 MeV dynamic range, and the colliding laser is in the green (*i.e.* $h\nu_L = 2.4\ eV$, second harmonic of the Yb:Yag laser wavelength). Therefore we can use equations 1,2 and 3 to evaluate the energy of the back-scattered photons as a function of the electron kinetic energy: the result is shown in Fig. 2.

Clearly, the relative red shift $\Delta$ given by Compton recoil is small, always below 3%, but it is not negligible with respect to typical bandwidths of interest, which are below 0.5%. As further discussed in section 1, a careful treatment of the scattering process has to be performed in order to assess the impact of the Compton recoil red-shift on the spectral characteristics of the Gamma photon beam. In particular, this inherently quantum effect cannot be neglected and the simulation tools adopted will have to be capable of describing it - although the full classical description is still quite a good approximation for most cases (see section 1), and it is still the best method to evaluate non-linear effects in the spectrum of the emitted gamma ray beam, and its induced bandwidth broadening.

### 0.2.3. Thomson/Compton Sources as electron/photon Colliders

There are several ways to evaluate analytically the characteristics of the photon beam produced by the Thomson/Compton backscattering of the two colliding beams: in this introduction for simplicity we will adopt a description based on the luminosity concept, as for colliders. In section 1 of this document a complete treatment of the radation emission mechanism is presented.  Here we want to derive simple formulas which, although approximate, delineate the fundamental scaling laws of the gamma beam parameter behavior, and help illustrating the basic design criteria for the electron beam and the laser beam characteristics, which in turns drive the choices about the linac technology and the laser system structure.

Adopting a collider-like description of the Compton source, we can say that the flux $N_\gamma$ (photons/s) of the $\gamma$-beam is simply given by

$$N_\gamma = \Sigma_{TH} \frac{n_{el} n_{ph} f}{4\pi \sigma_{TH}^2} \quad (4)$$

where $\Sigma_{TH} = 0.6 \cdot 10^{-24} \, cm^2$ is the Thomson cross-section, $n_{el}$ is the number of electrons in the scattering bunch, $n_{ph}$ is the number of photons in the incident laser pulse, $f$ is the repetition rate of the collisions (*Hz*) and $\sigma_{TH}$ the rms spot size of the two colliding beams at the collision point. This simple formula is valid whenever we can neglect the diffraction of both the electron beam and the laser beam across the interaction (focal) plane. That is equivalent to state that the beta-function of the electron beam, equal to $\beta^* = \frac{\gamma \sigma_{TH}^2}{\varepsilon_n}$, must be larger than the electron bunch length $\sigma_z$, i.e. $\sigma_z \leq \beta^*$, and the same condition for the laser pulse, i.e. its Rayleigh range, $Z = \frac{\pi W_0^2}{\lambda_L}$, should be larger than the laser pulse length, i.e. $c\sigma_t \leq Z$ (to avoid luminosity degradation due to the well-known hour-glass effect, or beam diffraction). Our machine will operate in this regime, as explained in details later on: typical beam spot sizes adopted are about $\sigma_{TH} \approx 15 \, \mu m$ with rms normalized emittances of about $\varepsilon_n \approx 0.5 \, \mu m$, which becomes $\beta^* = 225 \, mm$ at $\gamma = 500$, and $W_0 \approx 25 \, \mu m$, implying $Z \approx 3.9 \, mm$, so the use of laser pulses shorter than 10 ps.

The quantity $L_{TH} = \frac{n_{el} n_{ph} f}{4\pi \sigma_{TH}^2}$ is also known as the luminosity of the collision between the two beams, which gives rise to the Thomson back-scattered $\gamma$-beam. We can express $N_\gamma$ also in terms of the electron bunch charge $Q$ and the energy $U_L$ carried by the laser pulse, getting

$$N_\gamma = 2.1 \cdot 10^8 \frac{U_L[J] Q[pC] f}{h\nu[eV] \sigma_{TH}^2[\mu m]} \quad (5)$$

It is also useful to consider the efficiency of the photon production process, which is expressed by the ratio of the number of photons generated over the number of electrons in the bunch, i.e. the number of photons produced per electron

$$n_\gamma = 34 \frac{U_L[J]}{h\nu[eV]\sigma_{TH}^2[\mu m]} \quad (6)$$

Considering typical values for the electron beam and the laser beam as used for our ELI-NP-GS machine (and illustrated further on), i.e. a bunch charge of $Q = 250 \, pC$, a laser pulse energy of $U_L = 0.4 \, J$, an interaction spot size of $\sigma_{TH} = 15 \, \mu m$, we obtain a gamma ray flux of $N_\gamma = 3.9 \cdot 10^9 \, s^{-1}$ at f=100 Hz, and a number of photons produced per electron of $n_\gamma = 0.025$. This last value implies that 2.5 % of the electrons in the bunch will scatter a photon, thus being affected by the energy and momentum loss (typically a few MeV's, up to 20, vs. an initial energy in the range 200-700 MeV) in such a way to be lost from the electron beam after collision (this point will be further discussed in section 1 and well-illustrated in Ref. 54 and 57).

Since $N_\gamma$ represents the total number of scattered photons per second over the entire solid angle and the entire energy spectrum (from the minimum energy $\nu$ up to the maximum one $\nu_\gamma$, at the Compton edge), we should better consider the beam of photons which is emitted within a small solid angle and within a narrow bandwidth. The bandwidth $bw$ scales nearly like the square of the normalized collimation angle $bw \approx \gamma^2\theta^2$, and also the number of photons emitted within the collimation angle $\theta$ has a similar scaling. So we have

$$N_\gamma^{bw} = 3.5 \cdot 10^8 \frac{U_L[J]Q[pC]f}{h\nu[eV]\sigma_{TH}^2[\mu m]} bw \quad \text{and} \quad n_\gamma^{bw} = 56 \frac{U_L[J]}{h\nu[eV]\sigma_{TH}^2[\mu m]} bw \quad (7)$$

Defining $P_{las} \equiv U_L[J]f$ as the effective laser power available for collisions at the interaction point, we can rewrite as

$$N_\gamma^{bw} = 3.5 \cdot 10^8 \frac{Q[pC]P_{las}}{h\nu[eV]\sigma_{TH}^2[\mu m]} bw \quad (8)$$

Assuming a collision spot size $\sigma_{TH} = 15 \mu m$, which is almost the smallest achievable in a Thomson/Compton Source producing photons up to 20 MeV (for reasons related to the electron and laser beam quality and how they affect the γ-beam bandwidth as discussed further below), and considering a reference bandwidth $bw = 1\%$, we obtain

$$N_\gamma^{bw} = 6.5 \cdot 10^3 Q[pC]P_{las} \quad \text{and} \quad n_\gamma^{bw} = 0.001 \cdot U_L[J] \quad (9)$$

for the case of a collision laser with $0.5 \, \mu m$ wavelength. In order to produce high photon fluxes $N_\gamma^{bw}$ we need large collision laser powers $P_{las}$, while in order to achieve a good efficiency $n_\gamma^{bw}$ in generating photons we need a large energy per pulse $U_L$. Although the photon flux scales like the inverse square of the collision spot size $\sigma_{TH}$, there are limitations in the amount of focusing that can be applied to the electron (and the laser) beams in order to increase the luminosity, hence the photon flux. These are due to the focusing capability of the final focus (telescopes) systems for the two beams, but even more crucially, to the impact on the bandwidth due to the high rms transverse momentum typical of a tightly focused electron beam. This can be simply viewed by expressing the rms bandwidth $\frac{\Delta \nu_\gamma}{\nu_\gamma}$ of the gamma ray photon beam in terms of the collimation angle $\vartheta$ and the properties of the two colliding beams: the rms normalized transverse emittance $\varepsilon_n$ of the electron beam and its energy spread $\frac{\Delta \gamma}{\gamma}$, together with the bandwidth $\frac{\Delta \nu}{\nu}$ of the collision laser. The relative rms bandwidth $\frac{\Delta \nu_\gamma}{\nu_\gamma}$ of the $\gamma$-beam is approximately given by:

$$\frac{\Delta \nu_\gamma}{\nu_\gamma} \cong \sqrt{(\gamma \vartheta)^4 + 4\left(\frac{\Delta \gamma}{\gamma}\right)^2 + \left(\frac{\varepsilon_n}{\sigma_{TH}}\right)^4 + \left(\frac{\Delta \nu}{\nu}\right)^2} \quad (10)$$

which clearly shows the compromize we have to achieve between focusing (small collision spot size $\sigma_{TH}$) and small bandwidth. Since the photon flux within the bandwidth ($N_\gamma^{bw}$) scales like the collimation angle squared ($\gamma^2 \theta^2$), this formula clearly points out the critical importance to generate low emittance and energy spread electron beams together with high quality small bandwidth laser pulses in the collision point, as a compulsory condition to allow using large collimation angles together with small collision spot sizes to maximize the photon flux within the required bandwidth. Typically, this collimation angle is between 200 micro-rad at low energy (1 MeV) and 40 micro-rad at high energy (20 MeV), producing a very collimated and high brilliance $\gamma$-beam which is emerging by a 10 $\mu m$ wide source at very small emitting angles (as typical of X-ray beams from synchrotron radiation in undulators or FELs). The strategy to designing a Linac capable to produce high brightness and high phase space density electron beams, so to drive a Compton Source like the one for ELI-NP-GS, is largely discussed in [13].

Our careful analysis of the EuroGammaS machine, carried out with complete start-to-end simulations as illustrated in sections 1 and 2, will show that the best compromise is for a high brightness (low emittance, 0.4 $\mu m$, low energy spread, 0.05%) electron beam carrying 250 pC per bunch in bunch trains of nearly 30 bunches per RF pulse, focused down to spot sizes of about 15 $\mu m$, against a laser pulse carrying 0.2 J of energy, with an overall repetition rate of 100 Hz, coupled to a laser recirculator that allows to boost the effective repetition rate of collisions up to 3.2 kHz.

In order to clarify the choice about the Linac and Laser system technolgy, we now consider two examples of Compton Sources, one based on a CW electron beam as typical of a Super-Conducting Linac coupled to a CW Fabry-Perot laser cavity, and one based on a room temperature high gradient RF Linac coupled to a laser recirculator. This second paradigma will be the one chosen in our EuroGammaS proposal, and illustrated in details in this document: we can anticipate that the main reason for such a choice is the overall constraint on the cost of the machine. The CW SC Linac case is more performant in terms of the gamma ray beam performances, but largely exceeds the boundary on the cost imposed by the ELI-NP-GS Tender.

In order to set up the best laser and beam parameters for the two examples we need to consider another general constraint on the electron beam power, *i.e.* average electron beam current (assuming a maximum handlable beam power of 0.7 MW due to radio-protection and beam power limitations): at 0.7 GeV maximum electron energy this implies a maximum electron beam average current $I_{av}$ of 1 mA. On the other side we consider a maximum achievable laser power in a Fabry-Perot cavity of 1 MW (present state of the art is nearly 100 kW, but many R&D projects are aiming at a ten-fold upgrade of the performances within the next few years). Typical Fabry-Perot cavities frequencies are around 100 MHz (they may be operated up to 1 GHz but not much below 100 MHz). Running at 100 MHz with 1 MW of stored power, the Fabry-Perot cavity provides laser pulses carrying 10 mJ at each collisions. Taking $f = 100 \ MHz$ implies an electron beam carrying 10 pC bunch charge with 1 mA average current.

As far as the second example is concerned, the electron beam expected to be generated by our EuroGammaS proposed room temperature RF would deliver 250 pC electron bunches in trains of about 30 bunches, nearly 10 ns spaced, at a repetition rate of the RF pulses up to 100 Hz. Its average current is $I_{av} = 0.75 \ \mu A$, implying a very moderate beam power of 525 W at 0.7 GeV. The effective repetition rate of such an electron beam is $f = 3 \ kHz$ (30x100). The collision laser provides 0.2 Joule of pulse energy at a repetition rate of 100 Hz, that matches the repetition rate of the RF pulses feeding the accelerating sections of the Linac (for a CW SC Linac the temporal overlap is guaranteed anyway). Because each laser pulse carries about $10^{19}$ optical photons, of which only $1.3 \cdot 10^6$ per collision will be back-scattered by the electrons to become $\gamma$ photons (see Eq.5), the laser pulse is basically unperturbed by the collision, so that it can be recirculated by a proper optical device, named laser recirculator, in order to collide with all the 30 electron bunches delivered in each RF pulse, with a proper round-trip time that matches the 10 ns time separation of the electron bunches.

We will present a comparison between these two examples assuming an electron beam quality as follows: rms norm. transverse emittance $\varepsilon_n = 0.5 \ \mu m$, rms energy spread < 0.1%, laser rms bandwidth < 0.1% (rms pulse duration of the electron and laser beam shorter than 1.5 ps). Using eq.7 for a target bandwidth set to 0.5%, we find for the SC-CW electron beam coupled to the Fabry-Perot laser cavity:

$$N_\gamma^{bw}[SC-CW] = 3.2 \cdot 10^{10} \ ph/s \quad \text{and} \quad n_\gamma^{bw}[SC-CW] = 5.2 \cdot 10^{-6} \quad (11)$$

while for the ELI-NP EuroGammaS beam coupled to the laser recirculator:

$$N_\gamma^{bw}[ELI-NP] = 5.1 \cdot 10^8 \ ph/s \quad \text{and} \quad n_\gamma^{bw}[ELI-NP] = 1.0 \cdot 10^{-4} \quad (12)$$

These values represent a real upper bound for the performances of a Thomson/Compton $\gamma$-beam Source. The ELI-NP EuroGammaS case is less performant in terms of photon flux, but quite more efficient in terms of photon production per electron, with an electron beam power of only 525 W compared to 0.7 MW of the SC-CW case. The spectral density $SP$ achievable at 10 MeV energy for the gamma ray photon beam would be about $SP \approx \frac{5.1 \cdot 10^8}{0.5 \cdot 10^{-2} \cdot 10^7} \approx 10^4 \, ph/s \cdot eV$.

On the basis of all these considerations, we anticipate our choices for the machine parameter set specifying the physical characteristics of our EuroGammaS machine. These are summarized in the following Tables, and will be deeply motivated by the discussion developed in sections 1, 2, 3 and 4. In section 8 we finally summarize the overall EuroGammaS machine Parameter Set, with extensive description of the anticipated performances.

**Table 1. Summary of Gamma-ray beam Specifications**

|  |  |
|---|---|
| Photon energy | 0.2-19.5 *MeV* |
| Spectral Density | 0.8-4·10$^4$ *ph/sec.eV* |
| Bandwidth (rms) | ≤ 0.5*%* |
| # photons per shot within FWHM bdw. | ≤ 2.6·10$^5$ |
| # photons/sec within FWHM bdw. | ≤ 8.3·10$^8$ |
| Source rms size | 10 - 30 $\mu m$ |
| Source rms divergence | 25 - 200 $\mu rad$ |
| Peak Brilliance ($N_{ph}/sec \cdot mm^2 mrad^2 \cdot 0.1\%$) | 10$^{20}$ - 10$^{23}$ |
| Radiation pulse length (rms, *psec*) | 0.7 - 1.5 |
| Linear Polarization | > 99 *%* |
| Macro rep. rate | 100 *Hz* |
| # of pulses per macropulse | ≤ 32 |
| Pulse-to-pulse separation | 16 *nsec* |

**Table 2.    Electron beam parameters at Interaction Points: general characteristics**

| all values are rms | |
|---|---|
| Energy (MeV) | 80-720 |
| Bunch charge (pC) | 25-400 |
| Bunch length ($\mu$m) | 100-400 |
| $\varepsilon_{n\_x,y}$ (mm-mrad) | 0.2-0.6 |
| Bunch Energy spread (%) | 0.04-0.1 |
| Focal spot size ($\mu$m) | > 15 |
| # bunches in the train | $\leq$ 32 |
| Bunch separation (nsec) | 16 |
| energy variation along the train | 0.1 % |
| Energy jitter shot-to-shot | 0.1 % |
| Emittance dilution due to beam breakup | < 10% |
| Time arrival jitter (psec) | < 0.5 |
| Pointing jitter ($\mu$m) | 1 |

**Table 3.    Yb:Yag Collision Laser beam parameters**

|  | Low Energy Interaction | High Energy Interaction |
|---|---|---|
| Pulse energy (*J*) | 0.2 | 2x0.2 |
| Wavelength (*eV,nm*) | 2.3,515 | 2.3,515 |
| FWHM pulse length (*ps*) | 3.5 | 3.5 |
| Repetition Rate (*Hz*) | 100 | 100 |
| $M^2$ | $\leq$ 1.2 | $\leq$ 1.2 |
| Focal spot size $w_0$ ($\mu m$) | > 28 | > 28 |
| Bandwidth (*rms*) | 0.1 % | 0.1 % |
| Pointing Stability ($\mu rad$) | 1 | 1 |
| Sinchronization to an ext. clock | < 1 *psec* | < 1 *psec* |
| Pulse energy stability | 1 % | 1 % |

**Table 4.    Laser beam Recirculator parameters**

|  | Low Energy Interaction | High Energy Interaction |
|---|---|---|
| Distance between the two Parabolic Reflectors | 2.38 *m* | 2.38 *m* |
| Collision Angle | 7.5° | 7.5° |
| beam waist $w_0$ | 28 $\mu m$ | 28 $\mu m$ |
| rotation at IP of linear laser polarization (along 32 passes) | $\geq$ 1° | $\leq$ 1° |
| integrated luminosity over 32 passes | > 90 % | > 90 % |
| Mirrors parallelism default | $\leq$ 10 $\mu rad$ | $\leq$ 10 $\mu rad$ |
| Mirrors alignment tolerance | $\leq$ 10 $\mu m$ | $\leq$ 10 $\mu m$ |
| Sinchronization to an ext. clock | < 1 *psec* | < 1 *psec* |

# 1. Technical Design for ELI-NP Gamma Beam System

## 1.1. Theory and scientific background

### 1.1.1. Thomson Cross Section and TSST code

The classical model describing the interaction between electron and radiation, based on the fundamental laws of electromagnetism [14] has been widely analyzed in the framework of the development of Thomson sources [15]-[19] and already cross checked versus experiments [20-[23].

When the interaction takes place between laser pulses few picosecond long in the linear or moderately non-linear regime and ultrarelativistic electron bunches, the total scattered radiation intensity can be evaluated as sum of the distributions produced by the single particles. The i-th electron, while entering the laser pulse, experiences longitudinal and transverse ponderomotive forces that respectively lower its longitudinal momentum and induce quivering and secular transverse motion. In the far zone the spectral-angular distribution of the emitted photons is obtained by using the well-known relation involving the Fourier transform of the retarded current [14]:

$$\frac{dN_i}{d\nu d\Omega} = \alpha \nu \left| \int dt \, \underline{n} \times (\underline{n} \times \underline{\beta}_i(t)) e^{i\omega(t - \underline{n} \cdot \underline{r}_i(t)/c)} \right|^2 \qquad (1)$$

where $\underline{n}$ is the unit vector in the scattered direction (*i.e.* the direction of the observer), $\underline{\beta}_i(t)$ is the normalized velocity of the i-th electron, $\underline{r}_i(t)$ its position, $\alpha = 1/137$ is the fine structure constant and cgs units are used throughout. An analytic description of the electron trajectory is possible only in the case of flat-top laser pulses when transverse ponderomotive forces are negligible. For long pulses and small angles, the paraxial approximation, valid when $\sigma_z \gg \lambda_0$ and $|\theta_i| < 2w_0 (\beta_i)_z / \sigma_z$ ($\lambda_0 = c/\nu_0$ is the wavelength, $\sigma_z$ the longitudinal r.m.s. dimensions, $w_0$ the waist diameter of the laser and $\theta_i$ is the polar angle of the i-th electron velocity), yields useful relations for the scattered radiation distribution [18] which are strictly valid for a flat-top laser profile and when transverse ponderomotive forces can be neglected. A generalization to non-flat top pulses (as in the case of ELI-NP-GS settings) can be made by slicing the laser pulse in a sequence of sub-pulses of time duration $1/\nu_0 \ll T_s \ll \sigma_z/c$, in which the amplitude variation is negligible, and by solving the motion of the particles in each slice. The distribution of the scattered photons distribution of the s-th flat-top sub-pulse:

$$\left[ \frac{dN_i}{d\nu d\Omega} \right]_s = \alpha (|V_{\theta,s}|^2 + |V_{\phi,s}|^2) \qquad (2)$$

can be then generalized to the whole pulse by combining all contributions:

$$\frac{dN_i}{d\nu d\Omega} = \alpha (|V_\theta|^2 + |V_\phi|^2)$$

where $V_\theta = \sum_s V_{\theta,s}$, $V_\phi = \sum_s V_{\phi,s}$, and, in each slice, the terms $V_{\theta,s}$ and $V_{\phi,s}$ are computed according eq. 27 in [18]:

$$V_{\theta,s} = T_s \, \text{sinc}(\frac{\Delta\omega_n T_s}{2}) e^{i\phi_s} \sum C_{\theta,s} \qquad (3)$$

$$V_{\phi,s} = T_s \, \text{sinc}(\frac{\Delta\omega_n T_s}{2}) e^{i\phi_s} \sum C_{\phi,s} \qquad (4)$$

In Eqq. [3] and [4] sinc(x)=sin(x)/x, $\Delta\omega_n = (\omega - n\omega_0)$, $\phi_s$ is the phase of the central oscillation in the slice and $C_{\theta,s}$, $C_{\phi,s}$ are complex terms involving Bessel J functions (see Eqq. 31 and 32 in [18]) in the paraxial limit. A sum over the harmonic number n accounts for the non-linear effects. In the absence of correlation between the electrons, and in the moderate non-linear regime, the generalization to an electron beam can be made by summing over the whole beam the contribution by the single electrons. Eqq. [3] and [4] generalize the computation of the scattered radiation in the linear and nonlinear regime reported in [18] to non-flat-top laser pulses that eventually diffract while interacting with the electron bunch. The use of the principle of the slicing for estimating analytically the contributions of the single particles is at the basis of the semi-analytic TSST (Thomson Scattering Simulation Tools) code and turns out to be an accurate and fast procedure for radiation computation.

### 1.1.2. Beam interaction

As regards the laser system adopted, a Gaussian pulse has been supposed, with the normalized electric field $A_L$ given by the expression:

$$A_L = \frac{E_L e^{-\frac{\xi^2+\eta^2}{2\sigma_t^2(1+\frac{\varsigma^2}{Z_R^2})} + -\frac{(\varsigma - ct\cos\alpha)^2}{2\sigma_z^2} - i\phi}}{h\bar{\nu}_0 (2\pi)^{3/2} \sigma_t^2 \sigma_z \sqrt{1+\frac{\varsigma^2}{Z_R^2}}}$$

$$\phi = \frac{\xi^2+\eta^2}{2\sigma_t^2(\frac{Z_R}{\varsigma}+\frac{\varsigma}{Z_R})} - artg(\frac{\varsigma}{Z_R})$$

where ξ, η, ζ are proper coordinates of the laser beam, connected with the laboratory frame by:

$$\xi = x$$
$$\eta = y\cos\alpha - z\sin\alpha$$
$$\varsigma = y\sin\alpha + z\cos\alpha$$

with the geometry represented in Fig. 3.

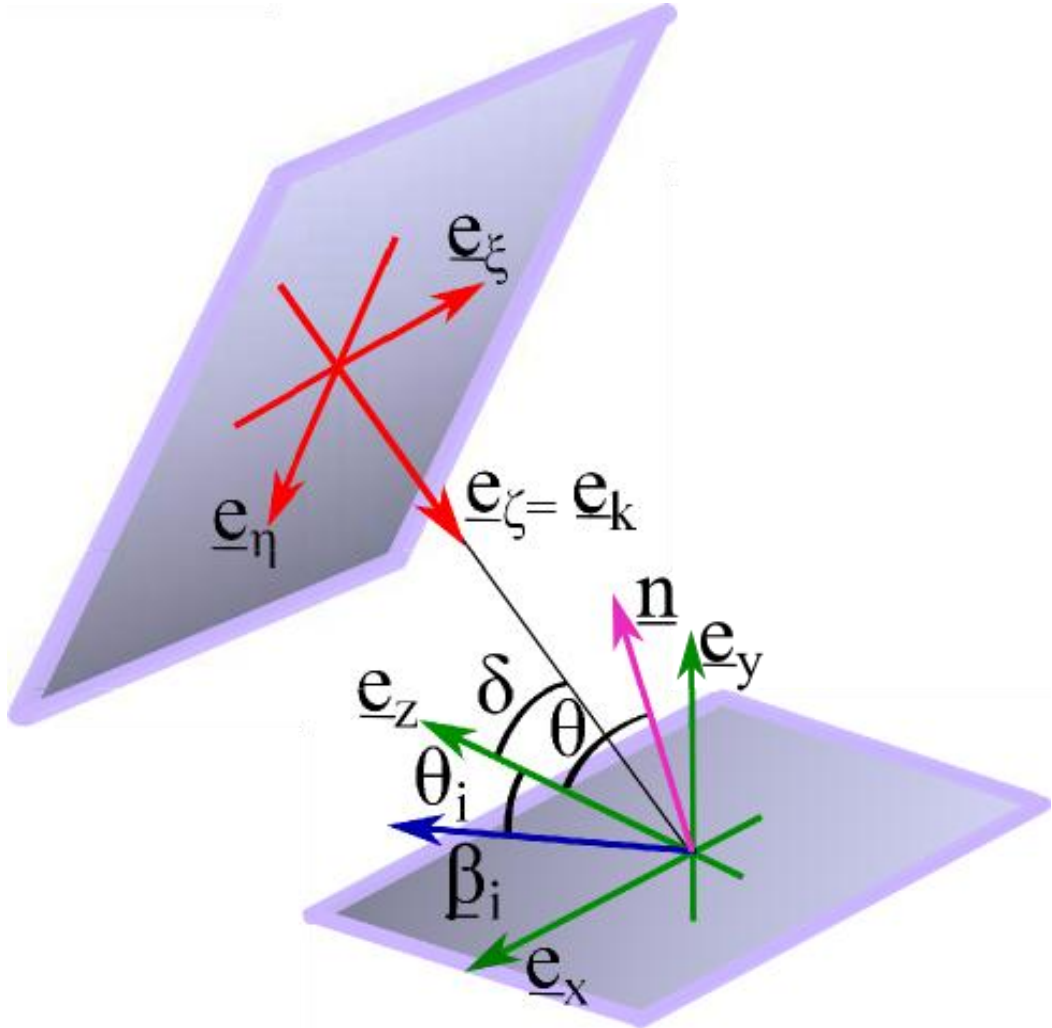

**Fig. 3.** Geometry of the laser-electron interaction. $\underline{e}_k$ is the unit vector of the laser wave vector, $\underline{\beta}_i$ is the electron normalized velocity, $\underline{n}$ is the scattered radiation direction, α=π-δ is the angle between $\underline{e}_k$ and the axis of the electron beam $<\underline{\beta}_i>=\underline{e}_z$, θ the angle between the electron beam axis and $\underline{n}$, and $\theta_i$ the angle between the i-th electron velocity $\underline{\beta}_i$ and the beam axis.

Moreover, $E_L$ is the maximum energy delivered by the laser, $Z_R$ is the Rayleigh length, and $\sigma_t$ the transverse r.m.s. dimensions of the laser field profiles.

The geometry of the interaction is described in Fig. 3. The axis of the electron beam $<\underline{\beta}_i>$ coincides with the $\underline{e}_z$, and the laser beam wave vector forms an angle α with $\underline{e}_z$.

## 1.2.  The Quantum model

### 1.2.1.  Compton Cross Section

The Compton model permits to evaluate the final condition of the electron-photon system after the scattering in all those conditions where the recoil of the electron cannot be disregarded. From the momentum and energy conservation laws it follows that the electrons present, after the interaction, a diminished Lorentz factor given by:

$$\gamma = \gamma_0 - \frac{h}{mc^2}(\nu_p - \nu_0) \qquad (5)$$

where

$$\nu_p = \nu_0 \frac{1 - \underline{e}_k \cdot \underline{\beta}_0}{1 - \underline{n} \cdot \underline{\beta}_0 + \frac{h\nu_0}{mc^2}(1 - \underline{e}_k \cdot \underline{n})} \qquad (6)$$

is the Compton frequency of the scattered photon. In these formulas, the index 0 refers to the coordinates before the scattering, $\underline{n}$ is the direction of the scattered photon and $\underline{e}_k$ is the unit vector of the direction of the incident photon of the laser.

A very useful expression is given by the wavelength:

$$\lambda_p = \lambda_0 \frac{1 - \underline{n} \cdot \underline{\beta}_0}{1 - \underline{e}_k \cdot \underline{\beta}_0} + \frac{h}{mc\gamma_0} \frac{1 - \underline{e}_k \cdot \underline{n}}{1 - \underline{e}_k \cdot \underline{\beta}_0} \qquad (7)$$

where the classic and quantum contributions appear clearly separated.

The photon-electron scattering is described by a cross section deduced in the rest electron frame by Klein and Nishina [24], and revisited in [25-28], [12].

The starting point is the Dirac equation:

$$i\hbar \frac{\partial \psi}{\partial t} = \hat{H}\psi$$

with: $\hat{H} = \hat{H}_0 + \hat{H}_{int}$.

where ψ is the wave function, and $\hat{H}$ the Hamiltonian formed by the unperturbed $\hat{H}_0$ and the interaction part $\hat{H}_{int}$.

Denoting with primes the quantities in the electron rest frame, the operator $\hat{H}_{int}$ contains the quantum radiation field:

$$\underline{\hat{A}} = c\underline{e}_0 \sqrt{\frac{\hbar}{v'_L}} (e^{i(\underline{k}'_0 \cdot \underline{r}' - 2\pi v'_0 t')} \hat{a}_0 + e^{-i(\underline{k}'_0 \cdot \underline{r}' - 2\pi v'_0 t')} \hat{a}_0^+)$$
$$+ c\underline{e}_p \sqrt{\frac{\hbar}{v'_p}} (e^{i(\underline{k}' \cdot \underline{r}' - 2\pi v'_p t')} \hat{a} + e^{-i(\underline{k}' \cdot \underline{r}' - 2\pi v'_p t')} \hat{a}^+)$$

where $\hat{a}_0(\hat{a})$ and $\hat{a}_0^+(\hat{a}^+)$ are creation (annihilation) operators and $\underline{e}_0$, $\underline{e}_p$ the polarization vectors relevant respectively to the incident and scattered photons.

The linear transition probability is given by: $w_{n,m} = \frac{2\pi}{\hbar} \rho \left| \sum_{n'} \frac{H_{m,n'} H_{n',n}}{E_m - E_n} \right|^2$

with ρ the density of the states with energy $E_m$, and $H_{m,n'}$ and $H_{n'n}$ the interaction matrix elements projected on the common states given by the product of the spin states, the momentum and the radiation eigenfunctions. In the electron reference frame the differential transition probability $dw_{n,m}/d\Omega$ turns out to be:

$$\frac{dw_{n,m}}{d\Omega} = c \frac{r_0^2}{4} \frac{v'^2_p}{v'^2_0} \left[ 4(\underline{e}'_p \cdot \underline{e}'_0)^2 + \frac{h\Delta v'}{mc^2} (1 - \underline{n}' \cdot \underline{e}'_k) \right]$$

and the cross section:

$$\left(\frac{d\sigma}{d\Omega}\right)' = \frac{r_0^2}{4} \frac{v'^2_p}{v'^2_0} \left[ 4(\underline{e}'_p \cdot \underline{e}'_0)^2 + \frac{v'_p}{v'_0} + \frac{v'_0}{v'^{-2}_p} \right]$$

with $\underline{e}'_0$ and $\underline{e}'_p$ the photon polarizations before and after the scattering. In the laboratory reference frame, the differential Compton cross section dσ/dΩ can be deduced by Lorentz transforming frequencies, polarizations and integration element over the solid angle according to the expressions:

$$v'_p = v_p \gamma_0 (1 - \underline{\beta}_0 \cdot \underline{n})$$
$$v'_0 = v_0 \gamma_0 (1 - \underline{\beta}_0 \cdot \underline{e}_k)$$

$$\underline{e}'_0 = \underline{e}_0 (1 - \underline{\beta}_0 \cdot \underline{e}_k) + \underline{\beta}_0 \cdot \underline{e}_0 (\underline{e}_k - \frac{\gamma_0}{1 + \gamma_0} \underline{\beta}_0)$$

$$\underline{e}'_p = \underline{e}_p (1 - \underline{\beta}_0 \cdot \underline{n}) + \underline{\beta}_0 \cdot \underline{e}_p (\underline{n} - \frac{\gamma_0}{1 + \gamma_0} \underline{\beta}_0)$$

$$d\Omega' = \frac{d\Omega}{\gamma_0^2(1-\underline{\beta}_0 \cdot \underline{e}_k)^2}$$

obtaining:

$$(\frac{d\sigma}{d\Omega}) = \frac{r_0^2}{2\gamma_0} \frac{{v'}_p^{\,2}}{{v'}_0^{\,2}} \frac{X}{(1-\underline{\beta}_0 \cdot \underline{n})^2} \qquad (8)$$

with:

$$X = \frac{v_p}{v_0} \frac{1-\underline{\beta}_0 \cdot \underline{n}}{(1-\underline{\beta}_0 \cdot \underline{e}_k)} + \frac{v_0}{v_p} \frac{1-\underline{\beta}_0 \cdot \underline{e}_k}{(1-\underline{\beta}_0 \cdot \underline{n})} + (e0 \cdot ep - \frac{c(\underline{\beta}_0 \cdot \underline{e}_p)(\underline{e}_0 \cdot \underline{p})}{hv_p(1-\underline{\beta}_0 \cdot \underline{n})} + \frac{c(\underline{\beta}_0 \cdot \underline{e}_0)(\underline{e}_p \cdot \underline{p})}{hv_0(1-\underline{\beta}_0 \cdot \underline{e}_k)})^2 \qquad (9)$$

Finally for the double differential cross section we obtain:

$$(\frac{d^2\sigma}{dvd\Omega}) = \frac{r_0^2}{2\gamma_0} \frac{{v'}_p^{\,2}}{{v'}_0^{\,2}} \frac{X}{(1-\underline{\beta}_0 \cdot \underline{n})^2} \delta(v_p - v_0 \frac{1-\underline{e}_k \cdot \underline{\beta}_0}{1-\underline{n} \cdot \underline{\beta}_0 + \frac{hv_0}{mc^2}(1-\underline{e}_k \cdot \underline{n})}) \qquad (10)$$

where the factor X, after averaging over the polarizations of the photons, becomes:

$$X = \frac{v_p}{v_0} \frac{1-\underline{\beta}_0 \cdot \underline{n}}{(1-\underline{\beta}_0 \cdot \underline{e}_k)} + \frac{v_0}{v_p} \frac{1-\underline{\beta}_0 \cdot \underline{e}_k}{(1-\underline{\beta}_0 \cdot \underline{n})} + (1 + \frac{mc^2}{hv_p\gamma_0(1-\underline{\beta}_0 \cdot \underline{n})} - \frac{mc^2}{hv_0\gamma_0(1-\underline{\beta}_0 \cdot \underline{e}_k)})^2 \qquad (11)$$

Another delta function multiplies the cross section guaranteeing the conservation of the total momentum between the initial and the final configuration.

### 1.2.2.    Beam Interaction

We carry out the calculations with the same electron beam density as used in the classical scheme.

As regards the laser system, the same Gaussian pulse as the classical case has been supposed, with the photon density represented by modulus squared of the laser electric field:

$$\frac{dN_L}{d\underline{x}_0 d\underline{k}_0} = \frac{E_L F(\underline{k}_0) e^{-\frac{\xi^2+\eta^2}{\sigma_t^2(1+\frac{\varsigma^2}{Z_R^2})} - \frac{(\varsigma - ct\cos\alpha)^2}{2\sigma_z^2}}}{h\overline{v}_0 (\pi)^{3/2} \sigma_t^2 \sigma_z \sqrt{1+\frac{\varsigma^2}{Z_R^2}}}$$

Where $\underline{x}_0$ ad $\underline{k}_0$ are laser photon coordinates and momenta, $Z_R$ is the Rayleigh length, $E_L$ is the maximum energy delivered by the laser, $F(\underline{k}_0)$ the dependence on the photon momenta, $\sigma_t$, $\sigma_z$ the transverse and longitudinal r.m.s. dimensions of the laser field profiles and $\bar{\nu}_0$ the central frequency value of the laser spectrum.

The number of emitted photons for frequency $\nu$ and solid angle $\Omega$ units can be evaluated in the laboratory frame as:

$$\frac{dN}{d\nu d\Omega} = h \int \frac{d\sigma}{d\nu d\Omega} \frac{dN_e}{d\underline{x} d\underline{p}} \frac{dN_L}{d\underline{x}_0 d\underline{p}_0} d\underline{x} d\underline{p} d\underline{k}_0 (1 - \underline{\beta}_0 \cdot \underline{e}_k) c dt \qquad (12)$$

The distribution of the electrons can be represented, in turn, by a sum of delta functions:

$$\frac{dN_e}{d\underline{x} d\underline{p}} = \sum_i \delta(\underline{x} - \underline{x}_i(t)) \delta(\underline{p} - \underline{p}_i(t)) \qquad (13)$$

With $\underline{x}_i(t)$ and $\underline{p}_i(t)$ coordinates of the i-th electron and the index i runs over all electrons. Once that equation (13) is inserted in (12) the number of emitted photons is:

$$\frac{dN}{d\nu d\Omega} = h \sum_i \int \frac{d\sigma}{d\nu d\Omega} \frac{dN_L}{d\underline{x}_0 d\underline{p}_0} d\underline{x} d\underline{p} d\underline{k}_0 (1 - \underline{\beta}_0 \cdot \underline{e}_k) \delta(\underline{x} - \underline{x}_i(t)) \delta(\underline{p} - \underline{p}_i(t)) c dt \qquad (14)$$

and the delta functions permit to solve the integrals in d$\underline{x}$ and d$\underline{p}$. Since the cross section depends only on the electron momenta which are constant in time while the photon distribution depends on coordinates, one has:

$$\frac{dN}{d\nu d\Omega} = h \sum_i \int \left[\frac{d\sigma}{d\nu d\Omega}\right]_{\underline{p}=\underline{p}_i} \left[\frac{dN_L}{d\underline{x}_0 d\underline{p}_0}\right]_{\underline{x}=\underline{x}_i(t)} d\underline{k}_0 (1 - \underline{\beta}_0 \cdot \underline{e}_k) c dt \qquad (15)$$

The distribution of the laser photons can be evaluated from the shape of the potential vector of the laser. We assume a simplified photon distribution depending only on the frequency f :

$$F(\underline{k}_0) d\underline{k}_0 = A_{l\nu}(f) df = \frac{1}{\sqrt{\pi} \sigma_\nu} e^{-\frac{(f-\bar{\nu}_0)^2}{\sigma_\nu^2}}$$

where diffraction and curvature of the phase surfaces of the laser are disregarded with the assumption of a very large Rayleigh length.

The df integral can be solved by using the delta function involving frequency $\delta(\frac{mc^2}{h}(\gamma-\gamma_0)+(\nu_p-f))=\delta(g(\nu_p,f))$:

$$\frac{dN}{d\nu d\Omega}=h\int df \sum_i \left[\frac{d\sigma}{d\nu d\Omega}\right]_{\underline{p}=\underline{p}_i} A_{L\nu}(f)(1-\underline{\beta}_0\cdot\underline{e}_k)\int [A_L]_{\underline{x}=\underline{x}_i(t)} cdt \quad (16)$$

According to the linear hypothesis, the unperturbed electron orbits can be assumed as:

$$\underline{x}_i(t)=\underline{x}_{0i}+c\underline{\beta}_{01}t \quad (17)$$

$$\underline{p}_i(t)=\underline{p}_{0i} \quad (18)$$

and the time integral evaluated analytically, giving :

$$\frac{dN}{d\nu d\Omega}=\frac{E_L}{\nu_0\pi^2\sigma_t\sigma_\nu}\int df\sum_i\left[\frac{d\sigma}{d\nu d\Omega}\right]_{\underline{p}_i}\delta(g(\nu_p,f)e^{-\frac{(f-\bar{\nu}_0)^2}{\sigma_\nu^2}}(1-\underline{\beta}_0\cdot\underline{e}_k)\frac{e^{\psi(\underline{x}_{01}\cdot\underline{\beta}_{0,1})}}{\sqrt{\sigma_z^2(\beta^2_{0ix}+\beta^2_{0iy})+\sigma_t^2(1+\beta_{0iz})^2}}$$

with: $g(\nu_p,f)=\delta(\frac{mc^2}{h}(\gamma-\gamma_0)+(\nu_p-f))=\dfrac{\delta(f-\nu_0)}{\left|\dfrac{dg(\nu_p,f)}{df}\right|_{f=\nu_0}}$ and:

$$\left|\frac{dg(\nu_p,f)}{df}\right|_{f=\nu_0}=\frac{\left[(1-\underline{\beta}_0\cdot\underline{e}_k)-\dfrac{h\nu_p}{mc^2\gamma_0}(1-\underline{n}\cdot\underline{e}_k)\right]}{(1-\underline{\beta}_0\cdot\underline{n})(1-\underline{\beta}_0\cdot\underline{e}_k)}$$

$$\Psi(\underline{x}_{0i},\underline{\beta}_{0i})=-\frac{x_{0i}^2+y_{0i}^2}{\sigma_t^2}-\frac{z_{0i}^2}{\sigma_z^2}+\frac{\left[\dfrac{x_{0i}\beta_{0iy}+y_{0i}\beta_{0ix}}{\sigma_t^2}+\dfrac{z_{0i}(1-\beta_{0iz}\cos\alpha)}{\sigma_z^2}\right]^2}{2\left[\dfrac{\beta_{0ix}^2+\beta_{0iy}^2}{\sigma_t^2}+\dfrac{(1-\beta_{0iz}\cos\alpha)^2}{\sigma_z^2}\right]}$$

Finally we get:

$$\frac{dN}{d\nu d\Omega} = \frac{E_L(1-\underline{\beta}_0 \cdot \underline{e}_k)e^{\psi(\underline{x}_{01},\underline{\beta}_{0,1})}}{\overline{\nu}_0(2\pi)^{5/2}\sigma_t\sigma_\nu} \frac{\left[\sum_i \left[\frac{d\sigma}{d\nu d\Omega}\right]_{p_i} \frac{e^{-\frac{(f-\overline{\nu}_0)^2}{\sigma_\nu^2}}}{\left|\frac{dg(\nu_p,f_L)}{df}\right|}\right]_{f=\nu_0}}{\sqrt{\sigma_z^2(\beta^2_{0ix}+\beta^2_{0iy})+\sigma_t^2(1+\beta_{0iz})^2}}$$

with:

$$\nu_0 = \nu_p \frac{1-\underline{n}\cdot\underline{\beta}_0}{1-\underline{e}_k\cdot\underline{\beta}_0 - \frac{h\nu_0}{mc^2}(1+\cos\theta)}.$$

Quantum effects can be weighted from the Compton formula, Eq. 7. A first regime, almost completely classical, occurs when $\lambda_0(1-\beta_0\cos(\theta+\theta_i)) \gg h/(mc\gamma_0)(1+\cos(\alpha+\theta_i))$, that, for $\theta, \theta_i$ and $\alpha$ much less than 1 can be written as $\lambda_0\left(\frac{1}{4\gamma_0^2}+\frac{\beta_0(\theta+\theta_i)}{2}\right) \gg \frac{h}{mc\gamma_0}\left(2+\frac{(\alpha+\theta_i)^2}{2}\right)$.

This is the usual Thomson classical regime, valid if $\gamma < mc^2/(8h\nu_0)$ and where the shift of the frequency is due only to Doppler effect. Acceptance and emittance effects extend the validity range of this regime.

If the incident radiation is a laser pulse with energy $h\nu_0$ ranging between 0.1 eV ($CO_2$ laser) to 10 eV (harmonics of the Ti:sapphire laser), the Thomson regime can extend up to electron beam energy of the order of several GeV. But if the relative bandwidth of the emitted radiation $\Delta\nu p/\nu p$ is required to be very narrow by the particular foreseen application, the red shift of wavelength due to quantum effects must be taken into account already when $(8h\nu_0)/mc^2 \sim \Delta\nu p/\nu p$.

If for instance $\Delta\nu p/\nu p \sim 10^{-3}$ the quantum shift on frequency begins to be manifest already at $\gamma \sim 100$.

The quantum model has been applied to the electron beams whose parameters are summarized in Table 1 for typical cases of ELI-NP-GS, obtained by accelerating the electron beam in a high-brightness C-band linac driven by a S-band photo-injector [29].

For the benchmark of the numerical code based on the model aforementioned, we have hypothesized as laser system the second harmonics of a IR laser ($\lambda_0$ =523,5 nm), with an energy of 0.5 J, a temporal rms duration of 4 ps and a waist diameter of 35 $\mu$m. The interaction is supposed head-to-head for sake of semplicity, with the angle $\delta$ defined in Fig. 3.

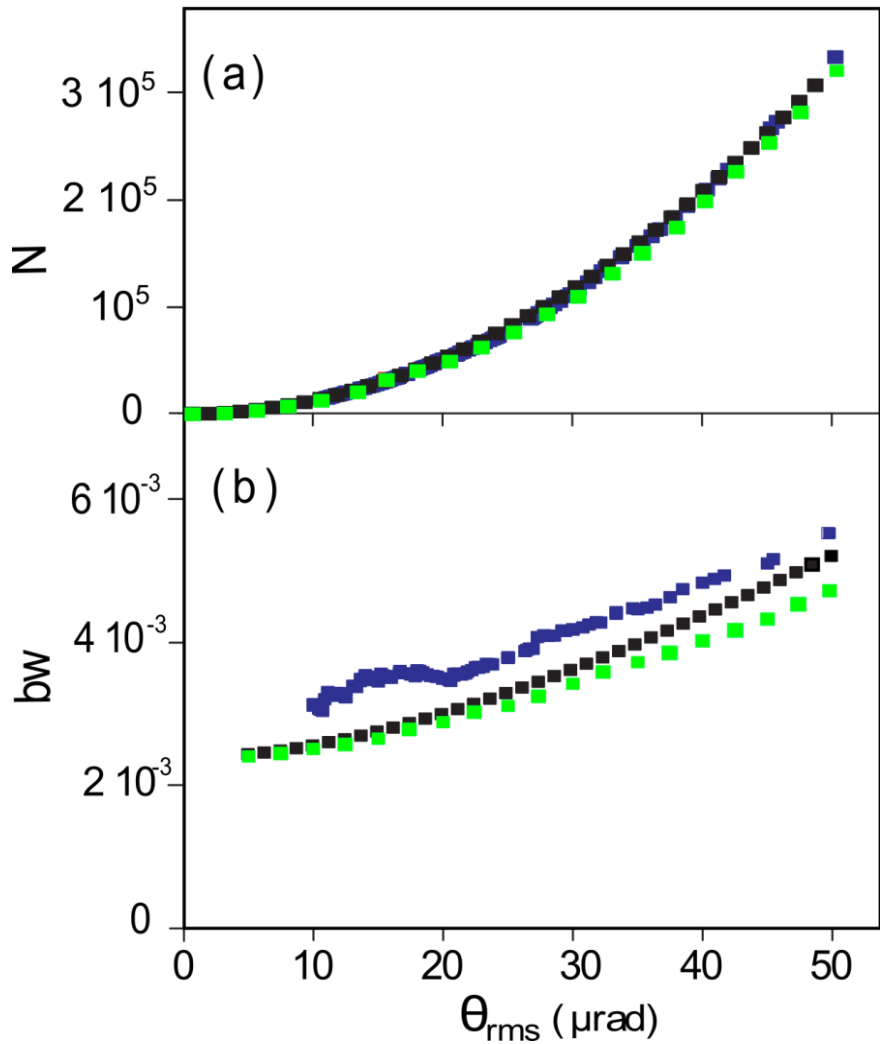

**Fig. 4.** Comparison of the total number of photons N (a) and relative bandwidth bw (b) vs the rms acceptance angle θrms(μrad) of the data obtained by the CAIN code (blue), the Compton model (red) and the classical non-linear theory TSST (black). The electron beam used is the working point A (Table 1). The total number of photons is almost exactly the same. As regards the bandwidth, differences within 20% were found.

In Fig. 4 we have compared the total number of photons (a) and relative bandwidth (b) vs the rms acceptance angle θrms(μrad) obtained by three numerical codes based on different models: CAIN (blue squares), which is a well-known quantum Monte Carlo code [30, 31] bench-marked for Compton sources in [9], a semi-analytical quantum code based on the linear Compton model described previously [12] (red squares) and the upgraded version of the classical non-linear code TSST (black squares) [12, [18] in the case of an electron beam with total charge of 250 pC, energy 360 MeV, energy spread 234 KeV, emittance 0.6 $\mu m$.

While for the total number of photons the agreement presented by the three different models appears striking, differences within 20%, are shown by the values of the bandwidth.

A photon number of order $10^5$ for a bandwidth of few $10^{-3}$ is obtained in a cone of acceptance $\theta_{rms}$ of about ten μrads.

In particular, for a bandwidth at the nominal value $3 \cdot 10^{-3}$, $1.5 \cdot 10^5$ photon are collected in about 25 µrad.

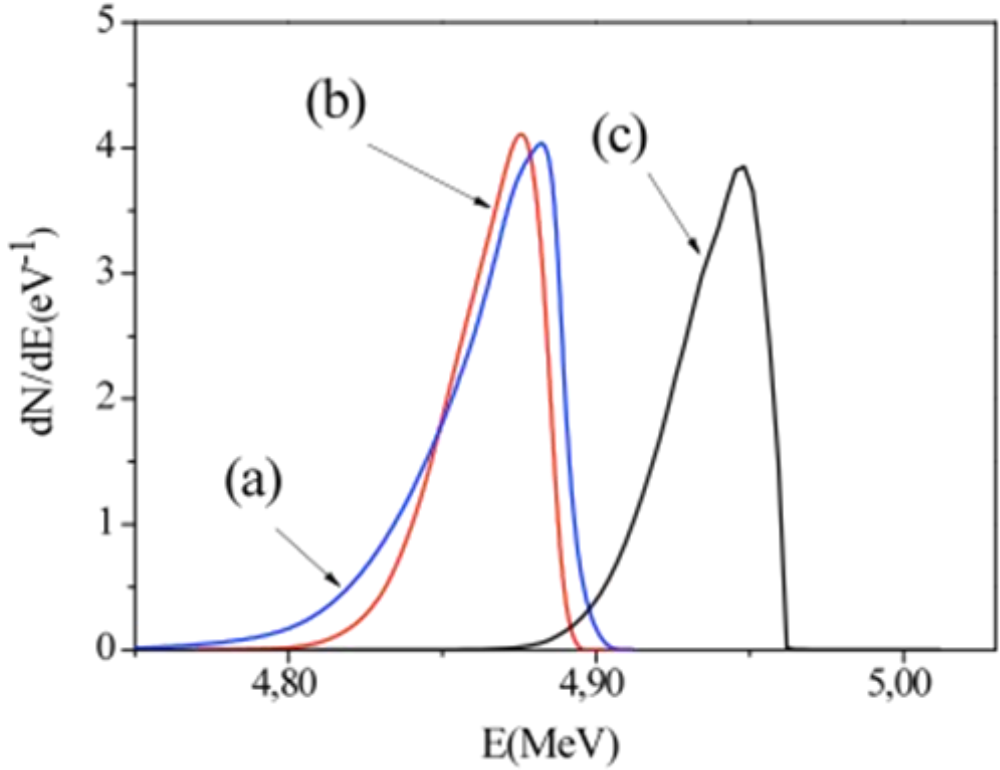

**Fig. 5.** Spectra of the rays. (a) CAIN (b) Quantum model (c) Classical treatment in the case of beam (A) and for the laser parameter of Table 1 and interaction angle α=π ; rms acceptance angle $\theta_{rms}$ = 25µrad

The spectra produced by the three codes are presented in Fig. 5 for the same beam of Fig. 4 and $\theta_{rms}$ = 25µrad. We can see here that the differences on the bandwidth values, shown in Fig. 4 (b), have to be attributed to the discrepancies in the tails, the rms evaluation being very sensible to them.

As can be seen, the quantum effects introduce a redshift in energy, that, in the present regime, can be quantified in $h\Delta\nu \sim 3.1 \cdot 10^{-5} \gamma^3 (h\nu_0)^2$=67 keV. The classical calculations give therefore a larger value of the mean frequency, while the shape, the peak and the width of the spectra are very similar for all cases, confirming that the classical approach can give approximate but still valid predictions as far as the average red-shift due to quantum Compton electron recoil is taken into account with analytical a priori estimations.

The simple model based on the luminosity concept permits to deduce a very useful scaling law for the total number of photons emitted in a cone of rms emi-angle $\theta_{rms}$. Assuming, in fact, an electron beam with circular transverse section of radius $\sigma_x$; the photon number for a single shot scales as:

$$N = \frac{\sigma_T}{\pi \sigma_x^2} N_e N_L \Psi^2 = \frac{0.35 \cdot 10^9 E_L Q \Psi^2}{h\nu_L \sigma_x^2} \qquad (19)$$

where $\sigma_T$ = 0.67 barn is the Thomson cross section, $\Psi = \gamma_0 \theta_{rms}$ is the acceptance of the system, $N_e$ the total number of electrons in the beam and $N_L$ the total number of photons in the laser pulse, and in the right hand term the energy of the laser is in Joule, the charge Q in pC, the factors $h\nu$ in eV and the lengths in μm. Inserting the data used in Fig. 7, and $\sigma_x$ = 13.5μm, Eq.19 fits very accurately the results of Fig. 8. The bandwidth, deduced from the Compton relation, scales with the quadratic sum of contributes due respectively to the acceptance $\Psi$, to the normalized emittance $\varepsilon_n$, to the energy spread, to the laser natural bandwidth, diffraction and temporal profile:

$$\left[\frac{\Delta \nu_p}{\nu_p}\right]_\Psi \approx \Psi^2$$

$$\left[\frac{\Delta \nu_p}{\nu_p}\right]_{\varepsilon_n} \approx \left[\frac{\varepsilon_n}{\sigma_x}\right]^2$$

$$\left[\frac{\Delta \nu_p}{\nu_p}\right]_\gamma \approx \left[\frac{2\Delta\gamma}{\gamma}\right]$$

$$\left[\frac{\Delta \nu_p}{\nu_p}\right]_d \approx \left[\frac{M^2 \lambda_L}{2\pi w_o}\right]^2$$

$$\left[\frac{\Delta \nu_p}{\nu_p}\right]_{\nu_L} \approx \left[\frac{\Delta \nu_L}{\nu_L}\right]^2$$

$$\left[\frac{\Delta \nu_p}{\nu_p}\right]_{\sigma_z} \approx \left[\frac{a_0^3/3}{1+a_0^2/2}\right]$$

The relative bandwidth has the expression:

$$\frac{\Delta \nu_p}{\nu_p} = \sqrt{\Psi^4 + \left[\frac{\varepsilon_n}{\sigma_x}\right]^4 + \left[\frac{\Delta \nu_p}{\nu_p}\right]_\gamma^2 + \left[\frac{\Delta \nu_p}{\nu_p}\right]_L^2} \qquad (20)$$

where in the last term all contributions of the laser are taken into account.

Once that the bandwidth have been fixed at the required value $\left[\frac{\Delta \nu_p}{\nu_p}\right]_r$ the corresponding acceptance is:

$$\Psi_r^2 \approx \sqrt{\left[\frac{\Delta\nu_p}{\nu_p}\right]_r^2 - \left[\frac{\varepsilon_n}{\sigma_x}\right]^4 - \left[\frac{\Delta\nu_p}{\nu_p}\right]_\gamma^2 - \left[\frac{\Delta\nu_p}{\nu_p}\right]_L^2} \qquad (21)$$

and the spectral density in the bandwidth in a single shot defined as:

$$S = \frac{N}{\sqrt{2\pi}h\Delta\nu_p}$$

turns out to be:

$$S_r(eV^{-1}) = \frac{0.35 \cdot 10^9 E_L Q \Psi^2}{h\nu_L \sigma_x^2} \frac{1}{\sqrt{2\pi}h\nu_p \left[\frac{\Delta\nu_p}{\nu_p}\right]_r} \qquad (22)$$

Fig. 6 and Fig. 7 show the numerical optimization of the spectral density with respect to the transverse dimension $\sigma_x$ of the electron beam (green hexagons, left axis), obtained by changing the final focusing system, for low (360 MeV) and the high (720 MeV) energy cases, together with the acceptance angle on the right. As can be seen in the Figures, the spectral density, once fixed the required bandwidth at the value $\left[\frac{\Delta\nu_p}{\nu_p}\right]_r = 3 \cdot 10^{-3}$, shows a maximum for a value of $\sigma_x$. For beams characterized by high focusing, the term $\left[\frac{\Delta\nu_p}{\nu_p}\right]_{\varepsilon_n}$ dominates in the bandwidth, thus limiting the collection of the photons to a small acceptance angle. Lower focusing decreases the transverse momenta of the electrons in the interaction region and permits to increase the acceptance. A further relaxation in the focusing leads to a low density beam with radius larger than the laser waist and to a diminishing spectral density. In this example the maximum value of photons $S_{r,max} = 9$ eV$^{-1}$ occurs for $\sigma_x(S_{r,max}) = 17\mu m$ for the lower energy beam, while $S_{r;max} = 3$ eV$^{-1}$ for $\sigma_x(S_{r,max}) = 16\ \mu m$ in the higher energy case. In Fig. 6 and Fig. 7, the validation of Eq. 22 (solid lines) is also shown.

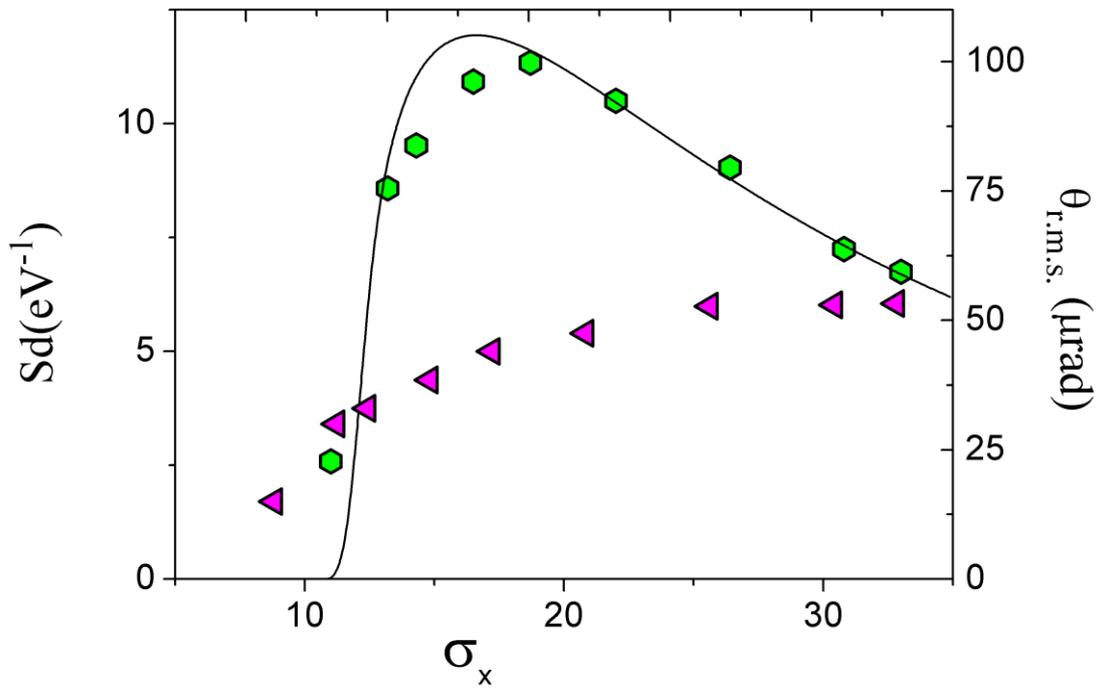

**Fig. 6.** Spectral density Sr(eV$^{-1}$)vs rms electron beam transverse dimension σx(μm) (green hexagons, left axis) and corresponding acceptance angle (stars, right axis) in the case of the beam at 360 MeV of Fig. 4.The solid line is the plot of Eq.22 for the same value.

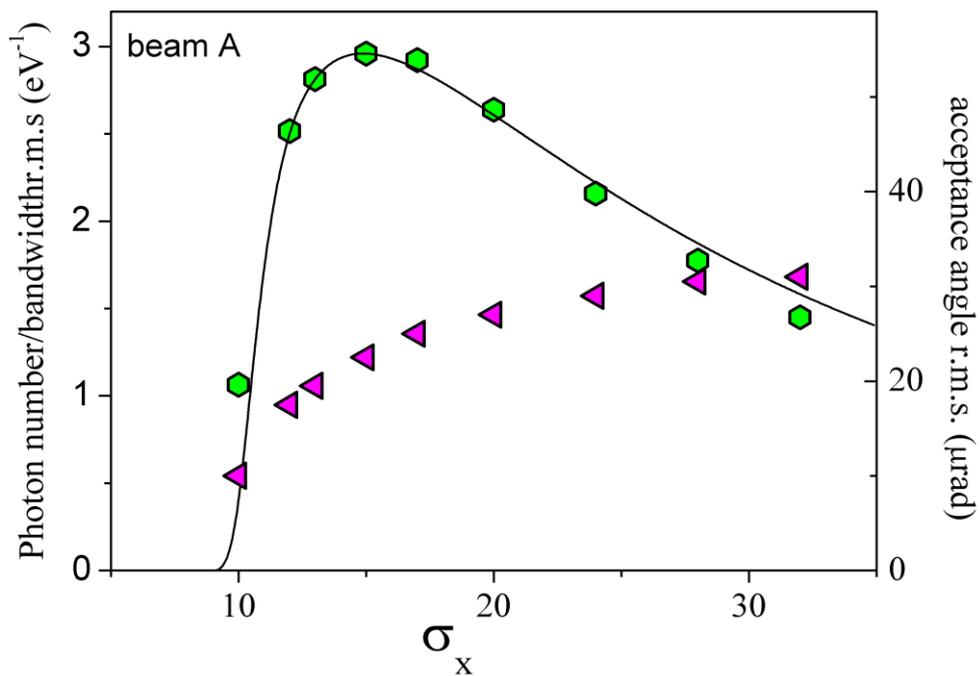

**Fig. 7.** Spectral density Sr(eV$^{-1}$)vs rms electron beam transverse dimension σx(μm) (green hexagons, left axis) and corresponding acceptance angle (stars, right axis) in the case of a beam at 720 MeV. The solid line is the plot of Eq.22 for the same value.

All these numbers represent the values obtained in a single shot.

An upgrade of the photo-injector, now under study, permits the operation at 100 Hz, enhancing by an extra factor $F_{pj} = 10^2$ the photon number per eV and per second.

A system of recirculation of the laser pulse producing a train of 32 bunches provides moreover a multiplication factor nearly equal ($F_r$ =32), that finally permits to increase the photon number per eV and per second in a bandwidth of $5·10^{-3}$ by an overall factor $F_{pj}F_r$ = 2 - 3 $10^3$. The number of photons become respectively 3.6 – 5.4 $10^4$ for the low energy case, and 1.2 – 1.8 $10^4$ for the high energy case.

As regards the non-linear effects due to strong laser energy, we carried out an analysis based on the numerical solution of the trajectories of the electrons under a realistic laser profile and on their insertion in Eq. 1. The complete discussion of these simulation results is presented in [31].

The non-linearity, in general, induces a series of distortions in the spectrum appearing in sequence with increasing laser energy: a shift in the spectrum towards lower energies, a broadening in the bandwidth, the rising of side bands, the growth of harmonics, the enhancement of the background, the superposition, shift and merging of all harmonics. In the case of the parameters of ELI-NP-GS, the only effect that appears is a weak shift of the frequency peak, which is however much lower than quantum shift. In Fig. 8 the non-linear shift for the case $a_0$ = 0.03 (curve (b)) and $a_0$ = 0.3 (curve (c)) are compared with the quantum one (curve (a)). For the typical values of the ELI working points we have considered (600<<1500, $a_0$ ~0.03), the quantum shift exceeds by more than one order of magnitude the non-linear one, and even an $a_0$ ten times larger do not change substantially this order. We conclude that bandwidth broadening due to non-linear effects, for the range of parameters of the present EuroGammaS proposed machine, are quite small compared to other relevant effects due to electron and laser beam quality.

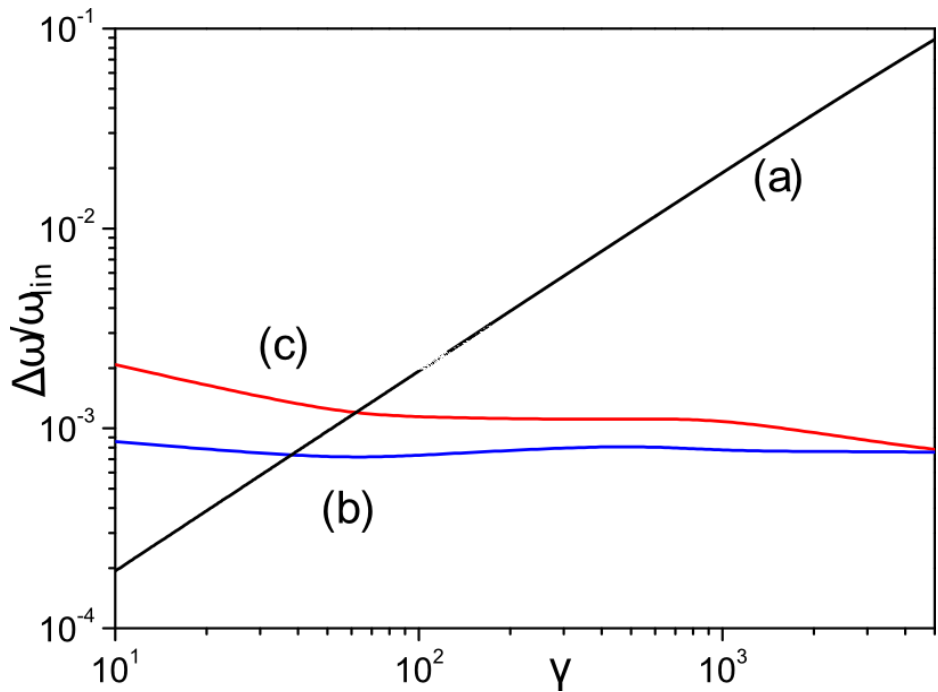

**Fig. 8.** Normalized frequency shift as function of $\gamma$. (a) Quantum model. (b) $a_0$ = 0.03; (c)$a_0$ = 0.3.

As a final remark on the physics of Compton sources in the energy range of ELI-NP-GS, we analyzed the dynamics of the electron beam through the collision with the laser pulse, in order to understand what happens to the electron phase space distributions and what kind of process rules their evolution. A complete discussion can be found in [32, 33]. Here we summarize a typical result, shown in the following figure.

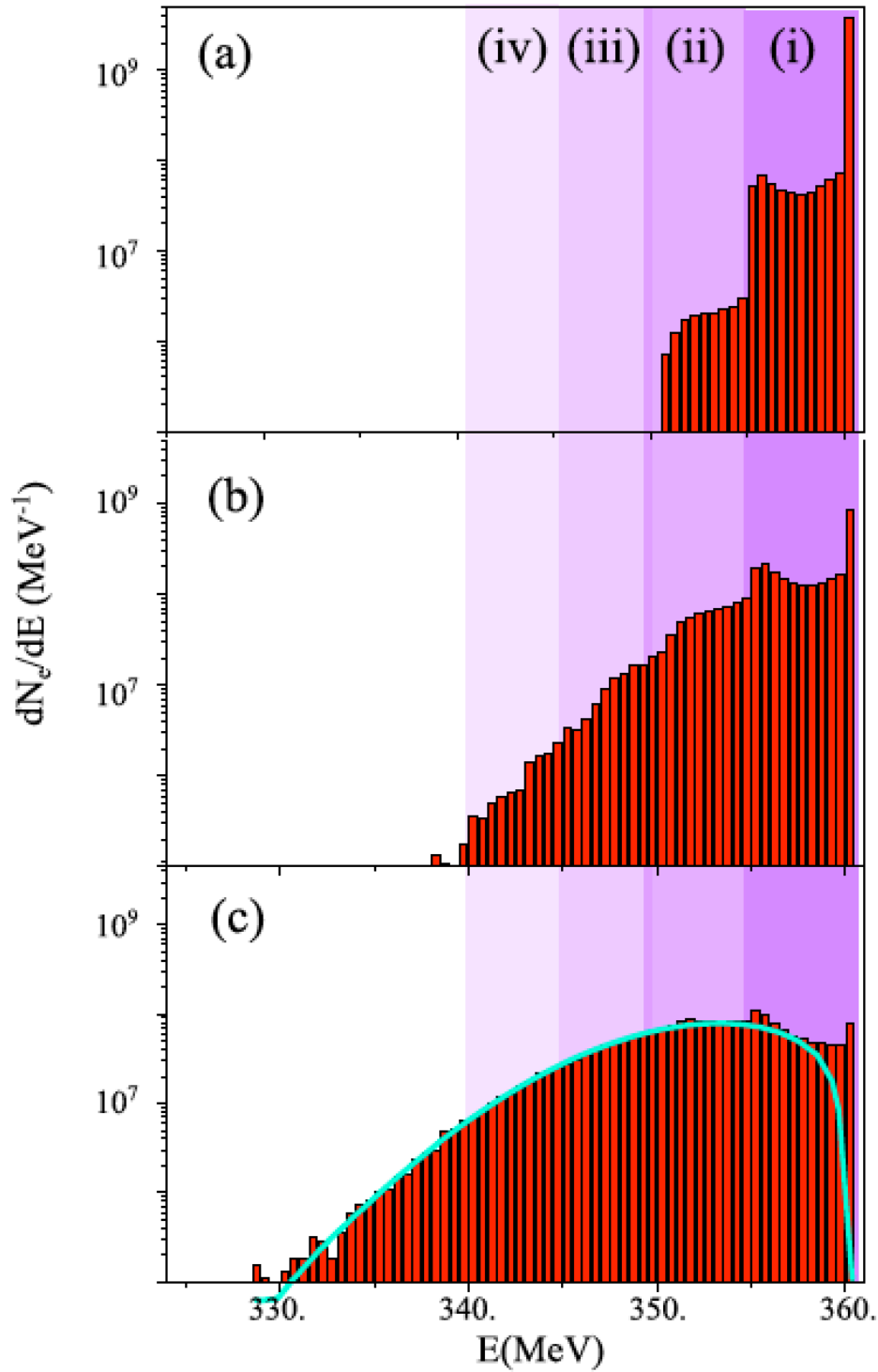

**Fig. 9.** Energy Spectrum of the electron beam colliding at 360 MeV against the Yb:Yag collision laser, to produce 5 MeV gamma ray photons, after the collision, for the case of a 0.33 ps long laser pulse (upper diagram), 3.33 ps long laser pulse (centre diagram) and 8.32 ps (lower diagram), showing diffusion of the electron energy distribution due to multiple back-scattering events

Here we plot the energy spectrum of the electron beam simulated for a typical ELI-NP-GS configuration in the EuroGammaS machine by means of the code CAIN, for different laser pulse lengths (0.33 ps, 3.33 ps and 8.32 ps, from top to bottom diagram, for an invariant laser parameter $a_0$=0.046), after the collision with the laser pulse and emission of the gamma ray beam (the energy spectrum before collision is a very thin Gaussian distribution with relative rms energy spread of 0.05%, absolute energy spread of about 200 keV).

The effect of multiple scattering of the electrons within the laser pulse against laser photons is evident, with formation of fringes in the electron energy distribution, up to a complete diffusion of the longitudinal phase space (lowest diagram), that can be described by a Chapman-Kolmogorov master equation for Markov random-walk phenomena with a suitable transition probability (see [52] and [55] for further details). It is clear from these results that the transport of the electron beam after a first collision with the laser to another collision point is not compatible with the preservation of high phase space quality of the beam itself, preventing to design a Linac lay-out based on multiple interaction points at different energies served by the same electron bunch propagated through the interaction points and re-accelerated from one to the next. Furthermore, the beam halo generated in the collision can produce beam losses on the beam pipe if transported and re-accelerated to a following Linac section, so to have significant impacts on the background radiation production.

## 2. Accelerator description

### 2.1. Beam physics

#### 2.1.1. Photo-injector Beam Dynamics: generating high Phase Space Density Beams

The expected quality of the emitted gamma photon beams needs the generation of a high quality electron beam. In particular from the analytical relation (Eq. 22) of section 1, showing how the spectral density in a fixed bandwidth is affected by electron beam and laser parameters, one can retrieve that the requirement of a spectral density > $10^4$ ph/sec.eV in a narrow rms bandwidth < 0.3% corresponds in the proposed collision scheme to an electron beam with a rms energy spread < 0.1% and a phase space density $(Q/\varepsilon_n^2) > 10^3$ ,where $\varepsilon_n$ is the rms normalized total projected emittance, as extensively discussed in Ref. 58. These specifications are somehow different from those required by FEL driver photo-linacs that aim at maximum beam brightness and minimum slice emittance. Both experiments and simulations [37], [38], [39], [40] show that, independently on the RF frequency band of the linac, it is very difficult to satisfy at the same time the above requirements: in fact a high spectral density requires a long beam with consequent increase of energy spread due to RF curvature, while achieving a reduction of the energy spread by starting with a short pulse is paid with an unavoidable increase of emittance. A way to overcome this problem is to start at the photocathode with a bunch long enough to control the emittance growth due to space charge in the gun region and then to reduce the bunch length by applying the "velocity bunching" technique [41] in the first accelerating section placed after the gun. This technique consists in injecting a not yet relativistic beam in a RF structure with a phase near to the zero crossing of the field: the beam slips back up to the acceleration phase undergoing a quarter of synchrotron oscillation and is chirped and compressed.

In our proposal this concept is applied to an accelerator based on a hybrid layout consisting in an S-band photo-injector followed by a C-band linac. The S-band segment is similar to SPARC and consists in a 1.6 cell RF gun with a copper photocathode and an emittance compensation solenoid, which is followed by two SLAC-type 3 meter long travelling wave sections. A gentle (compression factor <3) velocity bunching in the first accelerating section, as routinely done at SPARC, will allow to inject, without emittance degradation in the C-band booster, a beam short enough to reduce the final energy spread. The emittance is controlled by using a solenoid embedding the RF compressor as in SPARC.

From the approximated relation $\left.\frac{\Delta\gamma}{\gamma}\right|_{RMS} \approx 2(\pi f_{RF})^2 \sigma_t^2$ a 0.83 ps (280 µm) bunch length at the injector exit is required to achieve an energy spread around 0.05% at the end of the C-band linac.

This solution for the accelerator has several advantages: it reduces the risk factor being based on a conservative design of the photo-injector and at the same time is compact by using the C-band for the next part of the accelerator. In addition it can gain profit from the possibility to do some useful experimental tests (in single bunch) on the SPARC facility, as the matching between the S-band photo-injector with a C-band linac : in fact a third 3 meter long S-band structure actually present at SPARC is going to be removed and replaced with two C-band linacs for an energy upgrade from 170 to 240 MeV.

The operation of the S-band photo-injector for ELI gamma source has been optimized by extensive computer simulations based on the TSTEP [42] code, an updated version of the macroparticle code PARMELA [43], able to take into account the space charge effects that mainly affect the beam dynamics in this part of the accelerator. Calculations include a thermal emittance of 0.9 mm-mrad/(mm rms), the measured value for a S-band RF gun operating with a field of 120 MV/m [44]. The accelerating gradient in simulations is 21 MV/m, typically employed during SPARC operation, although a gradient of 23-25 MV/m could be reached. A number of 40K macroparticles has been used, giving a good statistic.

Computations explored different operating points in a range of electron beam charge 20-500 pC. Two of these points are presented here in detail:

a "reference" working point based on a 250 pC electron beam: in this point the phase space density results maximized and, for this reason, it has been selected for start-to-end simulations [45] aimed at meeting gamma source requirements a "commissioning" working point based on a low charge (25 pC) beam to be used in the early stages of the machine setup.

In the simulations for these two working points the beam starts with a transversally uniform distribution and the longitudinal distribution shown in Fig. 10, whose parameters have been found to minimize the final emittance.

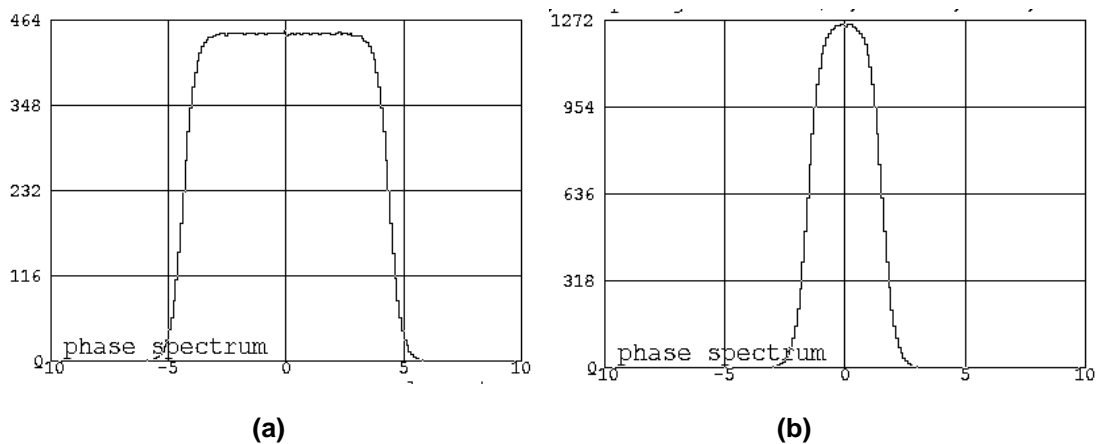

(a)  (b)

**Fig. 10.** Photocathode laser pulse shape used in beam dynamic simulations. (a) Reference working point,Q=250 pC. (b) Commissioning working point,Q=25pC.

The plots in Fig. 11 show the details of the computed dynamics of the reference beam: it exits from the photo-injector with a projected emittance of 0.4 mm-mrad beam and a rms length of 280 µm achieved by applying a RF compression factor of 2.5; the energy spread induced by the compression is foreseen to be recovered working off-crest in the following C-band linac.

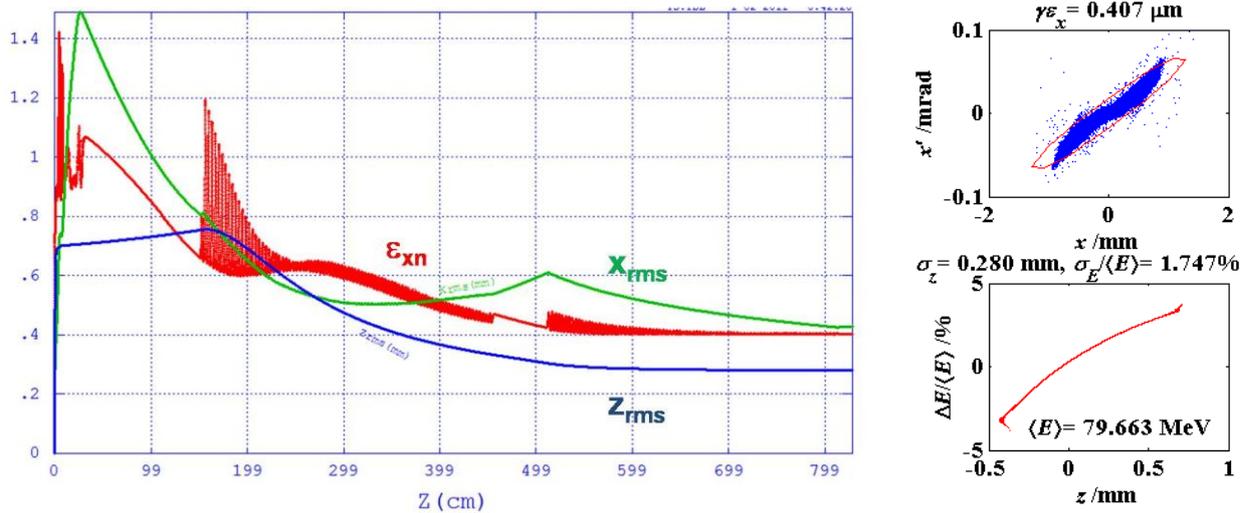

Fig. 11.  TSTEP output for the reference working point, Q=250pC. (top) Evolution of emittance, transverse and longitudinal envelopes in the S-band photo-injector (bottom) Transverse (top) and longitudinal (bottom) phase space at the photo-injector exit.

As the total projected emittance is affected by the mismatching of the tails of the longitudinal distribution the sensitivity of this parameter to the rise time of the photocathode laser pulse has been studied and the results are shown in Fig. 12: a final emittance of 0.4 mm-mrad is compatible with a rise time ≤ 1 psec and a tolerance of 10% in the emittance growth requires a rise time not larger than 1.5 psec.

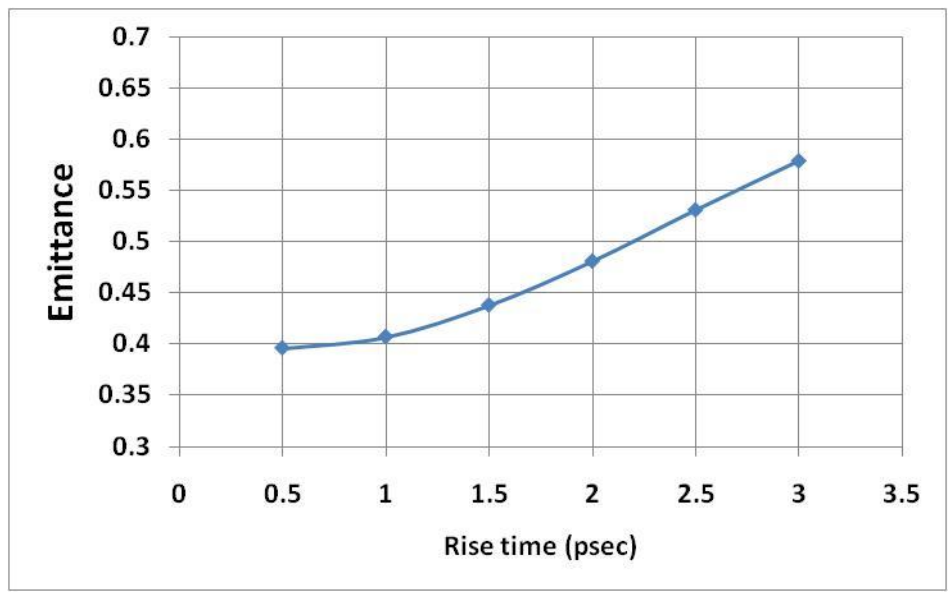

Fig. 12.  Sensitivity of projected emittance to the rise time of the photocathode laser pulse for the reference working point

The need to get a short rise time may result in the formation of longitudinal ripples. However also if the computations results presented here correspond to a plain laser pulse (Fig. 10) some ripple is sustainable in the realistic pulse. In fact, due to the compensation of longitudinal irregularities induced by space charge, as the beam goes through the gun and drifts, the temporal oscillations transform in energy oscillation and then the energy ripples are soon suppressed in the linac with no consequences on emittance for a peak-to-peak ripple up to 20%.

A low charge (25 pC) working point has been also optimized that can be usefully employed in phase of commissioning and setup of the machine. It has been splitted in two operation modes, with and without RF compression. In absence of RF compression (first section RF phase on crest) it has been optimized requiring the same bunch length of the reference working point (that means same final energy spread at high energy); TSTEP results are shown in Fig. 13: at the exit of the photo-injector a beam with a 280 µm rms length and a final emittance of 0.2 mm-mrad is obtained, operating the first section on crest with the solenoid embedding it switched off, which simplify the initial setting of the accelerator.

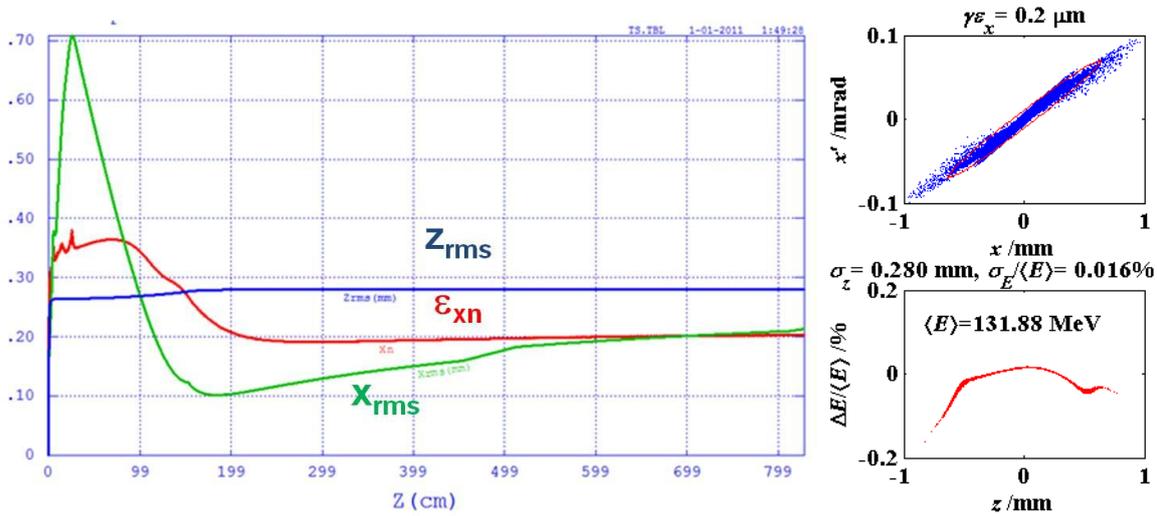

**Fig. 13.** TSTEP output for the commissioning beam, Q=25pC. Operation mode: no RF compression (a) Evolution of emittance, transverse and longitudinal envelopes in the S-band photo-injector (b) Transverse (top) and longitudinal (bottom) phase space at the photo-injector exit.

In absence of compression the final energy reaches 131 MeV and the energy spread for this no-chirped beam is very low (0.016%) that means that the off-crest operation in the following C-band sections is not needed.

If we apply to this low charge beam the velocity bunching with the same compression factor used for the reference working point, we obtain at the exit of the photo-injector a beam with a very short length (corresponding to a foreseen energy spread, at high energy, around 0.01%) without emittance degradation respect to the "on-crest" operation (Fig. 14).

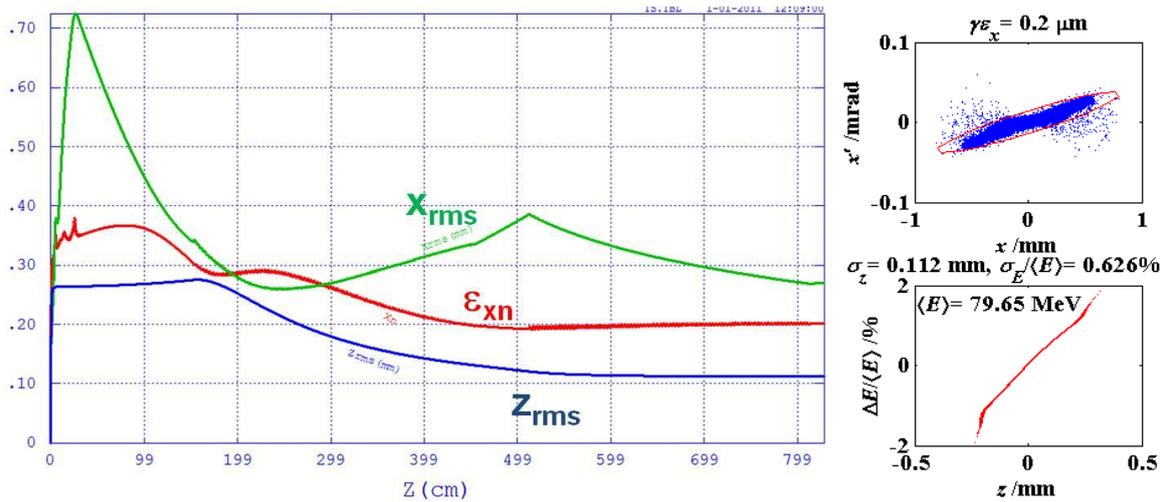

**Fig. 14.** TSTEP output for the commissioning beam, Q=25pC. Operation mode: RF compression on (a) Evolution of emittance, transverse and longitudinal envelopes in the S-band photo-injector (b) Transverse (top) and longitudinal (bottom) phase space at the photo-injector exit.

The characteristics of the two working points presented here are summarized in Table 5.

**Table 5.** Electron beam parameters for the S-band photo-injector in the hybrid scheme

|  | Reference beam (Compression factor = 2.5) | Commissioning beam (On crest operation) | Commissioning beam (Compression factor = 2.5) |
|---|---|---|---|
| Charge (pC) | 250 | 25 | 25 |
| Photocathode Laser pulse length (FWHM) (ps) | 8.5 | 3 | 3 |
| Photocathode Laser rms spot size (µm) (uniform transverse distribution) | 250 | 150 | 150 |
| Output energy (MeV) | 79.7 | 132 | 79.6 |
| Output RMS Energy spread (%) | 1.75 | 0.02 | 0.63 |
| Output Normalized RMS Projected Emittance (mm-mrad) | 0.4 | 0.2 | 0.2 |
| Output RMS bunch length (µm) | 280 | 280 | 112 |

The particle distributions, as computed by TSTEP at the end of the photo-injector (where the space charge effects are frozen), have been used as input for ELEGANT, the code used to propagate the beam up to the interaction point.

Another option has been also explored potentially able to improve the performance of the photocathode gun by optimizing the spatial distribution at the cathode. In fact, according to some recent numerical and experimental results achieved at LCLS, a spatial-cut Gaussian distribution could be more advantageous than the uniform one . Usually a spatial uniform distribution with a defined radius on the cathode is obtained by cutting with an iris a laser beam much bigger than the iris size, resulting in most of the laser beam being cut by the small iris. According with LCLS experience the use of a truncated Gaussian laser profile with a ratio $\sigma x/r \sim 1$ ($\sigma x$=rms size of the generating gaussian laser distribution, r=iris radius) could have two benefits: a significant improvement in laser transmission and a reduction in emittance. This last benefit is associated to a more linear behaviour of the transverse space charge force for σz/r~0.1 (a condition usually occurring when the bunch is immediately released from the cathode).

Some beam dynamics simulations have been done to explore the possibility to improve the ELI photocathode gun operation by using this approach. For the reference 250 pC working point the uniform spatial distribution used in the simulations discussed above has been compared with a truncated Gaussian spatial profile (Fig. 15) by keeping the same beam diameter on the cathode (1 mm) and optimizing the rms size of the generating laser Gaussian distribution. The longitudinal pulse profile has been kept equal to the optimized shape of Fig. 10.

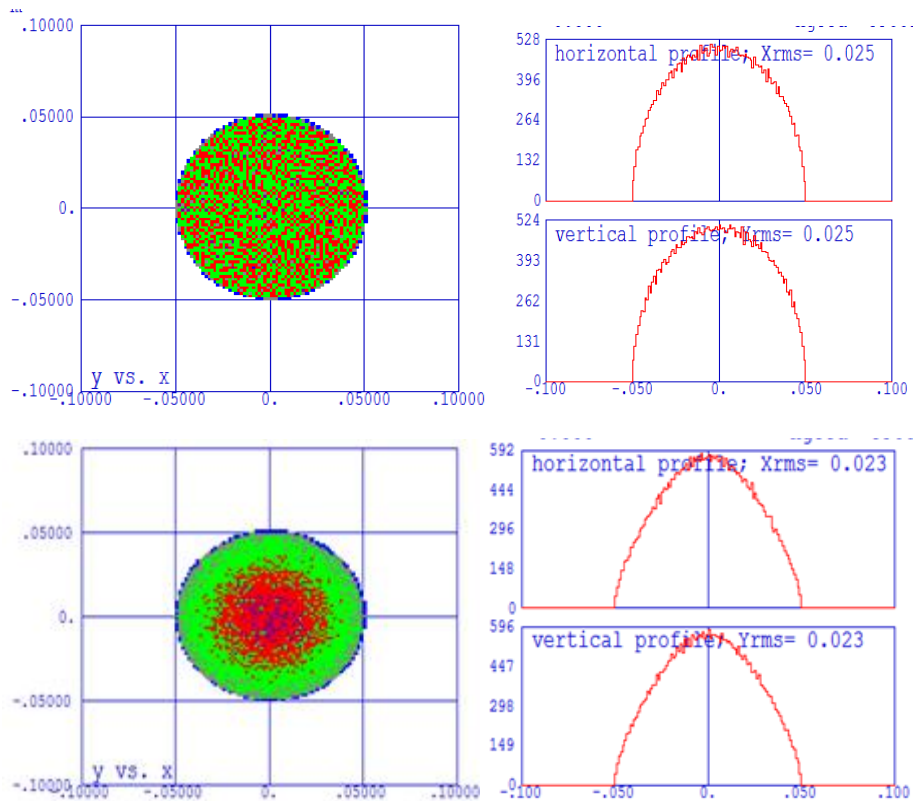

**Fig. 15.** Initial transverse distributions used in computations. Top: uniform, Bottom: Truncated Gaussian with σx=0.4 mm. Left: spot size right: horizontal and vertical projections

The rms size of the two distributions in Fig. 15 are very close, 250 μm and 235 μm, being the second one obtained by truncating a Gaussian distribution with σx=0.4 mm by a 0.5 mm radius iris. Therefore the effect of the thermal emittance is negligible. TSTEP beam dynamics computations show an reduction in the final emittance of about 33% (down to 0.31 mm-mrad) for the truncated Gaussian distribution respect to the pure uniform one (0.407 mm-mrad) with the same final bunch length (Fig. 16).

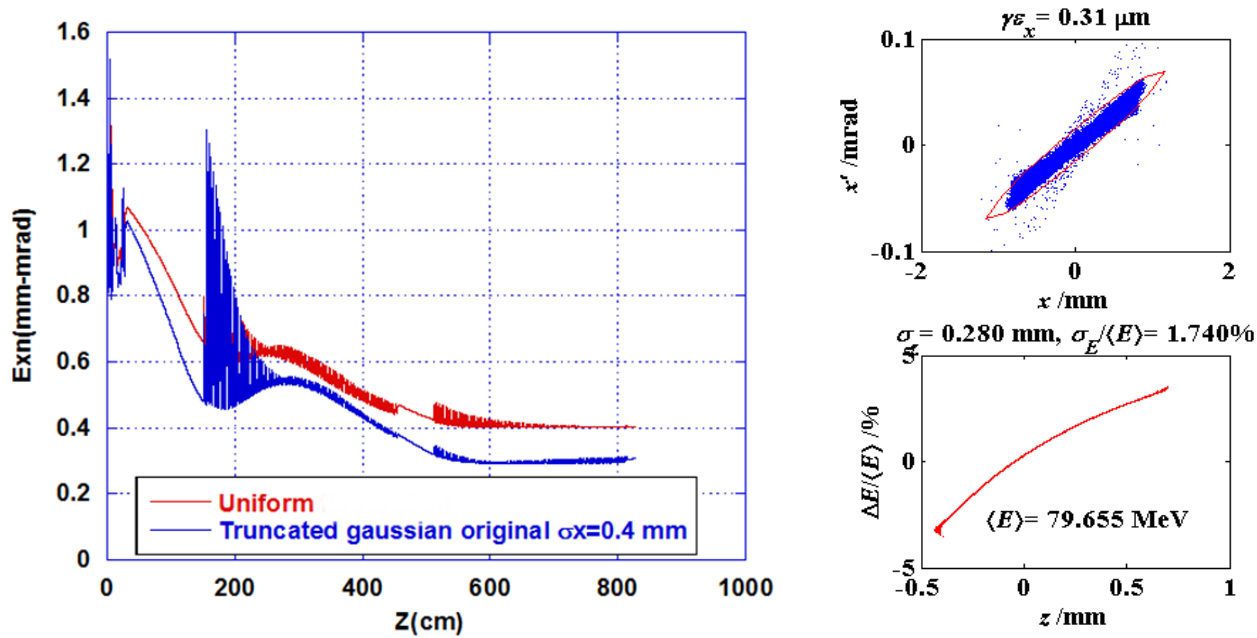

**Fig. 16.** Q=250 pC. Left: computed emittance evolution in the ELI S-band photoinjector for a uniform and a truncated Gaussian with σx=0.4 mm. Right: computed output phase space for the truncated Gaussian distribution

So this unconventional pulse shaping could improve the quality of the final beam, with the significant benefit to double at least the laser transmission (0.8 instead of 0.4-0.3) through the iris, loosening the required laser power.

This extra-budget could be also spent to increase the beam charge, instead to reduce the emittance. So other simulations have been done optimizing the beamline parameters (phase and magnetic fields) to find the maximum charge value compatible with the same final emittance of the reference 250 pC working point in the case of a Gaussian-cut distribution. As it is shown in Fig. 17 this upper limit has been found to be 390 pC, with the same bunch length of the reference working point. The only differences consist in a small increase in energy spread and a reduction of the final energy of about 1 MeV (see for comparison Fig. 11). The charge increase corresponds to an increase in the number of photons of the gamma source of a factor 1.56.

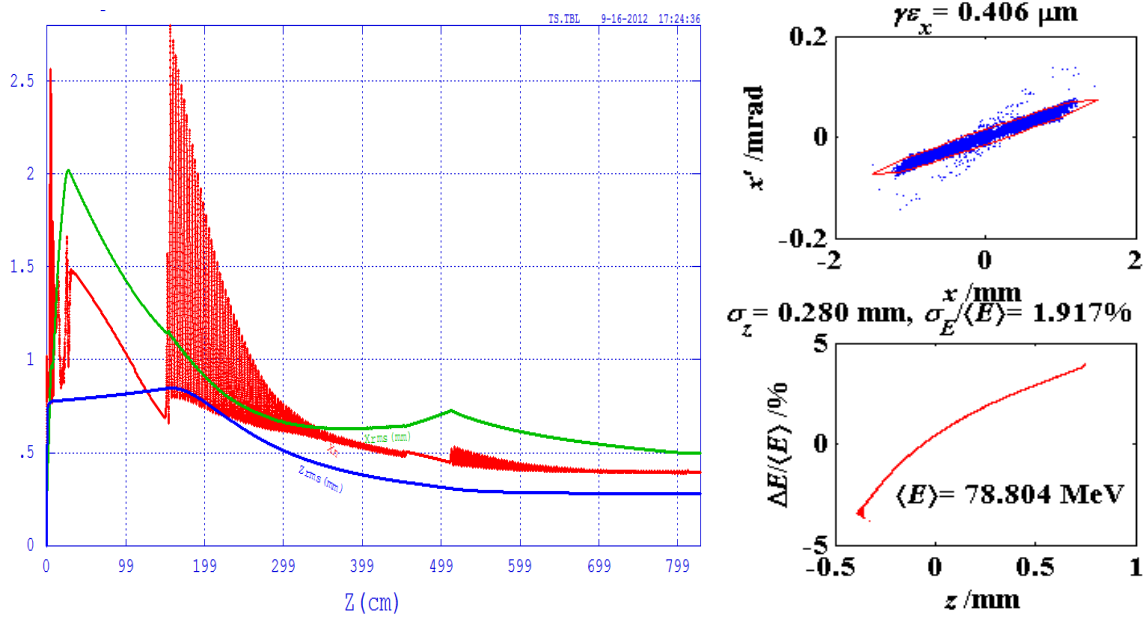

**Fig. 17.   Q=390 pC and initial Gaussian truncated distribution with σx=0.4 mm. Left: emittance and envelopes evolution in the S-band photoinjector. Right:  Output computed phase spaces**

All simulations presented here have been done with a perfect symmetric transverse Gaussian-cut distribution, demonstrating how the overall efficiency could be improved adopting this type of spatial laser profile respect to the uniform one. However it is necessary to remark that in the practical operation the improvement in emittance could be wash out by transverse beam asymmetries. Therefore this option requires a special effort in phase of commissioning to control the laser and cathode uniformity.

## 2.2.   RF Linac Beam Dynamics: preserving emittance and controlling energy spread

### 2.2.1.   The Linac lattice

The RF Linac that follows the photoinjector is based on twelve C-band accelerating sections, operating at 5.7 GHz with an average accelerating gradient Eacc=35 MV/m, for a maximum final energy of about 800 MeV on crest. Two beamlines are provided for delivering the electron beam at the two foreseen Compton interaction points : one at E=280 MeV and one at E=600 MeV respectively identified as Low and High energy Interaction Points, as it is shown in Fig. 18 where the schematic layout of the gamma-ray source is reported.

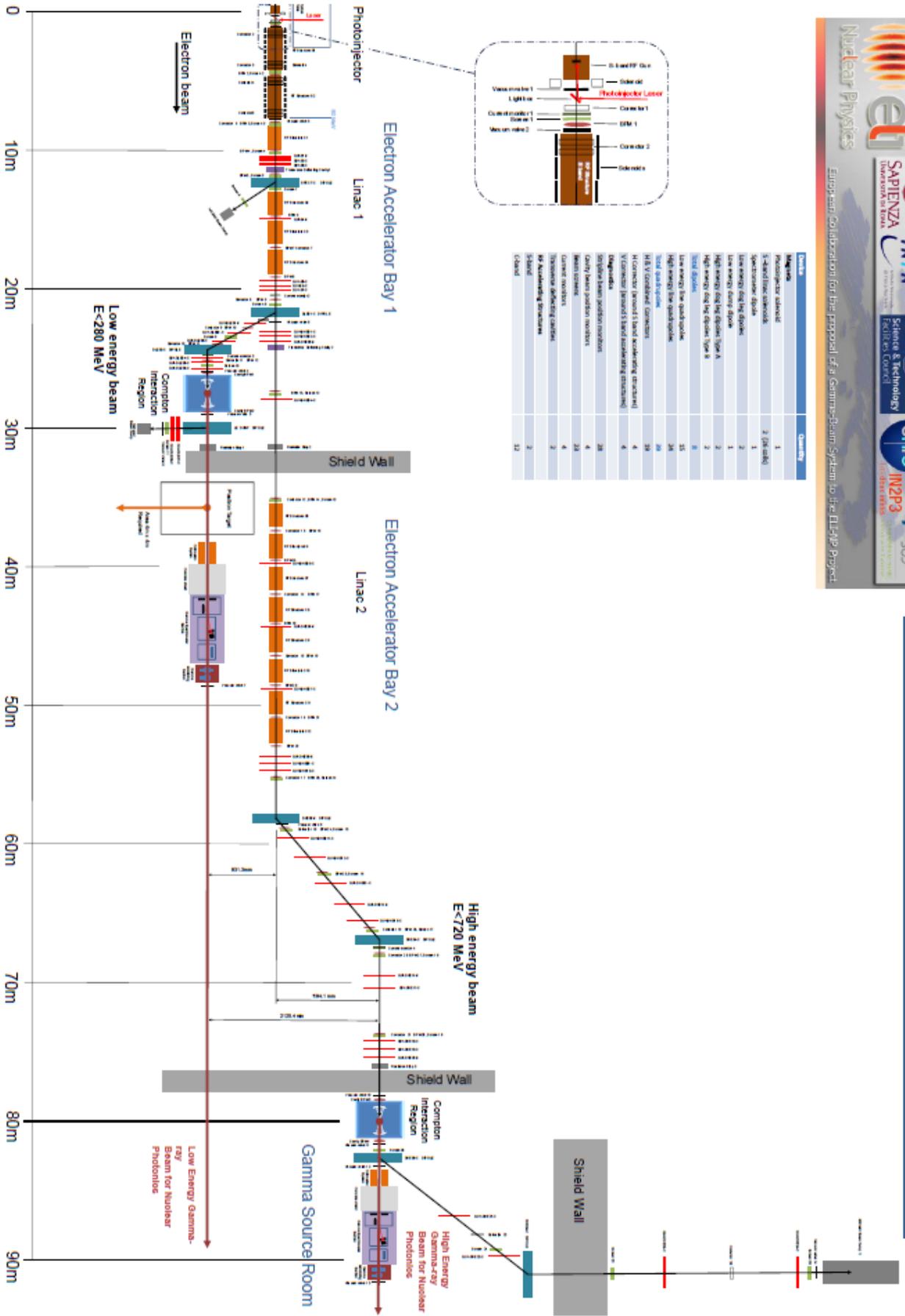

**Fig. 18.** Gamma-ray source schematic layout

Downstream the photoinjector a first linac section, Linac1, with 4 C-band 1.8 m long accelerating structures is foreseen to bring the electron beam energy up to a maximum energy of 320 MeV on crest, it is followed by a dogleg DL1 that provides on right side an off axis deviation of about 90 cm. This branch delivers the electron beam to the low energy Compton interaction point avoiding in this way the bremsstrahlung radiation contribution. The straight extension downstream the Linac1 exit brings the electron beam at the entrance of the Linac2 where the beam energy can be raised up to a maximum energy of 800 MeV on crest. A pulsed magnet is foreseen at the first dogleg entrance to split up the incoming electron bunches between the two Compton Interaction points. After the Linac2 a second dogleg DL2 delivers the beam to the high energy Compton interaction point with a further off axis deviation of 120 cm while the downstream doglegs DL3 and DL4 guide the electron beam to the high energy dumper located outside the accelerator bunker.

A repetition rate of 100 Hz is foreseen together with the possibility of accelerating up to 40 bunches in the same RF pulse with a 15-20 ns spacing to raise the effective repetition rate up to a 4 kHz value. The C-band accelerating structure design has been developed at LNF (see 3.3.4) and they consist of 102 cells for a total length of L=1.80 m. A full scale prototype of the structure is foreseen to be tested and fully characterized with the beam at SPARC (LNF) in spring 2013.

For the Linac lattice the considered FODO cell consists of two C-band sections with a phase advance of 58° as shown in **Error! Reference source not found.**.

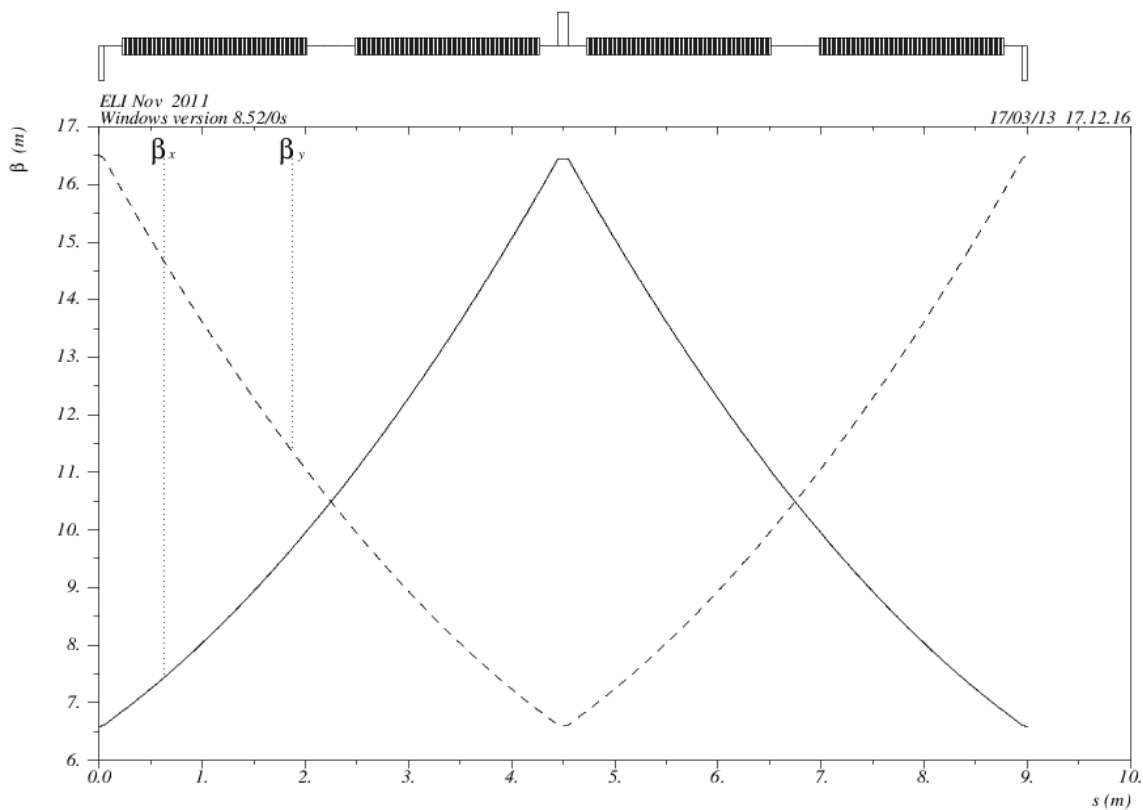

**Fig. 19**     Twiss parameters of the FODO lattice for the C-band linac, the considered phase advance is 58°

The lattice foreseen for the linac takes into account a diagnostic section after the first C-band structure, where the electron beam energy is E~120 MeV, to perform the 6-D phase space characterization just at the exit of the photoinjector by means of a RF-deflector combined with the adjacent spectrometer and of the quadrupole scan technique. After the DL1 dogleg a quadrupole triplet provides the electron beam transverse focusing for the interaction with the laser.

In Fig. 20 the Twiss parameters of the low energy beamline are reported from the photoinjector exit down to the low energy interaction point.

The straight beamline downstream the Linac1 injects the electron beam in the Linac2 of the high energy beamline that provides the desired acceleration and focusing for the high energy Compton Interaction point; in Fig. 21 the high energy beamline Twiss parameters are shown from the photoinjector exit down to the high energy interaction point.

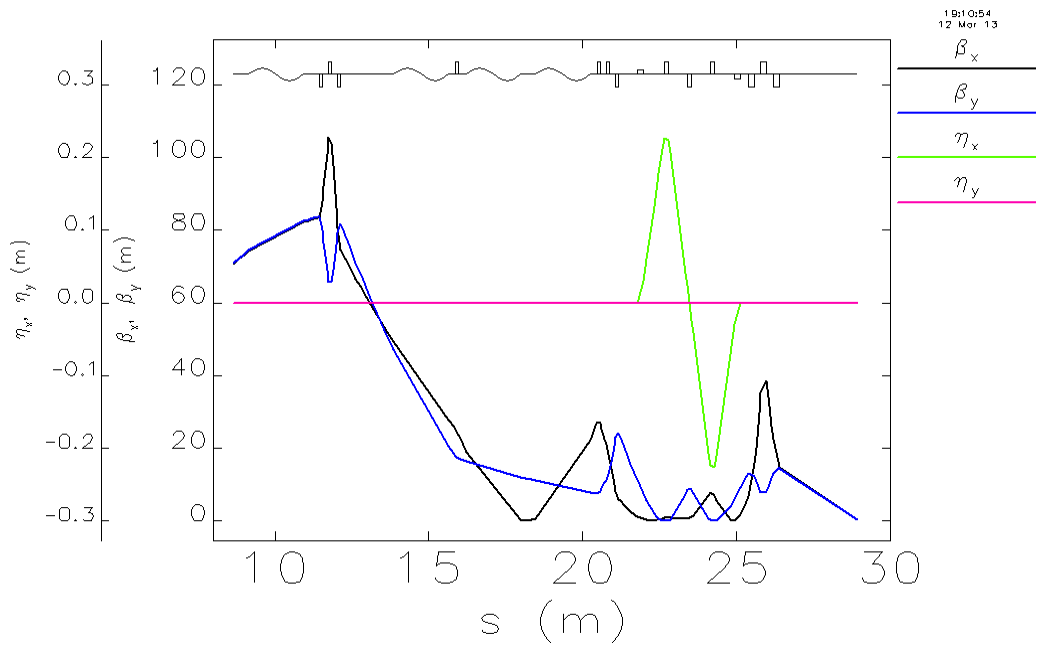

**Fig. 20.** Twiss parameters of the low energy beamline from the photoinjector exit down to the low energy Compton Interaction point

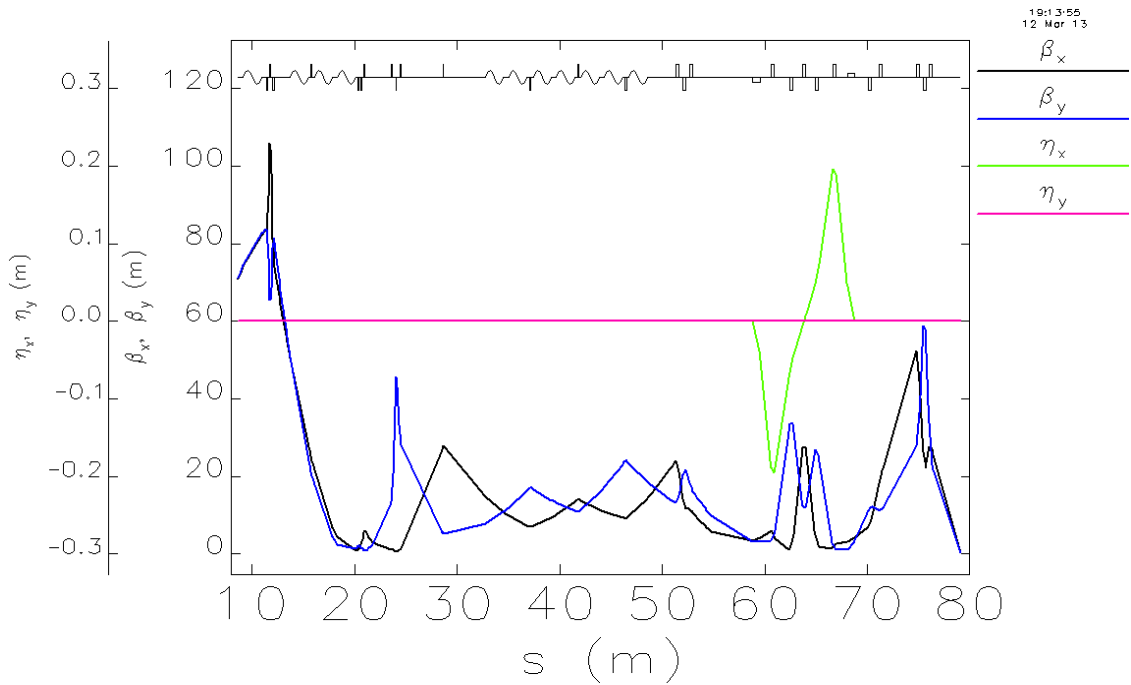

**Fig. 21.** Twiss parameters of the high energy beamline from the photoinjector exit down to the high energy Compton Interaction point

In Table 6, the main lattice parameters are reported for the two beamlines.

**Table 6.** Gamma-ray source electron beamline parameters up to the Interaction Point

| Beamline | $E_{in}$ (MeV) | $E_{out}$ (MeV) | $\varepsilon_{nx-in}$ (μ-rad) | $\varepsilon_{nx-out}$ (μ-rad) | $\sigma_{\delta-in}$ % | $\sigma_{\delta-out}$ % | Off axis deviation (cm) | $R_{56}$ (mm) |
|---|---|---|---|---|---|---|---|---|
| Linac1 | 80-130 | 280 | 0.4 | 0.4 | 1.7 | 0.08 | 0 | |
| DL1-H/L | 280 | 280 | 0.4 | 0.5 | 0.08 | 0.08 | 93 | -1.9 |
| Linac2 | 280 | 600 | 0.5 | 0.5 | 0.08 | 0.05 | 0 | |
| DL2-H | 600 | 600 | 0.5 | 0.5 | 0.05 | 0.05 | 120 | 5.3 |

The doglegs DL3 and DL4 provide the electron beam transport from the interaction point down the high energy beam dumper located in a dedicated area outside the linac bunker according to the safety requirements. In Table 7 the four dogleg parameters are reported.

**Table 7.    Gamma Source dogleg parameters**

| Dogleg | Total length (m) | Number of dipoles | Dipole length (m) | Dipole angle (rad) | Off axis deviation (cm) |
|---|---:|---:|---:|---:|---:|
| DL1 | 3.2 | 2 | 0.20 | 0.300 | 93 |
| DL2 | 9.2 | 2 | 0.70 | 0.130 | 120 |
| DL3 | 12.0 | 2 | 0.70 | 0.245 | 200 |
| DL4 | 15.0 | 2 | 1.30 | 1.571 | - |

### 2.2.2.    Simulation results

In order to optimize the linac and transfer lines design computer simulations have been performed by means of the code Elegant [57] tracking the beam macroparticles obtained with the Tstep [43] code from the photoinjector exit down to the two interaction points.

The code is able to take into account the wake fields generated by the electron beam inside the accelerating structure together with the coherent and incoherent synchrotron radiation effects in the bending magnets.

For the C-band structure the asymptotic values of the longitudinal and transverse wake functions have been calculated according to [58, 59]:

$$W_{0\parallel}(s) \approx \frac{Z_0 c}{\pi a^2} \exp\left(-\sqrt{\frac{s}{s_1}}\right) \text{ (V/Cm)}, \qquad s_1 = 0.41\frac{a^{1.8}g^{1.6}}{L^{2.4}}$$

$$W_{0\perp}(s) \approx \frac{4Z_0 c s_2}{\pi a^4}\left[1 - \left(1+\sqrt{\frac{s}{s_2}}\right)\exp\left(-\sqrt{\frac{s}{s_2}}\right)\right] \text{ (V/Cm}^2\text{)}, \quad s_2 = 0.17\frac{a^{1.79}g^{0.38}}{L^{1.17}}$$

where $Z_0$ is the free space impedance, $c$ is the light velocity, $a$ is the iris radius, $L$ is the cell length, $b$ and $g$ are the cavity radius and length respectively of the considered pill box model reported in Fig. 22.

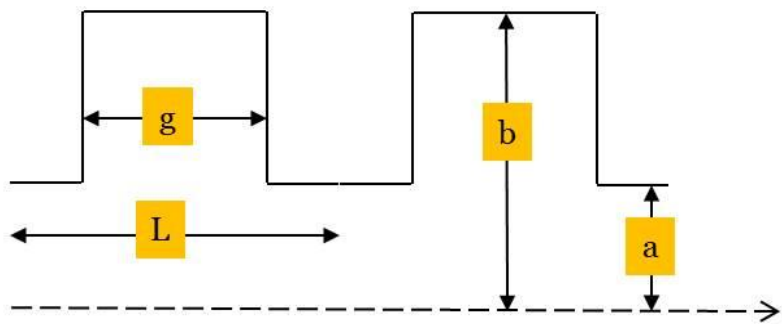

**Fig. 22.    Pill box cavity model considered for the wake feld calculations : *a* is the iris radius, *L* is the cell length and *b* and *g* are the cavity radius and length.**

The obtained results for the longitudinal and transverse cases are reported in Fig. 23 where they are compared with the results obtained for the SLAC type S-band structures.

In the Elegant code the Coherent Synchrotron Radiation effect is taken into account according to references [60] and [61].

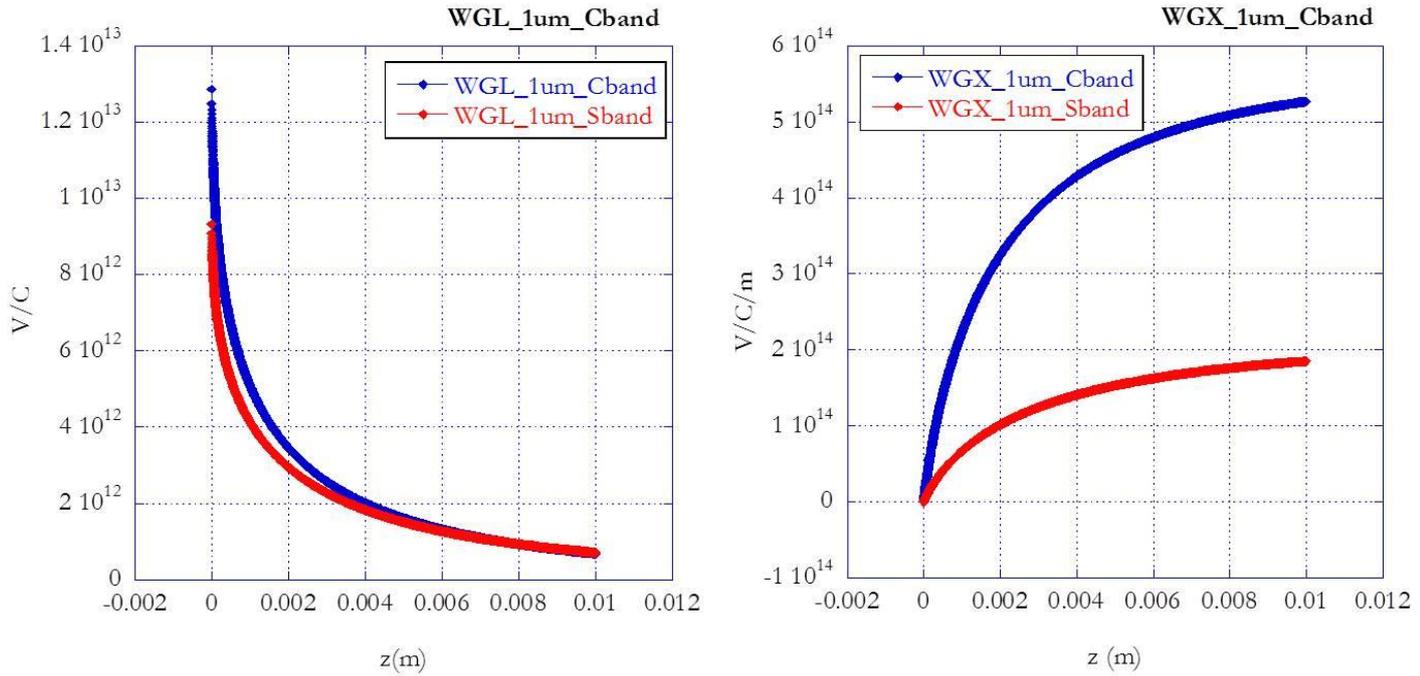

**Fig. 23.** Longitudinal and transverse short range wake field curve integrated over one cell for the S-band accelerating structure (red curve) and for the C-band structure (blue curve)

The considered electron beam as obtained at the exit of the photoinjector has the following characteristics: charge Q=250 pC, longitudinal rms length $\sigma_z \approx 280$ µm, energy $E \approx 80$ MeV, energy spread $\sigma_\delta \approx 1.8$ %, and transverse normalized emittance $\varepsilon_{n\ x-y} \approx 0.4$ µrad, and it is referred as the reference working point (see chapter 2.2.1). The simulated tracking through the linac down to the IP's of the Np=40 kparticles describing this electron beam give the results reported in the Table 8:

**Table 8.** Electron beam parameters at the low and high energy Interaction Point

| Beamline | E (MeV) | $\varepsilon_{nx}$ (µ-rad) | $\sigma_\delta$ % | $\sigma_x$ (µm) | $\sigma_y$ (µm) |
|---|---|---|---|---|---|
| Low energy IP | 280 | 0.4 | 0.09 | 18 | 18 |
| High energy IP | 600 | 0.4 | 0.05 | 10 | 10 |

In Fig. 24 the transverse and longitudinal beam distribution are reported for the two beams, while in Fig. 25 the energy spread and energy distribution are reported together with the current distribution along the bunch.

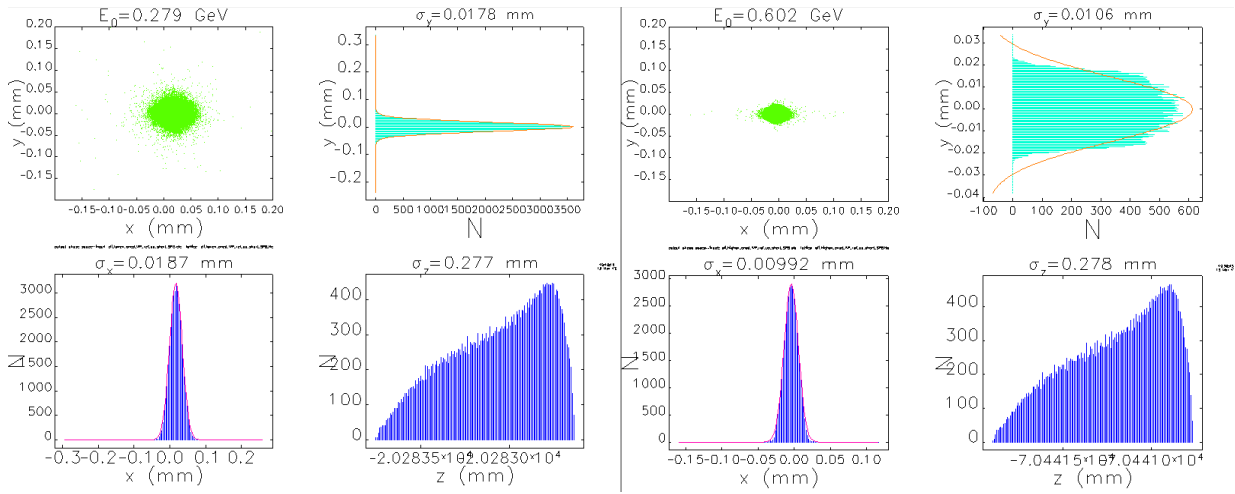

**Fig. 24.** Transverse beam size and distribution plus the longitudinal one for the reference working point electron beam at the low (left) and high (right) energy interaction point

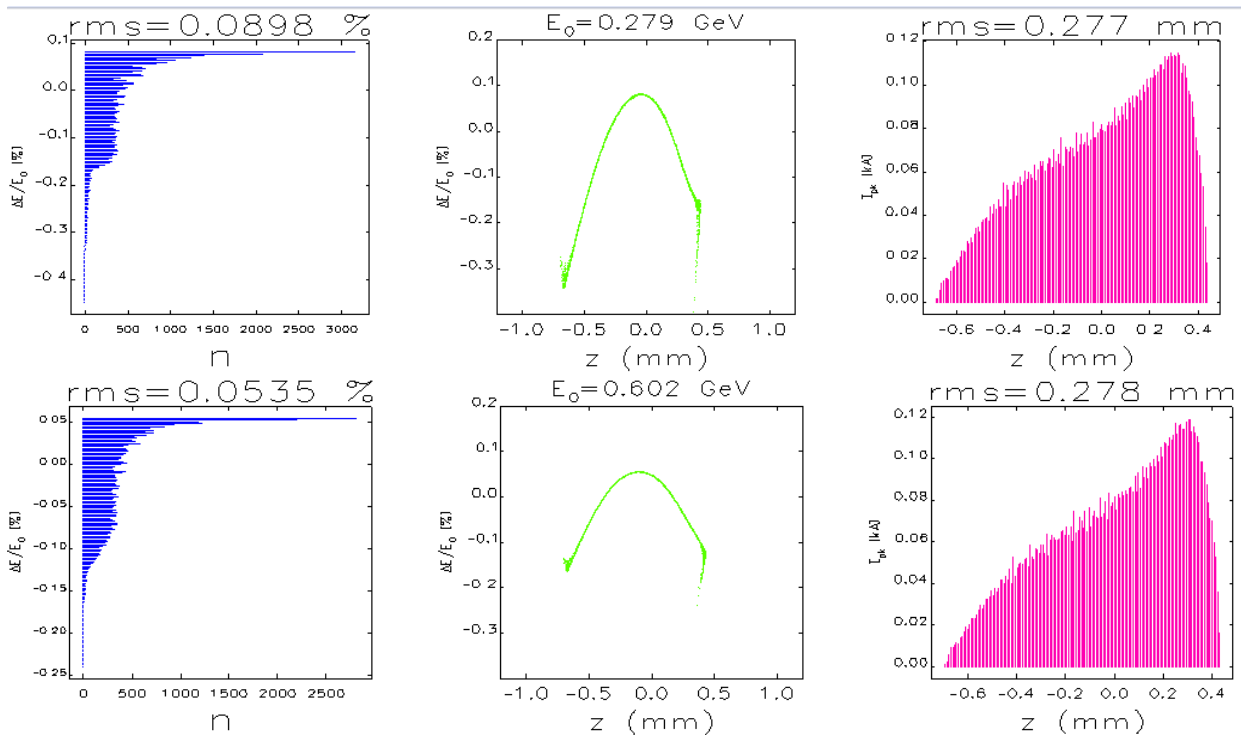

**Fig. 25.** Energy spread, energy and current distribution of the reference working point electron beam at the low (above) and high (bottom) energy interaction point

### 2.2.3. Lattice error sensitivity

In order to explore the lattice sensitivity to the machine errors a sample of 100 machine runs has been analysed where the errors of each lattice elements such as accelerating structures and quadrupole magnets are taken into account for both the low and high energy beamlines. The jitter values distribution is calculated according to the latin hypercube scheme and to the following table of jittering values:

**Table 9.  Jittering values for the lattice elements as considered in the simulation runs**

| Element | $|\Delta x|$ (µm) | $|\Delta V|$ (kV) | $|\Delta \varphi|$ (deg) | $|\Delta k/k_{max}|$ | $|\Delta B/B_{max}|$ |
|---|---|---|---|---|---|
| Acc. sections | 80 | 300 | 1 | - | |
| Quadrupole | 80 | - | - | 5. E-4 | |
| Dipole | 80 | | | | 1. E-3 |

The matrix of the latin hypercube that factorizes the previous jitter values is calculated via a Matlab script that applies also a normal random distribution of minus and plus sign to the jitter matrix. The jitters have been applied separately to the low and high energy lattices lattice to establish the most dangerous contributions; in Fig. 26 to Fig. 30 the overall probability distribution are reported for the beam size and energy spread values at the IP taking into account each jitter contribution separately. In Fig. 31 and Fig. 32 the result obtained with the total error contribution are reported.

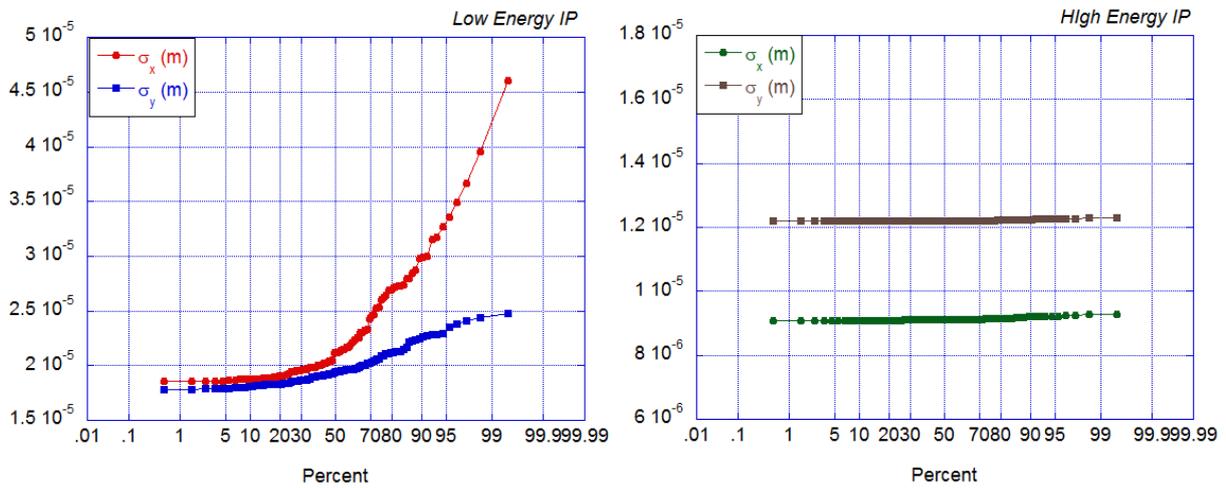

**Fig. 26.**  Beam size probability distribution for a random transverse misalignment of ± 80 µm of all the RF and magnetic elements at the Low Energy IP (left) and at the High Energy IP (right)

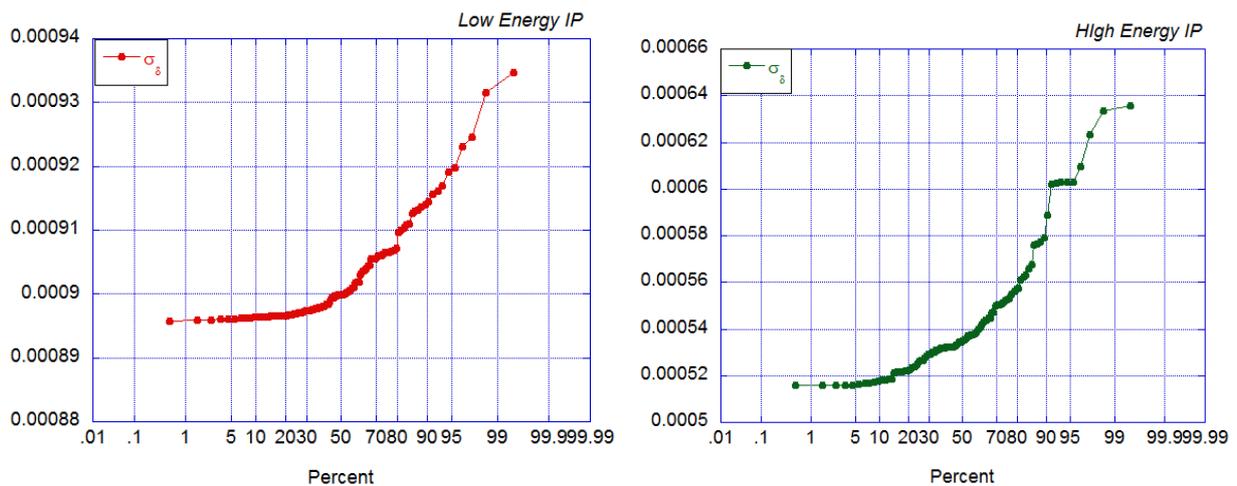

**Fig. 27.**  Energy spread probability distribution for a phase error of 1 deg at the Low Energy IP (left) and at the High Energy IP (right).

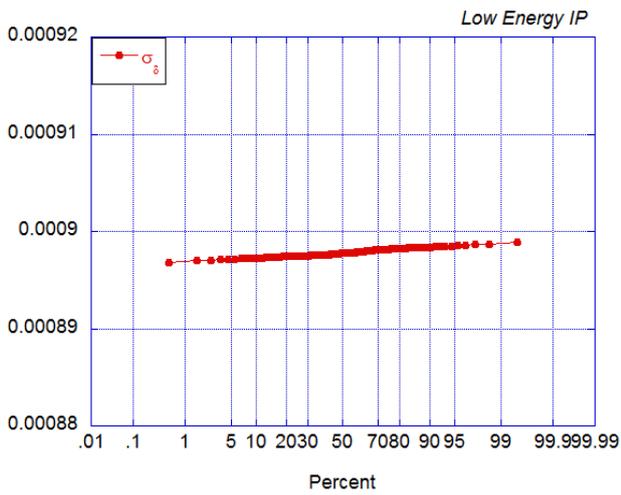
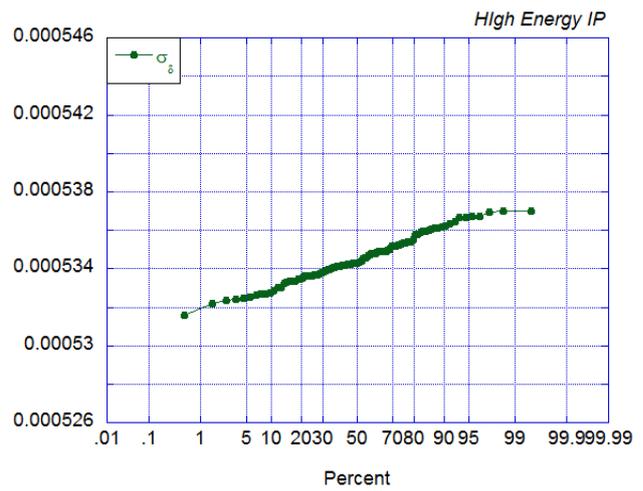

Fig. 28. Energy spread probability distribution for ± 300kV error in the accelerating gradient at the Low Energy IP (left) and at the High Energy IP (right)

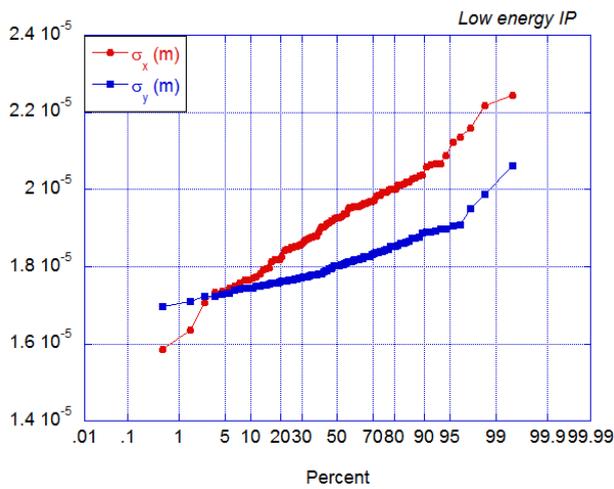
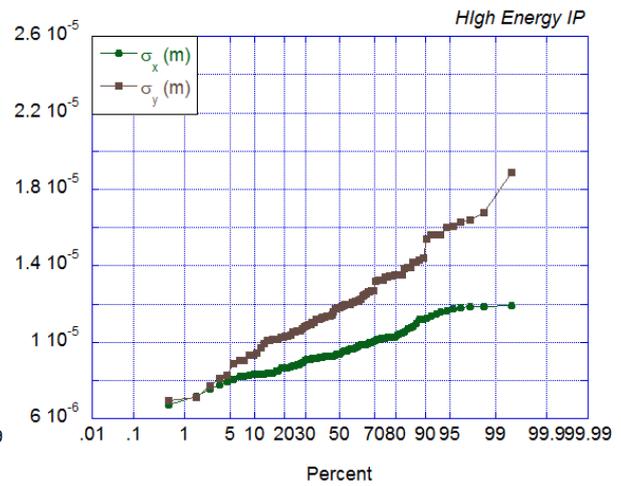

Fig. 29. Beam size probability distribution for a quadrupole focusing relative error Δk/kmax= ± 5 E-4 at the Low Energy IP (left) and at the High Energy IP (right)

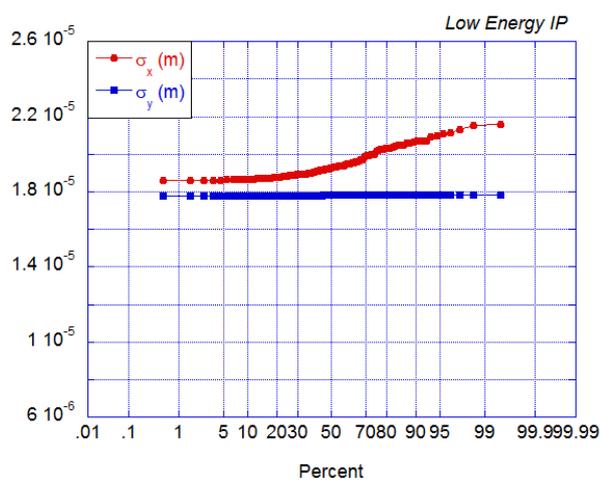
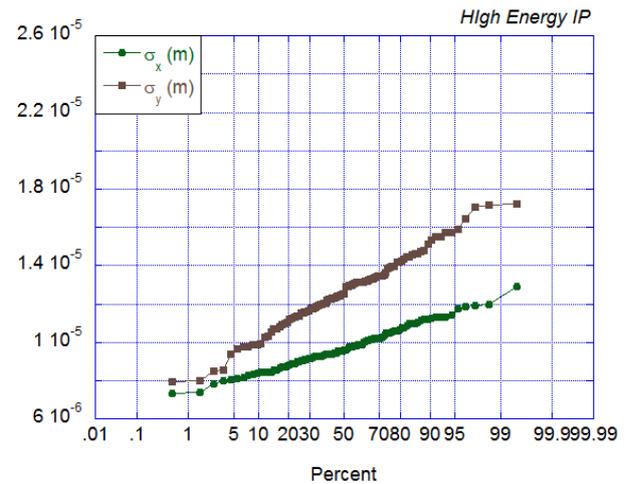

Fig. 30. Beam size probability distribution for a dipole strength relative error ΔB/Bmax= ± 1E-3 at the Low Energy IP (left) and at the High Energy IP (right)

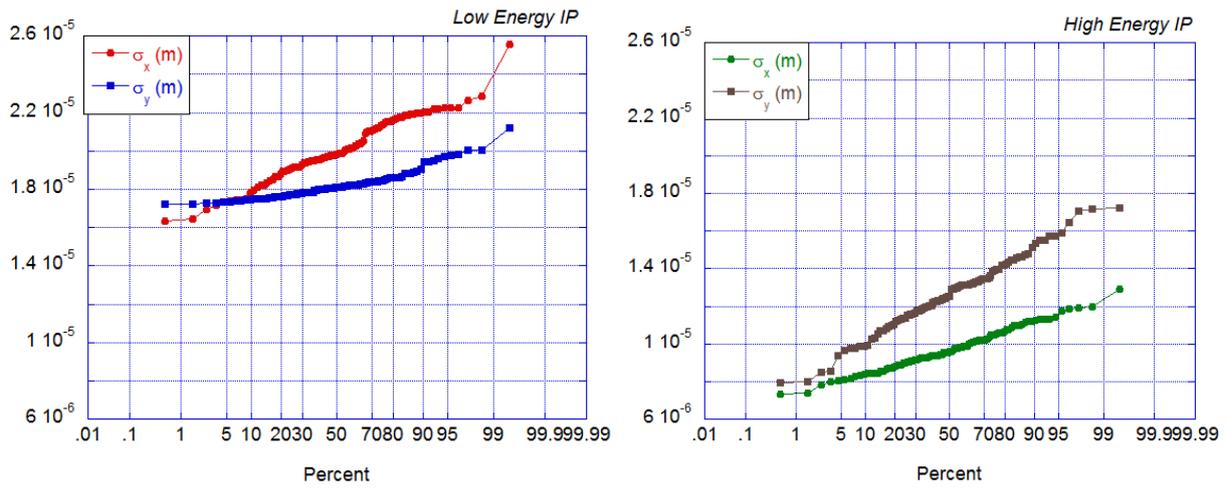

**Fig. 31.** Beam size probability distribution with all the previous errors considered at the Low Energy IP (left) and at the High Energy IP (right)

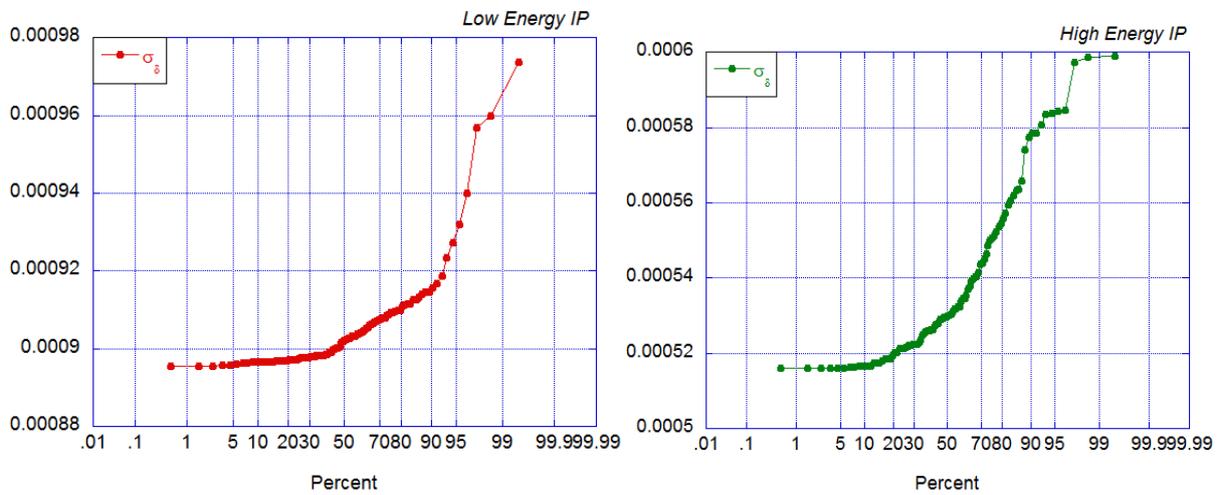

**Fig. 32.** Energy spread probability distribution with all the previous errors considered at the Low Energy IP (left) and at the High Energy IP (right)

From the obtained results it can be seen that when all the mentioned errors are considered the whole span of the beam transverse size is in the range of 10 µm around the nominal value in both planes for the low and high energy beamlines.

## 2.3. Multi-bunch Issues: Beam Loading and Beam Break-up

The passage of electron bunches through accelerating structures excites electromagnetic wakefield. This field can have longitudinal and transverse components and, interacting with subsequent bunches, can affect the longitudinal and the transverse beam dynamics. Those related to the excitation of the fundamental accelerating mode are referred as beam loading effects and can give a modulation of the beam energy along the train while transverse wakefields, can drive an instability along the train called multibunch beam break up (BBU).

### 2.3.1. Beam Loading

The main effect of the beam loading is the decrease of the accelerating field gradient in the structure since the effective field can be assumed as a superposition of the RF field and induced wakefield. If VRF is the RF accelerating voltage in the structure and VB is the induced voltage by longitudinal wakefield, the total accelerating voltage is VT=VRF+VB. Several approaches can be found in literature to study and compensate the beam loading effects [34]. The approaches are different in case of standing wave (SW) structures (like the RF gun) and travelling wave (TW) ones (like the S-Band or C-Band sections).

Other longitudinal higher order modes (HOM) can be also excited in the structure and can contribute to the beam energy perturbation. We will focalize, nevertheless, on the contribution given by the fundamental mode that is the dominant one.

#### 2.3.1.1 BL in the gun and its compensation

The RF gun we want to use for the ELI-NP injector is a 1.6 cell RF gun working in S-Band at 2.856 GHz. The main RF parameters are discussed and calculated in section 3. When a Gaussian bunch of total charge q passes through the gun the beam-induced voltage on the fundamental mode is given by:

$$V_{B\_GUN}(t) = -\omega_{RF}\frac{R}{Q}qe^{-\frac{\omega_{RF}^2\sigma_\tau^2}{2}}e^{-\frac{2Q_L}{\omega_{RF}}t}\cos(\omega_{RF}t) \quad (1)$$

where Q and QL are the unloaded and loaded quality factors, R and ωRF are the shunt impedance and the angular frequency of the gun working mode, σ$_\tau$ is the bunch length. According to the beam loading theorem, the beam voltage induced on the bunch itself is given by:

$$V_{B\_GUN\_SELF} = \frac{V_{B\_GUN}(0)}{2} = -\omega_{RF}\frac{R}{2Q}qe^{-\frac{\omega_{RF}^2\sigma_\tau^2}{2}} \quad (2)$$

If we consider a train of n equispaced identical bunches with distance ΔT, the total beam loading voltage experienced by the n-th bunch is given by:

$$V_{B\_GUN}\big|_n = -\omega_{RF}\frac{R}{Q}qe^{-\frac{\omega_{RF}^2\sigma_\tau^2}{2}}\left(\frac{e^{-\frac{\omega_{RF}}{2Q_L}(n-1)\Delta T}-1}{1-e^{\frac{\omega_{RF}}{2Q_L}\Delta T}}+\frac{1}{2}\right) \quad (3)$$

The beam induced voltage assuming the ELI beam parameters (q=250 pC, 31 bunches and ΔT=16 ns) and the gun parameters of Table 14 is given in Fig. 33.

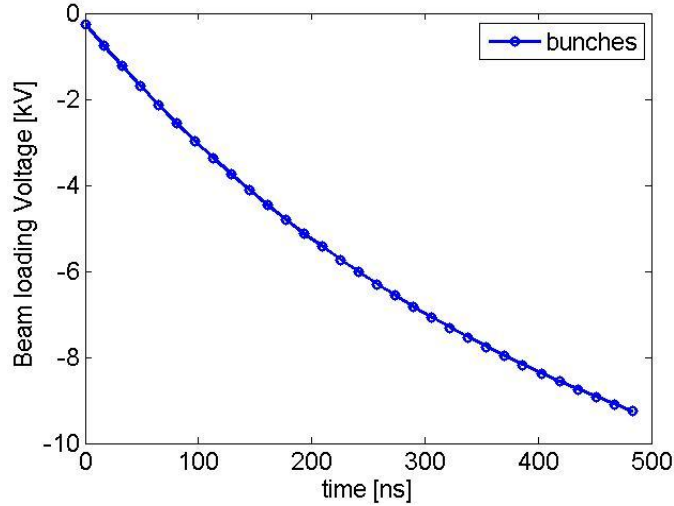

**Fig. 33.** Beam induced voltage in the gun (q=250 pC, 31 bunches and ΔT=16 ns)

The total voltage (VTOT_GUN) experienced by each bunch is given by the superposition of the beam loading voltage and of the external accelerating voltage (VRF_GUN):

$$V_{TOT\_GUN}\big|_n = V_{B\_GUN}\big|_n + V_{RF\_GUN}\big|_n \quad (4)$$

If we assume a simple excitation of the gun with a constant RF input pulse we have that:

$$V_{RF\_GUN}\big|_n = \sqrt{2P_{IN}R}\frac{2\sqrt{\beta}}{1+\beta}\left(1-e^{-\frac{\omega_{RF}}{2Q_L}[t_{inj}+(n-1)\Delta T]}\right)\cos(\phi) \quad (5)$$

where $P_{IN}$ is the incident input power in the gun, β is the coupling coefficient, Φ is the injection phase $t_{inj}$ is the injection time of the first bunch.

The energy spread due to the beam loading is, in the ELI case, very small (about 0.1%) while the energy spread due to the cavity filling time can give a much higher contribution if the injection time is not chosen properly. In principle, if one fixes the "nominal" accelerating voltage (Vnom) and the coupling coefficient β it is possible to choose the input power PIN and the injection time $t_{inj}$ to have a perfect compensation between the beam loading and the exponential increase of the field in the cavity. In most practical cases, nevertheless, and also in the ELI case, if we adopt this criterion to perfectly compensate the beam loading we risk to obtain large injection times ($t_{inj}$), that means long RF pulses and, therefore high average dissipated power in the 100 Hz operation and higher Breakdown Rate Probability. Alternatively, we can tolerate a bunch train residual energy spread relaxing the other RF parameters. The best working point for the gun and also the choice of the coupling coefficient $\beta$ has, therefore, to be done taking into account several figures of merit as the residual bunch energy spread, the average dissipated power into the cavity, the required input peak input power to reach $V_{nom}$ and the RF total pulse length. These figures of merit in the case of a flat input pulse are given in Fig. 34 to Fig. 36 considering the ELI parameters and the gun parameters of Table 69 and three different coupling coefficients. In the figures we have considered the case of $\phi=0$ and a "nominal" accelerating voltage of $V_{nom}$=5.97 MV corresponding to a peak field in the cathode equal to 120 MV/m.

The plots put in evidence that with a larger coupling coefficient we can reduce the input pulse length and therefore the average dissipated power. On the other hand we have to increase the incident input power to reach the required energy.

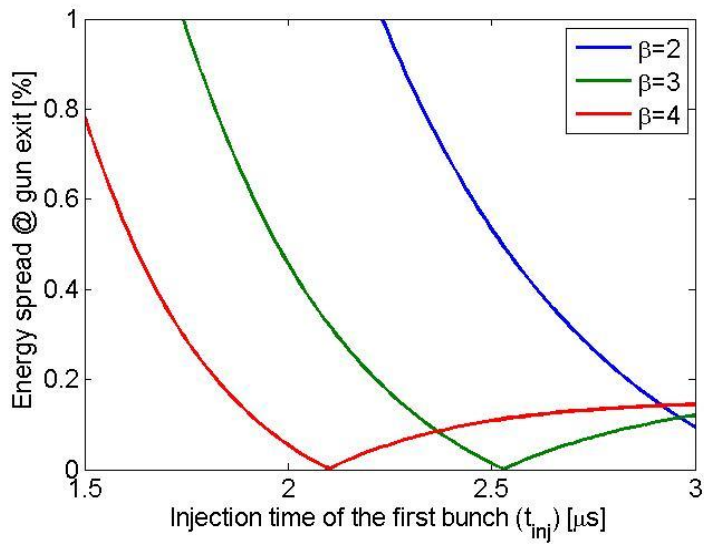

**Fig. 34.** Energy spread of the bunch train at the exit of the gun as a function of the injection time and for different coupling coefficients (q=250 pC, 31 bunches and ΔT=16 ns)

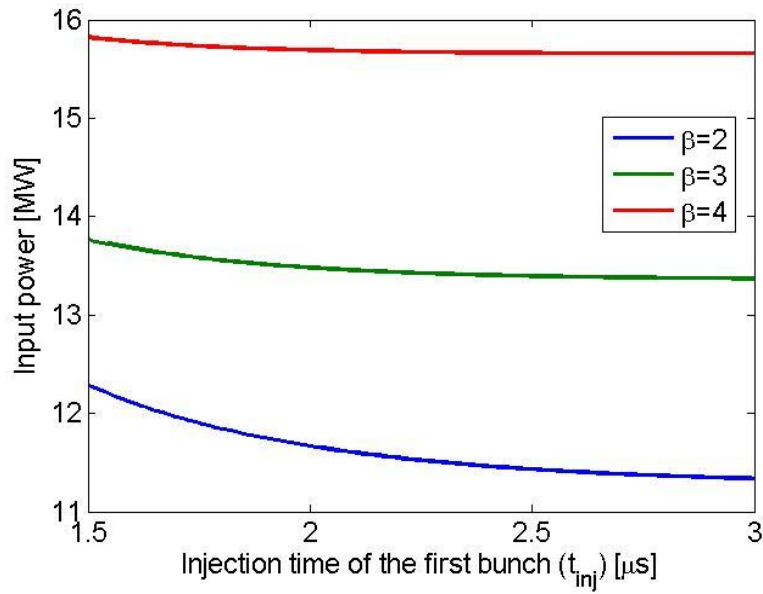

**Fig. 35.** Peak input power (to have an average energy of the bunch train equal to the nominal energy) as a function of the injection time and for different coupling coefficients

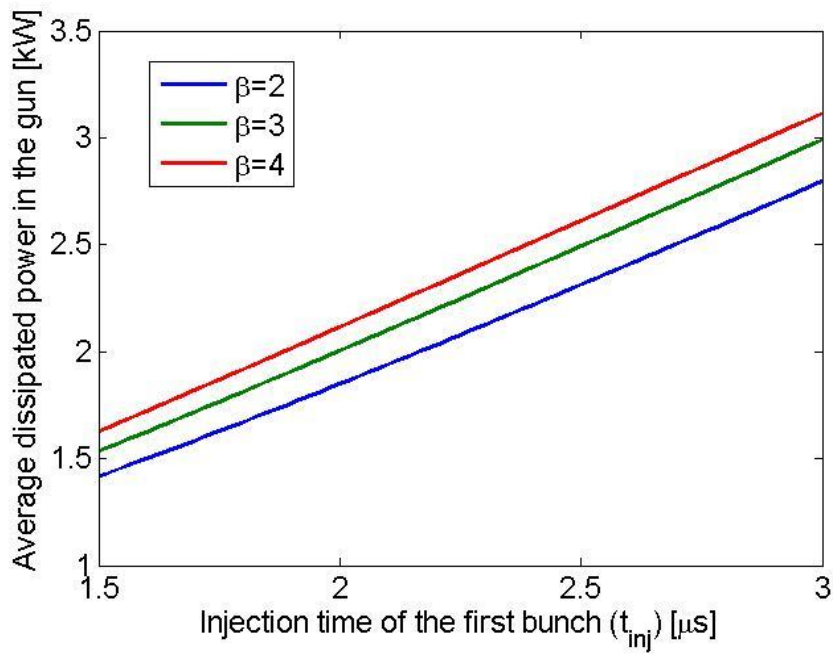

**Fig. 36.** Average dissipated power in the gun as a function of the injection time and for different coupling coefficients

Also not uniform input pulse profiles can be adopted to have more flexibility in the beam loading compensation allowing, also, reducing the RF pulse length and the average dissipated power. As an example an input pulse with the profile given in Fig. 37 can be considered. In this case there is a step in the input power corresponding to the injection time of the first bunch. One can play with both the injection time and with the ratio between the two levels (α) to compensate the beam loading as given in Fig. 38 to Fig. 40.

From all previous results it is possible to conclude that the beam loading in the gun can be compensated in different ways and that a large coupling coefficient (>3) allows reducing the input pulse length and therefore the average dissipated power in the 100 Hz operation. On the other hand if we increase the coupling coefficient we can need a too much high input power and one can introduce a strong perturbation in the accelerating mode configuration giving multipole field components. A good compromise between these requirements can be β=3

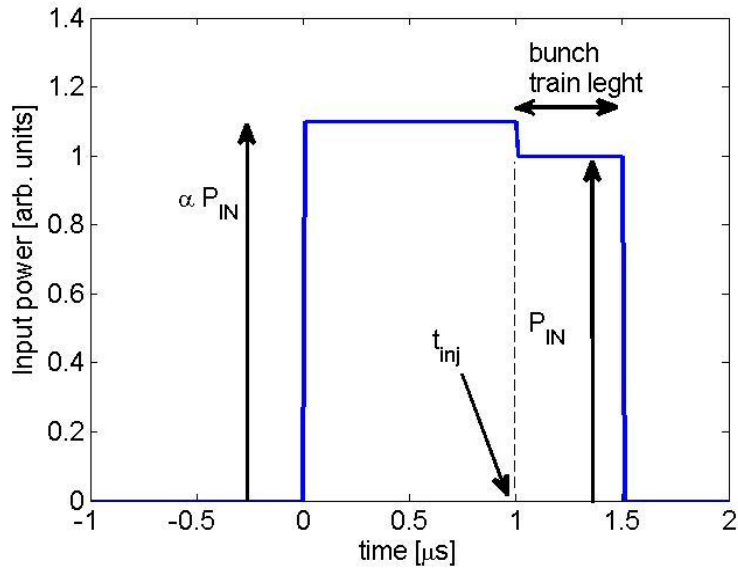

**Fig. 37.    Two step input pulse profile**

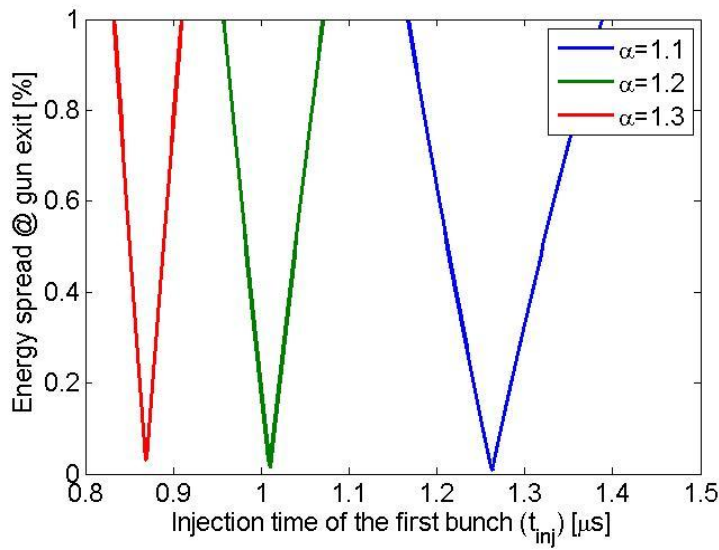

**Fig. 38.    Energy spread of the bunch train at the exit of the gun as a function of the injection time assuming a two steps input power profile with different α coefficients (q=250 pC, 31 bunches and ΔT=16 ns)**

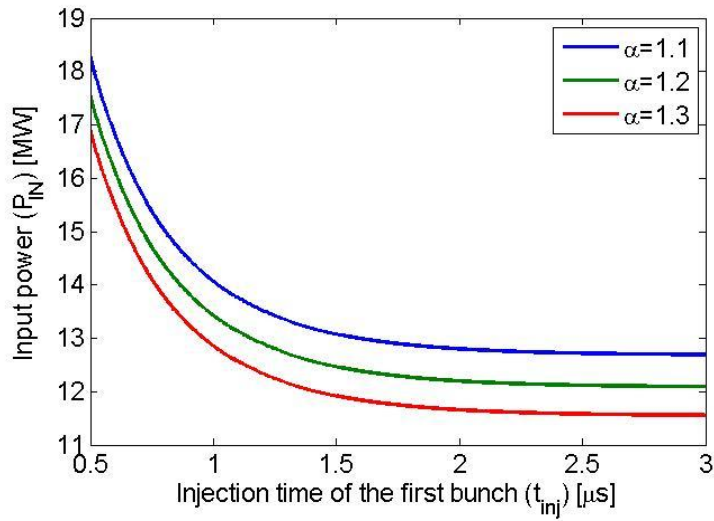

**Fig. 39.** Input power $P_{IN}$ (to have an average energy of the bunch train equal to the nominal energy) as a function of the injection time assuming a two steps input power profile with different α coefficients (q=250 pC, 31 bunches and ΔT=16 ns)

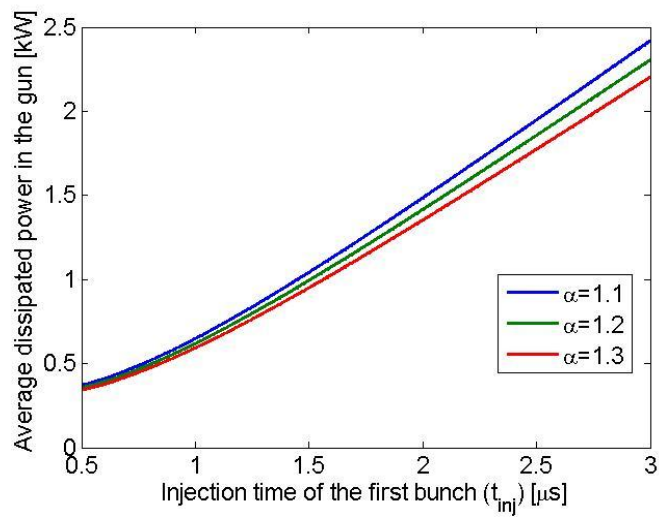

**Fig. 40.** Average dissipated power in the gun as a function of the injection time assuming a two steps input power profile with different α coefficients (q=250 pC, 31 bunches and ΔT=16 ns)

### 2.3.1.2 BL in the TW structures and its compensation

A complete discussion on beam loading in TW structures and possible compensation techniques can be found elsewhere in several published papers [35-37].

When a single bunch passes through a TW structure it absorbs energy from the accelerating field and the field intensity subsequently decreases. The perturbation on the accelerating field travels along the structure from the beginning to the end with a velocity that is essentially equal to the group velocity of the structure and with a phase velocity that is the beam velocity. The phenomenon can be also viewed from another point of view, whereby the beam, whilst interacting with the periodic structure, excites the accelerating mode whose envelope propagates with the group velocity from the beginning to the end of the structure, with a phase velocity equal to the beam velocity. The total accelerating field from subsequent bunches is exactly the superposition of this field and the external RF field. This simplified treatment neglects the dispersive effect related to the fact that the beam excites the field at all frequencies of the pass-band accelerating mode. All field components propagate into the structure with different group and phase velocities. A complete treatment has been done for other structures [35-43].and demonstrated that this simplified approach gives the main contribution to beam loading contributions.

The described mechanism is simplified in Fig. 41 for a constant impedance structure showing that each bunch generates a beam loading field and, if the bunch train is longer than the filling time of the structure, the beam loading field reaches a steady state condition.

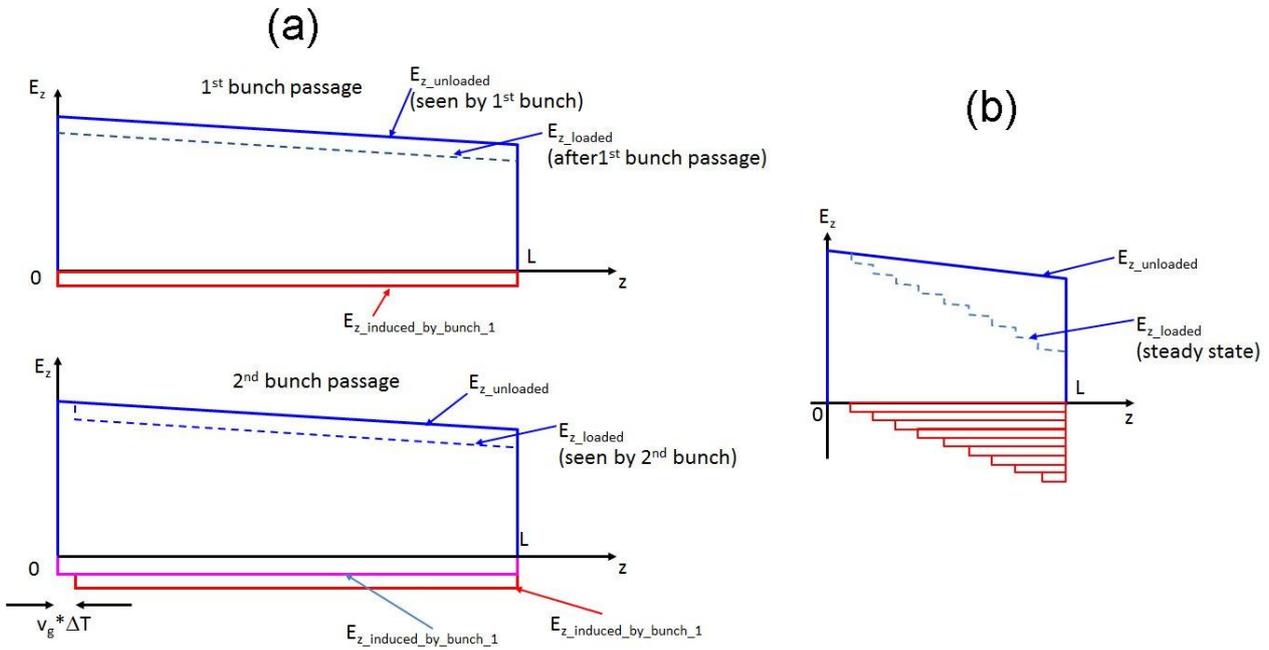

**Fig. 41.** Schematic view of the beam loading process in TW structures

For the ELI linac sections there are two different type of TW cavities proposed: S-Band cavities in the injector which are constant gradient (CG) and C-Band cavities in the linac booster that are quasi-constant gradient (QCG). Moreover, for the S-Band cavities the beam train is shorter than the filling time whilst for the C-Band sections the beam train exceeds the filling time by more than a factor two.

The external voltage given by the RF generator and the beam induced voltage in CG structures can be analytically calculated in the frequency domain [35] and are given by the following expressions:

$$V_{RF}(\omega) = E_{in}(\omega) \frac{\omega_{RF} L}{(1-e^{-2\tau})(j\omega + \omega_{RF}/Q)Q} \left(1 - e^{-2\tau Q/\omega_{RF}(j\omega + \omega_{RF}/Q)}\right) \quad (6)$$

$$V_{BEAM}(\omega) = -\frac{\omega_{RF} i(\omega) r L}{j2\omega Q(j\omega + \omega_{RF}/Q)Q} \left(1 - \frac{\omega_{RF}}{(1-e^{-2\tau})}\right)\left(1 - e^{-2\tau Q/\omega_{RF}(j\omega + \omega_{RF}/Q)}\right) \quad (7)$$

Where *r* is the shunt impedance of the structure per unit length (considered constant along the structure), *Q* is the quality factor per unit length, $\omega_{RF}$ is the angular excitation frequency, τ is the total attenuation of the structure, *L* is the structure length, *i(ω)* is the Fourier transform of the beam current, $E_{in}$ is the Fourier transform of the accelerating field profile at the beginning of the structure, related to the input power by: $E_{in}(\omega) = \sqrt{P_{in}(\omega)2\alpha(0)r}$ with *α(0)* being the attenuation constant at the beginning of the linac structure.

In the same way it is possible to calculate the previous quantities of beam induced voltage in a CI structure by the formulae:

$$V_{RF}(\omega) = E_{in}(\omega) \frac{1 - e^{-(j\omega/v_g + \alpha)L}}{(j\omega/v_g + \alpha)} \quad (8)$$

$$V_{BEAM}(\omega) = -\frac{i(\omega)r\alpha}{j\omega/v_g + \alpha} \left( L + \frac{e^{-(j\omega/v_g + \alpha)L} - 1}{j\omega/v_g + \alpha} \right) \quad (9)$$

In this case the attenuation *α* and the group velocity $v_g$ are constant.

The parameters for the S band and C band structures which prove useful for beam loading calculations are given in Table 10 and Table 11 respectively.

Knowing the beam current profile i(t) and the input power profile P(t), it is possible to calculate the total voltage by anti-transforming the previous quantities. For example, if we consider the rectangular input power profile of Fig. 42 for the S-band case and a uniform beam is injected after one filling time, we obtain the results given in Fig. 43 for the external RF voltage $V_{RF}$(t) and for the beam induced voltage $V_{BEAM}$(t). The superposition of the two effects gives the train energy spread of Fig. 44a (blue curve).

Similarly it is possible to calculate the beam induced energy spread in the C-Band structures. The result is given in Fig. 44b (blue curve).

**Table 10. Parameters of the S-Band TW accelerating structures useful for beam loading calculation**

| Structure type | Constant gradient, TW |
| --- | --- |
| Working frequency (fRF) | 2.856 GHz |
| Structure length | 3 m |
| Nominal RF input power (PIN) | 40 MW |
| Average accelerating (Eacc) | 22 MV/m |
| Quality factor (Q) | 13000 |
| Shunt Impedance per unit length (r) | 55 MΩ/m |
| Filling time (τF) | 850 ns |
| Structure attenuation constant (τ) | ~0.5655 |

**Table 11. Parameters of the C-Band TW accelerating structures useful for beam loading calculation**

| Structure type | Quasi-constant gradient, TW |
|---|---|
| Working frequency ($f_{RF}$) | 5.712 GHz |
| Structure length | 1.8 m |
| Nominal RF input power ($P_{IN}$) | 40 MW |
| Average accelerating ($E_{acc}$) | 33 MV/m |
| Quality factor ($Q_0$) | 8800 |
| Shunt Impedance per unit length | 74.5 MΩ/m |
| Normalized group velocity | 0.021 ($v_g/c$) |
| Filling time ($\tau_F$) | 290 ns |
| Field attenuation constant ($\alpha$) | 0.321/m |

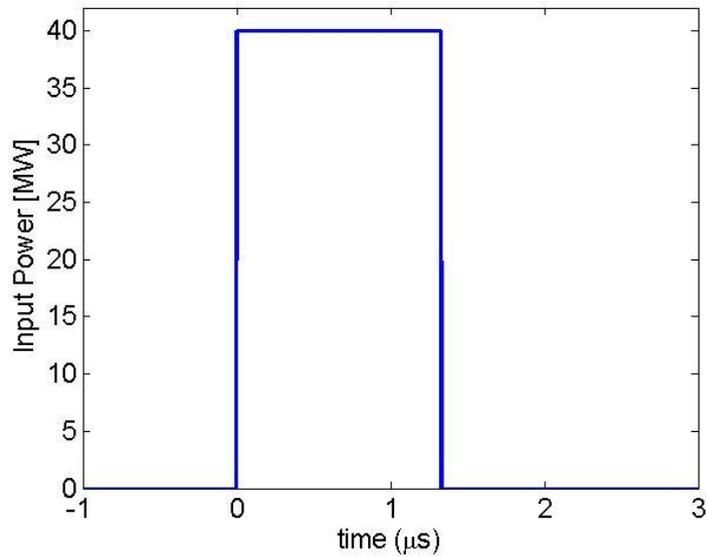

**Fig. 42.** Input power at the entrance of the S-Band structure

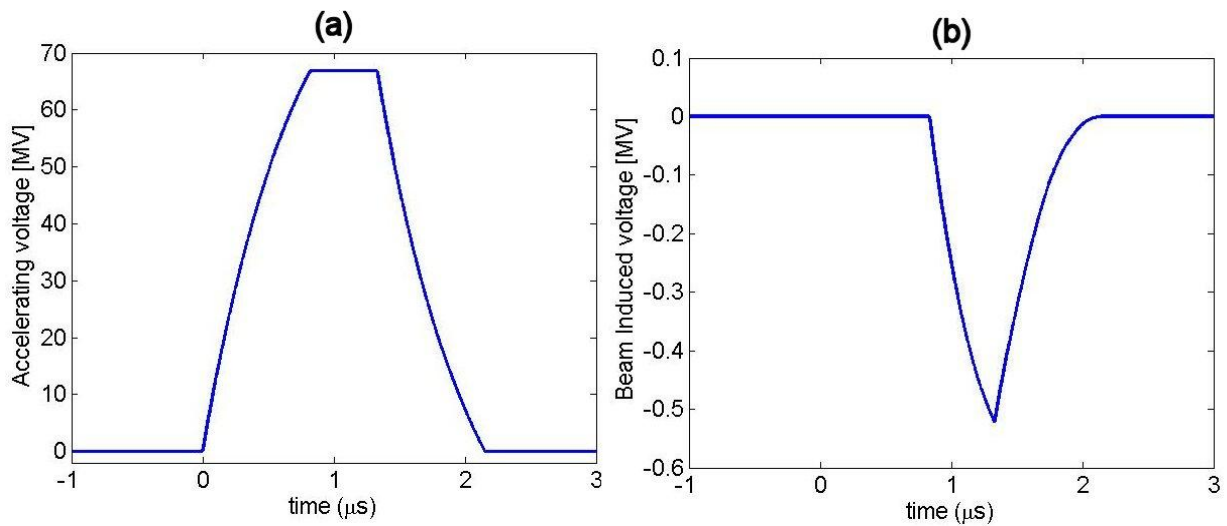

**Fig. 43.** (a) external RF voltage and (b) beam induced voltage in the S-Band TW structure

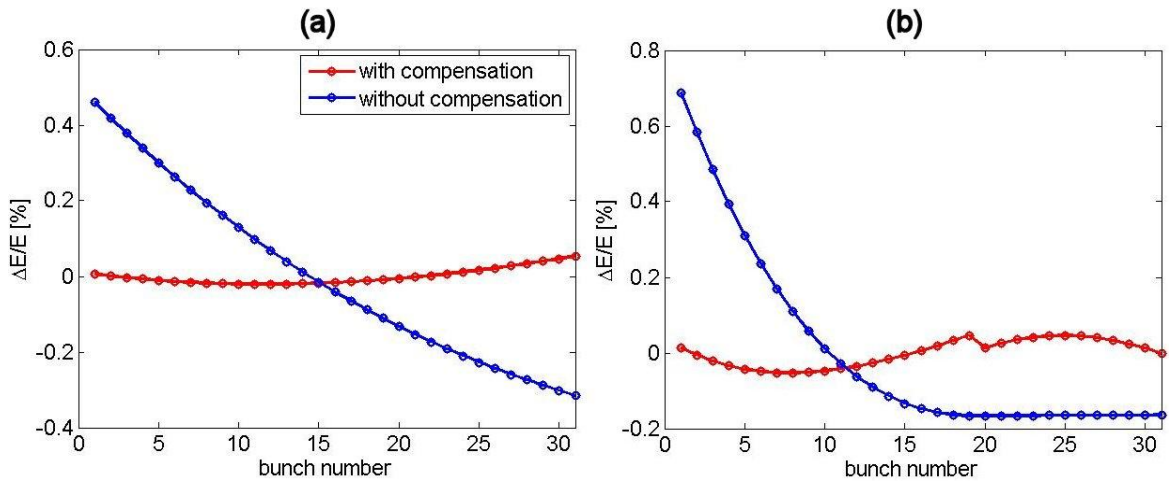

Fig. 44.  Energy spread induced by beam loading in the S-Band (a) and C-Band structures (b) with and without compensation

The simplest way to compensate the beam loading effect is to appropriately shape the RF input power to compensate the energy spread induced in the beam train. One possibility is shown in Fig. 45. After the first bunch is accelerated the input power can be slightly increased to compensate the beam induced voltage. The step in power also partially compensates the beam loading induced by the second bunch since the power discontinuity propagates into the TW structure with a group velocity that is equal to the emptying velocity of the beam induced field.

The input power profiles necessary to compensate the beam loading for the S-Band and C-Band structures are given in Fig. 46. Whilst for the case of the S-band injector sections, the bunch train is shorter than one filling time of the structure, in the C-Band case the bunch train is longer and a doubling in power is necessary to compensate the associated beam loading. This is due to the fact that after one filling cycle the first step in power (necessary to compensate the beam loading after the first bunch passage) comes out from the structure and it is necessary to reintegrate the power for a correct beam loading compensation. The energy spreads at the end of each S-Band or C-Band structure with the compensation of beam loading are given in Fig. 44 (red curves).

In conclusion the beam loading in the TW structures of ELI can be compensated with a simple modulation of the input power. The calculations neglect the dispersion effect of the structure and also the finite bandwidth of the klystron. The compensation technique can, nevertheless be improved by an appropriately shaped low level RF input power pulse with the required energy spread along the bunch train (following the ELI parameter specifications) being maintained. Moreover the proposed beam loading compensation is a "local" compensation that means that the energy spread is compensated in each section.

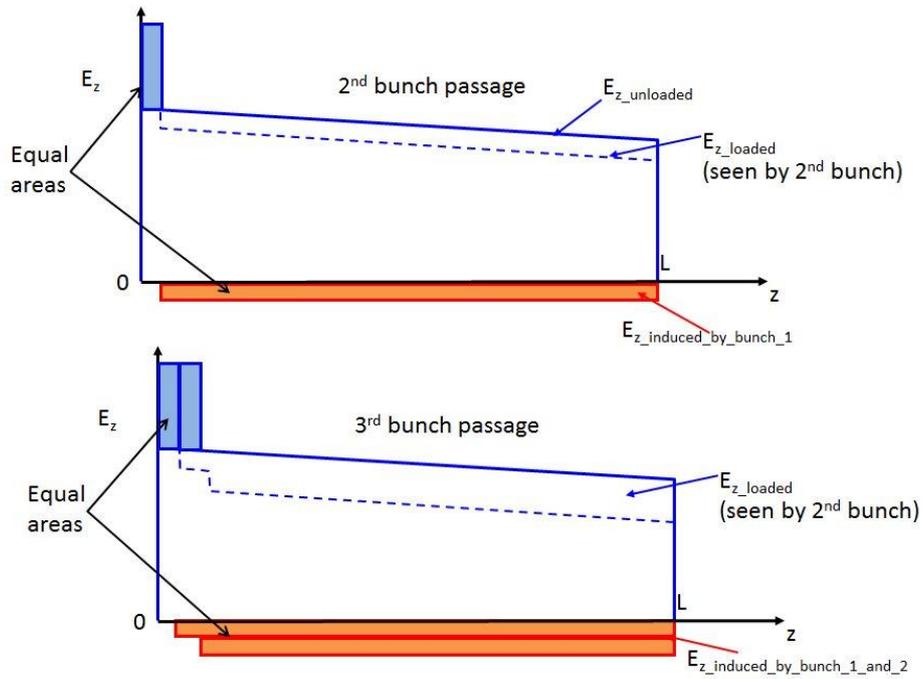

Fig. 45. Simple proposed method to compensate beam loading in TW structures

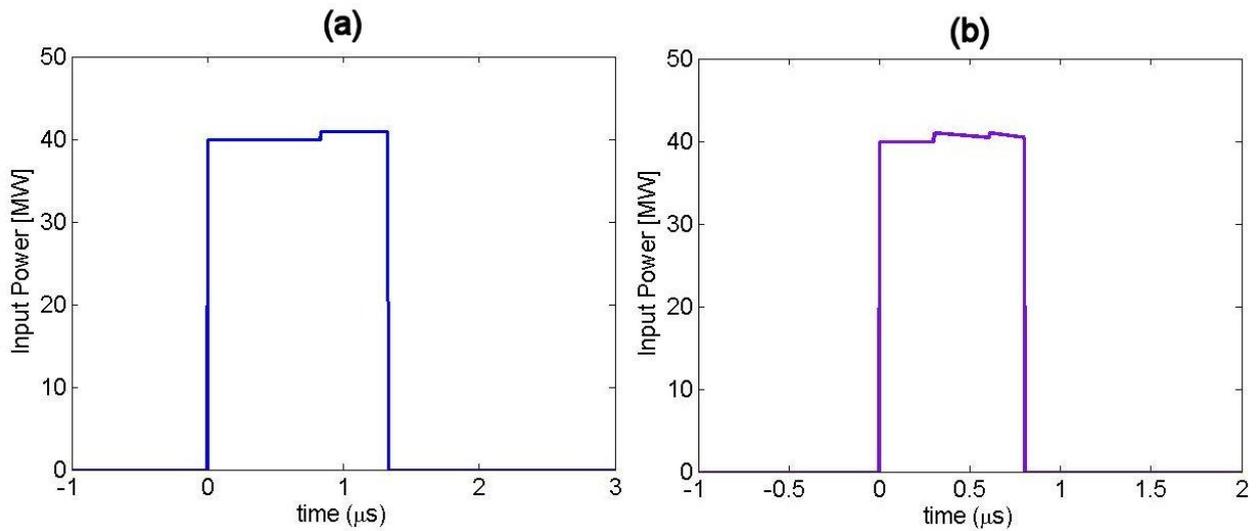

Fig. 46. Input power profiles to compensate the beam loading in the S-Band (a) and C-Band (b) structures

### 2.3.2. Beam Break Up

As pointed out in the previous section, a charged particle beam, travelling across accelerating structures, generates self-induced electromagnetic fields, called wakefields. Of particular concern in a linac accelerating intense bunches, is an instability driven by the transverse wakefields. This instability, excited by off-axis beam trajectories, can develop within a single bunch or along a train of bunches, in which case it is called multibunch beam break up (BBU).

Off-axis beam trajectories arise due to a variety of errors like offset at injection, misalignment of focusing magnets and misalignment of accelerating sections.

As a bunch in a beam pulse is displaced from the axis, transverse deflecting dipole modes are excited. The trailing bunches are then deflected by the wakefield forces whether they are on-axis or not. The angular deflections transform into displacements through the transfer matrices of the focusing system and these displaced bunches will themselves create wakefields in the downstream cavities of the linac. The subsequent bunches will be further deflected leading to a beam blow-up [44].

The beam can excite several dipole modes at different frequencies but, in general, the contribution of the dipole modes at the lower frequency dominates. To describe the effects of such excited modes, one introduces the transverse wake function [44] per unit length and displacement. As example the transverse wake per unit length for the C-Band structures of the SPARC linac [46, 47] obtained by the ABCI code [48] is given in Fig. 47. In the same plot it is reported the transverse wake due to the first dipole mode contribution. From the plot it is easy to note that the contribution of the first dipole dominates the transverse wake field.

For this reason in all our calculation we have considered only the effect of this first dipole.

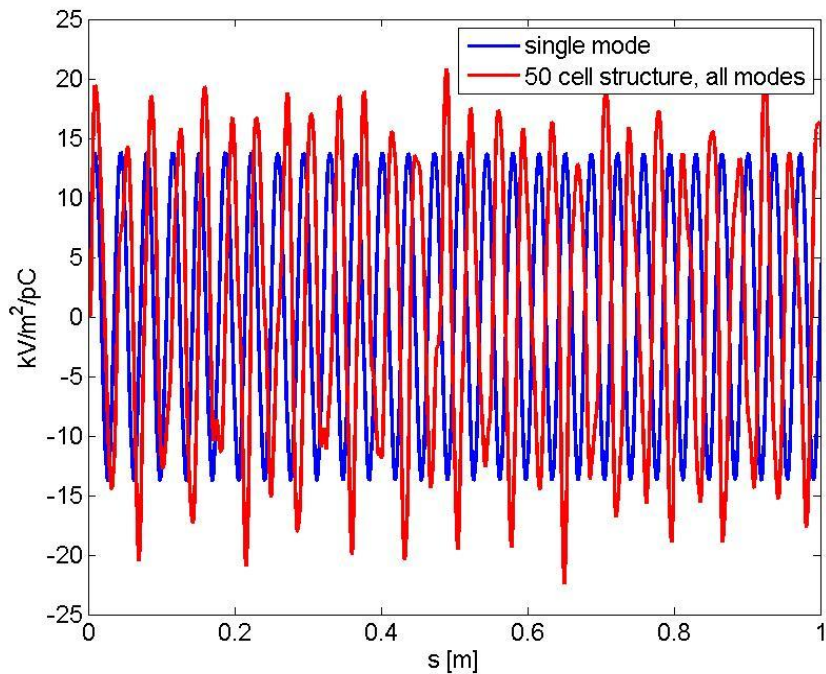

**Fig. 47.** Wakefield calculated by ABCI in the case of SPARC C-Band structure (red curve) and first dipole mode contribution (blue curve)

### 2.3.2.1 BBU study in the C-band LINAC

The analysis of the multibunch BBU in the ELI C-Band booster linac has been performed by using two independent approaches:

1) an analytical study in which the bunches are considered to be rigid macroparticles, like delta-functions, and assuming all bunches injected with the same initial offset $x_0$ [44];

2) a simple tracking code which uses essentially the same assumptions of the analytical approach, but which allow more flexible analysis (contribution of several resonant modes, different initial offsets of the bunches, different charge per bunch, not uniform β-function along the linac…);

In the Mosnier analytical approach, bunches are considered as macroparticles of charge Q without internal structure, spaced by period T (which is an integer number of RF period of the accelerating mode) and injected in the linac with the same initial offset $x_0$.

All the cells of the TW structure are assumed identical and only the first dipole mode is excited by the beam. The transverse optical β-function is considered constant along the linac. This analytical approach allows obtaining the normalized transverse position at the Linac exit for each bunch in the train, i.e.:

$$\frac{x(n) - x(\infty)}{x_0 \sqrt{\frac{\beta_{fin}}{\beta_{in}}} \sqrt{\frac{\gamma_{in}}{\gamma_{fin}}}} \quad (10)$$

With n being the bunch number, x(∞) the steady state solution reached for long (rigorously infinite) trains, $\beta_{fin}$ the β-function at the end of the linac, $\gamma_{in}$, $\gamma_{fin}$ the relativistic factors at the entrance and at the end of the linac. This approach allows also obtaining the so-called asymptotic solution valid for large BBU strength [9].

A dedicated tracking code has been also developed. In this case bunches are modelled as macroparticles, only the fundamental dipole mode is considered and the β-function is considered constant along the linac. Moreover, it is possible to easily extend the code by taking into account the contribution of several HOMs, by considering different initial displacements of the bunches, different resonant frequencies of the HOMs in the different cells of the accelerating structure and a not uniform β-function.

The code allows obtaining the transverse positions and angles of bunches at the exit of the LINAC.

The C-Band structures are constant-impedance structures, which means that all cells have the same dimensions. The parameters of the first dominant dipole mode of a single cell are given in Table 12. In the table we have defined the transverse shunt impedance of the dipole mode $R_T$ as:

$$R_T = \frac{\left| \int_0^{L_c} F_T e^{j\omega_{res} z/c} dz \right|^2}{2P_{diss}} \quad [\Omega] \quad (11)$$

where FT is the Lorentz transverse force, $L_c$ is the single cell length, $P_{diss}$ is the dissipated power in the cell. The transverse shunt impedance defined in (11) is related to the transverse impedance per unit length $R$ (used in the Mosnier approach) by the relation reported in Table 11.

These parameters have been used to evaluate the BBU effects in the linac using both the analytical solution and the tracking code. In the calculations we have considered an injection energy at the entrance of the linac (Einj) equal to 80 MeV, an average β-function of 5 m and an average accelerating gradient of 30 MV/m.

With these parameters, the Mosnier approach gives the normalized displacement at the exit of the linac reported in Fig. 48. In the same plot it has been also reported the solution obtained by the tracking code. The corresponding normalized Courant-Snyder Invariants ($I_n$) of each bunch at the exit of the linac for an initial displacement at the entrance of the linac equal to 500 µm are given in Fig. 49. For each bunch we have defined In as follows:

$$I_n = \frac{\frac{1}{\beta}(x_n - x_1)^2 + \beta(x'_n - x'_1)^2}{\gamma} \quad (12)$$

Where $x_n$, $x'_n$ are the output position and angle of the n-th bunch. This quantity can be directly compared with the nominal normalized emittance of bunches (~0.4 mm mrad). The result clearly shows that the BBU strongly increase the final projected emittance of the train and cannot be tolerated.

**Table 12.** Parameter list used in the BBU estimate of C-Band structures

| | |
|---|---|
| RT/Q | 26 Ω |
| Q | 11000 |
| fres | 8.398 GHz |
| R/Q (=ωres/c*RT/Q/Lc) | 0.26 MΩ/m2 |
| Einj | 80 MeV |
| <Eacc> | 30 MV/m |
| <β> | 5 m |

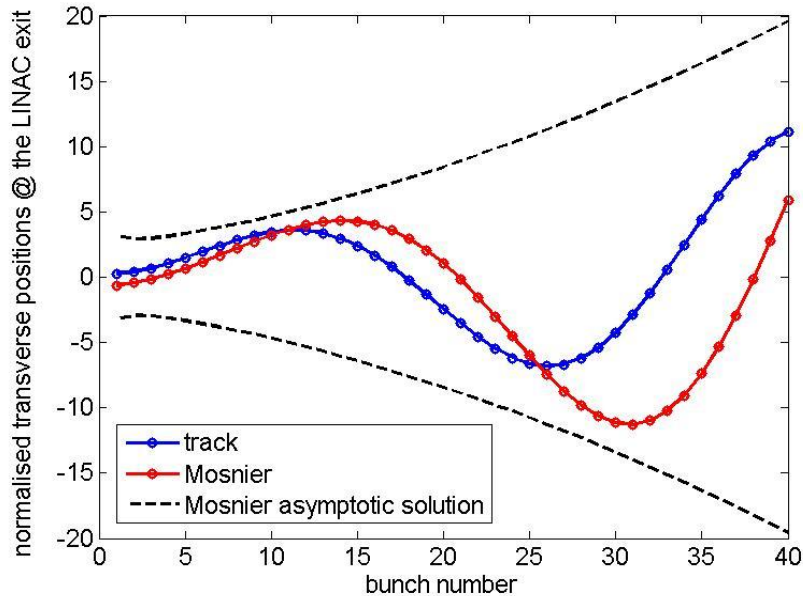

**Fig. 48.** Normalized transverse position at the exit of the linac due to the BBU obtained by the Mosnier analytical approach and by the tracking code

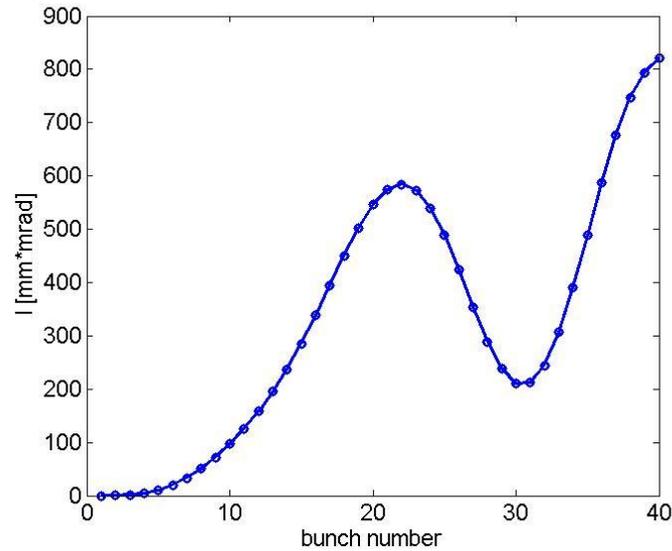

**Fig. 49.    Normalized Courant Snyder Invariant at the exit of the linac for an initial displacement of all bunches of 500 µm**

Several solutions can be adopted to mitigate the BBU, like the damping and/or detuning [49]. For the ELI structures we have proposed to adopt the strong damping solution similar to the X-Band CLIC structures [50]. The design criteria of the C-Band structures are given in section 3. Each cell of the structure has four waveguides that allows the excited HOMs to propagate and dissipate into loads. To estimate the maximum quality factor necessary to avoid the BBU instability we have calculated the maximum In as a function of the first dipole mode quality factor, for different injection errors amplitude and for different average β-functions. Moreover in the calculation we have considered a perfect build up mechanism of all transverse wakes to have a maximum estimation of the BBU effects. The results are given in Fig. 50 and Fig. 51, and clearly show that, with a quality factor below 100, the maximum $I_n$ is below $10^{-2}$ that means that the BBU is completely cancelled. Similar considerations can be done for the other dipole HOM at higher frequencies.

In conclusion, the design of the C-band sections has been addressed to reach an overall quality factor of the dipole modes below 100 to avoid BBU instability. This value has been considered as a design parameter for the C-Band structures.

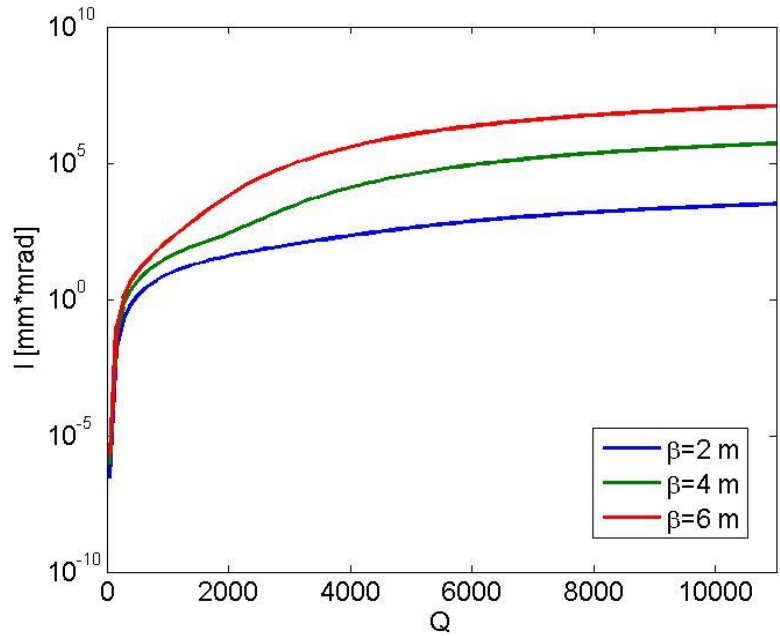

**Fig. 50.** Maximum value of the normalized Courant Snyder Invariants at the end of the C-Band linac as a function of the first dipole mode quality factor and for different β-functions (initial displacement of all bunches: 500 µm)

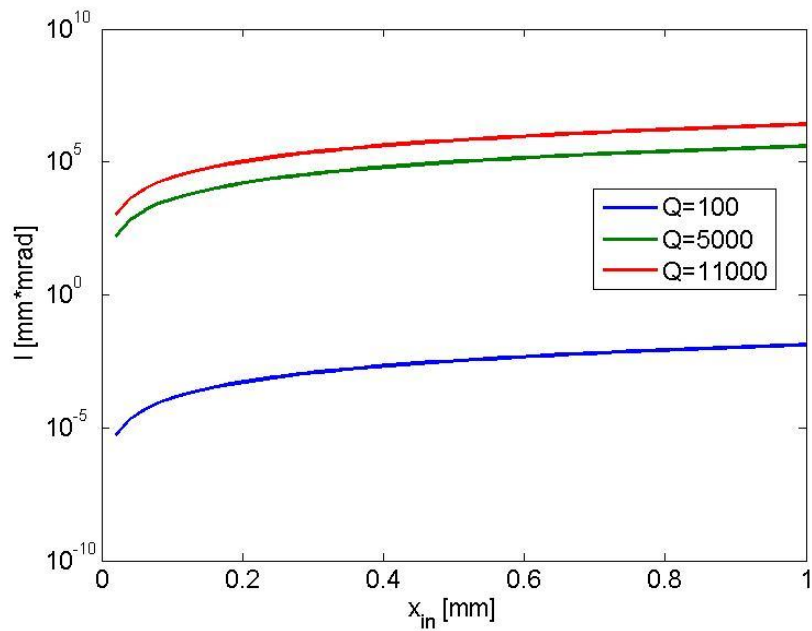

**Fig. 51.** Maximum value of the normalized Courant Snyder Invariants at the end of the C-Band linac as a function of the injection errors and for different Q-factors of the first dipole modes

### 2.3.2.2 BBU Study in the S-Band injector

The accelerating structures we want to use in the injector after the RF gun are two S-Band travelling wave cavities of the SLAC-type. Such cavities are undamped structures and a careful analysis of the possible induced BBU instabilities have to be done. Even if, in fact, the transverse wakefiled per unit length scales with the third power of the working frequency, the beam rigidity in the injector is much lower than in the booster and a transverse voltage induced by the beam can have strong effects in the beam dynamics. We were encouraged in the use of S-Band undamped structures even because experimental measurements of induced BBU in S-Band linacs have been done in other experiments and did not put in evidence strong instabilities effects. The ELI case is, nevertheless, a particular case in which a total very small projected emittance have to be preserved and also small transverse perturbations can results in a decrease of the luminosity at the IP.

The analysis of the multibunch BBU in the ELI S-Band injector has been performed by using two independent approaches for the transverse wakefield calculation:

1) calculating the dipole modes frequencies and transverse impedances with a mode matching technique and reconstructing the total transverse wakefield by a superposition of the different modes;

2) calculating the transverse wakefield of each TW structure by the electromagnetic code GdFidL.

The wakes have been, then, inserted in the tracking code illustrated in the previous paragraph and the Courant Snyder Invariants at the end of the injector have been calculated for all bunches.

The two different approaches for the wake calculation have advantages and disadvantages. The first one is an analytical approximated technique based on the mode matching technique that allows calculating the properties of the single resonant modes while, in principle, GdFidL can give the exact transverse wake of the structure. On the other hand, nevertheless, with the first approach, we can calculate different transverse wakes assuming small errors in the calculated frequencies of the dipole modes and a statistical analysis of BBU can be performed. The errors in the frequencies calculation are due to the approximations in the wake calculation itself.

The transverse wakefield calculated with the analytical approach is given in Fig. 52. The corresponding CS Invariants of all bunches at the end of the injector assuming an initial energy (after the gun) of 5 MeV and an energy at the end of the S-Band sections of 80 MeV, are given on Fig. 53 assuming an initial error for all bunches at injection of 200 μm and an average β-function of 3 m.

The maximum CS Invariant along the train has been also calculated considering errors in the frequencies of the dipole modes. The distribution of such maximum CS invariants is given in Fig. 54 (a) considering random errors of ±0.5% in the frequencies and calculating the distribution over 5000 cases. The corresponding probability function is given in Fig. 54 (b). From this last plot it is possible to conclude that the probability to obtain an increase in the projected emittance of 10% for the worst bunch of the train is below 10%.

The transverse wakefield calculated with the simulation code GdFidL is given in Fig. 55. The CS Invariants along the train obtained by using the tracking code are given in Fig. 56 (assuming an initial error for all bunches at injection of 200 μm and an average β-function of 3 m). In this case the number of bunches has been fixed to 32 because the wake has been calculated (for computing time limitations) up to 150 m.

Also this type of analysis gives results that show a negligible emittance increase.

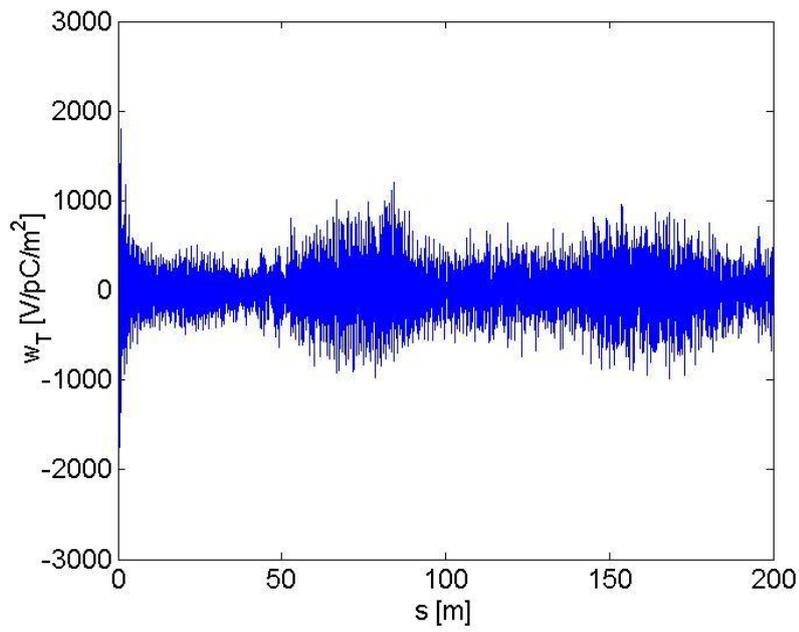

**Fig. 52.    Transverse wakefield calculated using the mode matching technique**

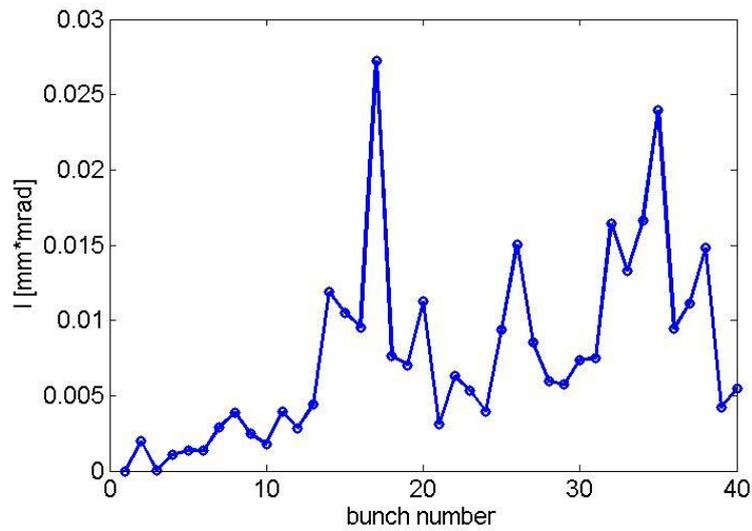

**Fig. 53.    Normalized Courant Snyder Invariant at the exit of the injector for an initial displacement of all bunches of 200 µm and using the transverse wake given by the mode matching technique**

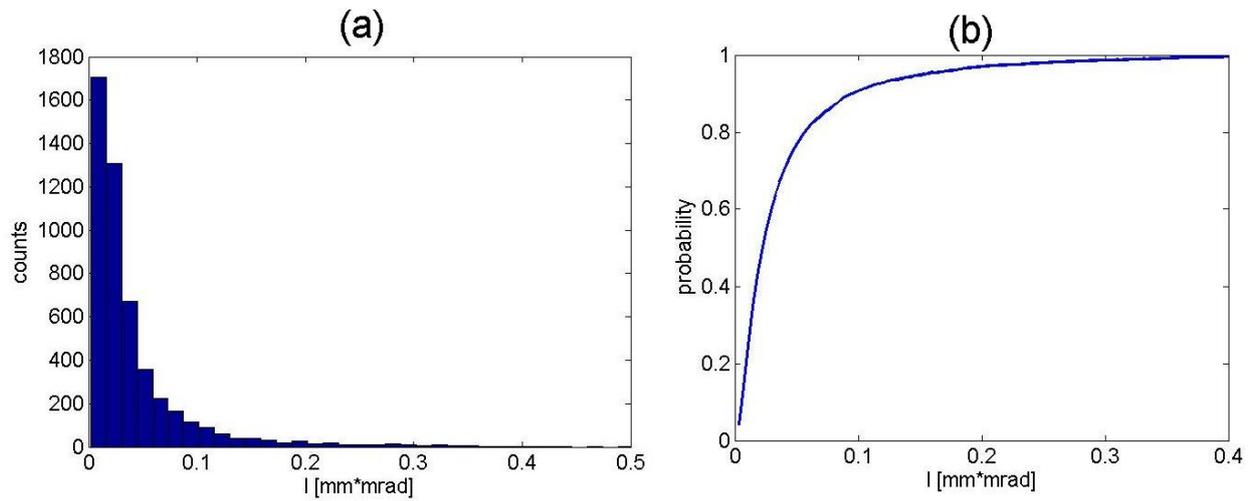

Fig. 54. Distribution of the maximum normalized Courant Snyder Invariants at the exit of the injector for an initial displacement of all bunches of 200 μm and using the transverse wake given by the mode matching technique. The distribution has been calculated considering 5000 different cases with random variation of the resonant frequencoies of dipole modes of ±0.5% with respect to the theoretical ones

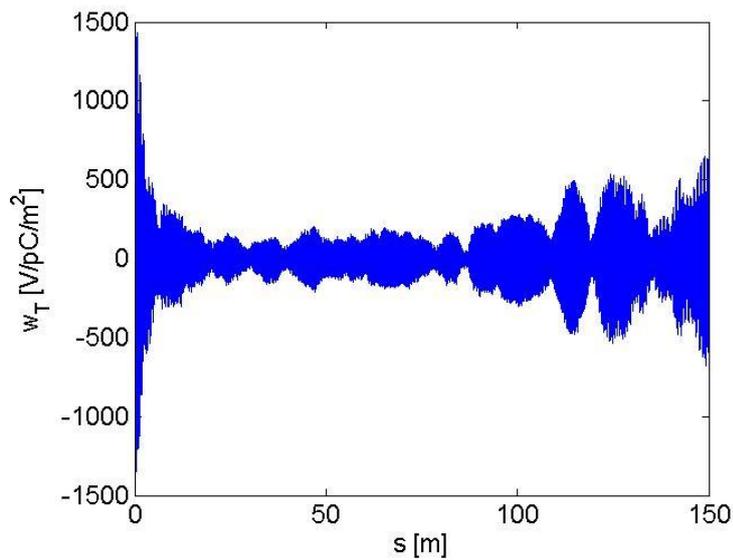

Fig. 55. Transverse wakefield obtained using GdFidL

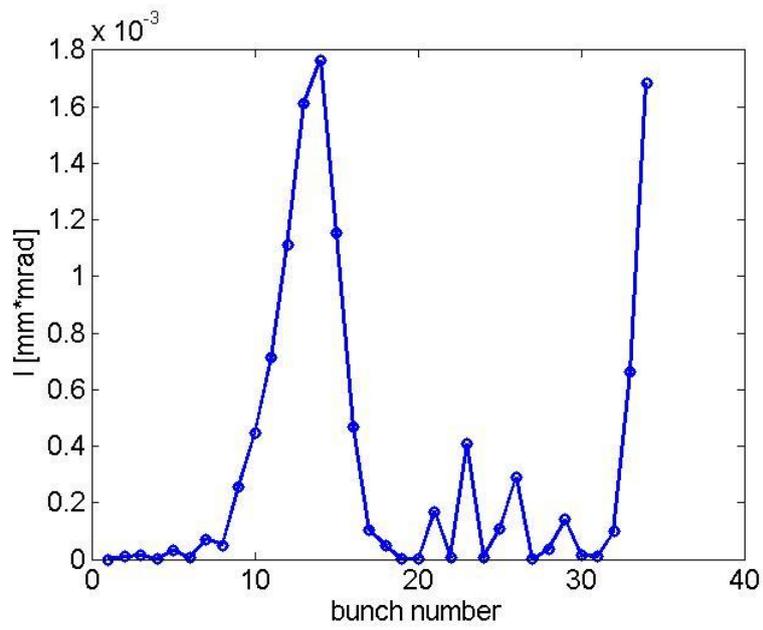

**Fig. 56.** Normalized Courant Snyder Invariant at the exit of the injector for an initial displacement of all bunches of 200 µm and using the transverse wake given by GdfidL

# 3. Accelerator System

## 3.1. S-Band Accelerating structures

The electron beam is generated in the laser-driven photocathode 1.6-cell standing-wave (SW) RF cavity and then accelerated to about 80 MeV (in the velocity bunching configuration) by two SLAC type [51] travelling wave (TW) structures. The structures are 3 m long, are Constant Gradient (CG) units and operate in the $2\pi/3$ mode at 2.856 GHz. They are made up of a series of 86 RF copper cells, joint with a brazing process performed in high temperature, under vacuum furnaces. The cells are coupled by means of on-axis circular irises with decreasing diameter, from input-to-output, to achieve the constant-gradient feature. The RF power is transferred to the accelerating section through a rectangular slot coupled to the first cell. The power not dissipated in the structure (about 1/3rd) is coupled-out from the last RF cell and dissipated on external load.

To meet the severe emittance requirements for the injector, the single-feed couplers of the pre-injector accelerating structures will be replaced by a dual-feed design to minimize the multipole field effects generated by the asymmetric feeding, which induces transverse kicks along the bunch, causing beam emittance degradation.

The industrial companies, which can develop the accelerating structures, are only a few in the world. The fabrication is a complex task that requires specialized know-how, availability of very advanced equipment and facilities, top-level organization.

The maximum achievable accelerating gradient is the most important parameters of such devices. The will operates at an average gradient of 22 MV/m. Nevertheless, it requires the use of selected materials, precise machining, high-quality brazing process, surface treatments and cleaning, ultra-pure water rinsing, careful vacuum and RF low power tests. The Fig. 57 shows an S-band accelerating structures installed in the Frascati SPARC injector that will be very similar to the ELI one. The Table 13 gives the main parameters of the sections.

In order to maintain the structure tuned to the $2\pi/3$ mode, that guarantees the necessary cumulative energy gain for the beam particles, the accelerating sections are kept at very constant temperature ($\Delta T = \pm 0.1°C$) by means of regulated cooling water systems.

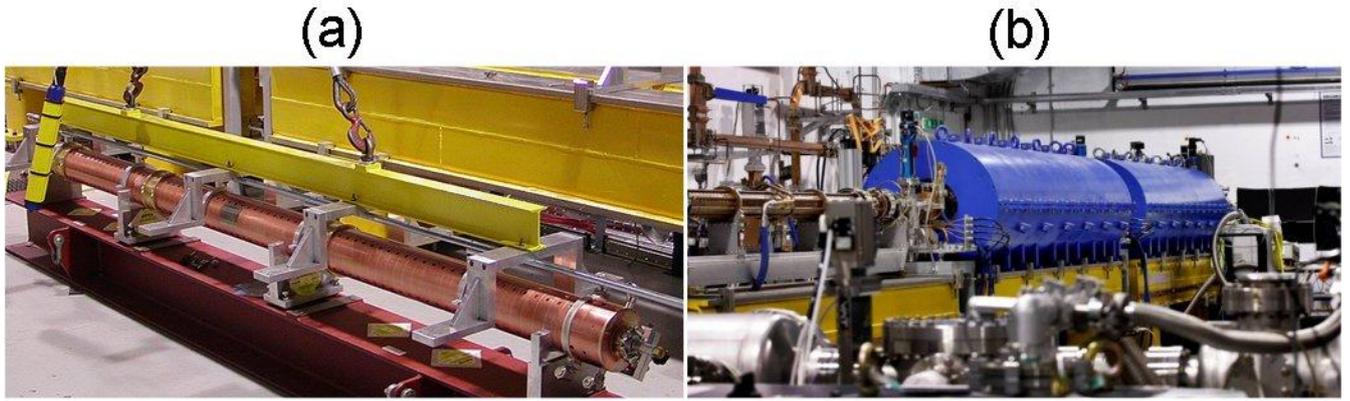

**Fig. 57.** S-band accelerating structure: (a) S-band accelerating structures during installation in the Frascati SPARC injector; (b) structures installed in SPARC with solenoids around

**Table 13.** Technical specifications of the S-Band accelerating sections

| Structure type | Constant gradient, TW |
|---|---|
| Working frequency ($f_{RF}$) | 2.856 [GHz] |
| Number of cells | 86 |
| Structure length | 3 m |
| Working mode | $TM_{01}$-like |
| Phase advance between cells | $2\pi/3$ |
| Max RF input power ($P_{IN}$) | 40 MW |
| Average accelerating ($E_{acc}$) | 20-22 MV/m |
| Quality factor ($Q_0$) | ~13000 |
| Shunt Impedance per unit length | 53-60 MΩ/m |
| Phase velocity | c |
| Normalized group velocity | 0.0202-0.0065 ($v_g/c$) |
| Filling time ($\tau_F$) | ~850 ns |
| Structure attenuation constant | 0.57 neper |
| Operating vacuum pressure (typical) | $10^{-8}$-$10^{-9}$ mbar |
| Max RF input pulse length ($\tau_{IMP}$) | 1.5 μs |
| Pulse duration for beam acceleration ($\tau_{BEAM}$) | <600 ns |
| Rep. Rate ($f_{rep}$) | 100 Hz |
| Average dissipated power | ~3.5 kW |

## 3.2. Photo-injector

### 3.2.1. RF-Gun

The RF GUN of the ELI LINAC will be a 1.6 cell gun of the BNL/SLAC/UCLA type [53, 54]. With respect to the original design of such type of GUN, the ELI gun will implement several features recently integrated in the new gun recently developed for the SPARC photo-injector. This gun has the following improvements:

1) the iris profile has an elliptical shape and a larger aperture to simultaneously reduce the peak surface electric field, to increase the frequency separation between the two RF gun modes (the working π mode and the so called 0-mode) and to increase the pumping speed on the half-cell;

2) the tuners realized by an hole in the full cell and a movable metallic cylinder have been substituted by two deformation tuners on the outer wall of the full cell;

3) the coupling window between the rectangular waveguide and the full cell have been strongly rounded to reduce the peak surface electric field and, therefore, the pulsed heating;

4) the cooling pipes have been improved and increased in number to guarantee a better gun temperature uniformity;

5) the structure have been realized without brazing but using special gaskets in order to maintain hard copper (instead of soft copper resulting from brazing process at high temperature) to reach higher accelerating field with lower BDR.

The structure simulated by HFSS [55] with main dimensions is given in Fig. 58. The mechanical drawing of the gun is given in Fig. 59.

The final main gun parameters are reported in Table 14. In particular the coupling coefficient has been chosen equal to 3 according to the considerations of beam loading compensation discussed in section 2.

A detailed thermal analysis has been performed to investigate the impact of the 100 Hz operation. According to the considerations reported in the paragraph on beam loading compensation, the average dissipated power in the gun, is, in the worst case, 2 kW but can be reduced below 1 kW with a dedicated beam loading compensation. Thermal simulations has been performed using ANSYS [55] assuming an average dissipated power of 1 and 2 kW and the same cooling pipes distribution of the new SPARC gun. The results are given in Fig. 60, where the temperature distribution on the gun is plotted assuming a cooling water temperature of 32 deg. From the plot it is possible to note that (except two hot spots in the coupler window) the temperature of the gun resonant cells is uniform within less than 3 deg in the 1 kW case and less than 5 deg in the 2 kW case. A dedicated cooling of the cathode have to be implemented to reduce its temperature and several solutions can be adopted. The corresponding deformations have been also calculated and in the resonant cells they are within 30 µm in the worst 2 kW case. Other cooling pipes can be easily added to increase the temperature uniformity of the gun and to reduce the hot spot in the coupler region.

**Table 14. RF gun parameters**

| Parameters | Value |
|---|---|
| $f_{res}$ | 2.856 GHz |
| $Q_0$ | 15000 |
| $E_{surf\_peak}/E_{cathode}$ | 0.85 |
| Coupling $\beta$ | 3 |
| Shunt impedance (R) | 1.78 M$\Omega$ |
| Filling time $\tau_F$ | 550 ns |
| Frequency separation between the 0 and the $\pi$-mode | 38 MHz |
| Pulsed heating (2 $\mu$s RF pulse length) | 60 deg |

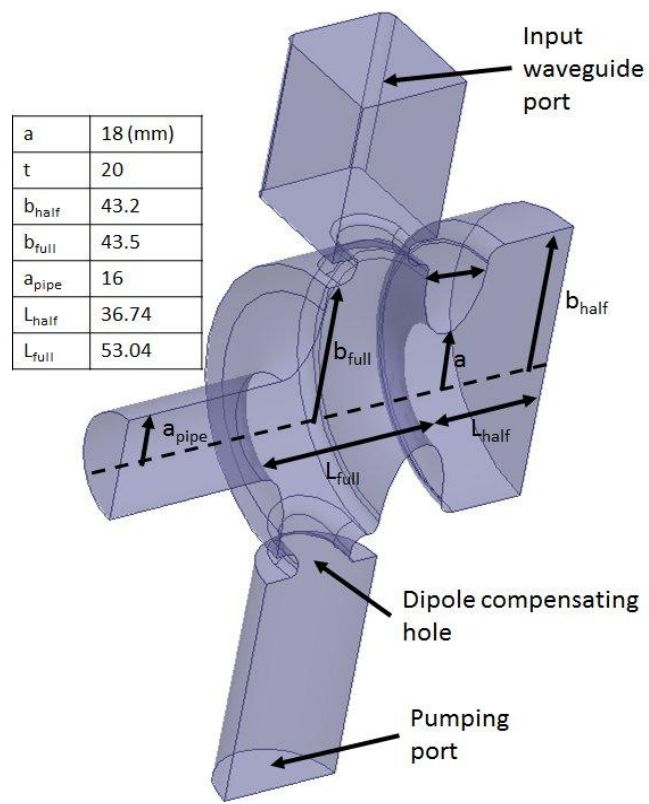

| a | 18 (mm) |
|---|---|
| t | 20 |
| $b_{half}$ | 43.2 |
| $b_{full}$ | 43.5 |
| $a_{pipe}$ | 16 |
| $L_{half}$ | 36.74 |
| $L_{full}$ | 53.04 |

**Fig. 58.** HFSS simulated structure with main dimensions

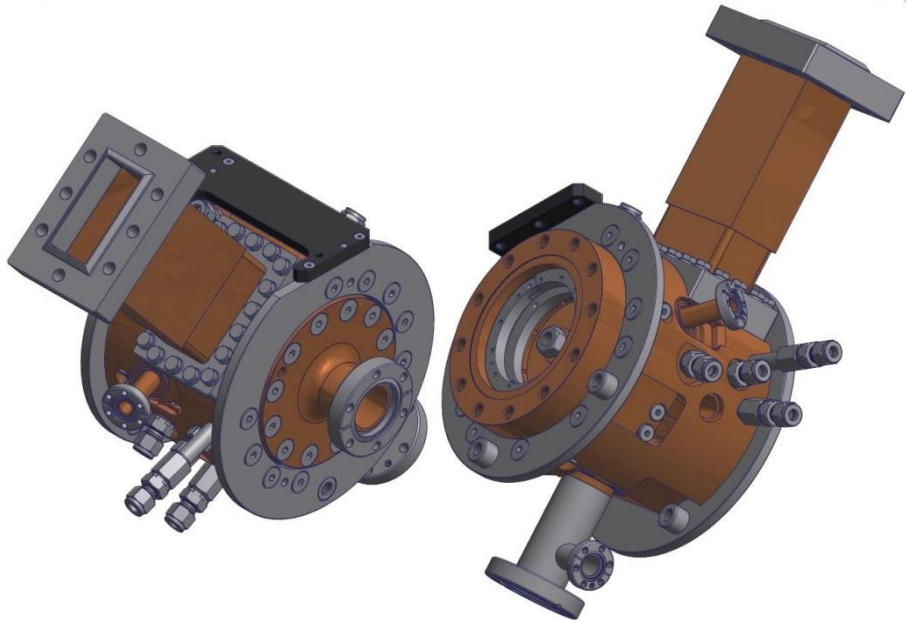

**Fig. 59.** Mechanical drawing of the gun

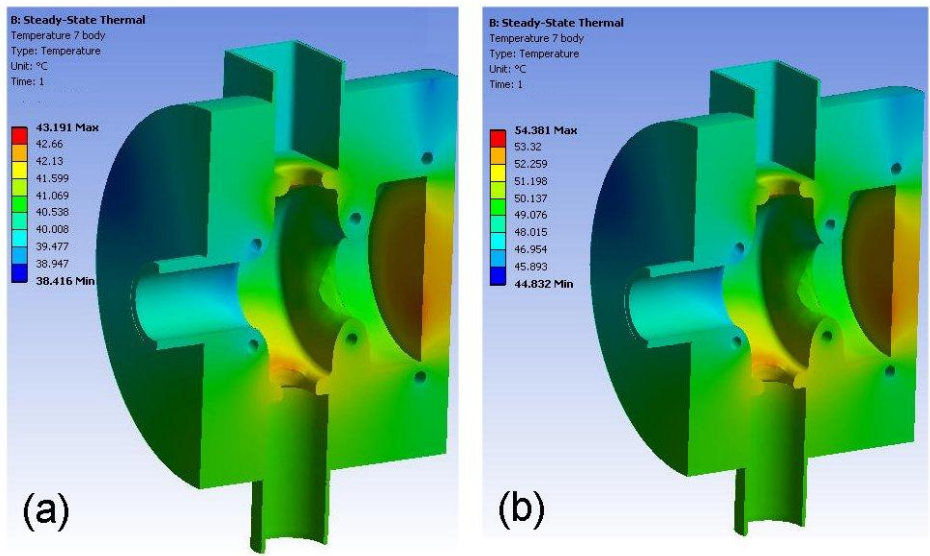

**Fig. 60.** Temperature distribution on the gun assuming a cooling water temperature of 32 deg: (a) 1 kW average dissipated power; (b) 2 kW average dissipated power

The RF-gun, Fig. 61, represents probably the most delicate part of the whole system. In fact this structure should operate with very high electric fields gradients (about 120 MV/m) and high repetition rates (up to 100 Hz) thus, in order to decrease as much as possible the probability of discharge events, the vacuum of this device has to be kept in the range of few $10^{-9}$ mbar during operation. Such low pressure is also required in order to keep the cathode surface free from contaminants, which have been demonstrated to alter the photoemission properties of the cathode.

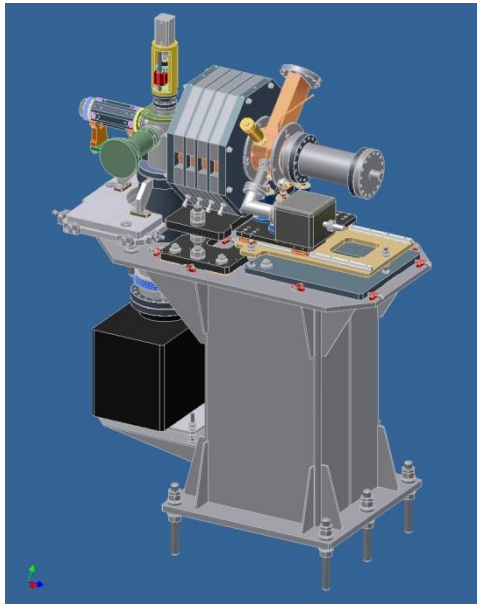

**Fig. 61.     ELI-NP RF Gun**

On the RF Gun two vacuum pumps will be installed: one 50 l/s sputter ion pump will be connected to the RF coupler and one on the compensating port.

Moreover, pumping stations with a 100 l/s sputter ion pump, to pump down the vacuum chambers equipped with diagnostic tools for the characterization of electron beam, are foreseen just before RF accelerating sections and between them.

The gun assembly (RF-gun, waveguide, diagnostic chambers and ion pumps) must be completed with a suitable backing system in order to perform an *in situ* degassing with the aim to remove all the residual water vapors that otherwise cannot be sufficiently eliminated avoiding the reaching of the requested vacuum level.

### 3.2.2.     RF Gun Solenoid

The Gun solenoid, together with the Gun design, defines the emittance compensation process and the final emittance performance at the linac exit. The design consists in four coils, embedded and separated by iron armature, that can be powered independently. In this way it is possible to shape the field profile and moves the field peak around the central position. This solenoid design, as tested at Sparc_Lab, makes it possible to power the coils with alternate signs (e.g. +-+- or ++--), that gives a better compensation for alignment errors and multipolar components. In Fig. 62 are shown the solenoid before the installation and the magnetic fields as computed by Poisson-Superfish FEM code.

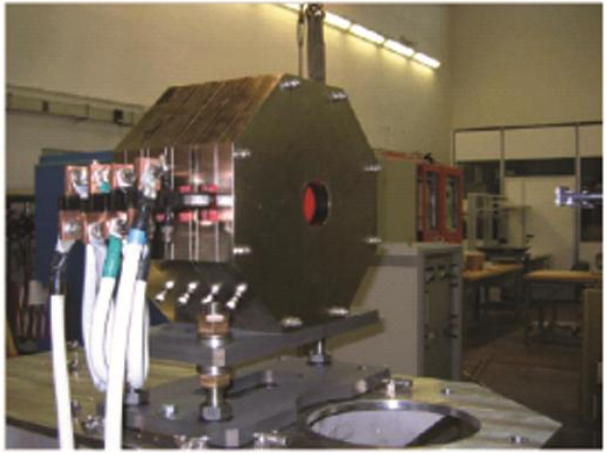 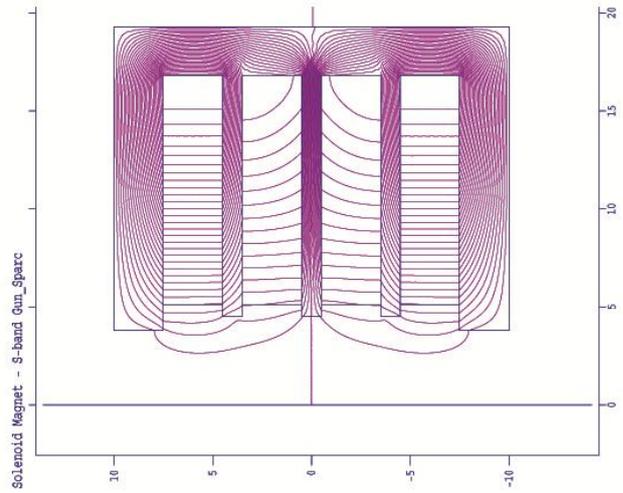

**Fig. 62.** On left the Sparc Gun Solenoid, on right the field distribution, in Sparc configuration ++--, as computed by Poisson-Superfish code

## 3.3. Linac

The ELI-NP Injector, whose layout is shown in Fig. 63, has a vacuum system that is quite simple. The complex is made of one RF Gun with a photocathode, two accelerating sections, a system of RF waveguides and three klystrons.

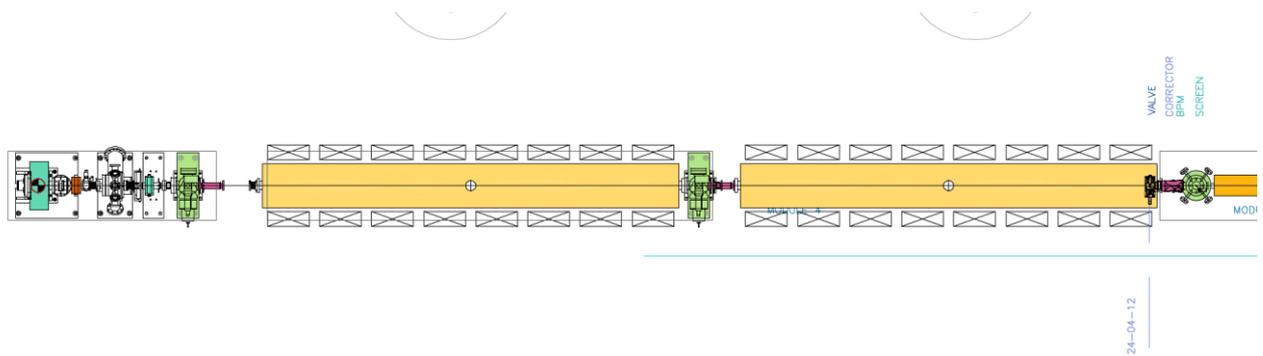

**Fig. 63.** ELI-NP Injector Layout

The performance of the injector depends strongly on the vacuum pressure. Special care must be adopted for the design of the RF Gun vacuum system, because of the very high pollution sensitivity of the photo cathode.

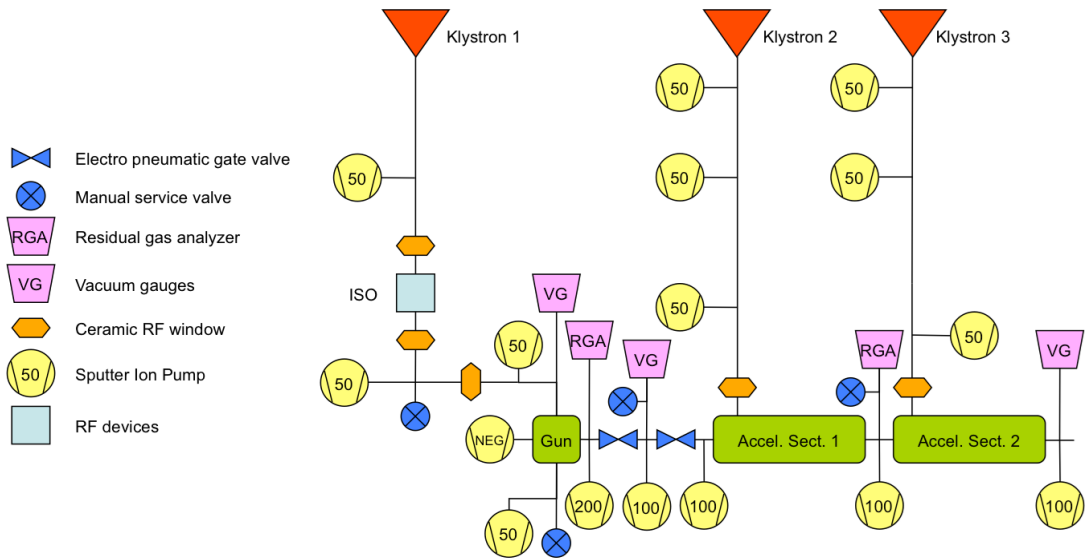

**Fig. 64.**   **ELI-NP Injector Vacuum Layout**

The vacuum system of the ELI-NP Injector, shown in Fig. 64, can be divided into three subsystems that differ mainly on vacuum requirements. Based on our past experience, the vacuum levels requested for the operation of the ELI-NP Injector will be the following:

**Table 15.   Vacuum levels of the ELI-NP Injector**

| RF-gun | ~$10^{-10}$ mbar |
|---|---|
| RF accelerating section | ~$10^{-9}$ mbar |
| RF Waveguide | ~$10^{-8}$ mbar |

### 3.3.1.    Vacuum chambers and beam pipe

The vacuum chamber of the transfer lines sections that are foresee in between each accelerating structures will be made using stainless steel pipe, Fig. 65. Only sputter ion pumps will be adopted.

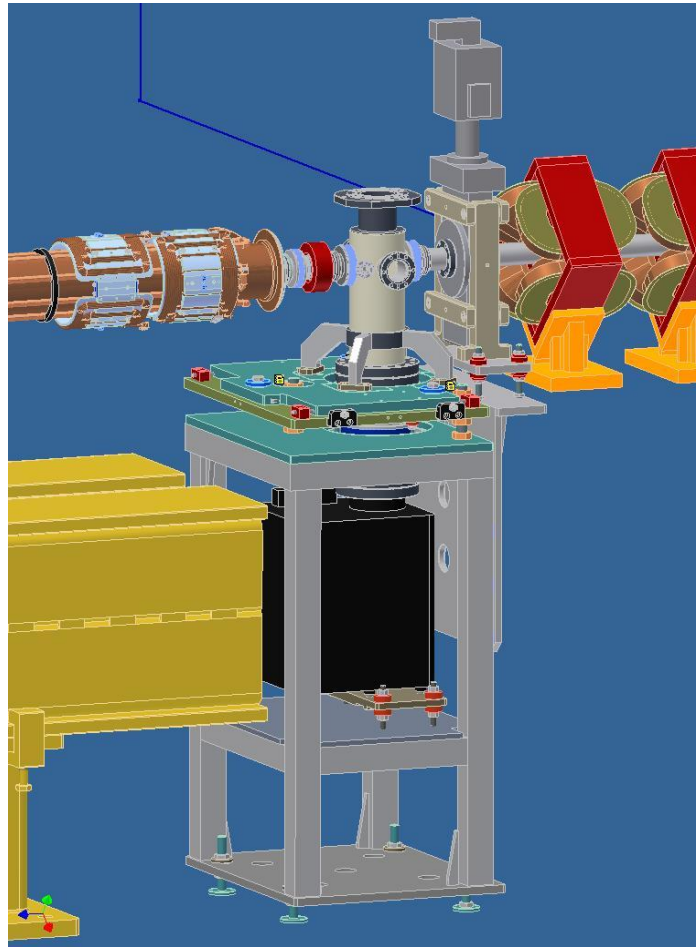

**Fig. 65.     Injector standard pumping station**

### 3.3.2.    RF Waveguide

The radio frequency, needed to drive each accelerating section, flows trough a complex system of copper wave guides operating under vacuum at a pressure of about $10^{-8}$ mbar. The vacuum is obtained by means of a certain number of 50 l/s sputter ion pumps placed along the wave guides every about 10 meters. The RF connecting flanges will be of LIL type.

### 3.3.3.    Vacuum diagnostics and devices

Standard commercial vacuum diagnostics can be adopted on the ELI-NP Injector. Vacuum gauges must be UHV compatible and must be able to read from atmospheric pressure down to $10^{-11}$ mbar. Moreover, in order not to disturb some beam diagnostic devices, UHV gauges must be of cold cathode type. Few residual gas analyzers are foreseen to monitor the residual gas composition near some critical points of the Injector, the RF Gun for example.

A suitable number of vacuum gate valves are foreseen to separate the gun from the RF accelerating sections and the injector itself from the LINAC.

### 3.3.4. Accelerating structures

**3.3.4.1 C-Band accelerating structures for the booster**

The linac booster is composed of 12 TW disk loaded structures working in C-Band (at 5.712 GHz). Each structure is 1.8 m long and the field phase advance per cell is 2π/3. The design of the structure follows the criteria we have adopted for the SPARC C-band sections [46]. In particular, the dimensions of each cell have been optimized to simultaneously obtain:

a) the lowest peak surface electric field on the irises;

b) an average accelerating field of 33 MV/m with an available power from the klystron of 40 MW (conservative assumption);

c) the largest iris aperture compatible with the previous points in order to better pump the structure, in order to reduce both the wakefields contribution and the filling time of the structure itself. This last point is also related to the needed input pulse length to fill the structure and, therefore, to the breakdown rate probability (at least from X Band) in the sense that a reduction of the pulse length strongly reduces the breakdown rate.

Since ELI operation is multi-bunch, the structures have been designed with an effective damping of the HOM dipoles modes in order to avoid beam break-up effects, as widely illustrated in section 2. Several possible schemes for dipole modes damping can be found in the literature [3]. The solution that we have adopted for the ELI structures is based on a waveguide damping system and is very similar to the design adopted for the CLIC structures at CERN [50]. Each cell of the structure has four waveguides that allows the excited HOMs to propagate and dissipate into loads. To simplify the fabrication, these structures have constant impedance that means all irises have equal dimensions. For this reason, the accelerating field has a maximum at the beginning of the section and a minimum at the end due to the dissipation factor along the structure. A possible design with quasi-constant gradient is also possible with minor modifications of the geometry. In this case the irises are modulated along the structure to partially compensate the field attenuation.

**Table 16. Main parameters of the C-Band TW accelerating structures**

| Structure type | Constant impedance, TW |
|---|---|
| Working frequency ($f_{RF}$) | 5.712 [GHz] |
| Number of cells | 102 |
| Structure length | 1.8 m |
| Working mode | $TM_{01}$-like |
| Phase advance between cells | c |
| Nominal RF input power ($P_{IN}$) | 40 MW |
| Average accelerating ($E_{acc}$) | 33 MV/m |
| Quality factor ($Q_0$) | 8800 |
| Shunt Impedance per unit length | 67-73 MΩ/m |
| Phase velocity | c |
| Normalized group velocity | 0.025-0.014 ($v_g/c$) |
| Filling time (τ) | 310 ns |
| Structure attenuation constant | 0.58 neper |
| Operating vacuum pressure (typical) | $10^{-8}$-$10^{-9}$ mbar |
| Max RF input pulse length ($\tau_{IMP}$) | 0.8 μs |
| Pulse duration for beam acceleration ($\tau_{BEAM}$) | <500 ns |
| Iris half aperture (a) | 6.8-5.8 mm |
| Rep. Rate ($f_{rep}$) | 100 Hz |
| Average dissipated power | 2.3 kW |

The input and output couplers have a symmetric feeding. In the input coupler the power is divided in two by an external splitter as shown in Fig. 66. Round edges allow reducing the pulsed heating of the surfaces. The damping system and interface geometries have been designed using HFSS [53].

The mechanical drawings of the structure are shown in Fig. 66. Further details of the single cell are highlighted in Fig. 67. Each cell has four tuners and eight cooling pipes to sustain the 100 Hz operation. Special silicon-carbide (SiC) absorbing tiles, which have been optimized to avoid reflections, are integrated into the structure.

The final structure parameters are given in Table 16. To verify the performance of the damping system, GdFidL [52] simulations have been performed. The result of the transverse wake simulation is given in Fig. 67 and shows a reduction of the quality factor of the dipole mode below 100, which is within the specification of the BBU instability study.

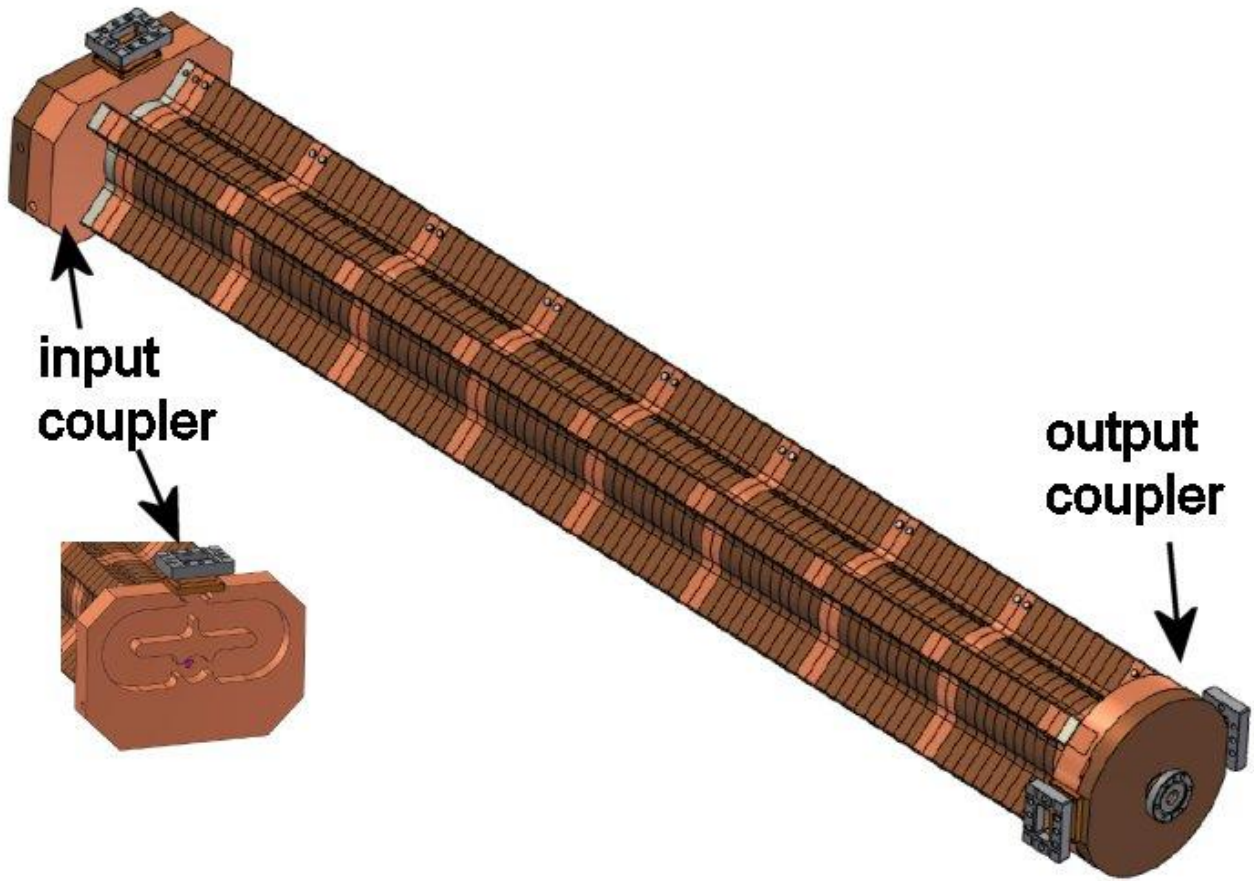

**Fig. 66.	Mechanical drawings of the C-band structure with details of couplers**

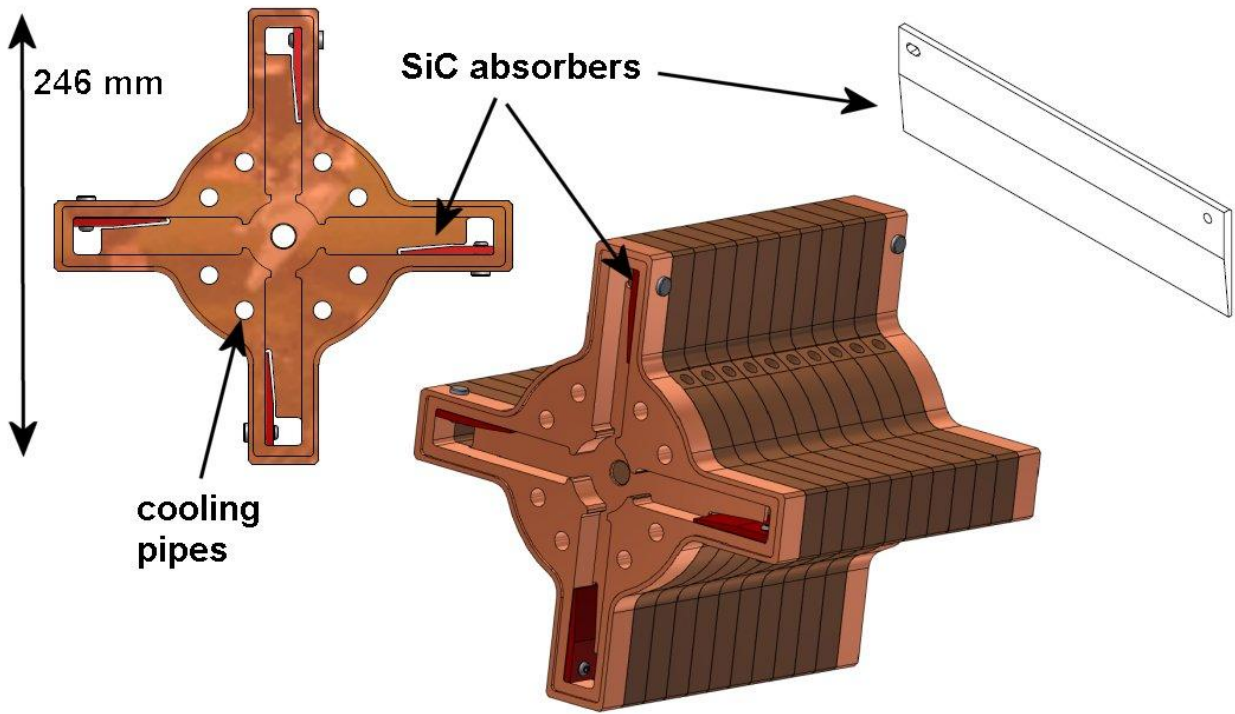

**Fig. 67.	Detail of the TW cell with absorbers**

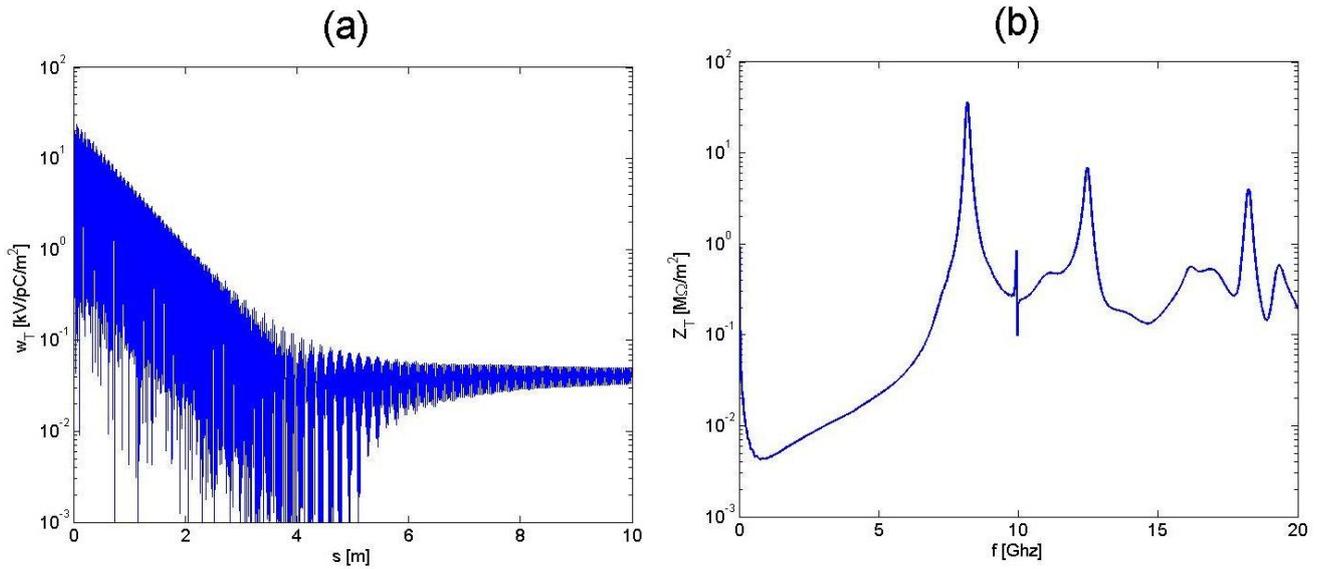

**Fig. 68.** Transverse wake per unit length (a) and impedance (b) obtained with GdFidL (20 cells plus 2 couplers, σz=5 mm, mesh step 500 μm)

### 3.3.5. RF Power and distribution

The RF power is generated by means of high power multi-megawatt microwave sources operating at 2856 MHz or 5712 MHz and transmitted to the accelerating structures with a network of copper rectangular waveguides [13, 34].

The microwave sources consist of commercially available power klystrons, supplied by pulsed high voltage modulators. The klystrons are driven by solid-state pulsed amplifiers (SSA). The SSA's are also available on the RF market while the pulsed modulators must be generally tailored, on a case by case basis, to the specifications of the klystrons.

The Fig. 69 shows the schematic layout of the ELI-NP RF system.

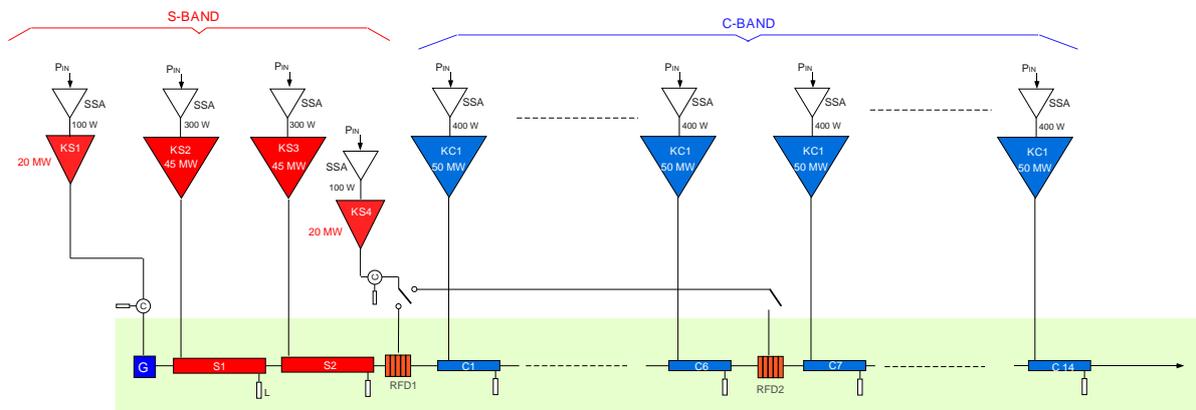

**Fig. 69.** Layout of the ELI-NP Linac

The linac complex consists of an S-band pre-injector followed by a C-band 700 MeV injector. The pre-injector comprises an RF-gun and two traveling wave, constant gradient, 3 meter long accelerating sections.

Each RF element is supplied by an individual klystron because the use of pulse compressors is not allowed as, in order to accelerate trains of multi-bunches, the entire length of the RF pulses must be available. The pre-injector stage is completed by a 10 MW S-band klystron to feed two deflecting cavities for beam longitudinal diagnostics, selectable with a waveguide RF switch. The RF deflectors are positioned as shown in Fig. 69. The pre-injector output energy can range from 70 to 120 MeV. The lowest energy value occurs when the phase of the first S-band section is set for the velocity bunching configuration.

The successive power system consists of a series of 14 RF stations to feed as many C-band accelerating structures.

The following Table 17 summarizes the most significant parameters of the high power klystrons needed for ELI. The S-band klystrons can be developed by the three main companies operating in the high power RF market sources i.e. Thales (F), CPI (US) and Toshiba (JP). The C-band klystron is produced, up today, solely by Toshiba. These systems are routinely operated in several accelerator laboratories.

**Table 17.  Significant parameters of high power klystrons**

| Power station | K1 | K2, K3 | K4 | K5 to K18 |
|---|---|---|---|---|
| Klystron (°) | TH2128 | TH2100L | TH2163A | E37202 |
| Frequency (MHz) | 2856 | 2856 | 2856 | 5712 |
| Output power (MW) | 15 | 60 | 10 | 50 |
| Pulse width (µsec) | 4 | 2.5 | 2.5 | 1 |
| Rep. rate (Hz) | 100 | 100 | 100 | 100 |

(°) only as a reference

The klystrons are powered by pulsed high voltage modulators. Recently, the need to reduce their maintenance costs and size has meant that began to spread the use of modulators completely or almost totally made of solid-state technology.

In the standard case, the modulators consist essentially of a HV power supply charging a LC line which operates as a pulse forming network (PFN) when a HV switch discharges the line through a 1/n transformer. The HV switches are usually vacuum tubes that must be replaced regularly every 10÷15 khours. The secondary winding of the transformer, immersed in insulating mineral oil, is connected to the klystron cathode and provides the HV pulses. The principle diagram of a PFN modulator is shown in Fig. 70. The charging units operate in the range 25-50 kV

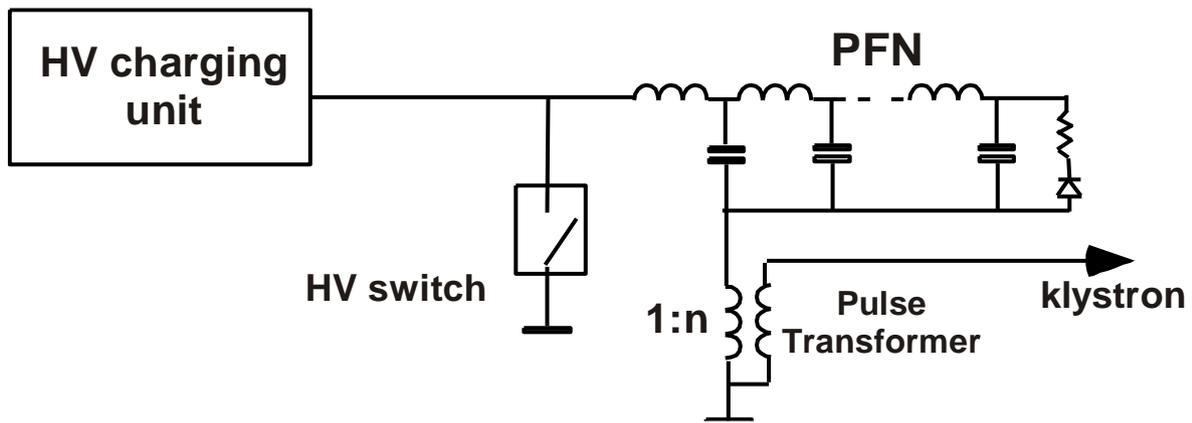

Fig. 70. Layout of a PFN modulator

A series of solid-state switching boards replaces the line-type PFN in the solid-state modulators. The current pulses generated by triggered IGBT's are summed on the primary of a 1/n transformer. Similarly to the PFN case, an HV pulse is produced on the secondary winding. The voltages on the primary are much lower respect to the PFN modulators for a better safety and reduced size of the system. The long-term reliability of the solid-state solution is still matter of some concern because these systems are in use only recently in particle accelerators. Nevertheless, the solid-state modulators offer other important advantages as lower operating costs, excellent stability and repeatability of amplitude and phase of the output voltage/current pulses. The Fig. 71 shows the basic concept of a solid state modulator.

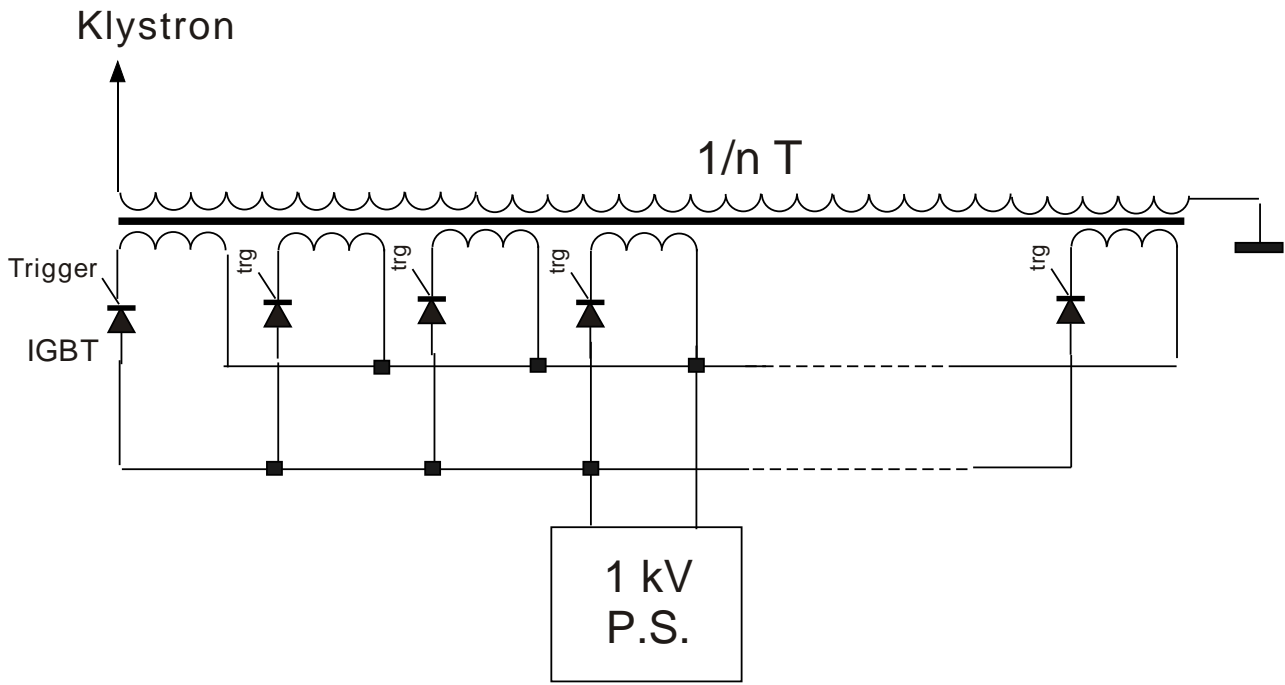

Fig. 71. Conceptual scheme of a solid state modulator

The following Table 18 gives the significant parameters of the pulsed modulators needed for the linac.

Table 18. Significant parameters of pulsed modulators for Linac

| Klystron | K1 | K2, K3 | K4 | K5 to K18 |
|---|---|---|---|---|
| Beam Voltage (kV) | 280 | 350 | 170 | 360 |
| Beam current (A) | 290 | 410 | 125 | 320 |
| FWHM Pulse width (µsec) | 5 | 3 | 3 | 3 |
| Max rep. rate (Hz) | 100 | 100 | 100 | 100 |
| Average wall plug power (kW) | 41 | 43 | 6.5 | 35 |

The RF power will be distributed to the cavities with a system of copper rectangular waveguides. The S-band section will use WR284 guides. WR187 is instead the standard adopted with the C-band system. With the exception of the K1 and K4 channels, in which the waveguides will be pressurized with 2 atm SF6, other waveguides operate in ultra-vacuum at $< 10^{-8}$ mbar because of the very high peak power of the transmitted RF power. The use of SF6 in K1 and K4 channels is required because the RF Isolators contains ferrite materials that cannot operate in vacuum. The isolators protect the klystrons against the power reflected by the RF gun and the RF deflectors that are standing wave cavities.

The next two figures show, with more details, the RF layout of the S-band and C-band channels. The waveguide network comprises an RF monitor at each klystron output and each accelerating structure input. A further directional coupler could eventually be added at each output load of the accelerating sections to monitor the RF power not dissipated through the sections. The monitors are 60 dB directional couplers commercially available from the RF market. The Linac vacuum is separated from the waveguide system with RF ceramic windows, also these commercially available, at each accelerating section input. In the K1 and K4 channels, where the SF6 is used as insulating medium, a double window system is adopted to avoid that, in case of a window break, the SF6 enters the linac pipe. The vacuum of the waveguides is maintained at a pressure lower than $< 10^{-8}$ mbar with vacuum ion-pumps distributed every 5÷6 m in the S-band and 1.5÷2 m in the C-band system . The pumps are connected to the waveguides by means of T-units and the pumping takes place through some slits on the narrow side of the guides or through an RF notch-filter from the same narrow side.

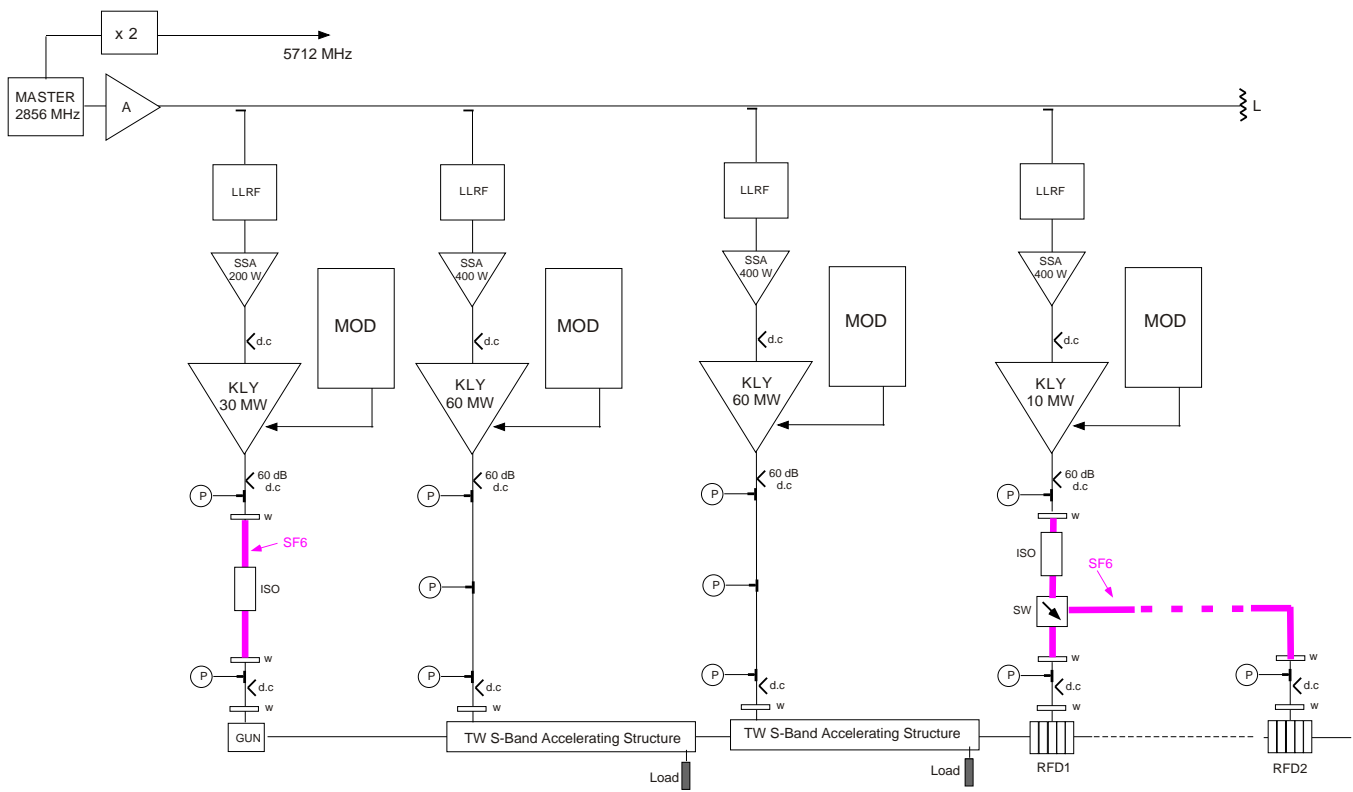

**Fig. 72.** Scheme of the S-band sector of the ELI Linac

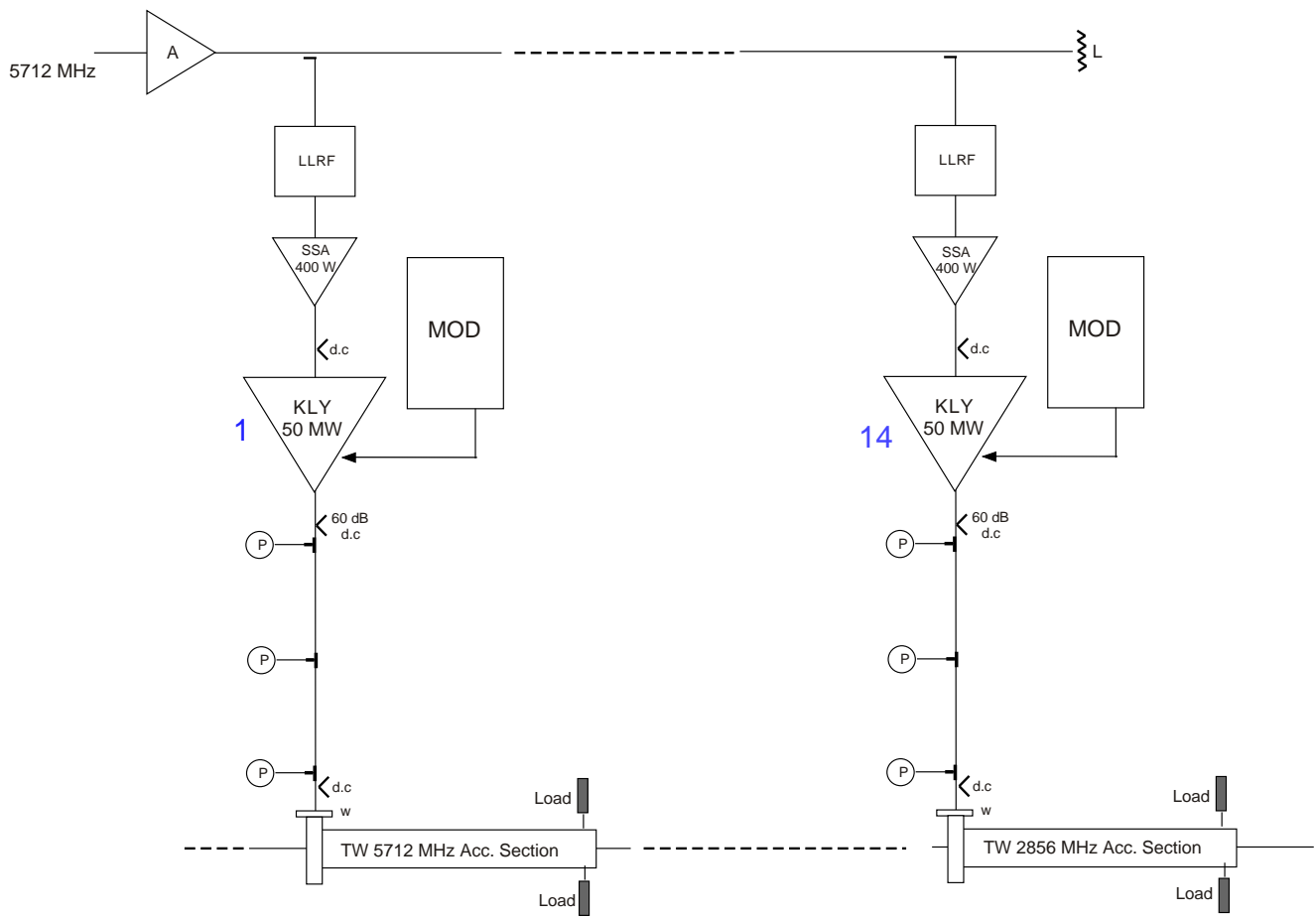

**Fig. 73.** Scheme of the C-band sector of the ELI Linac

The RF losses in copper WR284 and WR187 waveguides are approximately 0.02 dB/m and 0.035 dB/m respectively that mean about 5 and 8% of power loss with 10 m. long guides. Therefore the RF power routinely available for the accelerating structures is 51 MW in S-band and 41.6 MW in C-band, since it is good practice to operate the klystrons 10% below their nominal maximum power.

## 3.4. Magnets and Power Supplies

### 3.4.1. General Remarks

The magnets and power supplies are a core technical system for the ELI-NP project. It is essential that they achieve their specifications in terms of field strength, quality, stability and reproducibility in order for the full project to meet its' overall aims and ambitions. Fortunately the project team has considerable expertise and experience in the design and delivery of similar systems and so is qualified to carefully assess all of the magnet parameters required to ensure that they are achievable with robust, reliable design solutions. Examples of our expertise include the design and delivery of magnets for synchrotron light sources (Diamond), free electron lasers (SPARC), novel FFAG accelerators (EMMA), and other low energy test facilities (ALICE, EBTF). We have expertise in all types of accelerator magnet (dipoles, quadrupoles, sextupoles, undulators, wigglers) employing electromagnet (DC, AC, pulsed) and permanent magnet technologies. Fig. 74 highlights some recent examples. The project team has expertise in magnet design using state of the art modelling codes, mechanical and electrical engineering, and survey and alignment techniques. In addition we have developed state of the art magnet measurement laboratories (see Fig. 75) in order to assess the field strength and quality achieved by the manufactured devices. Our magnet expertise is in demand internationally and we are often asked to take on challenging magnet projects by the wider community (e.g. CLIC, ILC, SwissFEL, LHeC, etc).

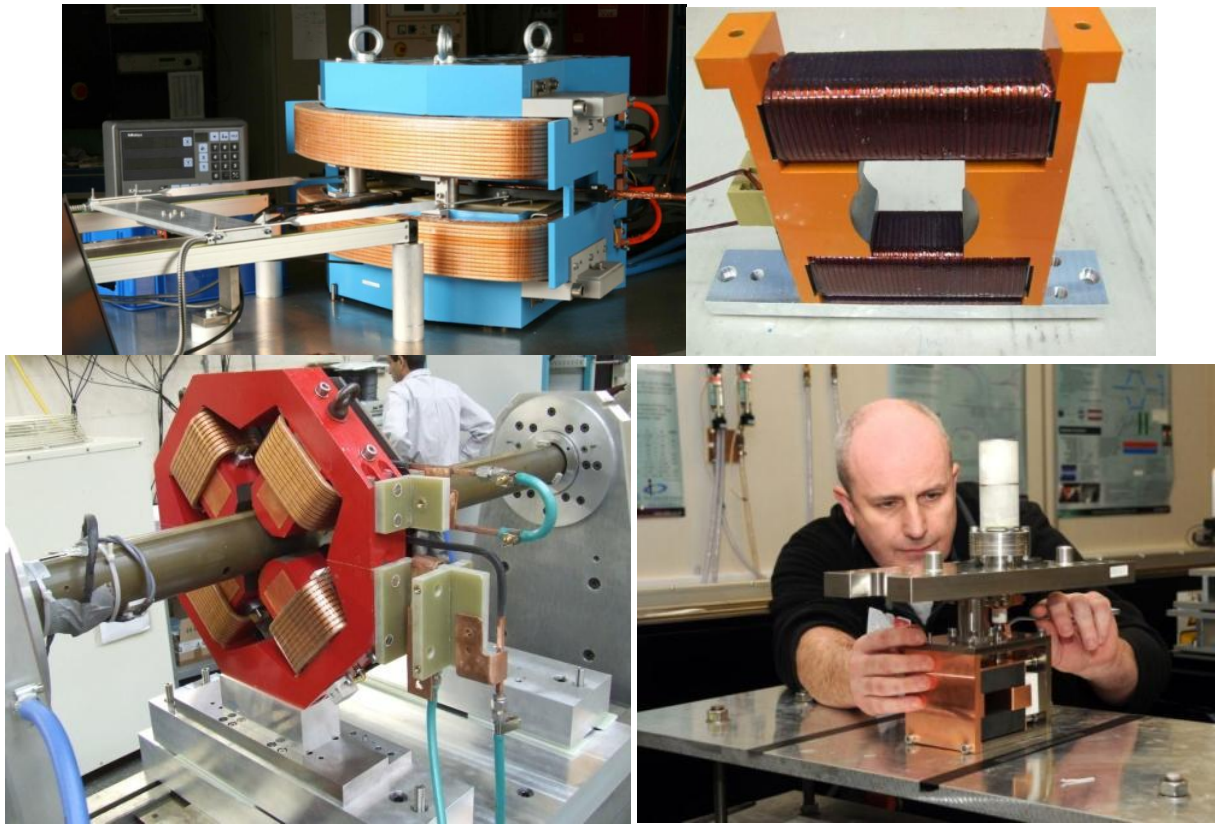

**Fig. 74.** Examples of recent magnet designs which have all been delivered and operated successfully (top left ALICE dipole, top right EMMA corrector, bottom left ALICE quadrupole, bottom right EMMA kicker)

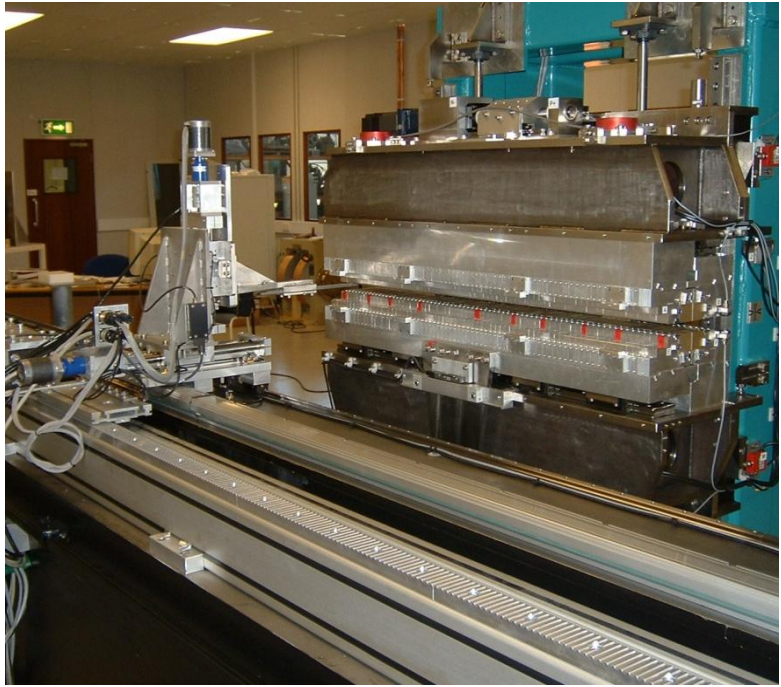

**Fig. 75.    The ASTeC magnet measurement laboratory at Daresbury. This facility can assess magnet quality using a wide variety of techniques (Hall probe, flipping coil, rotating coil, stretched wire, pulsed wire)**

For the ELI-NP project we have carefully assessed every magnet type and confirmed that it is feasible in terms of field strength and quality using standard accelerator magnet technology and that the power supply requirements and water cooling requirements are reasonable. In general, all of the magnets, except the low strength correctors, will require steel yokes with accurately machined pole faces designed to maximise the field quality in an optimal manner and they will be powered by high current coils with integral water cooling. Such solutions are now routine for magnet manufacturers and accelerator magnets are highly reliable during facility operations in all accelerator laboratories. The magnets will all be excited by direct current (DC) except for one dipole which will operate at 50 Hz AC in order to switch the electron beam between the two branches.  The designs shall ensure that the upper and lower halves of dipoles and quadrupoles correctly mate and align on reassembly after separation to insert the vacuum vessel. Every individual magnet shall be subject to rigorous electrical, thermal, and mechanical testing during manufacture and detailed magnetic measurements once assembled.

### 3.4.2.    Gun and Linac section

The photo-injector gun will employ a sophisticated solenoid similar to that used on the SPARC gun except the solenoid for ELI-NP will be upgraded with respect to the SPARC design as it will include the coil winding connections on all four opposing sides to be quadrupole symmetric in the iron boundaries, instead of only two ports (dipole symmetry). This improved design of solenoid has already been successfully adopted for both the FINDER and FERMI projects.  Fig. 76 shows the existing SPARC solenoid design.

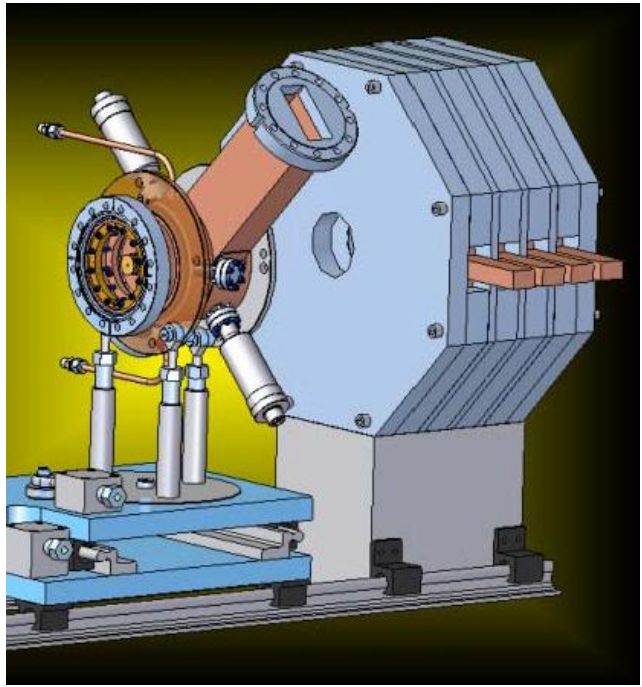

**Fig. 76**     **SPARC photo-injector solenoid design with coil access on two sides only. The ELI-NP will be improved by having coil access on four sides to improve the symmetry**

The linac solenoids will also be based upon the successful experience of SPARC. In the SPARC facility there are 13 independent coils mounted on each S-band linac. Experimental experience from SPARC has shown that such a large number of independent coils is not required in practice to generate beams of excellent quality. For ELI-NP the system will be simplified, reducing the number of coils to five but still covering the full length of the linac. Fig. 77 illustrates the SPARC linac solenoids.

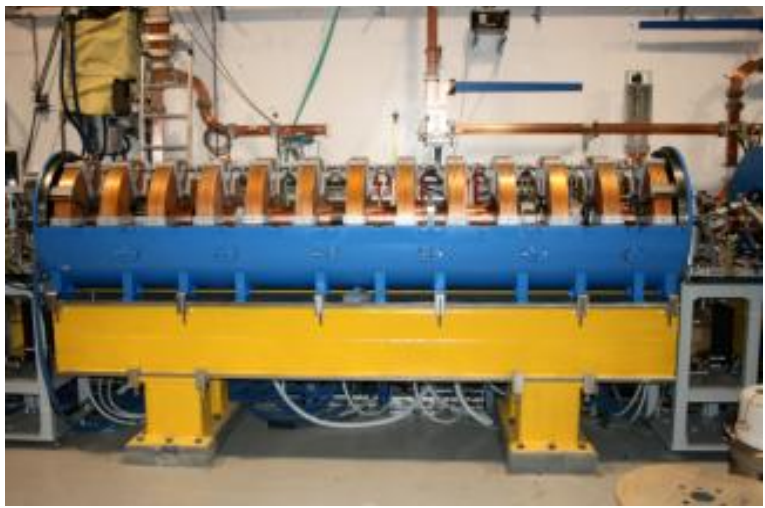

**Fig. 77.**     **Photo of one of the SPARC linacs showing the 13 independent solenoid coils**

The linac section also requires a number of quadrupoles and a dipole for the energy spectrometer. The quadrupole strength is relatively modest and the use of water cooling in this case is marginal. Detailed design will be carried out in the future on these magnets and if possible they will be cooled using natural convection or perhaps indirect water cooling. The dipole will require direct water cooling and it has been estimated that the temperature rise of the magnet coils will be limited to 10°C with a water pressure drop of 4 bar. The power supply will need to be rated to 150 A, 20 V, with the dipole consuming around 2.2 kW at the required field level of 0.84 T.

### 3.4.3. Low Energy Transport

The low energy transport section (360 MeV) contains some of the most challenging magnets within ELI-NP. This is because of the tight space constraints and also because of the need to switch the beam between the two beamlines at 100 Hz, requiring a rapidly alternating dipole field. Several of the quadrupoles are specified for 30 T/m gradient which is relatively high given the 4 cm magnet aperture that is required. Design calculations suggest that each quadrupole will consume ~500 W, and require a power supply rated at around 80 A and 10 V. Direct water cooling will be essential with a pressure drop of ~3.5 bar being sufficient. The three dipoles required for the dogleg and spectrometer/dump line are broadly similar to that required by the linac section, although the slightly higher field level required implies an increased power consumption of ~2.8 kW and a current level of ~170 A.

The switching dipole is required to change polarity at 100 Hz so that the electron beam is shared between the two transport lines. The transport system is designed so that a dipole field of ±1.05 T is needed to deflect the electron beam alternately down each beamline. Therefore a dipole operating at 50 Hz AC will fulfil the requirements of ELI-NP. The design of a 50 Hz AC magnet requires a number of additional issues to be taken into account. Fortunately these are well understood and, whilst the field level is relatively high for such a magnet, none of these issues are expected to affect the feasibility of the device. One of the key challenges that must be tackled is the induced eddy current effects which cause power losses in the coil and the steel. At such high frequencies it is essential that the coil is wound from narrow strands of copper conductor twisted to form a wire and then made into a cable with suitable transposition to ensure each strand couples the same total flux. Water cooling will be incorporated into the cable by embedding a separate cooling pipe. To minimise eddy losses in the steel the magnet must be laminated. These laminations will be made of high silicon content steel (non-grain oriented) and be 0.35 mm thick. Preliminary calculations of a technically feasible solution give a magnet inductance of ~0.01 H, requiring a 50 Hz power supply capable of 400 A and 1100 V peak.

### 3.4.4. High Energy Transport

The high energy transport section (720 MeV) contains a number of dipoles and quadrupoles, most of which are relatively straightforward. The quadrupoles are about twice the length of the low energy ones but have lower gradients in general. As a result the power supply ratings are rather similar, with current of ~80 A and voltage of ~10 V. The power loss per magnet will be ~600 W and the pressure drop around 3 bar. The dipoles before the interaction point and in the spectrometer dogleg are also about twice the length of the low energy ones and have power supply ratings of up to ~200 A and ~25 V. The two dipoles which take the high energy beam to the dump are more challenging as each must bend the beam by 45° in a relatively confined space. To achieve this, the magnetic field is 1.5 T and the length of each dipole is 1.3 m. These parameters can be achieved with a 450 A, 50 V power supply, with power loss per magnet of 19 kW. In this case the temperature rise in the coil can be maintained at 20°C with a pressure drop in the coils of 5.3 bar.

### 3.4.5. Summary

All of the magnets and power supplies for ELI-NP have been assessed and found to be feasible. In the next phase of the project detailed 3D magnet design simulations would be carried out by experienced magnet designers within the consortium with a strong track record in this area. The detailed pole shapes and other important mechanical features would be optimised and then the exact coil arrangements (conductor cross-sections, cooling channel size, number of turns per pole, etc) would be decided in close cooperation with power supply experts to create the most cost effective and efficient solutions. Table 19 summarises all of the magnet types required and their key parameters.

**Table 19.    Summary of the magnet parameters for ELI-NP**

| Type | Location | Strength | Length | Aperture | |
|---|---|---|---|---|---|
| Solenoid | Gun | 0.31 T | 19.5 cm | 7.6 cm | |
| Solenoid | Linac | 0.05 T | 60 cm | 30 cm | |
| | | | | | |
| **Type** | **Location** | **Gradient** | **Length** | **Aperture** | |
| Quadrupole | Linac | 12 T/m | 12 cm | 4 cm | |
| Quadrupole | LE dogleg | 30 T/m | 12 cm | 4 cm | |
| Quadrupole | LE IP | 30 T/m | 14 cm | 4 cm | |
| Quadrupole | LE Dump | 12 T/m | 12 cm | 4 cm | |
| Quadrupole | HE Lines | 16 T/m | 30 cm | 4 cm | |
| | | | | | |
| **Type** | **Location** | **Strength** | **Length** | **Aperture** | **Bend Angle** |
| Dipole | Linac | 0.84 T | 20 cm | 4 cm | 250 mrad |
| Dipole | LE Dogleg | 1.05 T | 20 cm | 4 cm | 170 mrad |
| Dipole (50Hz) | Switching | 1.05 T | 20 cm | 4 cm | ±170 mrad |
| Dipole | LE Dump | 1.0 T | 30 cm | 4 cm | 245 mrad |
| Dipole | HE before IP | 0.52 T | 70 cm | 4 cm | 150 mrad |
| Dipole | HE Dogleg | 0.85 T | 70 cm | 4 cm | 245 mrad |
| Dipole | HE Dump | 1.5 T | 130 cm | 4 cm | 45° |

## 3.5.   Diagnostic

### 3.5.1.    General considerations

High performance diagnostic is mandatory in order to achieve high brightness in high repetition rate machine. The properties of the single bunch, as well as the characteristics of the whole train of bunches must be measured. In our machine the beam is manipulated in the phase space in order to obtain the best performance in term of emittance and bunch length. The so called 'invariant envelope' scheme has been adopted. These choices put additional requests to the beam diagnostic, for instance the measurement of the beam envelope in the region downstream the gun up to an energy of about 80 MeV.

The emittance has a paramount importance for such a machine. We plan to measure it in three different locations with quadrupole scan technique, and in one other branch with multi-screen system [48]. In the following we are going to describe also the technique to measure the emittance for the single bunch in the pulse train.

Beam trajectory is measured by means of BPM striplines in the entire machine, but interaction points where we consider the use of cavity BPM, with a resolution of few $\mu$m in order to properly align the position of the electron beam with respect to the laser beam.

The bunch length is monitored by means of Radio-Frequency-Deflector, and also in a parasitic way using Coherent Diffraction or Synchrotron Radiation.

### 3.5.2. Optical diagnostic

The optical diagnostic is based on the measurement of the beam size. Envelope measurement is needed in order to check the correct emittance compensation in the invariant envelope scheme. Quadrupole scan measurements will be used to retrieve the emittance value. We need about 23 vacuum chambers, AISI316L. In Fig. 78 you can find the basic layout.

The quotes and all the technical details can be found in Fig. 79.

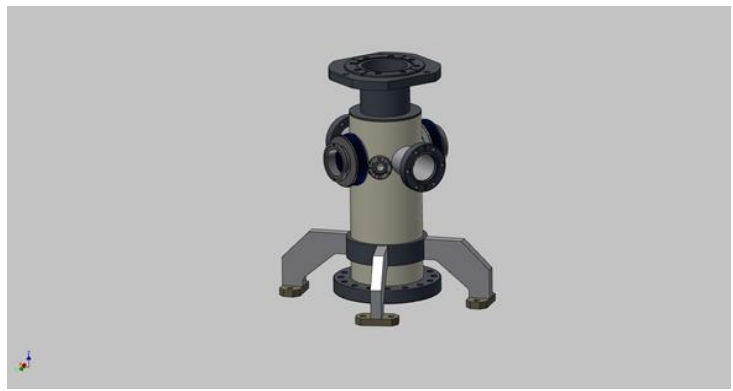

**Fig. 78.** Schematic Drawing of the vacuum chamber

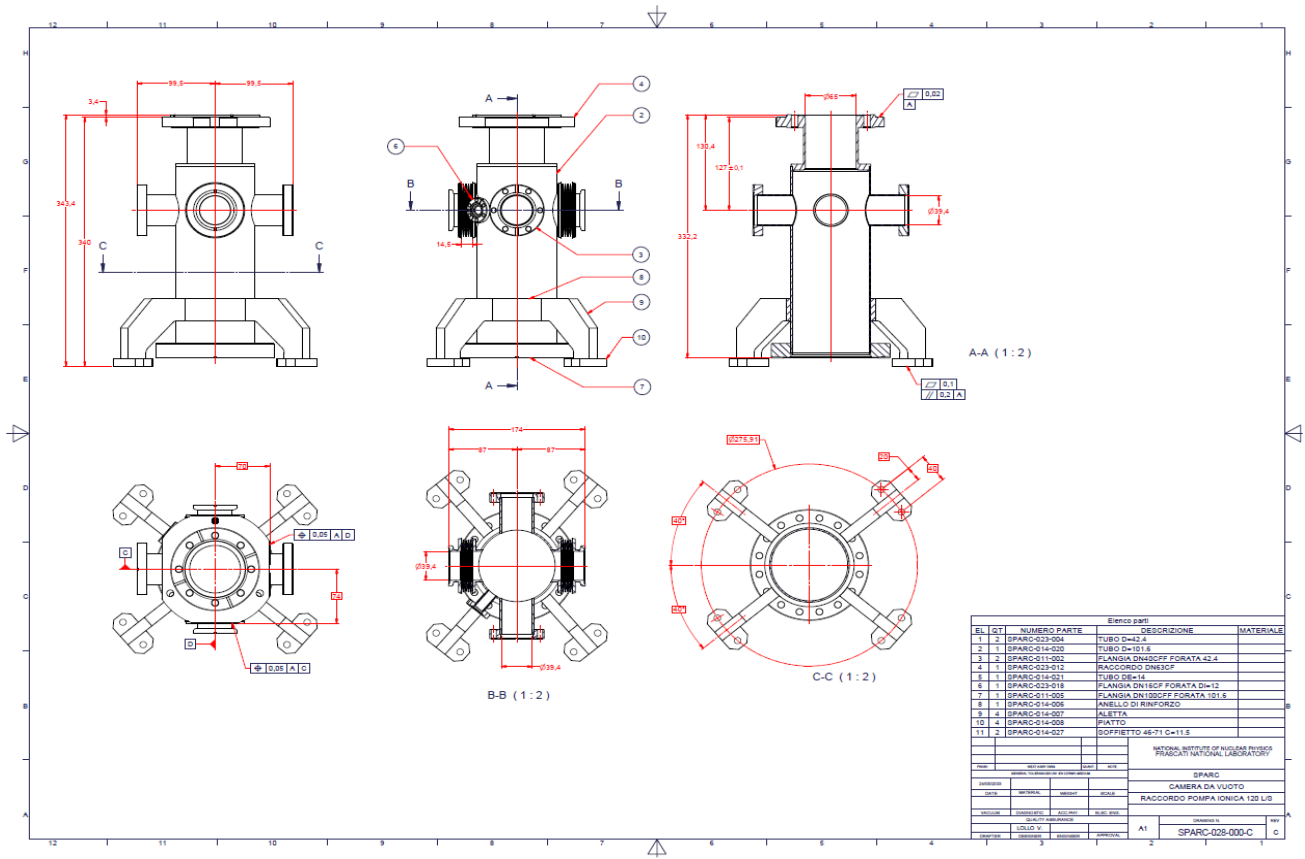

**Fig. 79.    Drawing of the vacuum chamber**

Every chamber is equipped with a mechanical actuator, with UHV feedthrough to drive inside a screen holder. In Fig. 80 a sketch view of the mover is shown, while in Fig. 81 there is a complete drawing.

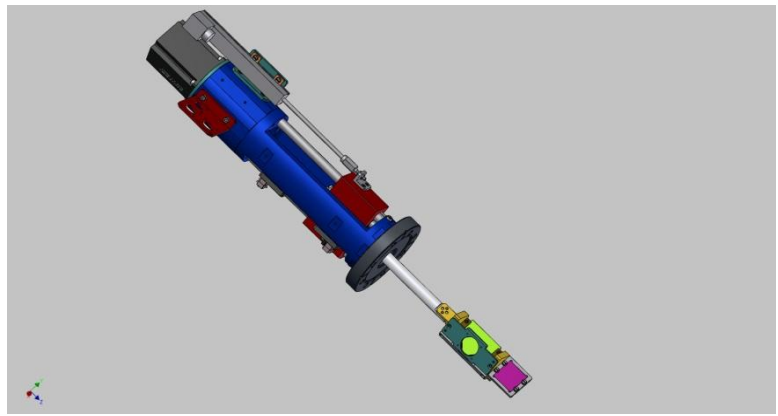

**Fig. 80.    3D View of the actuator with the screen holder**

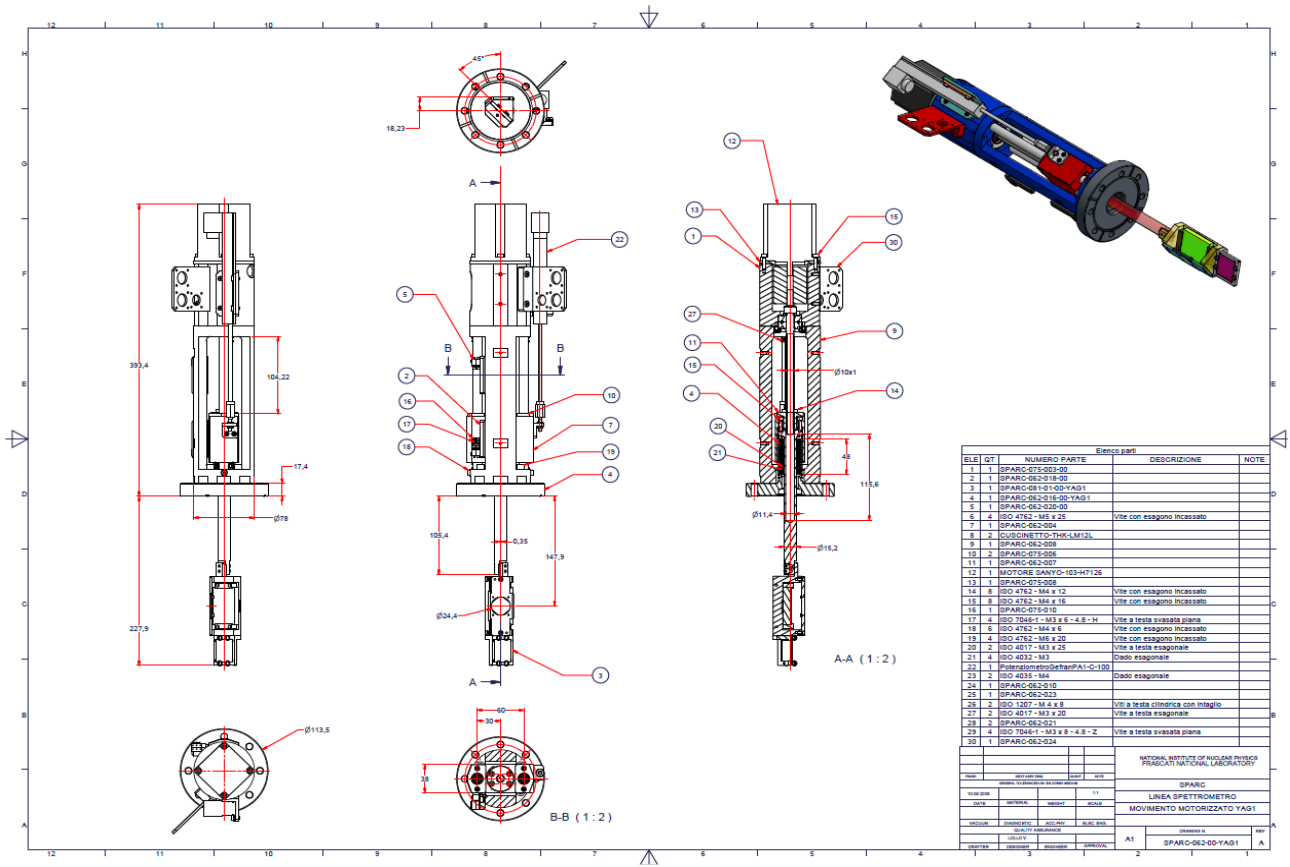

**Fig. 81.**     Complete drawing of the actuator with the screen holder

The mover is equipped with a linear potentiometer (Gefran series PK with 100 mm useful electrical stroke range), a stepper motor (for instance a Sanyo Denky 103H7126-6610 or equivalent). The microstepping capability is not required. A YAG:Ce crystal for single bunch and low charge operation and a silicon aluminated OTR (Optical Transition Monitor) are placed at the end of the mover. A calibration screen is included in the screen holder.

Outside of the vacuum we plan to install a CCD camera, likely a Basler, model scA640-74gm/gc. The holder of the camera includes a mirror that reflects down the light, coming out from the vacuum window.

For time resolving beam size measurement, useful to monitor change in transverse properties between microbunches, an ICCD (Intensified CCD camera) will be used. The fast gate (about 2 ns) of the MCP (multi-channel plate) allows the exposition of the CCD to be tailored on the single bunch of the train. In such a way time resolved measurements will be possible, in particular the emittance measurement. It is really important in a multi bunch machine to keep under control this parameter for every bunch in the train.

### 3.5.3.    BPM striplines

The striplines type beam position monitors are composed by stainless steel AISI 316L electrodes of length 146 mm and width 14.3 mm, mounted with a π/2 rotational symmetry at a distance of 4 mm from the vacuum chamber, and forming a transmission line of characteristic impedance $Z_0=50$ Ω with the beam pipe. Matched loads at the end of the stripline electrodes have been preferred to reduce the loss factor and to avoid unwanted reflections to the detection electronic. Integration of the matched resistive load inside the vacuum chamber allowed to halve the number of UHV feedthroughs and to reduce pickup size for installation inside quadrupole magnets.

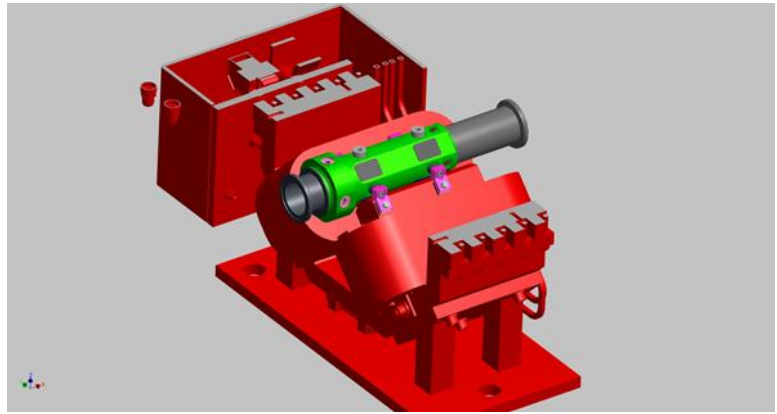

**Fig. 82.    3D View of the Stripline BPM embedded in a quadrupole**

In Fig. 82 there is a schematic sketch of the BPM stripline mounted inside a quadrupole, while the technical drawing is shown in Fig. 83. The signal readout will be provided using Libera modules, produced by Instrumentation Technologies, a standard in such a field.

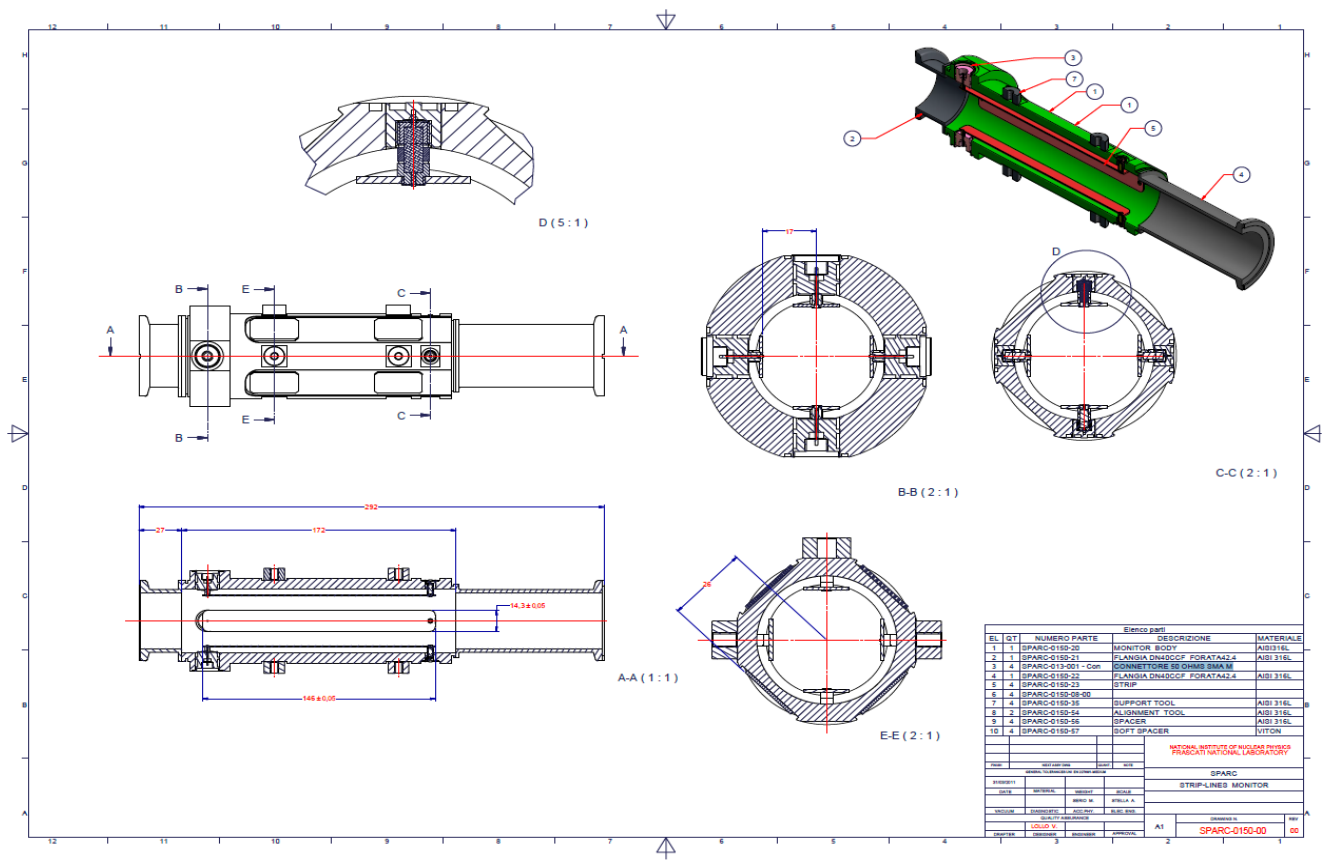

**Fig. 83.    BPM stripline**

## 3.5.4. Cavity BPM

The Cavity BPMs have been developed for the SwissFEL linac (C-band accelerating structures). 4 Cavity BPMs will be used to align the beam, at micrometer level, in the interaction regions. The principal parameters are reported in the following table:

**Table 20. Cavity BPM parameters**

| Beam pipe diameter | 16 mm |
|---|---|
| Working mode frequency | 3.3 GHz |
| Loaded Q | 40 |
| Sensitivity | 7 V/nC/mm |
| TM010 freq | 2.252GHz |
| TM210 | 4.308 GHz |

In Fig. 84 there is a sketch of the Cavity BPM structure, with the two resonant cavities, one for position measurement, and the other for reference calibration, highlighted. In Fig. 85, a picture of a cavity BPM is shown.

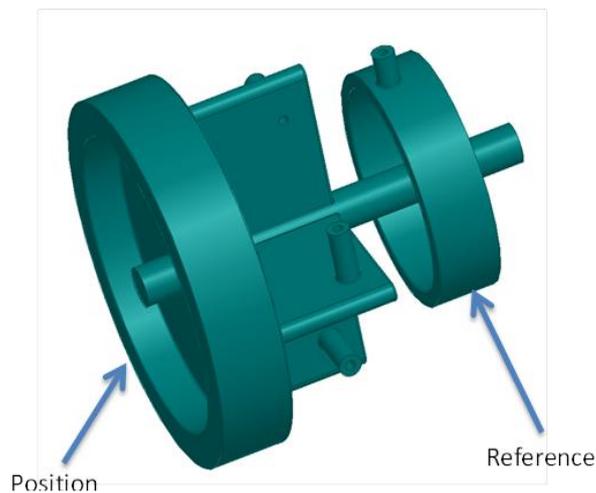

**Fig. 84.    Sketch of a cavity BPM**

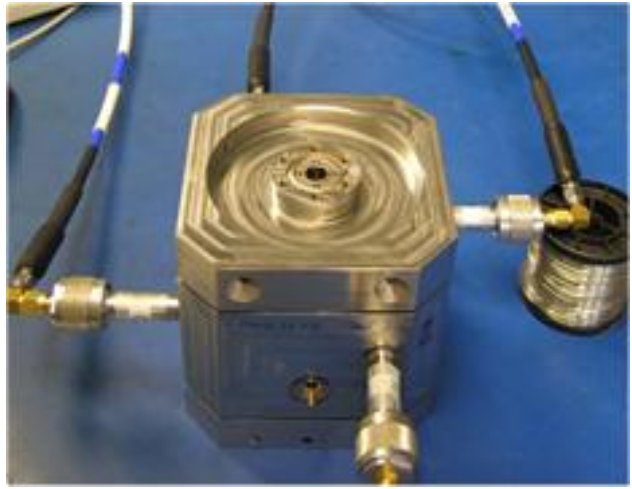

**Fig. 85.    Picture of a cavity BPM**

### 3.5.5.    Toroids

In-flange FCT (Fast current transformer) can be mounted in the beam line. They have short axial length, included a ceramic gap vacuum-brazed on kovar. They do not require bellows, wall current bypass or electromagnetic shield. They are UHV compatible.

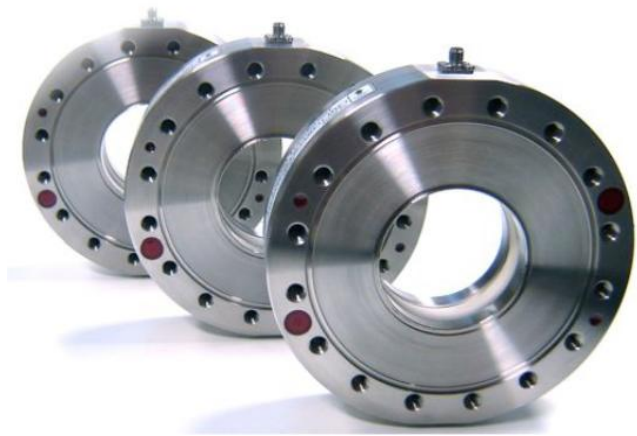

**Fig. 86.    FCT**

We plan to use FCT (series –WB) produced by Bergoz instrumentation. The readout can be done with a scope, or digital board on an industrial PC.

### 3.5.6.    Online Bunch Compressor Monitor

In order to have online measurement of the bunch length, we plan to install a diffraction radiation target. This not intercepting device allows the beam transport, but at the same time produces coherent diffraction radiation. The setup is shown in Fig. 87. The radiation is extracted from the vacuum pipe through a z-cut quartz window and guided using two 90 deg off-axis parabolic mirrors. One is placed in front of the z-cut quartz with the focus on the source plane (i.e. the plane of the radiation target). The second one is used to focus the radiation onto the FIR-THz detector.

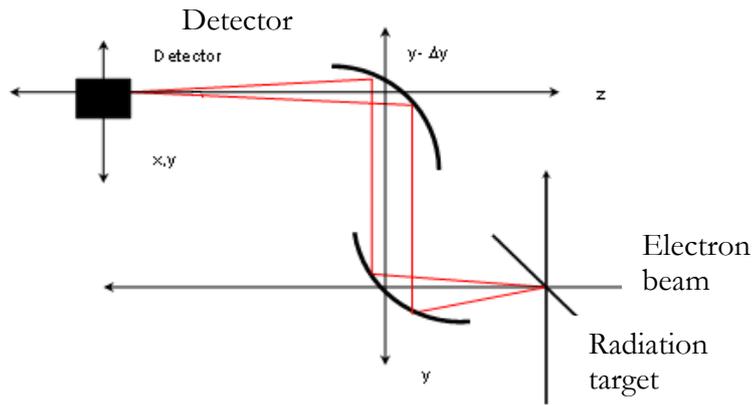

**Fig. 87.   Experimental setup**

Both transmission and technical characteristic of the z-cut quartz window are summarized in the following plot (Fig. 88) and table (Table 21).

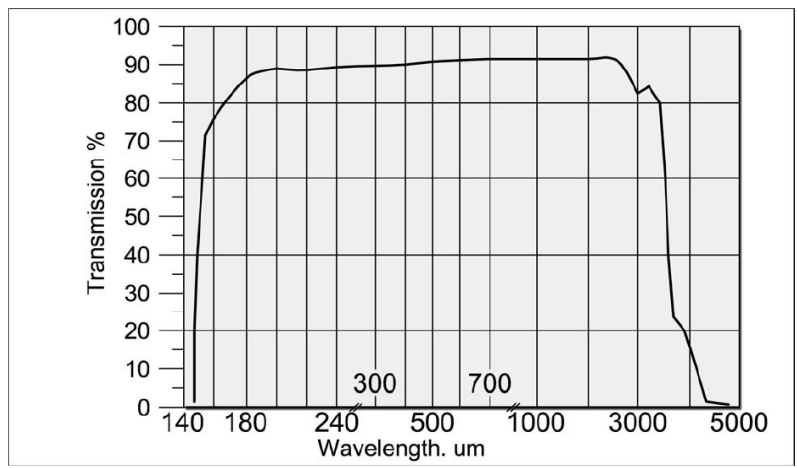

**Fig. 88.   Transmission curve of the z-cut quartz window**

**Table 21.   Parameter of the quartz window**

| Type | Quartz Natural Z-Cut Zero Length Viewport |
|---|---|
| Flange | NW63CF |
| Flange OD | 114 mm |
| View Diameter | 63 mm |
| Seal Type | Bond |
| Max Temp. Deg C | 120 |
| Flange material | St. Steel 304L |
| Weld Ring | Kovar |

The technical drawing of the 90 deg off-axis parabolic mirrors are reported in Fig. 89, while the specifications are shown in and Table 22.

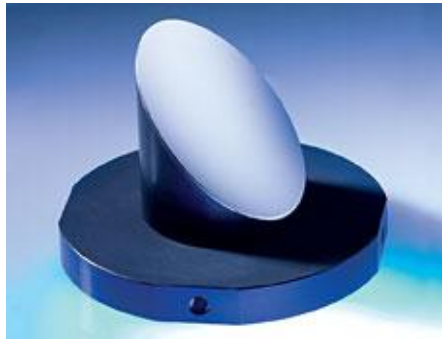 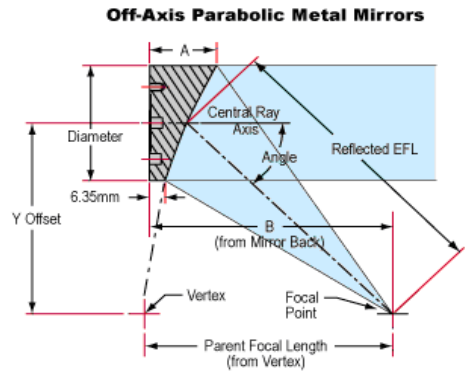

**Fig. 89.**     90 deg off-axis parabolic mirror. Left: picture. Right: schematic drawing

**Table 22.**     Specifications of the off-axis parabolic mirror

| Type | 90° Off-Axis |
|---|---|
| Diameter (mm) | 76.2 |
| Diameter Tolerance (mm) | +0.00/-0.38 |
| Focal Length Tolerance (%) | ±1 |
| Surface Accuracy (λ) | 1/4 RMS |
| Effective Focal Length EFL (mm) | 177.80 |
| Y Offset (mm) | 177.8 |
| A (mm) | 83.8 |
| B (mm) | 41.7 |
| C (mm) | 7.6 |
| Surface Roughness (Angstroms) | <175 RMS |
| Substrate | Aluminum 6061-T6 |
| Coating | Protected Aluminum |

The radiation signal in the millimeter and sub-millimeter wavelength range is detected by a pyroelectric detector based on LiTa03 crystal. At square wave modulation of radiation, the saw-tooth signal voltage from the detector is proportional to the intensity of radiation.

**Table 23.    Specifications of the pyro-detector**

| Operating spectral range | 0.02-3 THz |
|---|---|
| Typical Responsivity | 1000 V/W |
| Dynamic Range | 1 µW - 10 mW |
| Optimal modulation frequency | 5-30 Hz |
| Noise level | 1.0 mV |
| Power supply voltage (2 AA batteries) | 2 x 1.5 V |
| Power supply current | 0.14 mV |
| Operating battery lifetime | > 10,000 hours |
| Dimensions | 135 x 40 x 30 mm |
| Operating spectral range | 0.02-3 THz |

### 3.5.7.    The Gun sector and S-band structures

In the area just downstream the cathode several parameters need to be measured. The beam imaging is mandatory, both to center the laser spot on the cathode and to determine the beam size in order to properly match the 'invariant envelope' emittance compensation scheme. A Yag:Ce screen, mounted at 90 degrees with respect to the beam line is followed by a mirror to deliver the radiation emitted outside the vacuum pipe. Such a mounting prevents any problem related to the depth of field when using an imaging detector as CCD because the optics focuses on the back of the flat YAG:Ce screen.

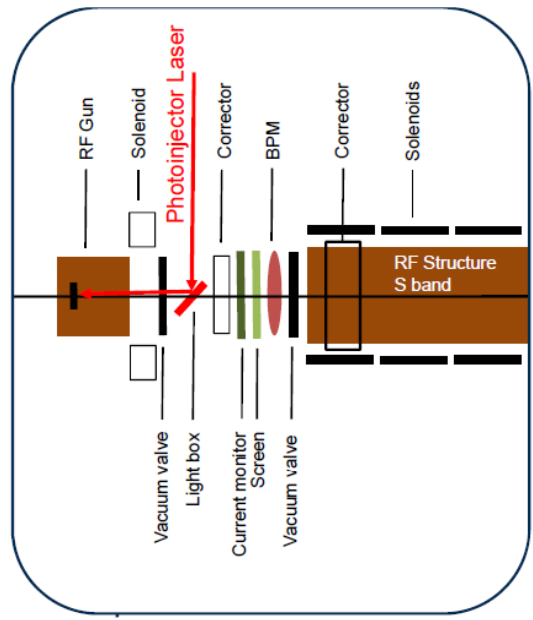

**Fig. 90.    Sketch of the cathode and low energy (<6 MeV) region**

This screen can be also used to measure the beam energy at the exit of the gun sector. This measurement is really important to control that the gradient inside the gun is at expected level. The corrector deflects the beam center on the screen depending on the beam energy. So varying the current in the corrector and measuring the beam center on the screen results in the determination of the beam energy.

The current monitor is needed to properly adjust the beam charge, varying for instance the laser power. It is also possible to obtain the so-called phase scan, changing the RF injection phase and recording the charge value, identifying in such a way the proper injection phase.

Finally we placed in this area a stripline BPM to control the orbit in the critical phase of the injection in the first travelling accelerating section.

After the first S-Band structure another diagnostic chamber is placed just to measure the beam size to match the invariant envelope scheme. As in the gun sector a Yag:Ce screen is mounted on an actuator and imaged by a ccd camera. Again a stripline BPM is available in this area to monitor the beam trajectory.

### 3.5.8. Module 4-5

Following the two S-band structures another BPM (to properly inject in the following C-band structure) and a screen, to measure the beam size, are placed. The region downstream the C-band 1 is totally dedicated to the beam diagnostic. The main parameters to be measured are: emittance (whole train and single bunch), energy, energy spread, bunch length and longitudinal phase space.

The first screen in this module is the third needed to measure the beam size in order to check the working of the invariant envelope scheme. The Quads QUA 1,2,3 are placed to be used in the quadrupole scan emittance measurement. A transverse deflecting cavity (TDC) will be used to measure the bunch length [49], imaging the streaked beam on the following screen flag.

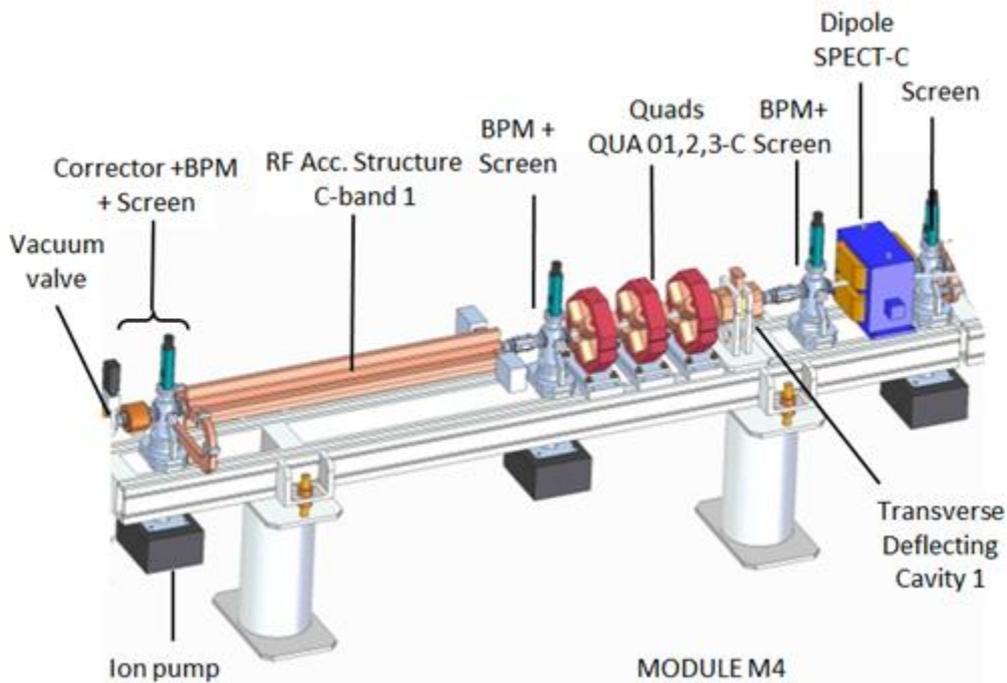

**Fig. 91.** Module 4 layout

The Dipole SPECT-C is placed to be used as a spectrometer together with the following screen in the dogleg region (see local beam dump module 5). Not only the energy can be measured but also the energy spread, a fundamental parameter in this sector because the working point foresees the use of velocity bunching The Dipole LE1 is placed to be used as a spectrometer together with the following screen in the dogleg region. Not only the energy can be measured but also the energy spread, a fundamental parameter in this sector because the working point foresees the use of velocity bunching [50] in the S-Band accelerating structure. The compensation of the energy spread is done in all the following C-band structures. At this point it is crucial to compare the value with the expectation from simulation to validate the velocity bunching phase. The use of the TDC together with the dipole allows also the measurement of the longitudinal phase space in a single shot. in the S-Band accelerating structure. The compensation of the energy spread is done in all the following C-band structures. At this point it is crucial to compare the value with the expectation from simulation to validate the velocity bunching phase. The use of the TDC together with the dipole allows also the measurement of the longitudinal phase space in a single shot.

The bunch length can be also measured online using a Coherent Diffraction Radiator (CDR). A metallic screen with an aperture of about 5 mm is placed in the beam line. The electromagnetic extension of the radial field travelling with the bunch is larger than this size. So the beam goes through the hole while part of the field touches the screen. The Diffraction Radiation emitted is coherent at wavelength longer than the bunch length. The radiation can be extracted by a z-cut quartz window and propagated to a pyroelectric detector. From the analysis of the radiation integral a relative bunch length monitor can be obtained, operating in a parasitic way.

The layout of module 5 is shown in Fig. 92.

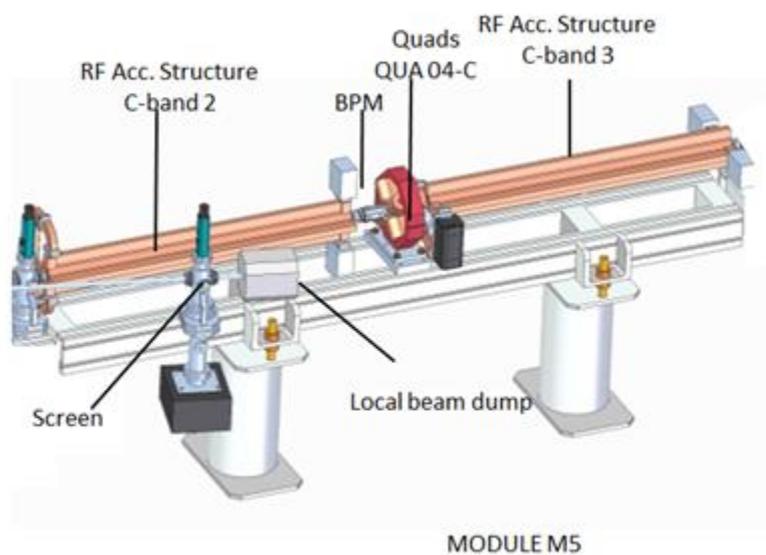

**Fig. 92.    Module 5 layout**

The diagnostic following the module 4 is quite standard. The main requirement is only to check the correct orbit inside the accelerating structure, using BPM measurements.

In module 5 (see Fig. 92) there are BPMs between C-band accelerating structures, and sometimes screens to measure also the beam size, even if this measurement is not fundamental at this point.

### 3.5.9. Module 6

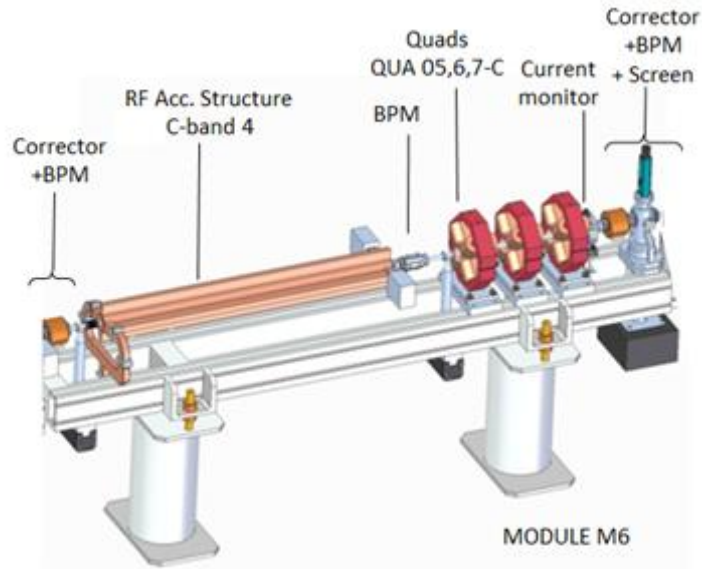

**Fig. 93.** Module 6 layout

The module 6 contains just the quadrupole to perform the emittance measurement at the end of the low energy region acceleration.

### 3.5.10. Module 7

The module 7 is the dogleg for the low energy branch. After that accelerating has been taken place, and before the beam extraction for the low energy branch, a complete characterization of the electron beam properties is needed.

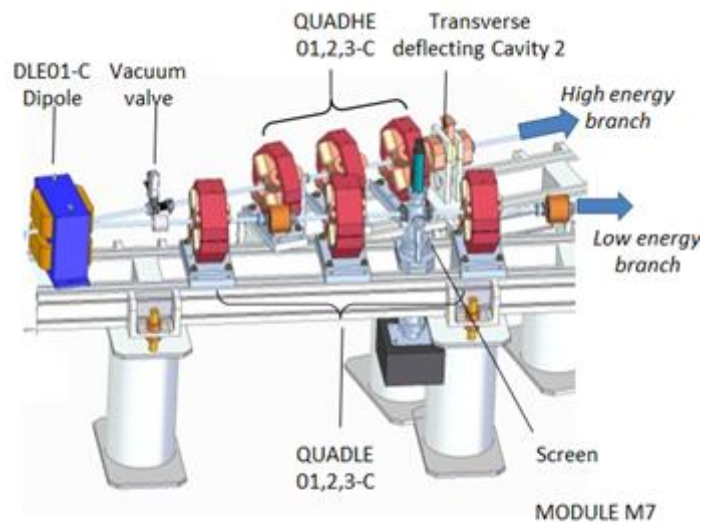

**Fig. 94.** Module 7 layout

The emittance, as well as the energy spread, the bunch length, and the longitudinal phase space can be measured in the same way already mentioned in section 3.5 (module 4/5). At this point the process of the compensation of the energy spread induced by the velocity bunching is completed and so the validation using the bunch length measurement and the energy spread is mandatory. A wall current monitor is also implemented in this area to check that all charges has been properly transported.

### 3.5.11.     Main Linac High Energy

In the high energy region linac the diagnostic plays a minor role. The trajectory is the only important parameter to be checked.

The module 9 is a simple duplication of module 5. After that accelerating has been taken place, and before the beam extraction for the low energy branch, a complete characterization of the electron beam properties is needed.

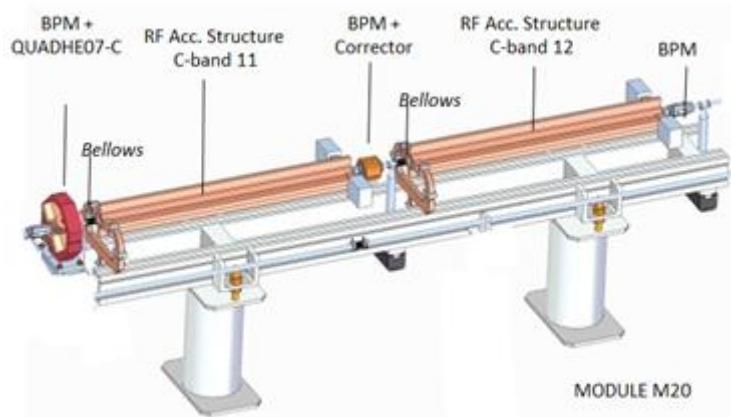

**Fig. 95.     Layout of typical section in high energy region**

In Fig. 95, the typical layout of the diagnostic in this area is shown. All the modules are the same. In front of every C band structure there is a BPM to measure the beam position. Every 2 or 4 C-band structure a screen can be placed to check the correct matching of the beam.

At the end of the whole line there is a triplet of quadrupole that can be used to measure the emittance at full energy in the downstream screen.

### 3.5.12.     High-Energy Doglegs

The high energy shallow angle dogleg downstream locates the beam in line with experimental end station fo ELI-NP-GS accelerator bay. Being a dispersive arm – like the low-energy module 7 – can be used to measure the energy and the energy spread. A beam position monitor is needed before the quadrupoles to properly set the beam on the magnetic axis.

The high energy dogleg angle is shallow and so the dogleg is long and therefore spans several support module (girder) lengths. It is shown in Fig. 96 below.

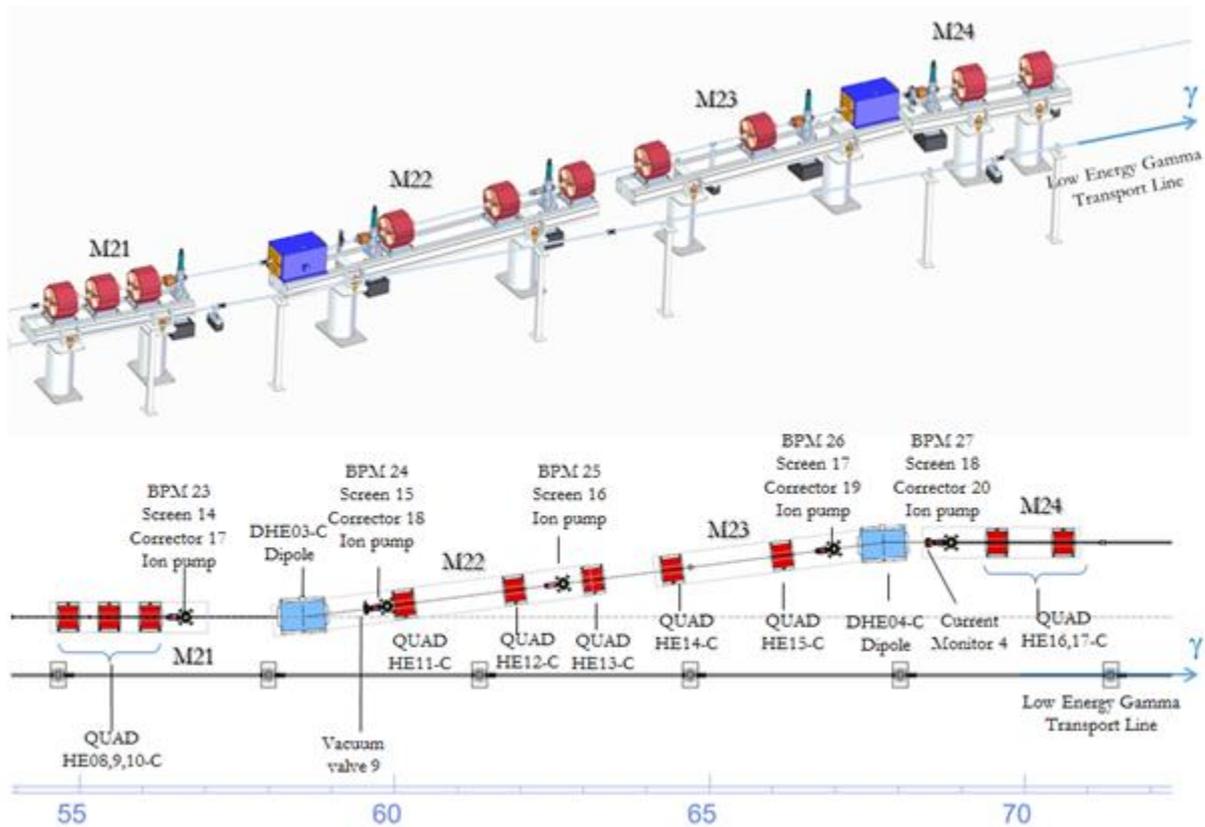

**Fig. 96.  Module 21-24 layout**

Two screens are placed in every dogleg. One is dedicated to the energy measurement, the other to check the beam spot before the last dipole in order to control that the beam is properly matched in the line.

### 3.5.13. Interaction regions

The two interaction regions (IR) will be supplied by the same instrumentation. A couple of cavity BPM will be placed, one before the IR and one after. In the IR a mechanical actuator will drive into the beam line two OTR screens, mounted at 45 degrees with respect to the beam line, and at 90 degrees with respect of each other.  A CCD camera, outside the vacuum chamber will image the two OTR screens as shown in Fig. 97, one at time.

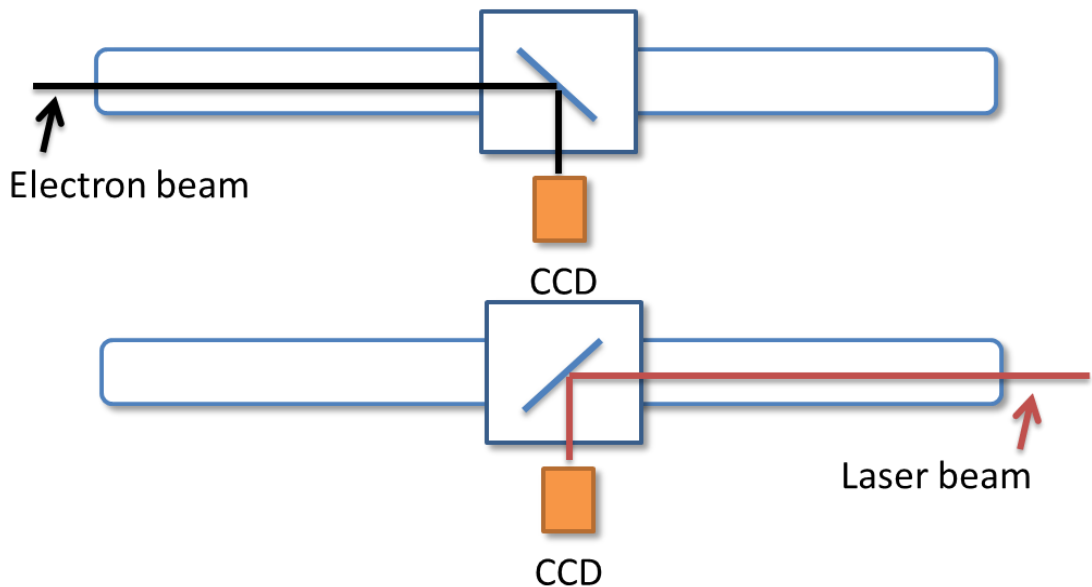

**Fig. 97.**   Schematic Layout of the optical diagnostic in the IR

In one of them we'll image the beam, in the second an attenuated laser pulse. A commercial lens with a magnification of 1:1 will be used for the camera, so we can achieve a resolution of about 10 µm. In such a way we can establish the position of the laser beam and correct the position electron beam in order to overlap the two beams in space. A Hamamatsu photodiode (example model G4176-03 MSM) can be used in the same position of the CCD to synchronize both the OTR signal coming from the beam and the light coming from the laser pulse.

## 3.6.  Vacuum system

### 3.6.1.  Introduction

The vacuum system of the ELI-NP project can be divided into a number of subsystems that require different vacuum specifications. In general, however, the design is based upon the extensive experience of the design team in delivering successful vacuum systems for facilities such as SPARC at INFN, Rome, Italy; Diamond Light Source near Oxford, UK and ALICE at Daresbury, UK.

**Table 24.**   Mean vacuum working pressure for each accelerator subsystem

| Subsystem | Dynamic Vacuum Level |
|---|---|
| RF Gun and S-Band Linac | $10^{-8}$ Pa |
| C-Band Linac 1 and 2 | $10^{-6}$ Pa |
| High and Low Energy Interaction Regions | $10^{-7}$ Pa |
| High and Low Energy Gamma Spectrometers | $10^{-6}$ Pa |
| Gamma Beamlines (Transfer Lines to Experimental Stations) | $10^{-6}$ Pa |
| Positron Target Region | Future Upgrade |
| RF Wave Guides | $10^{-6}$ Pa |

The range of vacuum requirements falls comfortably in the ultra-high vacuum region and therefore the general design principles for this vacuum regime will be adopted. The Vacuum Quality Assurance Documents[1] developed at Daresbury Laboratory, for modern accelerator applications, provide a good example of design principles that will be adopted for ELI-NP.

As this project will involve a number of different partners it is essential that integration of the vacuum systems, delivered by the partners, is carefully addressed. This function has been identified and will be the role of a specific work package for integration of the whole project including vacuum systems. Additionally it has been decided that the procurement of the vacuum equipment for the complete accelerator will be the responsibility of a single point of contact.

Key vacuum challenges for ELI-NP include:

- Achieving low pressure without extensive in-situ bakeout.
- Maintaining a contamination free environment, including some particle controls, particularly in the region of the RF-gun.
- Providing sufficient differential pumping where vacuum specifications vary by more than one order of magnitude between sections.
- Providing sufficient pumping for conductance limited beam pipes.

### 3.6.2. General Design Objectives

In any accelerator based project, it is inevitable that detailed consideration of the vacuum system comes some way down the line of the design process. The major reason for this is that a relatively detailed understanding of the mechanical layout of the machine and of the design of individual vacuum vessels/components is required before any final analysis of the pumping requirements can be made. At this stage in the ELI-NP project there are some areas where more detailed engineering is still required to finalise the design, this may result in some adjustments to the vacuum system but these are not expected to be significant.

The vacuum levels in an electron accelerator are such that it is not in fact pressure that is important per se; rather it is the number density of gas molecules inside the vacuum envelope. Number density plays an important role in two general types of process, viz., scattering and ionisation, and impact rate on surfaces.

For the ELI-NP project the basic accelerator system can be considered as a single pass machine and therefore a modest number density of gas molecules inside the vacuum envelope can be tolerated. Typically pressures in the region of $1 \times 10^{-6}$ Pa are expected to be sufficient to meet both the accelerator physics and engineering requirements. Vacuum requirements in the RF-gun are more stringent and will be considered separately. Although the pressure levels are generally quite modest, cleanliness criteria will still be stringent and will follow standard UHV practice. It is recommended that all components receive a full UHV cleaning procedure followed by a vacuum bake to 250oC for 24 hours before installation (ex-situ) where possible.

---

[1] These documents are available from STFC Daresbury Laboratory, Vacuum Science Group, contact joe.herbert@stfc.ac.uk

Differential pumping may be required to minimise gas transfer between various parts of the machine. This will be determined during the modelling stage of the project and is not expected to increase costs above that specified in the tender.

### 3.6.3. Vacuum Requirements for Specific Areas

#### 3.6.3.1 RF-gun and S-band Linac

The RF-gun vacuum represents the most challenging part of the accelerator vacuum system. This structure will operate with a very high electric field gradient (~120 MV/m) and high repetition rates (100 Hz) thus in order to decrease as much as possible the probability of discharge events the vacuum of this device has to be kept in the range of a few $10^{-8}$ Pa during operation. This low pressure is also required to keep the cathode surface free from contaminants that alter the photoemission properties of the cathode such as oxygen containing species.

The vacuum levels will be achieved using two 50 l/s sputter ion pumps connected to each of the two RF couplers. Additionally, to increase the pumping speed for hydrogen, carbon monoxide and carbon dioxide, a Non Evaporable Getter (NEG) cartridge pump will be used, connecting to a flange at the rear of the gun. The diagnostic chamber will be pumped by a 200 l/s sputter ion pump before injection into the accelerating structures. It will be necessary to carry out in-situ bakeout of this part of the accelerator vacuum system in order to reduce the outgassing sufficiently to achieve the low pressures required in this area. An ancillary mobile pumping system will be used for this task.

In order to establish that the components of the gun, etc., are in fact sufficiently clean prior to final assembly, a high specification partial pressure analyser will be used.

#### 3.6.3.2 Gamma Spectrometer

Although the required vacuum levels in this region are modest (~$10^{-6}$ Pa) there are some challenges that influence the vacuum system design. These include the following:

- Polimeric components are required in the vacuum envelope (plexiglass-PS).
- O-ring seals are required for feedthrough components.
- Large volume vacuum enclosures.

To deal with the above challenges a pumping system has been designed consisting of three modular oil free stations comprising scroll, turbomolecular and ion pumps with integral isolation valve system. The vacuum system is designed to allow some bakeout at a minimum of 100oC.

#### 3.6.3.3    Interaction Region

The requirements for vacuum in this region are modest with the basic requirement being to produce a working pressure suitable for providing a clean environment and one that will prevent gas phase particle interaction with the laser beam. Pump down will be done using the mobile pumping carts with sputter ion pumps providing the working pressure. Bakeout will not be possible in this region and so pre-cleaning of components will be essential. To allow for the vacuum transition between this and the other regions either side some differential pumping will be provided using sputter ion pumps.

#### 3.6.3.4    Positron Target Region

This will be a future upgrade for ELI-NP and not included in the scope of the initial project.

#### 3.6.3.5    Gamma Beamlines

Basic vacuum requirements will be provided using mobile carts and then sputter ion pumps. No special requirements for this region.

### 3.6.4.    Vacuum System Design Principles

#### 3.6.4.1    General

In general, the vacuum system design of ELI-NP is based on well-tried and well understood design principles. It is not a particularly demanding design, except for the RF-gun, but sufficient experience in operating guns of this type is well documented [56, 57].

Knife edge sealing is the default standard for bakeable vacuum systems of the type required for ELI-NP. However, in recent times, the reliability of VAT seals has been proven on a number of machines and it may be that their conformal inner surface geometry may be chosen to control the impedance characteristics of ELI-NP as the final detailed design is completed. To ensure that all vacuum equipment is compatible with each region of the accelerator complex it is important that flange connections are clearly specified. Unless otherwise specified all vacuum equipment purchased under work package 6 will use the ConFlatTM Flange (CF) for connecting to the vacuum system. This is the most common type of flange connection for ultra-high vacuum systems and uses the knife-edge principle to achieve an all-metal vacuum seal.

Use of clean laminar flow hoods should be adopted during installation to prevent significant particle contamination in the accelerator although more attention to particle control will be required for those parts that constitute the RF-gun vacuum system. These procedures are recommended good practice for the build and installation of this type of accelerator. It should be noted that it is not sufficient simply to ensure that initially the sensitive areas of the machine (RF-gun) are themselves particulate "free". Experience shows that particulates readily migrate through machines of similar construction, so that if there is a source of particulates somewhere, undesirable particulates will eventually turn up where they are not wanted. To reduce the migration of particulates through the machine it is important that the location of let-up and pump-down ports is carefully selected, essentially setting up a gas flow away from the areas sensitive to particulates. Additionally the rate of gas flow during both let-up and pump-down in the region close to the RF-gun should be restricted to prevent turbulence.

### 3.6.4.2 Vacuum Pumping

It is expected that ELI-NP will be conductance limited in most places, which will limit the pressures that can be achieved in the machine using a reasonable number of vacuum pumps. As usual, a number of iterations of machine layout and calculation of pressure distributions will be required before a satisfactory final scheme of vacuum pumping can be determined. At this stage an estimate has been made based upon previous experience. It is anticipated that at most the position of pumps may need to be modified to meet the vacuum specification.

Since the whole machine is sensitive to hydrocarbon contamination to a greater or lesser degree, rough pumping will not use any oil-sealed pumps. Pre-pumping will use proven scroll pumps for a good balance of pumping throughput and costs, and high vacuum pumping will use clean turbomolecular pumps. These pump sets will be mounted on roughing carts that can be moved into position when required. Similar pumping schemes have been used successfully in the past such as for the ALICE facility at Daresbury Laboratory [58].

Main UHV pumping will be by sputter ion pumps, supplemented by NEG cartridge pumps in the RF-gun area. The ion pump power supplies will be located in the main control rack room away from the accelerator complex. To reduce costs, power supplies with multiple outlets will be used. The disadvantage of this approach is that the pressure indication for each individual pump is compromised such that it cannot be relied upon as a true indication of pressure. However, this approach reduces costs significantly and has been used successfully elsewhere.

### 3.6.4.3 Pressure measurement

Adequate pressure measurement performance will be obtained by using Pirani Gauges and Inverted Magnetron Gauges supplemented by information from ion pump power supplies and residual gas analysers.

A Residual Gas Analyser (RGA) will be placed on the RF-gun vacuum system and at a few strategic locations throughout the accelerator complex. Otherwise, RGA facilities will be mounted on the mobile roughing carts where they will be used mainly for leak testing and monitoring of initial cleanliness of the systems.

The total pressure gauge controllers will be located in the main control rack room and so long cables will be required to connect these to the gauge heads, the correct specification of cable will be specified for this purpose to minimise noise and interference of the gauge operation.

For the residual gas analyser the main electronic unit will be located close to the analyser head but sufficiently decoupled (via an RF cable extender) to avoid unwanted radiation damage or high magnetic fields.

**3.6.4.4      Valves**

Gate valves are required to provide sectorisation of the complete ELI-NP vacuum system for practical reasons and for machine protection. Dividing the machine into discrete vacuum sections makes it easier to install and commission whilst at the same time simplifying maintenance and breakdown interventions. At the same time it is important to keep the number of gate valves at a minimum to reduce both costs and impedance.

In most circumstances gate valves will be all-metal sealed and RF lined to reduce impedance.

Roughing valves (right angled valves) and let up valves will be situated at convenient points consistent with the requirements for minimising particulate transport mentioned earlier. These will be an all-metal construction and for pump-down will be sized as DN63.

**3.6.4.5      Bakeout**

Most of the machine will not be baked in situ, although all warm-bore parts of the machine will be baked prior to installation as part of the conditioning and cleaning process. However, the system is designed such that in-situ bakeout may be used if necessary such as to improve vacuum levels after interventions. One exception to this is the RF-gun vacuum system, this does require in-situ bakeout to at least 150o C, suitable facilities will be purchased and installed for this.

**3.6.4.6      Vacuum Control System**

A full system of vacuum controls will be installed on ELI-NP, providing monitoring, automation, alarms and safety protection. Specific details of the control system are included in the controls section of this document (see section 3.7).

# 3.7.   Control System

In a complex system like ELI-NP an unique control system will manage the whole machine, from the gun to the radiation beam-lines. This means that the control system must be able to execute commands on all the active elements and control all the diagnostic devices, giving the needed information to people operating the accelerator. Furthermore, it has to be easy to upgrade the system substituting old elements or introducing new ones.

### 3.7.1.      General Description

As above mentioned the control system should guarantee and simplify machine operation. In general the main operations in an accelerator control system are:

- data taking,
- display of information,

- data analysis,
- command execution,
- storage,
- automatic operation
- alarms managing

The control system could be made of processors distributed on the machine area with a three levels architecture.

- First level: at this level we find the console with their human interface to allow the operator to control the machine; an electronic logbook to share information within the collaboration; a database to store all information coming from the accelerator; web tools to help the management of the control system and to share some information outside the collaboration;
- Second level: at this level we find the front-end CPU's that execute commands and make all the information about the status of the machine available to the first level. Meanwhile it automatically saves data from its various elements in two ways: on value changes, or at fixed time intervals;
- Third level is the acquisition hardware where we find the appropriate acquisition board or the secondary field bus to acquire data from the real element.

The interconnection bus between the levels is an appropriate communication bus.

### 3.7.2. Hardware

Each distributed CPU has to control only a certain type of elements. This simplifies the number and type of acquisition board assigned to the front-end processor. The CPU used will depend on the element to be controlled. We plan to implement in our system PXI boards, industrial PC's and real time processors.

At console level we need the maximum flexibility in terms of number of screens and possible remote connections. At this level, we plan to use small form factor PCs with at least 3 monitors each.

A storage and backup facility for the whole system software and data is foreseen.

Each CPU must be interconnected via an Ethernet network with the necessary bandwidth, at least 1 Gbit, and the required number of connections.

### 3.7.3. Software

In order to reduce the time of development of the ELI-NP control system, we propose to use well known Rapid Application Development (RAD) software. It is very important that experts from various fields can develop their own control windows based on the control system. Among the software to be integrated into the control system, due to their diffusion, there are Labview and Matlab.

### 3.7.4. Service Programs

The ELI-NP collaboration involves different national and international research institutions. Some services are necessary to make all the information about the machine status and the work progress available to all involved people. The old system based on a paper logbook where the operator writes the data and glues pictures can be useful but may not be available to remote researchers. One good example is the logbook developed in SPARC. This logbook is on PostgresSQL database. The integration of the logbook system in the user interface helps the users in inserting entries and data in simple and homogeneous way.

During machine operation automatic data saving is mandatory. This mechanism can be useful both for machine maintenance and for offline analysis.

Acquired data will be recorded in a database with a possibility to communicate via TCP/IP.

### 3.7.5. Elements

The ELI-NP machine can be divided in three parts: the LINAC, the Laser and the radiation lines. Each part of the machine has its elements to be controlled or displayed.

### 3.7.6. Laser

To allow the optimization of the laser on the cathode and in the interaction point, its remote control is essential. In order to implement this we need to completely control the components of the laser apparatus. The main operation is the alignment of the light with the mirror that can be controlled by means of a motor while the laser light is acquired through a video camera. Other parts can be controlled by means of standard interfaces such as Ethernet or serial.

### 3.7.7. RF

The RF section includes controls for the high power and the low power apparatus. The modulator can be controlled with Ethernet or serial interface with an appropriate protocol.

Signal monitoring and synchronization can also be controlled via Ethernet interface or using a demodulation board and digitizer cards in an PXI or compactPCI chassis, where data analysis and device control are accomplished. The signal apparatus can be seen as a custom multi-channel digital scope, able to display in the control room all the demodulated signals coming from the RF structures placed along the whole machine.

Other devices such as attenuators, phase shifters, amplifiers and so on will be controlled with serial or Ethernet interfaces.

### 3.7.8. Magnets

In the accelerator we have different kind of magnets such as solenoids, correctors, dipoles and quadrupoles. Magnet control means controlling their power supplies. We decided to use as much as possible the same interface between the control and the power supply.

The following specifications will be required: Ethernet, RS232 or RS485 interface; the well-defined communication protocol Modbus as standard interface protocol.

### 3.7.9. Vacuum

In the whole machine we need to display and control vacuum through different kinds of apparatus.

#### 3.7.9.1 Pumps

The most widely used pumping systems for ultra-high vacuum (UHV) will be a combination of ion sputter pumps (briefly ion pumps) and titanium sublimation pumps (TSP). Control units for these pumps usually include a serial (RS-232 or RS-485) interface for remote control and a set of logical switches for alarm output to the control system. The brand and model of these controls, as previously indicated, should be the same for all the machine, in order to avoid duplication of software and spare parts. For particular sections (those requiring gas differential pumping) turbomolecular pumps with scrolling fore pumps will be required whose control units are again equipped with serial ports and switches. Turbo and scroll pump combinations will be also required for first stage of evacuation of each section, but in this case a remote control should not be needed.

#### 3.7.9.2 Vacuum gauges

The vacuum gauges allow an accurate measure of the vacuum in different point of the machine. The most widely used ion gauges controls are equipped with serial ports or a FieldBus equipped with analog and digital channels. Unfortunately, at the time of this writing, these gauges are designed for industrial applications, and the required pressure range needed for radiation beam lines applications (in the $10^{-9}$ Pa scale) is not achieved, so the serial approach would be chosen. Some sections will require low vacuum gauges, such as thermocouples or Pirani heads, also generally equipped with control units having serial interfaces.

#### 3.7.9.3 Valve controls

All the valves should be electro-pneumatic gate type. The control units, generally home-made, will include a solid state switch for valve opening/closing and a couple of logical switches for valve status monitoring. All the valve control units will be logically (and in some case physically) connected to the vacuum gauge, ion pump and cooling system controls in order to avoid valve opening in unsafe vacuum/cooling conditions and to close them automatically in case of vacuum/cooling alarms. Valves belonging to the switching mirror chambers will be also connected to the chamber position encoders, so that valves may be only opened on safe beam path conditions. Also, the correct opening and closing sequence will be ensured in order to avoid radiation directly hitting the front-end valves without the beam stoppers inserted. All these conditions imply hardware controls for the safety-related situations and realtime software controls in the other cases.

One special valve, designed for protecting the accelerator and the undulator from sudden venting of the beam lines, is a fast valve, which has its own sensor and control unit.

### 3.7.10. Diagnostic

A part of the machine parameter (emittance, bunch length and energy) will be measured with images. The use of a versatile camera system is strategic in the realization of the corresponding diagnostic. The rapid evolution in the image acquisition systems allows us to choose the camera and its own interface in a wide variety of products. The IEEE1394 or GvisionEth protocol gives us the possibility to interface different cameras with different specifications without changing the acquisition program. The cameras are acquired by different distributed personal computers that send data through a TCP/IP channel. The data transfer structure will be well defined to allow full integration of all cameras inside the control system.

Another important component in the diagnostic is the control of motors to move flags and slits to allow the acquisition of the beam image. The flag movement is allowed by the stepper motors.

The beam position and charge monitors depend on the pickup that we use. In general for this two kind of diagnostic we can use analog acquisition board with appropriate signal conditioning.

### 3.7.11. Motion and positioning

Most of the chambers holding optical elements namely mirror gratings target and slits, coupled with suitable position encoders. Depending on the accuracy required for positioning, stepper motors or CC motors will be used, coupled with optical encoders or potentiometric transducers respectively. In both cases, motor embedded encoders will not be used, relying on the motor reproducibility and on the external encoders. Motor control units are currently available with the most common interfaces such as Ethernet, CANopen, Profibus-DP, serial ports and so on. The latter three are also generally available for potentiometric transducers. Each motor will be equipped with a couple of switches (end of travel) to be used for both motor stopping and alarm generation.

## 3.8. Timing and Synchronization

### 3.8.1. General overview

To fulfill the synchronization requests for the ELI-NP gamma production project we designed a system using the state-of-the-art technology used also in recent photo-injectors for FELs.

As mentioned before in this document, the relative time jitter of electron bunches respect to the photon pulses should be less than 500fs RMS. This can be achieved in a large scale infrastructure by means of an optical signal reference distribution system. This kind of system is described in the following paragraphs and most parts of it are nowadays commercially available. A general scheme of the system is reported below in Fig. 98.

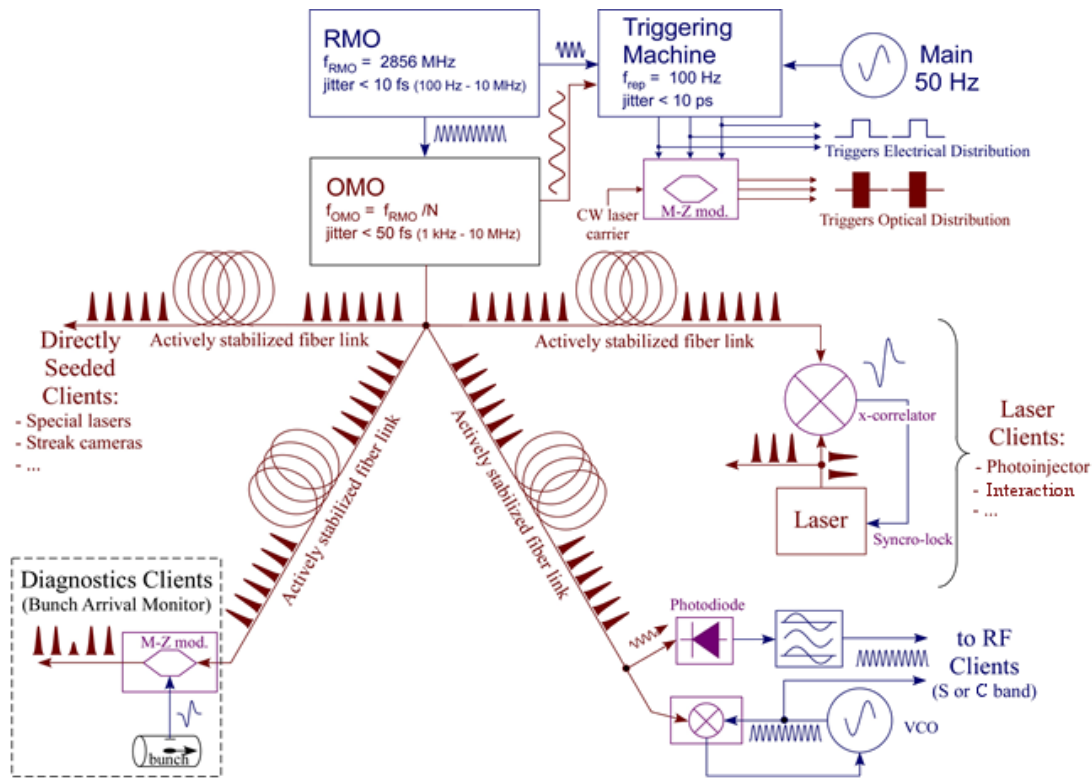

**Fig. 98.    Timing and synchronization system layout**

The timing and synchronization system consists of three main parts:

- Timing generation and distribution. An ultra-stable reference signal generated in a central timing station will be distributed to the various clients through actively stabilized links. Due to the remarkable link lengths, an optical reference will be distributed to exploit the fibre-link low attenuation and the large sensitivity obtainable by optical based timing detection. The required stability of each link is ≈ 70 fs over any time scale.

- Client synchronization. Each individual client (laser systems, RF power stations, beam diagnostics hardware...) has to be locked to the local reference provided by the timing distribution systems. The lock technique depends on the particular client (laser optical cavity PLL, multi-pulse structure generation inside the macro-pulse, RF pulse-to-pulse or intra-pulse phase feedbacks,...).

- Client triggering. Together with a continuous reference signal, low repetition rate trigger signals (100Hz or less) must be provided to some clients, which contain essentially the information on the timing of the macro pulses needed to prepare all the systems to produce and monitor the bunches and the radiation pulses (laser amplification pumps, klystron HV video pulses, beam diagnostics, ...). The triggering system is a coarser timing line (≈10ps stability) that can be distributed either optically (through fibre-links) or electrically (through coaxial cables). Slight variations in the listed frequencies are possible without changing the system architecture.

## 3.8.2. Reference Master Oscillator

The reference signal is originated by a Reference Master Oscillator (RMO) which is a µ-wave crystal oscillator with ultra-low phase noise characteristics. The role of this device is to provide a reliable reference tone to an Optical Master Oscillator which is a highly stable fiber-laser that encodes the reference timing information in the repetition rate of short optical pulse in the IR spectrum.

The RMO guarantees the long term stability of the OMO, and, through the OMO locking system, imprints its low-frequency noise figure to the whole facility timing line.

The state of the art low-noise µ-wave oscillators can provide pure sine tones with phase jitter at few fs level over a spectral range from 10Hz to 10MHz. As an example, the phase noise SSB spectra of sapphire oscillators commercially available from Poseidon Scientific Instrumentation (PSI) are reported in Fig. 99.

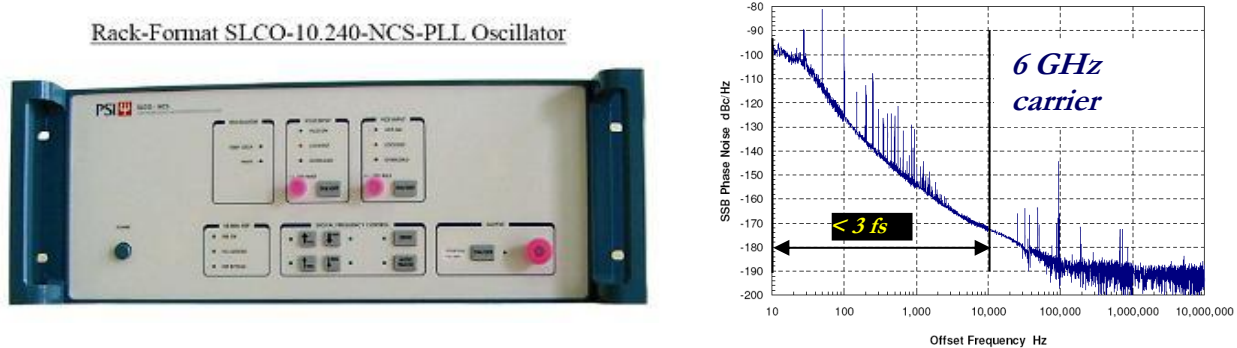

**Fig. 99.**   Ultra-low phase noise sapphire oscillator

## 3.8.3. Optical Master Oscillator

The timing reference will be encoded in an optical signal before being distributed over the whole facility through a glass fiber network. The µ-wave to optical conversion is accomplished by mode-locking a low noise fiber laser (the facility Optical Master Oscillator – OMO) to the RMO. The OMO to RMO synchronization is obtained by a locking system, which consists in a PLL scheme controlling the path length of the fiber laser cavity by stretching the fiber with piezo-controllers driven by the relative phase error between the two oscillators. This is a standard technique to synchronize also in-air laser oscillators to external references, with the piezos controlling the position of one or more mirrors in this case. Due to the limited frequency response of piezo-controllers, the locking loop gain rolls off typically around 5 kHz. Above this cut-off frequency the OMO retains its typical noise spectral properties, while below the cut-off frequency the OMO phase follows the RMO one, and the spectra of the two oscillators result to be very similar. However, the intrinsic phase noise spectrum of a good fiber laser oscillator above the locking cut-off frequency is comparable or even better respect to that of a µ-wave reference oscillator.

The overall phase noise of fibre laser OMO fairly locked to a high-class RMO can be as low as ≈100fs in a wide spectral region spanning from 10Hz to 10MHz. The specifications of the OMO are reported in Table 25.

**Table 25.   Typical OMO specifications**

| Pulse width | τpulse | < 200 fs |
|---|---|---|
| Wavelength 1 | λ1 | 1560 nm |
| Wavelength 2 | λ2 | 780 nm |
| Pulse rep rate | Frep | ~80 MHz |
| Pulse energy | Epulse | > 2 nJ (~ 180 mW) |
| Phase jitter | τrms | < 100 fs rms<br>(SSB Δf > 1 kHz) |
| Amplitude jitter | (Δ A/A)rms | < 0.05 % rms |
| Synchrolock BW | Fcutoff | > 5 kHz |
| Phase jitter relative to reference | τrel | < 10 fs rms<br>(dc – 1 kHz) |

The OMO pulse repetition rate specified in Table 25 is around 80MHz, corresponding to the 1/N of the linac S band RF frequency 2856MHz. One can choose an optimal number depending on the manufacturer availability and to the system requirements. A convenient choice could be 79.33MHz that corresponds to the S band RF/36.

Fibre lasers with characteristics close to the OMO specifications are already available on the market, needing only a limited customization to fully meet our requirements.

### 3.8.4.    Reference Signal Generation and Distribution

The distribution of the synchronization reference for ELI-NP must be realized using multi fiber optic channels. The source of the temporal reference signals for all the sub-systems is foreseen to be generated by an optical master oscillator (OMO) situated at a median position of the facility infrastructure. The channels extend from the source to the several end users up to 300 meters distance. The choice of optical waveguide channel to distribute the synchronization signals is motivated by a number of advantages of the optical link respect to the electrical cables. In fact the optical fibers offer THz bandwidth, immunity to electromagnetic interference and very low attenuation. These properties make the optical fiber a mandatory technology when one aims to synchronize a large scale facility at sub 100fs level.

The layout of the synchronization source system is schematized in Fig. 100. The OMO consists of a fiber optic oscillator generating soliton pulses. The cavity is designed to operate at repetition around 80MHz as explained before. In a fiber oscillator the active medium is a portion of the fiber doped by Erbium and is, generally, excited through fiber-coupled diode. In the cavity, a portion of the optical path is in free space to allow the active control of the optical length. The fact that the cavity and the output coupling are realized within an optical waveguide determines a net improvement in the stability respect to a free space propagation laser. The output spectrum is centered at 1560 nm, because it is the standard wavelength for the telecom applications and we can use low cost commercially available technology.

The soliton passive mode locking OMO offer the possibility to produce and distribute a short pulse over long distance. After the oscillator, the signal level needs to be augmented using an erbium doped fiber amplifier (EDFA) to achieve a proper power. At this point a fiber beam splitter will be used then to divide the signal to the different synchronization channels.

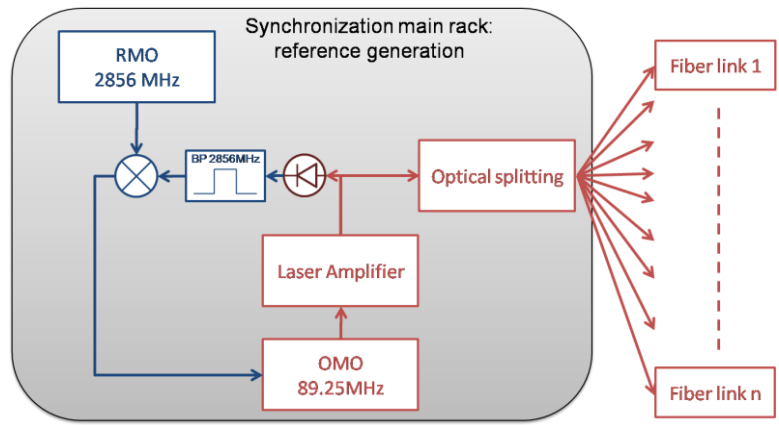

**Fig. 100.** Schematics of the synchronization reference source and distribution

The oscillator plus amplifier chain will be synchronized using a phase lock loop respect to the 2856 MHz μ-wave reference synthesizer. At the output of EDFA the signal recorded from a fast photodiode (with the necessary bandwidth > 3GHz) will be sent to a narrow band pass filter at the 2856 MHz and the result signal is then mixed with the reference to measure the pulse to pulse phase noise and generate the feedback toward the OMO. The oscillator repetition rate is kept constant using a piezoelectric actuator to drive a cavity mirror on the optical path of the cavity in air. The present status of the technology allows a jitter much lower than 100 fs rms. Since possible time jitters between the RF oscillator and the OMO are common mode disturb for all the injector subsystem, it is important to stress that the synchronization of the OMO is required in order to keep stable operating frequency on long term scale.

The optical signal will be sent to several fiber links with equalized optical length. The fiber optics will to be length-stabilized in order to deliver synchronous pulse to the end users. In fact the time of propagation in the fiber is in general affected by temperature drift and acoustic noise. The stabilization is achieved sending back a pulse from the end of the fiber and optically compare it with another pulse from the oscillator as schematized in Fig. 101. To reflect part of the power a Faraday mirror can be used. In details the pulse to send through the fiber is divided with a polarization beam splitter: part of the pulse is sent to the fiber and the other is used as a reference for the optical mixing with the retroreflected pulse. The comparison between the reference and the other pulse is carried out with cross-correlator. The error signal will be used to drive a piezo path length stabilization unit and a stepper actuator to move trombone. The feedback aims to keep constant the pulse's propagation length.

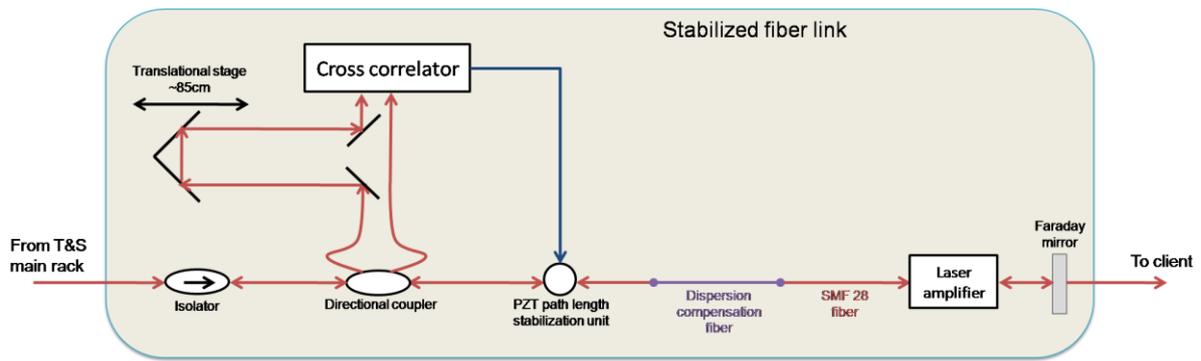

**Fig. 101.** Layout of a path length stabilized optical link

A balanced cross-correlator is preferred for suppressing the laser amplitude noise and operates on a linear response of the optical mixer. This technique demonstrated a stabilization better than 10fs on 300 meter channel. For low demanding end user the optical channel can be stabilized using electronic techniques based on high-speed photo detectors and microwave mixers. The limited timing resolution limits the stabilization to 100fs level. The cross-correlation or the electronic mixing requires more than half of the input pulse energy is not sent to the user. This set the required power per channel at level of 50mW.

The propagation over hundred meters fiber induces a pulse broadening due to several mechanisms. The pulse broadening reduces the synchronization resolution for the end user and cross-correlator. Modal dispersion introduces a optical pulse broadening an unwanted effect that can be overcame with single mode fibers. A possible single mode fiber is the SMF-28. In fact this fiber is characterized by very limited attenuation: at 1560 nm the energy loss is 0.25db/km that corresponds to an attenuation over 300 meters of less than 2% of power. Unfortunately this fiber shows a quite large dispersion 18ps/(nm*km) that corresponds to a pulse broadening up to more than 130ps for the typical oscillator bandwidth (25 nm) over 300 meters. To compensate the normal dispersion of the SMF-28 it is possible to use a fiber characterized by anomalous dispersion, named dispersion compensating fiber (DCF). A proper length DCF can completely balance the dispersion introduced by the normal dispersion fiber and produce a transform limited pulse to the end user and for the cross-correlation.

The price to pay for using a hybrid typology of fibers, are larger losses. In fact, the typical attenuation found in a DCF is three times higher than in standard fiber and other losses are induced by the splicing losses between the two fibers. If one includes also the losses for match the fiber from the trombone delay line, the transmitted signal results to be very attenuated (up to 10 dB). To recover the energy losses at the far end of the optical link it is advisable to use an EDFA. The amplifier assures a good level of power for the final user and the cross-correlation. Insignificant time jitter is expected from the EDFA also because it is in the stabilized loop.

For better environment housing the OMO, the amplifier, the splitter and the fiber stabilizer should be controlled at better of ± 0.1 °C rms with a moderate air flow velocity. The stabilization of temperature and an optical table to dump possible acoustic noise is required when a total sub-100fs are the target stabilization level. Moreover the optical fiber must be housed in a temperature controlled pipe (±1 °C) to reduce the thermal effect during the propagation.

Finally the exposure of the optical link to radiation and particles induce a fiber darkening and therefore needs to be minimized.

### 3.8.5. Synchronisation of Laser Clients

In the ELI-NP photo-injector, a precise synchronization of the photocathode drive laser is necessary to have a fixed and stable time-of-arrival of the photons on the cathode with respect to the 2856MHz and 5712MHz RF field. This condition is very important to guarantee stability and shot-to-shot reproducibility of crucial beam parameters such as bunch charge, energy, energy spread, beam emittance, bunch time of flight and time of arrival across the linac, and to ensure a proper matching condition in the accelerator.

Other laser systems will operate at ELI-NP facility for interaction and for electron beam diagnostic. All the above laser sources are required to be synchronous with the RF field and the electron bunch. In particular the interaction laser is specified to be synchronized with the electron bunch time of arrival within 500fs RMS.

The RF frequency for the ELI-NP accelerator has been chosen to be the standard SLAC S-band of 2856 MHz, but typical laser oscillator repetition rates are between 70 and 100 MHz, so that a Nth S band RF sub-harmonics should be chosen for ELI-NP laser oscillators (36th sub-harmonic corresponding to f=79.33MHz could be a reliable choice). A proper technique has to be adopted to compare the low and high frequencies to retrieve the phase error that is used to implement active synchronization loops. In fact, whenever various clients whose intrinsic repetition rate is the $N^{th}$ sub-harmonic of the reference frequency are directly synchronized to the reference, their relative timing can differ by an integer number of RF periods. To avoid that, the synchronization has to be carried out also at the fundamental client frequency, to make different systems virtually overlap in time and space.

The synchronization between laser and RF fields has to be achieved at two different levels:

a) by controlling that the laser emission happens at the time when the high power RF fills all the accelerating cavities (coarse timing – device triggering) and
b) by phase-locking the laser optical oscillator with a the reference signal also at the Nth reference sub-harmonic, for real sub-ps synchronization (fine timing – device synchronization).

All the ELI-NP laser system have the same topology: an high repetition rate oscillator which provide pulse rate around 80 MHz with few nJ energy and a low repetition rate 10-100 Hz high energy amplifier. Also an intermediate frequency around 50MHz should be provided to the systems to generate the intra-pulse train with 15-20ns separation. For more detail see the TDR chapter dedicated to the lasers. A first fine synchronization is carried out mainly at the laser oscillator level. The following amplification and manipulation introduce practically only a fixed delay. Fluctuations and drifts in the time of arrival of the laser on the cathode are mostly due to acoustic vibrations and/or thermal gradients, leading to slow drifts of the laser optical path that can be compensated by shifting the phase of the laser oscillator reference signal. Slow drifts among different laser systems can be compensated with motorized optical delay lines.

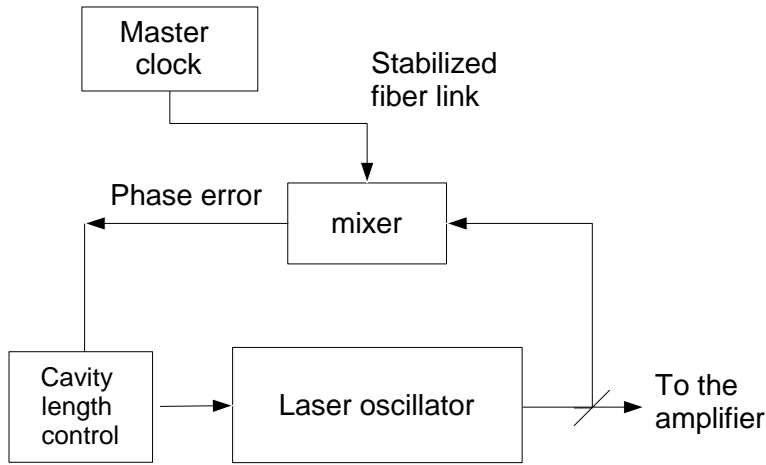

**Fig. 102.  Schematic of the laser oscillator phase locked loop**

### 3.8.6.  Laser systems synchronization

All the ELI-NP laser systems have the same topology: an high repetition rate oscillator which provide pulse rate around 80 MHz with few nJ energy and a low repetition rate 10-100 Hz high energy amplifier. Also an intermediate frequency around 50MHz should be provided to the systems to generate the intra-pulse train with 15-20ns separation. For more detail see the TDR chapter dedicated to the lasers. A first fine synchronization is carried out mainly at the laser oscillator level. The following amplification and manipulation introduce practically only a fixed delay. Fluctuations and drifts in the time of arrival of the laser on the cathode are mostly due to acoustic vibrations and/or thermal gradients, leading to slow drifts of the laser optical path that can be compensated by shifting the phase of the laser oscillator reference signal. Slow drifts among different laser systems can be compensated with motorized optical delay lines.

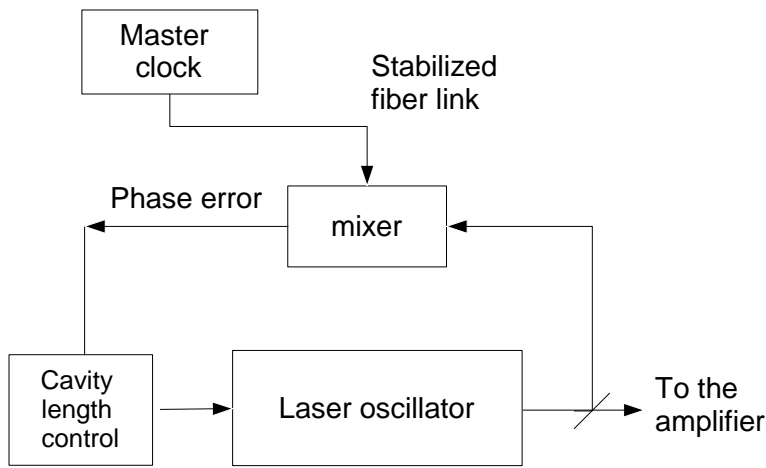

**Fig. 103.  Schematic of the laser oscillator phase lock loop**

The repetition rate of the laser oscillator is equal to c/2L, where c is the speed of the light and L is the optical length of the cavity. In our case by design the cavity length L should give a laser pulse frequency around 80MHz, Nth sub-harmonic of the linac S band RF frequency. The chosen frequency is also a multiple of the facility repetition rate (100 Hz nominal).

The synchronization is achieved through an active feedback loop driven by the laser to reference phase error. In Fig. 103 it is reported the conceptual scheme of the synchronization loop. The reference signal is generated by an optical master oscillator (OMO) and transmitted to the end users by a stabilized fibre optic link. The phase error can be measured either with electronic or optical mixing. The electronic mixer requires two photodiodes, not drawn in the figure, as opto-electrical transducers: one placed at the laser oscillator exit and the other terminating the fibre link. The signals from the photodiodes can be filtered with a < 1 % bandwidth bandpass filter centred at 2856 MHz, in order to reject all the harmonics but the Nth and drive the mixer directly at the linac RF frequency. The use of an electronic harmonic loop (i.e. working at a frequency higher than the pulse repetition rate) increases the phase detection sensitivity. The only requirement is that the photodiode bandwidth extends to the chosen loop operational frequency. The intensity of the laser beam on the photodiode has to be optimized to reduce the amplitude-to-phase conversion present typically in the photodiode response near saturation.

As already mentioned, the ambiguity associated to the use of an harmonic loop derives from possibility of locking the laser at any of the N intermediate temporal positions corresponding to the harmonic periods within a period of the laser repetition frequency. This is not relevant whenever the laser has only to be locked to the RF field because the laser-to-RF synchronization is always guaranteed. But since the laser has to be synchronized also with other 80MHz lasers, the use of a fundamental loop (i.e. working at the laser repetition frequency) in addition to the more sensitive harmonic loops is necessary.

Electric phase noise measurement can be performed with a resolution of ≈ 0.01 degree, which corresponds to ≈ 10fs at 2856MHz.

Optical mixing is obtained by mean of cross-correlators. In this kind of devices one pulse from the laser oscillator and one from the OMO have to overlap inside a BBO (Beta Barium Borate $BaB_2O_4$) non-linear crystal at the fundamental repetition rate. It is possible to use either the fundamental wavelength at (1560 nm) or the second harmonic (780 nm) of the OMO for the optical mixing with the wavelength from the laser oscillator. The frequency sum pulse energy is proportional to the correlation signal calculated for the relative delay τ. Using a standard DC photodiode, it is possible to measure the pulse to pulse time jitter recording the variation of intensity of the sum optical frequencies. To reduce the effect of input amplitude jitter the cross-correlator is normalized to the intensity of the two input pulses. Recently it has been proposed a balanced cross-correlation configuration suppressing the effects of laser amplitude noise, showing also a linear characteristic response around the zero crossing.

The main advantages of the optical mixing respect to the electronic one are the intrinsic higher resolution and the need of less sophisticated acquisition and manipulation electronics (lower frequencies, 0-100 MHz). The cross-correlation allows resolving jitter with resolution that depends on the diode sensitivity and the input pulse length. The pulse from the oscillator is of the order of 100fs rms and the signal coming out the fiber can be compressed to the same duration. This results in a resolution on the scale of the fs or even lower.

The measured phase noise error is used to drive actuator to active control the oscillator's cavity length. In general high frequency piezo-electric transducers (PZT) are associated to a lower frequency stepper motor driven optical delay line. One cavity mirror is mounted on the PZT for high resolution frequency control. It has a typical range of few μm for voltages up to 100 V corresponding to a regulation range of few hundred Hz. The PZT frequency response extends up to 5-10 kHz. For long term stability the stepper motor can be used to slowly adjust another mirror that changes the optical path. This broadens the range of frequency correction of the PZT and allows operating the piezo-motor at its optimal DC bias voltage. The commercial active synchronization units guarantee the locking for temperature drifts of ≈ 10 degrees in a frequency range of 10 kHz.

Another fine synchronization has to be made to achieve the repetition rate in the laser amplification path. This repetition rate around 50MHz should be provided to the laser systems isolating the Mth sub-harmonic of the 2856MHz S band frequency. A convenient choice could be 57.12MHz (M=50), that yields to a pulse separation of 17.5ns, meeting the required repetition rate. The maximum limit of 40 pulses can be maintained since 17.5ns x 40 = 700ns, so the pulses can be accelerated also in the C band RF macro-pulse that lasts 1μs. Every laser system has its own method to lock the rep. rate to that frequency, as described in the laser section of this TDR.

The $N^{th}$ and $M^{th}$ sub-harmonics should be synchronous with the 100Hz repetition rate of the machine timing in order to ensure that the pulse train is placed in the correct macro-pulse region and the operation is then shot-to-shot repeatable.

### 3.8.7. Synchronization of the RF clients

The problem of keeping all the RF fields interacting with the beam always well synchronized among them and to the machine reference is obviously crucial to obtain and preserve the beam quality.

The RF driving signal for all the power sources (S band and C band) will be locally extracted from the OMO reference transported to each station by stabilized phase links. The optical to electrical conversion will be accomplished by photodiodes with proper bandwidth and/or Sagnac loop (a PLL with an electro-optical phase detection system insensitive to laser amplitude jitter).

Samples of the RF voltages will be available all along the linac from directional couplers placed near the accelerating sections input/output coupling ports, and from RF probes located inside the standing wave cavities (RF gun, SW RF deflectors...) to monitor the resonant fields inside. Directional couplers will be placed also near the output window of the klystrons to directly monitor the forward and reflected power on the tubes.

The low-level RF control will allow to demodulate some of the monitored RF pulses to work out amplitude and phase of the RF voltages, and monitor them on a pulse-to-pulse base at the rate of 100 Hz. Drifts and slow fluctuations of the monitored signals will be detected, and consequent actions to restore the optimal RF synchronization conditions will be done, essentially moving some variable phase shifters placed in the low-level hardware or in the waveguide network for independent phase control of different network arms. A sketch of the low-level control for a standard ELI-NP RF power station is shown in Fig. 104.

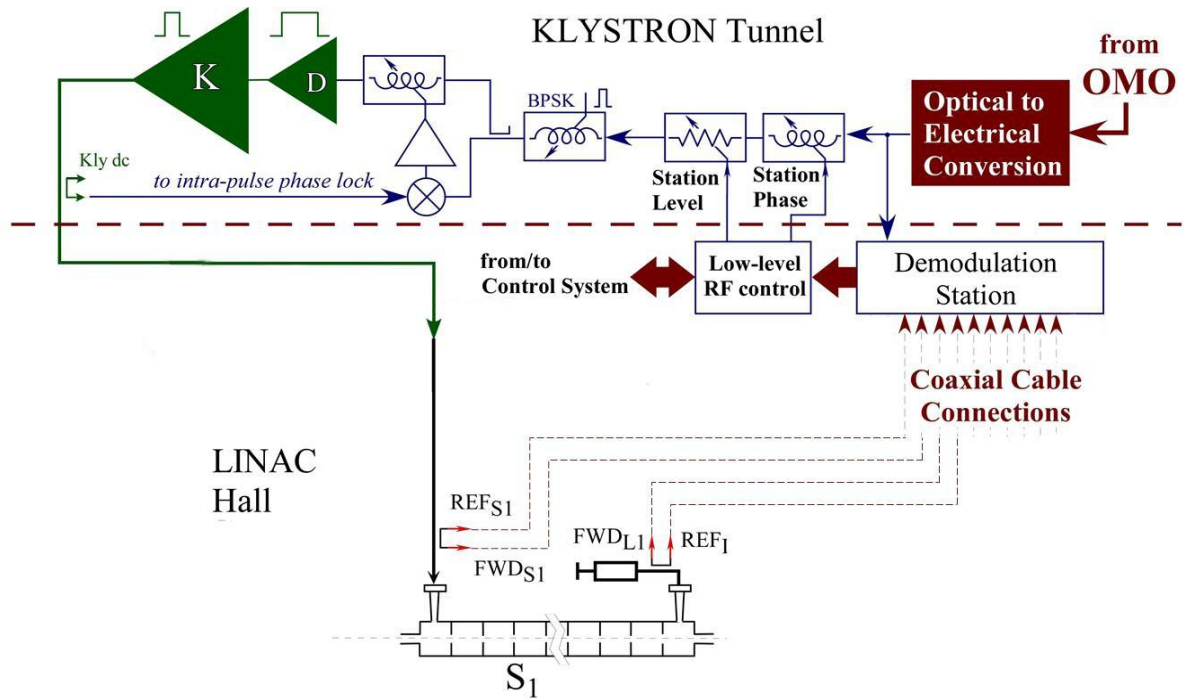

**Fig. 104.    Sketch of the ELI-NP single power station**

The reference RF signal to drive the whole RF chain is obtained from optical-to-electrical conversion of the OMO signal. The same reference is used to demodulate various RF pulses sampled over the network, and the whole station is re-phased in real time on the base of the measured values. The coaxial connections bringing the sampled signals to the demodulation board are out of the loop, so that any length variation will produce an error in the phase readout and a corresponding deterioration of the level of synchronization. Measures have to be taken to avoid these effects, such as reduction of the cable lengths, real time continuous calibration of the cable delays and use of thermally stabilized ducts to house and guide the cables along their paths.

Being 100 Hz the maximum repetition rate, phase stabilization loops based on pulse-to-pulse acquisition are intrinsically bandwidth limited to 50 Hz as a consequence of the sampling theorem. in order to overcome this limitation, an innovative intra-pulse phase lock scheme has been positively tested at SPARC and will be implemented in the ELI-NP low-level RF control to push the RF synchronization to the best achievable specifications.

The intra-pulse phase lock schemes could be not necessary for solid state HV klystron supply. It is a fast feedback loop with a response time of ~1µs, capable to reach its regime condition within the time duration of each individual pulse (~2.5µs for S-band and 1µs for C band modulators). Phase jitter extending its bandwidth beyond 50 Hz (power supply ripples, acoustic vibrations, etc.) is effectively reduced below 100fs with this approach, as measured at LNF in the SPARC_LAB facility. However, the loop path length has to be reduced to the minimum in order to obtain the largest bandwidth in stable conditions. For this reason the loop can only encompass klystrons and RF driver amplifiers, while most of the waveguide network, SW cavities and TW sections cannot be included and the phase noise entering the system at that level can only be cured by slower pulse-to-pulse feedback loops.

### 3.8.8. Timing distribution

The timing signals distribution could be made by means of electrical coaxial cable or optical fiber. In both cases a system based on event generation, distribution and reception could be used. A main station will be placed in the tunnel and its function is to generate the machine events for all the subsystems. Then the signals will be sent through the preferred medium. The target receiver will decode the information real time and generate the proper trigger signals for the relative subsystem. Such implementation for timing control and distribution is commercially available and it was yet designed, engineered and tested with success. It is used in most of the light source around the world (LCLS, APS, Swiss Light Source, Diamond). A sketch of principle is reported in Fig. 105.

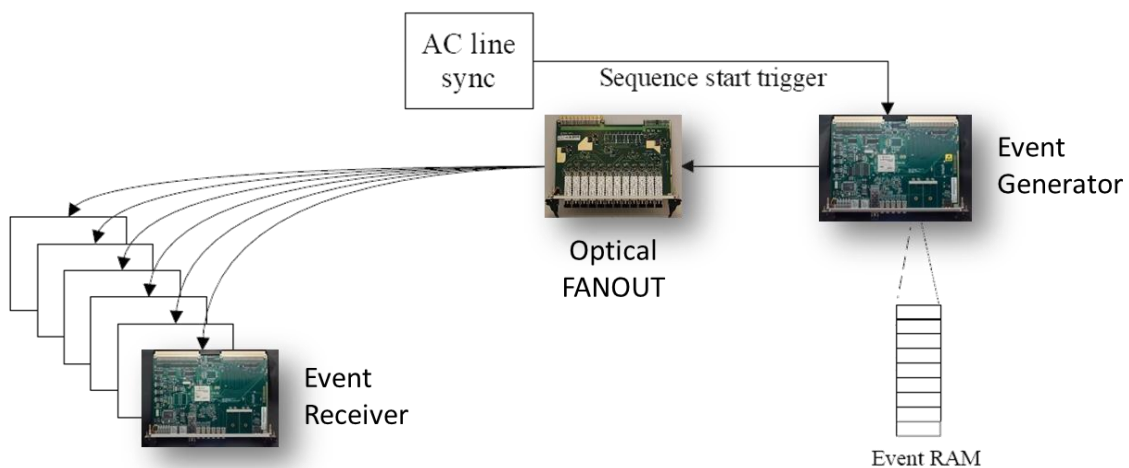

**Fig. 105.   Layout of the timing signals distribution**

Every event could be created at the 100Hz machine repetition rate or even at higher frequency, if necessary (multiple delayed events arriving to the same client). Every event could be tagged by a sort of complex status word. This allows a cross check of data collected from different devices, also when the machine is not running. The control software is also available in the market (written in EPICS standard) and could be easily integrated in the most common control system architectures used in modern accelerators.

## 3.9. Layout, Integration, Survey and Alignment

### 3.9.1. Accelerator Layout & Engineering

The layout of the whole accelerator is shown below in Fig. 106 and Fig. 107 as a 2D plan and a 3D isometric view, in isolation from the building. The engineering layout of the accelerator has been greatly driven by the constraints of having to fit within a pre-designated building layout whilst ensuring that the lattice design is ultimately able to achieve low and high energy beam delivery within specification. The accelerator in context of the building and associated technical services and infrastructure is addressed further on in chapter 6.

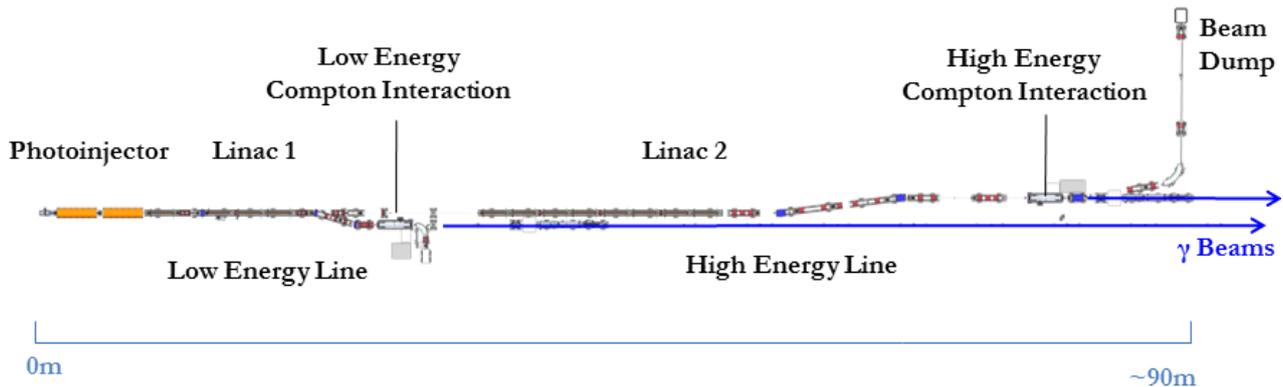

Fig. 106.    Accelerator Layout – Plan View

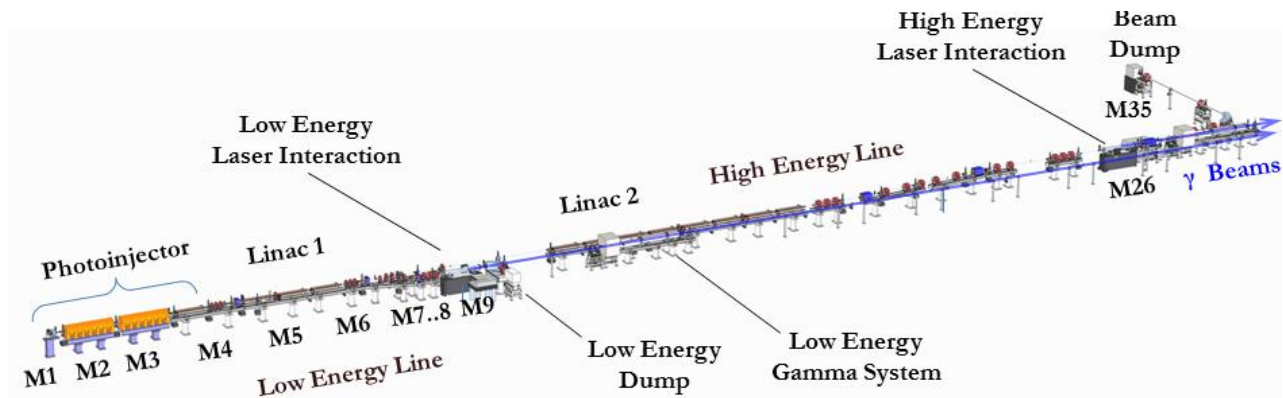

Fig. 107.    Accelerator Layout - Modules M1 – M28

Note that in Fig. 107 the low energy accelerator (linac 1) is indicated as a series of module elements M1 – M10. The high energy linac continues the series up to M35 (the high energy dump). This 'modular system' is key to a strategy that should ensure fast and efficient installation and integration of the accelerator at the Măgurele site and is discussed in the next section on integration.

The module element trains make up the primary sections of the accelerator. These are;

(a) Photo-injector: Modules M1 – M3 comprising the photocathode gun and first two accelerating S-Band RF structures and solenoids, the overall design of which is closely based on the existing proven SPARC S-Band injector at Frascati;

(b) Low Energy Line: Linac 1, consisting of 3 modules (M4 → M6) supporting four C-Band RF structures modules between them, accelerates the beam to low energy operation at 280MeV. This part of the linac runs from 8m downstream of the photocathode origin to ~22m at which point it 'doglegs' 287mrad (~16.5°) into the final leg of the low energy line with a center offset ~0.93m from the photoinjector. Beam bunches can be switched via the dipole at this point between the straight-on line (forming the start of the high energy linac) and the low energy branch line modules (M7-M10)

The low energy 'dog-leg' branch feeds into the Compton interaction point in the laser recirculator (module M9) where the low energy gamma beam is generated. Immediately beyond the laser interaction point the electron beam is directed horizontally through a 90o bending dipole to a beam dump fully contained within accelerator bay 1 shield wall. The gamma beam propagates through the dump dipole magnet 2 via a small hole in the shield wall into the Gamma beam characterisation system (modules M11-14) located in accelerator bay 2 downstream. Note that to transport the characterised gamma beam into the experimental areas much further downstream requires a train of evacuated pipes ~50 – 60m long that does not form part of the scope of the accelerator system described in this TDR;

(c) High Energy Line: consisting of the straight-on branch dogleg into Linac 2, where a further eight C-Band RF structures accelerate the beam to the full energy of up to 720MeV and into the high energy Compton laser interaction point ~ 80m from the photocathode source. Before the high energy gamma beam characterisation system, the electron beam is diverted horizontally via a shallow 9o angle directly to a further 81° dipole bending magnet that ultimately steers the beam to a prescribed high enrgy dump in the energy recovery room. The gamma beam from the interaction point passes through a gamma conditioning system (collimator..etc) located ~10m downstream and very similar in construction to the low energy gamma system. The Gamma characterisation (or demonstration) systems are described further in chapter 5 of the TDR.

The biggest challenge in engineering a workable layout of the accelerator has been the low energy line. This is because the maximum distance for the interaction point and turn-off into the local beam dump is fixed into the building design layout at just less than ~33m from the inside wall edge of the accelerator hall (or ~31m from the photocathode zero).

Such a contingent design space requires an extremely compact engineering solution in order to stay within the very demanding overall length constraint and reach achievable low energy. This involved many space saving engineering work-arounds in the low energy line - e.g. expansion bellows welded directly to chambers, use of ISO-KF vacuum flange joints (instead of all Conflats) and the strategic omission of matching all BPMs with complimenting YAG screens in parts of linac 1.

Further aspects of the layout of the accelerator are described below in sections on integration and buildings and infrastructure. In addition, a comprehensive pictorial breakdown of all components and module train that comprise the full accelerator and the gamma characterisation system is shown in appendix D.

### 3.9.2. Accelerator Systems & Integration

A wide range of accelerator components and systems must be integrated together to ensure a smooth design, planning, construction and commissioning process for ELI-NP.

The methodology for construction proposed is to assemble the accelerator as series of modules as shown in Fig. 107, the different types of which are shown in Fig. 108 below.

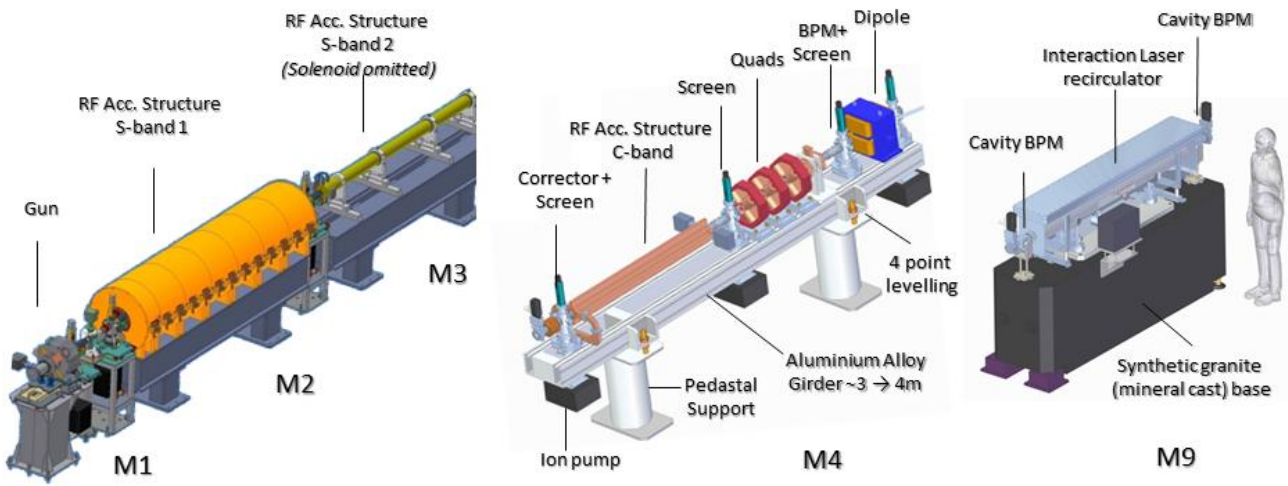

**Fig. 108.** Close-up of Accelerator Module Types; Photo-injector (M1 – M3), linac girder (typical M4 to M25) and laser recirculator (M9 & M26)

The **photo-injector modules** (M1-M3) are based on proven design already successfully operating at SPARC, INFN Frascati.

The **laser interaction modules** M9 & M26 will be commissioned by LAL / Alysom. Because of the more acute sensitivity to thermal and vibration stability the interaction recirculator devices will be mounted on a large mass (4-5Tonne) mineral cast block (synthetic polymer granite) with low thermal expansion and high damping characteristics. Such blocks have original application in ultra-precision machining and metrology test beds and in sensitive applications in other accelerator facilities (e.g. hard X-ray monochromators) and have been shown to typically give better thermal and vibration performance of between x3–10 in comparison with standard steel frame structures.

All **other accelerator modules** will consist of specific trains of linac components; magnets, diagnostics, RF structures, etc. mounted on common aluminium girder of approximately 4m in length. In all there will be 23 such girder modules and it is proposed that each is assembled, aligned and tested off-line (at STFC Daresbury Laboratory) prior to installation in the ELI-NP building. This methodology will allow a significant amount of accelerator construction to be achieved in parallel to the building construction, enabling the earliest opportunity to install within the accelerator hall and so meet the ELI-NP commissioning milestones.

Many vacuum joints, motor control and diagnostic and key stage acceptance testing can be proven earlier in the construction process, thereby reducing the amount of work and time involved in installation, testing and commissioning on site at Măgurele. For the girder mounted C-Band RF structures it is intended to follow the testing approach proposed by SwissFEL [62] whereby the first initial batch of structures will be fully tested at low level and high power RF off-line before assembly but thereafter will be dimensionally inspected and low level 'tune' characterised only. Thus, final full power tests of the whole ensemble of RF structures will only be carried out during the phased commissioning at Măgurele.

Accelerator components will be accurately pre-aligned on their support girders which are themselves mounted on sand filled pedestals, bolted and grouted to the accelerator floor plinth to provide a stable support structure. To maintain extremely good levels of mechanical and thermal stability of the accelerator magnets, RF structures, diagnostic devices and optical components, careful attention to detail will be required in the structural and thermal management of the facility. Further details on stability requirements are covered later in this chapter and those influencing the building design and construction described in Chapter 6.

### 3.9.3. Stability

ELI-NP-GS must function as an integrated facility from its photo-injector, accelerating systems, focusing magnets, precise electron beam alignment through to electron–laser collisions to routinely produce high energy gamma beams.

To maintain accelerator component positional accuracy to the order of a few microns or better it is essential that measures are taken to control both mechanical and thermal stability. Both long and short term stability issues as well as high frequency motion must be addressed and suitably managed.

The overall timing jitter from source to electron–laser interaction needs to be in the order of 100 fs. Quantitative evaluation and control of mechanical vibration and thermal drift is therefore essential.

#### 3.9.3.1 Ground Vibration

Fig. 109 below shows typical ground vibration displacement power spectral densities (PSD) in µm2/Hz from 0.01-100Hz for several accelerator sites around the world [63-64].

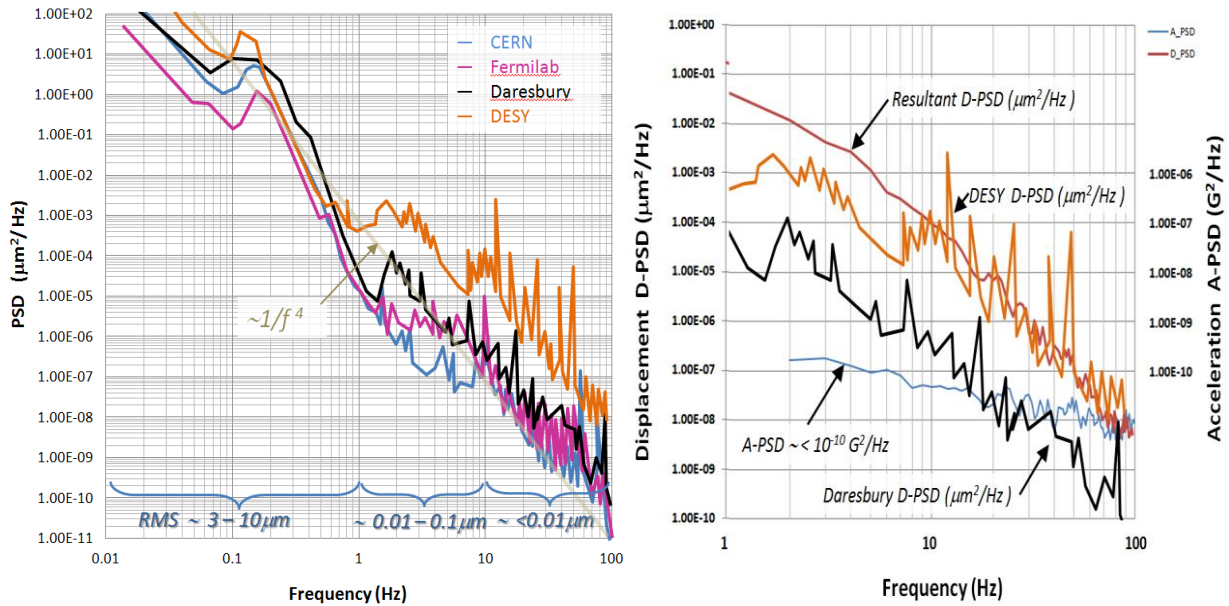

Fig. 109. (a) Left: Ground noise spectrum (PSD) at several sites with RMS values. (b) Right: a generic acceleration A-PSD at $<\sim 10^{-10}$ G2/Hz (blue line) predicted for ELI-NP and transformed to a displacement D-PSD (red line) - for comparison with two other sites (Daresbury & DESY)

From 1–100 Hz vibration spectrum is driven by 'cultural noise' – typically local, man-made noise arising from sources such as road traffic (often around 10 Hz) and technical equipment (pumps, compressors etc). As can be seen from the figure this varies by orders of magnitude between various sites. As frequency increases movement across smaller distances comes increasingly uncorrelated 'random' vibration. Fortunately the PSD profile falls off strongly with frequency as $1/f^4$ so it is also reasonable to set 100 Hz as an upper limit of significance.

Between 1-100Hz the RMS residual random 'fuzzy' motion at the ground level will be in the approximate range ~0.02 – 0.1µm, depending on site. This appears small but a typical supporting mount can easily amplify this by a factor ~ x 10 – 100 dependent on frequency response and eigen-modes of the design. Thus motion of at least several microns is not unusual occurrence.

For ELI-NP the effects of ground vibration are primarily of twofold concern;

(a) Acclerator magnet lattice stability: Small movements of magnetic elements of the accelerator, particularly the quadrupoles, introduce a series of electron beam transverse motion and angular "kicks" in the trajectory ultimately giving rise to problems with gamma photon beam stability. For many different types of light source an upper limit in overall rms motion is that it should not typically exceed ~10% of the electron beam size in either plane. For ELI-NP with a spot size of ~10 – 30um (as listed in the parameter table) this equates to a quadrupole motion ~1- 3um. However other work [65] has shown that the overall rms contribution is increased by factor $\sqrt{N_Q}$ where $N_Q$ is the number of quadrupoles. For ELI-NP high energy line $N_Q =28$ up to the interaction point, so that the upper limit for any individual quadrupole becomes ~1- 3um/$\sqrt{28}$ ~ 0.2-0.6um.

(b) Focus stability at laser interaction: Angular deflections of <1urad in the laser recirculator will exceed the pointing stability required for the multipass laser system to maintain an acceptably low distortion within the 25um focal spot width. Over the half length of the ~2.8m recirculator such an angular deviation corresponds to ~ 1um in transverse deflection. In addition to angular tolerances the focussing mirror optic surface is typically specified to at least λ/4. At the laser energy of 2.4eV, λ~0.5um, for which a λ/4 would ideally imply a limit of <<0.2um to be maintained. In this case it would appear that the laser interaction could potentially dominate limits for acceptable vibration.

In either case it can be appreciated that a higher level of overall rms vibrations of 0.1um (as indicated in Fig. 109 (left) by √ (integration under PSD curve) over 1- 100Hz along with amplification Q factors of supports of only ~x10 has the potential to exceed acceptable levels of motion. In the case of the laser recirculator the added susceptibility to angular deviation over such long structures means that in practical terms additional measures must be taken to minimise vibration effects – hence the specific use of large granite mass block in these locations that would be uneconomic to employ in the remainder of the accelerator.

The accelerator hall and laser room floors have been designed by ELI-NP as a monolithic slab supported by anti-vibration mounts, custom sized for the various loading at various locations around the facility. At the time of writing the building specification points to the behavioural characteristics under earthquake load. However there is no measured ground vibration data under normal conditions reported for the Măgurele site (of a type as shown in Fig. 109 (left)) or a predictive spectral estimate of the finished anti-vibration floor under nominal conditions.

The technical description in the tender for the ELI-NP-GS accelerator makes reference to <1 um RMS movement up to 100Hz. If measured from 0.01Hz to 100Hz this is an acceptably low value – as can be evidenced from a quick reference to RMS values summed over the frequency range in Fig. 109 (left).

However if this upper limit on RMS is taken as applying to non-correlated motion – i.e. for frequencies >1Hz and up to 100Hz then using the same reference it can be seen that 1um is in fact an unacceptably high value of noise – by almost an order of magnitude. Thus it is apparent that the tender specifies an RMS value that is of little use without the spectral distribution.

Without this key information it is difficult to provide any form of quantitative assessment of the anticipated ground vibration on the proposed accelerator design. Without more definitive reports we reference a presentation [66] by the ELI-NP civil engineering working group that states that the ground acceleration PSD will be < 10-10 G2/Hz over the 1-100Hz spectrum. We have taken this and transformed it into a displacement PSD in um2/Hz – which is shown in Fig 4.4b. This indicates that such a level of vibration is at the top of end of noisy sites as surveyed in this Figure. However, it is still classified as a relatively quiet background and based on simulation studies and actual measurements at Daresbury of a girder design similar to that proposed for ELI-NP we are confident that a maximum overall rms of <<1um can be achieved on the girder support.

In addition to ground borne vibration, more localised vibration sources can also cause problems unless mitigated;

- Potential noise sources mounted on the girder or close by sensitive equipment – e.g. such as water flow channels into magnets. In general, experience has shown that in most cases careful design can mitigate such vibration to be substantially less than the effects arising from ground vibration.
- Noisy plant and machinery – such as rotary vacuum pumps, water and aircon pumps ...etc. Small nearby pumps should be properly mounted on correctly balanced AV elastomer mounts with connecting vacuum or fluid hoses coupled via flexible bellows to large mass to dissipate vibration. Larger service plant equipment must be carefully mounted in a similar way but also located some distance away so that attenuation through intervening subsoil is significant. Studies suggest [67] that for even large plant, such as Megawatt size cryogenic compressors, distance can reduce vibrations to an acceptably quiet level

### 3.9.3.2 Diffuse Motion & Settlement

Over very long timescales and distance, Shiltsev [68] has shown that displacement can be described as being diffusive – a 'random walk' with limits governed by the ATL law. For the ELI-NP site, ground displacements could potentially run into >mm over years to decades. In addition, for the first few years after construction another more critical but systematic motion exists due to concrete shrinkage and ground settlement. These relatively large displacements arise from the initial ground disturbance during excavation and re-compaction of the surrounding area over time and also from diurnal and seasonal changes in ground water.

To minimize settlement, suitable construction methods, already employed at other accelerator sites across the world, should be adopted by ELI-NP such as substantial piling of the sub-basement floor. Heavy vibro-compaction of disturbed earth should also be conducted during construction.

ELI-NP have designed a monolithic concrete construction for the accelerator and laser room floor slabs that is supported on custom sized anti-vibration pads.

Specific additional concerns over settlement at ELI-NP arise over the use of this approach. This is because viscoelastic materials are known to undergo non-linear creep effects over time and whilst such levels of creep (mm per year?) would not be a problem on conventional buildings (where these devices are typically employed), in the case of ELI-NP-GS this potentially could be very significant – especially so if differential rates point to point across the floor were significant diverse. We note that no other accelerator (that we are aware of) has been constructed in this way and therefore experience in this is practically nil. If creep settlement is high then this may requiring more frequent resurveys and beam based realignments than at other facilities. This possibility is should be further evaluated during the tender phase.

Correcting for diffuse motion & residual settlement (both defined by characteristically large displacements, hundreds of μm, over long periods) is best addressed by classifying them as alignment, rather than stability issues. These are examined in section 6.

Due to the thickness of the floor slab and shield wall construction, settlement in the concrete structures as they set and cure over the initial months and years is also of especial importance and this is also briefly addressed in the survey and alignment sections.

### 3.9.3.3 Thermal Stability

On the time scale of months to hours (and sometimes minutes) another very important class of instability exists – that of thermally induced displacement.

Within the accelerator hall and technical rooms various localized heat sources (mostly from equipment such as vacuum pumps, magnets, certain instrumentation racks etc.) will give rise to temperature gradients unless adequately controlled. This control is achieved through the use of a distributed heating, air-conditioning and ventilation (HVAC) system (Section 6).

Temperature variations induce quadrupole magnet and BPM displacements, and hence trajectory changes, as described above for ground motion. The ELI-NP Air conditioning (HVAC) system has been specified to deal with this issue and has been specified at 22 ±0.5°C in the accelerator hall and 22 ±1.0°C for the experimental areas. Relative humidity has been specified at 40 – 60%.

Temperature stability of magnets can be improved and thermal load on the HVAC system reduced by running all of the magnet coils as water cooled even for low current coils where air cooling would normally be sufficient. In the case of ELI-NP-GS proposal this is included in the magnet specification.

The HVAC temperature control also defines the differential movement of the building and floor slab in the horizontal plane. For nearly uniform changes the alignment of the accelerator does not change but the lengths along the beam axis will. Over the 80m length from gun to high energy interaction point a 0.5ºC change in temperature will correspond to a change of ~0.3mm or 1000fs , considerably higher than the timing jitter budget and could therefore give rise to certain problems in timing and synchronisation of the RF and interaction laser pulse. In practice such thermal drifts would be likely smaller than indicated and over long time scales (hours, days). Therefore it could be compensated for – e.g. by measuring in situ using a permanent laser interferometer and correcting in the timing system.

### 3.9.3.4 Stability Conclusions

A broad overview of ELI-NP stability issues has been presented here, which should be addressed further during the detailed design phase of the project. Overall, it is encouraging to note that many of the most challenging stability problems faced by ELI-NP are of the same order or less than the tolerance requirements at other accelerators or Free Electron Laser facilities, which have been accommodated for by careful design.

# 4. Lasers

The laser beam characteristics have to fulfill the requirements delineated in the introduction. However, strong constraints are put on the choice of the optical system and on the lasers system: they must be reliable in order to be routinely operated and no R&D, or at least a very low risk R&D, is required for their developments.

A major attention was paid on mitigating the risks and simplifying the overall architecture, by exploiting at its maximum the photons produced by the interaction point lasers, recycling them at the interaction point using a re-circulating geometry, keeping also reasonable requirements on the photoinjector laser. Moreover, the choice for using 2 identical lasers, one alone for IP1, and a combination of the two lasers for IP2, allows to reduce the complexity and allows for easier maintenance.

In this chapter, both the photoinjector laser and the interaction point (IP) laser systems are presented. Following the specifications given in the introduction, the photoinjector laser will produce a sequence of trains made of 32 laser pulses at 100Hz repetition rate with a ~10ps pulse duration in the UV range, whereas the high energy IP laser will produce one pulse with 0,5J pulse energy, at 515nm and 3.5ps pulse duration, achieved by frequency doubling a 0.4J pulse at 1030nm. Each pulse will then be used 32 times in a recycling architecture, reducing by the same factor the requirements on the pulse energy/average power.

This chapter is thus organized as follows: the technology chosen for the IP laser systems is presented in section 4.1. The description of the optical re-circulator is given in section 4.2. Finally, the photocathode laser is presented in section 4.3.

## 4.1. IP Lasers

The parameters of the IPs lasers combine at the same time high energy and high average power, which reduces strongly the range of technologies available. Firstly, diode pumped technology is the technology of choice to achieve high average power with reduced cooling needs, together with maintenance issues. Secondly, the Joule-class energy is only achievable with bulk lasers, in contrast to fiber laser technology.

Combining energy and average power have only been achieved by Ytterbium lasers, thanks to their superior thermal behavior, higher energy storage capability [69] and laser diode pumping compatibility. Various approaches have been developed recently to address the thermal issues, from the Thin-Disk technology to the cryogenic cooling technology. Even if simple water cooling appears attractive, the thin disk technology [70] suffers from severe limitations for energy scaling together with power scaling: the already low gain drops down when enlarging the beam spot size, ending with a complex and poorly efficient multi pass architecture limited today to 500mJ in picosecond regime [71]. Working at cryogenic temperature, even if increasing the complexity of the cooling system, simplifies strongly the laser architecture, by improving significantly the gain and thermal properties of the laser gain medium [72]. This technology allowed to demonstrate recently 1J at 100Hz in the picosecond regime (1,5J before compression) using "thin-disk" active mirrors [73, 74], close to the requirements of the high energy IP laser system. Very good beam quality (M2<1.2) and power stability (~3%) were also recently reported [74].

The architecture proposed aims to deliver a highly stable system, based on one part on state-of-the-art commercial products that proved their stability and reliability in highly demanding industrial applications, and on the other part on solutions taking benefit of 15 years of experience on ultrafast diode-pumped laser technology.

In order to optimize the performance reliability of the system, the global architecture is declined as follow :

- A diode-pumped femtosecond Yb oscillator, synchronized to a common reference signal
- A fiber-based compact stretcher
- A regenerative amplifier based on Yb:KYW
- A booster amplifier based on cryo-cooled Yb:YAG
- A main amplifier based on cryo-cooled Yb:YAG
- A deformable mirror
- A high efficiency compressor
- A frequency converter

The first elements are standalone products, delivered in high quantity (>100) to scientific and industrial customers worldwide, proven for their stability and reliability. A specific development needs to be addressed for the main amplifier system, for which cryo-cooling is identified to be the best solution in terms of stability and beam quality of the laser source.

The cryogenic technology is already used by Amplitude Co. in commercial ultrafast lasers to improve the thermo-optical properties of Ti:Sapphire crystals on high average power systems [75], and is currently transferred to Ytterbium laser technology [69]. As an example, a 100W class laser was demonstrated by Amplitude Systèmes in collaboration with Institute of Optics Graduate School, based on diode-pumped Yb:CaF2 crystal..

Therefore, the technology selected by Amplitude to fulfill the requirements for the IP lasers is cryogenically cooled Yb:YAG in a chirped pulse amplification architecture [76] [77], followed by a frequency conversion stage to achieve 515nm wavelength.

### 4.1.1. IP Laser architecture

#### 4.1.1.1 General architecture

The laser architecture consists in a chirped-pulse amplification architecture, composed of an oscillator, a stretcher, several amplifiers and a compressor, subsequently frequency doubled in a conversion stage.

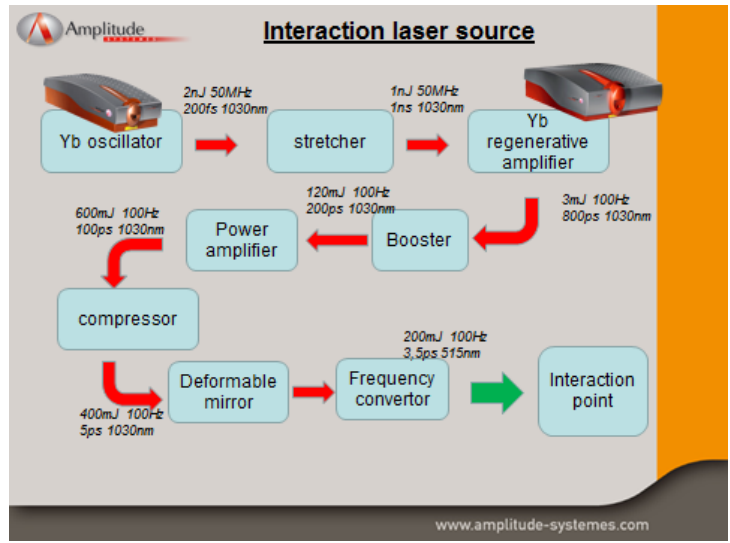

**Fig. 110.** Functional drawing of the IP laser system.

### 4.1.1.2 Oscillator

The oscillator is a standard product from Amplitude Systemes, from the t-Pulse series, a diode-pumped ultrafast oscillator delivering 1W average power at 1030nm, with 200fs pulse duration, at a repetition rate of 50MHz [78]. The repetition rate can be factory set to the required value, around 62MHz, with a precision of 100Hz and a tunability of +/-2kHz, to compensate for potential drift of the reference signal frequency.

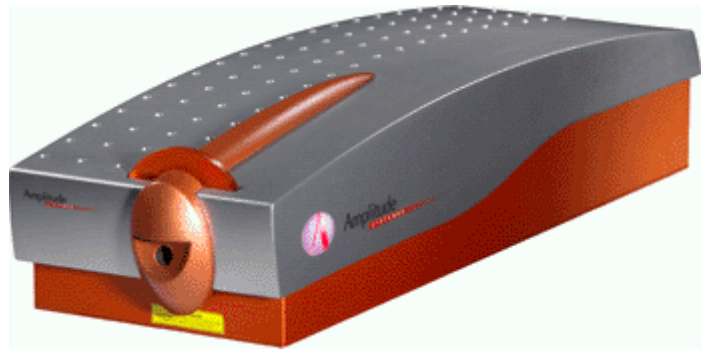

**Fig. 111.** An Amplitude Systèmes t-Pulse series diode-pumped ultrafast oscillator

The laser cavity is actively stabilized and synchronized to an external electronic reference: an electronic feedback loop activates a piezoelectric actuator in the cavity for fast (kHZ) feedback and an additional translation stage for slow feedback. This active stabilization allows to achieve a timing jitter lower than 500fs integrated on 10Hz to 1MHz frequency range. Please note that the feedback loop electronics can also be done on any error signal, such as a cross-correlation optical measurement done with the OMO used as a secondary reference clock of the gamma beam machine. The cavity is compact and sealed, ensuring stability and environmental isolation.

**Table 26. Specifications of the oscillator**

| OSCILLATOR | |
|---|---|
| Output energy | > 10 nJ |
| Repetition rate | 62,08 MHz |
| Repetition rate accuracy | 100 Hz |
| Timing jitter (100-10MHz) | <500 fs |
| Central wavelength | 1030 +/-2 nm |
| Spectral bandwidth | >5 nm |
| Pulse duration | <250 fs |
| Beam diameter (1/e²) | < 1 mm |
| Beam divergence | < 2 mrad |
| Spatial mode TEM00 | (M² < 1.3) |
| Polarization | > 100:1 (horizontal) |
| Noise | < 0.1 % rms |
| Power stability | +/- 1% |
| Size | 600x200 mm |

The oscillator requires a water cooling system with a 100W cooling power, at a controlled temperature of 20°C typically, with a 0.1°C regulation. The electronics driver is fully controllable through RS232.

#### 4.1.1.3 Stretcher

The stretcher is a compact system based on chirped fiber Bragg grating technology. Due to a precise control of the Bragg grating density along the fiber, this technology gives access to a customized spectral phase profile, allowing to adjust second order and third order of dispersion to compensate exactly for the dispersion of the compressor to be used. The typical dispersion of such a component is about 200ps/nm per piece, on a bandwidth of 6nm.

An Yb:fiber amplifier is used to subsequently amplify the pulses by one order of magnitude in order to compensate for the losses in the stretching part.

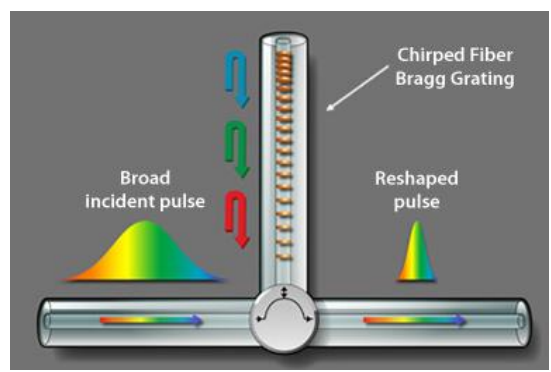

**Fig. 112.    Fiber Bragg grating based stretcher**

**Table 27.    Specifications of the stretcher**

| STRETCHER | |
|---|---:|
| Stretching ratio | > 200ps/nm |
| Spectral aperture | > 6nm |
| Efficiency | > 30 % |
| Energy after stretcher | > 500 pJ |

#### 4.1.1.4    Regenerative amplifier

The regenerative amplifier is the one used in the commercial product s-Pulse HP², based on diode-pumped Yb:KGW crystals, delivering up to 3mJ energy at 100Hz [79]. This amplifier can be adjusted at 1029nm wavelength, which allows a perfect spectral overlap with the following Yb:YAG amplifiers.

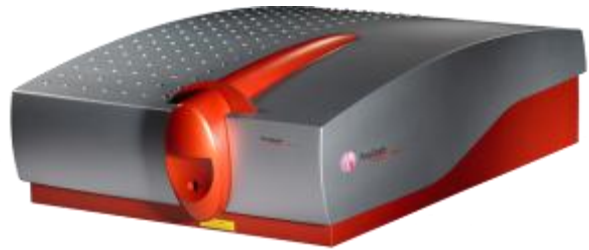

**Fig. 113.    A regenerative amplifier**

Moreover, in such amplifier, the amplified bandwidth is about 4nm, which is beneficial to act against spectral narrowing in the following amplifiers. It reduces also the dispersion required in the stretcher to avoid nonlinearities during the whole amplification process. A high pulse-to-pulse stability and pointing stability is ensured by the integration of the regenerative cavity in a highly rigid metallic, sealed structure.

**Table 28.    Specifications of the regenerative amplifier**

| REGENERATIVE AMPLIFIER | |
|---|---:|
| Output Energy | > 3mJ |
| Central wavelength | 1029+/-1nm |
| Spectral bandwidth | > 4nm |
| Repetition rate | 1 to 1000 Hz |
| Spatial profile (M²) | <1.3 |
| Footprint | 750x500 mm |

The control electronics consists of a power supply, and a timing unit, integrated in a single, 19" rack (weight 9.8Kg). The laser can be controlled from the front panel but also via a RS232 protocol. The laser software is user-friendly and is programmable via RS232.

#### 4.1.1.5    Booster amplifier

The booster amplifier consists in a multi pass amplifier based on a diode-pumped cryo-cooled Yb:YAG, amplifying the pulse energy from 3mJ to 120mJ after 10 passes. The pulses experience a significant spectral narrowing effect during amplification, ending with 1nm bandwidth, thanks to the broad bandwidth and high energy injected from the regenerative amplifier. The stretched pulse end with a pulse duration of 200ps on this 1nm spectral bandwidth. The cooling unit, operating at approximately 100K, is based on a closed loop system with minimized vibrations.

**Table 29.    Specifications of the booster amplifier**

| MULTIPASS BOOSTER AMPLIFIER | |
|---|---|
| Pumping power | 500 W |
| Pumping time | 1 ms |
| Repetition rate | 100 Hz |
| Output Energy | >120 mJ |
| Efficiency | > 25 % |
| Number of passes | 10 |
| Spectral bandwidth | 1 nm |
| Footprint | 1200x800mm |

#### 4.1.1.6    Main amplifier

The main amplifier consists in a multi pass amplifier based on one Yb:YAG crystal cooled by a second cooling system. The energy is amplified from 120mJ to 600mJ, in a 6 pass angular multiplexed architecture. The final spectral width is 0,5nm. The crystal is pumped by a fiber coupled laser diodes arrangement emitting at 940nm, allowing for reliable operation and easy replacement. The cooling power is also reduced due to the typical 50% electrical-optical efficiency of the laser diodes.

The cryo-cooling allows for high gain and simple amplifier architecture, together with an excellent beam quality due to enhanced thermo-mechanical properties of Yb:YAG at cryogenic temperature. The cryo-cooling system is a standard close-loop commercial product, already used in many high-power ultrafast lasers based on Ti:Sa amplifier.

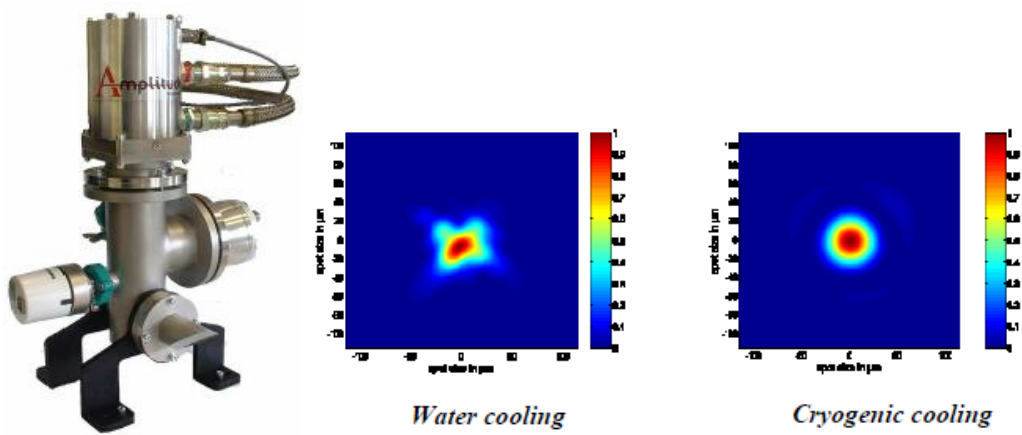

**Fig. 114.    (left) Main amplifier with cryo cooling module. (Right) Typical focal spot of the laser output beam**

**Table 30. Specifications of the main amplifier**

| MULTIPASS MAIN AMPLIFIER | |
|---|---|
| Pumping power | 2 kW |
| Pumping time | 1 ms |
| Repetition rate | 100 Hz |
| Output Energy | >600 mJ |
| Efficiency | > 30 % |
| Number of passes | 6 |
| Spectral bandwidth | 0,5nm |
| Footprint | 1500x1000mm |

#### 4.1.1.7 Deformable mirror

In order to correct for the wavefront distortions due to amplification, a deformable mirror is considered before the compression stage, in order to optimize the conversion efficiency of the second harmonic generation in the nonlinear crystal by increasing transverse pulse quality.

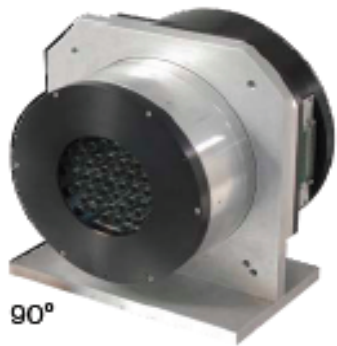

**Fig. 115.    High power deformable mirror**

#### 4.1.1.8 Optical compressor

The compressor is composed of highly efficient diffractive gratings, based on multilayer dielectric mirrors, allowing for recompressing the pulses to 5ps pulse duration with an overall efficiency of 65%, ending to pulse energy of 400mJ.

At this level of peak power (80GW), the nonlinear effects in the air are still negligible in the compressor. Therefore the compressor does not require to be placed in a vacuum chamber.

**Table 31.  Specifications of the optical compressor**

| COMPRESSOR | |
|---|---:|
| Input Energy | > 600mJ |
| Output Energy | > 400 mJ |
| Efficiency | 65 % |
| Beam diameter | >15 mm |
| Spectral Aperture | >2nm |
| Footprint | 1500x600mm |

#### 4.1.1.9    Frequency converter

The frequency converter consists in a LBO nonlinear crystal, well known for its availability in large sizes and high crystal quality, together with its high efficiency with a high damage threshold, and limited aging issues. While the alignment to adjust the phase-matching angle is done mechanically, the stability of the conversion efficiency is ensured by temperature stabilizing the nonlinear crystal. A telescopic system is needed in order to adjust the beam size in the crystal for best energy conversion and beam quality. Two dichroic mirrors are finally installed in order to remove the infrared part, and keep only the visible part of the beam.

**Table 32.  Frequency converter for Interaction Point 1 (low energy)**

| FREQUENCY CONVERTER ON IP1 | |
|---|---:|
| Output energy | >200mJ |
| Pulse duration | <3,5 ps |
| Spectral bandwidth | <0,3 nm |
| Beam quality (M²) | <1,2 |
| Efficiency | > 50 % |
| Polarization state | linear |
| Footprint | 500x500mm |

For the second Interaction Point (IP2), the compressed IR beam of laser room 1 is reflected before the frequency converter and transported to the room 2 through vacuum tubes, including relay imaging system in order to reduce beam pointing fluctuations issues (responsibility of ALSYOM). The two compressed infrared beams are then combined with a polarizer, and converted in a nonlinear crystal used in type 2 configuration. The synchronization of the two pulses is adjusted using a phase shifter (electronic resynchronization of the two separate oscillators) between IP1 laser and IP2 laser.

Two dichroic mirrors are also used after the nonlinear crystal, to separate the green part and the IR part of the beam.

**Table 33.     Frequency converter for Interaction Point 2 (High energy)**

| FREQUENCY CONVERTER ON IP2 | |
|---|---|
| Output energy | >400mJ |
| Pulse duration | <3,5 ps |
| Spectral bandwidth | <0,3 nm |
| Beam quality (M²) | <1,2 |
| Efficiency | > 50 % |
| Polarization state | linear |
| Footprint | 500x500mm |

#### 4.1.1.10     Laser diodes for pumping

All the laser diodes used in the IP lasers are based on telecom-class technology, emitting at 940nm or 980nm (InGaAs). They are well known for their long lifetime (2000h routinely), and high electrical to optical efficiency (>50%). All the laser diodes are fiber coupled, providing circular homogenized beams, and allowing simple replacement procedures.

They are conductively cooled (passive cooling) with water, with a typical temperature of 20°C, with a stabilization of +/-0.1°C.

#### 4.1.1.11     Diagnostics

Several diagnostics are placed throughout the laser chain, for synchronization requirements, and alignment issues. Photodiodes, CCD cameras, are placed for continuous operation, whereas a spectroscope, power meter/energy meter, wavefront analyzer and auto-correlator are used for checking.

## 4.2.     The Laser beam recirculator

The purpose of the re-circulator is to provide an effective average power enhancement of the laser beam at the IPs. However, the optical system constituting the re-circulator must not degrade the γ-ray spectral width produced by the ELI-NP machine. It means that laser beam – electron beam crossing angle must be constant. The re-circulator must also provide a stable and small enough laser beam waist to optimize the γ-ray flux.

Two classes of optical devices can meet these requirements: the optical resonators [80], [81] and the multi-pass re-circulator.

### 4.2.1.     Optical resonators

Stable optical resonators, or Fabry-Perot cavities (FPC) [80], are made of a set of mirrors of which at least one is concave. The power enhancement factor inside a FPC is given by $G=F/\pi$, where F is the FPC finesse which depends on the mirror reflection coefficients. However, to take benefit from the power gain, one must keep the cavity at resonance while the electron beam is sent inside the FPC [81]. The FPC length must indeed be controlled to an accuracy better than $\Delta L=\lambda/F$ (or equivalently the laser frequency at a relative accuracy $\Delta v/v \sim \lambda/(FL)$ where L is the FPC round trip optical path). For example, even a moderate gain G=100 leads to $\Delta L \sim 1.7$nm for $\lambda \sim 500$nm (i.e. $\Delta v/v \sim 10^{-10}$ for L of the order of few meters). A strong feedback between the laser frequency and the cavity length is therefore needed, even for low finesse FPCs (e.g. see [82] for more details).

FPCs have been widely used in continuous wave (cw) regime and more recently also in pulsed regime [83] using the spectral properties of passive mode lock laser oscillator [84]. At present, the state of the art [85] is: 72kW average power inside a FPC, FWHM=2ps and frep=78MHz. That is ~2GW peak power and 900µJ pulse energy which is more than 2 orders of magnitudes below of what is needed for ELI-NP.

Increasing the pulse energy would lead, a priori, to an increase of instability sources. Among them, the radiation force induced by the momentum transfer of the laser beam photons on the mirrors may lead to sizable effects [86] (note that the best laser cavity feedback bandwidth is at present of the order of 10 MHz [87], that is noise sources of period approximately below 1µs cannot be eliminated or dumped...).

In order to reduce the effect of radiation pressure, one may increase the laser pulse duration, but at the same time the electron-laser crossing angle should be reduced. This consideration leads to the use of donuts FPCs [88] (i.e, the focusing mirrors are drilled at the center to let the electron beam passing through) for which the low loss Laguerre-Gauss modes (i.e. 'donuts' modes) are expected to be used. However, since a four-mirror FPC is needed to provide both stability and small enough laser beam [89], elliptical donuts beams should be considered. These modes being not eigenmodes of the Helmotz propagation equation [90], only higher order Hermite-Gauss will in fact resonate in four-mirror donuts FPC. Work on Hermite-Gauss FPC is indeed progressing in the context of the gravitational wave interferometers but, still some technical issues remain to be solved [91].

Finally, in order to reduce the average power circulating inside the cavity, one should operate the laser amplifier in pulsed mode (i.e one should switch the laser amplifier during a given time widows at a frequency of 100Hz). However, very few experiments have addressed this problem (only one to our knowledge [92(a)]). This technique is not yet mature and requires further R&D [92(b)].

Active FPCs have also been considered [93][94] but, because of the presence of an active material inside the resonator, the re-circulation of a high peak power pulse is expected to induce destructive non-linear effects.

In conclusion, although some of us have already build and operated a four-mirror FPC on an accelerator to produce γ-rays [82][95], a two-mirror FPC to measure the HERA electron beam polarization [96][97] and demonstrated high finesse F~30000 (i.e. gain G~10000) FPC locking in picoseconds regime [98], we decide not to choose the FPC technology for the ELI-NP machine because of a too high technological risk. We therefore considered the 'brute force' method, namely the multi-pass re-circulator.

### 4.2.2. Multipass recirculator

Among the possible multi-pass re-circulator geometries, one must choose those having a fixed focus point. This is the case of confocal re-circulators (see Fig. 116) which have been indeed studied for Compton scattering in [99][100]. However, the planar geometry considered in these references leads to a variation of the IP crossing angle pass after pass. Herriot cells [101] have also been used to focus a laser beam onto a molecular flow [102] but the focus area is, by construction, much wider than what is needed in the case of the ELI-NP machine.

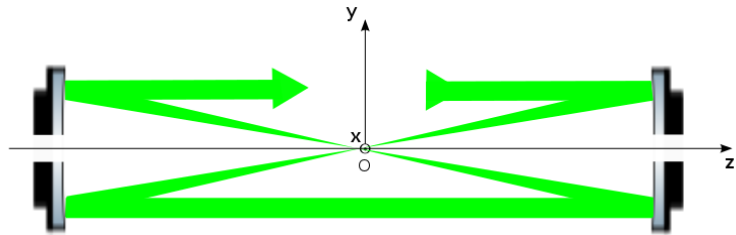

**Fig. 116.** Schematic view of a confocal multi-pass re-circulator. The point labeled 'O' designed the interaction point (IP)

One must thus consider 3D extensions of the planar confocal re-circulator of Fig. 116.

We found two solutions.

The first solution that we studied simply consists on mapping two rings with spherical (or parabolic) mirrors (see Fig. 119). The equivalent lens sequence is also shown in Fig. 117. We made an experimental 5 passes test setup using 9 (one inch) concave mirrors of radius of curvature 2m mounted on commercial mounts. A stable cw λ=1064nm laser beam was used and beam waist at the focus point was 40μm (corresponding to 20μm for λ=534nm).

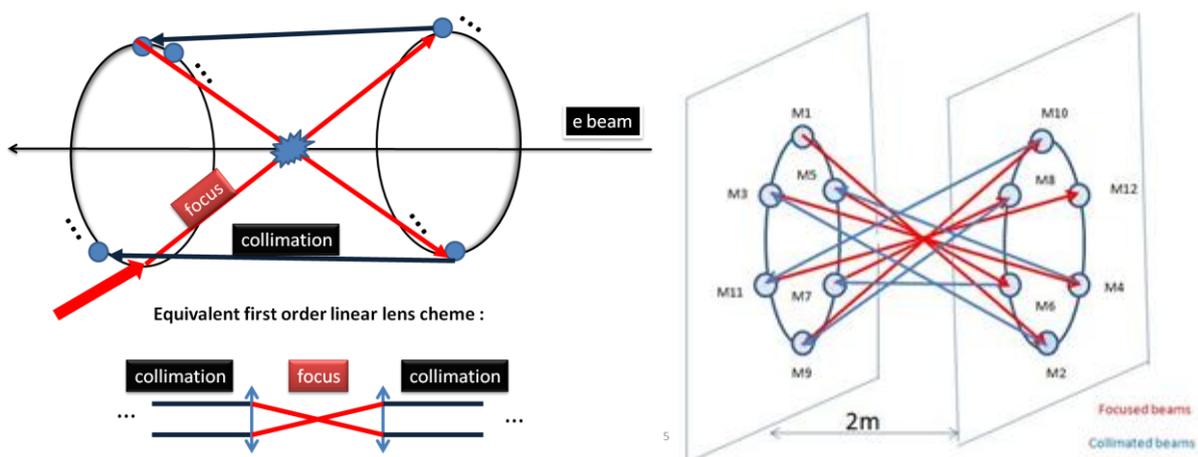

**Fig. 117.** Top left: schematic view of the first multipass setup. Circular concave mirrors (small discs) are located on two coaxial rings. Bottom left: equivalent linear lens sequence. Right: experimental test setup (Mi are spherical concave mirrors of ROC=2m)

Although the mirrors rings were located 2m apart, we were able to align them by hand by simply imaging the focus plane with a thin parallel plate. Fig. 118 shows the superposition of the 5 spots. Although this system is simple and can be made robust, it has three drawbacks: one a priori needs to align all mirrors one after the other leading a large number of tilt actuators (3 per mirror); the implementation of a synchronization system can be an issue; by construction, off axis reflections occur after each pass, thus inducing beam ellipticity and astigmatism. We estimated numerically an increase of the beam waist and ellipticity of the order of 1-2µm per pass.

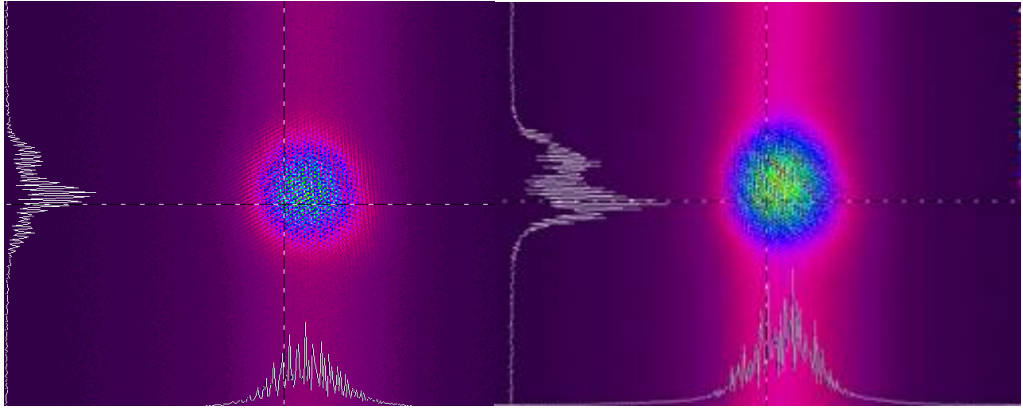

**Fig. 118.** Image of focus point at the 1rst pass (left) and 5th pass (right). Interference patterns are induced by the highly coherent cw laser beam (Innolight co.). The waist has been measured to $\omega_0$;50µm (for $\lambda$=1064nm)

The second solution that we end up with cures the weaknesses of the first solution. Namely, it avoids astigmatism, it reduces the number of alignment's degree of freedom and it allows synchronizing all optical paths with the electron bunches. A detailed description of this system can be found in [103].

### 4.2.3. Recirculator design

Our system minimizes the number of alignment degrees of freedom and the optical aberrations accumulated along the passes while providing a fixed electron-laser beams crossing angle. A schematic drawing of the system is shown in Fig. 119. Two parabolic mirrors focus the laser beam at the IP and re-circulate it parallel to the electron beam axis (= the z.axis).

To keep constant the crossing angle, several interaction planes must be considered. Each interaction plane must contain the z axis so that one must design an optical system capable to switch the optical path from one plane to another.

To switch from one interaction plane to the next one, two parallel flat mirrors are used (see Fig. 120). Such a pair of parallel mirrors forms an optical invariant and is well known to provide an emerging beam parallel to the entrance one whatever its orientation. The re-circulating laser beam axis is thus always parallel to the symmetry axis z when impinging the parabolic reflectors, thereby reducing the optical aberrations. As shown in Fig. 119, the mirror pairs are located on a circular helical in order to maximize the number of passes.

In terms of number of optical surfaces optimization, one could have only inserted one mirror pair each two re-circulation passes (i.e two successive bunch crossings in a single plane). However, as described below, the re-circulator round-trip must match a sub-harmonic of the RF frequency. Besides, as also shown in section 3, the distance between the two parabolic reflectors must be fixed within a few micrometers so that it cannot be used to synchronize the re-circulator to the RF clock. We are then forced to insert a mirror pair for each re-circulation pass and, in addition, to play on their orientations in order to vary the round-trip lengths independently the one from the other. This the way the re-circulation round trip frequency will be lock to the RF clock.

To optimize the number of re-circulating passes, the optical paths undergo a 4π rotation inside the re-circulator. Consequently one can show that the number of passes has to be a multiple of 4 (see Fig. 121).

From an optical point of view, if the mirror pairs of Fig. 120 are parallel 'enough', one only needs to align the incoming laser beam parallel to the electron (z) axis with the proper beam diameter. In addition, the two parabolic reflectors must also be aligned in a confocal geometry. Therefore very few adjustment screws must be implemented on the device.

Nevertheless, it is clear that a good operation of the system relies mainly on the mirror pair parallelism on the parabolic mirror alignment and on the stability of direction and size of the incident laser beam. Although robust and precise experimental tools enabling the laser beam stabilization are available, the mirror pair construction procedure must be carefully addressed. Since this is the crucial technical point of the re-circulator, we describe the alignment procedure in the next section.

The re-circulator round trip length must also be synchronized to the RF clock. From the study described in 4.2.5 one sees that a precision of ~100fs is required on the re-circulator round-trip period (i.e. ~30µm on the round trip length). The synchronization procedure is described below.

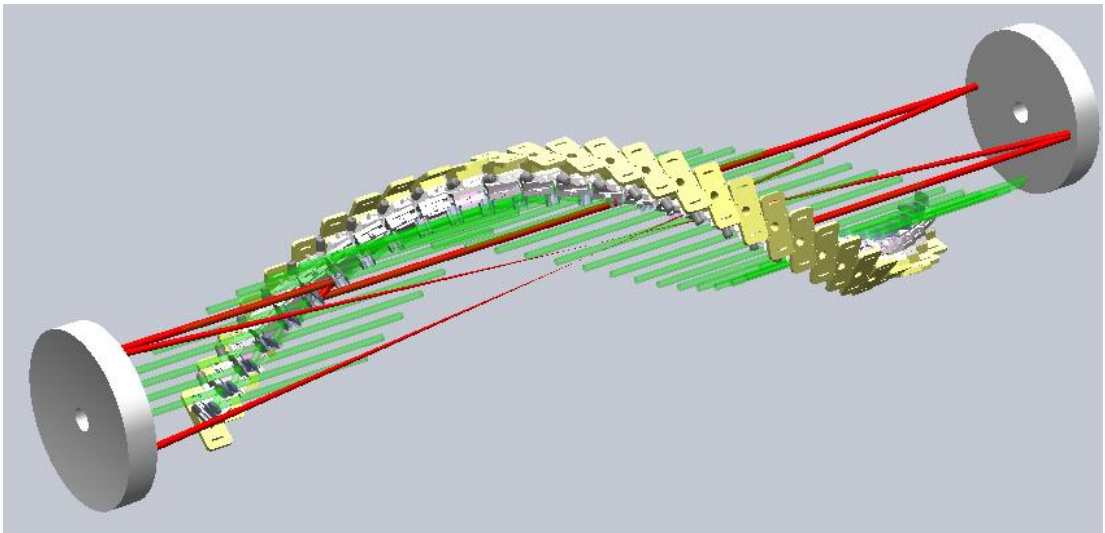

**Fig. 119.**     Schematic view of the re-circulating principle

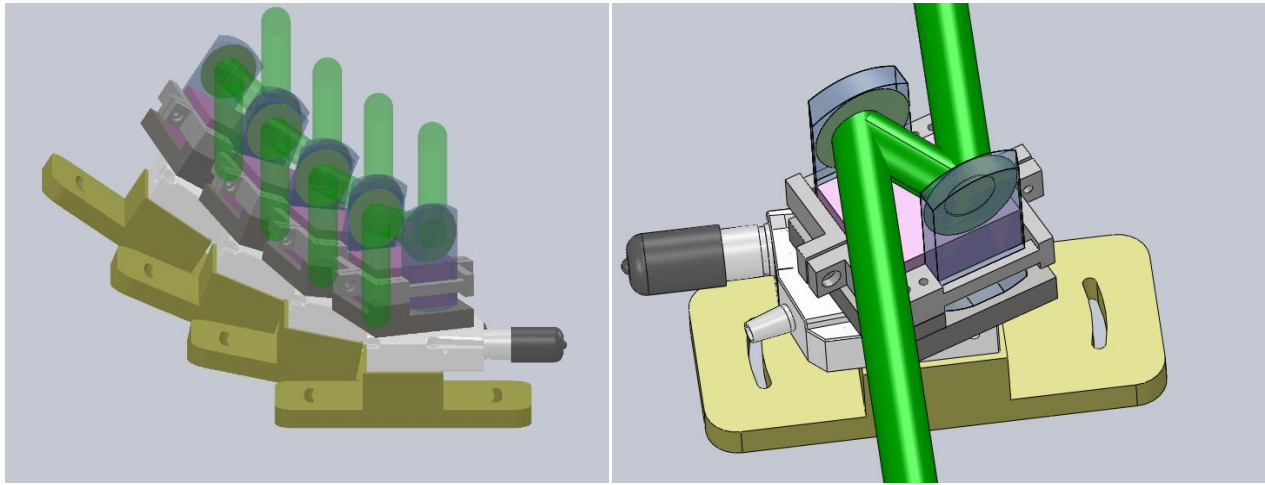

**Fig. 120.** Schematic view of the motorized mirror pairs used in the re-circulator

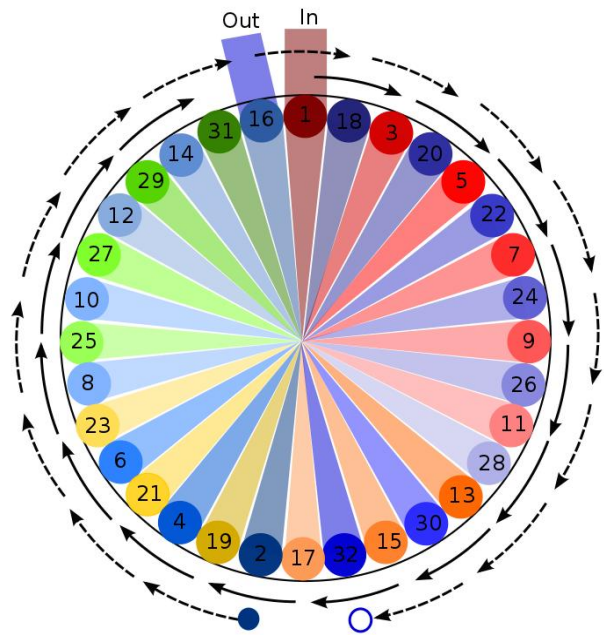

**Fig. 121.** Schematic front view of the re-circulator together with the pass numbers

### 4.2.3.1 Description of the mirror pair precise alignment

The alignment method must be based on interferometric or auto-collimation measurement. The auto-collimation procedure [104] is described in Fig. 122: the image of an illuminated object, located in the posterior focal plane of the collimator lens, is projected to infinity and reflected by a mirror. The image is picked up by a light-sensitive receiver. A slight alteration of the angle between the optical axis of the autocollimator and the mirror causes a deviation which can be determined very precisely.

The procedure that we will use to align the multi-pass re-circulator is sketched in Fig. 122. It is an iterative procedure where the mirror parallelism is adjusted sequentially so that the adjustment of the $n^{th}$ mirror pair can compensate for the residual misalignment of the n-1 previous ones. We checked, using the Code V simulation described below, that this procedure strongly reduces the impact of the mirror pair misalignments on the re-circulator performances.

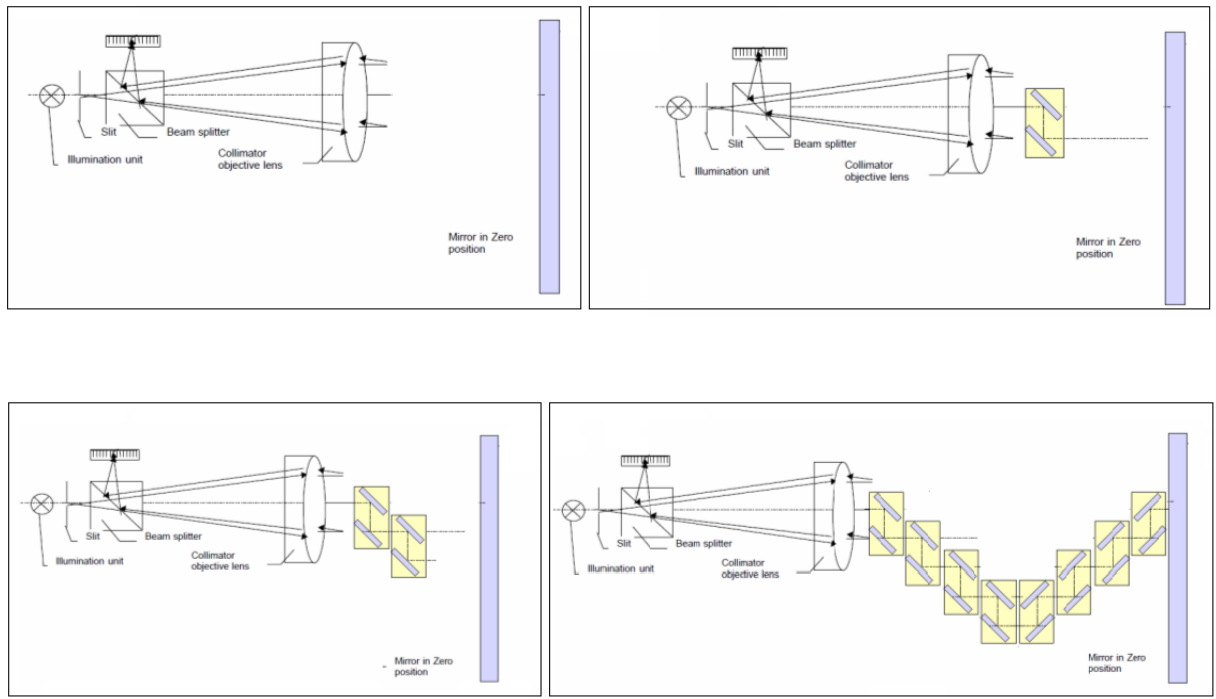

**Fig. 122.    Illustration of the mirror pairs parallelism measurement.**

### 4.2.3.2     Fine alignment tools

To optimize the performances of the re-circulator, the 32 laser spot crossing the electron beam axis should overlap within a few micrometers accuracy at the IP. Due to the geometry of the re-circulator an optical imaging with high spatial resolution and long working distance is required. Moreover, to isolate each round trip laser focal spot, an area detector with an ultra-fast shutter is needed. A nanosecond time gated optical intensifier and a large high sensitivity coupled charged device (CCD) compose this detector. A drawing of the optical imaging system is given in the figure below.

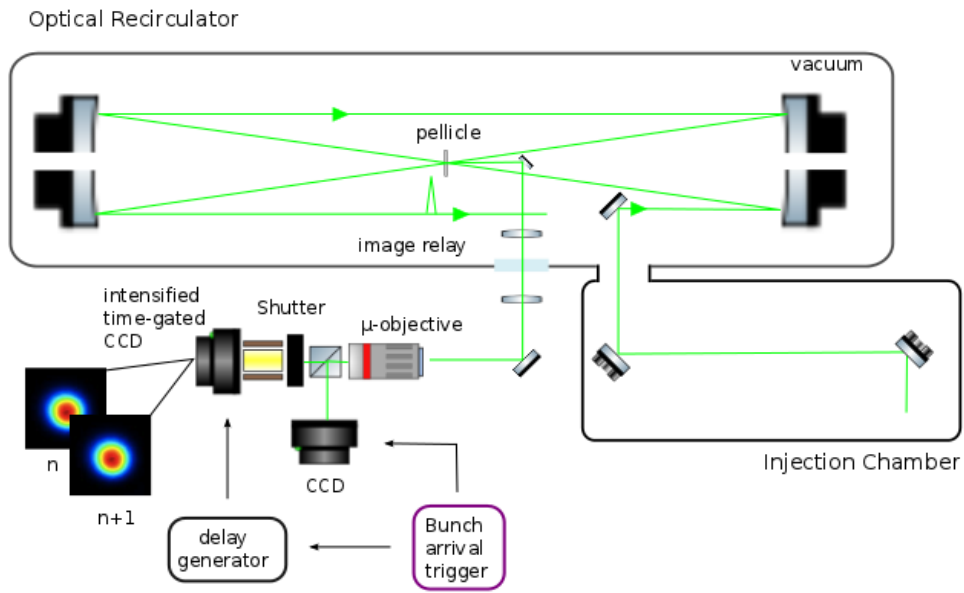

**Fig. 123.** Scheme of the optical system set on the re-circulator environment (not on scale)

In order to image the beam without blocking it, a motorized translation stage is used to insert an ultra-thin pellicle. The pellicle thickness is about 2µm and both faces are AR coated for 515nm. The very low reflection level allows the use of an intensified CCD and the record of all roundtrip focal spot. The transverse beam walk-off due to refraction in the thin pellicle is about 80nm/round trip and negligible for a fine spatial tuning of the re-circulator. The pellicle is placed perpendicularly to the electron beam axis. The image is then transported with a 2f-2f or 4f combination imaging lens doublet outside of the vacuum chamber to the object plan of a high aperture and magnification microscope objective. A non-polarizing beam splitter is inserted to send the beam toward two kinds of detectors: an intensified CCD and a large field of view high dynamic CCD camera that provides all together the 31 round-trips images and ensures a preliminary control of light flux for intensified CCD activation. The intensified CCD is triggered on RF pulses via delay generator computer controlled to switch the acquisition time window to record successively the 32 round-trips.

#### 4.2.3.3    Re-circulator synchronization

We are considering to different technique for the synchronization of the re-circulator round trip period. The first one uses the IP optical imaging system of Fig. 125. The second one uses non-linear optical parametric amplification in a non-linear crystal inserted at the interaction point. Both methods make necessary the use of an external femtosecond oscillator lock-on the OMO pulse train (see section 3.8).

**Alternative 1**
As shown later in Fig. 138, the electron bunches and the re-circulating laser pulses have to overlap in time within a few hundred of femtosecond. Thanks to the motorized parallel plan mirrors, all the round-trip lengths are independently tunable for adjusting the electron bunch/laser pulse synchronizations. The first method that we can use is thus based on the existing nanosecond-gated optical imaging device devoted to the fine spatial alignment of the re-circulator (see Fig. 123). It is based on the contrast measurement of an interference pattern obtained by coherent time combining of two laser pulses at the IP. One pulse is addressed to represent an electron bunch of the train on the axis of the re-circulator, the other pulse is injected on the re-circulator.

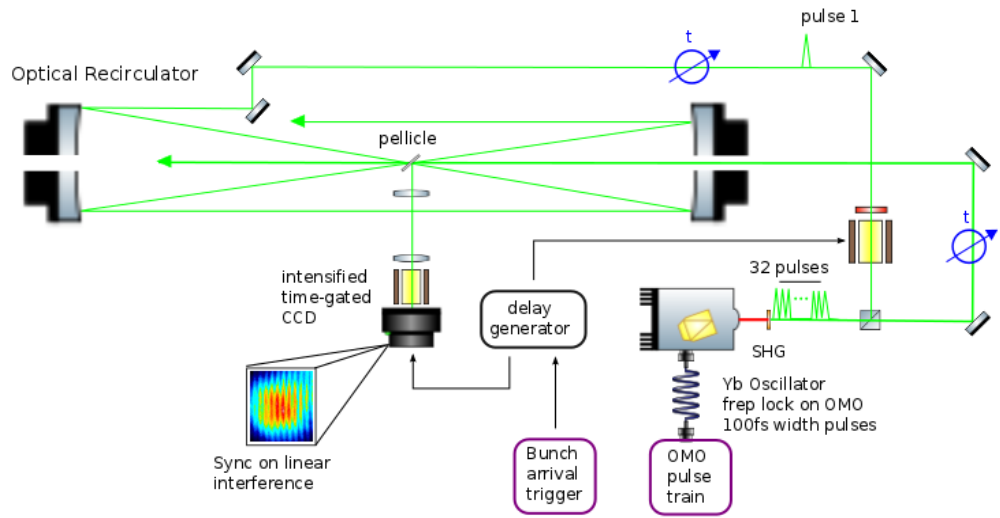

**Fig. 124.** Re-circulator synchronization principle: the 1030nm femtosecond oscillator is locked (optical mixer) on the distributed machine RF clock (i.e., the OMO). The 515nm radiation is created by the second high harmonic generator (SHG) module. An electronic synchronization of the moving time gate for iCCD and the electro-optic modulator (EOM) are used to pick up the pulse injected in the re-circulator

The re-circulator synchronization is then based on a frequency doubled Yb optical laser oscillator locked to the distributed machine RF clock (see Fig. 127). The laser beam is split into two beams. The first beam is injected on the electron axis and used as reference of the synchronization of the re-circulator. The second beam is sent into an electro optic modulator used as a pulse picker.

The synchronization of the re-circulator is realized by fringes contrast optimization for each pass. Pass after pass, the contrast is extracted from images taken by the nanosecond-gated optical imaging system. Then, after post processing, all the mirror pair are rotated to adjust laser beam path length of each trip in the re-circulator. The synchronization is performed when the electron beam is off and the IP laser blocked. The timing accuracy depends on the synchronization laser pulse duration. The synchronization laser pulse FWHM is 100fs.

**Alternative 2**

To reach the 100fs precision on the re-circulator round trip period, we will use a technique inspired from the optical synchronization distribution method described in section 3.8. The method principle is depicted in Fig. 125. A non-linear crystal will be inserted at the waist position of the circulating laser beam (*i.e.* at the focus points of the two parabolic reflectors). Then, by launching a 100fs femtosecond laser beam synchronized to the RF clock on the electron beam axis, one produces signal usually called 'idler' from the non-collinear parametric coupling between the 'pump' beam and the re-circulating 'signal' laser beam inside the crystal [105]. By measuring the idler intensity for each re-circulation pass one can thus optimize the re-circulation round trip durations.

A non-linear crystal thickness lower than 10µm must be used to limit the perturbation induced on the re-circulating beam. To provide a Type I phase matching one must also send the pump beam along the longitudinal z-axis and send the signal beam into the re-circulator. Since this experiment will be done using low power laser beam we are free to choose the pump and signal beam wavelengths. We can re-circulate a 1550nm wavelength beam since the mirror coating for 515nm is also highly reflecting for 1550nm. The idler beam will come out along a direction satisfying the phase matching condition $\vec{k}_p = \vec{k}_s + \vec{k}_i$ [106]. The smallest $\lambda_p$ the closest to the z-axis will come out the idler beam. By choosing the second harmonic of Ti:Sapphire laser amplifier $\lambda_p$ =400nm, the idler beams emission angles will be below 2°, that is inside the cone delimited by the optical path of the re-circulating laser beam. A simple imaging optics can then be implemented around the z-axis between the IP and le parabolic mirror $PM_0$ to focus all idler beams onto a single photo-detector. With this technique, inspired from [107] and those described in section 3.8, a time synchronization better than 100fs could be achieved. Note that a similar technique, based on four-wave mixing and described in [108], led to synchronization better than 100fs.

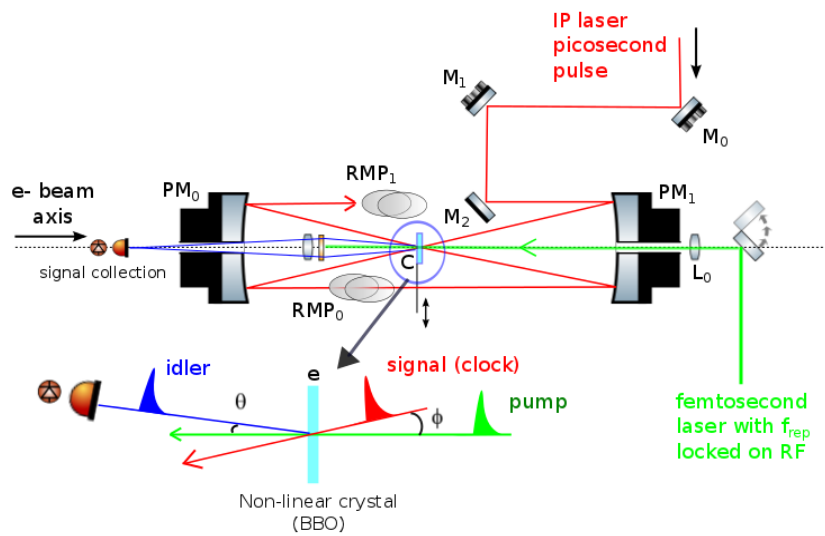

**Fig. 125.** Schematic drawing of the method foreseen to synchronize the re-circulator round-trips

During the machine operation, the slow control of the gamma ray flux will provide information on the re-circulator/RF synchronization and will be slowly adjusted by rotating the mirror pairs ($RMP_i$ in Fig. 125).

#### 4.2.3.4 Optical performances: aberrations

The optical astigmatism and ellipticity induced pass after pass are estimated using the Code V software [109]. For the re-circulator geometry given by the parameters of Table 36, we obtain negligible ellipticity and astigmatism for all optical passes. Fig. 126 shows the laser beam spots on one of the two parabola of the re-circulator. The spots are perfectly Gaussian and equidistantly spaced on a circle. The beam propagation is done using the very precise Code V beam synthesis propagation.

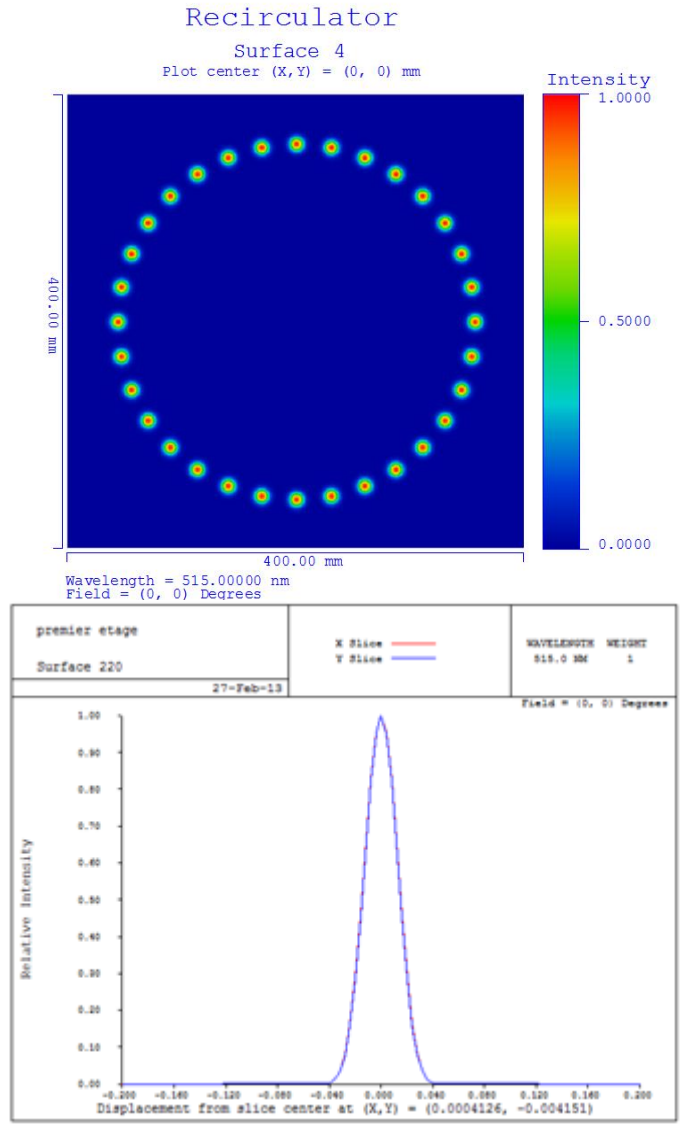

**Fig. 126.** Top: intensity profile of the laser beam (from Code V beam synthesis propagation) on one of the parabolic reflector. The laser beam spots map a circle on the parabola. Bottom: transverse projections (x and y axes) of the laser beam intensity after 32 passes at the IP of the re-circulator (calculation from Code V beam synthesis propagation). The x and y distributions lie on each other

#### 4.2.3.5  Optical performances: laser beam polarization Transport

We model a typical high reflectivity mirror coating made of $N_{DL}$ $SiO_2/Ta_2O_5$ double layers using the formalism of [110]. Fig. 127 shows the $s$ and $p$ wave transmittance $T_s=1-|r_s|^2$, $T_p=1-|r_p|^2$ as a function of the incident angle and $N_{DL}$. Sizable differences are observed for large incident angles. The phase shift between the $r_s$ and $r_p$ exactly matches $\pi$ for a perfectly aligned system.

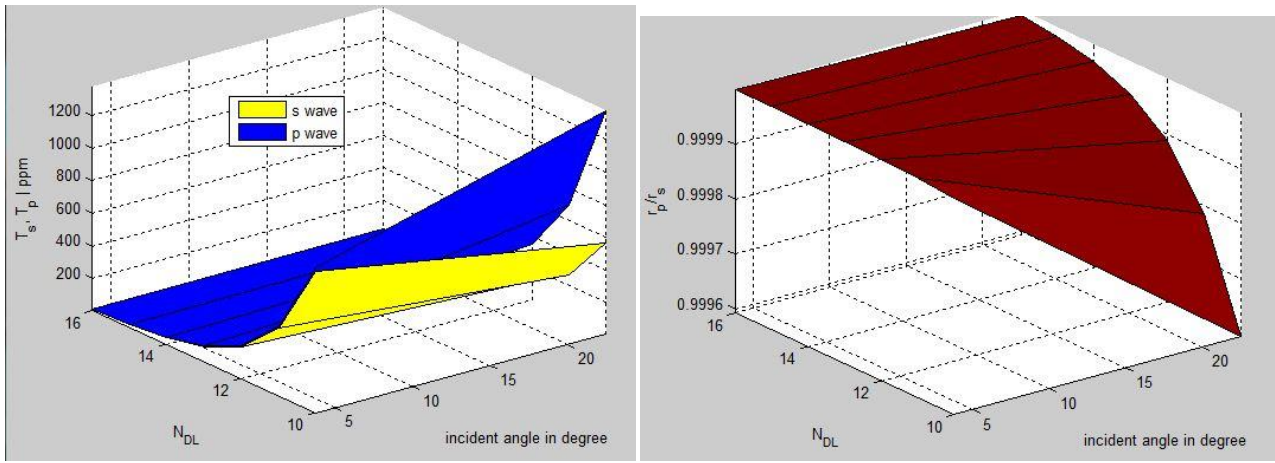

**Fig. 127.** Left: s and p wave transmittances as a function of the incident angle and the number of double layers. Right: reflection coefficient ratio $r_p/r_s$ as a function of the incident angle and the number of double layers.

For 1"arc misalignment incident angle one obtains a phase shift of $\pi$ - 1μrad for an incident angle of 22.5° and for all $N_{DL}$. One can then safely assume that no noticeable polarization ellipticity is induced during the optical path (as will be shown later on in this chapter, a few arc second is the typical angular alignment tolerance of our system).

For an incident linear polarization, one thus observes a rotation of the polarization direction at the IP induced by the ratio $r_p/r_s$. This ratio is shown in Fig. 127. To calculate the pass by pass polarization rotation, we use the OSLO code assuming various values of $r_p/r_s$. A typical result is shown in Fig. 128.

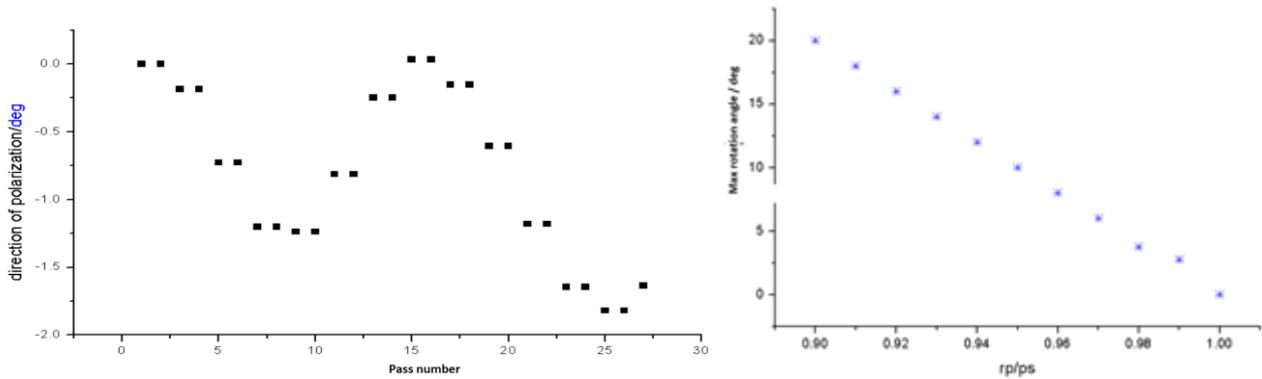

**Fig. 128.** Left: variation of the orientation of a launched linear polarization at the IP as a function of the pass number for $r_p/r_s$ =0.99. Right: maximum variation of the linear polarization orientation occurring during the optical passes as a function of $r_p/r_s$.

Assuming the $r_p/r_s$ variation range of Fig. 127 one then gets a maximum rotation of the linear laser beam polarization between 0.01° and 0.1° and a negligible ellipticity.

### 4.2.4. Optimization of the re-circulator geometry

To optimize the re-circulator geometry, we do consider the beam parameters corresponding to the production of 2MeV and 10 MeV gamma ray beams. We choose the same geometry for both IPs because they will be operated quasi-simultaneously. The only possible difference between the two IPs is the laser pulse energy. The laser beam characteristics are given in Table 36 below.

### 4.2.4.1 Optimization of gamma-ray beam luminosity

For a fixed beam quality e-beam and Laser system, the Thomson Scattering output from a single laser-bunch interaction depends on the free parameters of Table 34.

**Table 34.    Free parameters of the Thomson scattering output**

| | |
|---|---|
| $\alpha_{Dev}$ | Deviation from backscattering |
| $W_0$ | Pulse minimum waist |
| $\sigma_b$ | Bunch rms transverse size |
| $\Psi_{Acc}$ | Normalized acceptance rms angle = $\gamma * \theta_{Acc}$ rms, being $\theta_{Acc}$ = ½ $\theta$ max |

Deviation from backscattering ($\alpha_{Dev}$) angle controls pulse-bunch overlapping which reduces as $\alpha_{Dev}$ increases; pulse minimum waist ($W_0$) is linked to both intensity (and nonlinear effects which introduce blurring) and overlapping with the e-bunch. Bunch rms transverse size ($\sigma_b$) is mainly linked to overlapping but it controls a portion of intrinsic energy spread since tranverse momentum spread is inversely proportional to $\sigma_b$. Normalized acceptance angle $\Psi_{Acc}$, finally, is the key parameter for the determination of the number of collected photons and the correlated energy spread.

The energy spread of the collected photons can be roughly estimated with leading terms as

$$\left(\frac{\delta E}{E}\right)_{rms} \cong \frac{1}{4}\Psi^2_{Acc} + 2\frac{\delta\gamma}{\gamma} + \left(\frac{\varepsilon_n}{\sigma_b}\right)^2 + \left(\frac{\delta E}{E}\right)_{Laser} + \frac{1}{2}a_0^2 + ...$$

Since for a fixed scattering angle the contribution

$$\left(\frac{\delta E}{E}\right)_{UNcorrelated} \cong 2\frac{\delta\gamma}{\gamma} + \left(\frac{\varepsilon_n}{\sigma_b}\right)^2 + \left(\frac{\delta E}{E}\right)_{Laser} + \frac{1}{2}a_0^2$$

is completely uncorrelated, we can refer it as the Uncorrelated energy spread *(dE/E)$_{Uncorrelated}$*, which is the lowest energy spread we can extract from our laser and bunch interaction. Moreover, since peak energy depends also on the scattering angle, different angles within the acceptance cone will receive gamma rays having different energies, thus producing a *correlated* energy spread contribution *(dE/E)$_{Correlated}$* =1/4 $\Psi^2_{Acc}$.

Photon flux within the acceptance cone depends on $\Psi_{Acc}$ through

$$N_{Acc} \propto \Psi^2_{Acc} \frac{1 + \frac{1}{4}\Psi^2_{Acc} + \frac{1}{6}\Psi^4_{Acc}}{(1 + \frac{1}{4}\Psi^2_{Acc})^2} \cong \Psi^2_{Acc} \quad \text{if} \quad \Psi_{Acc} << 1 \ .$$

In order to make a full scan in the Thomson Scattering interaction output quality a 4D matrix would be needed. Once the goal energy spread is fixed, however, the dependence of the Spectral Density on the best $\sigma_b$ is weak: this allows a simplification of the scan that has been performed as follow.

**Step1**. Having fixed a 0.5% goal energy spread find the best (trial) $\sigma_b$ ($\sigma_{b\ TRIAL}$) for a pre-scan with coarse-grained grid in the 3D sub-matrix ($\alpha_{Dev}$, $W_0$, $\Psi_{Acc}$) .

**Step2**. Fix $\sigma_b = \sigma_{b\ TRIAL}$ and scan ($\alpha_{Dev}$, $W_0$, $\Psi_{Acc}$) with a fine-grained grid

**Step3**. Working point for ($\alpha_{Dev}$, $W_0$, $\Psi_{Acc}$) is chosen by maximizing the global TS output

**Step4**. Refine $\sigma_b$ with a final scan with ($\alpha_{Dev}$, $W_0$, $\Psi_{Acc}$) from previous step.

### 4.2.4.2 2MeV working point

The procedure in **Step1** gave a rough estimation of the bunch size as 18 μm. Simulations in **Step2** produced results summarized in Fig. 129.

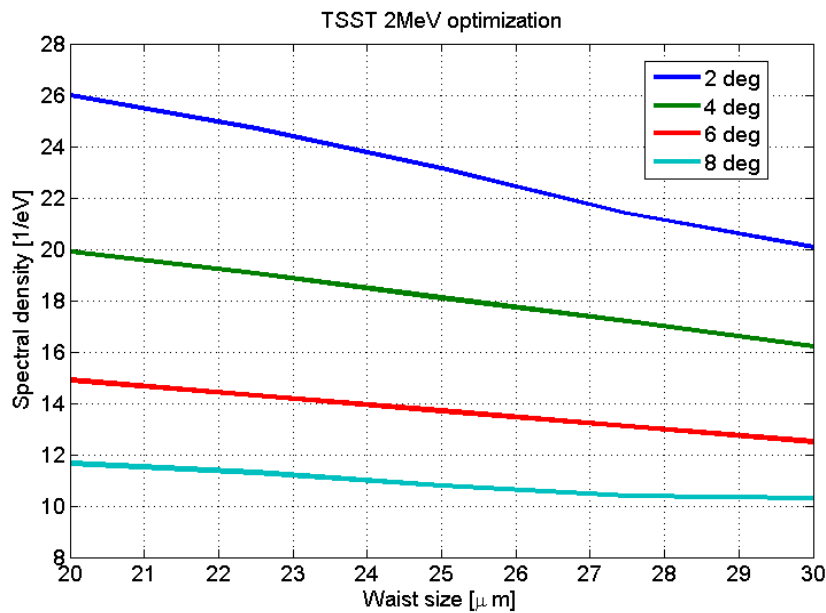

**Fig. 129.    Results of Step2**

### 4.2.4.3 10MeV working point

The procedure in **Step1** gave now a rough estimation of the bunch size as 20 μm. Simulations in **Step2** are finally shown in Fig. 130**.**

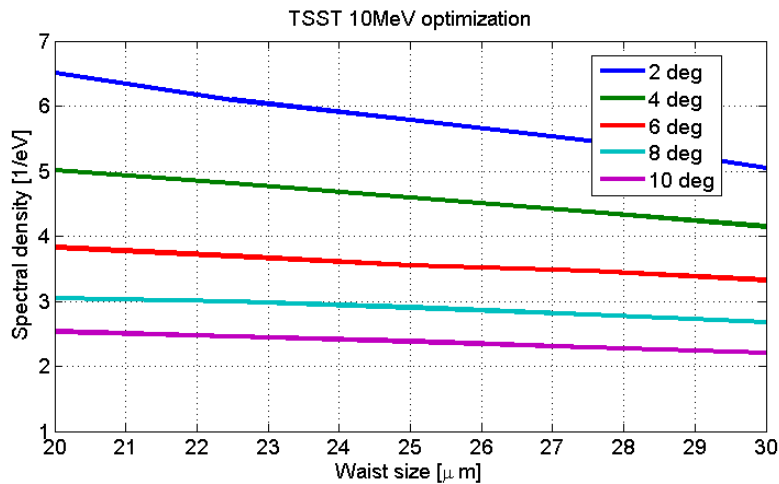

Fig. 130.     Simulations in Step2

#### 4.2.4.4     Constraints for the optimization

In optimizing the re-circulator we have to take into account the following constraints:

1. *Laser fluency damage threshold of reflective surfaces*: the laser pulse energy (400mJ at maximum) and FWHM (3.5ps) impose a constraint on the beam spot size on mirrors and therefore on geometry (focal length, optics' size).

2. *RF wavelength and re-circulator path length matching*: the optical length of each round trip inside the re-circulator must be matched to the distance between two bunches of electrons if the system is perfectly aligned.

3. *RF pulse duration*: the total re-circulation path duration may be less or equal to the RF duration (600ns).

4. *Azimuth ordering of the optical passes:* the laser beam spots on the parabola describe a circle. The spots must be a multiple of four and equally spaced on this circle (see Fig. 126).

5. *Maximum number of passes=32.* This limit is givenby the photoinjector specifications.

6. *Tolerances on optics manufacturing, mechanical stroke and thicknesses of mirrors and structures*: with the objective of a realistic re-circulator design we have to take into account the tolerances provided by the optical and mechanical manufacturers. We restrict ourselves to the tolerances on the distance between the mirrors of mirror-pairs (=0.1mm), on the focal length of the parabola (=0.025mm) and the possibility to vary by ±200fs the optical path using the mirror-pairs for synchronization purpose. We also account for a minimum space (=7mm) between two adjacent mirror-pairs and for mirror and structure thicknesses (Th=2x1cm), that is $D_M = 7mm + 2 \times 1cm = 27mm$(see Fig. 131). More details can be found in [103].

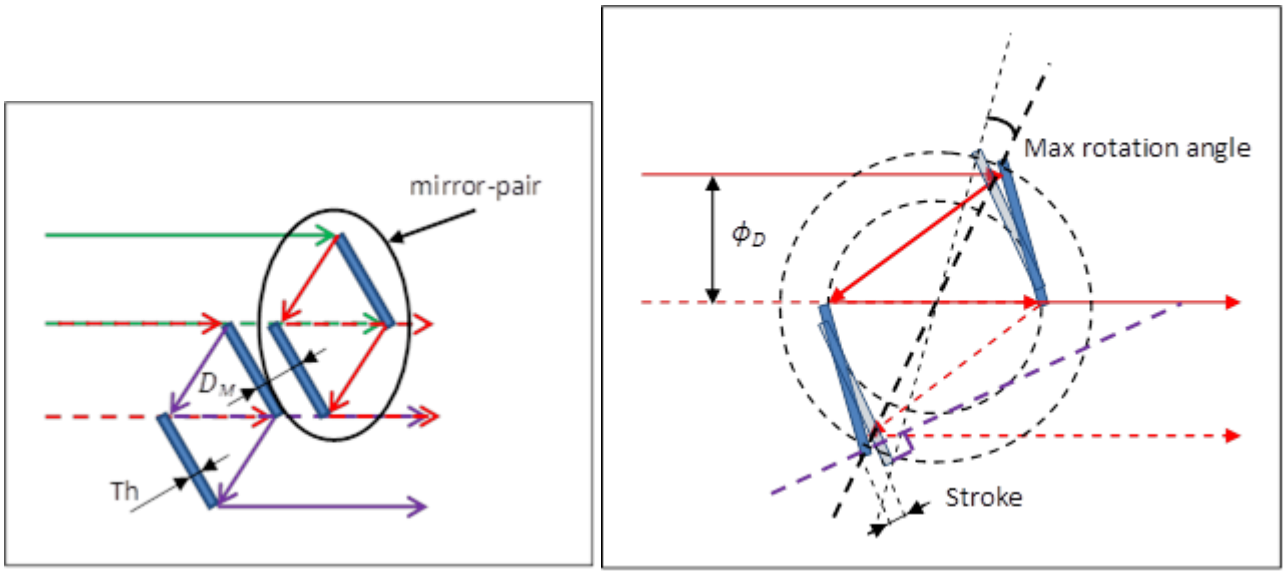

**Fig. 131.** Left: In plan scheme of the mirror-pair disposition (unrolled view of 2 mirror pairs). Right: Scheme of the mirror pair's stroke

### 4.2.4.5 Optimization

The re-circulator length should be in the order of 2.5m. We assumed a realistic value for the mirror coating damage threshold $F_0$=20J/cm² for FWHM=10ns. We propagate it down to the picoseconds range using the empirical law derived in [110]: $F_{max}=F_0(\text{FWHM[ns]}/10)^{1/2}$. It is important to notice that we are considering a conservative damage threshold value. Table 35 shows damage thresholds of various mirror coating manufacturers. From this table one sees that the $F_0$ value that we chose is a factor of 2 below the performances of these manufacturers. In addition, extrapolating the REOSC-SAGEM result for 0.5ps pulse duration, a factor of 10 higher could be even be envisaged at 3ps since the scaling low of [110] starts to break down in this region. Nevertheless, one must keep in mind that the damage threshold also depends on the number of laser pulses [111] and on the laser beam diameter [112].

**Table 35.    Damage thresholds values for various manufacturers**

| Manufacturer | Damage threshold |
|---|---|
| REOSC-SAGEM [113] | ~45J/cm²@1057nm, 3ns<br>[~2J/cm²@1057nm, 0.5ps]<br>R ~ 99.6% |
| REO [114] | ~20J/cm²@1064nm, 3ns<br>R ~ 99.99% |
| Laser Components [115] | ~15J/cm²@532nm, 10ns<br>R ~ 99.8% |
| CVI [116] | ~9.7J/cm²@532nm, 10ns<br>R ~ 99.9%<br>200 multiple shots |
| Layertech | 0.5-1J/cm²@800nm, 1 ps<br>R>99.9% |

*Nota: The manufacturer list is not exhaustive but just representative on the high damage threshold coating commercially available. The damage threshold values are taken from the web sites*

With

$$F = \frac{2U_{geo}}{\pi \omega^2}$$

And assuming $F = F_{max}$, we obtain

$$\omega = \sqrt{\frac{2U_{geo}}{\pi F_{max}}}$$

Where $\omega$ is the laser beam radius on the parabolic mirrors and $U_{geo}$ is the laser beam energy considered for the optical design of the re-circulator (it gives the maximum energy that can be launched inside the re-circulator). Using this expression, we obtain the distance $D$ between the two parabolic mirrors:

$$\omega \approx \frac{\lambda}{\pi \omega_0} M^2 \frac{D}{2} \Rightarrow D = \frac{2\pi \omega_0}{\lambda M^2} \sqrt{\frac{2U_{geo}}{\pi F_{max}}}$$

Where $\omega_0$ is the beam waist at the IP, $\lambda$ the laser wavelength and where $M^2$ describes the laser beam wave front quality. Since the circulating laser beam is collimated between the two parabolic mirrors, we further fix the diameter of the mirror pairs as $\phi_D = 2n\omega$ with $n = 1.8$. For the re-circulator geometrical design we also fix $U_{geo}$ = 400mJ (notices that, during the operations, we shall consider 200mJ and 400mJ at the first and second IPs respectively). For fixed $F_{max}$, $U_{geo}$ and $\lambda$ values, the distance $D$ is then solely determined by $\omega_0$.

The gamma ray flux or the Time Average Spetral Density (TASD) is directly proportional to the number of passes, which mainly depends on $\omega_0$ and on the crossing angle at the IPs. Our geometrical optimization of the re-circulator geometry is then done by changing only two parameters: the crossing angle and $\omega_0$.

Other important parameters are those related to the mirror-pair dimensions. These parameters are tuned in order to locate at equal distance the laser beam spots on the parabola circle.

To preserve the laser beam polarization all along the re-circulation, the angle of incidence on the mirror-pairs should be the lowest. We found that 22.5° was the good compromise between optical and geometrical constraints.

We performed the numerical studies using the Matlab software package. The top plots of Fig. 132 show the results of the TASD calculations (normalized to 5000γ/eV.sec) and the maximum number of passes as a function of the crossing angle and $\omega_0$ at the first IP (*i.e* for the 2MeV γ ray beam). The results for the second IP (*i.e.* 10 MeV γ ray beam) are shown in the top of Fig. 133. These figures lead us to choose $\omega_0$=28.3µm which is a good compromise between a high TASD value and a reasonable crossing angle. The TASD corresponding to $\omega_0$=28.3μm are shown in the bottom plots of Fig. 132 and Fig. 133 for the 2MeV and 10MeV γ ray beams respectively. Our final parameters are summarized in Table 36.

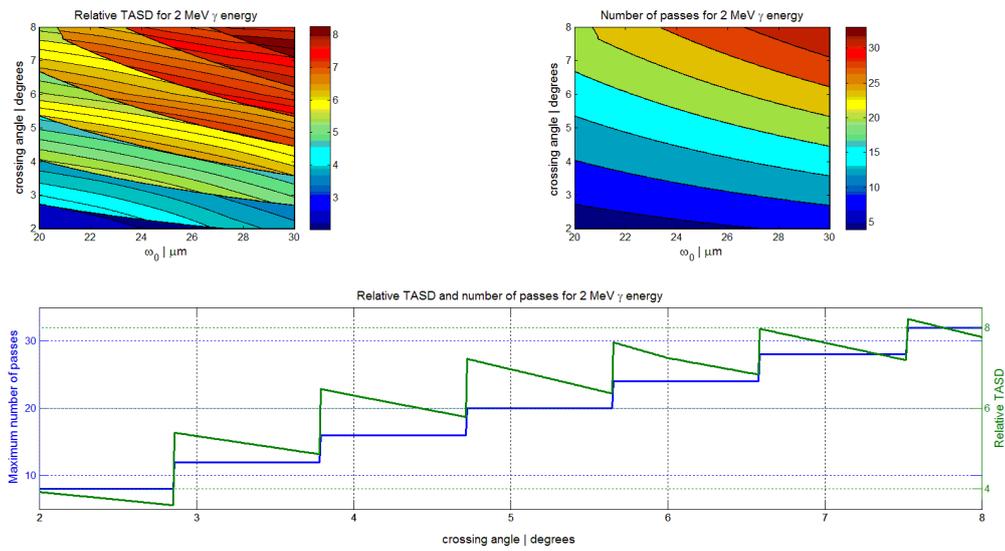

**Fig. 132.** Top left: TASD (=Time Average Spectral Density) as a function of the crossing angle and $\omega_0$. The TASD is normalized to 5000/s.eV. Top right: Maximum number of passes as a function of the crossing angle and $\omega_0$. The results correspond to the 2MeV γ ray beam (i.e. 1rst IP electron beam parameters). The angle of incidence on the mirror pairs are fixed to 22.5°, the laser pulse energy to U=200mJ and the pulse time width to FWHM=3.5ps. The bottom plot shows the TASD and the number of passes as a function of the crossing angle for $\omega_0$=28.3μm.

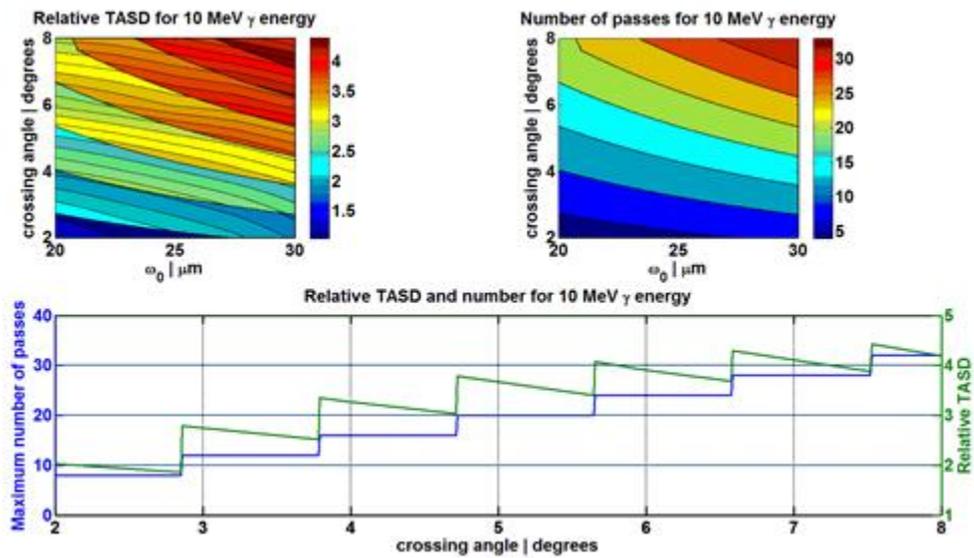

**Fig. 133.** Same as Fig. 132 but for the 10 MeV g ray beam (i.e. the second IP electron beam parameter) and for a laser pulse energy U=400mJ.

**Table 36.    Summary table of the chosen configuration**

|  | First IP<br>(2 MeV γ ray) | Second IP<br>(10 MeV γ ray) |
|---|---|---|
|  | **Laser beam** | |
| Laser pulse energy (in mJ) | 200 | 400 |
| Laser wavelength (nm) | 515 | |
| Laser time FWHM (ps) | 3.5 | |
| $M^2$ laser beam quality factor | 1.2 | |
|  | **Mirror pairs** | |
| Distance between mirrors in mirror pair (mm) | 40.1 | |
| Incidence angle on mirror pairs (deg) | 22.5 | |
| Transverse occupation of a mirror pair (mm) | 60.5 | |
| Distance between mirrors on corona (mm) | 30.7 | |
|  | **Re-circulator** | |
| Number of passes | 32 | |
| Distance between parabolic mirrors (mm) | 2378.9 | |
| Incidence angle at IP (deg) | 7.5 | |
| Corona radius (mm) | 156.7 | |
| Beam waist at IP (µm) | 28.3 | |
| Laser field radius on mirrors (mm) | 8.3 | |
| Flat mirrors diameter (mm) | 29.7 | |
| Conversion factor from waist on mirrors to flat mirrors diameter | 3.6 | |
| Relative TASD (divided by 5000 γ/eV.s) | 8.1 | 4.4 |

### 4.2.5.    Tolerances on the re-circulator alignment

The Code V optical simulation software is used to estimate the mechanical tolerances and we cross checked our results using an independent code, OSLO. We move all the alignment degrees of freedom independently and compute the beam characteristics at the IP for each misaligned configuration. Next, from the Code V output (waist sizes, waist positions and the beam direction for each pass) we compute the luminosity geometrical factor as described in [117] using a MATLAB [111] code (*i.e.* a triple overlap integral over *x,y,z* accounting also for the delay induced by the misalignments). The calculation is done for all passes. The TASD is finally compared to the nominal TASD factor corresponding to the perfectly aligned geometry.

#### 4.2.5.1    Injection: pointing instability

We parameterize the incident laser beam wave vector as:

$$\mathrm{k}_{inj} = k\bigl(\sin \delta\theta_{inj} \cos \varphi_{inj}, \sin \delta\theta_{inj} \sin \varphi_{inj}, \cos \delta\theta_{inj}\bigr)$$

with $k=2\pi/\lambda$ and $\delta\theta_{\mathrm{inj}} \ll 1$, $\varphi_{\mathrm{inj}} \in [0,2\pi]$.

This parameterization is supplied at the incident beam waist position, *i.e.* at 252.6mm in front of the parabolic reflector.

The relative TASD ($r_{TASD}$) loss is shown in Fig. 134. From this figure one set a tolerance of 0.5"arc (3μrad) on the injection beam pointing angle.

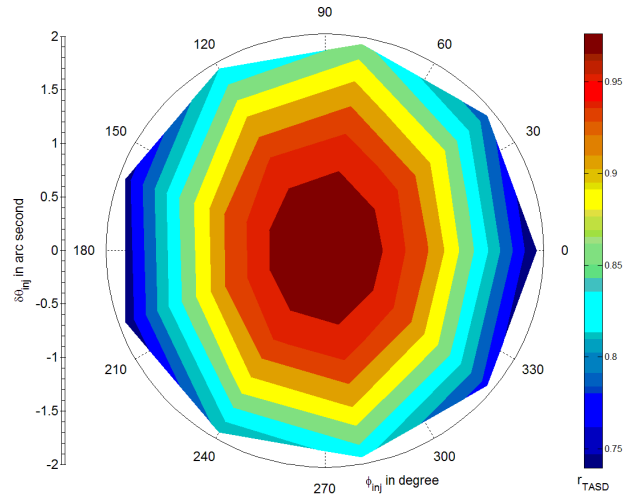

**Fig. 134.** Relative TASD loss induced by an injection beam pointing misalignment as a function of $\delta\theta_{inj}$ and $\varphi_{inj}$.

#### 4.2.5.2     Injection: waist mismatch

The incident waist is fixed to $\omega_{0,in}$ = 8.26mm and is located 252.6mm in front of the parabolic reflector. We varied the waist position from 0 to 1m and obtained negligible variation of the TASD.

Fixing the waist position the beam radius reaches $\omega_{para}$=8.26mm on the parabolic mirror. Then, from $\omega_{para} \simeq \lambda L/(2\pi\omega_0)$, where $\boldsymbol{\omega_0}$ is the beam waist at the IP, one gets $\Delta\omega_0/\omega_0 \simeq \Delta\omega_{para}/\omega_{para}$. From Fig. 132 and Fig. 133, one sees that a variation of a few microns (*i.e.* ~ $\omega_0/10$) doesn't affect significantly the TASD. However, depending on how much the fluence is closed to the damage threshold, a decrease of $\omega_{0,in}$ may induce mirror damage. For instance, if the fluency is chosen 10% below the damage threshold, then one must keep $\Delta\omega_{para} < \frac{\omega_{para}}{10} \sim 0.8mm$. This fixes the tolerance on the incident waist size.

#### 4.2.5.3     Parabolic mirror misalignment

We proceed as in the injection instability study. We parameterize the angular orientation of the parabolic reflector, *i.e.* the orientation of the normal direction $\hat{\boldsymbol{n}}$ at the center of the parabola, with two angles $\delta\theta_{para} \ll 1, \varphi_{para} \in [0,2\pi]$ such $\hat{n} = (\cos\varphi_{para}\sin\delta\theta_{para}, \sin\varphi_{para}\sin\delta\theta_{para}, \cos\delta\theta_{para})$. We obtain the relative TASD variation show in Fig. 135.

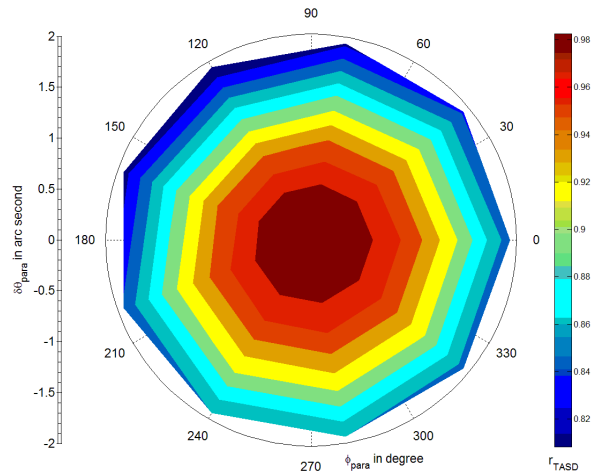

**Fig. 135.** Relative TASD loss induced by a parabolic mirror misalignment as a function of $\delta\theta_{para}$ and $\varphi_{para}$.

Next, we studied the spatial misalignments by moving the parabolic reflector along the x, y, z axes by $\Delta\rho\cos\zeta, \Delta\rho\sin\zeta\ and\ \Delta z$ respectively (with $\Delta\rho, \Delta z \ll 1$ and $\zeta \in [0, 2\pi]$). The corresponding losses are presented in Fig. 136.

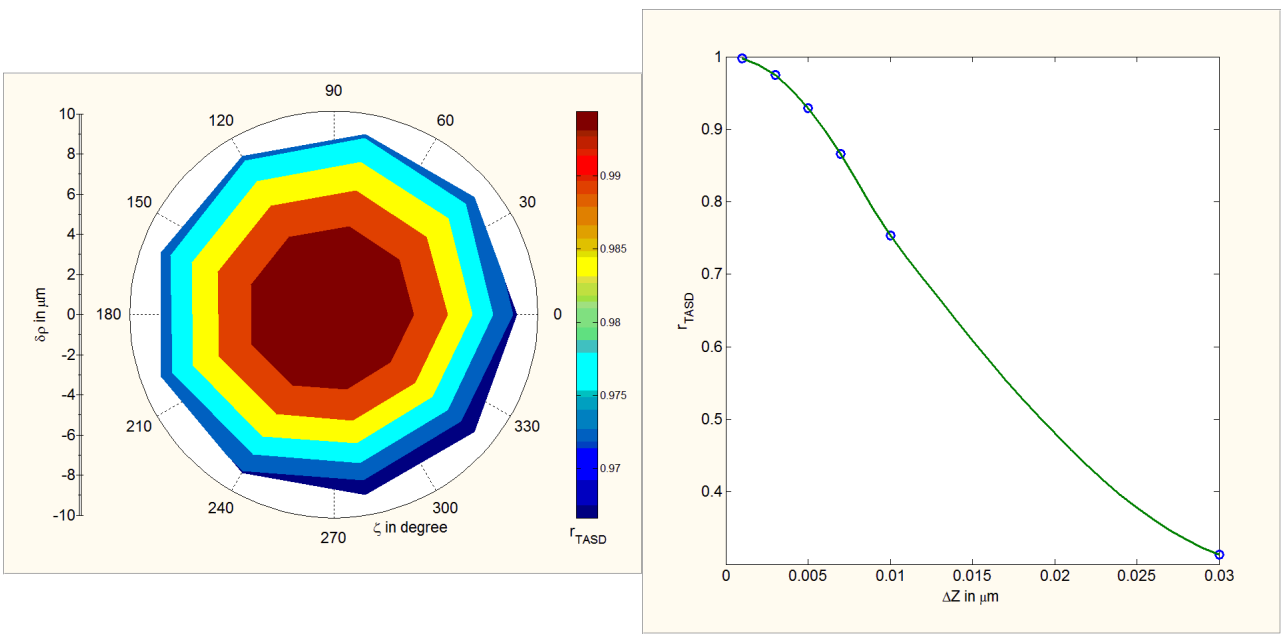

**Fig. 136.** Left: relative TASD loss induced by a parabolic mirror misalignment as a function of $\Delta\rho$ and $\zeta$. Right: relative TASD loss induced by a parabolic mirror misalignment as a function of $\Delta z$.

From these results, one sees that a tolerance of 3µrad and 3µm are required on the alignment of the parabolic reflectors. Note that one of the two reflectors can be rigidly fixed so that only the other one needs to be aligned. Note also that the losses induced by the z misalignment also include a cumulative timing delay effect.

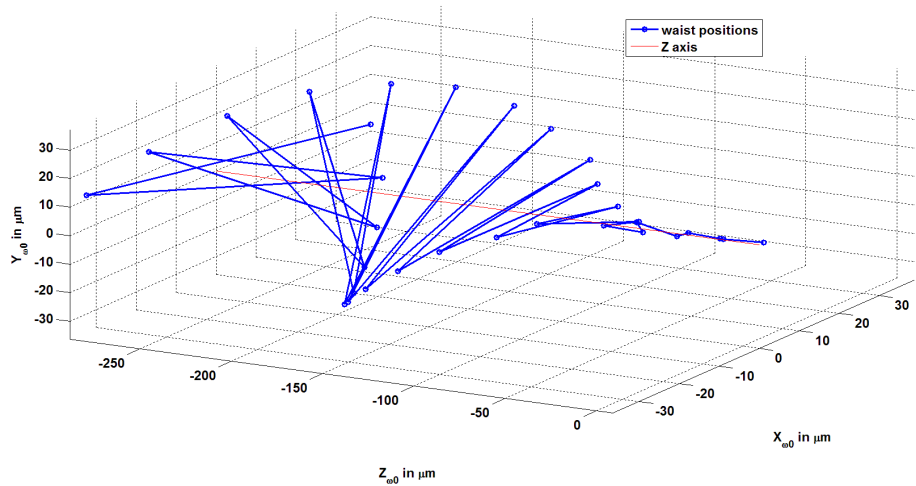

**Fig. 137.** Positions of the beam waist in space pass after pass for: Δz=35μm

Fig. 137 shows the variations of the beam waist position pass after pass for a longitudinal displacement of one parabolic mirror of Δz=35μm. As the waist position is progressively shifted along the z axis, the transverse displacement becomes comparable with the electron beam size. Although the z shifts are partially compensated by the delay accumulated along the path inside the misaligned re-circulator, the transverse displacement induces an irreducible loss. This plot shows that the length tuning between the two parabolic mirrors cannot be used for the re-circulator/RF clock synchronization.

A standard tolerance on the knowledge of the Radius of Curvature (ROC) of a parabolic mirror before manufacturing is ~0.5%. This means that, since our parabolic mirrors are set in a confocal geometry, the distance between the two mirrors is known a priori (i.e. before mirror manufacturing) with an accuracy of ~1mm. Since the re-circulator round-trip has to match an RF wavelength harmonic, such a 'large' uncertainty cannot be tolerated. We therefore have to include a length fine tuning system for each pass individually.

### 4.2.5.4     Mirror quality surface

From the above statements about the parabolic mirror alignment precision, it comes out that an important parameter is the mirror substrate and coating quality. Whereas surface roughness r.m.s. below 0.1nm is at present currently achieved by mirror substrate manufacturers, the homogeneity of the surface shape over large mirror areas is still an issue. In our case, we have to use parabolic mirrors of ~38cm diameter and a departure from a perfect parabola below the micrometer level all over the mirror surface. Such a high polishing performance has in fact been achieved, e.g., by numerical polishing techniques as reported in (46). In this reference, a maximum departure from a perfect parabola of about ±0.04μm has been obtained over a parabolic mirror diameter of 20cm and for ROC=2.5m. This is the surface quality that is needed for our re-circulator.

Another TASD loss may come from a mismatch between the focal lengths (*i.e.* the ROCs) of the two parabolic reflectors. In our case and within the paraxial approximation, the following expression for the waist of the n$^{th}$ pass is $\omega_{0n} \approx \left(\frac{ROC1}{ROC2}\right)^n \omega_{01}$. Defining $\delta_{ROC}$=1-ROC1/ROC2, one gets a decrease/increase of the luminosity by 5%-6% for a standard value |$\delta_{ROC}$|=0.5%. One would then need |$\delta_{ROC}$|<0.1% to limit this effect at the percent level.

#### 4.2.5.5    Mirror pair cylinder mount misalignment

No noticeable variations of the optical performances were observed when moving the orientation and the transverse position of the cylinder by 0.1° and 0.3mm respectively.

#### 4.2.5.6    Timing tolerance

Fig. 138 shows the TASD variation as a function of timing mismatch between electron bunches and the laser pulses. From this figure, one sees that a maximum delay of 400fs, and preferably even less, is needed to keep the TASD loss at the percent level.

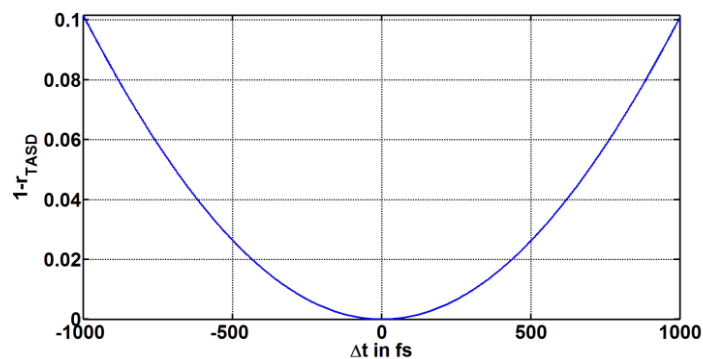

**Fig. 138.    TASD loss induced by a delay between the laser beam and the electron beam. The beam parameters of the high energy IP has been considered.**

In reality, when facing a misaligned re-circulator, the synchronization optimization leads to an increase of the TASD

#### 4.2.5.7    Mirror pair parallelism misalignment

We have simulated the mirror-pair alignment procedure sketched in Fig. 124. For a given measurement precision of the autocollimator device, we obtain, using Matlab Software, a series of random misaligned mirror-pairs that we include in the re-circulator Code V simulation. The correspondence between the autocorrelation measurement precision and the mirror-pair parallelism (defined by the angular difference between the normal directions of the two plane mirrors) is shown in Fig. 139.

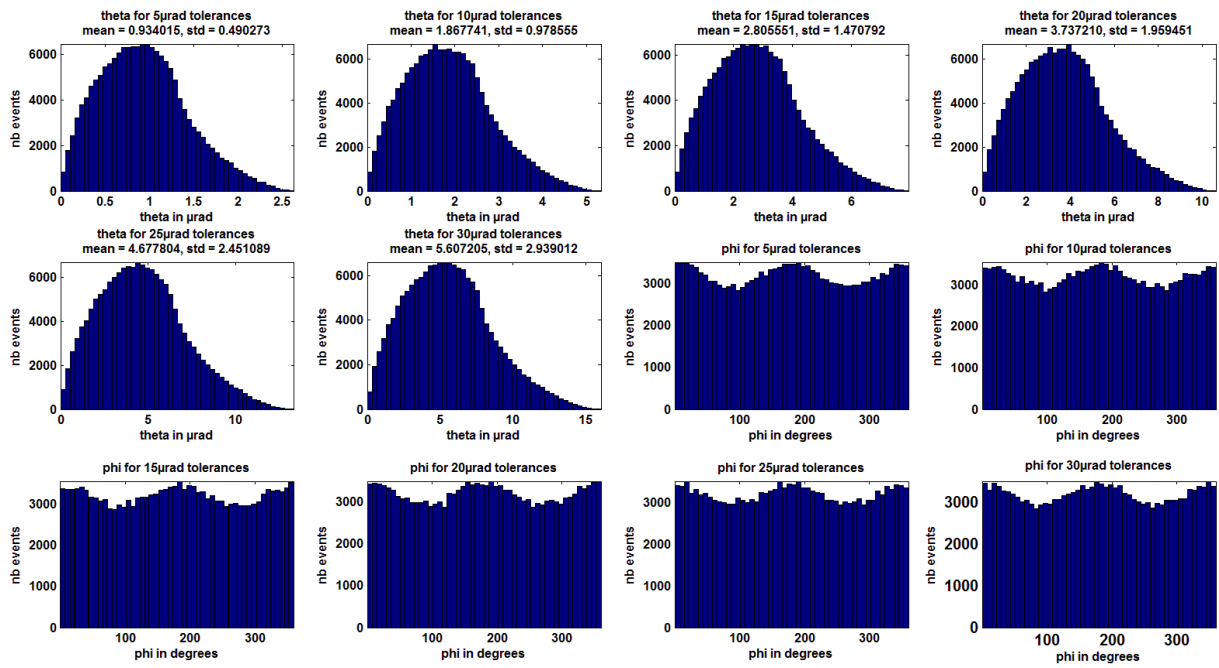

**Fig. 139.** Results of the random simulations of the mirror-pair parallelism misalignments. Six autocorrelation measurement precision have been considered: 5, 10, 15, 20, 25, 30 µrad. Both polar (theta) and azimuth (phi) angles describing the mirror pair parallelism are shown.

The results of the Matlab simulations for various autocollimator precisions are summarized in Fig. 140. In these simulations, the alignments of the injection and of the parabolic mirrors are assumed to be perfect.

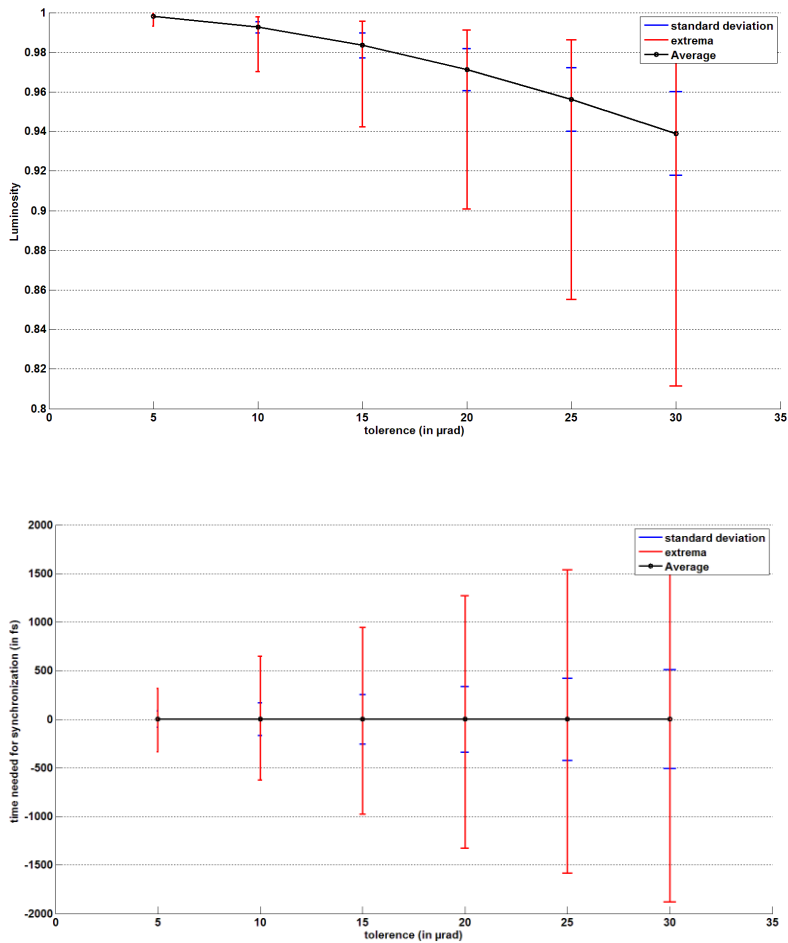

**Fig. 140. Results the Matlab simulations for six autocollimator precisions (5,10,15,20,25 µm). 5000 simulations have been performed for each autocollimator precision values. The main statistical estimators, the averages, the root mean square and the extrema, are shown as a function of the autocollimator precision for the TASD distributions (top) and the synchronization correction (bottom).**

When mirror pairs are not parallel to each other, the optical paths differ from the nominal design resulting in a synchronization mismatch between the electron bunches and the re-circulating laser pulse. One must then turn the mirror pairs in order to compensate for this mismatch. The bottom plot of Fig. 140 shows the synchronization correction delay that is needed to optimize the TASD (shown in the top plot). For 30µrad precision, the delay is ~2ps at most. This corresponds to a rotation of a mirror that is tolerable with the geometrical parameters that we have considered (see Fig. 131). The TASD losses for the 20 µrad precision are also tolerable with regard to the average and root mean square values of Fig. 140. It is to mention that when running the fine alignment optimization described in the next section, the maximum losses observed in this figure reduces by ~10%. We then choose to fix the tolerance on the mirror pair parallelism misalignment to 20µrad.

### 4.2.5.8 Re-circulator alignment procedure

The re-circulator is first aligned in two steps. First, the optical elements are fixed thanks to a high precision mechanical referencing of surface shapes. Then an optical pre-alignment is performed in order to reach a precision of ~20µm, 20µrad on the parabolic mirror and injection positions. Once the pre-alignment is done, we start an optimization procedure using the *simplex* method of Matlab. The optical observable that is minimized is based on the measurement of the intersection points of the laser beam with a plane located at the focus of the reference parabola. This is done by imaging the optical path on a thin plate inserted at the focus plane.

We simulated this procedure using Code V for the optical propagation inside the misaligned re-circulator and Matlab to define the misaligned geometries and to compute the optical estimator at each step of the *simplex* algorithm. A typical simulation result is shown in Fig. 141 where we randomly misaligned the parabolic mirror and the laser beam injection by ±20µm and ±20µrad. In this figure, the mirror-pair were also randomly misaligned assuming a residual precision of 20µrad in the collimator. From Fig. 141, it is clear that whereas a pre-alignment within ±20µm/µrad of the parabola and laser beam injection can lead to 70% of TASD loss, the use of the most simple and robust minimization algorithm, *simplex*, can next reduce this loss to a level below 5%.

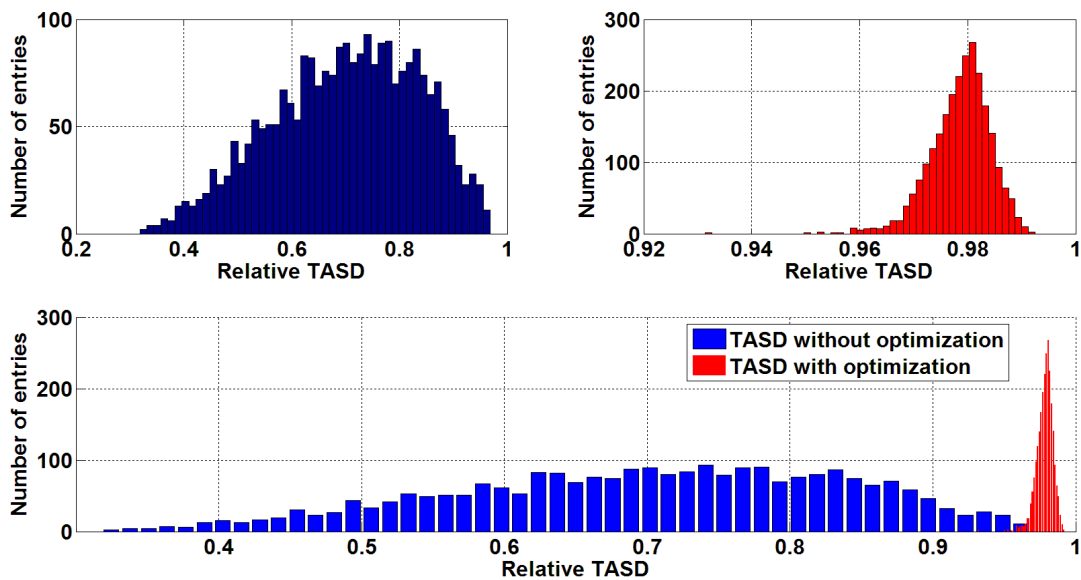

**Fig. 141.** TASD distributions after simulations (see text). Top left: before *simplex* optimization; Top right: after *simplex* optimization. The bottom plot shows the same distributions superimposed.

#### 4.2.5.9  Summary

Our optical system allows circulating 32 times a laser beam within a time window less than 600ns and without any noticeable aberration. However, to reach this level of very high optical quality, tight tolerances of the order of a few µm and µrad are required on the alignment of the optical elements. From our simulation studies, it ends up that a simple minimization algorithm based on a simple and robust optical observable successfully brings a re-circulator pre-aligned with an accuracy ~20µm; 20µrad down to the µm, µrad level.

### 4.2.6.  Laser beam losses

The effective beam losses due to the $M^2$ laser beam factor can be estimated as [118]

$$\delta \approx \frac{1 - e^{-2/M^2}}{1 - e^{-2}}$$

from this formula we see that the effective power lost at the IPs is only a few percents for our target value $M^2$=1.2. Using this $M^2$ value is justified by recent experimental results [74].

There are 4 mirror reflections per pass, which yield 124 reflections for 32 passes. The mirror coatings have a reflectance R and transmittance T such R+T=1 if T is much larger than the coating scattering and absorption losses (i.e. T>>10ppm for high quality coatings). For a given relative overall loss δ and a given number of passes $N_{pass}$, assuming that the parabolic and planar mirrors have the same coatings, one gets the following expression for the mirror transmission loss:

$$\delta = 1 - \frac{(1-T)}{N_{pass}} \sum_{n=1}^{N_{pass}} ([1-T]^m)^{n-1} = 1 - \frac{(1-T)}{N_{pass}} \frac{(1-(1-T)^{mN_{pass}})}{1-(1-T)^m}$$

That is, for m=4 and neglecting the absorption and scattering losses (that can be below 5ppm), we obtain: T=400ppm for 2.5% loss and T=1000ppm for 6%. These are reasonable/standard coating transmission values.

Diffraction on the edges of the re-circulating flat mirror pairs induces power losses. One can first estimate this loss, ρ, by computing the laser beam intensity contained on the first mirror surface which is supposed to act as an intensity spatial filter. Considering a circular mirror of radius nω, with ω the laser beam radius on the mirror, one gets ρ=exp(-2$n^2$), that is 300ppm for *n*=2 and 1000ppm for *n*=1.8. However, diffraction can also modify the laser beam intensity distribution at the IPs. We have model the laser beam transport using the FFT propagation technique [119]. We observe no noticeable modification of the intensity distribution at the IP after more than 30 passes for *n*=1.8.

In summary, with a good laser beam quality ($M^2$<1.2) and re-circulating mirrors of radii 1.8ω, the power loss after 32 passes is dominated by the mirror losses due to coating transmission. It can be easily kept below 5%.

### 4.2.7. Implementation in ELI-NP

The re-circulator will be mounted inside vacuum on a low thermal expansion material (e.g. invar, see Fig. 142). It could be located on a massive, thermal controlled, synthetic granite table. The whole system could thus be thermally controlled within 0.1°C. A schematic view of the re-circulator implementation is given in Fig. 143.

The re-circulator system for each IP is composed by: an optical re-circulator enclosed in the UHV chamber, a HV injection-chamber, the synchronization device (see previous section), the fine alignment imaging system and some laser beam diagnostics.

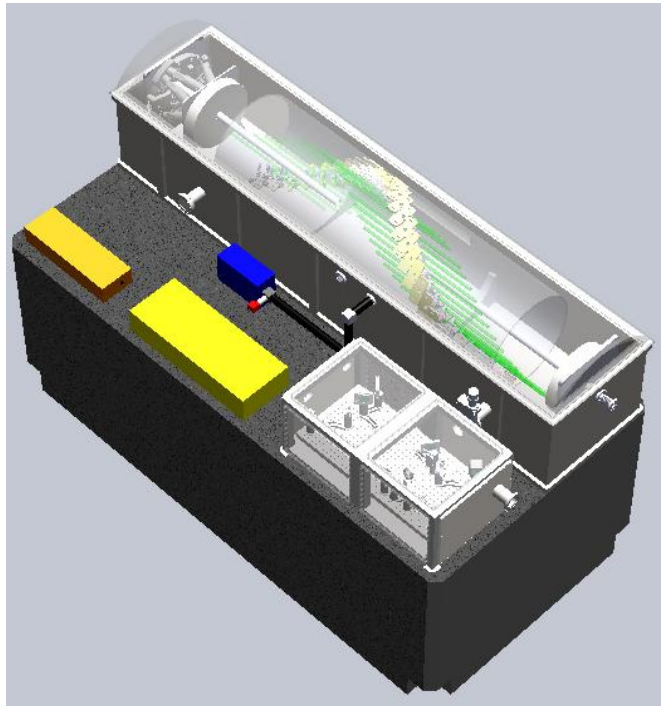

**Fig. 142.** Schematic view of the re-circulator mounted on a thermal controlled table

The injection chamber, connected to the re-circulator chamber, hosts the active beam stabilization which provides a beam pointing stabilization of ~3µrad and the injection steering mirror used to align the incident laser beam into the re-circulator. Motorized flip-mirrors are also implemented inside the chamber to inject the laser beams used to synchronize the re-circulator roundtrips on the RF clock.

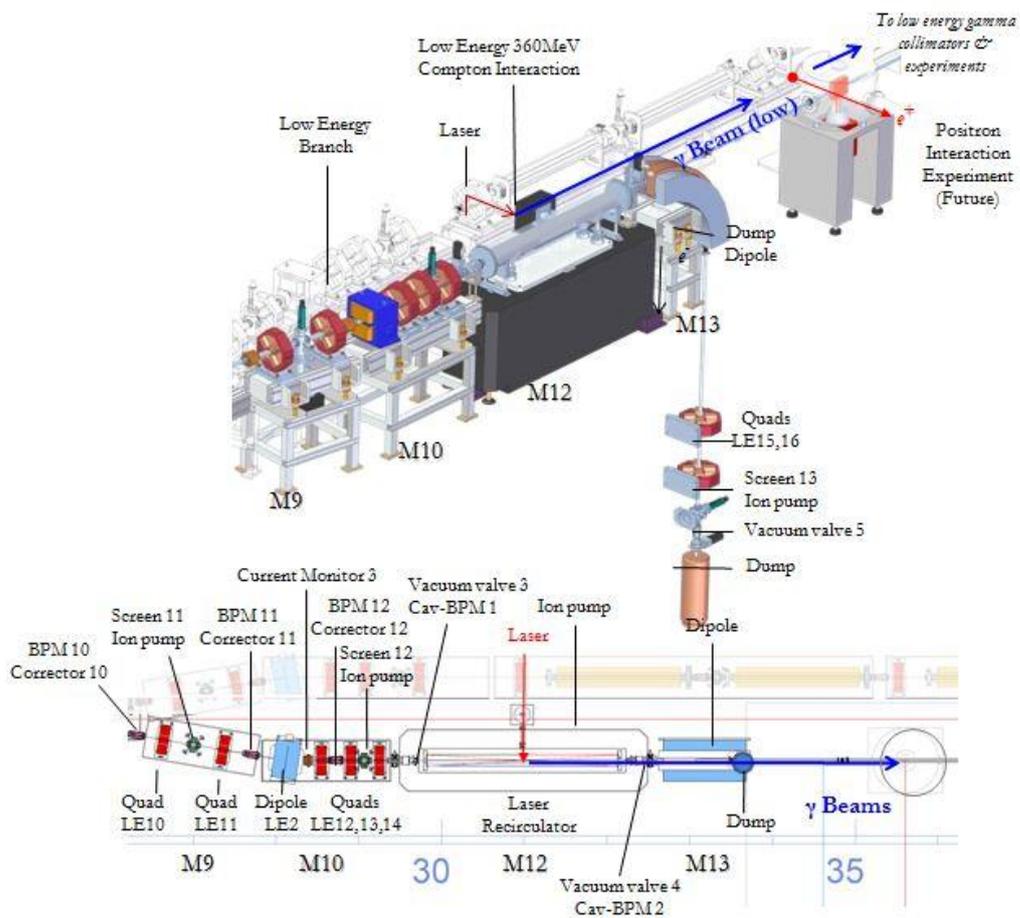

**Fig. 143.** Schematic view of the implementation of the re-circulator at the low-energy IP

## 4.3. Photocathode laser

The photocathode laser system task is to provide light pulses able to generate the requested electron-beam by photo electric effect on a metal surface (i.e. OFHC Copper). So it is possible to derive the related specs back from the e-beam ones from Table 37 hereafter.

**Table 37.** Main photocathode laser relevant parameters

| | |
|---|---:|
| Bunch charge (*pC*) | 25-400 |
| RF peak field at the cathode (*MV/m*) | 120 |
| RF Pulse duration for beam (*nsec*) | ≤ 600 |
| Accel. grad. in S-band Sections (*MV/m*) | 21 |
| Laser Pulse energy (*μJ*) at 266 nm | ≤ 150 |
| Bandwidth @ 266 nm (nm) | > 0.7 |
| Laser Pulse length (flat-top, *psec*) | 5-12 |
| Laser Pulse rise-time (*psec*) | < 1 |
| Laser focal spot size (*μm*) uniform | 100-400 |
| # pulses in the train | ≤ 32 |
| Laser pulses separation (*nsec*) | 16 |
| Laser Pulse energy jitter (%) | 2 |
| Time arrival jitter (*psec*) | < 0.5 |
| Pointing jitter (*μm*) | < 20 |

Main assumption in the photocathode scheme is that on a first order, the electron beam is a copy one to one of the laser beam shape (neglecting space charge), once taken in account the cathode quantum efficiency QE (i.e. ratio of #electrons and #photons). In the case of Copper, for a linear photoemission (one photon excites and eventually extracts one electron), one needs to overcome the material work function with each single photon. Given the average work function of Copper of 4.5-4.67 eV, this corresponds to illuminating the cathode with photons at wavelength lower than 267 nm. In the framework of large bandwith lasers like Ti:Sa it would be worth to have capability of a slight adjustment in wavelength tunability of plus-minus 5 nm around central wavelength. This means possibility to change the tuning of both amplifiers and oscillator, in a quite straightforward way. It is not necessary to have an online automatic control of such a parameter, since this would be done infrequently, in order to follow the cathode work function with photon energy.

### 4.3.1. Single pulse requests UV

Within the assumption of $10^{-5}$ QE for Copper, it is easy to retrieve the requested pulse energy in the UV for the highest charge requested (500 pC), that is 250 μJ.

Final laser transverse size and pulse length must reflect the requests of the electron-beam, given in Table 37. Bandwith specs on UV laser beam are given by the steepness requested to the bunch longitudinal distribution (<1 ps). We estimate 0.7 nm FWHM able to guarantee a very fast response, but moreover, to drive the harmonic process with a good efficiency, necessary to lower the requested energy coming out of the amplifier.

Longitudinal pulse shaping is a process to take in account, because the final pulse has to be in the ps range (quasi-flat longitudinal profile). We'll take in account a beam longitudinal manipulation mainly conducted directly in the UV, otherwise the harmonic generation would be affected by such a big difference in power in a detrimental way. For sake of efficiency, we'll take in account a shaping process made through the use of birefringent crystals, a technique which is now widely used for photoinjectors and well established (photo).

Principle is quite simple: a single short (polarized) pulse entering a birefringent crystal will be split in two pulses, each travelling with the different group velocity (here we will deal with uniaxial crystals [120]) related to the specific direction of the electric field. At the end of the crystal of length L, the two pulses will be split in time by the following amount: T=L/ve-L/vo, where the subscripts e/o stand for extraordinary and ordinary axes. In general after a stack of n crystals, we'll get a sequence of 2n pulse replica exactly split in two halves of (2n-1) pulses whose vector is lying on two orthogonal polarizations.

It is important to notice that in our scheme we are going to illuminate the cathode on a quasi-normal incidence. With such an assumption there is no difference between polarization vectors for photoemission (i.e. absorption of radiation by the cathode).

If properly designed, such a crystal stack may generate a train of pulses whose superimposed intensity resembles that of a square like pulse. More in detail we mention that if the sequence of pulses is split such that each successive replica has a change in polarization, then there won't be any interference between the pulses, and the overall intensity of the train will be the sum of the single replica intensity. Another option is to play with the interference (for example selecting a common plane for the train by projecting all replicas, whatever the polarization is). In such case, the interference between pulses will play a role, but this can be also an effect used mitigate ripples of intensity on the plateau of the pulse, when introducing some spectral phase orders (e.g. through an acousto-optic modulator). With such technique ripples below 2% RMS of maximum are possible.

Coming to the efficiency of the process. Losses are given by the reflectivity at each interface (absorption in the crystals are much lower and hence we'll neglect them). Given crystals with 2% losses (much better can be made through anti-reflection coatings) gives 88.5% efficiency, after a stack of 6 crystals. This brings for each single UV pulse, a required energy of 283 µJ at the entrance of the shaper. Obviously, UV transparent crystals such as a-cut BBO (beta barium borate) have to be employed.

### 4.3.2. Longitudinal pulse shaping

Amongst many strategies possible in achieving a laser pulse in the final UV wavelength, which is close to the specs: quasi flat-top with rise/fall time minor than 1 ps, we'll focus in this document about the use of birefringent crystals, for their ease of use, intrinsic efficiency, and successful use in other facilities.

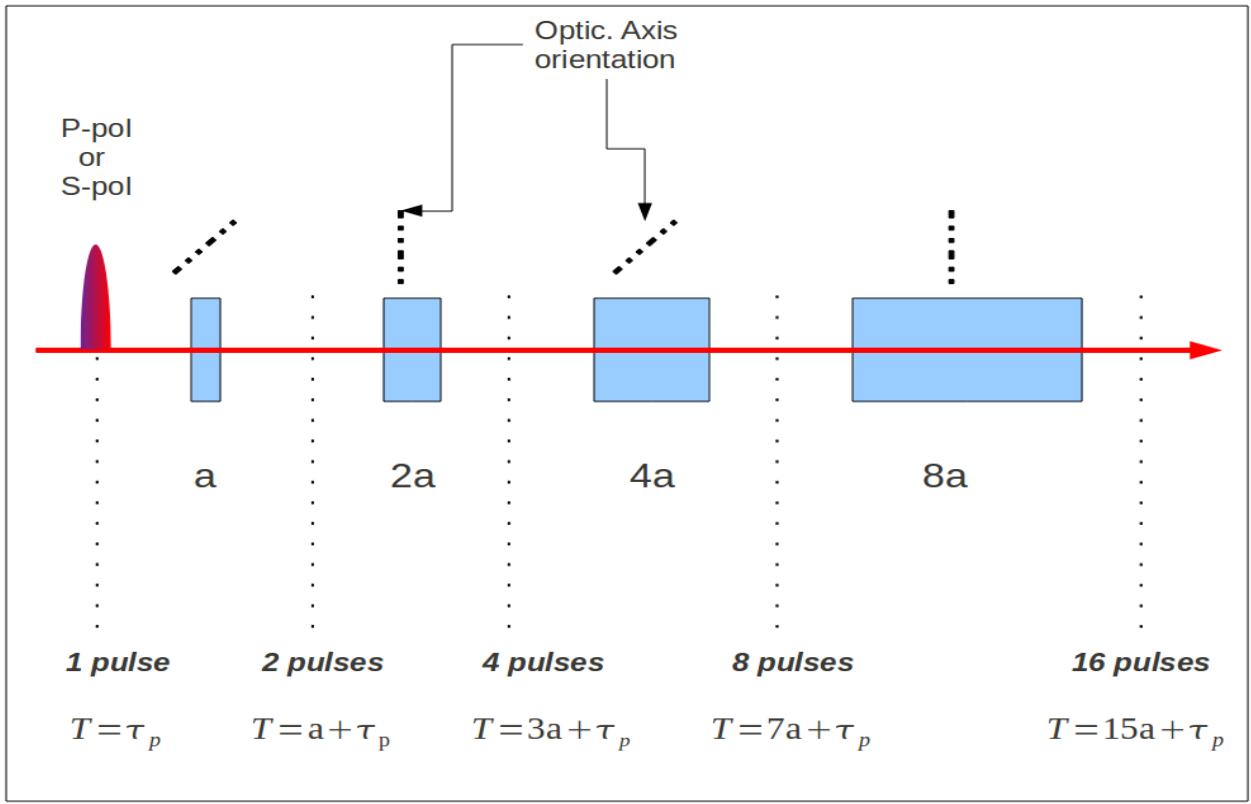

**Fig. 144.** Schematic lay-out of the sequence of birefringent crystals used to generate laser pulse replica, to be overlapped for the generation of flat-top laser pulses driving the photocathode

Let us take into account a scheme able to shape a single pulse so to appear as a longer quasi flat pulse. On a first approximation we'll forget about pulse enlargement due to dispersion and also differential enlargement of ordinary and extraordinary wave due to birefringence inside the crystals and walkoff angle induced by the same birefringence, eventually reporting afterwords about alignment tolerances and second minor corrections given by such effects.

Given a birefringent crystal, such as an uniaxial a-cut BBO (able to work with both first and third harmonic of a Ti:Sa laser line, if not coated for anti-reflection capabilities) we'll define the basic quantity **a**, as the difference in time, standing between the ordinary and extraordinary wave in propagating through the same crystal. In a very simple way we are able rto build pulse replica directly from a stack of crystals, whose lengths (we'll directly talk of pulse separation in time **a** which is the quantity we are interested in and which is proportional to the crystal length L) double that for each successive crystal (as reported in figure). In such a way we are able in a quite easy way to derive the overall length which contains all the pulses in the train, obviously taking care to fill up all the empty spaces between each pulse at the end of the stack, so to generate a real quasi flat top beam. In this framework we'll now take in account just the generation of all replica of the same heights, though the system allows for more flexibility and complex shapes. In this scheme, it is sufficient to enter the stack in p or s polarization. The first crystal optical axis has to be oriented at 45 degrees respect to the horizontal plane (i.e. respect to s or p polarization), and each successive crystal axis must be once again at 45 degrees respect to the previous, following the example outlined in figure. After the nth crystal we'll get a total number of pulses of $2^n$, for a total train length of $T = \mathbf{a}[2^n - 1] + t_{pulse}$. As a reference for calculations, given the difference in group velocities and possible crystal lengths compatible with a commercial manufacturing process, we can be able to span (having different lengths of crystals) **a** over a range that goes from hundreds of femtoseconds up to tens of picoseconds. It is useful to understand what is the longitudinal intensity distribution arising from such a scheme. In principle we should take in account interference of all single pulses in the electric field and hence deriving the intensity of such a superposition, eventually accounting for vector orientation able to give interference. Within the layout proposed, after the n-th stage, the first half of the pulses will share the same polarization orientation, and the second part will be normal to the first one. A polarizing optic placed at 45 degrees between the two polarization would restore the same interference on all the pulses within the train (otherwise only the first part will interfere within its pulses, and the second, in the same way). The interference is now sensitive to the relative phase shift between successive pulses, piled up thanks to different phase velocities of the ordinary and extraordinary wave. Such an interference will especially play a role in affecting the ripples on the top of the pulse. The strong advantage of such a shaping method is given by the fact that the rise and fall time of the overall pulse won't change, adding more crystals, being basically that of the single pulse used at the entrance of the device, plus minor correction due to pulse enlargement in crystal and air propagation (GVD etc.) for long enough pulses (hundreds femtoseconds). Moreover, further knobs can be used, giving further high order phase terms to the pulse, in order to minimize the ripples of the interference, for example by means of acusto optic modulators widely used in large bandwith Ti:sa systems nowadays. An even stronger reason is then given by the efficiency of the process. Losses are basically given just by reflection at the interfaces, which could be possible minimize to few percent (1%) through AR coatings (50% efficiency over 5 crystals).

Here is an example for a crystal stack of 4 elements, whose basic spacing **a** is 320 fs, hence 1.6 sigma in case of a gaussian pulse of 200 fs standard deviation (as here accounted). Here is an example of a perfectly matched interference between trasform limited pulses, and that of slight mismatches between successive pulses given by the relative delay/phase term for a 5 ps overall beam.

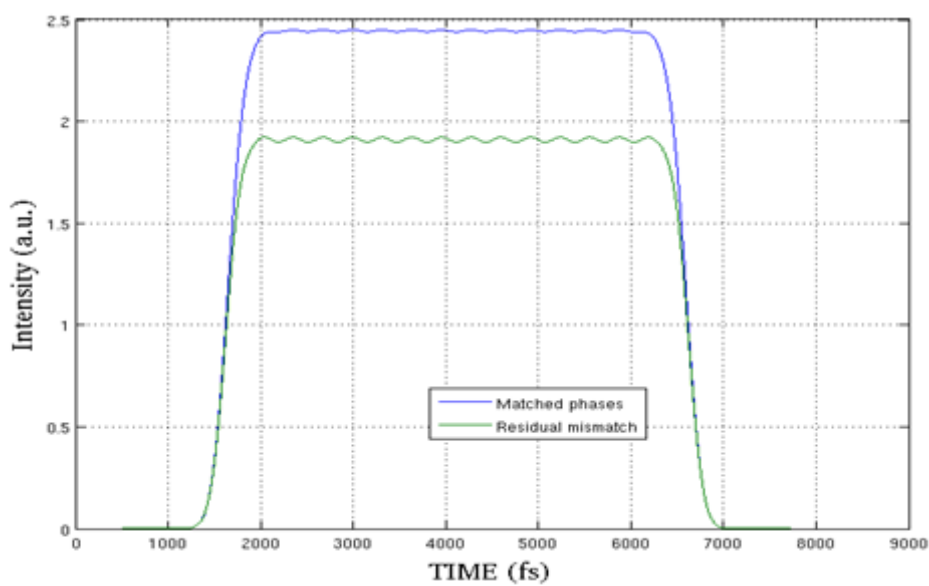

**Fig. 145.** Time profile of the quasi flat-top laser pulse longitudinal distribution, obtained with ideally matched crystals (blue line) and with slightly mismatched crystals (red line)

Results in terms of residual ripples and rise/fall time are satisfying, as shown in the figure below.

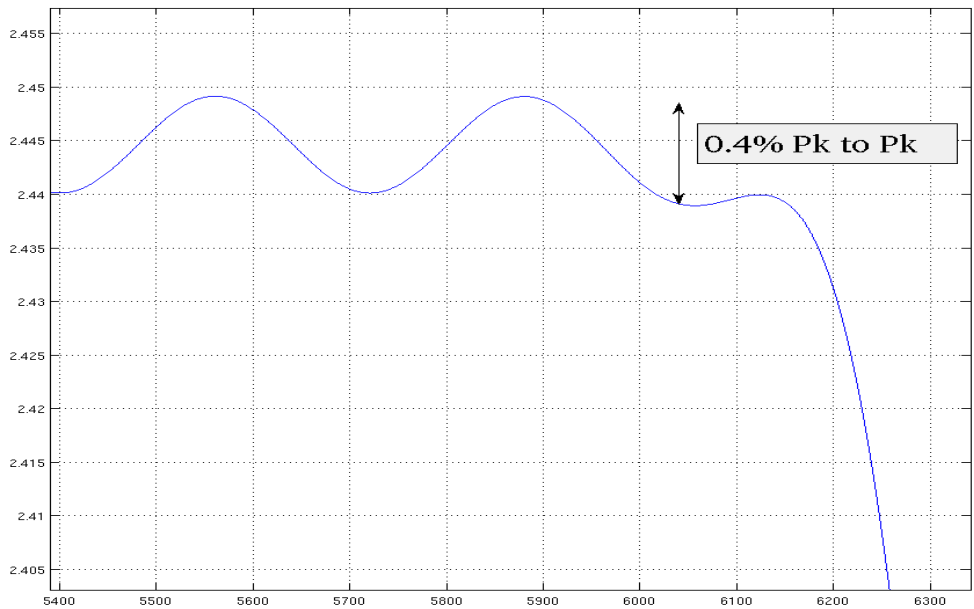

**Fig. 146.  Zoom-in of the ripples in the time profile distribution of the quasi flat-top laser pulse driving the photocathode**

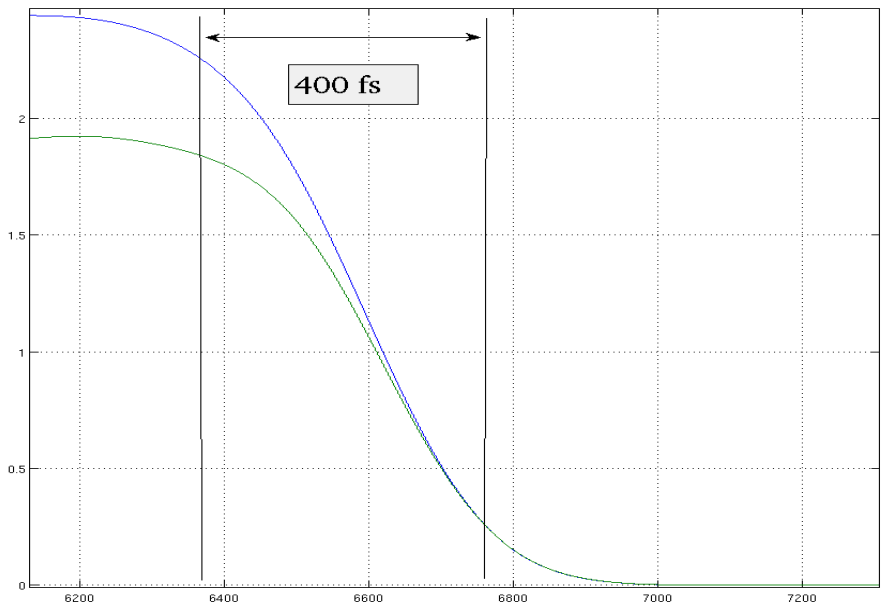

**Fig. 147.  Zoom-in of the rise-time region in the time profile distribution of the quasi flat-top laser pulse driving the photocathode**

Here we come to the bad side of this scheme, which could be the flexibility. In order to make a 12ps long pulse, here we show what comes out by using a 300fs standard deviation beam and a 5 crystals stack (32 pulses) with basic spacing **a** of 1.3 s, that means a different cristal stack for different pulse lengths. Slight adjustment might be done by changing the input pulse length, anyhow this is not completely a free parameter, because it is going to affect the efficiency in the harmonic process upstream and also because it is basically related to residual chirp in the beam coming from not perfect recompression of the stretched beam in the CPA system.

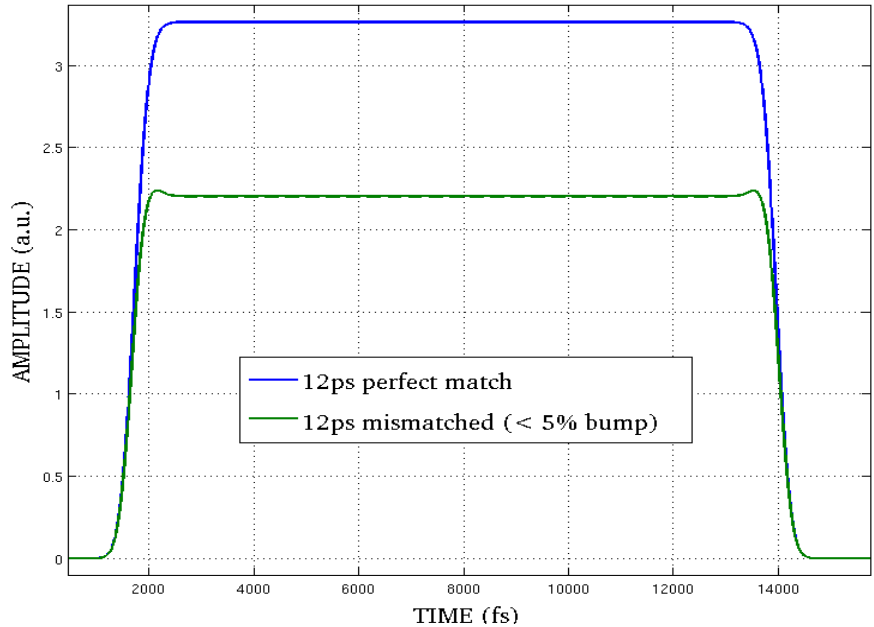

**Fig. 148.** Quasi flat-top time distribution of the laser pulse driving the photocathode with different stacking options

The effect of residual second order phase (or further orders) in the input beam have to be accounted, but do play a role in the interference of the beams, potentially causing deep modulation within the longitudinal distribution.

We also supply an example of shaping for nominal beam, 8.5 ps long, obtained starting with 200 fs standard deviation beam and **a** of 1.3 s and a 5 crystal stack in the same case of quite perfectly matched beam (0.8% peak to peak bumps) and slightly mismatched case (2%).

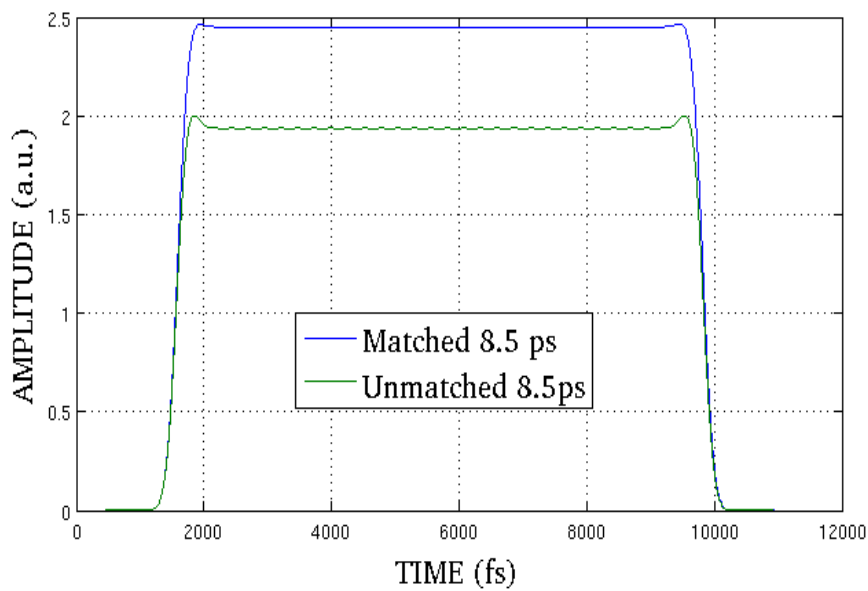

**Fig. 149.** Quasi flat-top time distribution of the laser pulse driving the photocathode with different stacking options

Using the same crystal stack but reversing the order of the elements, the overall resulting delay is the same, but we could be able to get a useful property out of this. In fact, in such a scheme, the polarization of successive pulses is at 90 degrees, hence, they are not able to interfere (in terms of resulting intensity, of course, wavevector is changed, but this is not affecting photoemission at normal incidence). Interference can be reestablished once again by selecting polarization in the middle of the two directions, going back to the previous cases. Anyhow simple envelope superposition is easier, but with less degrees of freedom, respect to the other case.

### 4.3.3. Single pulse requests IR

Once given the specs on the final energy in the UV beam, it is possible to eliminate the laser power at the fundamental wavelength. Calculations made on a few milli Joule energy level (3 mJ), for pulses of few hundreds of fs at 800 nm, employing a harmonic up-conversion to third harmonic, made up of a cascaded second harmonic generation ($2\omega=\omega+\omega$) and frequency sum mixing ($3\omega=\omega+2\omega$) with a double type I phase matching in BBO nonlinear crystals, both of 0.2 mm thickness staying within 25% efficiency, with a beam diameter of 5mm at the entrance of the harmonic generator.

### 4.3.4. Transfer Line to cathode

The fundamental point in the design of a high energy, ultra-short amplifier is the reliability of the performances. The transfer line will be used to transfer in the most efficient way the laser to the cathode. As the transfer line environment cannot be controlled as precisely as the laser room, it should not be used for beam manipulation. To limit the energy transported, the laser pulses will already have been converted to the UV avoiding the requirement for an in-vacuum beam line. Starting point is a gaussian mode with a RMS radius of 1.25 mm. We do consider a fixed magnification system which fills the distance (13 mt) up to the cathode and images an aperture (which will be probably a remotely controlled variable iris diaphragm) onto the cathode. Upstream we consider a variable system in which magnification might span to fill up requirements of the final spot size (100-400 µm RMS radius on cathode).

First element of the transfer line will be the iris. In such a way that the final cut distribution might be imaged with a scaling factor, no matter of which percent of the initial distribution is cut.

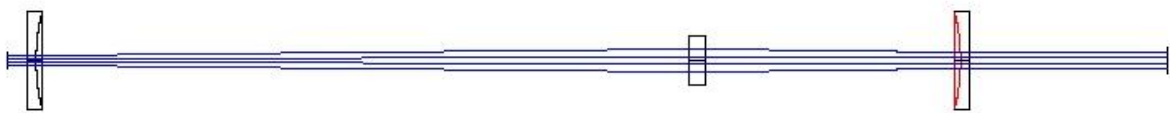

**Fig. 150.** First part of the photocathode laser transport line

Magnifications of the variable system will be between 0.4 and 1.6, upstream a subsystem of fixed magnification 0.2. The variable system will be realized through an afocal lens layout made up of three lenses of equal focal length in absolute value, and a commercially available two identical fused silica plano concave lenses and a biconvex lens of the same EFL (206 mm).

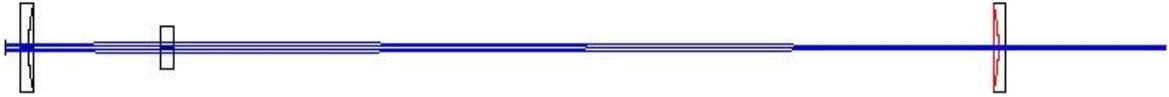

**Fig. 151.    Second part of the photocathode laser transport line**

The layout of the variable subsystem is the one shown in figure for the two extremal configurations (namely, 1.6 and 0.4):

The second and third lenses have to be moved in a defined pattern in order to recollimate the beam at the different magnifications foreseen. The overall span in the movement is 246 mm for the first lens and 140 mm for the second. 2 mt is the overall space needed by this system and it is desired to have it inside the laser room. In order to be conservative it might be thought to further magnify the entrance spot size and choose another magnification for the fixed system in order to decrease the fluence on the optics. The fixed magnification is given by a long focal length (2 mt biconvex) 1:1 imaging of the iris object, followed by a shorter 5:1 magnification of the first relay made by a 550 mm plano convex lens. With less than a mrad divergence, the beam will be always well inside the aperture of the lenses taken in account, of 2" diameter.

The low pointing jitter considering a plus minus 50 μrad reflects onto plus minus 1.3 μm of displacement on cathode. A detailed analysis shows that the relay system, as it is, has some off-axis aberrations (namely, coma and astigmatism), especially given by the 5:1 system that is not symmetric. For such a reason, the slight displacement is not really due to the chief-ray offset, but to the fact that the beam shape changes (not possible to notice that for such low field angles). Anyhow, it is possible to improve such performance by optimizing the shape of the 5:1 subsystem in a lens design which compensates for such aberrations (i.e. splitting the lens in more sub-elements and optimizing the curvature of the surfaces to reduce aberrations).

Effective distances must be then refined. In particular an imaging system like this goes through focus two times. The first focus is under control since the focal length of 2 mt is quite long to generate nonlinear effects in air. The second focus is right before the cathode, and hopefully in vacuum, but this has to be carefully checked through the final layout.

Efficiency of the shaping: there's about 96% of efficiency along the beamline, not considering the cut operated by the movable iris which is simply given by the integral of a normalized gaussian distribution up to the radius of the cut.

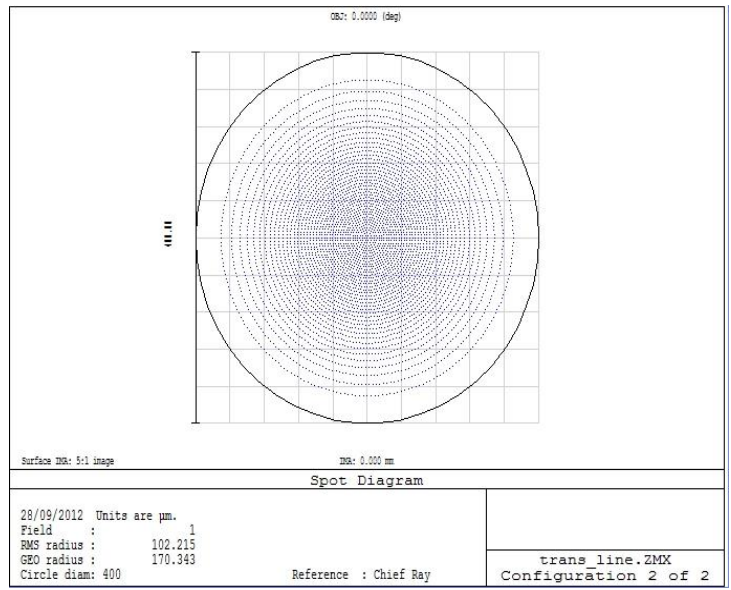

Fig. 152. First beam configuration on cathode (about 100 µm RMS)

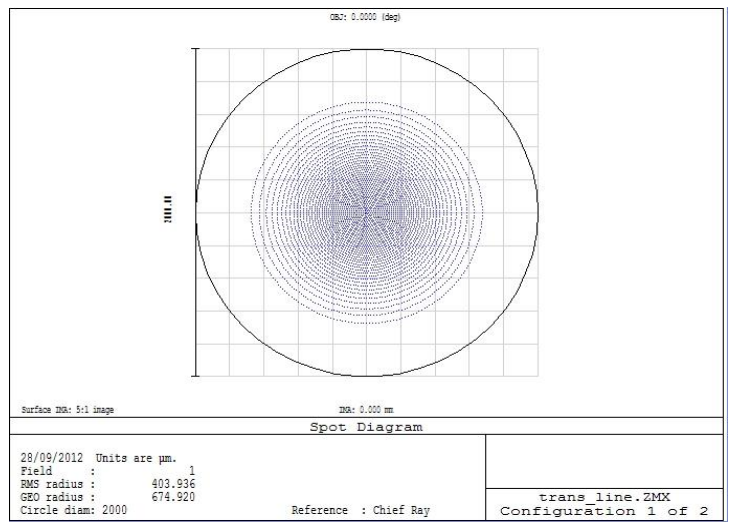

Fig. 153. Second beam configuration on cathode (about 400 µm RMS)

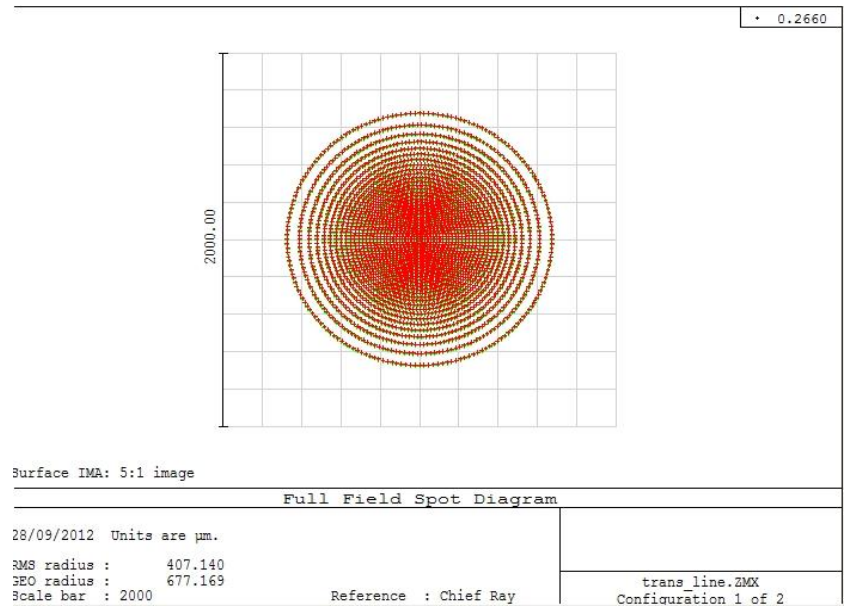

**Fig. 154.** Full spot diagram on cathode (ray tracing), considering full field distribution with possible pointing instability

### 4.3.5. Introduction to ultra-short high repetition rate laser systems

The fundamental point in the design of a high energy, ultra-short amplifier is the reliability of the performances. The system required for the photocathode is a Ti:Sa amplifier that should deliver 300mJ in the infrared at 100Hz. The system must be tunable between 790 to 810 nm to optimize the efficiency of the photocathode. After splitting the main pulse in 32 replicas the 800nm is converted to the third harmonic in order to reach an energy level of 150µJ for each individual pulse. The temporal and spatial shapes have to be close to square pattern.

To achieve these performances it is necessary to have a complete control over the temporal and spectral phase of the laser pulse.

Like for all complex machines it also important to have an easy and reliable control of the parameters and performances of the laser via an integrated computer interface. Thanks to the experience of Amplitude Technologies in the domain of 100TW class laser, supervision software with diagnostics can be integrated.

The next paragraphs will describe Amplitude Technologies strategy for the realization of a picosecond, high energy (>300 mJ), high repetition rate (100Hz) laser system.

#### 4.3.5.1 The Dazzler-Mazzler configuration: ultra-short temporal duration and wavelength tuning thanks to the active gain control

The Mazzler is an acousto-optic crystal used in the cavity of the regenerative amplifier. It is an "Acousto-Optic Filter Programmable Gain Control". The Mazzler helps to shape the amplitude spectrum in the cavity of the regenerative amplifier and operates as a spectral filter. In this way it is possible to fight the gain narrowing of the amplifier spectrum, and even to shape the spectrum for amplification in successive stages.

At the exit of the regenerative amplifier, the width of the spectrum can be more than 160nm and pulse down to 10fs can be produced (assuming a broadband oscillator).

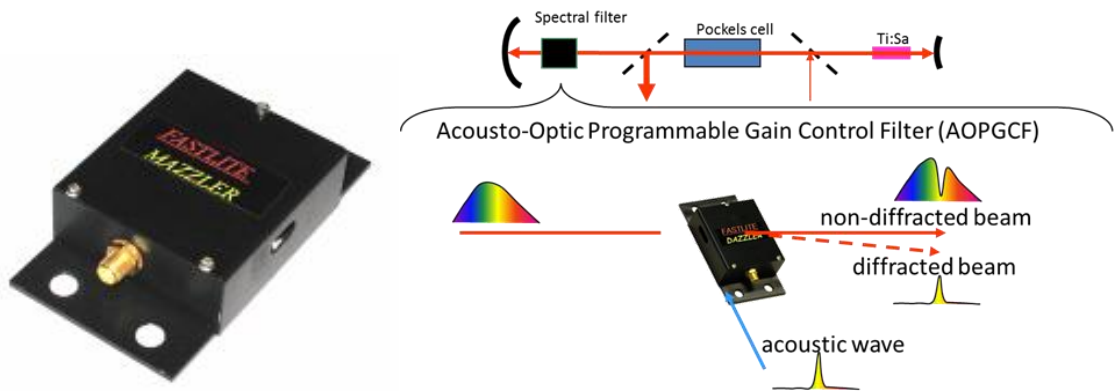

Fig. 155.    Configuration of the Mazzler in the cavity of the regenerative amplifier

#### 4.3.5.2    Wavelength tuning

The Mazzler plus Dazzler configuration allows controlling completely the spectrum over more than 110nm. The performances accessible thanks to this control are:

- Spectrum tunability: a square 15-30nm spectrum can be tuned over a 55nm spectral range (see picture below)
- Two-colors: 2 different parts of the spectrum can be amplified in the same time, allowing a "two-color" pulse.

The interesting point is that this tunability doesn't require any realignment of the system.

#### 4.3.5.3    High energy high repetition rate amplifiers design

The fundamental point in the design of high repetition rate amplifier is the management of the thermal lensing effect and the overlap between the pump beam and the amplified IR beam. At 100 Hz the thermal lens induced by the pumping is in in order of a meter, and then the propagation becomes an issue. To minimize this effect the Ti:Sa crystal is cooled down to low temperature (around 150K) in order to improve the conductivity of the material and then reduce the thermal lensing.

- Infrared beam propagation

The thermal stress induces lensing on the small diameter infrared beams propagating in the amplifier. It is fundamental to manage the beam propagation with passive optical systems in order to keep to right beam diameter in the crystal.

- Cryo-cooling

The cryo-cooling is essential to minimize the effect of the thermal load in the Ti:Sa crystal of the last amplifier. The temperature is kept around 150K in order to have the thermal lens as low as possible.

A heater is used on the crystal mount to maintain a constant temperature in order to avoid any a change of the thermals lens when the parameters of the amplifier are modified (pump level, rep rate…)

### 4.3.6. Detailed description of the system

#### 4.3.6.1 Oscillator

The oscillator used for the laser is a VITARA from Coherent Inc. The Base plate is thermally regulated for improved long term stability. To ensure the synchronization with the RF gun, a synchro lock AP has to be used.

**Table 38. Specifications of the oscillator**

| OSCILLATOR | |
|---|---:|
| Output energy @ 75 MHz | > 2 nJ |
| Beam diameter (1/e²) | < 1 mm |
| Beam divergence | < 4 mrad |
| Spatial mode TEM00 | ($M^2$ < 1.3) |
| Polarization | > 100:1 (horizontal) |
| Noise | < 0.1 % rms |
| Power stability | +/- 1% |

#### 4.3.6.2 Stretcher

The stretcher is an Offner configuration with rugged mechanical configuration and complete sealed cover to avoid unwanted perturbations and keep a very good beam pointing stability. Once the stretcher is optimized no further realignments are required (no setting on the mounts).

**Table 39.    Specifications of the stretcher**

| STRETCHER | |
|---|---:|
| Stretched pulse duration | > 250ps |
| Spectral aperture | > 130nm |
| Efficiency | > 40 % |
| Energy after stretcher | > 1.5 nJ |

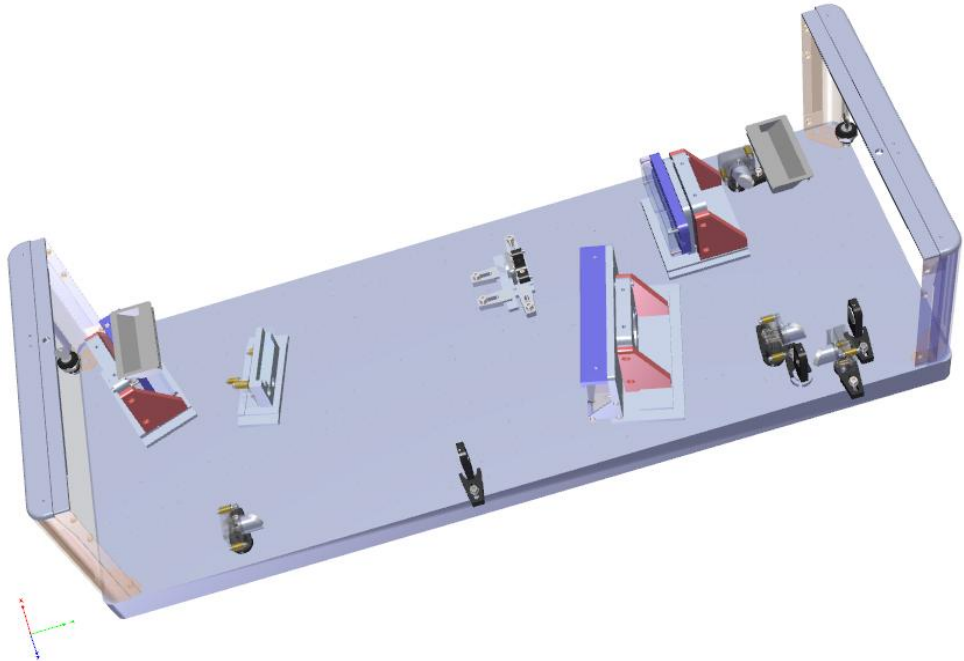

**Fig. 156.    Stretcher layout**

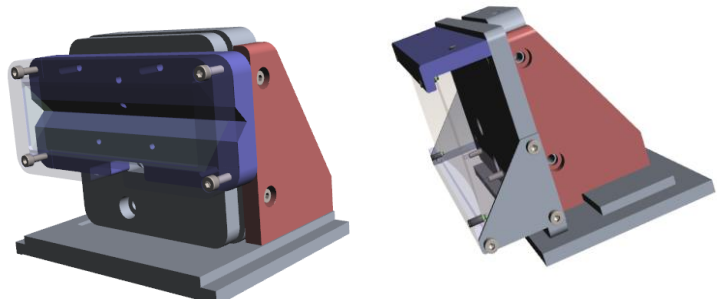

**Fig. 157.    Stretcher with rugged mounts for excellent long term stability and alignment free**

### 4.3.6.3    100 Hz Regenerative amplifier

The regenerative amplifier uses a Pockels Cell in order to seed the pulse coming from the stretcher and another one to extract the pulse. The timing control for these operations is done by the GenPulse. The amplifier is pumped with 10mJ. The cavity is a Z configuration with two concave mirrors in order to manage the thermal lens introduced by the pumping. No cryogenic cooling is necessary, water cooling is used to ensure a homogeneous temperature of the crystal and thus minimize the aberrations introduced by the thermal lens. The astigmatism introduced by the Brewster cut crystal is compensated by adjusting the angle of incidence of the concave mirrors.

A MAZZLER is added in the cavity and allows a complete control of the gain curve and permits to keep the spectral bandwidth of the seed pulse.

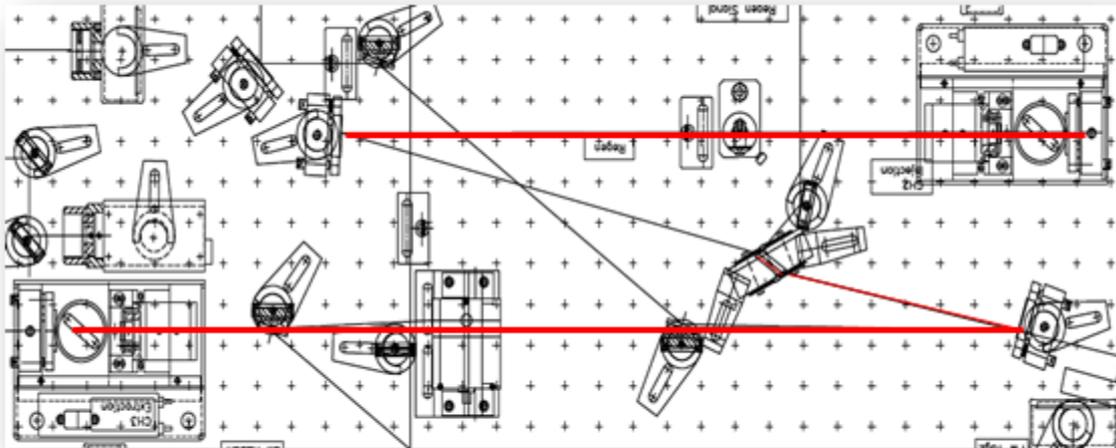

Fig. 158.    Optical Table lay-out

The main advantages of the regenerative amplifier compared to a multi-pass amplifier are a better beam quality (due to the filtering introduced by the cavity) and a better pulse to pulse stability (the extracted pulse is in the saturated regime). Another advantage is that spatial chirp is minimized by the cavity.

Table 40.    Specifications of the Regenerative Amplifier

| REGENERATIVE AMPLIFIER | |
|---|---|
| Pumping Energy | 10mJ |
| Output Energy | > 500µJ |
| Efficiency | > 5 % |
| Spatial profile | TEM 00 |

In order to improve the nanosecond contrast two Pockels cell will be installed before and after the regen.

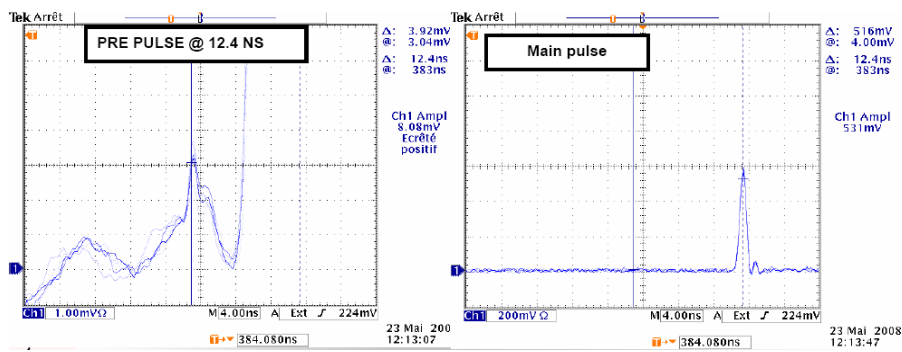

Fig. 159.    Contrast measured at 12ns >10-5 (typical 10-6)

- Pockels cell "pulse cleaner"
  A Pockels Cell is installed at the output of the regenerative amplifier. The precise adjustments on the delay and the way of switching the high voltage allow an improvement of the temporal contrast in the range of 1/1000.

- Pockels Cell "pulse picker", at 100 Hz

    A pulse picker is set up at the output of the oscillator in order to select the appropriate repetition rate from the pulse train coming out of the oscillator. The precise adjustments on the delay and the way of switching the high voltage allow a temporal contrast improvement in the range of 1/1000.

#### 4.3.6.4   100 Hz Multi-pass Pre-Amplifier MP1

A small pre amplifier is installed after the regenerative amplifier and allows reaching energy compatible with the next multi-pass amplifiers. To minimize the influence of the thermal lens, the crystal is pumped with only 20mJ of green coming from the Nd:YAG diode pumped laser (see later in this document).

**Table 41.   Specifications of the Multipass pre-amplifier**

| MULTIPASS PRE AMPLIFIER | |
|---|---|
| Pumping Energy | 20 mJ |
| Output Energy | >4 mJ |
| Efficiency | > 20 % |
| Number of passes | 5 |

#### 4.3.6.5   100 Hz Multipass Power Amplifier MP2

The MP2 amplifier is installed before the cryo cooled amplifier in order to minimize the final amplification factor. The pump energy, typically 100 mJ is enough to have around 25mJ from the amplifier. The crystal is pumped on both side to optimize the energy density and the heat deposition.

The extraction efficiency is around 25%.

**Table 42.   Specifications of the Multipass amplifier (MP2)**

| MULTIPASS AMPLIFIER 2 | |
|---|---|
| Pumping Energy | 100 mJ |
| Output Energy | >25 mJ |
| Efficiency | > 25% |
| Number of passes | 4 |

#### 4.3.6.6   100 Hz Multipass Power Amplifier MP3

The main amplifier is installed after AMP2 and allows for reaching the final energy. To reduce the thermal lens the crystal is cryogenically cooled. The pump energy depends on the required output energy, typically 1,2J are enough to have around 450mJ from the amplifier. The crystal is pumped on both side to optimize the energy density and the heat deposition.

Due to the gain control with the Mazzler we are able to keep a large and tunable spectrum at the output of the amplifier. The spectral shape is changed from a Gaussian (from the oscillator) to a super Gaussian shape.

**Table 43. Specifications of the Multipass amplifier (MP3)**

| MULTIPASS AMPLIFIER 3 | |
|---|---|
| Pumping Energy    1200 mJ | 1200 mJ |
| Output Energy     >450 mJ | >450 mJ |
| Efficiency    >35 % | >35% |
| Number of passes    3 | 3 |

#### 4.3.6.7 Compressor

The compressor uses a conventional Treacy two-gratings normal configuration with two holographic gratings. To be on the safe side concerning the damage threshold, a beam diameter of 30mm will be used at the entrance of the compressor. With a grating size of 140x120mm², the spatial and spectral bandwidths are enough to recompress the pulse down to 1ps.

Due to the low stretching factor the overall compressor size is kept to a minimum. This ensures the highest mechanical stability, required for a good beam pointing stability.

The compressor doesn't require any vacuum chamber.

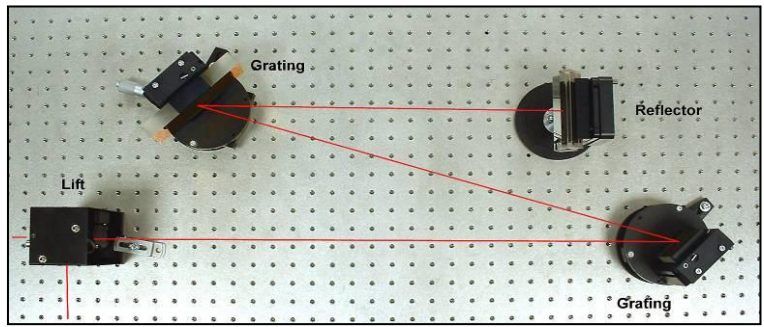

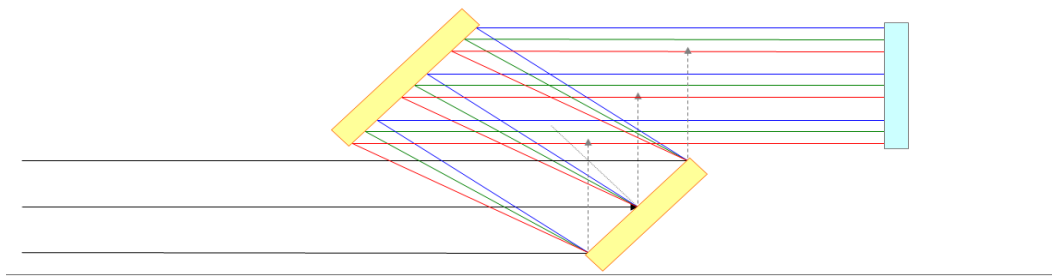

**Fig. 160.    Compressor**

**Table 44.   Specifications of the 100Hz Compressor**

| 100 Hz COMPRESSOR | |
|---|---|
| Input Energy | > 450 mJ |
| Output Energy | > 300 mJ |
| Efficiency | 70 % |
| Beam diameter | >30 mm |
| Spectral Aperture | >40nm |

#### 4.3.6.8   Pump lasers

Two type of pump laser will be used in the system:

- Centurion from Quantel, 20 mJ, 100 Hz @ 532 nm, diode pumping
- Nano TRL 250-100, > 100 mJ, 100 Hz @ 532 nm, flash lamp pumped.

The centurion is a very compact Nd:YAG diode pumped laser. The output energy reaches 20mJ at 532nm with a quasi-gaussian profile. Due to the diode pumping the short term stability is excellent and makes this laser an ideal source for pumping the front end of the femtosecond

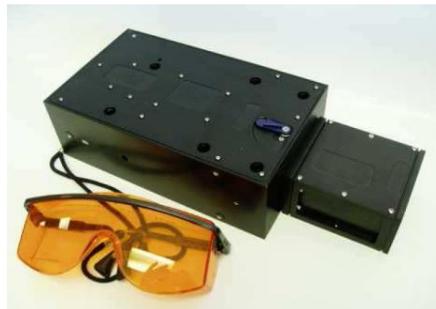

**Fig. 161.   Centurion Pump Laser**

This laser has been used in large quantities for a system installed with an accelerator at PSI (Paul Scherrer Institute).

The lifetime of the diodes is in the order of 1.5 Billions of shots (around 4000 hours).

For the power amplifiers (AMP 1, AMP 2 and AMP 3) a flash lamp pumped laser will be used. The LITRON NANO TRL 250-100 has been chosen for his rugged design ensuring a good reliability and long term stability. This laser has been made especially for industrial applications. The output energy is >100 mJ, 100 Hz @ 532 nm.

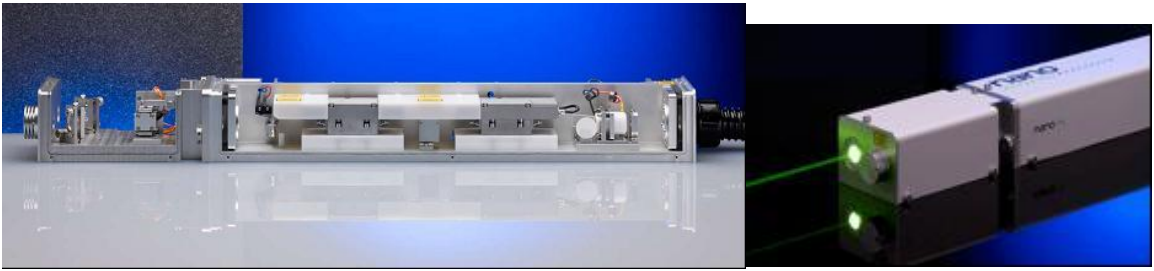

**Fig. 162.    LITRON NANO Laser**

The laser uses only 2 flashlamps with a lifetime around 250 Millions of shots (around 700 hours).

The Nano TRL 250-100 is cooled with water to water heat exchanger.

### 4.3.6.9    Generation of the 32 replicas

In order to generate the train of 32 pulses separated by 15ns, a passive system based on 50/50 beamsplitter and delay line will be used. The final recombination of the two 16 pulse train will be done by using a Pockels cell. This allows minimizing the overall losses of the system.

The device will be implemented before the compressor.

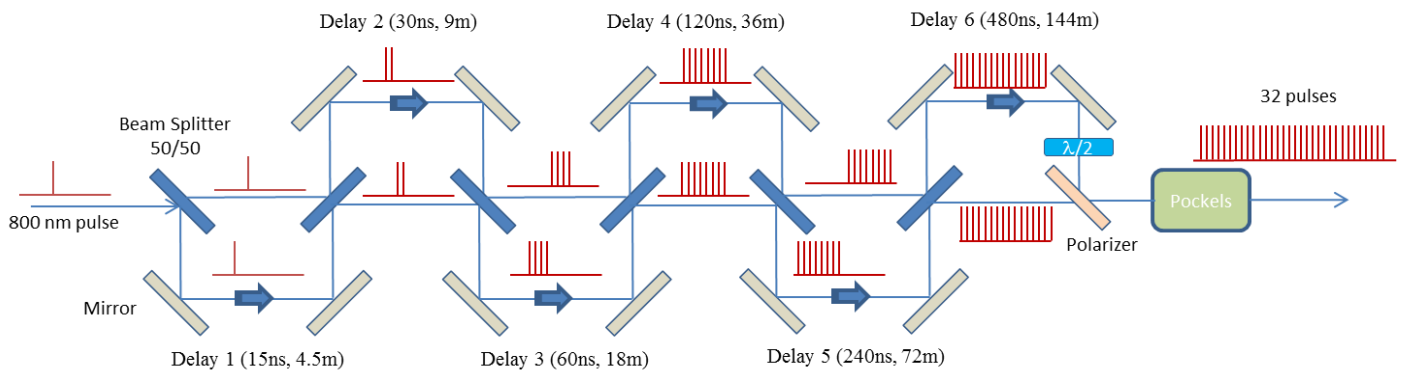

**Fig. 163.    Scheme for 32 Replica Generation**

GenPulse: An electro-Optic controller coupled to a digital delay generator

The GenPulse is the synchronization device developed by Amplitude-Technologies for all femtosecond systems. It uses the RF signal from the oscillator to generate the low frequency signals.

The GenPulse can control 4 Pockels cells. All delays can be modified independently with 500ps accuracy.

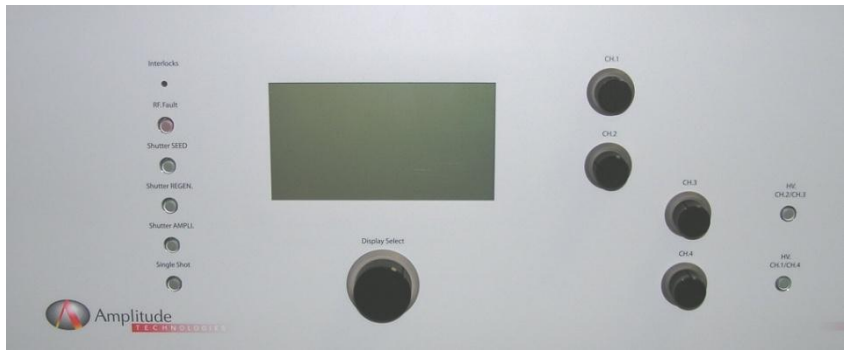

**Fig. 164.    GenPulse**

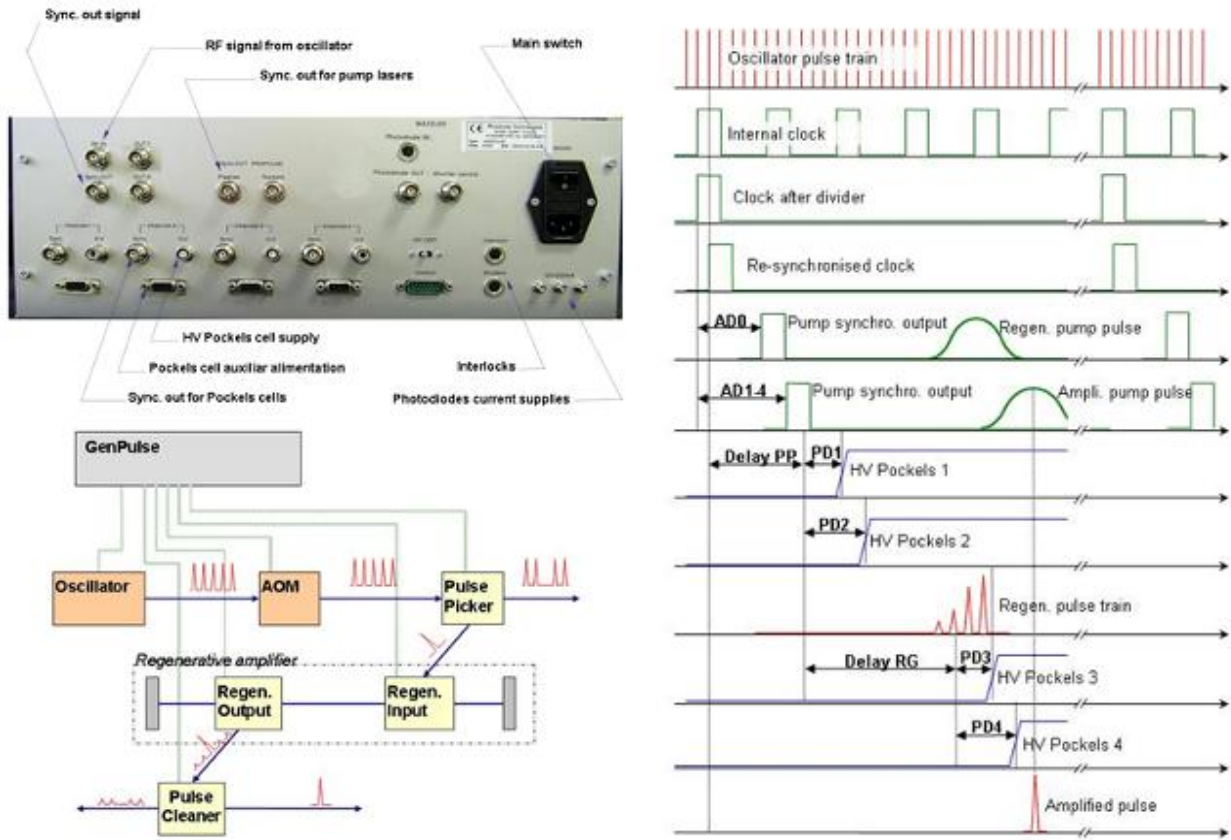

**Fig. 165.    Delay Line Scheme**

# 5. Gamma Beam Collimation and Characterization

## 5.1. Introduction

The ELI-NP gamma beam will be obtained by collimating the photons emerging from the Compton interaction region. This is a critical step to reach the required brilliance and requires a careful design of the collimation system. A precise energy calibration of the gamma beam and a continuous monitoring of the stability of its parameters during operation are also necessary features of the ELI-NP facility. A characterization system providing a prompt feedback on the energy spectrum, intensity, space and time profile of the beam is also essential for the commissioning and development of the machine.

Given the unprecedented characteristics of the beam, these tasks are extremely challenging. In this chapter, a description is given of the proposed collimators and detectors for the gamma beam, along with their technical implementation concepts. The majority of these devices will need a further work on the design finalization, Monte Carlo simulations and preliminary tests, in order to define the final specifications of the complete apparatus.The chapter is organized as follows: in the next section the collimation system is discussed; in Section 5.3 the physical requirements for the Gamma Beam Demonstration System (GaBX) are summarized and the proposed solutions outlined; in Section 5.4 the detector elements are described in detail; in Section 5.5 the expected detector performances based on Monte Carlo simulations are summarized and illustrated in detail; finally in  the integration with the machine is discussed in Section 5.6.

## 5.2. The gamma collimation system

The gamma energy bandwidth is one of the most challenging parameter to be achieved. Defined as the ratio of the r.m.s. to the maximum of the beam energy spectrum, the energy bandwidth should be in the range of of 3 % down to 0.5 %, resulting in a quasi-monochromatic gamma beam.

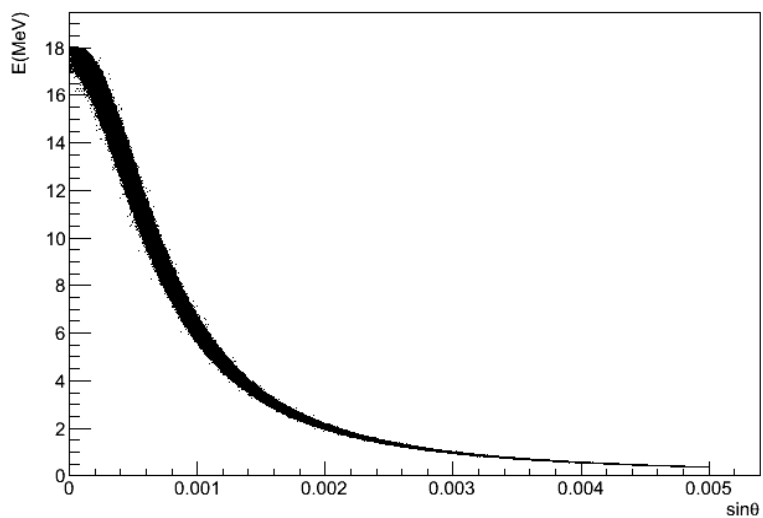

**Fig. 166.    Example of the angular dependence of the gamma energy**

Fig. 166 shows a typical energy distribution as a function of the angular divergence for the gamma emerging from the interaction point (IP) in the case of electron beam energies of 720 MeV, corresponding to a maximum photon energy of 18 MeV. In Fig. 167 it is shown the gamma energy distribution in the same case.

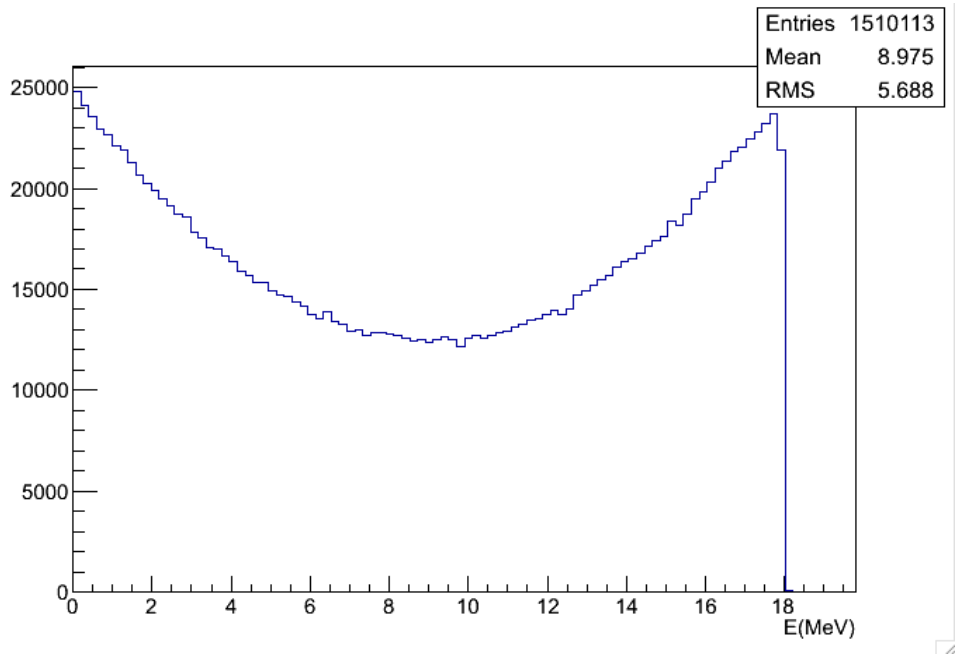

**Fig. 167.    Energy distribution for maximum gamma energy of 18 MeV**

By letting the photons propagate in the extraction line and by using appropriate collimators, the low energy part of the spectrum can be filtered out. The high energies are then selected and the bandwidth requirements can be achieved. Two different gamma beam extraction systems are considered, corresponding to the two different incident electron drive beam energies. In the case of the low energy (electron drive beam at 360 MeV) the mean energy of the gammas is about 2.4 MeV, and goes up to 9.0 MeV in the case of the high energy line (electron drive beam at 720 MeV). At those energy only a thick material with a high atomic number is able to stop the gammas.

In Table 45 a summary of the r.m.s divergences needed to obtain the 0.5 % bandwidth is shown for three typical values of gamma beam energy. Also, the corresponding diameters of collimation aperture, for a collimator placed at 10 m from the IP in the low energy case and 8.6 m in the high energy case, are listed.

**Table 45.    Collimation parameters for selected gamma beam energies**

|  | Low Energy | High Energy | High Energy |
|---|---|---|---|
| Energy (MeV) | 3.45 | 9.87 | 19.50 |
| Source rms divergence (μrad) | 100 | 50 | 40 |
| Collimator aperture (μm) | 2,000 (@ 10m) | 860 (@ 8.6 m) | 688 (@ 8.6 m) |

The main requirements for the collimation systems are:

1. Very low transmission of gamma photons (high density and high atomic number material);

2. Continuously adjustable aperture, in order to reach the correct energy bandwidth in the whole energy range;
3. Avoid production of secondary radiation (electromagnetic, neutrons…)

## 5.2.1. Collimation system

Various types of collimators and materials have been studied. The choice has been made on a dual slit collimator similar to the one already designed and assembled at INFN-Ferrara for a high energy X-ray experiment [121] (Fig. 168). It consists of an aluminium framework and two 20 mm-thick tungsten edges.

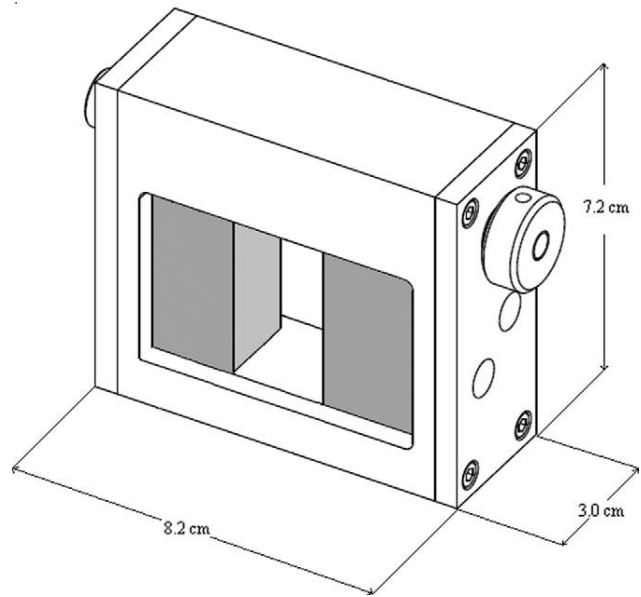

**Fig. 168.** Design of the dual slit collimator

The slit width can be adjusted by translating the tungsten edges. By adding several such collimators at different rotation angle, one can create a pin-hole collimator with a small enough angular acceptance and a significant thickness.

In particular for the two lines, low and high energy, of the ELI-NP facility, the collimator setup will be:
- Low energy (1-5 MeV): 12 tungsten slits with a relative rotation of 30° each.
- High energy (5-20 MeV): 14 tungsten slits with a relative rotation of 25.7° each.

The position of the collimation system along the beam-line will be at about 10 m of distance from the IP in the low-energy case and 8.6 m in the high-energy case. Each slit aperture will be adjustable from 0 to 20 mm . To minimize the misalignment two different mechanicals structures will be used. Each of them is composed of 6/7 dual slits related to each other by a rotation of 30°/27.5° in the case of low/high energy, respectively. The distance between the two frames can be adjusted considering mechanical constraint and alignment optimization.

A schematic drawing of the low energy collimation configuration, with a total of 12 slits, is shown in Fig. 169.

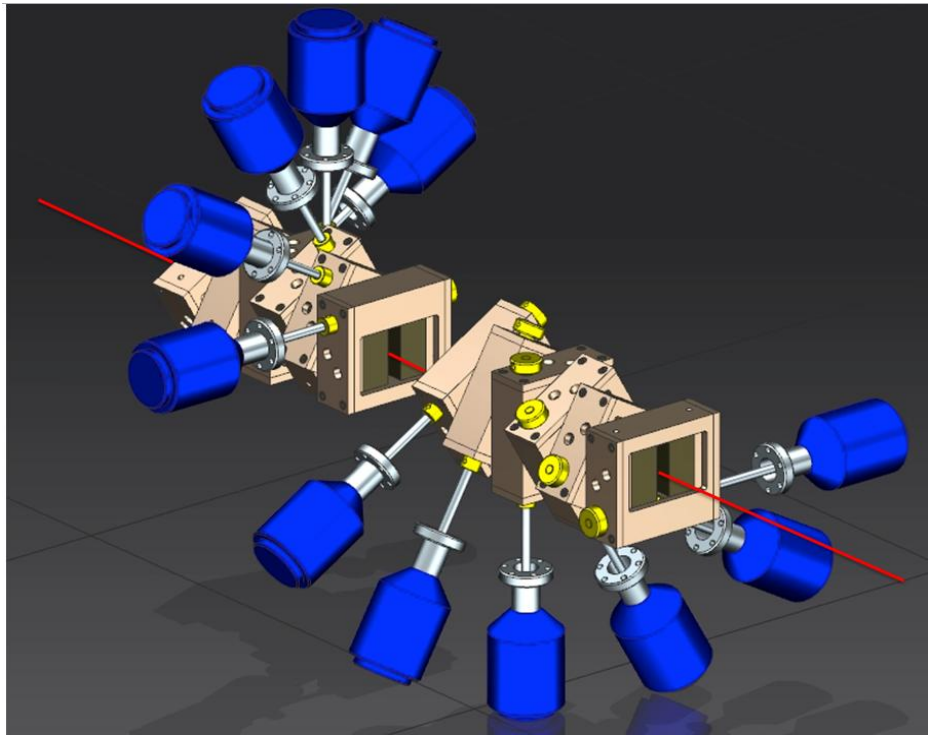

**Fig. 169.** Drawing of the configuration of low energy collimator made up of 12 tungsten adjustable slits with a relative 30° rotation each

In order to evaluate the transmission of these collimators setup a Monte Carlo simulation of the collimator assembly was set up using the MCNPX code. A ray-tracing technique ("radiography tally") was employed. Only the direct (source) contributions were made to the detector grid, neglecting all scattering events. A point source (at 10 m from the origin) was assumed to emit mono-energetic photons into a cone according to the parameters reported in Table 46. The geometry used is reported in Fig. 169. As it can be seen it is made up of 12 or 14 tungsten slits with an angular displacement of 30° or 25.7° for low and high energy, respectively. Each slit is composed by two blocks of tungsten of size 3.0 x 3.0 x 2.0 cm$^3$. The slit aperture is 0.96 mm for low energy case and 0.48 mm for the high energy case.

**Table 46.   Monte Carlo simulation parameters**

|  | LE | HE |
|---|---:|---:|
| Energy [MeV] | 4.6 | 18.5 |
| # of tugsten slits | 12 | 14 |
| Angular displacement [°] | 30 | 25.7 |
| Collimator gap [mm] | 0.96 | 0.48 |
| Detector distance [cm] | 36.1 | 42.1 |

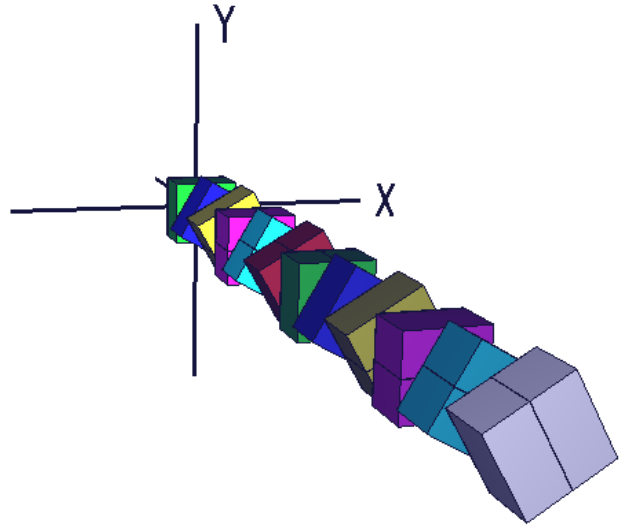

**Fig. 170.   Slits arrangement for MC simulation of transmission**

The results of these simulations are showed in Fig. 171, where the number of photon transmitted (not interacting) by the collimation system are represented as a function of position in grey levels or as a 3-d plot. It is possible to see the pattern due to the superimposition of different thicknesses of tungsten, notice that the scale is logarithmic.

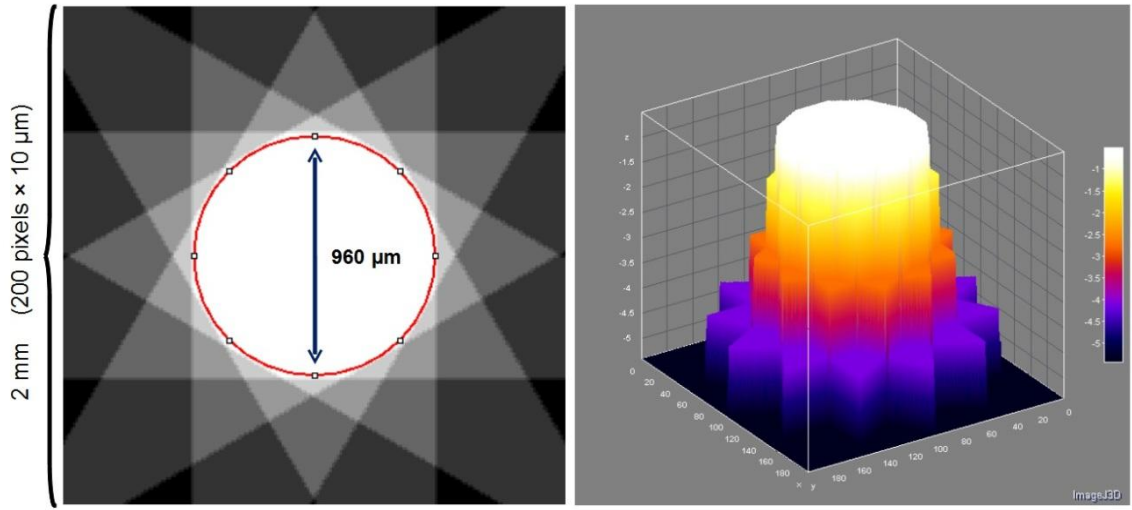
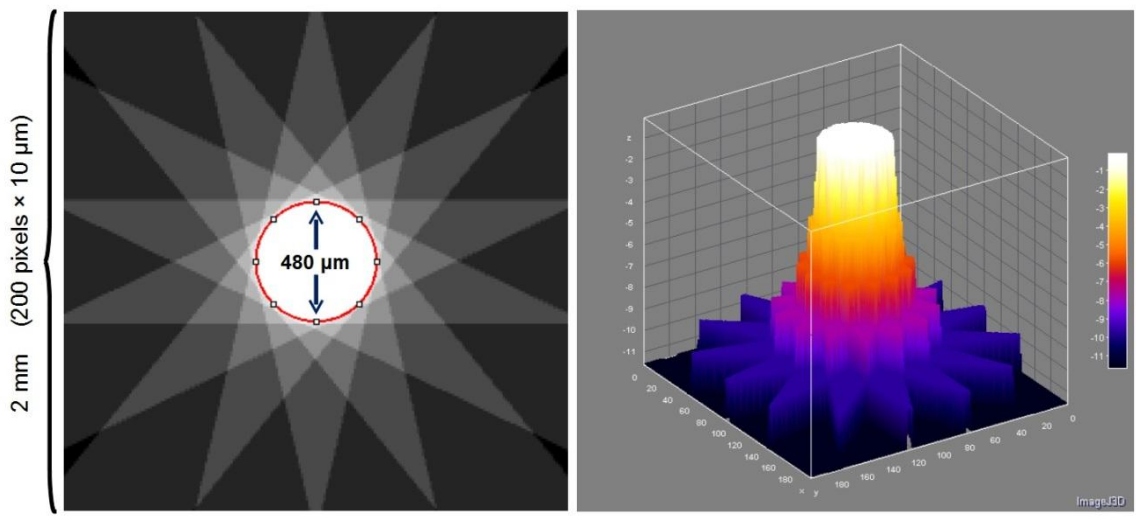

**Fig. 171.** Results of MC simulation for LE (4.6 MeV) on top, and for HE (18.5 MeV) on bottom

### 5.2.2. Collimation system stability

Let consider the low energy case. All the 12 dual slit collimators should be aligned both in space and in angle. To minimize the misalignment two different mechanicals structures will be used. Each of them are composed of 6 dual related to each other by a rotation of 30°. The distance between the two frames can adjusted considering mechanical constraint and optimization. The alignments accuracy for the collimation system structure requires 20 µrad in angle (with respect to the center of the structure) and 20 µm in position.

In the case of the high energy two structures of 7 collimators is considered. The collimators are related to each other by about 25.7°. The tolerances in position and in angle are the same than in the case of the low energy. To allow the mechanical support, stability and fine adjustment of the collimation system a set of movement will be implemented on the main support frame, as depicted by Fig. 172.

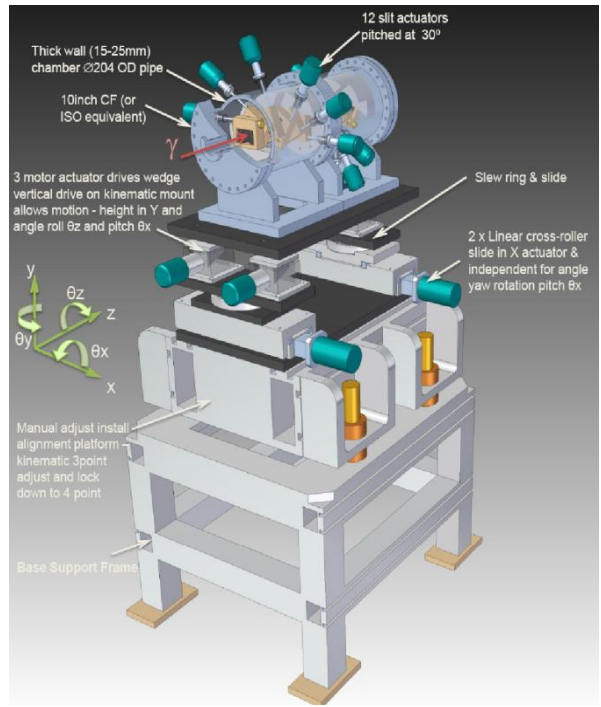

**Fig. 172.  Representation of the collimation system complete with framing and support**

### 5.2.3.  Background

In order to avoid that the scattered radiation produced by the interaction of the primary gamma beam with the collimation system interferes with the downstream detectors, a concrete block will be placed after the collimators exit. This block will be shaped as a cylinder of concrete material 100 cm long and with a radius of 100 cm. To evaluate the efficiency of this block in absorbing scatter radiation, being photons, electron, positrons and neutrons, a Monte Carlo simulation using FLUKA was performed. The gamma beam considered has energy of 20 MeV, a beam radius at source equal to 0.5 mm and a divergence of 200 rad. After propagating in vacuum for 10 m, the beam impinges on a cylindrical tungsten collimator (length = 28 cm, inner radius 0.024 cm, outer radius 0.5 cm). At a distance of 30 cm from the collimator exit is placed the parallelepiped concrete shielding 1 m long, with a square transverse section of 60x60 cm2 and an aperture radius of 1.45 cm. The simulation of 105 photons, comparable to the number of photons per each pulse, results in the collection of 1 electron, 2 positrons and 20 photons on a collecting screen: inner radius 1 cm, outer radius 20 cm.

## 5.3.  Requirements and overview of the Gamma beam characterization system

In order to assess the machine performance and monitor its operation, the GaBX gamma beam demonstration system is expected to measure the beam brilliance by accurately reconstructing its intensity and energy spectrum. According to the beam specifications described in the previous chapters, the required resolution on the energy measurement should not be worse than the expected bandwidth of 0.3% in a photon energy range between 1 and 20 MeV. The system must be able to cope with pulses of 2 to 6 x$10^5$ photons within a time window of 1-2 ps, and pulse trains separated by 15 ns. The capability of assessing the time dependence of the beam parameters within a macro-pulse, other than their long-term stability, is also a desirable property of the system that would allow providing feedback on the tuning of the laser recirculation system and of the interaction region.

To meet these requirements for the intensity monitoring, we foresee to use a luminometer based on thin detectors exposed to the beam, providing the integrated beam flux on two different time scales: fast (ns scale) to resolve single pulses and slow (ms scale) to monitor the integrated macro-pulse flux.

The spatial profile of the beam can be monitored with high resolution and minimal beam attenuation using thin scintillator screens readout by CCD cameras.

For measuring the gamma energy, the main difficulty comes from the very short length of the intense pulses, preventing to easily disentangle the response of single photons to any detector directly exposed to the beam line. Therefore, we propose to sample Compton interactions of single photons in a micrometric target, by accurately measuring the energy and position of the resulting electrons and photons off beam. The advantage of this technique is the minimal interference with the beam operation, making it an ideal tool for beam energy monitoring. From a sufficiently large amount of such measurements, the shape of energy spectrum can be precisely determined. On the other hand, the need to use detectors with very high energy resolution (either HPGe or LaBr crystals) prevents to perform this measurement on a time scale smaller than the macro-pulse length, resulting in a low rate of clean measurements (<< 100 Hz).

A complementary approach consists in performing a measurement of the total beam energy by absorbing the gamma pulses in a longitudinally segmented calorimeter. This approach relies on the high intensity and monochromaticity of the gamma beam: the longitudinal profile of the energy released by photons in a light absorber has a rather strong dependence on the incident photon energy in the range of interest, while the profile fluctuations are suppressed by the high number of photons. Once the gamma average energy is obtained from the longitudinal profile, the beam intensity is also measured at the same time from the total energy release. The advantage of this approach is that the full photon statistics can be exploited and, since fast detectors can be used, the measurement can be performed for every single pulse allowing it to be used during the machine commissioning and tuning to provide an immediate feedback on the beam energy and intensity and their variation within a macro-pulse.

The combination of the measurement performed by the Compton Spectrometer and the absorption calorimeter will make possible to fully characterize the gamma beam energy distribution and intensity with the precision needed to demonstrate the achievement of the required parameters.

Finally, an absolute energy scale calibration system must be included into the characterization system. Using appropriate targets, the detection of resonant scattering condition during a beam energy scan attests the beam energy very precisely for a number of values, providing accurate calibration for the other detectors components and the user experiments.

To summarize, the general concept of the characterization system consists of four basic elements:

- a beam position imager to spot the beam position for alignment and diagnostics purposes;
- a Compton spectrometer, to measure and monitor the photon energy spectrum, in particular the energy bandwidth;
- a sampling calorimeter for a fast combined measurement of the beam average energy and intensity, to be used als as monitor during machine commissioning and development;
- a resonant scattering spectrometer for absolute beam energy calibration and inter-calibration of the other detector elements.

Two similar beam characterization systems are foreseen for the two gamma beam lines, optimized for the different energy regimes.

All the detectors are contained in specific interaction chambers under vacuum, in direct connection with the gamma beam pipe in order to avoid transition windows.

## 5.4. Detector Design

### 5.4.1. The Beam Position Imager

The Beam Position Monitor has to display the location and uniformity of the beam. The use of the device is mandatory for checking the alignment of the collimation system with the source of the gamma emission (the point of interaction electron-light). The system is used also to control the size and uniformity of the field at the output of the collimation system. The typical size of the beams may vary between 0.1 mm to 10 mm for which the spatial resolution of the system will have to be between 0.05 mm and 0.10 mm. In Fig. 173 the conceptual drawing of the imager is shown: it is mainly composed of a scintillator screen, a lens system to focus onto a CCD camera the scintillator light distribution, a frame grabber and an acquisition system.

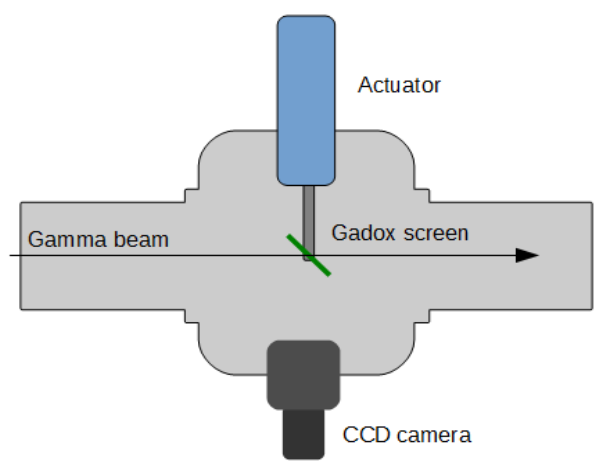

Fig. 173.    Sketch of the beam imager setup

## 5.4.2. The Compton Spectrometer

This detector is designed to precisely measure the energy and position of single electrons recoiling at small angles against incidents photons Compton–scattered on a thin target. At small recoil angles with respect to the beam direction, the electrons carry most of the incident photon energy and the sensitivity to polar angle, whose knowledge is limited by the beam size and the multiple scattering inside the target, is minimal. By also detecting in coincidence the scattered photon, whose position and energy can be predicted from the electron measurements, the background from pair production and Compton photons, as well as from electrons not fully contained in their detector, can be strongly suppressed. The principle of the spectrometer is illustrated in Fig. 174.

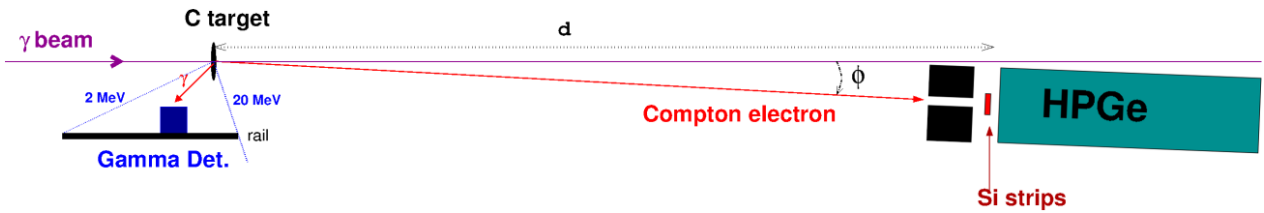

**Fig. 174.** Schematic view of the Compton spectrometer. The HPGe and Si strips detectors are used to measure the scattered electron at small angles. The scattered photon is measured by a movable device, to be placed according to the beam energy.

We discuss in this subsection the motivations driving the detector design and the optimization of its geometry, obtained from analytical and numerical computations using the cross sections for the relevant processes. The detector with the optimized configuration has then been simulated in detail using GEANT4 to anticipate their performances that are presented in Section 5.5.2.

We plan to use a low-Z target to minimize the relative contribution of pair production. We choose to use graphite as a practical solution, since foils of almost any thickness are available on the market.

To obtain the required resolution on the beam energy, the electron must be contained in an high-resolution homogeneous detector. We considered HPGe as best option, and LaBr crystal as an alternative. For both these technologies, the time needed for the measurement is not shorter than the macro-pulse length (600 ns) and any pileup occurring in this time window would spoil the measurement. This implies that the rate of useful measurement cannot be larger than the macro-pulse repetition rate (100 Hz), and the target for Compton scattering must have a very small thickness δ in order to keep the number of detected electrons per macro-pulse $N_{eMP}$ low:

$$N_{eMP} = N_{\gamma MP} \cdot \sigma_{Compton} \cdot \alpha \cdot \rho N_A \frac{Z}{A} \cdot \delta \sim 1$$

where $N_{\gamma MP}$ ~ 5x10$^6$ is the number of impinging photons per macro-pulse, the electron density $\rho N_A \frac{Z}{A}$ is 6.6 10$^{23}$ cm$^{-3}$ for graphite, and the Compton electronic cross section (~150 to 30 mb from 2 to 20 MeV) has to be multiplied by a realistic detector acceptance $\alpha$ ~ 1 %. From this simple calculation we get $\delta$ ~ 2-10 μm.

This is indeed an advantage, since such a micrometric target is almost transparent for the beam, and minimizes the multiple scattering of the emerging electron that, as shown in the Section 5.5.2, turns out to be the most limiting effect to the beam energy resolution.

Our simulations show that electrons in the energy range of interest can be contained with good efficiency using an HPGe (or LaBr) cylinder with a radius of 65 mm. To enhance the full energy peak of the measured electrons while reducing pileup, only particles entering the inner part of the detector can be selected using a lead collimator placed in front of the detector. The collimator has an inner hole of radius 30 mm, and a depth of 45 mm to effectively filter out electrons, positrons and photons outside the fiducial region. To measure the electron scattering angle a precise position measuring device is foreseen between the lead collimator and the energy detector. This also improves the identification of electrons against photon background. We plan to use a double-sided silicon microstrip detector, as described in more detail in Section 5.4.2.1.

The angle of the detector axis with respect to the beam direction must be the minimum allowed by the detector size, for several reasons:

- as already mentioned, the sensitivity of the reconstructed gamma energy on the electron polar angle $\Phi$ is minimal, being proportional to $|\cos(\Phi)\sin(\Phi)|$ for $\Phi \ll 1$;

- since the electron energy is maximal, the error on $\Phi$ due to multiple scattering, inversely proportional to the electron momentum, gets minimized;

- though the peak of the distribution of electron recoil angle depends on the incident photon energy (from about 160 to 24 mrad from 2 to 20 MeV), the purity of signals, defined as the fraction of signals in the detector that are due to Compton electrons, is minimal at the lowest possible angle for any energy between 2 and 20 MeV. This is shown on Fig. 175.

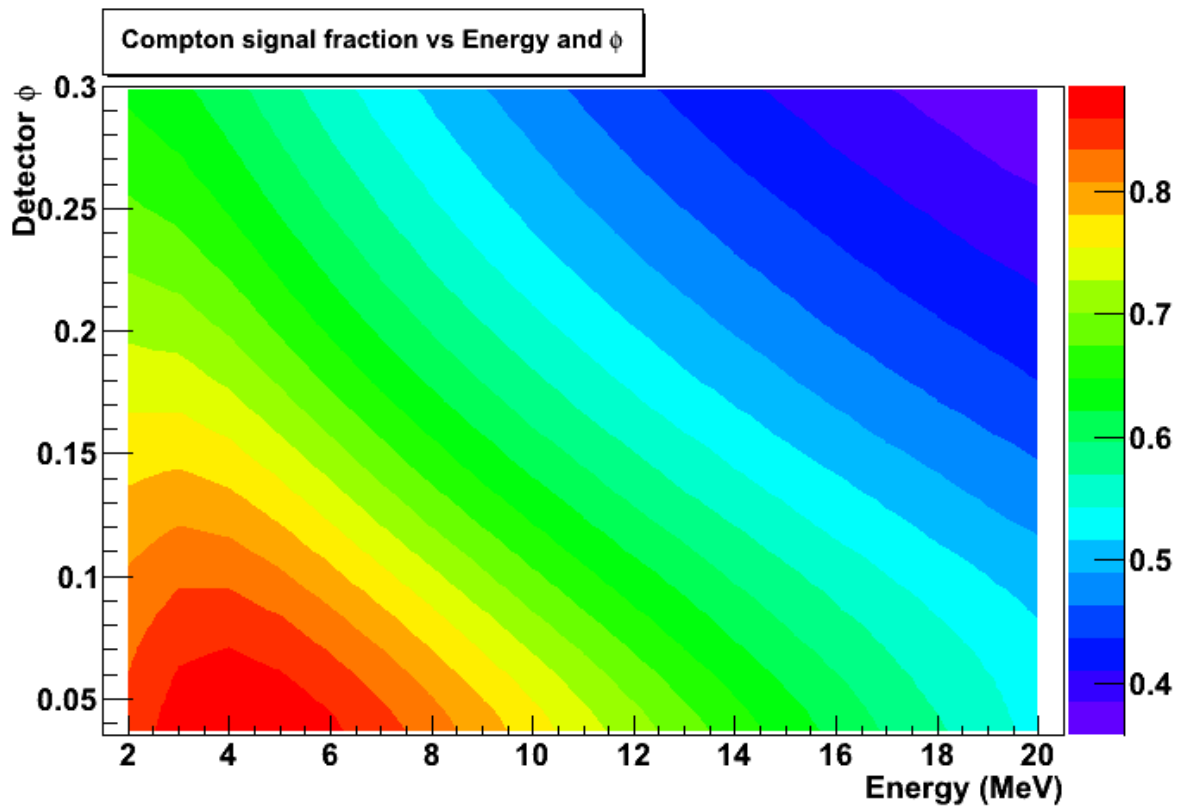

**Fig. 175.  Fraction of signal due to Compton electrons among the particles entering the detector as a function of the beam energy and the polar angle of detector axis.**

Finally, the distance *d* of the detector from the target along the beam direction is chosen as a compromise between conflicting requirements:

- a larger distance allows to reduce the polar angle for a fixed detector size, and the error on Φ due to the beam spot size;

- a smaller distance increases the acceptance, allowing to reduce the target thickness for a given rate, reducing the contribution of multiple scattering.

To optimize this parameter we compute, as a function of *d*, the target thickness needed to obtain a fixed rate (20 Hz) of isolated electron signals for 10 MeV gammas, placing the detector at the minimal practically possible angle of 80 mm/*d*. We obtain the resulting expected resolution on the gamma energy taking into account a resolution on the electron energy of 0.1%/ r.m.s., the uncertainty on the gamma position on the target (1 mm r.m.s. in the transverse directions), the effects of multiple scattering and energy loss of the electron inside the target. The results shown in appendix A2 for this optimization lead to the choice of d = 200 cm as a compromise between getting a reasonable rate and a good energy resolution. Finally, the rate and resolution expected from good signals using this simple calculation are shown as a function of energy in Fig. 176. Note that the calculation does not take into account the inefficiency of the electron measurements and of the detection of the scattered photon in coincidence, that will result in a reduction of the rate of useful signals, and we also assume that the background can be completely suppressed and/or subtracted. The numbers are thus intended as best limits from the principle of the method, while a more realistic evaluation of the detector performance will be given in Section 5.5.2.

We see that we can expect an important contribution on the energy resolution from multiple scattering. However, if larger beam intensities are reached, the same detector rate can be obtained with smaller target thickness, improving the achievable resolution down to the limit imposed by the electron energy measuring device.

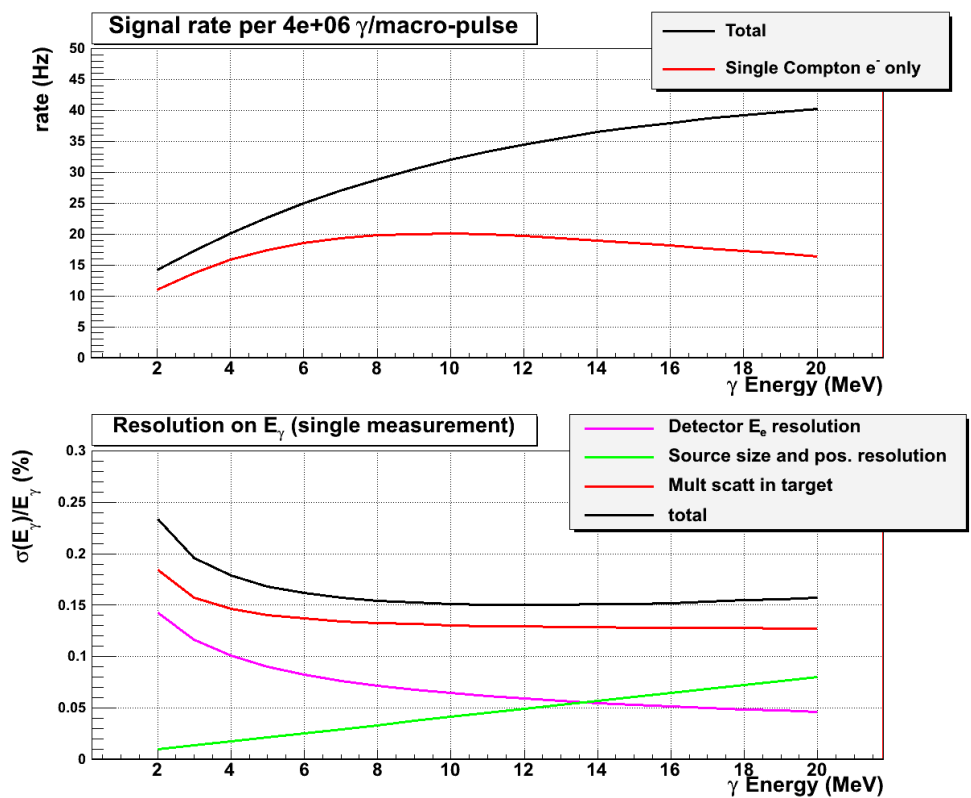

**Fig. 176.** Rate and beam energy resolution expected from the Compton spectrometer using a graphite target of 1.7 μm for a macro-pulse intensity of 4x10$^6$ photons. On the upper plot, the total rate of detector signals and the rate from single Compton electrons are shown. In the lower plot, the best achievable resolution from a single measurement on the primary gamma energy is shown.

In the following we discuss the proposed technologies for the electron and photon detection. The electron energy resolution must be as good as possible. The accuracy on polar angles estimation is limited by the uncertainty on the beam position, so that a resolution slightly better than 1 mm on the position measurements is needed. The Compton-scattered photons are detected using a detector close to the target in a position that depends on the beam energy. The detection of scattered photons does not improve the resolution on the primary gamma energy measurement, but provides a powerful veto against signals in the electron detector that are not due to a single fully stopped Compton electron.

### 5.4.2.1 The electron detection system

Two possible solutions are being considered for measuring the electron energy: an HPGe detector or a LaBr3(Ce) crystal. The HPGe solution is the preferred one given the better energy resolution achievable. On the other hand the LaBr3 crystals offer a faster time response and a lower price. The geometry of the detector is however independent of the chosen technology being the physical properties of Ge al LaBr3 very similar from the particle detection point of view (see Table 47). To fully contain electrons in the required energy range the detector geometry is designed as a cylinder with a radius of 65 mm and a length of 80 (40) mm for the high (low) energy line. As mentioned already, a cylindrical lead collimator with external radius of 65 mm, internal radius of 30 mm and thickness of 45 mm is placed in front of the detector. Given this geometry, a standard coaxial HPGe detector cannot be used due to the dead space around the inner electrode. Therefore a solution based on a stack of planar HPGe sensors is adopted, similar to the "Edinburgh Ge6 Array" described in [122]. We plan to use four (two for the low energy line) hyper-pure Ge sensors, each crystal being a 20 mm thick cylinder. An important feature is the use of ultra-thin (<~ 1μm) electrical contacts on the flat surfaces in order to minimize energy losses as the particles traverse from one crystal to the next. All the sensors are enclosed in a single cooling container with a thin (100 μm) beryllium entrance window. The window is needed to preserve vacuum and Be is chosen for its low atomic number a in order to minimize energy loss.

**Table 47. Physical properties of Ge and LaBr$_3$**

|  | Ge | LaBr$_3$(Ce) |
|---|---|---|
| density (g/cm$^3$) | 5.32 | 5.29 |
| average Z | 32 | 40.5 |
| average Z/A | 0.44 | 0.43 |
| dE/dx for m.i.p. (MeV/cm) | 7.3 | 6.9 |
| radiation length (cm) | 2.3 | 1.9 |
| Molière radius (cm) | 2.7 | 2.8 |

The front-end chip is a standard FET based charge amplifier. However care must be taken on the electrode total surface that might need to be segmented in order to minimize the noise.

The readout chain is based on a fast digitizer followed by a digital pulse shaping stage. A standard Germanium detector cooling system is foreseen.

An advantage of using LaBr3 instead, is that an homogeneous detector could be achieved. The crystal can be coupled with a photomultiplier tube [123] or, alternatively, with a solid state photo-detector such as an avalanche photo-diode or a silicon drift detector [124]. Since the LaBr3 is strongly hygroscopic, the crystal must be enclosed in a protection structure equipped with a beryllium entrance window, similar to the HPGe case, for not degrading the accuracy of the energy measurement.

The recoil angle of the electron must be measured with accuracy of <~ 1 mm in order to obtain, along with the energy measurement, the energy of the incident photon. The entrance position of the electron into the energy detector is precisely measured with a double sided silicon strip detector originally developed for the Pamela apparatus [125]. The sensor is 300 µm thick with 50 µm and 67 µm readout strip pitch respectively on the junction and ohmic side, with overall dimensions 5.33x7.00 cm2. The front-end chips used to readout the silicon sensor are the low noise VA1 Application Specific Integrated Circuits (ASICs) from IDEAS [126] which contain 128 charge sensitive preamplifiers connected to shapers and sample-and-hold circuits. A signal to noise ratio of ~50 for Minimum Ionizing Particles has been demonstrated [127], allowing both a very precise bidimensional reconstruction of the electron impact point (~10 µm), and a sufficiently accurate measurement of the energy released in the silicon sensor (with ~2 KeV resolution).

### 5.4.2.2 The photon detection system

The photon detector should have an adequate spatial resolution to allow for angular matching with the recoil electron. The proposed solution is based on LYSO crystals read out by APDs. The lutetium-yttrium oxyorthosilicate doped with Cerium is a scintillation material which provides high absorption coefficient, due to its high density, very short decay time and bright light output exceeding that of BGO, which is still commonly used for gamma-rays detection. The LYSO properties are listed in Table 48. We plan to use an array of 8 LYSO crystals with a transverse section of 5x5 cm2 and a depth of 2.5 cm, arranged in 2 layers covering a surface of 10x10 cm2, in order to ensure high photon detection efficiency in the needed range (0.2-0.9 MeV for beam energy of 2-20 MeV). The advantages of using monolithic crystals, when compared to a pixelated detector, are a simpler design, lower cost, larger sensitive volume, better energy resolution and comparable or better spatial resolution. Crystal blocks coupled to APDs have been widely studied in the field of medical applications, especially PET detectors.

**Table 48. LYSO properties**

| | |
|---|---|
| Density (g/cm$^3$) | 7.1 |
| Meltint point (C) | 2050 |
| Zeff | 65 |
| Hygroscopic | No |
| Wavelength of emission max. (nm) | 420 |
| Refractive index @ emission max | 1.81 |
| Decay time (ns) | 41 |
| Light yield (photons/keVγ) | 32 |

Statistical methods, also based on neural networks, have been developed in order to estimate the coordinates of entrance of incoming photons. Intrinsic spatial resolutions better than 1 mm have been achieved [128-130].

### 5.4.3. Target and Mechanics

The target thickness must be optimized according to the beam energy and intensity in order to avoid detector saturation while keeping an acceptable rate of useful signals. The target will be mounted on a revolver-like support to allow for different thicknesses and possibly also to host the targets for the resonant scattering measurements described in Section 5.4.5.

The photon detection system must be movable in order to match the proper detection angle as a function of the incident photon energy, and to optimize the efficiency for the resonant scattering measurement.

### 5.4.4. The Absorption Calorimeter

The total gamma absorption cross section might have a rather strong dependence on the incident photon energy as shown in Fig. 177. However, in the energy range of interest for ELI-NP, the energy dependence tends to saturate for heavy absorbers when pair production starts dominating the cross section, while a good sensitivity is maintained for low-Z materials, where the cross section is still decreasing with energy. The average longitudinal position of photons scattering inside a light calorimeter is thus expected to increase with energy. Moreover the average depth of the resulting electromagnetic showers in the detector also tends to increase with energy. The idea of this approach is to parameterize, with the help of detailed simulations, the expected profile of the energy release in a longitudinally segmented calorimeter as a function of energy. The average energy of the beam (and, with much lesser sensitivity, its bandwidth) can be measured by fitting the measured longitudinal profile against the parameterized distributions. Once the photon energy is known, the number of impinging photons is obtained from the total energy release, after correcting for the expected leakage that is also obtained from simulations.

Organic scintillators are natural candidates for being low-Z absorber and active detectors at the same time, also having the advantage of fast response time allowing to resolve pulses within a macro-pulse. However, the absorption of the very intense and collimated gamma pulses requires the detectors to be radiation hard. Our simulations suggest that, at 20 MeV gamma energy, doses up to $10^{-9}$ Gray per incident photon are released in the central mm2 around the beam direction. For any practical plastic scintillator, the performance is quickly degraded after about 10 Gray, i.e. a fluence of $10^{10}$ photons, corresponding to less than 1 hour of operation according to the beam specifications.

Therefore, we plan to adopt the scheme of a sampling device, where a passive plastic absorber is interleaved with thin detector layers, made of a hard-radiation and fast technology. We choose to use Si pixel planes. The sampling scheme is not expected to degrade too much the achievable resolution, since, due to the high beam intensity, the quantum fluctuations on the measured signals are not expected to be larger than the longitudinal fluctuations of the primary particle distribution.

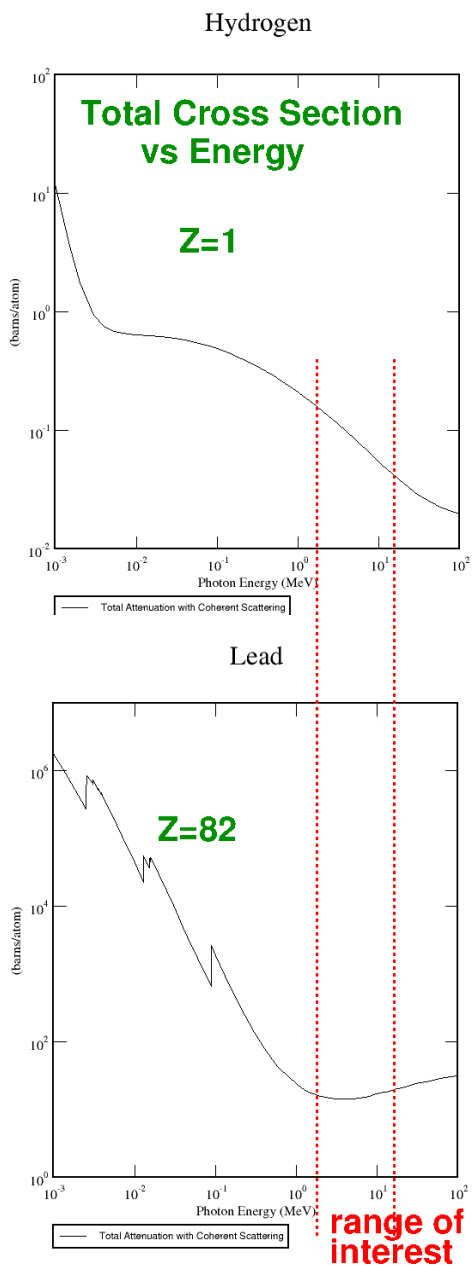

**Fig. 177.    Total photon cross section in hydrogen and lead, as a function of energy**

In order to optimize the detector layout, we simulated, using GEANT4, the energy release of photons in the 2-20 MeV energy range inside a large number (150) of detector elements made of 1 cm thick polyethylene blocks followed by a 1 mm thick Si layer. The distribution of the released energy and its fluctuation was studied as a function of the longitudinal and transverse position inside the detector. From this study, the absorber length was chosen to be 75 cm, and the transverse size to be 6x6 cm2. Simulations described in detail in appendix A2 suggest adopting a layout consisting in a stack of 25 identical detector elements.

### 5.4.4.1    The Detecting Layers

The active detections layers used for the calorimeter should have the following main characteristics:

- They should be made with detectors that are radiation hard, to avoid significant degradation of the response after several months of data taking, that correspond to doses ~ 50 KGy. We believe that this is a safe margin, since the absorption calorimeter will not be routinely used, but it will be inserted in the beam only during the commissioning phases;

- They should be fast enough to allow clear separation between the various bunches in the macro bunch. In this way we can provide important feedbacks on the tuning of the laser recirculation system and of the interaction region;

- They should allow a good signal over noise ratio, to precisely measure simultaneously the photon energy and flux.

Silicon sensors are the best candidates for this task. The technologies developed in the last 15 years for the LHC tracking systems allow us to have radiation hard detectors [131] that can safely sustain irradiation up to $10^{14}$ 1 MeV neutron/cm$^2$ and up to 100 KGy. The thickness of the sensors and the bias voltage applied can be chosen to have collection times on the electrodes of the order of 2-3 ns, allowing the measurement both the energy and the intensity of the single bunch.

We will use 25 6x6 cm$^2$, 200 µm thick silicon sensors (n-type bulk), with a proper implantation pattern (p-type implants) to match the required geometrical characteristics. Measuring the transverse distribution of the energy release is not strictly needed for the beam energy and intensity characterization. It is useful anyway, in order to detect the presence of beam halo, to verify the beam alignment, and to test the predictivity of the Monte Carlo simulations. For this we will implant 16 pads on a 4x4 matrix on each sensor. The readout of the pads will be done in a different way for the different layers, to optimize the number of readout channels, reducing or increasing the segmentation according to the importance of the various layers. The optimal segmentation that we found is the following:

- on the first layer we will readout all the 16 pads with 16 separate electronic channels, to have the possibility to finely analyze the first part of the shower produced by the incident photons;

- among the subsequent 24 layers, one every 6 will be readout using 2 electronics channels each, one for the 'inner core' (the 4 central pads, that will be bonded together) and one for the external ring (the 12 external pads, that will be bonded together);

- the remaining 20 layers will be readout with only one electronic channel, bonding all the pads together.

In this way we will reduce the overall number of electronic channels to 44, without sacrificing the physics performance of the absorbing calorimeter.

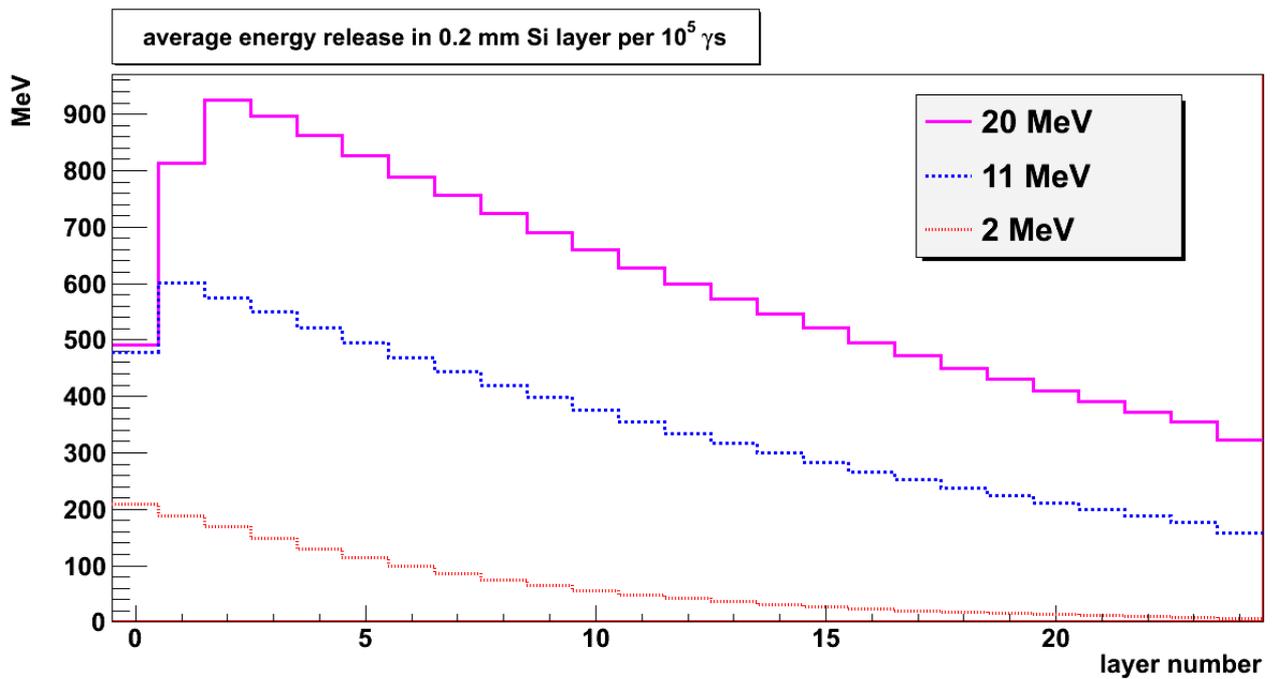

**Fig. 178.** Expected energy release in each of the 25 active Si layers for bunches of $10^5$ gammas of different energies.

To define the characteristics of the electronic channels (in particular noise and dynamic range), we have evaluated the total energy release that we will expect in every silicon layers for a single photon bunch. The results are shown in Fig. 178 for a typical bunch made of 105 photons. From the figure we see that the maximum energy release that we could expect in a single layer is 930 MeV for 20 MeV beam energy. To cope with such relatively large signals, we plan to use a fast spectroscopy shaping amplifier with reduced gain. The selection of the preamplifier has not yet been finalized; the proper solution will be found by taking into account both the technical characteristics and the cost. We are currently investigating also the possibility to develop a custom hybrid circuit, properly matched on the necessary technical specifications.

#### 5.4.4.2 Services and Mechanics

As the calorimeter will be in vacuum, a suitable cooling system for the readout electronics must be designed despite the simple open geometry of the system. When the calorimeter is being used, the gamma beam gets absorbed and it is not available for physics measurements. It must be therefore be retractable in order to allow routine operation of the gamma source.

### 5.4.5. Resonant Scattering Calibration system

The accuracy of the beam energy measurement through the calorimetric and Compton techniques requires a calibration of the absolute energy scale at the per mill level. We propose to achieve this level of precision with a Resonant Scattering Calibration System. The idea is to detect the gamma decays of properly selected nuclear levels of some suitable targets, when resonant conditions with the beam energy are achieved. While performing an energy scan of the beam, the resonant energies can be precisely identified through the detection of the corresponding decay photons. Resonance lines are usually very narrow allowing for an extremely precise calibration.

Nuclear resonant scattering is a process consisting in the resonant absorption of a gamma photon by a nucleus, followed by its de-excitation with emission of one or more photons, according to the decay sequence of the nuclear level. In the absence of Doppler broadening, the resonance cross section has a Breit-Wigner shape,

$$\sigma(E) = \pi \left(\frac{\lambda}{2\pi}\right)^2 \frac{2J_1+1}{2J_0+1} \frac{\Gamma_0 \, (\Gamma/2)}{(E-E_r)^2 + (\Gamma/2)^2} \qquad (1)$$

where λ is the wavelength of the incident photon of energy E, Er is the resonance energy, $J_0$ and $J_1$ are the nuclear spins of final state and excited state, respectively, Γ is the total decay width of the excited state, and $\Gamma_0$ is the partial width for electromagnetic (EM) de-excitation to the final state. The angular distribution of the emitted photons is described by a correlation function, W(θ), defined with respect to the direction of the incident photon. In many cases of practical interest, this distribution is roughly isotropic. The resonances in typical nuclei are narrow (widths of order $10^{-3}$ to $10^3$ eV). The resonance energy Er corresponds to isotope-specific energy differences between the final state and excited states of the target nuclei, with generally small corrections accounting for nuclear recoil.

The Doppler broadening of the resonance lineshape can be taken into account by introducing a modified expression [132] of eq. (1):

$$\sigma(E) = \pi^{3/2} \left(\frac{\lambda}{2\pi}\right)^2 \frac{2J_1+1}{2J_0+1} \frac{\Gamma_0}{\Delta} \, e^{\left(\frac{E-E_R}{\Delta}\right)^2} \qquad (2)$$

where $\Delta = E(2kT/Mc^2)$ is the Doppler width, with M mass of the nucleus, while k is the Boltzmann's constant, and T is the absolute temperature of the material.

The expected rate R of resonant γ production at ELI-NP depends on the target thickness T and target density *n* in the following way:

$$R = D \left(1 - e^{T/\lambda}\right) \qquad (3)$$

where D ≥ $10^4$ photons/s·eV is the design ELI-NP spectral density, and λ= 1/σ(E)·n is the radiation length.

As an example, for the decay to ground state of the 15.11 MeV level of $^{12}$C, shown in Table 49 [133], a 1 mm thick target of carbon, with n=$10^{23}$ atoms/cm$^3$, would produce R >$10^3$ γ's per second according to eq.(1), or an even greater value if the Doppler broadening is considered via eq.(3).

Table 49.   15.11 MeV $^{12}$C  level  decay to ground state

| Excited State Energy | Width to Ground State | $J_1$ | $J_0$ |
|---|---|---|---|
| 15.11  MeV | 38.5 eV | 1 | 0 |

Measurement of Nuclear Resonant Scattering may entail either to measure gamma-beam intensity attenuation at the resonant energies or to detect the emitted de-excitation photons. However the widths of most nuclear levels are typically some order of magnitude smaller than the ELI-NP beam bandwidth. This results in a large percentage of incident gammas not accessing the resonant level and does not allow performing easily attenuation measurements. For this reason a resonance monitor is here proposed, aimed to detect de-excitation gammas at large angles. This detector will unambiguously determine the establishment of resonance condition of the gamma beam with some properly selected nuclear levels. In resonance condition the gamma beam energy is known with a resolution simply given by the energy spread of the beam (<0.3% in the case of ELI-NP), since the intrinsic level width as well as the error of its energy value is usually negligible in comparison. This technique allows for a discrete number of beam energy points to be precisely measured, determining an on-site calibration of both the Compton Spectrometer and the Absorption Calorimeter.

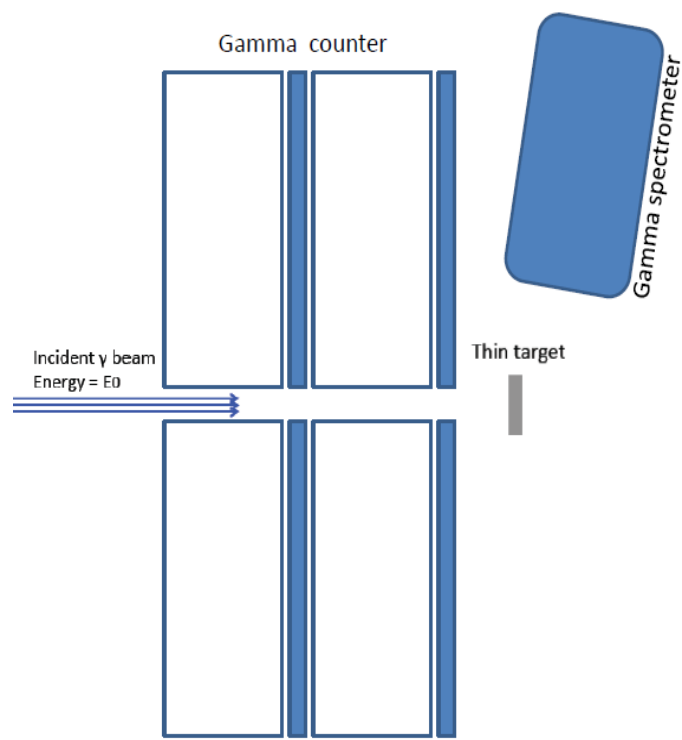

**Fig. 179.  Conceptual scheme of the Resonant Scattering Detector**

The Resonant Scattering Detector (RSD) should optimize all of the following characteristics:

- High efficiency in gamma detection;
- Fast response;
- Suitable granularity;
- Good energy resolution.

High efficiency, both intrinsic and geometrical, is necessary to shorten the time-consuming calibration procedure (consisting in a complete scan of beam energy, at smallest possible steps). Fast response and granularity, on the other hand, allow for pileup reduction. Eventually, an enhanced energy resolution allows unambiguous identification of energy levels: some levels, in particular, show a characteristic decay scheme, resulting in multiple energy peaks whose pattern represent a precise signature.

Energy resolution can, however, be disentangled from the other requirements. So a system made by two detectors is proposed (see Fig. 179): the first detector is a low-cost and fast-response Gamma Counter, covering backward angles with high intrinsic efficiency and suitable granularity, while the second one is a Gamma Spectrometer placed at large angle, reconstructing the energy spectrum produced by the resonant level decay.

The Gamma Counter allows for a fast beam energy scan, giving prompt information about the establishment of resonance condition, while the Gamma Spectrometer, which is a slower detector covering a much smaller solid angle, allows for the later identification of the level.

### 5.4.5.1 The Resonant Gamma Counter

This detector will cover almost the whole backward direction. A schematic view of its conceptual design is shown in Fig. 180. The detector has a 20 mm diameter hole to allow beam crossing, and is segmented in eight sectors housing the gamma counters. The whole system occupies a cylindrical volume of 120 mm diameter and 100 mm length or 50 mm length for the high-energy gamma line or the low-energy gamma line, respectively.

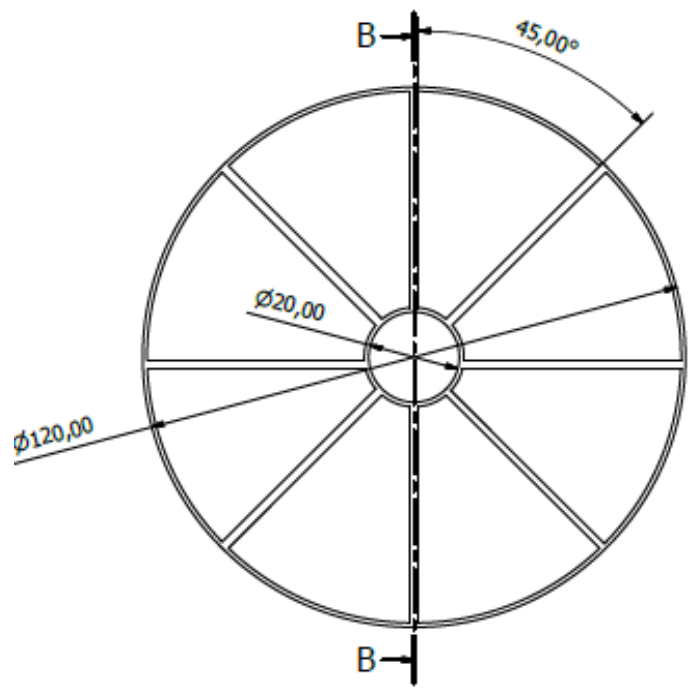

**Fig. 180.** Conceptual scheme of the resonant gamma counter

A possible choice for the eight gamma counters is a sandwich of lead converters and fast plastic scintillators. For the low energy gamma beam line a possible utilization of lead-loaded plastic scintillator will be studied. Attention has to be devoted to detector shielding to avoid problems arising when plastic scintillator are used in vacuum environment.

### 5.4.5.2 The Resonant Gamma Spectrometer

The Resonant Gamma Spectrometer (RGS) detector is aimed to clearly identify resonant excited levels, in particular when a complex decay pattern is expected as in the case of the 15.11 MeV level of 12C reported in Table 49. Reconstructed decay pattern may be used as calibration candles, thus allowing also allow a better identification of adjacent levels. A good energy resolution and an anti-Compton shield for background rejection are necessary to distinguish the peaks in the pattern.

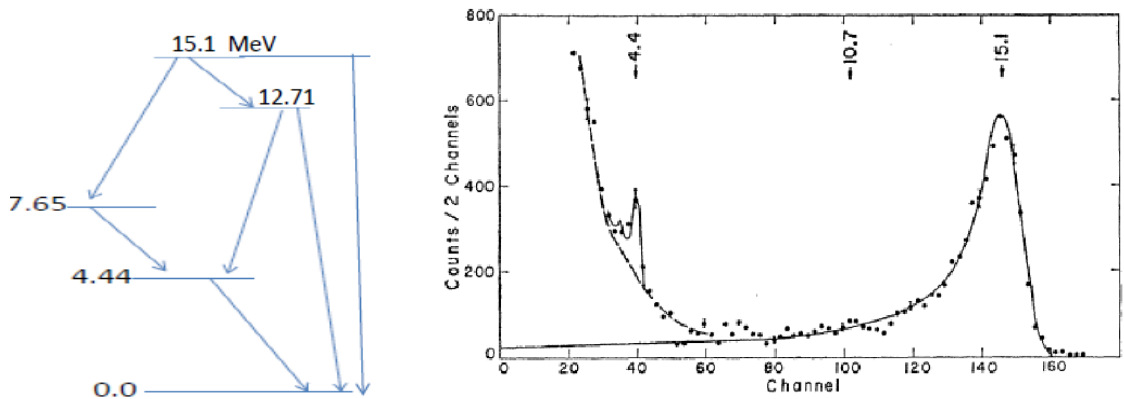

**Fig. 181.** Resonant spectrum of the 15.11 MeV level of $^{12}$C

In Fig. 181 [134] an experimental resonant spectrum of the 15.11 MeV level of $^{12}$C measured with a NaI detector is reported. It shows the difficulty to reconstruct the full peak pattern in the absence of an anti-Compton shield. For the low-energy line the Resonant Gamma Spectrometer would be a Ge(Li) detector, 30 mm diameter and 45 mm length, positioned just outside the vacuum chamber, at about 20 cm from the target. For the high-energy line a much larger detector is needed, while nuclei with lower density of levels will be used. So a less demanding energy resolution is needed and a NaI detector, 80 mm diameter and 150 mm long, will be used.

## 5.5. Expected Performances

### 5.5.1. Beam Intensity Profile

The performance of the imager has been evaluated estimating the signal produced by the light sensor of the device.

The average number of electrons produced in a pixel (20 µm x 20 µm) of a CCD coupled with s Gadox phosphor with a lens providing a magnification M = 1 and an image area surface up to 10 x 10 mm2 are reported in Table 50 (see appendix D for a detailed description of the method used to obtain these results).

**Table 50. Average number of electrons per CCD pixel per macro-pulse for two different beam divergence**

| Beam Energy [MeV] | 1.0 | 5.0 | 10.0 | 15.0 | 20.0 |
|---|---|---|---|---|---|
| average signal per CCD pixel ( 25 µrad) | 8.50E+02 | 5.44E+02 | 5.83E+002 | 6.18E+02 | 6.68E+02 |
| average signal per CCD pixel (250 µrad) | 8.50E+00 | 5.44E+00 | 5.83E+000 | 6.18E+00 | 6.68E+00 |

For a standard cooled CCD camera the typical dark current is <1 e-/pixel/s while the readout noise is in the range 20-30 e-/pixel: from these data the average signal for macro-pulse is larger than the readout noise signal for 25 µrad beam divergence, while it is lower than readout noise for 250 µrad beam divergence. To increase the lower signal it is possible to use the CCD binning modality that increases the pixel size summing the signal of neighbour pixels, or summing the signal of several macro-pulses.

### 5.5.2. Gamma Energy Spectrum from Compton Spectrometer

To evaluate the performance of the proposed Compton spectrometer a detailed simulation of the detector has been developed using Geant4, according to the layout resulting from the study of Section 5.4.2. As described in detail there, the spectrometer consists of a micrometric graphite target, a detector to measure the energy and position of the Compton scattered electron and a movable photon detector to be placed around the expected position of the recoil gamma that depends on the electron detection angle and on the beam energy.

Three different beam energies were simulated (5, 10 and 20 MeV), each simulation consisting of $3 \times 10^{10}$ events. The beam starts 25 m upstream the Compton spectrometer target with a x-y emittance ε = 20 µm x 40 µrad and is directed onto the target. In case of an interaction, the resulting particles are tracked along the detector and their identities, their positions together with the deposited energy in the different crossed volumes are recorded. A more complete description of the Monte Carlo simulation is reported in appendix B.

The number of interactions in the target per incident gamma as a function of the beam energy and type of interaction are reported in Table 51. As expected, with increasing beam energy the Compton rate decreases while pair production increases.

**Table 51. Probability of interaction in the target for the beam photon as a function of the beam energy and interaction type**

| Beam Energy (MeV) | 5 | 10 | 20 |
|---|---|---|---|
| Compton | 7.16 x 10$^{-6}$ | 4.44 x 10$^{-6}$ | 2.62 x 10$^{-6}$ |
| Conversions | 5.91 x 10$^{-7}$ | 1.20 x 10$^{-6}$ | 1.91 x 10$^{-6}$ |

*Nota: The graphite target has a depth of 1.7 μm.*

A fraction of these events will be recorded by the electron/recoil gamma detectors. The event rates in the electron detector are reported in Table 52 as a function of the beam energy, before and after selecting a matching signal in the Si strip and photon detectors.

**Table 52. Average number of vents rates in the electron detector per incident photon as a function of the beam energy and cut type**

| Beam Energy (MeV) | 5 | 10 | 20 |
|---|---|---|---|
| All | 3.31 x 10$^{-8}$ | 1.25 x 10$^{-7}$ | 2.49 x 10$^{-7}$ |
| 1Hit | 2.11 x 10$^{-8}$ | 0.81 x 10$^{-7}$ | 1.36 x 10$^{-7}$ |
| Signal | 1.72 x 10$^{-8}$ | 0.47 x 10$^{-8}$ | 0.39 x 10$^{-7}$ |

*Nota: "All" refers to all the energy depositions recorded. "1Hit" is the rate obtained when requesting 1 hit in the Si strip detector inside the fiducial volume defined by the collimator hole. Finally "Signal" is the rate corresponding to the additional request of a coincidence signal inside the gamma detector.*

Table 53 shows the effect of the required coincidences on the signal purity for 10 MeV beam energy. The request of a hit in the Si strip strongly reduces the events due to a Compton photon, while the detection of a recoil gamma in coincidence suppresses the background due to conversions inside the target. Photons surviving the selection are always accompanied by a good Compton electron, and are produced by electron interactions in the Si detector or at the border of the collimator. For all the simulated energies more than 99% of the selected events contain an electron generated by a Compton interaction in the target.

**Table 53. Recorded particle type as a function of the selection cuts for a beam energy of 10 MeV**

|  | e-(%) | e+(%) | γ(%) |
|---|---|---|---|
| All | 60.1 | 14.4 | 25.5 |
| 1Hit | 78.2 | 16.9 | 4.9 |
| Signal | 94.9 | 0.1 | 5.0 |

The ELI beam energy is obtained by measuring the electron scattering angle and its energy according to the formula:

$$E_\gamma = \frac{m_e}{\sqrt{\cos^2(\varphi)(1 + 2m_e/T_e)} - 1}$$

Where $m_e$ is the electron mass, $\varphi$ is the scattering angle and $T_e$ is the kinetic energy. The $\varphi$ angle is precisely measured by the Si strip detector.

The electron energy is measured in the HPGe detector, adding up the signal recorded in its four (two for the low energy line) elements. A tiny fraction of the electron energy is lost in passive components of the detector, namely the Be entrance window and the passive Ge layer providing the electrical contact in each HPGe detector.

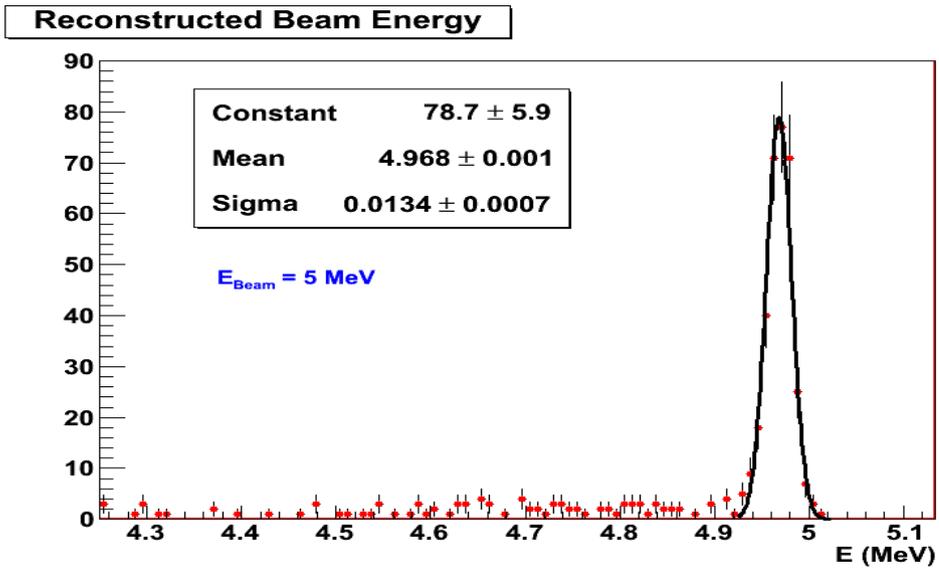

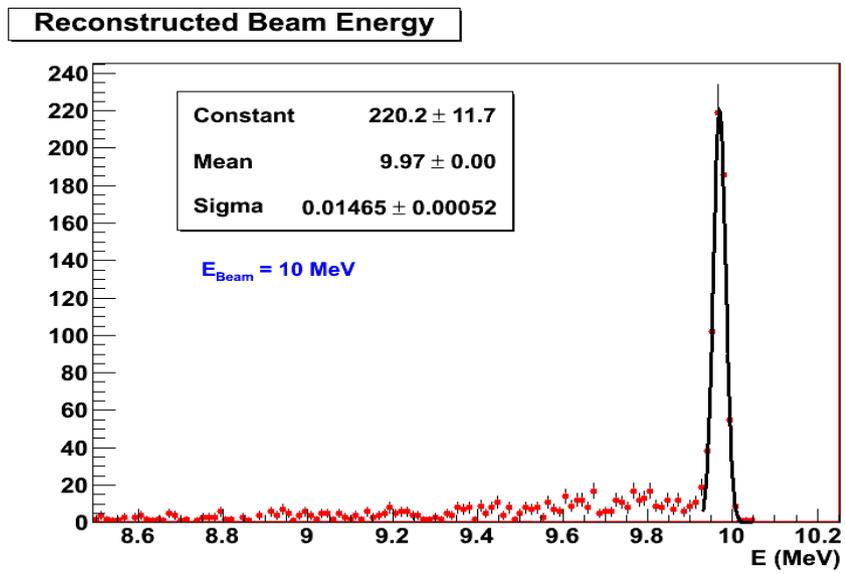

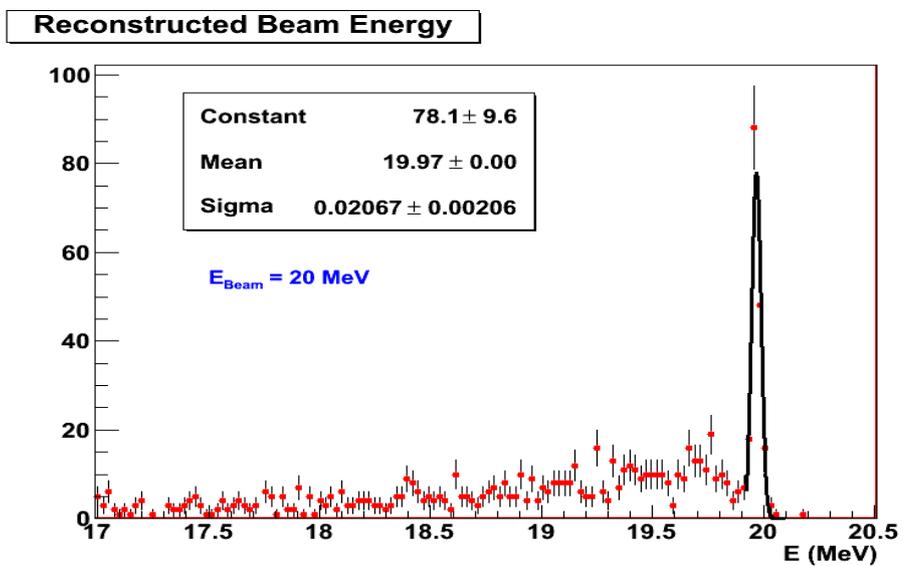

**Fig. 182.** Reconstructed beam energy by the Compton spectrometer for a simulated energy of 5, 10 and 20 MeV

The distribution of the reconstructed beam energy is shown in Fig. 182. To evaluate the expected resolution, the fill energy peak is fitted with a gaussian function. As shown in Fig. 182, the fitted mean values are shifted from the true values by about 30 keV, as expected from the average value of energy lost in the passive materials. The resulting resolution varies from 2 to 1.5 per mill from 5 to 20 MeV. The residual tails in the beam energy reconstruction can be further suppressed using the information from the impact position of the recoil gamma.

In conclusion, the simulations show that a clean sample of well reconstructed Compton interactions can be selected using the Compton Spectrometer, providing the beam energy with a resolution between 0.15 and 0.2 %. The expected number of useful signals per incident photons corresponds, for the nominal beam flux, to a rate of a few Hz for the whole range of beam energy. The spectrometer is thus able to provide a continuous monitoring of the beam energy during the routine operations of the ELI-NP facility with the required accuracy. The resolution function obtained from the simulations with a fixed beam energy, that can be verified using sources of monochromatic gammas, can be deconvoluted from the measured distribution. Therefore, after an adequate number of measurements, an accurate reconstruction of the time integrated energy spectrum of the ELI-NP gamma beam can be achieved.

### 5.5.3. Beam Energy and Intensity from Calorimeter

The optimized layout of the calorimeter described in Section 5.4.4, consisting of 25 elements made of 3 cm of plastic absorber and 0.2 mm Si planes, has been simulated in detail using GEANT4. The response has been studied simulating monochromatic photon pulses with energy between 2 and 20 MeV in steps of 0.1 MeV. For each energy value, the number of simulated events is $10^7$ (corresponding approximately to the total number of photons in a macro-pulse). The fraction of deposited energy in each layer with respect to the sum of the 25 signals is then parameterized as a function of energy, as shown in Fig. 183, as well as the fraction of sampled energy with respect to the total beam energy. Using this parameterization, the beam energy and its intensity are simultaneously obtained from a fit to the longitudinal energy profile. An example can be seen in Fig. 184 for $10^5$ photons (corresponding to less than one single nominal pulse) of 10 MeV.

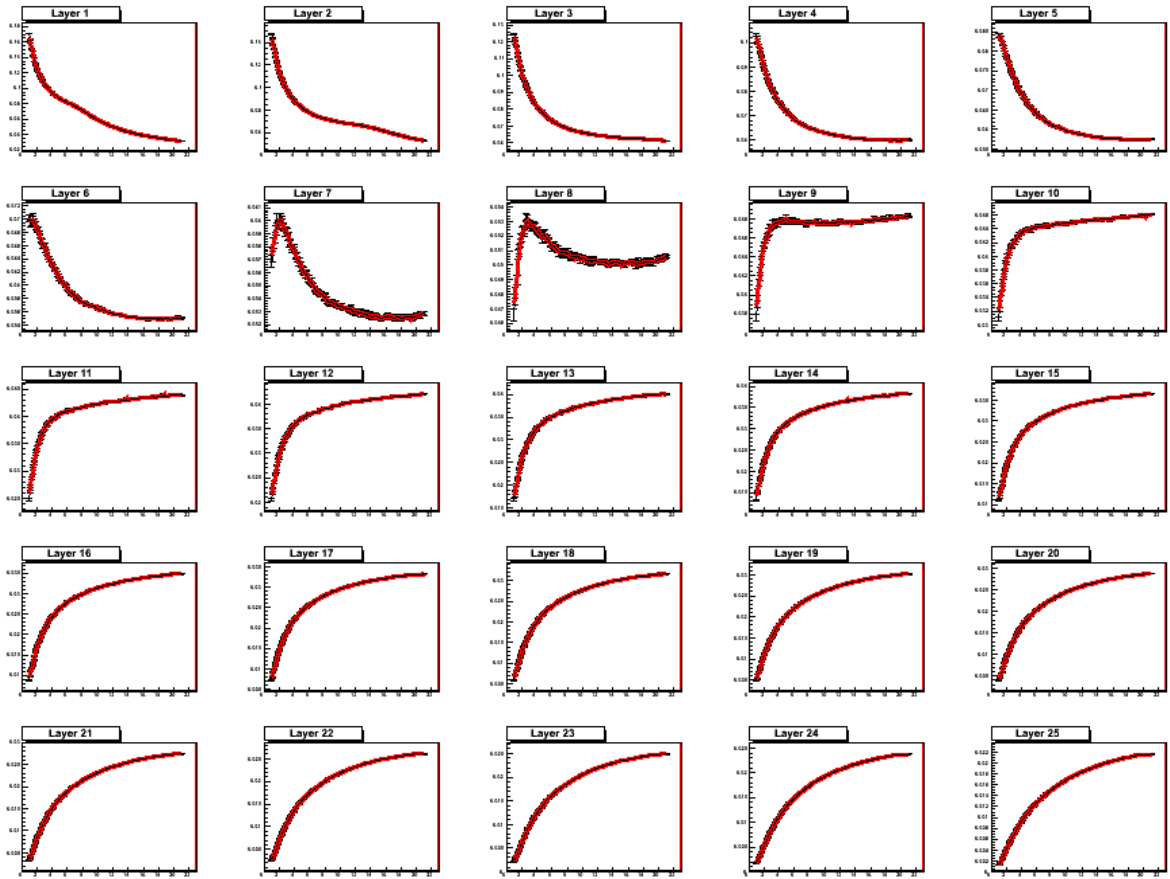

**Fig. 183.    Parameterization of the absorbed energy profile: the signal fraction in each of the 25 layers is shown as a function of gamma energy. The parameterized curves (in red) are obtained by smoothing the average values measured in simulated events.**

The energy resolution obtained with this method is shown in Fig. 185 as a function of the incident beam energy for a single gamma beam pulse of $10^5$ photons. The expected resolution is at the level of (1.6 – 3.2)% in the energy range of interest. The intensity is measured with similar precision, since the error on the beam energy is the main limitation to its accuracy. The statistical error on the energy measurement is thus well below 0.1% after integrating over a few seconds, and, assuming the macro-pulse reproducibility on this time scale, a detailed time profile of the macro-pulse can be obtained.

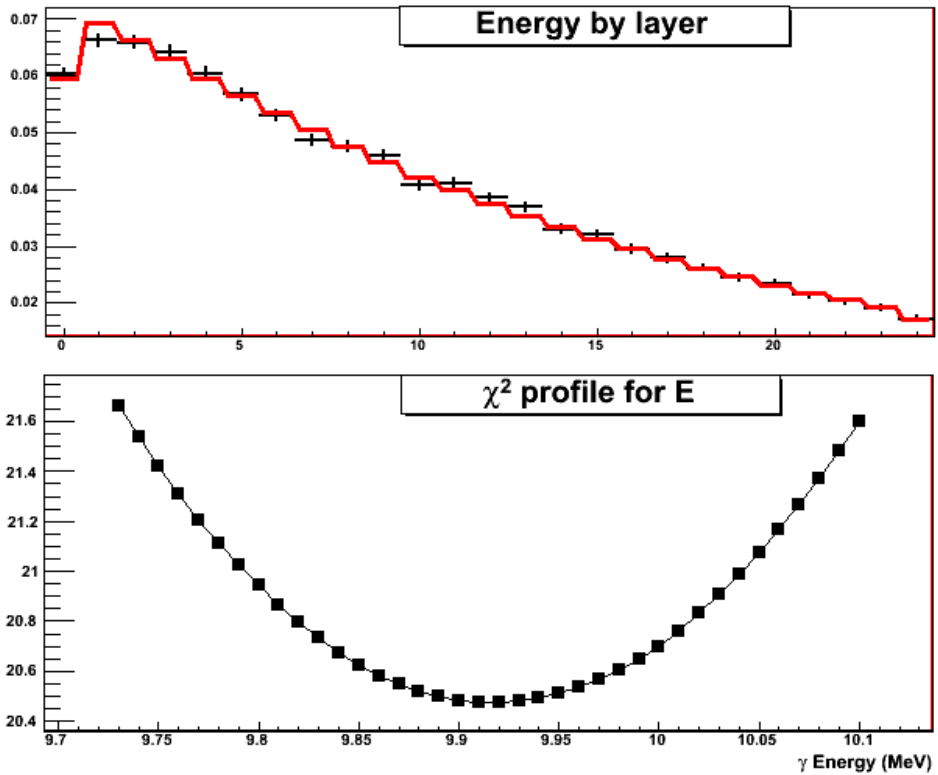

**Fig. 184.** Example of energy reconstruction from the longitudinal profile measured from a pulse of 10⁵ photons. The measured signal fractions, shown in black on the top plot, are fitted against the expected profile. The lower plot shows the fit χ2 scan as a function of the beam energy. The best fit in this case, shown in red, gives an estimation of (9.91 ± 0.18) MeV for a true energy of 10 MeV.

Systematic errors on the energy measurement can be induced by an incorrect modeling of the energy dependence of longitudinal profiles and by a miscalibration of the detector layers. The electromagnetic processes driving the gamma beam absorption are known very precisely and modeled in full detail within the GEANT4 framework. Errors on the effective sampling fraction of the active detectors, that is sensitive to the low energy cut in the simulation tracking, is common to all the 25 calorimeter layers, that we choose on purpose to be identical, and is thus not expected to affect the profile measurement.

The limiting systematic uncertainty is thus anticipated from differences in the response of the different layers, mostly due to a possible unaccounted non-linearity of the readout chain or a relative miscalibration of the readout channels.

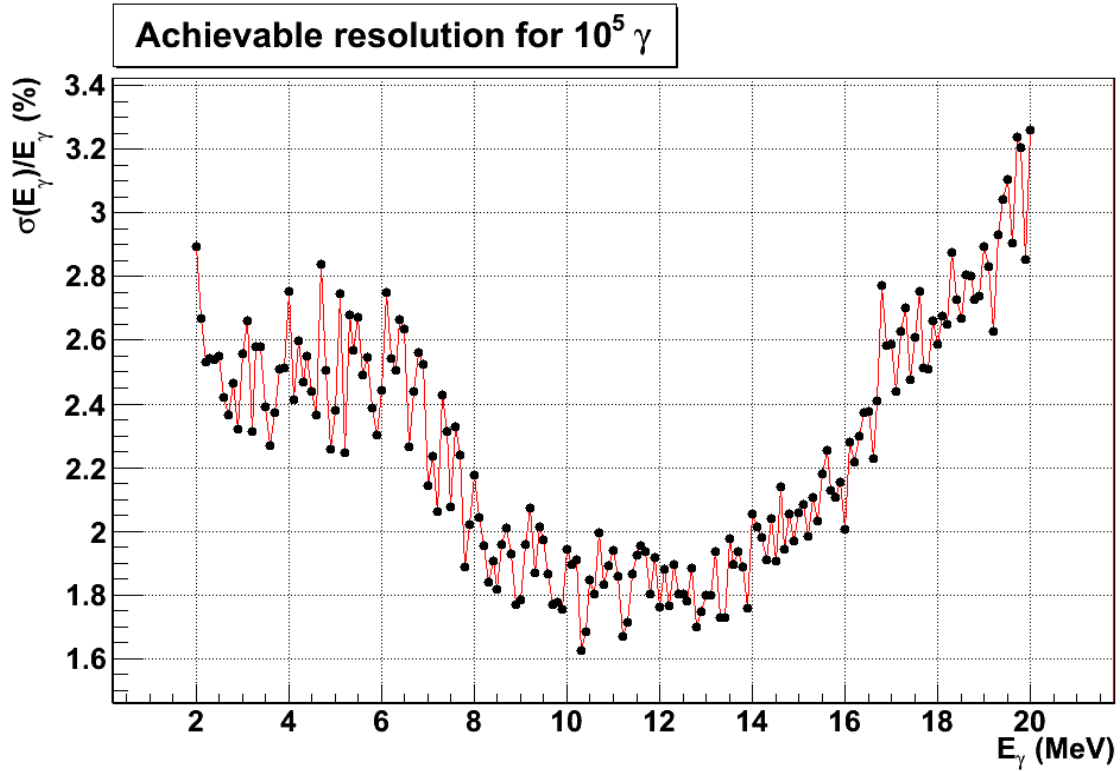

**Fig. 185.** Resolution of the calorimetric measurement of the beam energy expected from $10^5$ gammas. Only the statistical errors due to the fluctuations of longitudinal profiles are taken into account in this plot.

The effect of a non-linearity factor α, defined as

$$E_{rec} = E_{true} \cdot \left(1 + \alpha \frac{E_{rec}}{E_{max}}\right)$$

where $E_{max}$ ~ 10 GeV is the range of the readout chain, is estimated using the simulated events. To limit the resulting effect on the beam energy reconstruction below 0.2%, |α| is required to be kept below 0.5 %, a value achievable with a proper characterization of the readout electronics.

Using the simulated events, we also evaluated the effect on the beam energy reconstruction of a spread of 1 % (rms) in the calibration constants, a level which is easily attainable without an external calibration source. For each beam energy the same simulated event sample was analyzed multiple times after applying random biases to the response of each detector. The resulting spread on the measured beam energy goes from 0.5% at 2 MeV to 1% at 20 MeV.  It is therefore mandatory to reduce intercalibration errors to < 0.3%. We expect to reach this level of accuracy thanks to the resonant scattering calibration system that will provide very accurately the beam energy for several reference values, for which the measured profiles can be tuned to the shape expected from the simulation.

## 5.6. Integration with the machine

A representation of the entire collimation and characterization system of the gamma beam is shown in Fig. 186. It consists of three sections:

- **Gamma collimation**

  The gamma collimation section include a vacuum chamber (80 cm length) containing the collimating slits and mechanics. The chamber have to provide a vacuum level that can ranges from $10^{-7}$ to $10^{-5}$ mbar and needs to be interfaced with the beam pipe line by mean of two bellows, so as to permit X-Z-θ-φ movements for the alignment of the collimators. Downstream the collimation system a concrete block will be placed to absorb the secondary radiation produced in the interaction of the primary beam with the metal slits and pipes.

- **Gamma spectrometer**

  Gamma spectrometer consists of three different chambers providing a vacuum level of $10^{-7}$ -$10^{-5}$ mbar. The first one (150 cm length and diameter 60 cm) contains the target and gamma counter of the nuclear resonant spectrometer and the target and the Compton scattered photons detector mounted on a linear translator having a stroke of about 120 cm. Through an Al or Be window a nuclear gamma fluorescence detector is fixed to the chamber wall.

  The second and third chamber (max 180 cm length and diameter 20 cm and 80 cm length and diameter 20 cm respectively) are both connected off axes to the previous one with bellows and contain respectively the Compton electron detectors and its collimation system and the calorimeter container mounted on a shifting system.

- **Gamma beam profile imager**

  This section consists of a cylindrical chambers (20 cm high and diameter 40 cm) proving a vacuum level of $10^{-7}$ -$10^{-5}$ mbar, with flanges to connect motor screen for detector positioning, optical lens system and CCD detector.

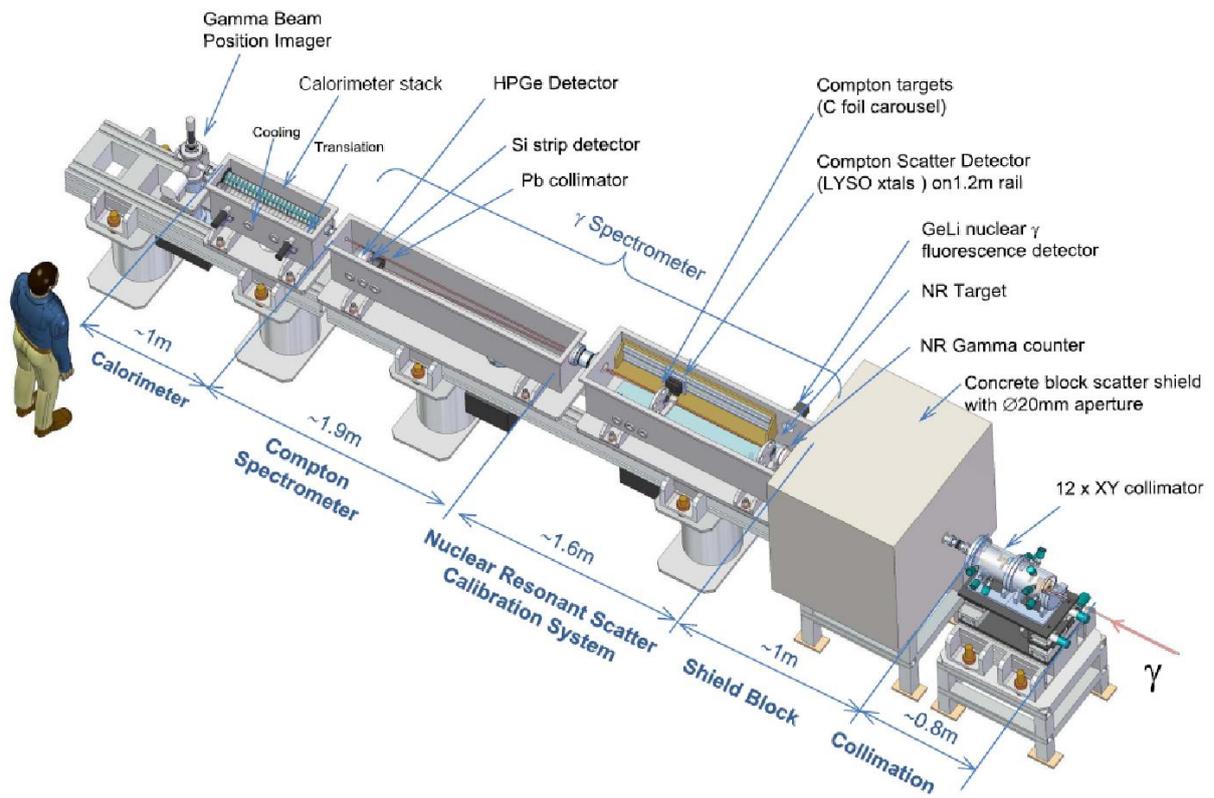

**Fig. 186.**   Drawing of the collimation and characterization system of the low-energy gamma beam line

# 6. Conventional Facilities

## 6.1. Building Layout & Infrastructure

### 6.1.1. General Overview

The building design plans for ELI-NP facility are effectively complete to a high level of detail. At the time of writing (early 2013) the details for the building contract are not in the public domain and enabling works have not yet begun. However, it is envisaged that there will be no opportunities to effect any significant changes to the plans based on accelerator design requirements and if there are any concessions these will only be permitted providing they have negligible impact on costs.

In this regard the technical design for the accelerator proposed here is unusual in that it must fit into a prescribed building layout with services infrastructure that have been devised without any specific reference to the accelerator in any particular way. Some of the challenges in matching a suitable lattice and engineering layout to fit within the predefined building layout are described in section 3.9.

Fig. 187 shows a 3D site illustration along with structural visualisations adapted from the building technical documentation which give a basic appreciation of the building form.

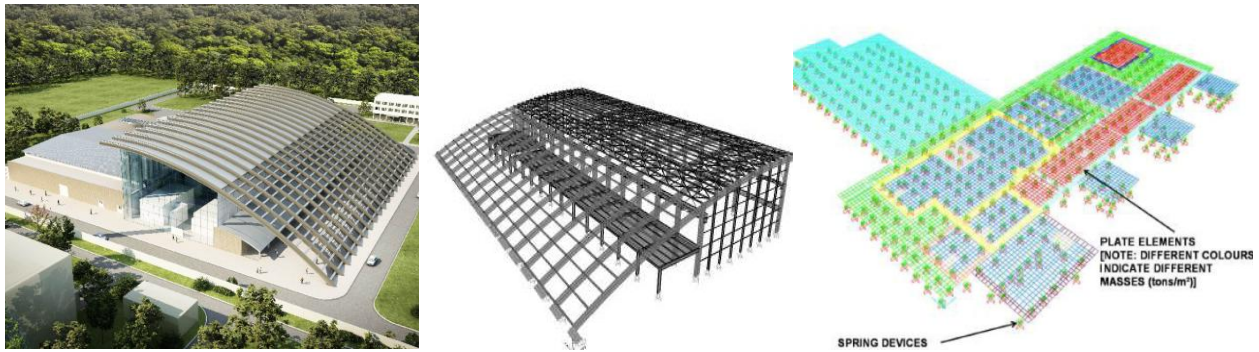

**Fig. 187.** Building Layout - left to right (a) 3D site visualisation; (b) Structural steelwork; (c) Anti-vibration support network in basement

Note that for the ELI-NP-GS system only the accelerator, experimental and technical areas are relevant so building layouts shown from herein do not include the high power laser areas. Fig. 187 highlights an unusual feature of the building design – not found in other such facilities – in that the entire accelerator, laser and experimental floors comprise a 1.5m thick concrete floor supported on a network of anti-vibration pad or building isolators. The support pad 'network' is also visible in side-elevation section in Fig7.3 below which shows 'T' pillar supports contained in a 3m height basement area.

At the time of writing we do not have information of the composition of the blocks that form the isolator pads or feedback regarding concerns over settlement or vibration transmissibility through such a floor. Some of these stability issues are addressed in section 3.9.

Fig. 188 shows the plan view of accelerator hall in the context of the wider building layout (without high power laser areas) with internal shielding concrete walls of the accelerator hall and various experimental rooms can clearly be seen in the plan view and also in the illustrative 3D isometric view of Fig. 190.

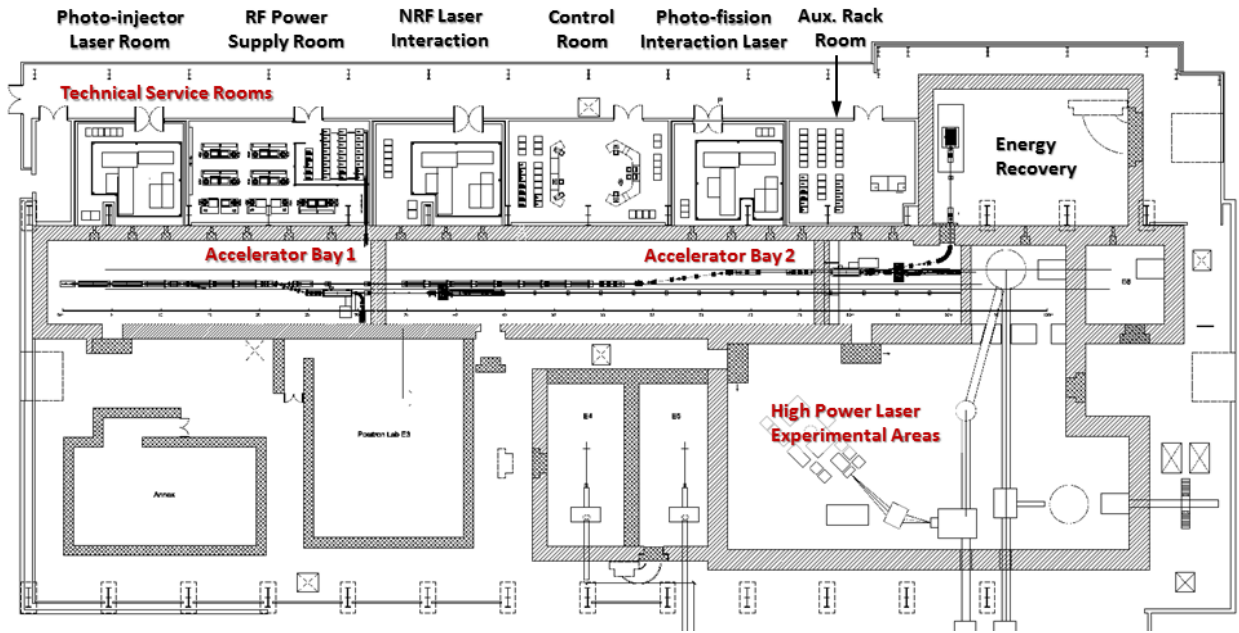

**Fig. 188.** Building Layout of the Accelerator Hall & Experimental Areas – Plan

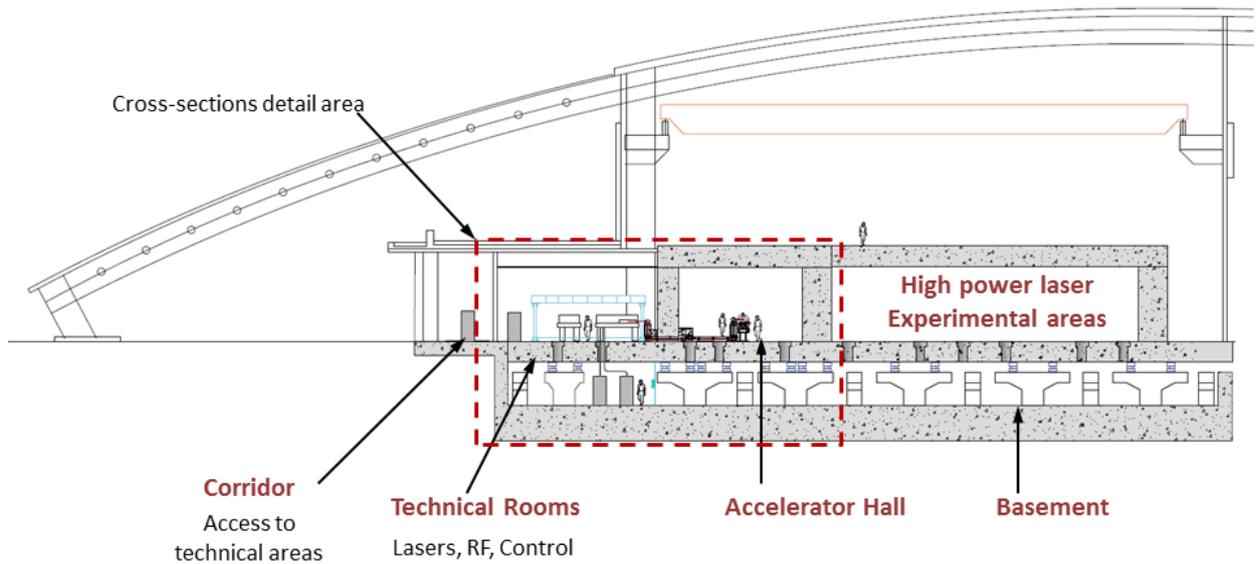

**Fig. 189.** Building Layout of the Accelerator Hall & Experimental Areas – Elevation Section

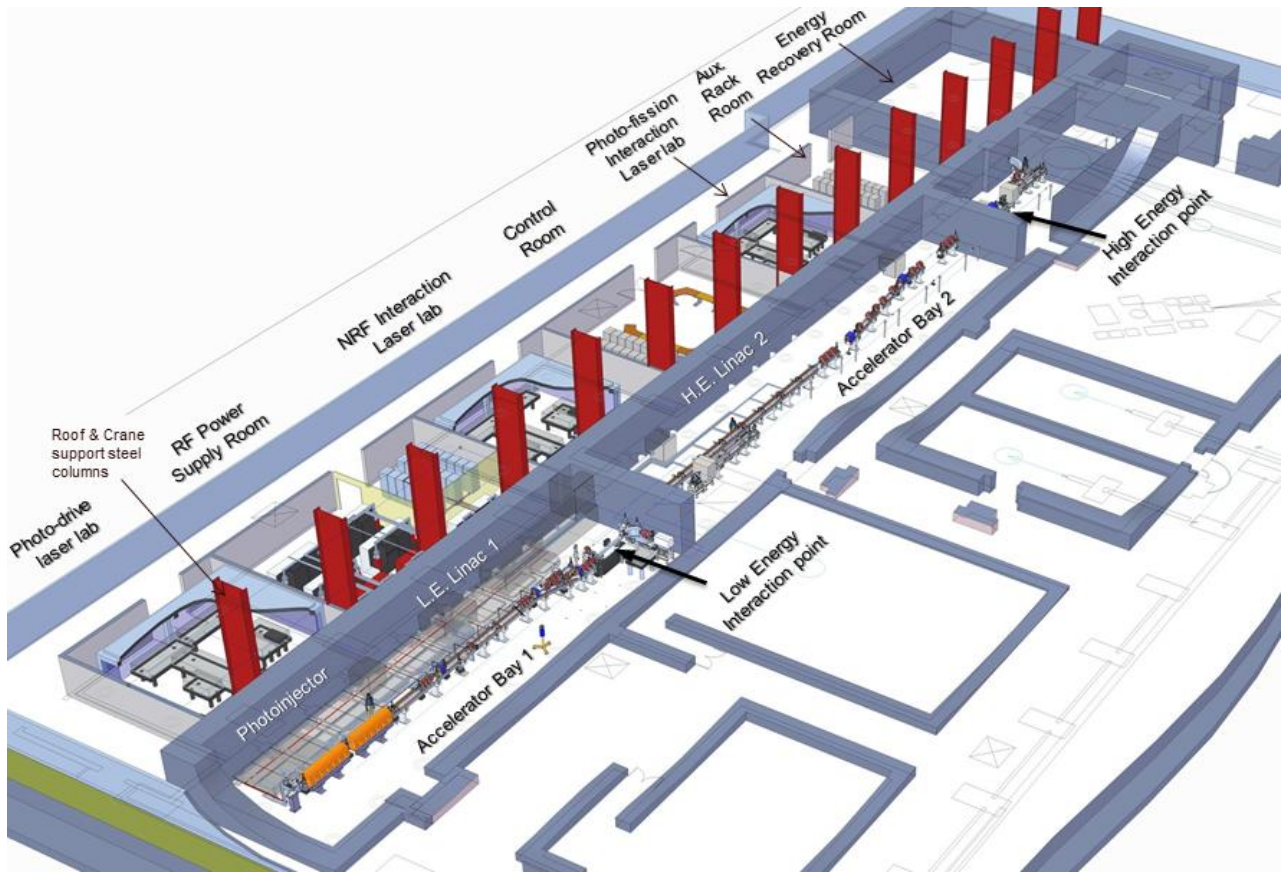

**Fig. 190.    Isometric 3D view of Building Layout of the Accelerator Hall & Experimental Areas**

It is important to note that in comparison to the ELI Whitebook and the building and accelerator tender we have assumed no significant change to the technical room allocation other than to provisionally assume a 0.5m extension of the room into the corridor. This moderate extension allows for idealised placement of all RF power supplies and a substantial rack enclosure into a single RF room #1 – reducing overall lengths of cable and waveguide runs. The RF room originally designated as RF Room #2 now acts as an auxiliary rack room for equipment far downstream with substantial space remaining for additional RF supplies should a higher energy accelerator (beyond the scope of this specification) ever be realized. This extension request is not a prerequisite and the detail design phase will determine whether it is of value in pursuing.  If the extension is not permitted under the building contract then alternative options exist that will slightly reconfigure the technical rooms from what is shown in this TDR with possibly some longer waveguide and cable runs being required.

Another option on building construction – that is permitted under the accelerator contract – and that we have assumed as prerequisite, is to move the shield walls between accelerator bays by up to ±1m. Thus, we have assumed that the end wall of accelerator bay 1 moves 1m downstream from nominal and the end wall of accelerator bay 2 moves upstream by 1m. All the drawing layouts shown in this TDR make this assumption implicitly.

### 6.1.2. Accelerator Hall

#### 6.1.2.1 General

Fig. 191 depicts a closer in plan view of the proposed 90m length of the ELI-NP-GS facility within the accelerator hall with more of the salient features are now visible.

Note that accelerator has been designed so that layout of the low and high energy gamma lines coincide with along the centrelines predetermined in the building layouts. The present design layout indicates the photo-injector line approximately central in accelerator hall resulting in ~4m space either side. Looking downstream, on the left hand facing side of the accelerator, the floor is reserved for services such as RF waveguides and photon transports emerging from the technical rooms, whilst on the right hand side it is proposed to keep the floorspace entirely clear for equipment transport access.

The services on the left side will be covered by treadplate flooring forming a walkway running the entire hall length at approx 0.4m height above the floor. The arrangement is similar to the SPARC facility at INFN Frascati. Fig. 192 also depicts the waveguides at floor level in a 3D visualisation with the flooring partially transparent for visualisation.

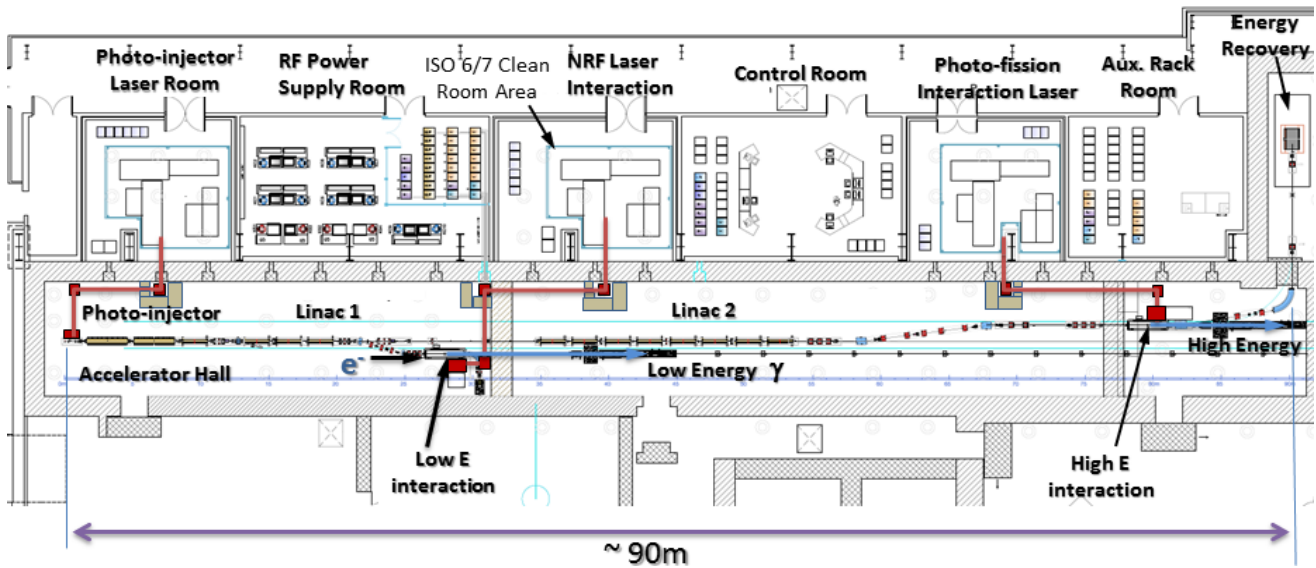

**Fig. 191.** Accelerator Hall and Technical Rooms Indicating Proposed Photon Transports (red)

#### 6.1.2.2 Low Energy Linac

The low energy linac1 has been described previously. The layout is apparent in Fig. 192.

The 'dog-leg' branch feeds into the Compton interaction point in the laser recirculator (module M9) where the low energy gamma beam is generated. Immediately beyond the laser interaction point the electron beam is directed horizontally through a 90° bending dipole to a beam dump fully contained within accelerator bay 1 shield wall.

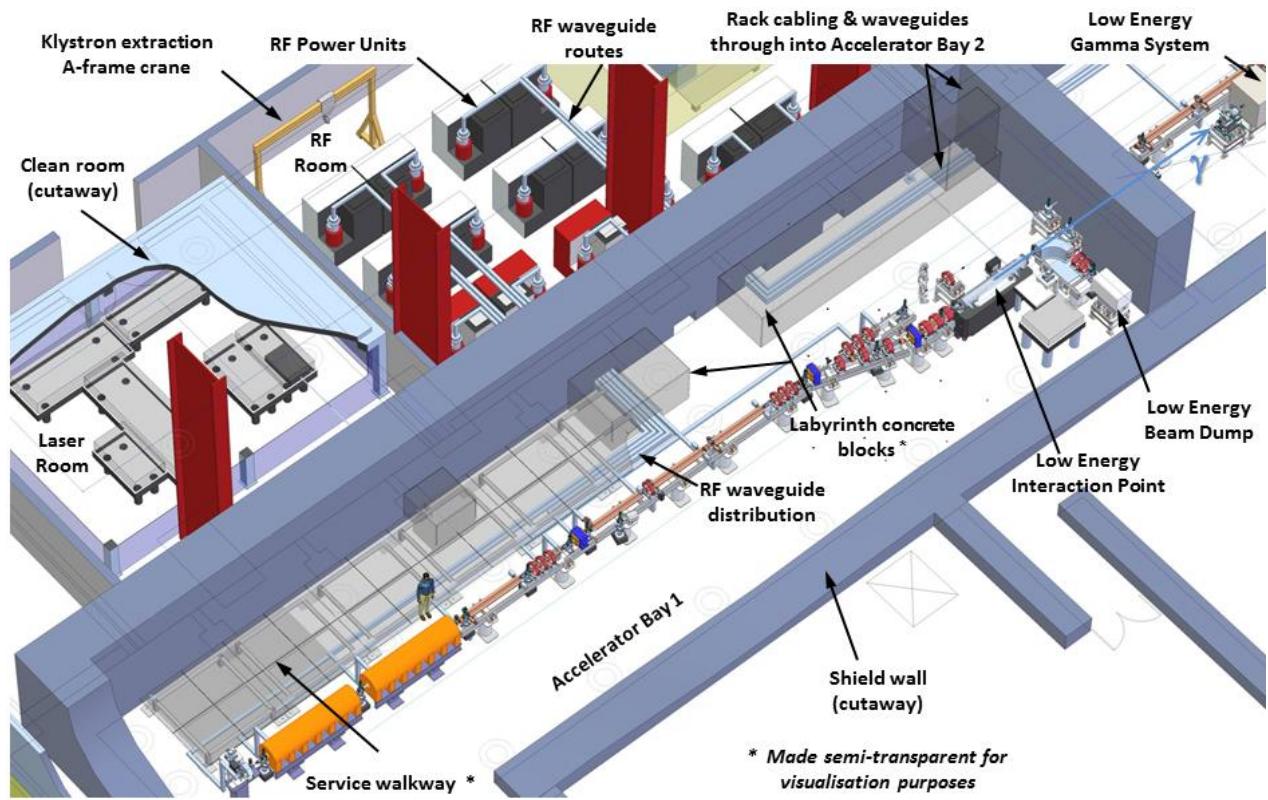

**Fig. 192.** 3D illustration showing low energy linac & proposed waveguide routes in Accelerator Bay 1

The compact area containing the laser recirculator, bending dipole and LE beam dump is shown pictorially in Fig. 193 with the downstream shield wall removed for clarity

The gamma beam propagates through the dump dipole magnet and into accelerator bay 2 via a small hole that can be cast into (or diamond drilled later through) the shield wall. The hole can be sealed via a PSS interlocked radiation shutter. Thus accelerator bay 1 and 2 can be separated radiologically and run in independent operation.

The gamma beam is conditioned and its properties measured and monitored by the gamma system installation. This consists of a collimator, spectrometer & calorimeter set that will confirm that the gamma beam produced meets the operational specification parameters of the facility. This is located ~10m beyond the interaction point (to maximize resolution) downstream of the shield wall in accelerator bay 2. It is described more fully in chapter 6 of this TDR and the low energy Gamma system can be seen in Fig. 194. The high energy Gamma system is essentially a copy of the LE system and detail is more apparent in close up of Fig. 195. Beyond the LE gamma system the beam will be propagated >60m to the experimental areas via beampipe – the supply and scope of which is not part of this TDR.

As described in chapter 3, the low energy (L.E.) line is highly constrained due to the fixed shield wall position. The design of the low energy beam dump is heavily influenced by these considerations and the number of siting options within the space available are very limited. A 1.2Tesla dipole bending magnet will result in a full 90o turn of the 280MeV electron beam in a short span (<0.8m radius) leaving an approximate 0.9m space before the shield wall – sufficient to fit the beam dump and local shielding.

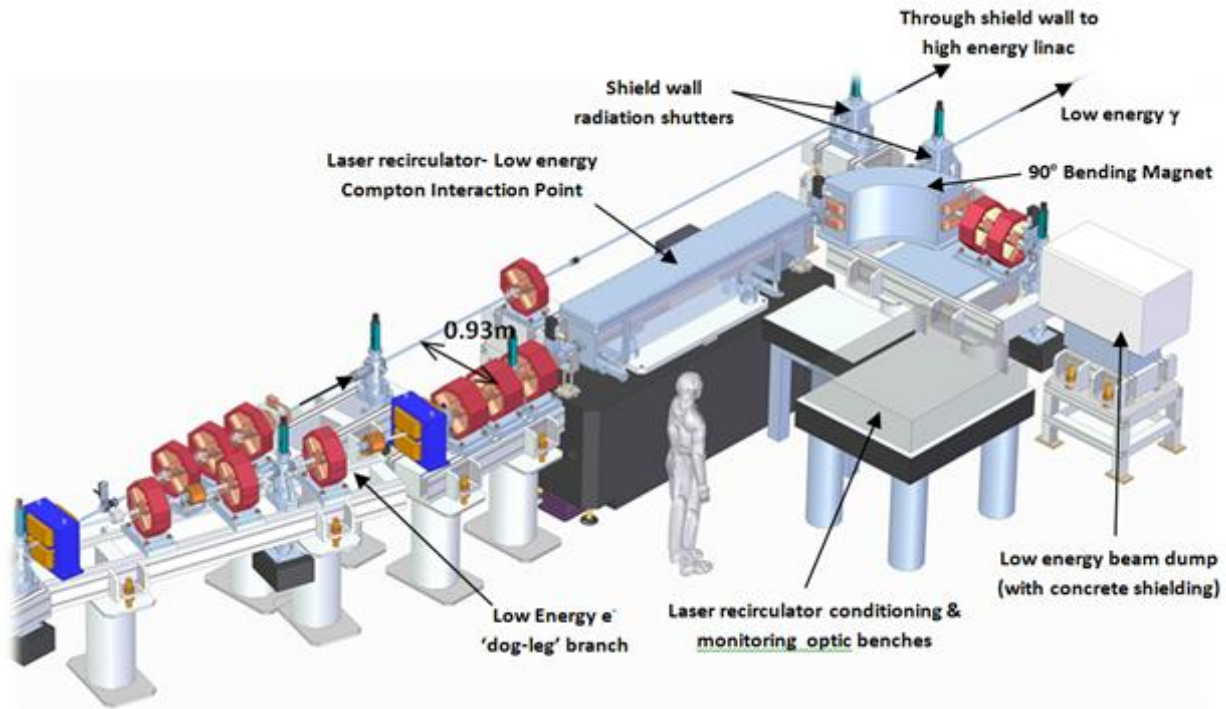

**Fig. 193.** Low Energy Interaction & Electron Beam Dump Area

It is likely that the dump may require some moderate cooling water as it will be designed for ~300Watt average heatload but calculations in the detail phase will show whether such services will be required or convective air cooling is sufficient.

The current design of the beam dump consists of a simple steel block 0.1m in width, surrounded by 0.5m concrete at the rear (and less at the sides). This shielding is derived from basic simple linear scatter calculations which predict sufficiently low dose rates beyond the shield wall with the beam at full energy and maximum 100Hz repetition rate. Further, more sophisticated, calculation and simulation work will be done to confirm this during detail design phase.

Note that in the event of failure of the power to the dump dipole we anticipate a failsafe mode to prevent the electron beam travelling downstream to the experimental area; a permanent magnet backstop be located beyond the positron area & upstream of the gamma collimator and will deflect the electrons to a local 'emergency' dump. This dump would be a simple uncooled mass designed to accept the beam momentarily. A hall-effect probe on the dipole would sense the magnet field and be interlocked to the machine to close the linac down quickly in the event of dipole magnet power failure.

#### 6.1.2.3 High Energy Linac & Gamma Systems

The high energy linac runs in Accelerator bay 2 (from ~ 33m to 76m from the gun) and is depicted in 3D isometric view in Fig. 194. No space restrictions exist as they did for LE linac and in fact there is sufficient space in the design to accommodate at least a further two accelerating RF structures should the energy require boosting as an upgrade option in the future.

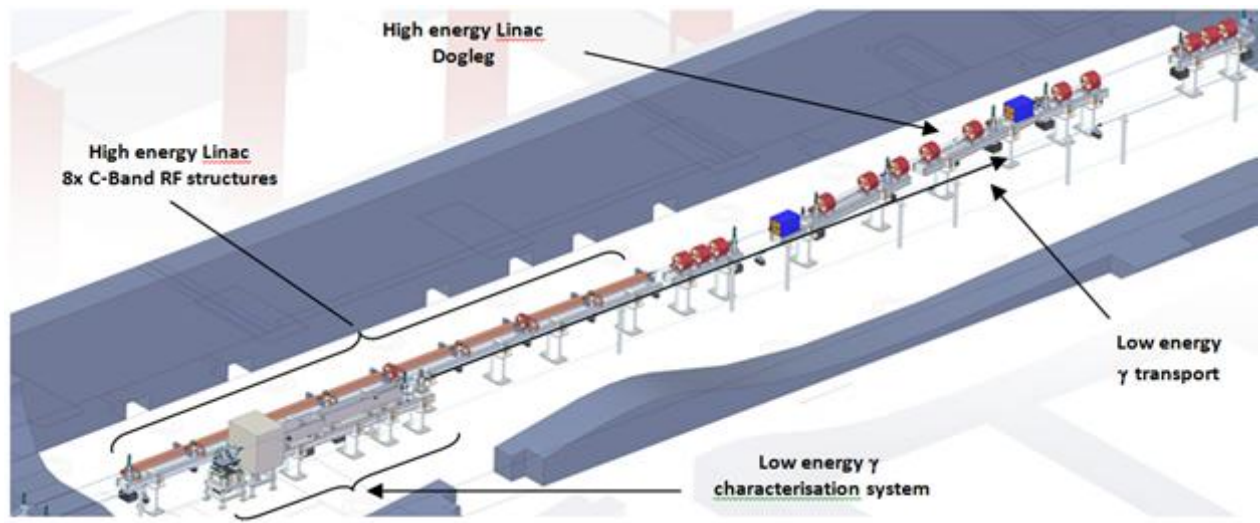

Fig. 194.    High Energy Linac & Low Energy Gamma System – Accelerator Bay 2

Downstream of 54m the full high energy beam is dog-legged horizontally by ~1.2m so as to more closely co-axially match the lines into the experimental areas as defined in the tender. In fact this offset is still ~0.3m short of the drawing layout as depicted in the tender. This slightly reduced offset is required on account of the relatively short space available in the end experimental area to house the Laser recirculator and Gamma system.

With an offset of 1.2m from straight a gap of 2.7m exists from the dogleg line to the energy recovery shield wall – just sufficient transverse space to accommodate the 9° angled line followed by the 81° bending magnet dipole into the HE energy recovery room. To increase the offset to the required 1.5m would require some considerable redesign and close re-examination of the magnet lattice.

The high energy gamma system is much closer to the interaction point source than in the LE case. This is because of the fixed building design which greatly restricts the space available in the end room. As a consequence the tolerances on the resolution and accuracy of the collimating slits will be more demanding for the HE Gamma system.

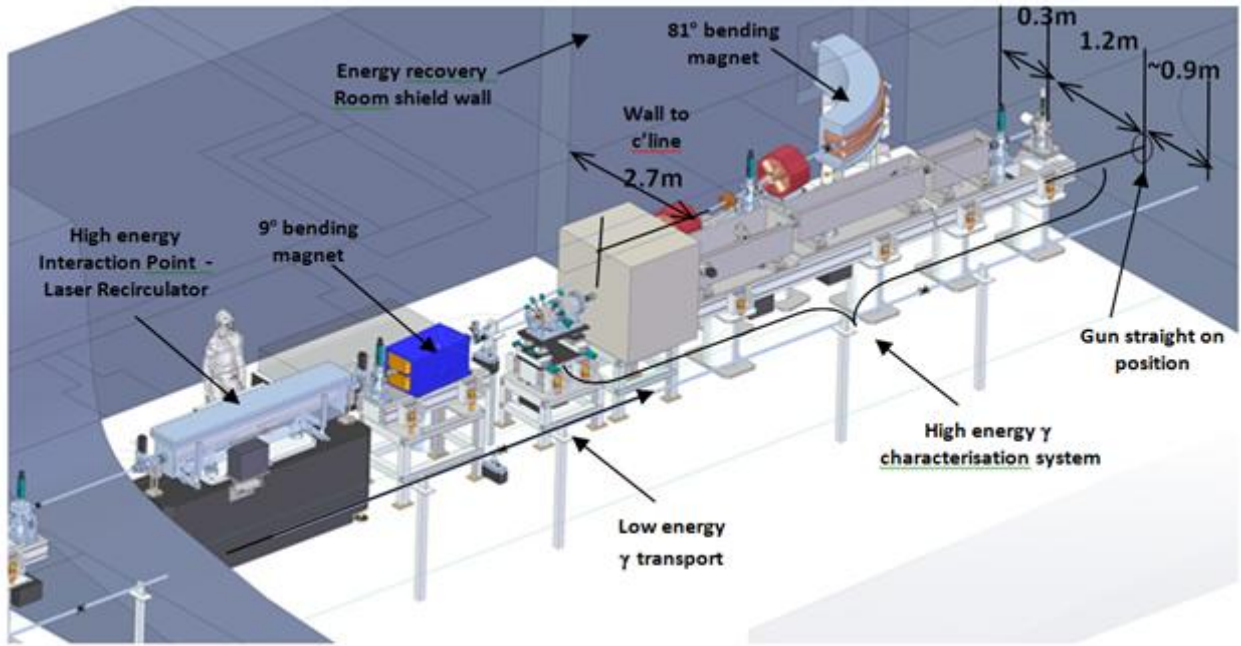

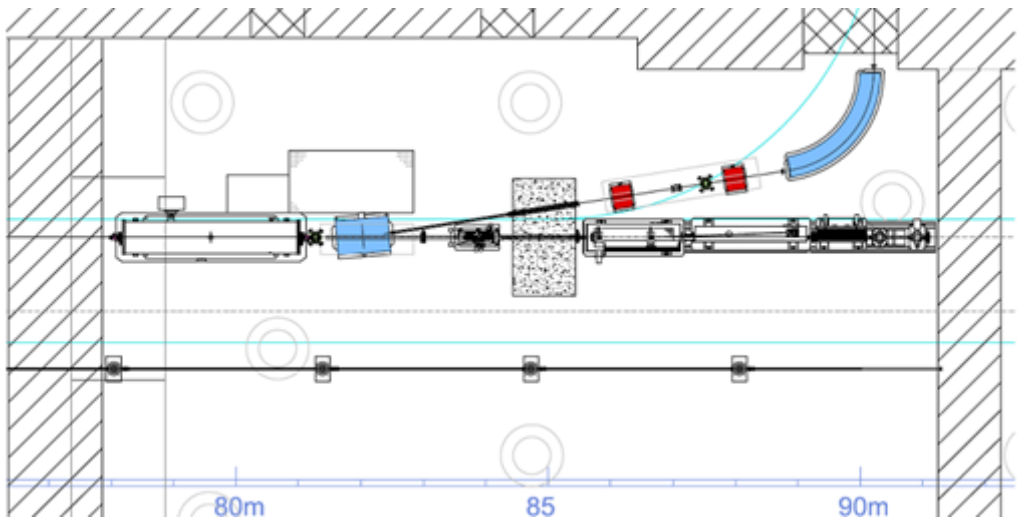

**Fig. 195.** High Energy Interaction & Gamma System (Isometric & Plan view)

### 6.1.3. Technical Rooms and services

#### 6.1.3.1 RF Power Supplies

The RF power supplies consist of 4 S-Band modulators/klystron units serving the photo-injector modules and 5 x dual C-Band units (with a reduced footprint of ~1.5x3.5m when compared with two single units) powering the RF structures in the LE & HE linacs. Together they occupy most of one of the technical rooms dedicated for RF use in the accelerator specification leaving the other RF room free for alternative use.

Fig. 196 depicts a close-up plan view of the RF room (of approximate size 10.8x18.3m). Note that part of the room has been reserved for racks (not necessarily all associated with RF power). At present it is assumed that the rooms are accessible (perhaps as supervised or controlled areas) when the machine is running. The 350kV Klystron heads will be shielded by several cm of lead so that X-ray radiation from the power supplies themselves will be reduced to acceptable levels. An interlock cage screen will also be erected to ensure that safe distance is maintained. It is assumed that radiations scatter into the RF rooms from the accelerator hall will be reduced to acceptable rates by the passing the waveguides through a labyrinth chicane.

Fig. 197 illustrates a section through the RF technical room (looking along beam direction) – a magnification of a similar cross section for laser rooms of Fig. 189. The arrangement shows a concept proposal for a mobile A-frame crane that would allow easy access to the klystrons heads should maintenance or removal for any other reason be required. The crane has a nominal 1 Tonne load capacity – which should be sufficient to remove the klystron shielding, the klystron unit and the solenoid - at least in staged operations.

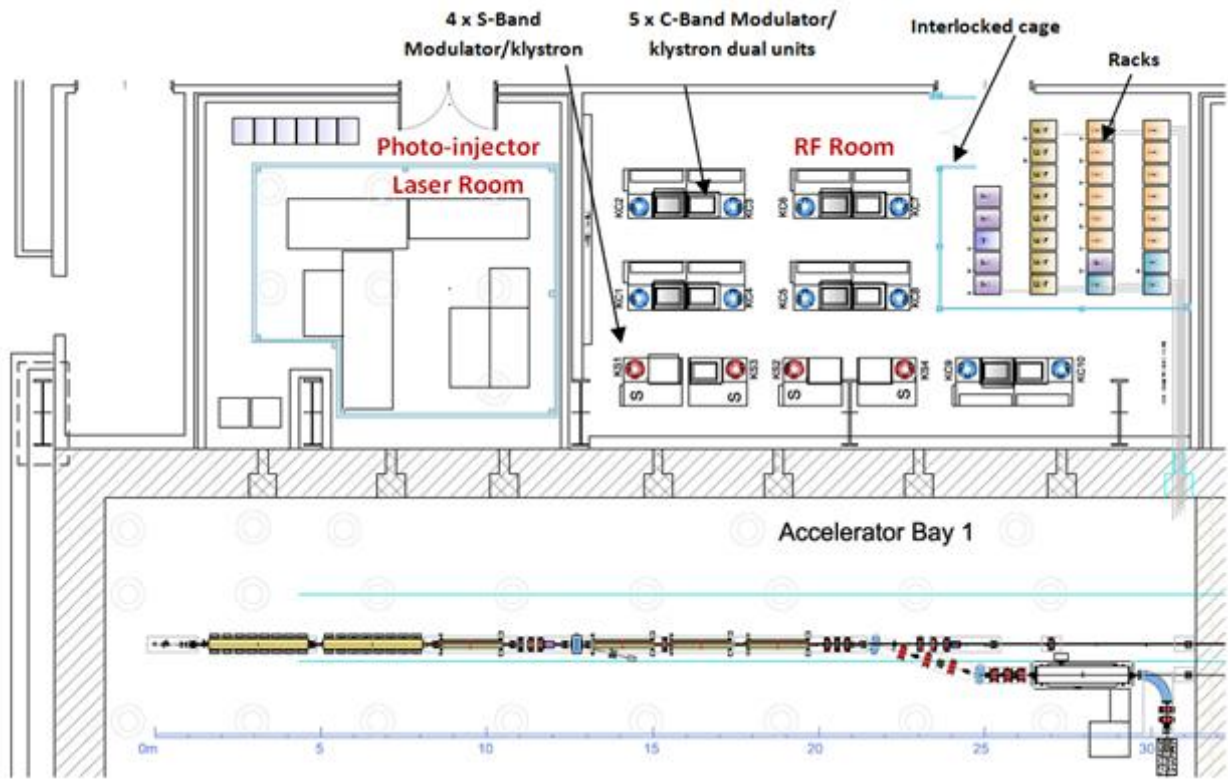

**Fig. 196.    Plan view RF + Laser Technical Room**

The heavy modulator units will be most likely be installed by being conventionally 'skated' into position. Inside the RF room it is provisionally envisaged that the waveguides, cooling and power cabling runs will be located overhead in central runs with respect to the RF power supplies so as to allow free access for the crane and extraction of any klystron without the necessity of removing all waveguides (see also Fig. 192 for 3D visual clarification).

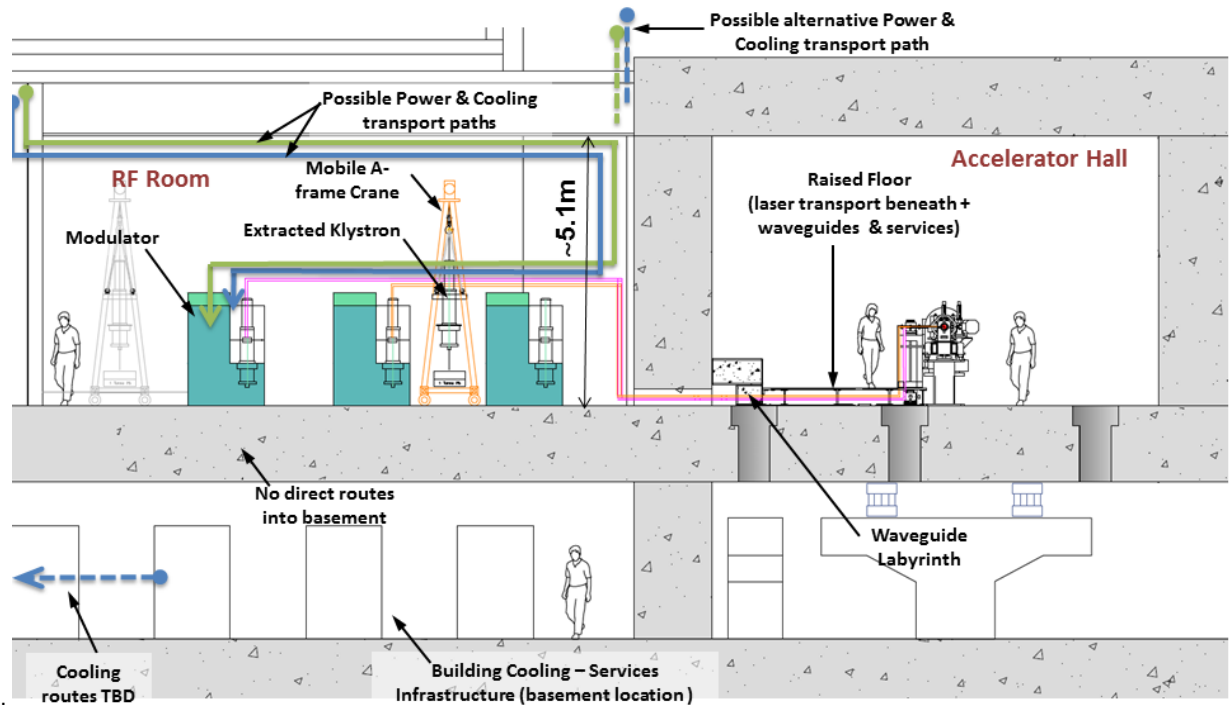

**Fig. 197.** RF Technical Room and Waveguide Distribution

Fig. 197 also indicates that the vibration damping pillar network does not extend under the RF rooms. Our present understanding is that building cooling services plant will be located here, in which scenario it would be obvious to minimise cooling runs by bringing the necessary Megawatts of cooling capacity directly up through the floor. Unfortunately, there does not appear to be any specific penetrations shown in these areas on the building plans. The present most likely route will be to bring pipes up through the corridor route and down into the room from there. Detail routes will be defined during detail planning phase.

#### 6.1.3.2   Waveguides

The proposal for the RF waveguide distribution in the accelerator hall is that it will drop down to low level in the RF room at the shield wall and emerge via a radiation shielding labyrinth into the accelerator hall at the same low level. Careful planning and design will be required in order to ensure that the RF distribution does not clash with the laser transports which will also run at low level. Alternative high level routes will also be considered during further detail planning and design phase.

The waveguides will be vacuum pumped to ~$10^{-8}$ mbar throughout the estimated 250m+ total distribution with only the ferrite circulators and waveguide switches housing a relatively low volume of SF6 (<6m$^3$). In this way no particular safety requirements as regards O² depletion alarms are anticipated in the accelerator hall due to RF services.

#### 6.1.3.3   Laser Rooms & Transports

Fig. 196 shows a 2D plan view of the Photocathode laser laser room layout adjacent to the RF room. It depicts a laser table layout based on the 100Hz 250mJ Ti:Sa ASUR Amplitude laser system at Marseille. The table area will be surrounded by either a soft wall or possibly panel sealed clean room environment most likely of ISO class 7 (class 10,000) - but this is yet to be confirmed. The other laser rooms will have similar layout. It is envisaged that each laser room will house its own dedicated laser racks (up to 6 or more) located outside of the softwall area in a walk around space – or possibly even in specific cabin enclosures. Rack distribution is discussed further in the next section.

The laser transports to the accelerator will most likely follow a path generically similar to RF waveguides in that they will run at low level in the accelerator hall under the services walkway. A labyrinth chicane tunnel, consisting of shielding blocks surrounding the transport pipe and turning mirror boxes, located either within the accelerator hall or in the laser room should provide protection from background scatter so that the laser rooms are accessible (under controlled conditions) when the machine is running. Periscope mirror systems will bring laser light from optic bench height in the laser rooms down and back up to the 1.4m height at the linac interaction points.

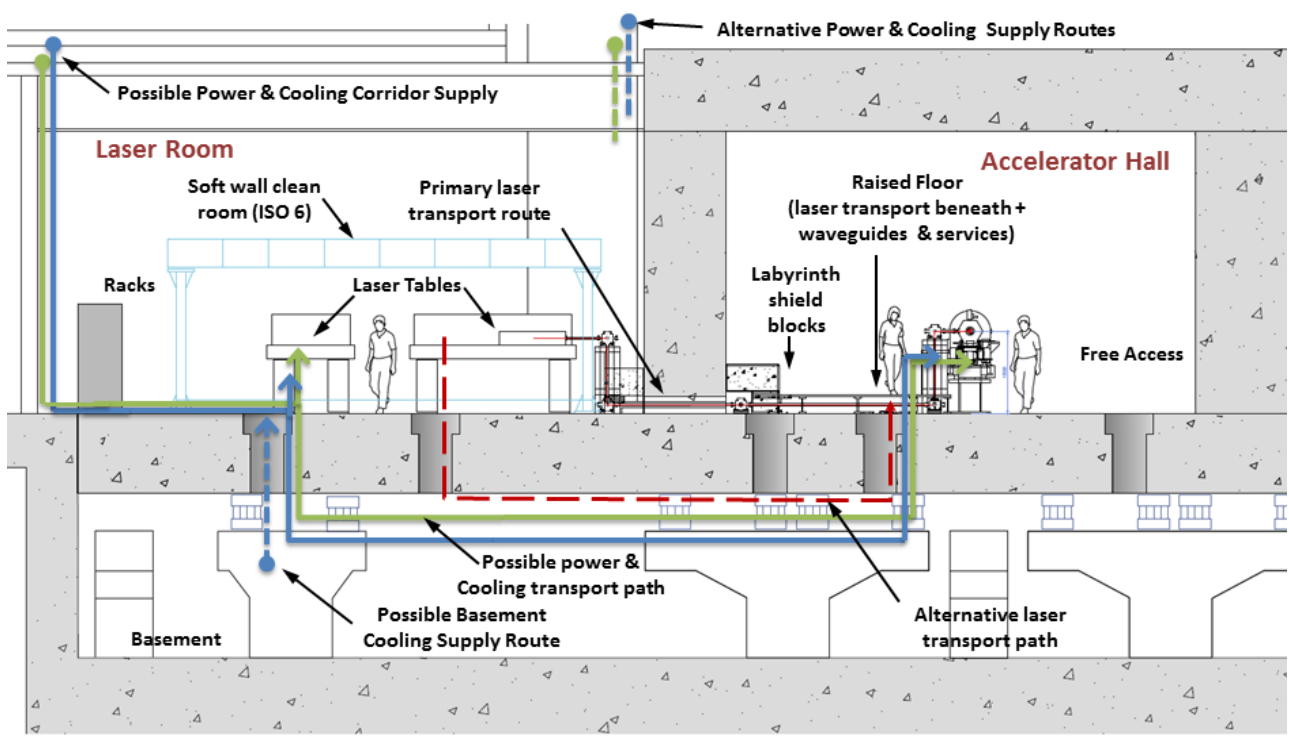

**Fig. 198.     Laser Room and Transport Lines**

The laser transports are shown generically in elevation cross-section in Fig. 198.

The anti-vibration pad network extends to below the laser rooms as does a network of penetration holes in the concrete floor. These may be used to transport cooling from plant in the basement into the laser rooms and also run cooling directly to the accelerator in a more general extended distribution. It is also proposed to route power and signal cabling from and to the laser rooms and the accelerator in a similar way - as depicted in Fig. 198. Detail routing schedules have not yet been done and the approach indicated will depend very much on the basement accessibility and other constraints defined by the facility managers at the Măgurele site. This must be determined early on in further technical negotiations during the detail design phase.

Note that an alternative route for the photon transports is also considered via the basement area - as indicated in Fig. 198 - but the preferred option is above the accelerator floor level.

The specific routes for each laser room are shown in the overview plan indicated in Fig. 191. Each route will take the minimum path length from the laser room location to the gun or LE/HE interaction point. Note that the LE interaction laser must pass through the accelerator bay 1→2 shield wall. It is anticipated that this could be done without periscope/ labyrinth given a sufficient long shielded straight-on length. However scatter calculations will be required to confirm this.

It is important to note that for HE operation both the NRF and the downstream Photo-fission laser rooms must be able to switch their outputs directly to the HE interaction point. For the NRF Laser Interaction lab to the HE point this necessitates a total transport length of over 40m – which may yet require the introduction of optic relays. Further details of the engineering of the laser optic transports are yet to be determined.

#### 6.1.3.4 Control Racks & Power supplies

The number of racks required for the various functions of the accelerator has been provisionally estimated and is listed in table below with a total of 80.

For rack working access we assume a 1m space between rack fronts (with in some cases 0.7m access behind). In this condition we observe that the total number of racks estimate for racks there is sufficient space to house them all within the technical rooms with some reserve capacity.

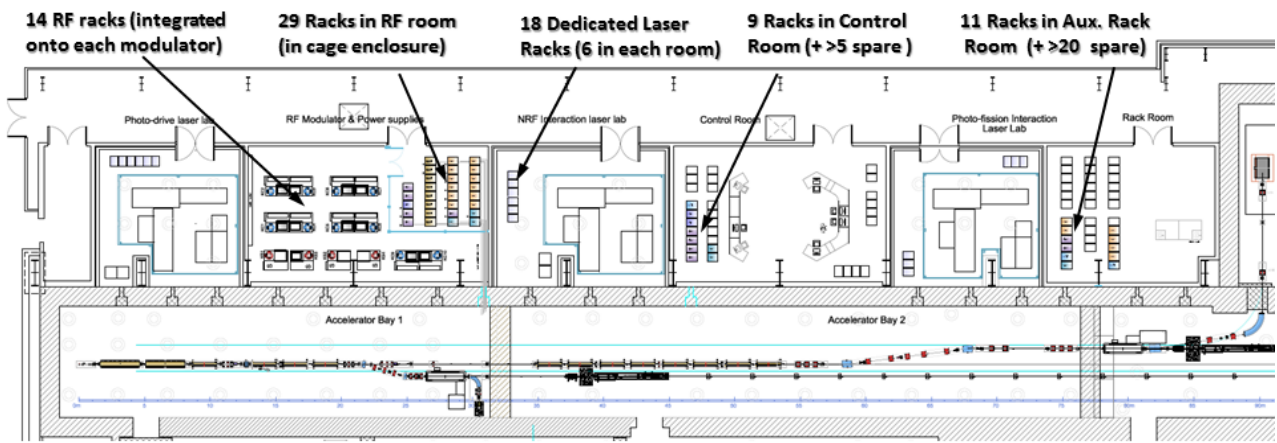

**Fig. 199.** Rack Locations

**Table 54. Rack numbers and distribution**

| Function | Quantity | Location |
|---|---|---|
| Laser | 18 | Laser Rooms (~6 in each room) |
| Diagnostics | 11 | RF room/Control Room/Aux Rack Room |
| Magnet powers supplies | 18 | RF room enclosure |
| Motion control | 3 | RF room/Control Room/Aux Rack Room |
| Vacuum | 5 | RF room/Control Room/Aux Rack Room |
| Control, monitoring and interlocks | 3 | RF room/Control Room/Aux Rack Room |
| LLRF & Synchronisation | 8 | RF room enclosure |
| RF power sources (1 per source) | 14 | RF room – integrated into modulator |
| Total | 80 | 30+ spare |

A tentative plan for rack distribution is shown in Fig. 199 above. As a first design pass it has been attempted to distribute the racks along the linac length approximately evenly. This has resulted in substantial number of racks operating with RF room 1. In addition 1/4 of the control room area has been assumed to be reserved for rack use with potential for considerable spare space. It is also feasible that a dedicated rack room could be constructed to house the most EMF sensitive of the racks including LLRF and synchronisation, and therefore would be EMF compatible with Faraday cage like construction and door seals. This room could also have specific high capacity air conditioning for higher power racks

The current plan indicates space for a spare rack capacity >30 populated mostly in the downstream auxiliary rack room, although 50 extra is feasible. As a provisional plan this large spare capacity is likely to be sufficient.

For magnet power supplies it is assumed that if sufficient separation space is available then air cooling will be acceptable. In this case a heatload of perhaps > 50 kW may be required to be handled by the air conditioning within the technical rooms. This needs to be confirmed against final building specification during detail planning at the earliest opportunity. The alternative is to provide water cooling to these racks.

## 6.2. General Infrastructure Requirements

### 6.2.1. Equipment Access

In general we foresee no particular requirement for an overhead gantry crane access in the accelerator hall or technical rooms. Access points into the accelerator area are assumed to be any of the three 2m wide doorways indicated (1,2,3) in Fig. 200 below.

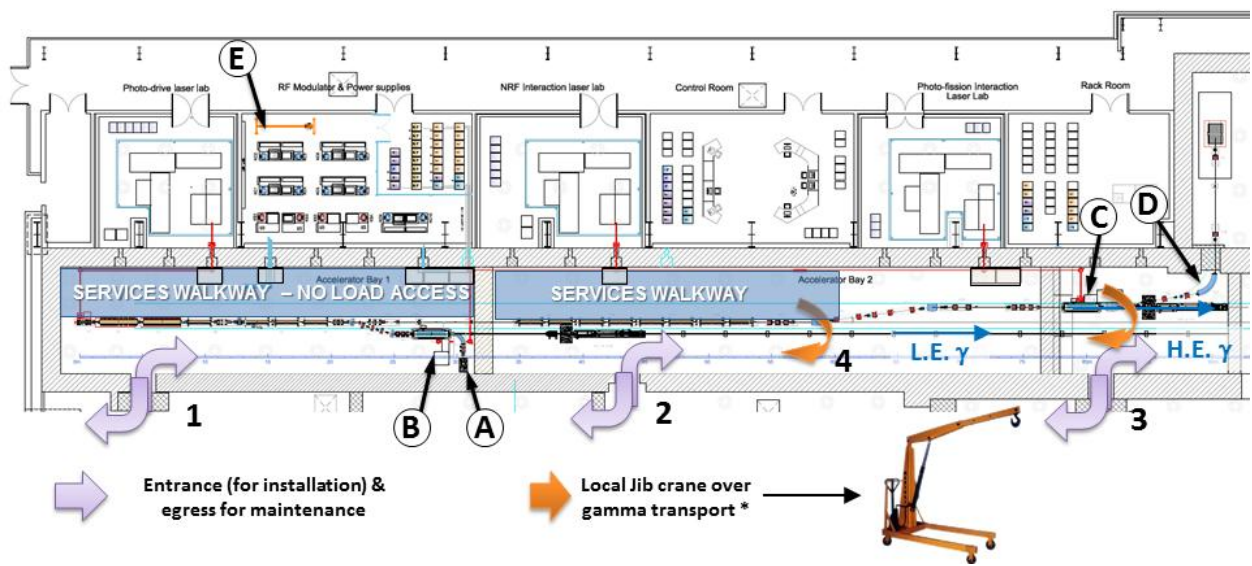

**Fig. 200.    Accelerator component installation and accessibility routes**

For the main accelerator hall it is envisaged that equipment can be 'skated' into position using inexpensive conventional industrial machine skate rollers. For the aluminum girder modules a low loading trolley system will be provided as part of the accelerator.

The heaviest item will be the granite block support for the laser recirculator. This could mass ~5T and it is envisaged that some form of powered pallet truck or forklift would be available to pull the block into position on heavy duty skates or lowloading bogie. Alternatively manual handling on an air-skate pallet may be employed.

Once the HE line is installed a mobile 1T capacity jib crane (engine hoist type) can be used to lift any module girder out and over the LE gamma transport line if maintenance cannot be done in-situ (4). An alternative is to close valves in the LE gamma line, let up to air and remove a beampipe so that the module can be skated out.

Once the accelerator is fully installed and operational some areas that may present particular difficultly of access for maintenance or repair removal could potentially be;

(A) Beam dump shielding removal may require laser table removal first. Alternative is to design local shielding using handle-able size concrete bricks.

(B) Note that lifting of the lid of the recirculator chamber will be done via jib crane (see 4) or local mechanism.  A mobile laminar air flow tent may have to be fitted over the recirculator. This may be softwall semi-permanent fixture or mounts fitted locally in the walls so it can be speedily erected using the jib crane.

(C) H.E. Interaction Laser Recirculator can be removed using jib crane – see general access note (4). The granite base itself may mass >5Tonne and can only be removed by removing a section of LE gamma transport  to skate out. (However this is a highly unlikely occurence).

(D) Removal of the dipole is probably best achieved by removing the end monitor module of the gamma section but lifting is also possible.

(E) Klystrons can be removed from a modulator for repair or replacement via an A frame crane (see also RF room figures). The overhead waveguide arrangement is configured so that the crane can travel to and access any klystron without need to be removed along the way. The klystron can be lowered from the crane in the area indicated onto a pallet or other low loading device and transported from the RF room .

### 6.2.2. Air conditioning

Loads on the building air conditioning (HVAC) have not yet been estimated and will depend on total electrical and cooling estimates as described below. Given that all major power items will be water cooled typically we anticpate loss to air to be relatively minor and given experience of other accelerators for this type and size of machine we would nominally expect heat loads to the HVAC in the accelerator bays to be < 20kW from the machine itself – including cable losses but excluding lighting and ancilliary (local pumping trolleys…etc) – which could raise this figure by >50%.

I higher figure is likely for the technical & rack rooms (dependent on whther racks are water cooled). A provisional conservative budget of ~ 50 - 100kW for this size facility can be considered as indicative prior to any detail estimates being done.

### 6.2.3. Electrical & cooling

Feasible routes for cooling and electrical distributions are tenetively indicated in Fig. 201. However these are indicative only & subject to further study & negotiation. In addition total capacities for electrical power and water cooling requirements are yet to be assessed in the following areas:

Cooling
- Magnets
- RF structures
- RF Klystron/Modulators
- Racks (notably magnet PS)

Electrical
- Magnets (130% rating)
- RF Klystron/Modulators
- Cooling Pumps (approx. electrical power required ~1/3 of total cooling load, – assumed efficiency ~0.3)
- UPS
- Earthing & Lighting

- Cabling
- Sub-distruibutions, cable trays

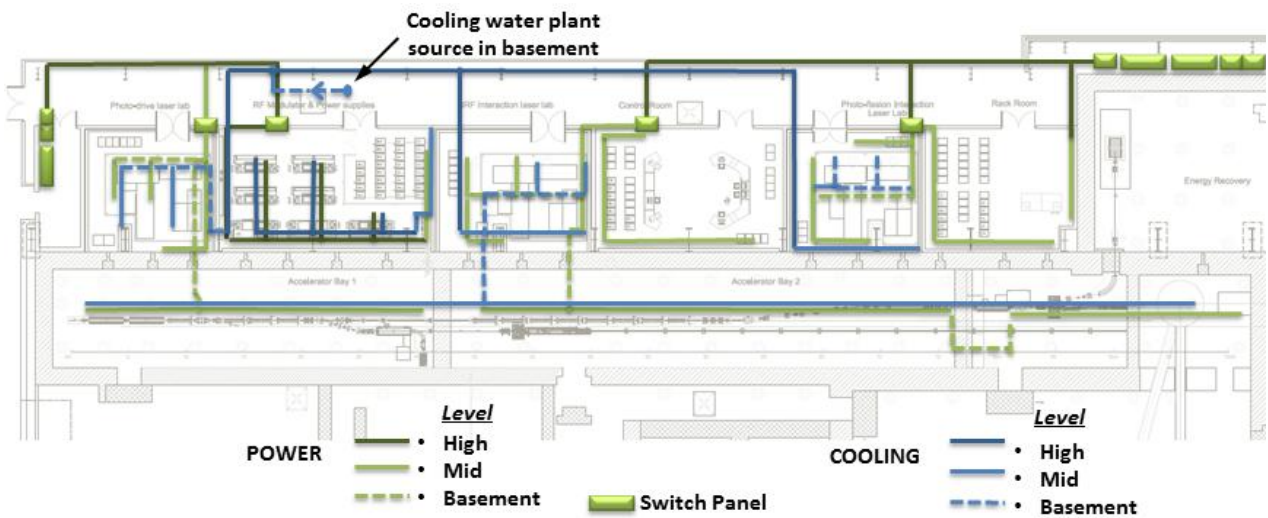

Fig. 201. Possible electrical sub-distribution cable & Water cooling routes

## 6.3. Radiation safety

### 6.3.1. Introduction

High-energy electron accelerators are complex devices containing many components. All facilities contain the same basic systems:

- Accelerators structures
- RF power components
- Vacuum system
- Magnetic system associated with steering and focusing the beam
- Water-cooling
- Etc

Prompt radiation and radioactivity induced by particle nuclear interaction in beam line elements and shielding structures represents the main radiation hazard of high energy accelerators.

The accelerator's design parameters are of crucial importance in the determination of the nature and magnitude of radiation source. The most important parameters are:

- Particle energy
- Beam power
- Target material
- Work load
- Beam losses

### 6.3.2. Operating parameter

The ELI-NP-GS project consists in an electron accelerator in which an electron beam is accelerated at the energies in the range 200 and 720 MeV GeV with an electron current of up to 1500 nC/s.

### 6.3.3. Radiation Protection

#### 6.3.3.1 Shielding outlines

The new machine general layout has been previously shown.

Using the previous operating parameters has performed calculations of shielding. Because of a great number of the precautions introduced, the results should be a conservative approximation of the doses actually expected. During the commissioning phase, the reliability of the assumptions made will be verified and, if necessary, additional precautions will be made.

For the calculation of low energy beam dump an energy of 360 MeV and an average current of 1500nC/s were considered.

#### 6.3.3.2 Shielding Design Criteria

The shielding design criteria have been base on the text of European Directives as well as the recent ICRP recommendations (ICRP 103) According previous documents the individual limits are 20mSv/y for radiation workers, and 1 mSv/y for the members of the public.

Moreover the definitions of controlled and supervised areas are useful as guidelines. A controlled area is every area where 3/10 of the limits recommended for radiation worker may be exceeded. A supervised area is one area where the overcoming of 1/10 of the previous limit may occur.

Taking into account the dose levels normally found around accelerators, the thickness of the shielding was calculated maintaining the doses, within the areas outside the shield frequented by the staff, 1-2mSv/year and 0.25 mSv/year within areas outside the shield frequented by members of the "public".

A shifting from these values could at most change the radiation classification of some areas. In normal working condition the dose rate outside shielding should not exceed a fraction of µSv/y.

### 6.3.4. Source Term

For shielding evaluation purposes, three components of radiation field which are produced when an electron beam, with an energy of hundred of MeV either a vacuum chamber wall or a thick target have to be considered.

### 6.3.4.1 Bremsstrahlung

Prompt photon fields produced by Bremsstrahlung constitute the most important radiation hazard from electron machines with thin shielding. Bremsstrahlung yield is very forward peaked, and increasingly so with increasing energy.

The following equation describes this behavior:

$$\theta_{1/2} = 100/E_0$$

where $\theta_{1/2}$ is in the angle in degrees at which the intensity drops to one half of that at 0º, and $E_0$ is the energy of the initial electrons in MeV. In order to evacuate the shield thickness a "thick target", usually a target of sufficient thickness to maximize bremsstrahlung production, was considered. Photon yield from a thick target as a function of angle consists of two components: sharply varying forward component, described in equation , and a mildly varying wide-angle component. Forward (or zero-degree) bremsstrahlung contains the most energetic and penetrating photons, while bremsstrahlung at wide angles is much softer.

The source term (per unit beam power) for bremsstrahlung at 90º is independent of energy.

### 6.3.4.2 Neutrons

Photons have larger nuclear cross-sections than electrons, so neutrons and other particles resulting from inelastic nuclear reactions are produced by the bremsstrahlung radiation. Neutrons from photonuclear reactions are outnumbered by orders of magnitude by electrons and photons that form the electromagnetic shower. However, some of these neutrons constitute the most penetrating component determining factor for radiation fields behind thick shielding.

**Giant resonance production**

The giant resonance production can be seen in two steps:

1. the excitation of the nucleus by photon absorption;
2. the subsequent de-excitation by neutron emission, where memory of the original photon direction has been lost.

The cross-section has large maximum around 20-23 MeV for light nuclei (mass number A<40) and 13-18 MeV for heavier nuclei.

The angular yield of giant resonance neutrons is nearly isotropic.

The giant resonance is the dominant process of photoneutron production at electron accelerators at any electron energy.

**Pseudo-deuteron production**

At photon energies beyond the giant resonance, the photon is more likely to interact with a neutron-proton pair rather than with all nucleons collectively. This mechanism is important in the energy interval of 30 to ~300 MeV, contributing to the high-energy end of the giant resonance spectrum. Because the cross-section is an order of magnitude lower than giant resonance, with the added weighting of bremsstrahlung spectra, this process never dominates.

**Photo-pion production**

Above the threshold of ~140 MeV production of pions (and other particles) becomes energetically possible. These pions then generate secondary neutrons as byproduct of their interactions with nuclei. While substantially less numerous than giant resonance neutrons, the photopion neutrons are very penetrating and will be the component of the initial radiation field from a target (with the exception of muons at very high energies) that determines the radiation fields outside very thick shields.

### 6.3.4.3 Muons

Muon production is analogous to e+/e- pair production by photons in the field of target nuclei when photon energy exceeds the threshold $2m_m c^2 \approx 211$ MeV.

Above a few GeV the muon yield per unit electron beam power is approximately proportional to electron energy $E_0$. Muon angular distribution is extremely forward-peaked, and this distribution narrows further with increasing energy. At energies of a few GeV adequate photon and neutron shielding will be also sufficient for muons.

### 6.3.4.4 Gas bremsstrahlung

The gas bremsstrahlung is produced by the interaction of the electron beam with residual low-pressure gas molecules in the vacuum pipe. Bremsstrahlung on residual gas is one of the main cause of beam loss in a storage ring and may represent a radiation hazard at syncrhrotron radiation facilities. This type of radiation has been thoroughly investigated at circular storage rings, where the beam current is much more intense. It is mainly in the straight section that a radiation problem could arise.

### 6.3.4.5 Induced Activity

Personnel exposure from radioactive components in the beam line is of concern mainly around beam lines, collimators, slots, beam stopper or beam dump, where the entire beam or a large fraction of the beam is dissipated continuously, while unplanned beam losses result from beam mis-steering due to inaccurate orbit adjustment or devices failure.

Beam losses induce activation in machine component as well as in

- the beam pipe      ($^{60}$Co, $^{54}$Mn, $^{51}$Cr, $^{46}$Sc, $^{22}$Na, $^{11}$C, $^{7}$Be)
- the cooling water  ($^{3}$H, $^{7}$Be, $^{15}$O, $^{13}$N, $^{11}$C)
- the air            ($^{15}$O, $^{13}$N, $^{38}$Cl, $^{41}$Ar)
- the concrete walls ($^{152}$Eu, $^{154}$Eu, $^{134}$Cs, $^{60}$Co, $^{54}$Mn, $^{22}$Na)

The activation of soil as well as the groundwater by neutrons and other secondary particles can have an environmental impact but at electron accelerators the radioactivity levels are generally low and absolutely negligible with the previous beam parameters.

### 6.3.4.6 Machine accesses

During machine operation the linac tunnel will be an excluded area.

During no operation periods the linac tunnel will be a controlled area, due to the possible activation of the machine structure.

The technical areas behind the shield will be classified as controlled or supervised areas.

The experimental areas will be a free access area. Only areas close to the front ends or at the end of the beam line will be classified.

In order to protect workers in the experimental areas, the electron beam will be dumped below the floor. A deflection of 45º is effected by electromagnets.

For additional safety permanent magnets and active radiation detectors interlocked with the beam will be used.

### 6.3.4.7 The operational radiation safety program

The purpose of the operational safety system program is to avoid life-threatening exposure and/or to minimize inadvertent, but potentially significant, exposure to personnel. A personnel protection system can be considered as divided into two main parts: an access control system and a radiation alarm system.

The access control system is intended to prevent any unauthorized or accidental entry into radiation areas.

The access control system is composed by physical barriers (doors, shields, hutches), signs, closed circuit TV, flashing lights, audible warning devices, including associated interlock system, and a body of administrative procedures that define conditions where entry is safe. The radiation alarm system includes radiation monitors, which measure radiation field directly giving an interlock signal when the alarm level is reached.

**Interlock design and feature**

The objective of a safety interlock is to prevent injury or damage from radiation. To achieve this goal the interlock must operate with a high degree of reliability. All components should be of high grade for dependability, long life and radiation resistant. All circuits and component must be fail safe (relay technology preferably).

To reduce the likelihood of accidental damage or deliberate tampering all cables must run in separate conduits and all logic equipment must be mounted in locked racks.

Two independent chains of interlocks must be foreseen, each interlock consisting of two micro switches in series and each micro switches consisting of two contacts.

Emergency-off buttons must be clearly visible in the darkness and readily accessible.

The reset of emergency-off buttons must be done locally.

Emergency exit mechanisms must be provided at all doors.

Warning lights must be flashing and audible warning must be given inside radiation areas before the accelerator is turned on.

Before starting the accelerator a radiation area search must be initiated by the activation of a "search start" button. "Search confirmation" buttons mounted along the search path must also be provided. A "Search complete" button at the exit point must also be set.

Restarting of the accelerator must be avoided if the search is not performed in the right order or if time expires.

The interlock system must prevent beams from being turned on until the audible and visual warning cycle has ended.

Any violation of the radiation areas must cause the interlocks system to render the area safe.

Restarting must be impossible before a new search. Procedures to control and keep account of access to accelerator vaults or tunnels must be implemented.

## 6.3.5. Electron Beam Dump

The electron beam will be dumped below the floor. The beam deflection of about 45 degrees in made using permanents magnet for additional safety.

Taking account all operating parameter as well as the foreseen beam loss (100%) in the beam dump and shielding design criteria, a dump of 30 cm of lead plus 30 cm of polyethylene surrounded by 300 cm of ordinary concrete is necessary.

In following figures are reported the ambient dose equivalent rate at 5 m from the source at 0° and 90° versus the thickness of concrete shield.

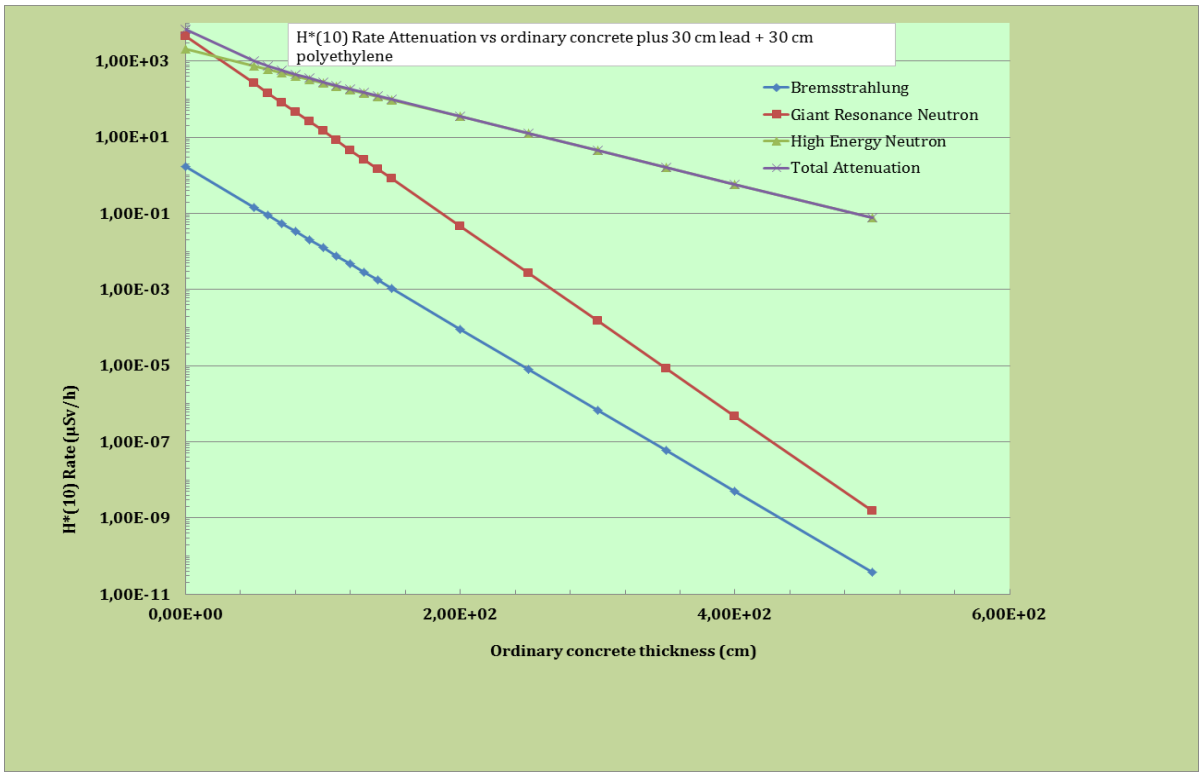

Fig. 202.    Ambient dose equivalent rate 0°

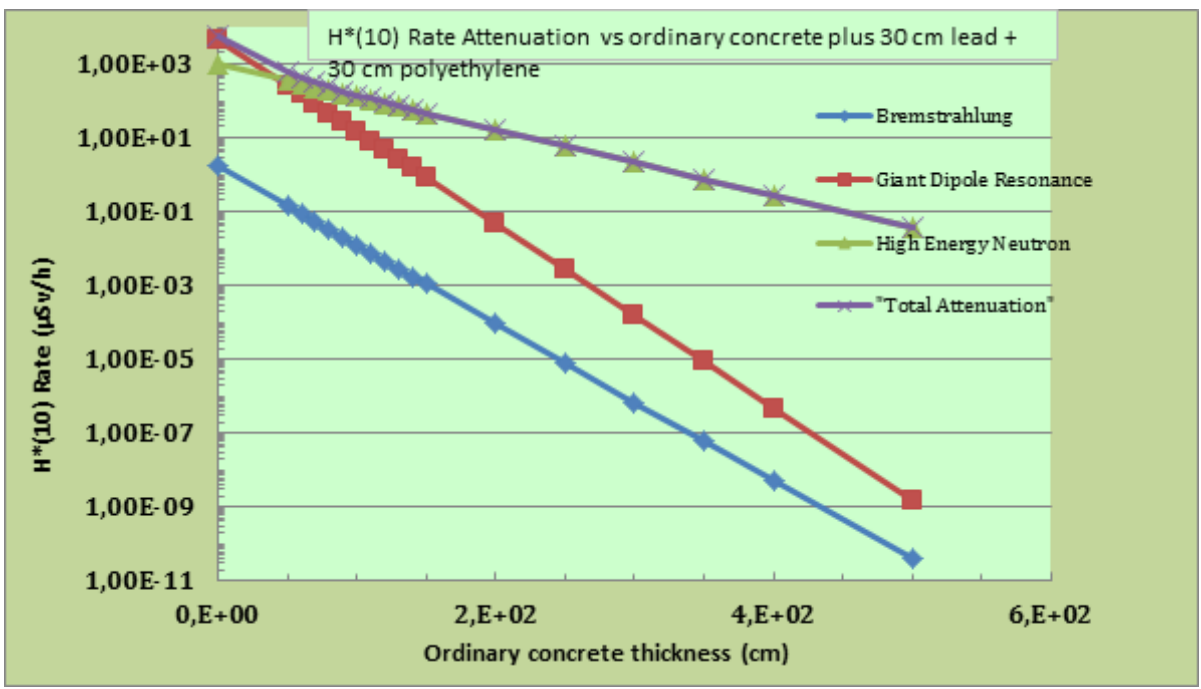

Fig. 203.    Ambient dose equivalent rate 90°

A cooling system for the lead of the beam dump has to be foreseen to avoid the melting of lead. A temperature higher than 606 Kelvin, melting point of lead, can be reached in an hour of continuous operation at about 500 Watt

### 6.3.6. Other Radiation Sources

The RF power sources for ELI are 55 MW S-band klystrons and 50 MW C-band klystrons.

The klystrons will arrive already shielded from factory.

Additional shield will be installed in order eliminate and/or to reduce radiation escape. An interlocked fence around klystrons is foreseen in order to reduce as low as possible the radiation level behind the fence.

## 6.4. Cooling system requirements

The project shall include: air conditioning of the technical halls, compressed air production and distribution, compressed technical gases storage and distribution, water cooling distribution for components (magnets, RF structures, laser equipment, power supply cabinets etc), three clean rooms, demineralized water treatment and production, and the control and supervision systems.

Chilled, cold (medium temperature) and hot water supply shall not be included as they shall be in charge of the local building facilities.

A first estimate of the thermal power could be drawn by the installed electrical power, of whom roughly 90% in water cooling and the remaining 10% in air conditioning.

**Table 55. Power estimation**

| System (WP) | RF STRUCTURES | RF POWER SOURCES | LASERS | MAGNETS and Power Supplies |
|---|---|---|---|---|
| **Location** | LINAC hall | LINAC hall and RF Modulators and Power Supply hall | Laser Labs (Clean Rooms) | LINAC hall and RF Modulators and Power Supply hall |
| **Components** | GUN S-Band<br>TW structures (S-Band)<br>TW C-band sections<br>S-Band deflectors<br>S-band loads<br>C-band loads | Modulators<br>Kly bodies<br>Kly collectors<br>Electromagnets | | Dipoles<br>Quadrupoles<br>GUN Solenoid<br>AC Solenoids |
| **Installed electrical power (kW)** | 76.3 | 765 | 65 | 188 |

#### 6.4.1.1 Water cooling specifications

The water for the cooling of accelerator components and for the air conditioning appliances shall be provided at the following levels of temperature and accuracy:

**Table 56. Water temperature level**

| RF POWER SOURCES, LASERS, MAGNETS and Power Supplies | 32 ±1°C |
|---|---|
| RF STRUCTURES | 30 ±0.1°C |
| Air Conditioning (chilled) | 7°C |
| Air Conditioning (hot) | 70°C |

The water shall be supplied at a pressure to be defined for each component, generally below 1000 kPa (10 bar). The flow required for every user has to be specified.

For the cooling of all the components of the accelerator the water shall be "ultrapure", corresponding to conductivity of less than 0.2 microS/cm, and with a low dissolved oxygen contents.

#### 6.4.1.2 Clean rooms

The halls housing the laser equipment shall be clean rooms of the level ISO 7 (352'000 part. <0.1 micron/m3 according ISO 14644-1, roughly corresponding to the class 10'000). If the level shall be maintained "at rest" or "in operation" has to be defined, as well as all the other details.

#### 6.4.1.3 Compressed air and gases

Oil free, dry (dew point 3°C) compressed air shall be produced and distributed in every technical hall at a pressure of 700±100 kPa. The air consumption shall be 50-100 m3/h.

Other technical gases (Helium, Nitrogen, Argon, Kripton, Xenon) will be stored and distributed in some of the technical halls, with small consumption rates.

# 7. Future upgrades

## 7.1. Full C-Band option

As it has been extensively described in the previous, the solution adopted for the accelerator consists in an hybrid scheme based on a photo-injector similar to the SPARC one, working in S-band, accelerating the electron beam to an energy around 80 MeV, followed by a sequence of C-band accelerating structures. A possible upgrade is to replace the S-band segment with a C-band photo-injector in a full C-band scheme.

The use of higher frequency RF gun (C-band or X-band), potentially, permit a better emittance compensation. In despite of these beam dynamic advantages, there are drawbacks coming from technological difficulties, which increase as the frequency rise. The smaller size of the cavities implies a worse power dissipation and an emittance compensation solenoid closer to the cathode, which due to higher field, might works with unfeasible current density or with a completely saturated armature. These problems justify the X-band gun projects, which are in 5+1/2 cells configuration than 1.6 cells.

This full C-band option has been carefully analyzed, optimizing a photo-injector consisting in a C-band RF gun, scaled from the SPARC S-band 1.6 cells RF gun and operating with a higher field of 170 MV/m. This field has been fixed at a safe value, far lower than $\lambda_{RF}$ linear scaling, avoiding unsustainable superficial electric fields and unfeasible emittance compensation solenoid (also Bz follows the $\lambda_{RF}$ linear scaling).

Downstream the gun, there are three 1.5 meters long sections, operating at a maximum field of 35 MV/m, similar to the sections composing the C-band booster. The optimization has been performed following the same criteria adopted for the S-band photo-injector: to start with a long bunch at the cathode and to apply the "velocity bunching" technique, in the first section, in order to get a 280 $\mu$m bunch length at the photo-injector exit, so to minimize the energy spread in the interaction point. As in the S-band photo-injector the emittance, during the compression, is controlled by the magnetic field of a solenoid embedding the RF compressor.

Two working points, at two different charge values, have been investigated: 250 pC (for comparison with the "reference" working point in the hybrid scheme) and 500 pC. In the calculations a thermal emittance of 0.9 mm-mrad/(mm rms) (the same value measured in S-band) has been used.

An extensive campaign of numerical optimization has been carried out based on the tracking code ASTRA [135] driven by the parallel genetic algorithm GIOTTO [136]. The optimization has been performed on a small 16 cores cluster by iterative runs of ASTRA over the free parameters of the problem, until best result is found, in terms of minimum emittance at the photo-injector exit. The genetic algorithm search has been carried out over the values of the cathode launching phase, the magnetic focusing solenoid field amplitude, the laser spot size at the cathode, its length (we assumed a flat pulse with a rise time of 1 psec and a uniform laser intensity distribution in the transverse plane), position, phases and gradient of the accelerating sections. The optimization included also the position of the gun solenoid. In Fig. 204 (left) the effect of this important geometrical parameter for a gun solenoid rescaled from the SPARC one, together with the new relative first section position, is shown: for each solenoid position, a roughly sweep of the first section position, around the rescaled values (from hybrid solution) has been done, and the emittance minimized by GIOTTO. The minimum emittance is obtained for a distance, between the solenoid entrance and the cathode, of 4.8 cm (pure S-band scaling case). Unfortunately this position is not compatible with the geometry of the RF gun that requires a distance not shorter than 6.5 cm. For this reason a new solenoid design has been done in the form of a short magnetic lens (Fig. 204 right), which moves the peak field closer to the cathode, in the way to compensate the shifted solenoid position. It is designed with only one coil and, with armature, measures 7.5 cm, whereas the scaled SPARC solenoid, composed by 4 coils is 10 cm long.

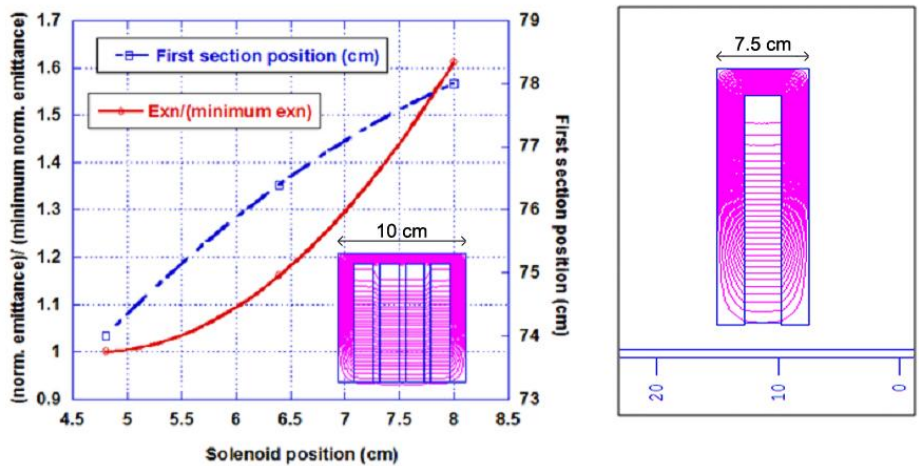

**Fig. 204.** FULL C-BAND photo-injector. (Left) Optimization of first section position and solenoid entrance position by using an emittance compensation solenoid on the gun scaled from the SPARC one. (Right) new gun emittance compensation solenoid design (POISSON code output [137])

In these conditions, with a solenoid entrance at 7.5 cm (very relaxed position from design point of view) the optimum position of the first section, by an accurate analysis, has been found to be 77 cm from the cathode. The plots in Fig. 205 and Fig. 206 compare the evolution of emittance and envelopes (transverse and longitudinal) along the photo-injector line and the exit phase spaces in the two working points that have been analyzed. In Table 57 the relative parameters are listed.

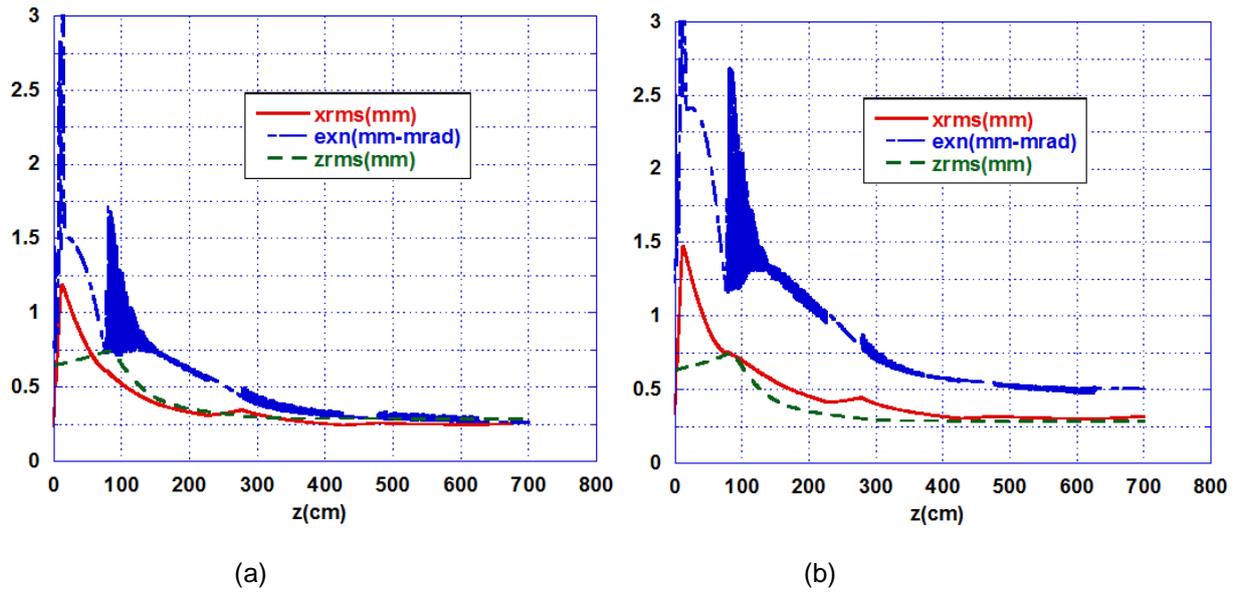

Fig. 205. FULL C-BAND photo-injector. ASTRA output: Evolution of emittance, transverse and longitudinal envelopes in the S-band photo-injector. (a) Q=250 pC (b) Q=500 pC

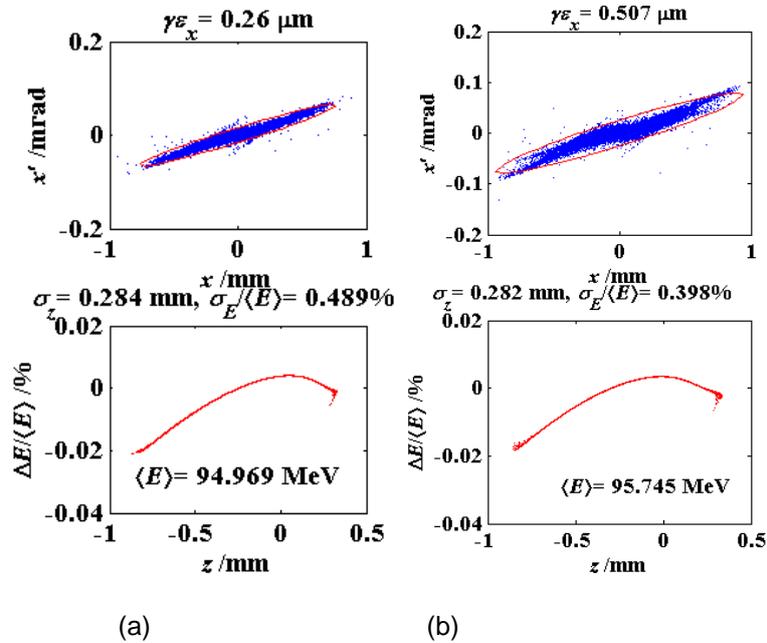

Fig. 206. FULL C-BAND photo-injector. ASTRA output: transverse and longitudinal phase space at the photo-injector exit; (a) Q=250 pC (b) Q=500 pC

For these setting, have been also checked the working conditions of the shorter solenoid: Fig. 207 shows a hysteresis curve portion, residual field on the cathode and coil current density per cm². In despite of the high magnetic field and high coil current, the working region shows achievable values.

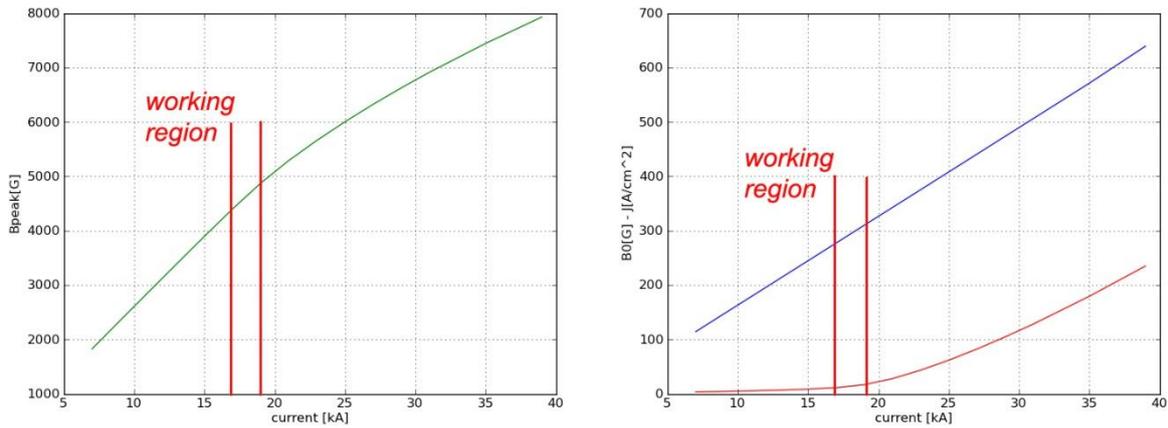

**Fig. 207.** On the left, in green, the portion of hysteresis curve around the solenoid working region, it shows a regime of light saturation. On the right, in blue, the current density per cm^2 and in red the residual magnetic field on the cathode

With this scheme for a charge of 250 pC, due to the higher field level on the cathode, the final emittance is reduced to less than 0.3 mm-mrad (against the value of 0.4 mm-mrad in S-band). At 500 pC the peak current is doubled, keeping a good control of the emittance that increases to 0.5 mm-mrad; but the point at 250 pC is preferred (as for the S-band photo-injector) from the point of view of phase space density.

**Table 57. Full C-band option: main photo-injector parameters**

|  | Reference beam | High charge wp |
|---|---|---|
| Charge (pC) | 250 | 500 |
| Photocathode Laser pulse length (FWHM) (ps) | 9 | 9 |
| Photocathode Laser rms spot size (µm) | 180 | 295 |
| Output energy (MeV) | 95.88 | 95.75 |
| Output RMS Energy spread (%) | 0.5 | 0.4 |
| Output Normalized RMS Projected Emittance (mm-mrad) | 0.26 | 0.5 |
| Output RMS bunch length (µm) | 283 | 282 |

Other advantages of this scheme, that is 1 meter shorter respect to the S-band photo-injector, are a reduced energy spread at the exit of the photo-injector easily recoverable in the following sections of the accelerator and a slightly higher energy at the photo-injector exit.

This upgrade seems promising for a phase-2 of the project, after a proper work of development on the C-band RF gun.

# 8. Parameter tables and Risk assessment

## 8.1. Parameter tables

**Table 58.    Summary of Gamma-ray beam Specifications**

|  |  |
|---|---:|
| Photon energy | 0.2-19.5 *MeV* |
| Spectral Density | $0.8\text{-}4\cdot10^4$ *ph/sec.eV* |
| Bandwidth (rms) | $\leq 0.5\%$ |
| # photons per shot within FWHM bdw. | $\leq 2.6\cdot10^5$ |
| # photons/sec within FWHM bdw. | $\leq 8.3\cdot10^8$ |
| Source rms size | 10 - 30 $\mu m$ |
| Source rms divergence | 25 - 200 $\mu rad$ |
| Peak Brilliance ($N_{ph}/sec\cdot mm^2 mrad^2\cdot 0.1\%$) | $10^{20} - 10^{23}$ |
| Radiation pulse length (rms, *psec*) | 0.7 - 1.5 |
| Linear Polarization | > 99 % |
| Macro rep. rate | 100 *Hz* |
| # of pulses per macropulse | $\leq 32$ |
| Pulse-to-pulse separation | 16 *nsec* |

**Table 59.    Electron beam parameters at Interaction Points: general characteristics**

|  |  |
|---|---:|
| all values are rms |  |
| Energy (MeV) | 80-720 |
| Bunch charge (pC) | 25-400 |
| Bunch length ($\mu$m) | 100-400 |
| $\varepsilon_{n\_x,y}$ (mm-mrad) | 0.2-0.6 |
| Bunch Energy spread (%) | 0.04-0.1 |
| Focal spot size ($\mu$m) | > 15 |
| # bunches in the train | $\leq 32$ |
| Bunch separation (nsec) | 16 |
| energy variation along the train | 0.1 % |
| Energy jitter shot-to-shot | 0.1 % |
| Emittance dilution due to beam breakup | < 10% |
| Time arrival jitter (psec) | < 0.5 |
| Pointing jitter ($\mu$m) | 1 |

**Table 60.  Yb:Yag Collision Laser beam parameters**

|  | **Low Energy Interaction** | **High Energy Interaction** |
|---|---|---|
| Pulse energy (*J*) | 0.2 | 2x0.2 |
| Wavelength (*eV,nm*) | 2.3,515 | 2.3,515 |
| FWHM pulse length (*ps*) | 3.5 | 3.5 |
| Repetition Rate (*Hz*) | 100 | 100 |
| $M^2$ | $\leq 1.2$ | $\leq 1.2$ |
| Focal spot size $w_0$ (*μm*) | > 28 | > 28 |
| Bandwidth (*rms*) | 0.1 % | 0.1 % |
| Pointing Stability (*μrad*) | 1 | 1 |
| Sinchronization to an ext. clock | < 1 *psec* | < 1 *psec* |
| Pulse energy stability | 1 % | 1 % |

**Table 61.  Laser beam Recirculator parameters**

|  | **Low Energy Interaction** | **High Energy Interaction** |
|---|---|---|
| Distance between the two Parabolic Reflectors | 2.38 *m* | 2.38 *m* |
| Collision Angle | 7.5° | 7.5° |
| beam waist $w_0$ | 28 *μm* | 28 *μm* |
| rotation at IP of linear laser polarization (along 32 passes) | $\geq 1°$ | $\leq 1°$ |
| integrated luminosity over 32 passes | > 90 % | > 90 % |
| Mirrors parallelism default | $\leq 10$ *μrad* | $\leq 10$ *μrad* |
| Mirrors alignment tolerance | $\leq 10$ *μm* | $\leq 10$ *μm* |
| Sinchronization to an ext. clock | < 1 *psec* | < 1 *psec* |

**Table 62.  Electron and Laser parameters for 3 selected collision examples (from STE simulations)**

| all quantities are rms | Low Energy Interaction | Low Energy Interaction | High Energy Interaction | High Energy Interaction |
|---|---|---|---|---|
| Energy (MeV) | 233 | 300 | 520 | 720 |
| Bunch charge (pC) | 250 | 250 | 250 | 250 |
| Bunch length (μm) | 280 | 280 | 280 | 270 |
| $\varepsilon_{n\_x,y}$ (mm-mrad) | 0.4 | 0.44 | 0.5 | 0.5 |
| Energy spread (%) | 0.1 | 0.08 | 0.05 | 0.05 |
| Focal spot size (μm) | 20 | 18 | 18 | 16 |
| RF pulse/bunch train rep rate | 100 | 100 | 100 | 100 |
| # bunches in the train / recirculator round-trips | 32 | 32 | 32 | 32 |
| emittance degradation due to beam break-up | <0.05 | < 0.05 | < 0.05 | < 0.05 |
| Laser pulse energy (J) | 0.2 | 0.2 | 2x0.2 | 2x0.2 |
| Laser pulse length  (psec) | 1.5 | 1.5 | 1.5 | 1.5 |
| Repetition Rate (Hz) | 100 | 100 | 100 | 100 |
| Laser focal spot size $w_0$ (μm) | 28 | 28 | 28 | 28 |
| Laser parameter $a_0$ | 0.02 | 0.02 | 0.04 | 0.04 |
| recirculator round-trip (nsec) | 16 | 16 | 16 | 16 |
| collision angle (deg) | 7.5 | 7.5 | 7.5 | 7.5 |

**Table 63.    Gamma-ray beam for 3 selected collision examples  (from Start-to-end simulations)**

| all quantities are rms | **Low Energy Interaction** | **Low Energy Interaction** | **High Energy Interaction** | **High Energy Interaction** |
|---|---|---|---|---|
| Energy (MeV) | 2.00 | 3.45 | 9.87 | 19.5 |
| Spectral Density ( ph/sec.eV) | 39,760 | 21,840 | 16,860 | 8,400 |
| Bandwidth  (%) | 0.5 | 0.5 | 0.5 | 0.5 |
| # photons per shot within FWHM | $1.2 \cdot 10^5$ | $1.1 \cdot 10^5$ | $2.6 \cdot 10^5$ | $2.5 \cdot 10^5$ |
| # photons/sec within FWHM | $4.0 \cdot 10^8$ | $3.7 \cdot 10^8$ | $8.3 \cdot 10^8$ | $8.1 \cdot 10^8$ |
| Source rms size ($\mu$m) | 12 | 11 | 11 | 10 |
| Source rms divergence ($\mu$rad) | 140 | 100 | 50 | 40 |
| Peak Brilliance ($N_{ph}$/sec·mm$^2$·mrad$^2$·0.1%) | $9.1 \cdot 10^{21}$ | $1.9 \cdot 10^{22}$ | $1.8 \cdot 10^{23}$ | $3.3 \cdot 10^{23}$ |
| Average Brilliance ($N_{ph}$/sec·mm$^2$·mrad$^2$·0.1%) | $2.9 \cdot 10^{13}$ | $6.2 \cdot 10^{13}$ | $5.9 \cdot 10^{14}$ | $1.1 \cdot 10^{15}$ |
| Rad. pulse length (rms, psec) | 0.92 | 0.91 | 0.95 | 0.9 |
| Linear Polarization (%) | > 99.8 | > 99.8 | > 99.8 | > 99.8 |
| Macro rep. rate (Hz) | 100 | 100 | 100 | 100 |
| # of pulses per macropulse | 32 | 32 | 32 | 32 |
| Pulse-to-pulse sep. ( nsec) | 16 | 16 | 16 | 16 |
| Contrast ratio 1$^{st}$ / 2$^{nd}$ harmonic | $1.5 \cdot 10^5$ | $8.5 \cdot 10^4$ | $7.0 \cdot 10^4$ | $4.4 \cdot 10^4$ |
| Luminosity @ (1,0.5)  psec  delay | (94,99) % | (92,98) % | (91,98) % | (85,96) % |
| Lumin. @ (5,2) $\mu$m  misalignment | (98,99) % | (96,99) % | (90,97) % | (87,95) % |

**Table 64.  Photo-injector and Photo-cathode drive laser specifications**

|  |  |
|---|---|
| Bunch charge (*pC*) | 25-400 |
| RF peak field at the cathode (*MV/m*) | 120 |
| RF Pulse duration for beam (*nsec*) | $\leq$ 600 |
| Accel. grad. in S-band Sections (*MV/m*) | 21 |
| Laser Pulse energy ($\mu J$) at 266 nm | $\leq$ 150 |
| Bandwidth @ 266 nm (nm) | > 0.7 |
| Laser Pulse length (flat-top, *psec*) | 5-12 |
| Laser Pulse rise-time (*psec*) | < 1 |
| Laser focal spot size ($\mu m$) uniform | 100-400 |
| # pulses in the train | $\leq$ 32 |
| Laser pulses separation (*nsec*) | 16 |
| Laser Pulse energy jitter (%) | 2 |
| Time arrival jitter (*psec*) | < 0.5 |
| Pointing jitter ($\mu m$) | < 20 |

**Table 65.  RF Structures : S-band system**

|  | RF Gun | 2 sections (3 m) constant gradient | 2 RF Deflectors |
|---|---|---|---|
| RF frequency (*GHz*) | 2.856 | 2.856 | 2.856 |
| # cells | 1.6 | 85 | 10 |
| Working mode | $TM_{01}$-like ($\pi$) | $TM_{01}$-like ($2\pi/3$) | $TM_{11}$-like ($\pi$) |
| Max RF inp. power (*MW*) | 15 | 40 | 15 |
| RF field (*MV/m*) | 120 (cath. peak) | 23 (acc. gradient) |  |
| Unloaded $Q_0$ | 15000 | 13000 | 18000 |
| Filling time (*nsec*) | 400-800 | 850 | 500-1000 |
| Shunt impedance (*M$\Omega$*) | 1.8 | 53-60 (*M$\Omega$/m*) | 5 |
| Max RF input pulse ($\mu sec$) | 2.5 | 1.5 | 2.5 |
| pulse duration for beam acceleration (*nsec*) | $\leq$ 600 | $\leq$ 600 | $\leq$ 600 |
| rep rate (*Hz*) | 100 | 100 | 100 |
| average dissipated power (*kW*) | 2.5 | 3.5 | 2.5 |

**Table 66.    RF Structures:  C-band system**

|  | 4+8 sections (1.8 m) constant impedance |
|---|---|
| RF frequency (GHz) | 5.712 |
| # cells | 102 |
| Working mode | TM01-like ($2\pi/3$) |
| Max RF inp. power (MW) | 40 |
| RF field (MV/m) | 33 |
| Unloaded Q0 | 9000 |
| Filling time (nsec) | 230 |
| Shunt impedance (M$\Omega$/m) | 74 |
| Max RF input pulse ($\mu$sec) | 0.8 |
| pulse duration for beam acceleration (nsec) | $\leq$ 500 |
| rep rate (Hz) | 100 |
| average dissipated power (kW) | 2.1 |

**Table 67.    RF Power Sources and Distribution Network:  S-band system**

|  | RF Gun RF Deflectors | Acc. sections |
|---|---|---|
| Frequency (*GHz*) | 2.856 ± 2 *MHz* | 2.856 ± 2 *MHz* |
| # Klystrons | 2 | 2 |
| RF pulses rep. rate (*Hz*) | 1-100 | 1-100 |
| P$_{out}$ (*MW*) | > 20 | 55 |
| RF pulse length ($\mu sec$) | 1.5 | 1.5 |

**Table 68.    RF Power Sources and Distribution Network:  C-band system**

|  | Acc. sections |
|---|---|
| Frequency (*GHz*) | 5.712 ± 5 *MHz* |
| # Klystrons | 12 |
| RF pulses rep. rate (*Hz*) | 1-100 |
| P$_{out}$ (*MW*) | 50 |
| RF pulse length ($\mu sec$) | 1 |

## 8.2. Risk table

We list in the following the risk factors associated to each target best value for the electron beam and the laser pulse and laser recirculator, in a range from 1 to 5, whre 1 stays for safe performance and 5 for risky.

**Table 73.   Risk Table on Electron Beam Performances**

|  | Target Best Value | Risk Factor |
|---|---:|---:|
| Energy (MeV) | 720 | 2 |
| Bunch charge (pC) | 250 | 2 |
| Bunch length ($\mu$m) | 270 | 3 |
| $\varepsilon_{n\_x,y}$ (mm-mrad) | 0.4 | 4 |
| Bunch Energy spread (%) | 0.05 | 4 |
| Focal spot size ($\mu$m) | 16 | 3 |
| # bunches in the train | 32 | 2 |
| energy variation along the train | 0.1 % | 4 |
| Energy jitter shot-to-shot | 0.1 % | 4 |
| Emittance dilution due to beam breakup | < 10% | 3 |
| Time arrival jitter (psec) | < 0.5 | 3 |
| Pointing jitter ($\mu$m) | 1 | 4 |

**Table 74.   Risk Table on Yb:Yag Collision Laser Performances**

|  | Target Best Value | Risk Factor |
|---|---:|---:|
| Pulse energy (*J*) | 0.2 | 4 |
| FWHM pulse length (*ps*) | 3.5 | 2 |
| Repetition Rate (*Hz*) | 100 | 2 |
| $M^2$ | 1.2 | 4 |
| Focal spot size $w_0$ ($\mu m$) | 28 | 3 |
| Bandwidth (*rms*) | 0.1 % | 3 |
| Pointing Stability ($\mu rad$) | 1 | 4 |
| Sinchronization to an ext. clock | 1 *psec* | 2 |
| Pulse energy stability | 1 % | 3 |

**Table 75.   Risk Table on Laser Recirculator Performances**

|  | Target Best Value | Risk Factor |
|---|---:|---:|
| beam waist $w_0$ | 28 $\mu m$ | 3 |
| max rotation at IP of linear laser polarization (along 32 passes) | 1° | 3 |
| integrated luminosity over 32 passes | 90 % | 5 |
| Mirrors parallelism default | 10 $\mu rad$ | 4 |
| Mirrors alignment tolerance | 10 $\mu m$ | 4 |
| Sinchronization to an ext. clock | 1 *psec* | 3 |

# 9. Appendix

## 9.1. Appendix A1: Compton spectrometer geometry optimization

In the Compton spectrometer the distance d of the electron detector from the target along the beam direction is chosen as a compromise between conflicting requirements:

- a larger distance allows to reduce the polar angle for a fixed detector size, and the error on Φ due to the beam spot size;

- a smaller distance increases the acceptance, allowing to reduce the target thickness for a given rate, reducing the contribution of multiple scattering.

To optimize this parameter we compute, as a function of d, the target thickness needed to obtain a fixed rate (20 Hz) of isolated electron signals for 10 MeV gammas, placing the detector at the minimal practically possible angle of 80 mm/d. We obtain the resulting expected resolution on the gamma energy taking into account a resolution on the electron energy of 0.1%/ r.m.s., the uncertainty on the gamma position on the target (1 mm r.m.s. in the transverse directions), the effects of multiple scattering and energy loss of the electron inside the target. The results are shown in Fig. 208 and Fig. 209. We choose d = 200 cm as a compromise between getting a reasonable rate and a good energy resolution. Finally, the rate and resolution expected from good signals using this simple calculation are shown as a function of energy in Fig. 176. Note that the calculation doesn't take into account the inefficiency of the electron measurements and of the detection of the scattered photon in coincidence, that will result in a reduction of the rate of useful signals, and we also assume that the background can be completely suppressed and/or subtracted. The numbers are thus intended as best limits from the principle of the method, while a more realistic evaluation of the detector performance are given in Section 5.5.2.

We see that we can expect an important contribution on the energy resolution from multiple scattering. However, if larger beam intensities are reached, the same detector rate can be obtained with smaller target thickness, improving the achievable resolution down to the limit imposed by the electron energy measuring device.

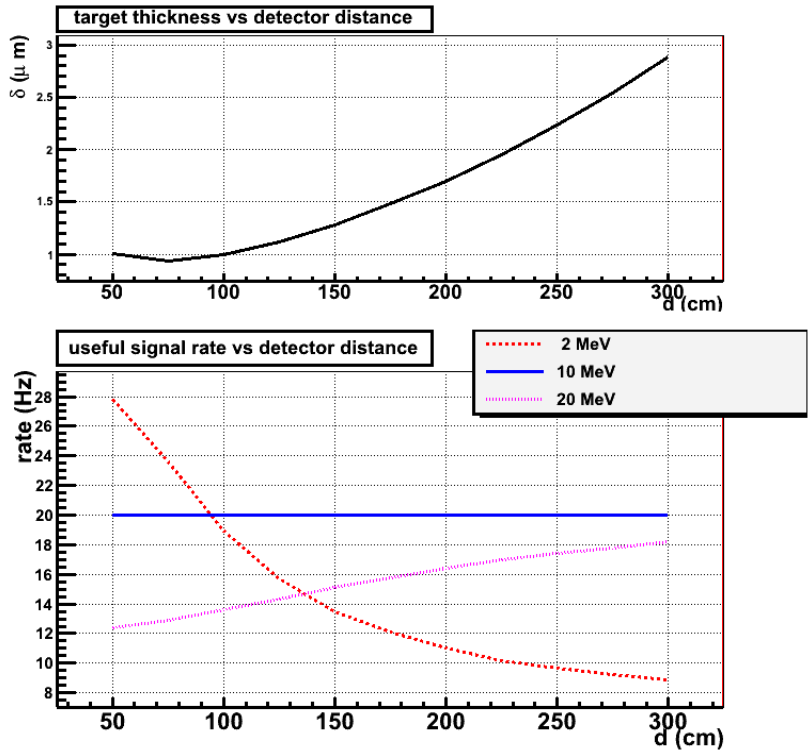

**Fig. 208.** On the top plot, the thickness of the target needed to get a rate of 20 Hz of useful signals for gamma energy of 10 MeV is shown as a function of the detector distance. On the bottom plots, the resulting rates for 2 and 20 MeV are also shown. For any value of d, the detector is placed as close as possible to the beam line.

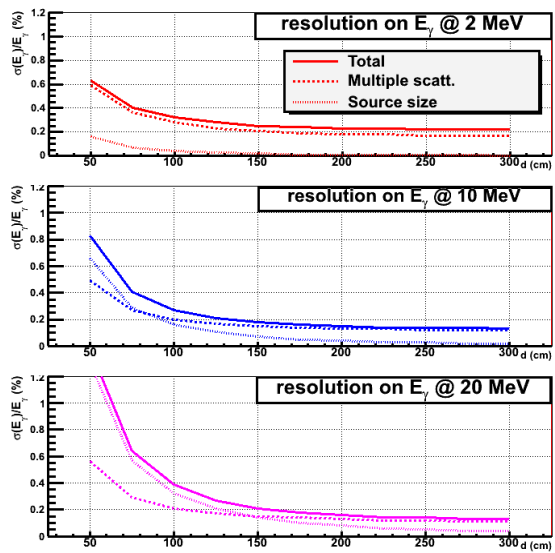

**Fig. 209.** Expected resolution on the photon energy as a function of the detector distance *d* for gamma of 2, 10 and 20 MeV, for a fixed rate of 20 Hz at 10 MeV. The contributions from multiple scattering and source size are also shown.

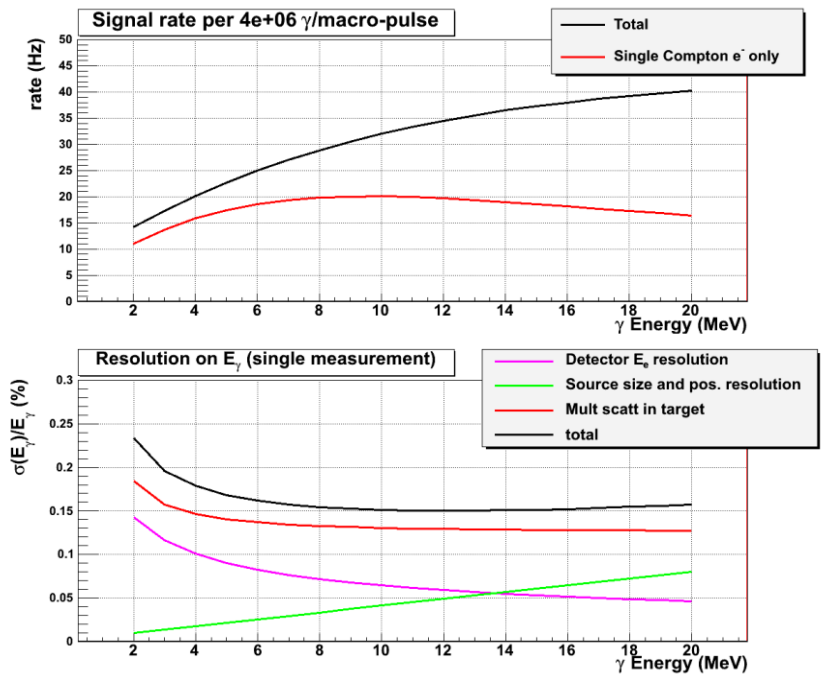

**Fig. 210.** Rate and beam energy resolution expected from the Compton spectrometer using a graphite target of 1.7 μm for a macro-pulse intensity of $4 \times 10^6$ photons. On the upper plot, the total rate of detector signals and the rate from single Compton electrons are shown. In the lower plot, the best achievable resolution from a single measurement on the primary gamma energy is shown.

## 9.2. Appendix A2: Evaluation of Compton spectrometer performances

To evaluate the performance of the proposed Compton spectrometer a detailed simulation of the detector has been developed using Geant4, according to the layout resulting from the study of Section 5.4.2. As described in most detail there, the spectrometer consists of a micrometric graphite target, a detector to measure the energy and position of the Compton scattered electron and a movable photon detector to be placed around the expected position of the recoil gamma that depends on the electron detection angle and on the beam energy.

Three different beam energies were simulated (5, 10 and 20 MeV), each simulation consisting of $3 \times 10^{10}$ events. The beam starts 25 m upstream the Compton spectrometer target with a x-y emittance $\varepsilon$ = 20 µm x 40 µrad and is directed onto the target. In case of an interaction, the resulting particles are tracked along the detector and their identities, their positions together with the deposited energy in the different crossed volumes are recorded.

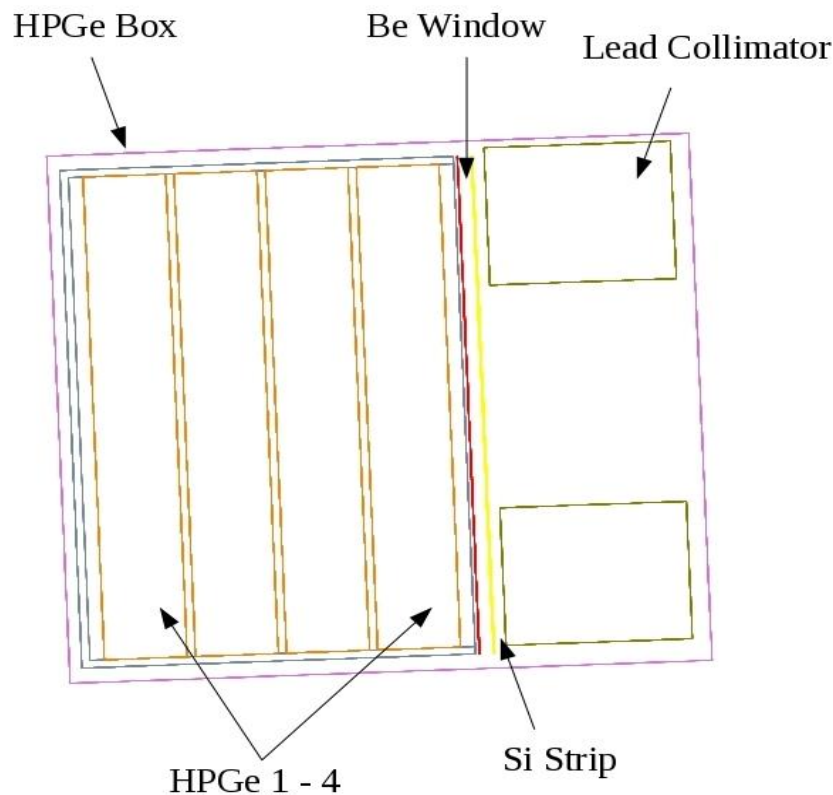

**Fig. 211.** **Lateral view of the electron detector for the high energy line. Particles are entering from the left.**

Interactions of photons and electrons were simulated using the Geant4 implementations of physics models developed for the PENELOPE code [138]

This code has been specifically developed for γ and e± transport in matter and great care was given to the description of the low energy processes including binding effects and Doppler broadening. Comparisons with experimental data showed that PENELOPE provides consistent results in the energy range from few keV up to about 1 GeV [139].

For the transport of e+- and gamma a range cut of 0.1 mm was used, with the exception of the thin volumes representing the target, the Be window and the HPGe contact layers, where the cut was lowered to 1µm. These range cuts are translated into an energy cut for each material depending on its physical properties.

For the simulation of the electron detector four (two for the low energy line) of the HPGe planar detectors have been placed in an aluminum case closed, at the incoming beam side, by a 100 µm Be window. The Ge blocks are separated by a distance of 2 mm. In order to simulate the ultra-thin contacts a passive Ge layer of 2 µm encloses each block. The energy resolution of the HPGe sensors was taken to be Gaussian and was simulated taking the standard 2.96 eV energy to produce an electron-hole pair, a Fano factor of 0.10 and a constant electronic noise of 500 eV.

Preceding the box housing the HPGe detectors, the silicon strip detector provides the impact coordinates of charged particles. The energy deposited by different tracking steps in this detector is merged if the hits are found within 50 µm distance from an initial seed. We assume to measure the deposited energy with 10% uncertainty.

The photon detector is made by an array of monolithic LYSO crystals, as described in Section 5.4.2. The scintillation process and the transport of scintillation photons to the APD light detectors were simulated. We used the crystal properties of LYSO type PreLude 420 from www.detectors.saint-gobain.com. The Geant4 Unified model was used to simulate the reflection of light at surfaces between dielectric materials. For the purposes of this work we coupled directly the crystal to the APD, assumed to be made of 50 µm of silicon and to have an efficiency of 56 %. This value accounts for a 75% quantum efficiency and a 75% geometrical efficiency. A 2 mm spatial resolution is assumed for the reconstruction of the impact coordinates.

Table 69 shows the number of interactions in the target per incoming gamma as a function of the beam energy and type of interaction. As expected, with increasing beam energy the Compton rate decreases while pair production increases.

**Table 69. Probability of interaction in the target for the beam photon as a function of the beam energy and interaction type**

| Beam Energy (MeV) | 5 | 10 | 20 |
|---|---|---|---|
| Compton | 7.16 x 10$^{-6}$ | 4.44 x 10$^{-6}$ | 2.62 x 10$^{-6}$ |
| Conversions | 5.91 x 10$^{-7}$ | 1.20 x 10$^{-6}$ | 1.91 x 10$^{-6}$ |

*Nota: The graphite target has a depth of 1.7 µm.*

A fraction of these events will be recorded by the electron/recoil gamma detectors. The event rates in the electron detector are reported in Table 70 as a function of the beam energy, before and after selecting a matching signal in the Si strip and photon detectors.

**Table 70. Average number of events rates in the electron detector per incident photon as a function of the beam energy and cut type**

| Beam Energy (MeV) | 5 | 10 | 20 |
|---|---|---|---|
| All | 3.31 x 10$^{-8}$ | 1.25 x 10$^{-7}$ | 2.49 x 10$^{-7}$ |
| 1Hit | 2.11 x 10$^{-8}$ | 0.81 x 10$^{-7}$ | 1.36 x 10$^{-7}$ |
| Signal | 1.72 x 10$^{-8}$ | 0.47 x 10$^{-8}$ | 0.39 x 10$^{-7}$ |

*Nota: "All" refers to all the energy depositions recorded. "1Hit" is the rate obtained when requesting 1 hit in the Si strip detector inside the fiducial volume defined by the collimator hole. Finally "Signal" is the rate corresponding to the additional request of a coincidence signal inside the gamma detector.*

Table 71 shows the effect of the required coincidences on the signal purity for 10 MeV beam energy. The request of a hit in the Si strip strongly reduces the events due to a Compton photon, while the detection of a recoil gamma in coincidence suppresses the background due to conversions inside the target. Photons surviving the selection are always accompanied by a good Compton electron, and are produced by electron interactions in the Si detector or at the border of the collimator. For all the simulated energies more than 99% of the selected events contain an electron generated by a Compton interaction in the target.

**Table 71. Recorded particle type as a function of the selection cuts for a beam energy of 10 MeV**

|  | e-(%) | e+(%) | γ(%) |
|---|---|---|---|
| All | 60.1 | 14.4 | 25.5 |
| 1Hit | 78.2 | 16.9 | 4.9 |
| Signal | 94.9 | 0.1 | 5.0 |

The ELI beam energy is obtained by measuring the electron scattering angle and its energy according to the formula:

$$E_\gamma = \frac{m_e}{\sqrt{\cos^2(\varphi)(1 + 2m_e/T_e) - 1}}$$

Where $m_e$ is the electron mass, $\varphi$ is the scattering angle and $T_e$ is the kinetic energy. The $\varphi$ angle is precisely measured by the Si strip detector. Given the detector location and the collimator hole dimension, the angle ranges between 0.25 and 0.55 mrad. With the assumed resolution of 50 µm on the impact position, we obtain a resolution on $\varphi$ of about 1 mrad, dominated by the uncertainty on the collision point due to the beam spot. This is illustrated in Fig. 212.

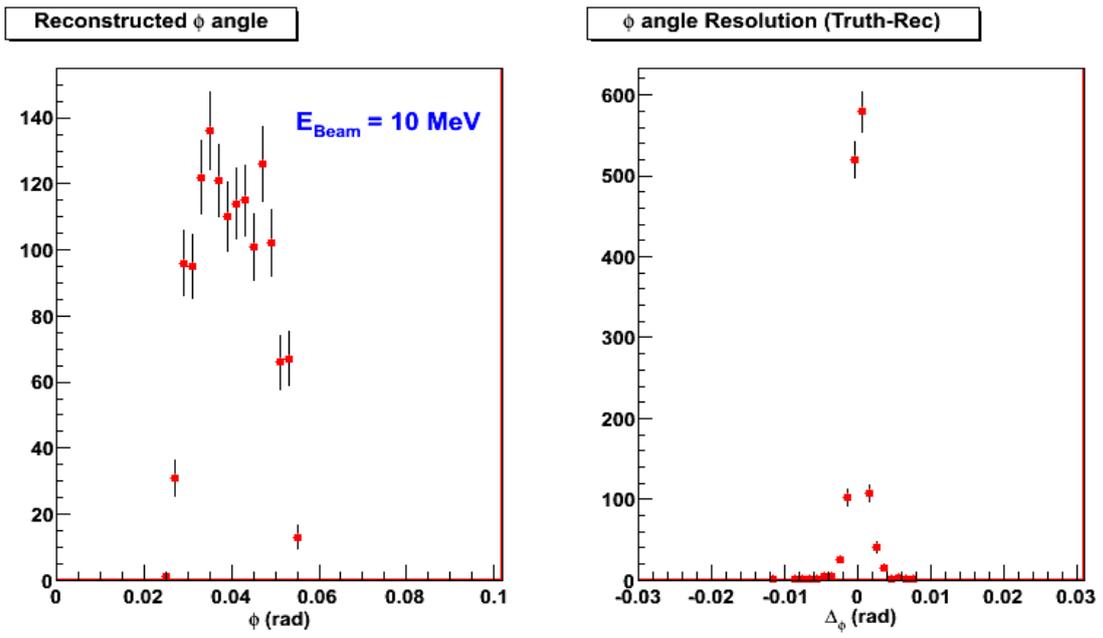

Fig. 212. Reconstructed electron angle by the Si strip detector (left plot) and resolution (right plot)

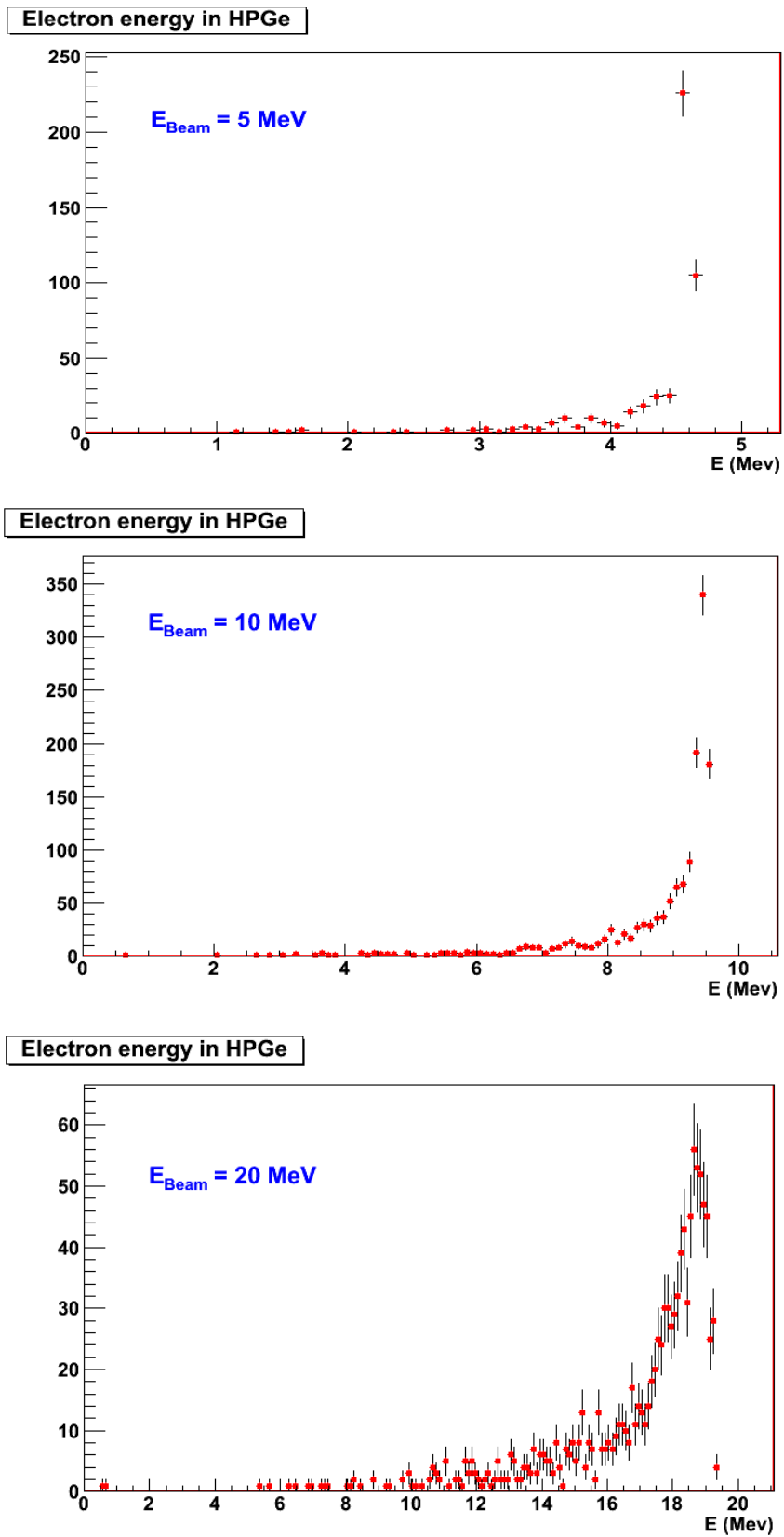

**Fig. 213.** Energy recorded inside the HPGe detector for a beam energy of 5, 10 and 20 MeV

The electron energy is measured in the HPGe detector, adding up the signal recorded in its four (two for the low energy line) elements. The reconstructed energy distribution for a beam energy of 5, 10 and 20 MeV is shown in Fig. 213. Other than the spread expected by the angular detector acceptance, we can notice the presence of a long low energy tail due to electrons not fully contained in the detector. The energy released in the position detector is measured event by event by the Si strips and amounts on average to about 100 keV (see Fig. 214).

A tiny fraction of the electron energy is lost in passive components of the detector, namely the Be entrance window and the passive Ge layer providing the electrical contact in each HPGe detector.

Fig. 215 shows the energy deposited in these components. The average energy loss obtained from the simulation, about 30 keV, can be applied to compensate for this effect. However, a fluctuation of about 20 keV is expected and much care has to be taken to minimize the thickness of material traversed by the electron in order to reduce this source of uncertainty.

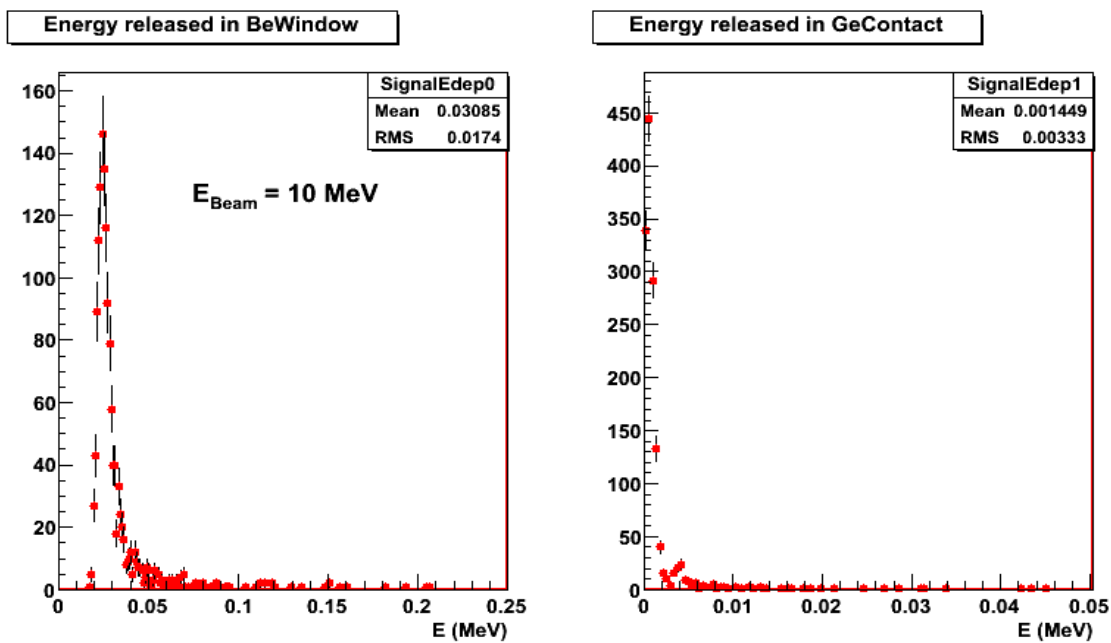

**Fig. 214.** Energy deposited by the electron inside the entrance window (left plot) and the Ge contacts (right plot) for a beam energy of 10 MeV

The distribution of the reconstructed beam energy is shown in Fig. 215. To evaluate the expected resolution, the fill energy peak is fitted with a gaussian function. As shown in Fig. 215, the fitted mean values are shifted from the true values by about 30 keV, as expected from the average value of energy lost in the passive materials. The resulting resolution varies from 2 to 1.5 per mill from 5 to 20 MeV.

The residual tails in the beam energy reconstruction can be further suppressed using the information from the impact position of the recoil gamma. Once the electron position and energy is known, the recoil photon angle θ is fixed by the kinematics of the Compton scattering:

Fig. 216 shows, for the three investigated energies, the distributions of the reconstructed scattering angles for the events of the selected signal sample. Events where the electron energy is not correctly reconstructed are expected to deviate from the kinematical relation

$$E = \frac{m_e}{\sqrt{cos^2(\varphi)(1+2m_e/T_e)-1}}$$

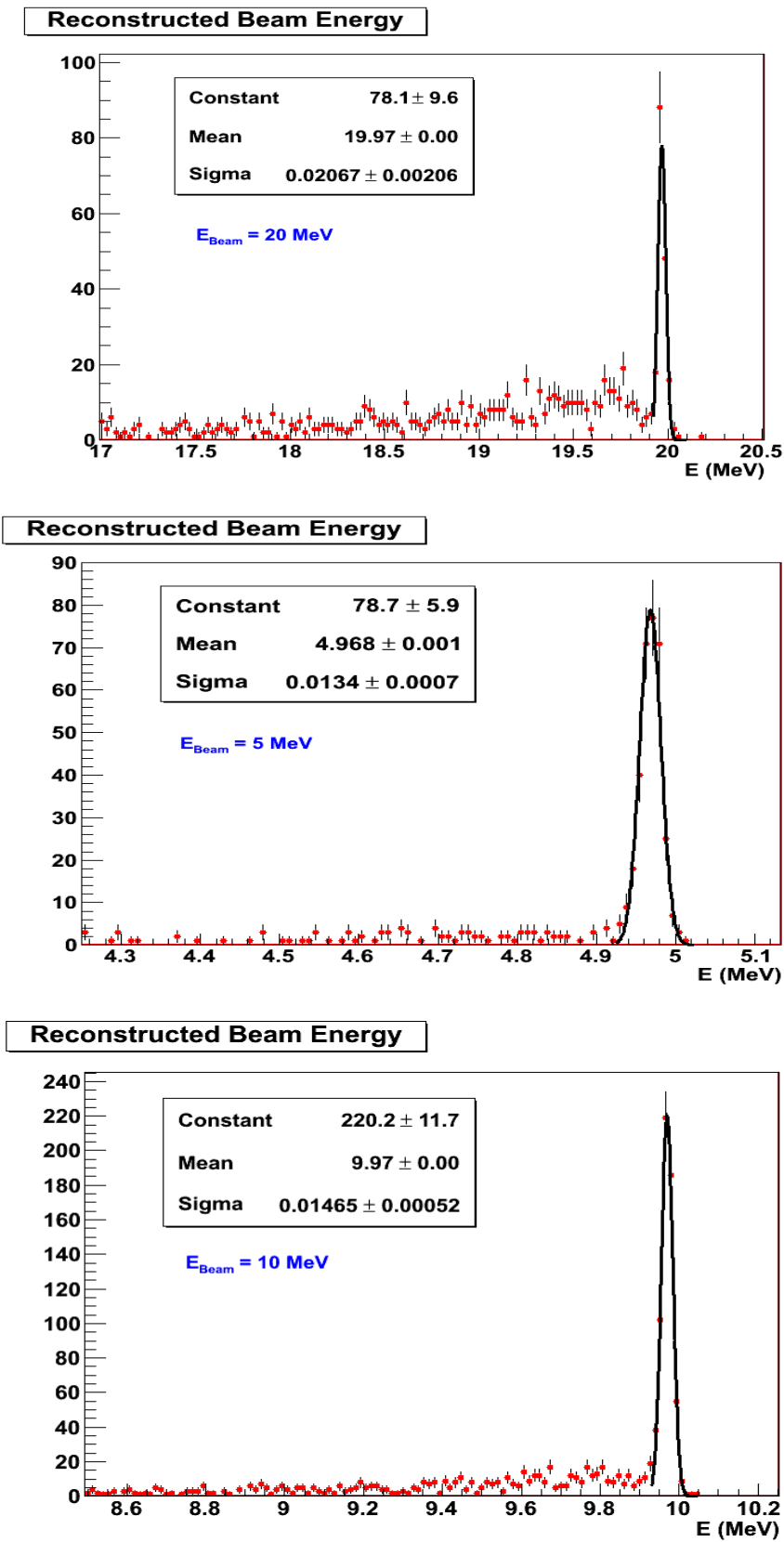

Fig. 215. Reconstructed beam energy by the Compton spectrometer for a simulated energy of 5, 10 and 20 MeV

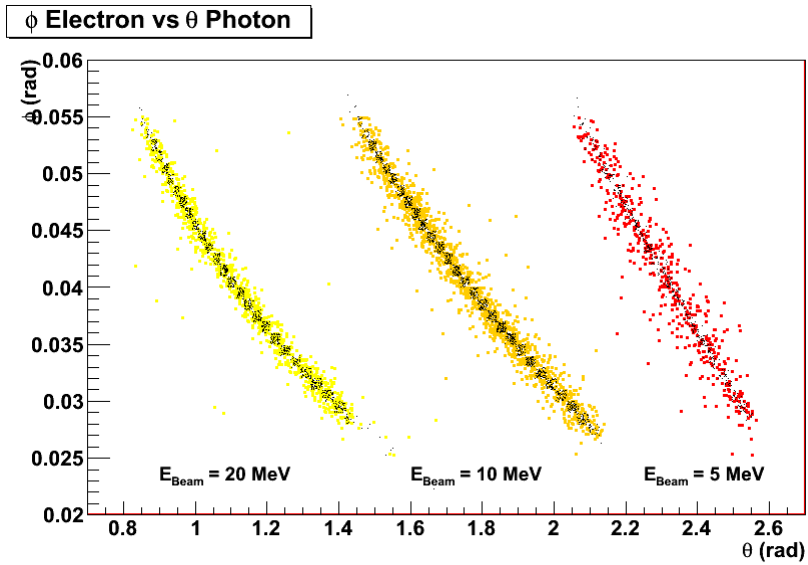

Fig. 216. Plot of the reconstructed scattering angles φ (electron) vs θ (photon) for the three investigated energies. On each plot is superimposed, in black color, the Montecarlo truth distribution.

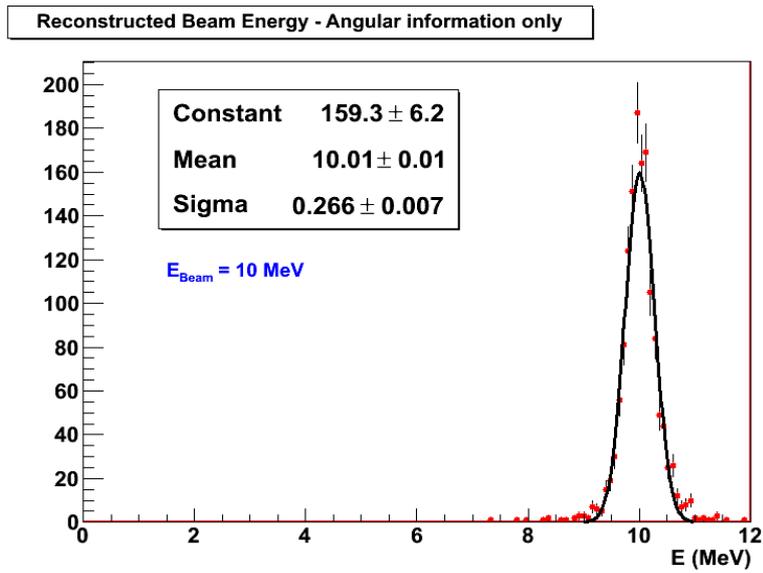

Fig. 217. Reconstructed beam energy using only the electron and recoil photon reconstructed angles φ and θ, for a 10 MeV simulated beam energy.

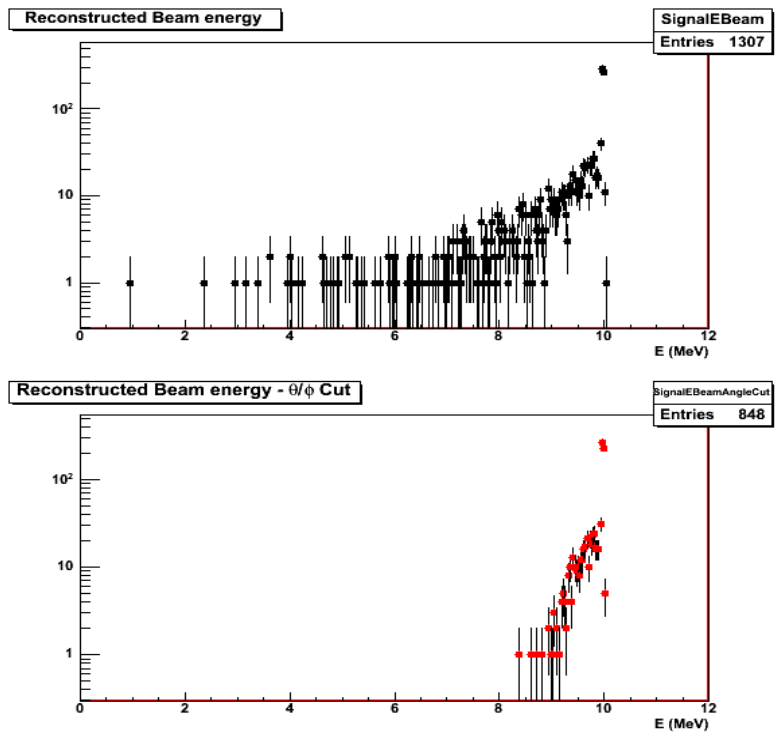

**Fig. 218.** Reconstructed beam energy (top plot), the long tail at low energies is due to escapers. This tail is suppressed when a constrain based on the value of energy determined by angles is applied (bottom plot).

To implement this kinematic check, we reconstruct the beam energy independently on the electron energy measurement, using only the reconstructed φ and θ. The distribution is shown on Fig. 217. The resolution of this measurement is much worse than the one based on φ and Te, but a requirement on the agreement between the two measurements can be used to reject cases where a relevant fraction of the electron energy escaped detection. We also notice that the distribution is centered around the correct value of the energy used in the simulation and can therefore be used also to cross-check the correction needed to compensate the losses in the passive materials for the electron energy reconstruction.

The effect of requiring the difference between the two values of reconstructed beam energy to smaller than 0.5 MeV results in suppression on the low energy tail, as shown in Fig. 218. A small improvement (a relative 7%) in the resolution of the full energy peak is also observed.

Another kinematic constraint which can be set from the precise measurement of the electron and photon positions is related to the azimuthal angle ψ that must be the same for the two particles. This is illustrated in Fig. 219. A cut on the difference of the ψ values can be used to reduce a possible residual combinatorial background that could be due to beam background (not included in this simulation).

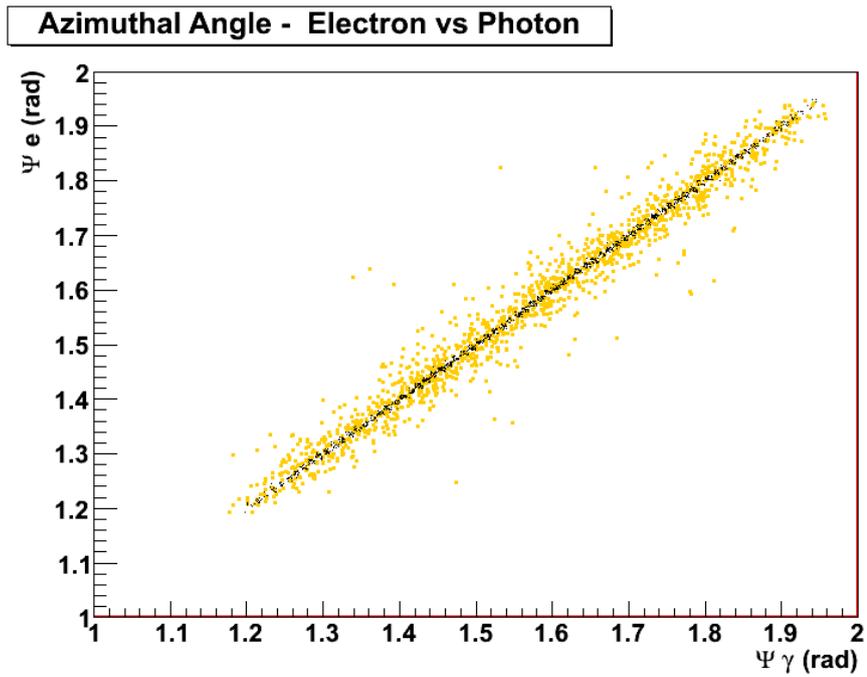

**Fig. 219.** Distribution of the azimuthal angle for the electron and photon. The Montecarlo true distribution, predicting equal values for the two particles, is superimposed in black.

In conclusion, the simulations show that a clean sample of well reconstructed Compton interactions can be selected using the Compton Spectrometer, providing the beam energy with a resolution between 0.15 and 0.2 %. The expected number of useful signals per incident photons corresponds, for the nominal beam flux, to a rate of a few Hz for the whole range of beam energy. The spectrometer is thus able to provide a continuous monitoring of the beam energy during the routine operations of the ELI-NP facility with the required accuracy. The resolution function obtained from the simulations with a fixed beam energy, that can be verified using sources of monochromatic gammas, can be deconvoluted from the measured distribution. Therefore, after an adequate number of measurements, an accurate reconstruction of the time integrated energy spectrum of the ELI-NP gamma beam can be achieved.

## 9.3. Appendix B: Evaluation of calorimeter performances

In order to optimize the detector layout, we simulated, using GEANT4, the energy release of photons in the 2-20 MeV energy range inside a large number (150) of detector elements made of 1 cm thick polyethylene blocks followed by a 1 mm thick Si layer. The distribution of the released energy and its fluctuation was studied as a function of the longitudinal and transverse position inside the detector. Fig. 220 shows the longitudinal profile of the energy released in the sampling detectors for different energies. Fig. 221 and Fig. 222 show the fraction of absorbed energy and, more importantly, its fluctuation as a function of the length and transverse size of the detector. From this study, the absorber length was chosen to be 75 cm, and the transverse size to be 6x6 cm2. We then evaluated the achievable resolution as a function of the distance between the readout active layers, in order to use the minimal amount of detectors. Fig. 223 shows that an active layer every 3 cm of passive material is a good compromise between the calorimeter resolution and the budget constraints limiting the number of active elements.

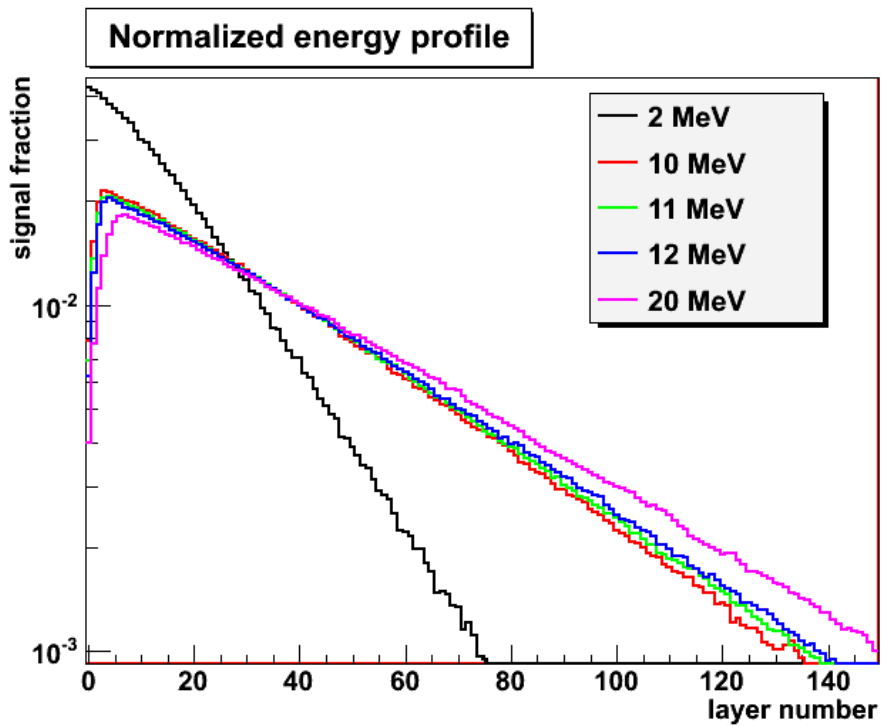

**Fig. 220.** Average longitudinal profile of the energy released by gammas of different energies. The profiles are obtained from $10^7$ simulated photons for each energy in an explorative setup with 150 active layers separated by 1 cm of plastic absorber.

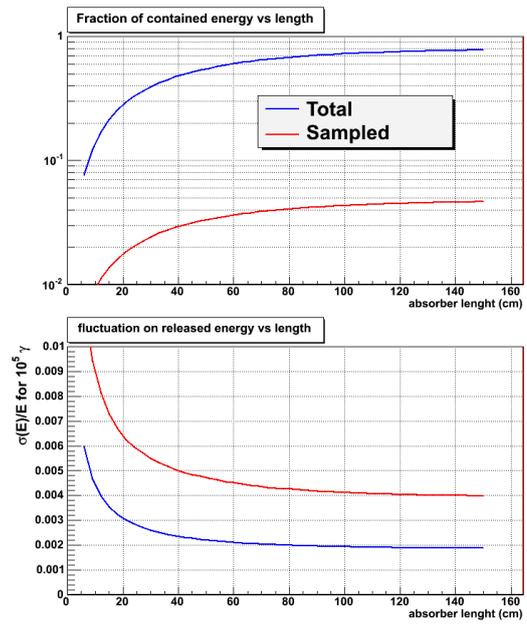

**Fig. 221.** The fraction of contained energy, in the full calorimeter and in the active devices, and its fluctuation as a function of the absorber length for $10^5$ photons of 11 MeV impinging the explorative setup described in the text.

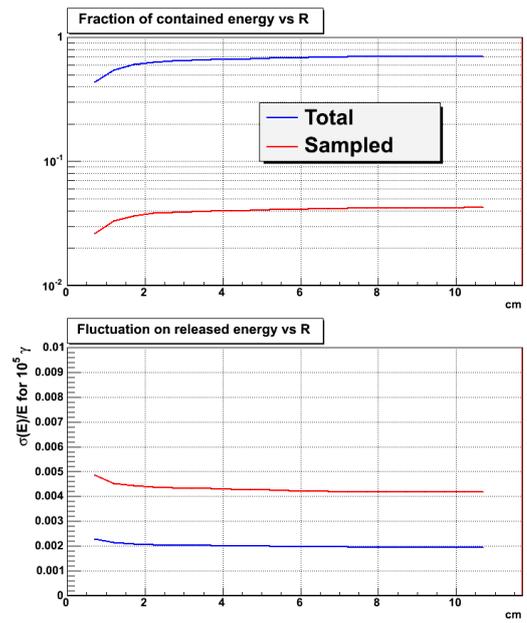

**Fig. 222.** The fraction of contained energy, in the full calorimeter and in the active devices, and its fluctuation as a function of the radial distance from the beam axis for $10^5$ photons of 11 MeV impinging the explorative setup described in the text.

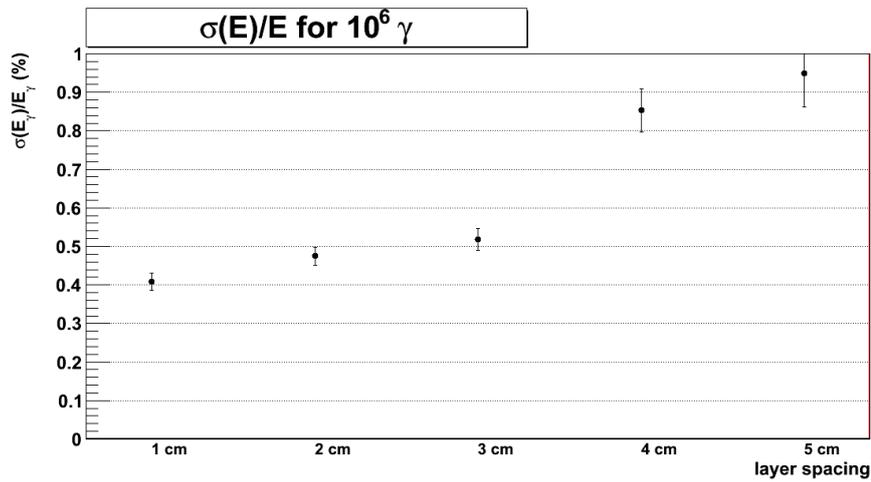

**Fig. 223.** Achievable resolution, for $10^6$ γ of 11 MeV energy, as a function of the spacing between active layers for a 75 cm long absorber

As shown in Fig. 224, a different layout of the detector planes, with a finer sampling around the maximum energy deposition (layout 2 in the figure), would allow to improve the resolution by 10%. However, as discussed in Section 5.5.3, the high beam intensity will suppress the statistical error to negligible level in a very short measuring time, so that we expect the accuracy of the measurement to be limited by systematic effects. Different sampling fractions along the calorimeter length would introduce non-linearities in the detection response (the average sampling fraction would become energy dependent) that would increase the systematic error on the detector modeling. Therefore, we prefer to adopt a layout consisting in a string of 25 identical detector elements.

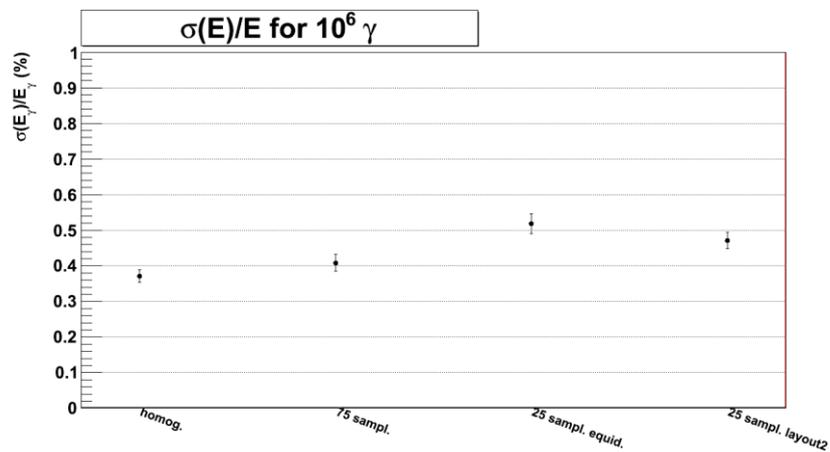

**Fig. 224.** Achievable resolution, for $10^6$ γ of 11 MeV energy, as a function of the spacing between active layers for a 75 cm long absorber

## 9.4. Appendix C: Beam imager performances

The performance of the imager has been evaluated estimating the signal produced by the light sensor of the device.

The average number of electrons produced in a pixel (20 μm x 20 μm) of a CCD coupled with s Gadox phosphor with a lens providing a magnification M = 1 and an image area surface up to 10 x 10 mm$^2$ are reported in Fig. 55.

According to the beam specifications, the transverse source image size is expected to have a FWHM in the range of 1.0 to 10 mm, for a source distance of 20 m and source rms divergence in the range 25-250 μrad.

To calculate the average number of e- produced in the CCD per incident gamma ray for the geometrical configuration described in Section 5.5.1, a commercially available scintillator screen "Kodal Lanex Fast" has been utilized in the Monte Carlo simulation by EGSnrc code. The scintillator is composed of terbium-doped gadolinium oxysulfide phosphor (Gd$_2$O$_2$S;Tb) with a 133 mg/cm2 mass thickness.

The equation 1 gives the total CCD signal for incident gamma:

$$n_{CCD}^{e^-} = \frac{E_\gamma \cdot \delta_a \cdot \delta_i \cdot \delta_t \cdot \delta_c \cdot \beta}{\varepsilon_k} \quad (1)$$

where:
- Eγ is the gamma ray energy;
- δa is the fraction of incident photon energy absorbed by the phosphor calculated by Monte Carlo;
- δi is the intrinsic gamma-ray to light conversion efficiency (0.15 for Gd2O2S;Tb);
- δt is the fraction of light that escape from phosphor (0.4 for Gd2O2S;Tb);
- δc is the lens system collection efficiency ($\delta_c = T_l \cdot [1 + 4 \cdot f^2 \cdot (1+m)^2]^{-1}$ : Tl is the lens system light transmittance, m is the magnification factor and f is the relative aperture of lens system);
- is the CCD quantum efficiency (tipically in the range 30%-50%);
- ε$_k$ is the mean energy of light photons created by phosphor (εk = 2.4 eV)

The results of Monte Carlo simulations for 1, 5, 10, 15, 20 MeV beams, with a magnification factor m = 1 for the lens system with a relative aperture f = 1.2 and transmission factor Tl=0.7, are shown in Table 72.

**Table 72. Total number of electrons produced in the CCD by single macropulse as a function of the beam energy**

| Beam Energy [MeV] | 1.0 | 5.0 | 10.0 | 15.0 | 20.0 |
|---|---|---|---|---|---|
| Deposited energy fraction per gamma | 2.725E-03 | 3.490E-04 | 1.870E-04 | 1.320E-04 | 1.070E-04 |
| # of electrons per gamma ($\beta$=0.3) | 7.439E-01 | 4.764E-01 | 5.105E-01 | 5.405E-01 | 5.842E-01 |
| # of electron per macropulse ($\beta$=0.3) | 2.98E+06 | 1.91E+06 | 2.04E+06 | 2.16E+06 | 2.34E+06 |

Starting from these results, it is possible to estimate the average number of electron per pixel that is a function of the image size and CCD pixel size. If the CCD pixel size is 20 µm, the dimensions of the images are in the range 50x50-500x500 pixels with magnification factor 1, while the average signal (number of electrons) per macro-pulse is given in Table 73.

**Table 73. Average number of electrons per CCD pixel per macro-pulse for two different beam divergence**

| Beam Energy [MeV] | 1.0 | 5.0 | 10.0 | 15.0 | 20.0 |
|---|---|---|---|---|---|
| Average signal per CCD pixel (25urad) | 8.50E+02 | 5.44E+02 | 5.83E+002 | 6.18E+02 | 6.68E+02 |
| Average signal per CCD pixel (250 urad) | 8.50E+00 | 5.44E+00 | 5.83E+000 | 6.18E+00 | 6.68E+00 |

For a standard cooled CCD camera the typical dark current is <1 e-/pixel/s while the readout noise is in the range 20-30 e-/pixel: from these data the average signal for macro-pulse is larger than the readout noise signal for 25 µrad beam divergence, while it is lower than readout noise for 250 µrad beam divergence. To increase the lower signal it is possible to use the CCD binning modality that increases the pixel size summing the signal of neighbour pixels, or summing the signal of several macro-pulses.

## 9.5. Appendix D: Accelerator Layout

The accelerator layout is depicted in pictorial detail. Fig. 225 gives an overview of the module sequence whilst subsequent figures (Fig. 226 and following) progress down the beam axis, depicting at each stage an isometric 3D visualisation of the components along with a 2D plan view (lower image) with scale bar (in blue) indicating position along from the source.

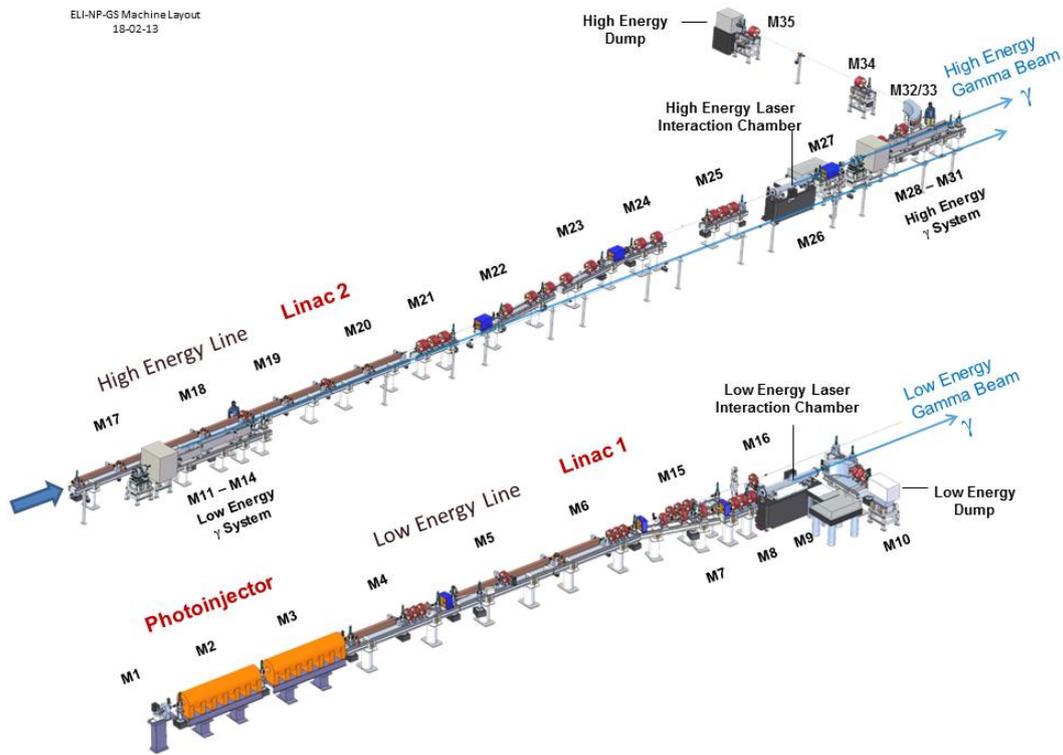

Fig. 225.    Accelerator Modules – Overview

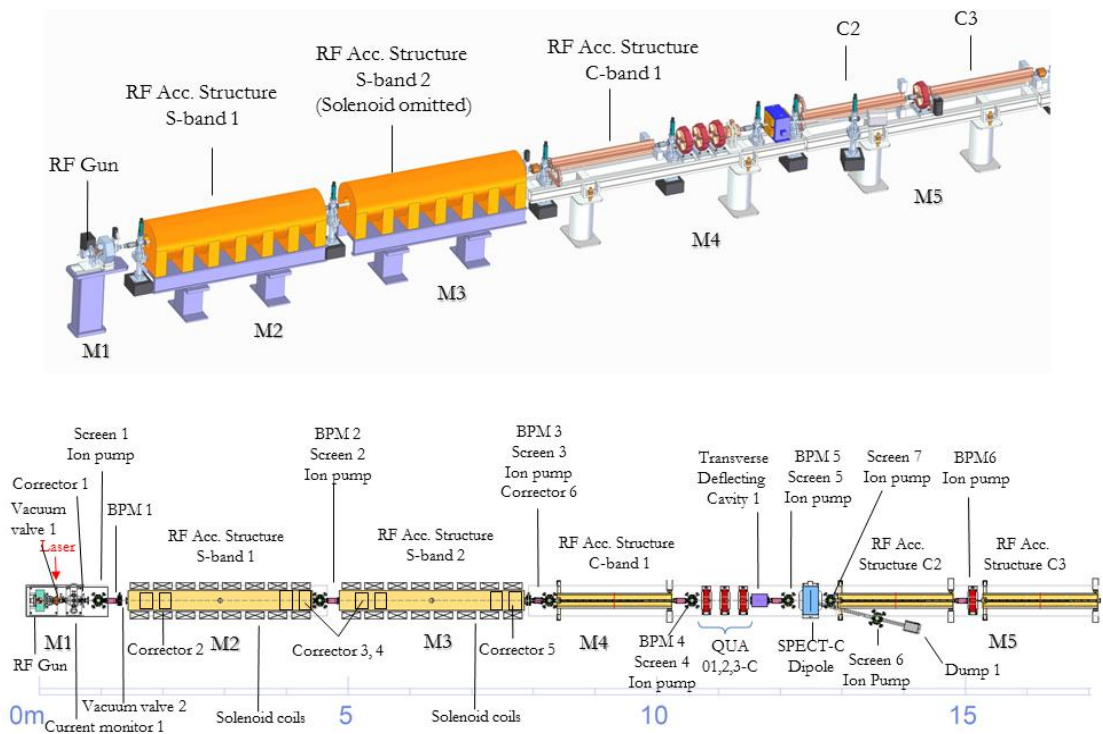

Fig. 226.    Accelerator Modules M1 – M5 (0 – 17m)

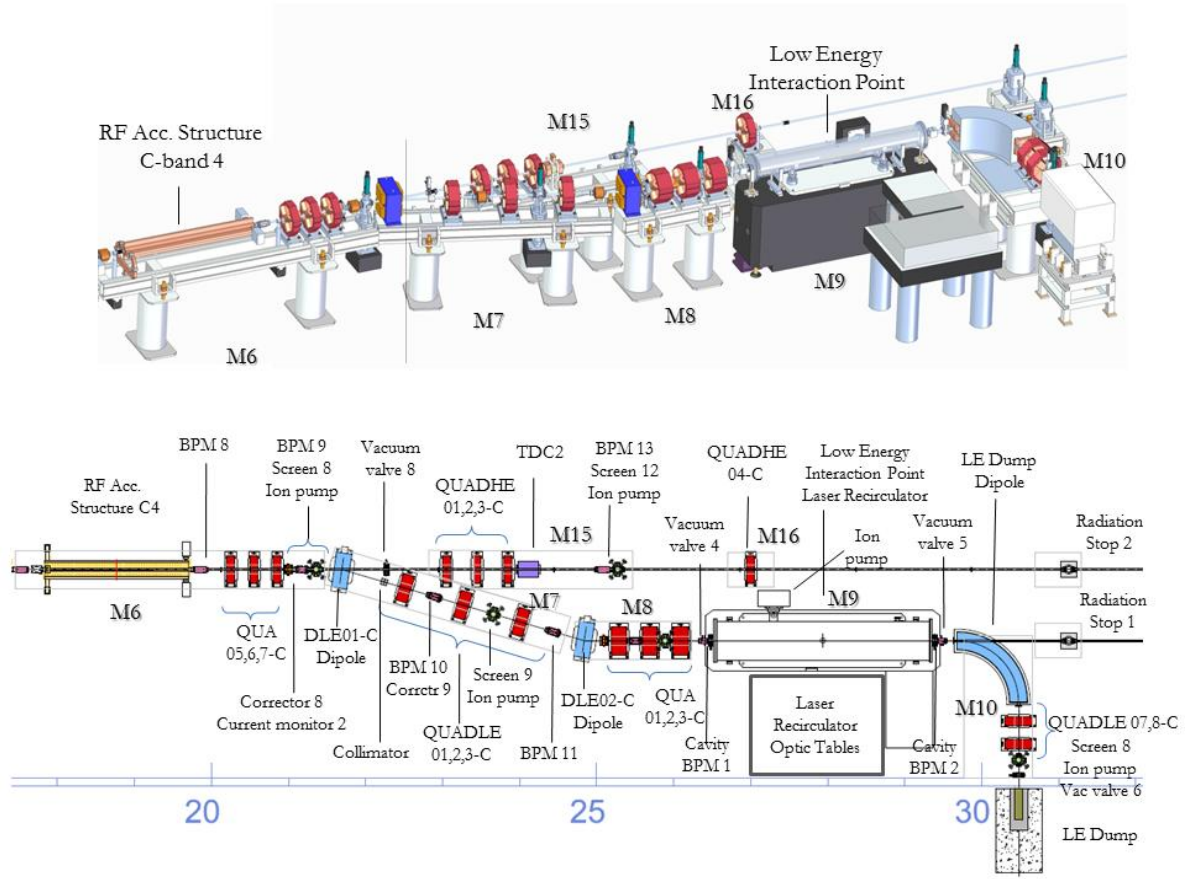

**Fig. 227.    Accelerator Modules M6 – M10 (17 - 32m)**

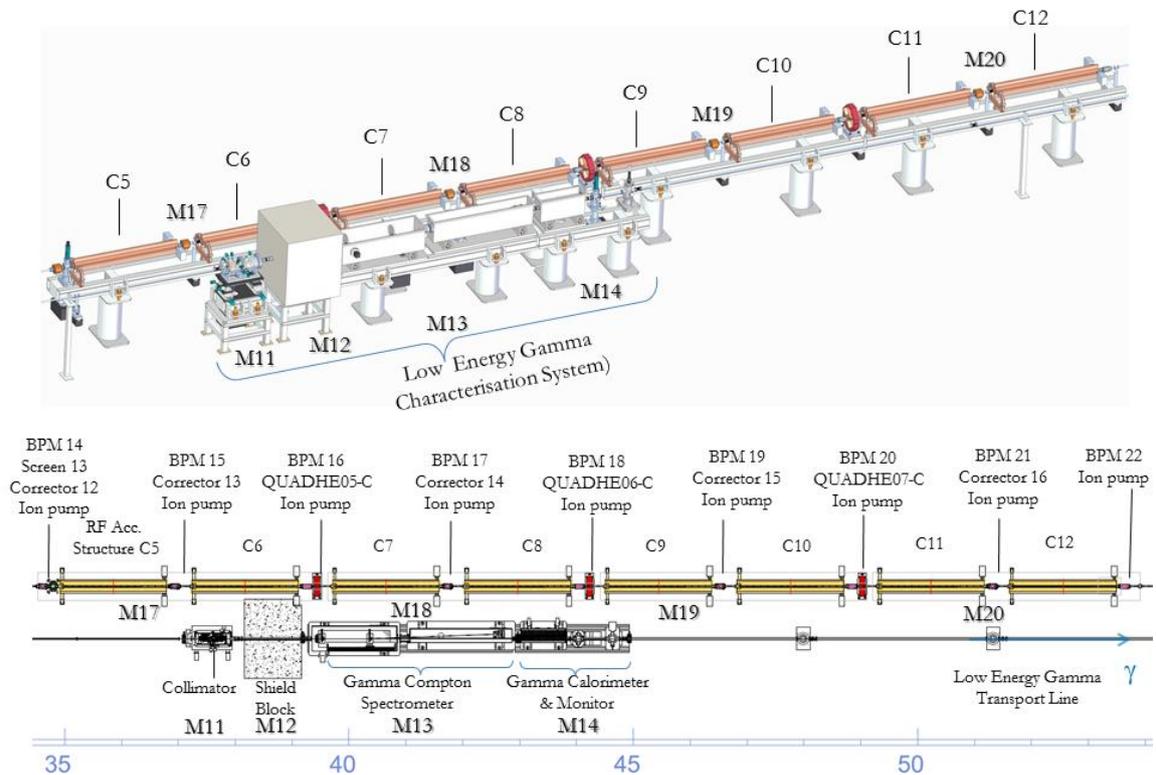

**Fig. 228.    Accelerator Modules M11 – M20 (34 - 54m)**

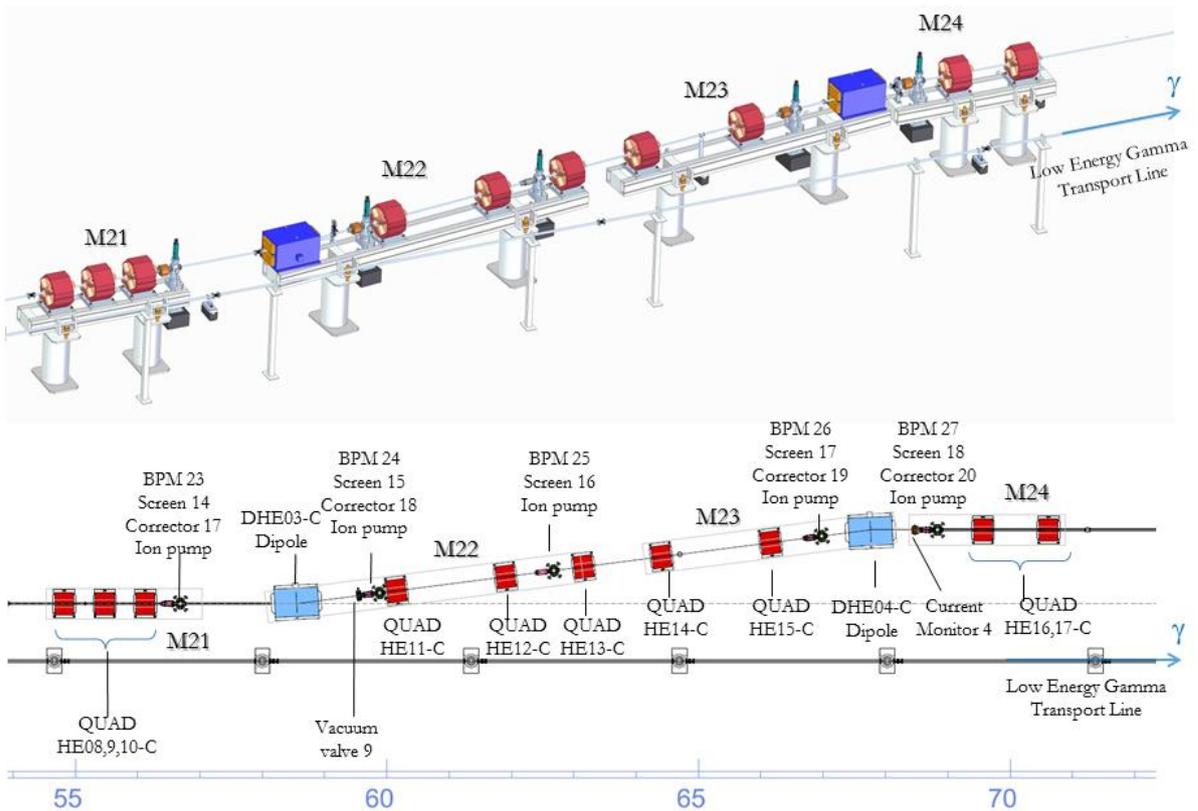

Fig. 229. Accelerator Modules M21– M24 (54 - 72m)

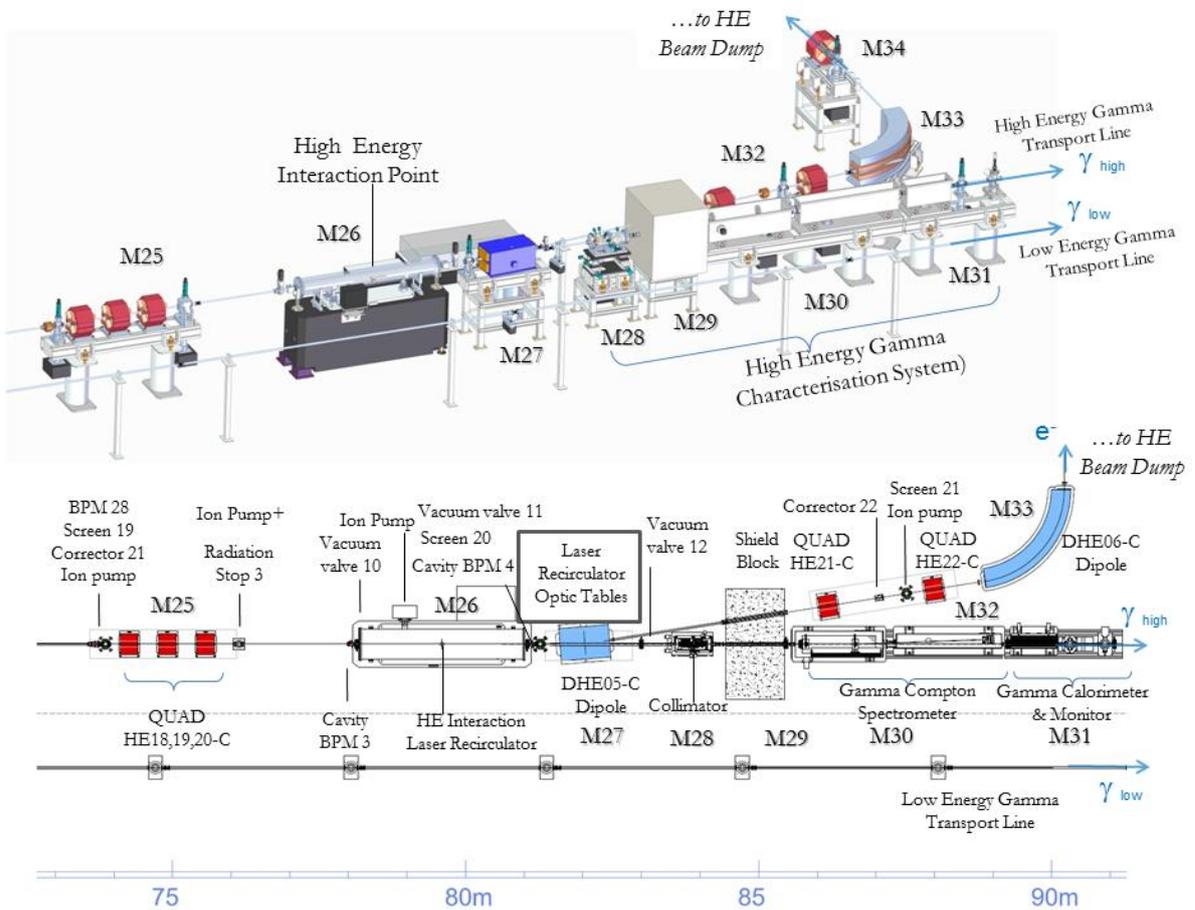

Fig. 230. Accelerator Modules M25 – M35 (73 - 90m)